\documentclass[a4paper,11pt]{book}
\usepackage[T1]{fontenc}
\usepackage{newpxtext,newpxmath}
\usepackage{fancyhdr}
\usepackage[inner=3cm,outer=2cm,top=3cm,bottom=2cm]{geometry}
\usepackage[linktocpage,backref=page, hyperfootnotes=false]{hyperref} 
\usepackage{makeidx}
\usepackage{fancybox}
\usepackage{titlesec} 
\usepackage{xcolor}
\usepackage[newcentury]{quotchap}
\usepackage{graphicx}
\usepackage{setspace}
\usepackage{amsmath}
\usepackage{tikz}
\usepackage[compat=1.1.0]{tikz-feynman}
\usepackage[sort&compress,numbers]{natbib}
\usepackage{braket}
\usepackage{wrapfig}
\usepackage{caption}
\usepackage{feynmp-auto}
\usepackage{float}
\usepackage{enumitem}
\usepackage{bm}
\usepackage{makecell}
\usepackage[dvipsnames]{xcolor}
\usepackage{emptypage}

\setlist[description]{itemsep=0.1cm, leftmargin=2cm}

\usetikzlibrary{arrows.meta}
\usetikzlibrary{positioning, shapes, arrows}

\usepackage{etoolbox}
\makeatletter
\patchcmd\NAT@citexnum{\let\NAT@last@num\NAT@num}{\MakeLinkTarget[cite]{}\Hy@backout{\@citeb\@extra@b@citeb}\let\NAT@last@num\NAT@num}{}{\fail}
\makeatother

\renewcommand*{\backref}[1]{}
\renewcommand*{\backrefalt}[4]{%
    \ifcase #1 (Not cited)%
    \or        (Cited on page~#2)%
    \else      (Cited on pages~#2)%
    \fi}

\hypersetup{
    colorlinks=true,
    linkcolor=blue,
    filecolor=magenta,
    urlcolor=cyan,
    citecolor=violet,
    pdftitle={Tesis},
    bookmarks=true,
    pdfpagemode=FullScreen,
}

\renewcommand{\headrulewidth}{0.4pt} 
\renewcommand{\chaptermark}[1]{\markboth{\thechapter.\ #1}{}}

\fancypagestyle{plain}{
  \fancyhf{}
  \fancyhead[LE]{\hspace{-2.5cm}\smash{\colorbox{blue!10}{\makebox[2cm][r]{\textcolor{blue}{\thepage}}}}\hspace{0.3cm}\leftmark}
  \fancyhead[RO]{\leftmark\hspace{0.3cm}\rlap{\smash{\colorbox{blue!10}{\makebox[2cm][l]{\textcolor{blue}{\thepage}}}}}\hspace{-0.3cm}} 
  \renewcommand{\headrulewidth}{0.4pt}
}

\fancypagestyle{empty}{
  \fancyhf{}
  \fancyhead[LE]{\hspace{-2.5cm}\smash{\colorbox{blue!10}{\makebox[2cm][r]{\textcolor{blue}{\thepage}}}}\hspace{0.3cm}}
  \fancyhead[RO]{\leftmark\hspace{0.3cm}\rlap{\smash{\colorbox{blue!10}{\makebox[2cm][l]{\textcolor{blue}{\thepage}}}}}\hspace{-0.3cm}} 
  \renewcommand{\headrulewidth}{0.4pt}
}

\makeindex 

\begin{document}

\pagenumbering{roman}

\begin{titlepage}
    \newgeometry{left=1in, right=1in, top=1in, bottom=1in} 
    \fancypagestyle{title}{
        \fancyhf{} 
        \renewcommand{\headrulewidth}{0pt} 
        \renewcommand{\footrulewidth}{0pt} 
    }
    \thispagestyle{title}
    \begin{center}
        \vspace*{0.5cm}
            
        \hrule
        \vspace{0.5cm}
        \huge
        \textbf{Optical excitations and disorder in two-dimensional topological insulators}
        \vspace{0.5cm}
        \hrule

        \vspace{0.5cm}
            
        \vspace{1.5cm}
        
        \large
        A thesis presented for the degree of\\
        Doctor of Philosophy by\\
        \Large
        \textbf{Alejandro Jos\'e Ur\'ia \'Alvarez}\\
        \vspace{0.5cm}
        \large
        Doctorate Programme in Condensed Matter Physics, Nanoscience and Biophysics\\

        \vspace{1cm}

        \begin{figure}[h]
            \centering
            \begingroup
            \tikzset{every picture/.style={scale=2}}%
            \input{images/torus}
            \endgroup
        \end{figure}

        \vspace{1.5cm}
        \large
        Director:\\
        \textbf{Juan Jos\'e Palacios Burgos}\\
            
        \vfill

        \vspace{0.1cm}

        \includegraphics[width=0.4\textwidth]{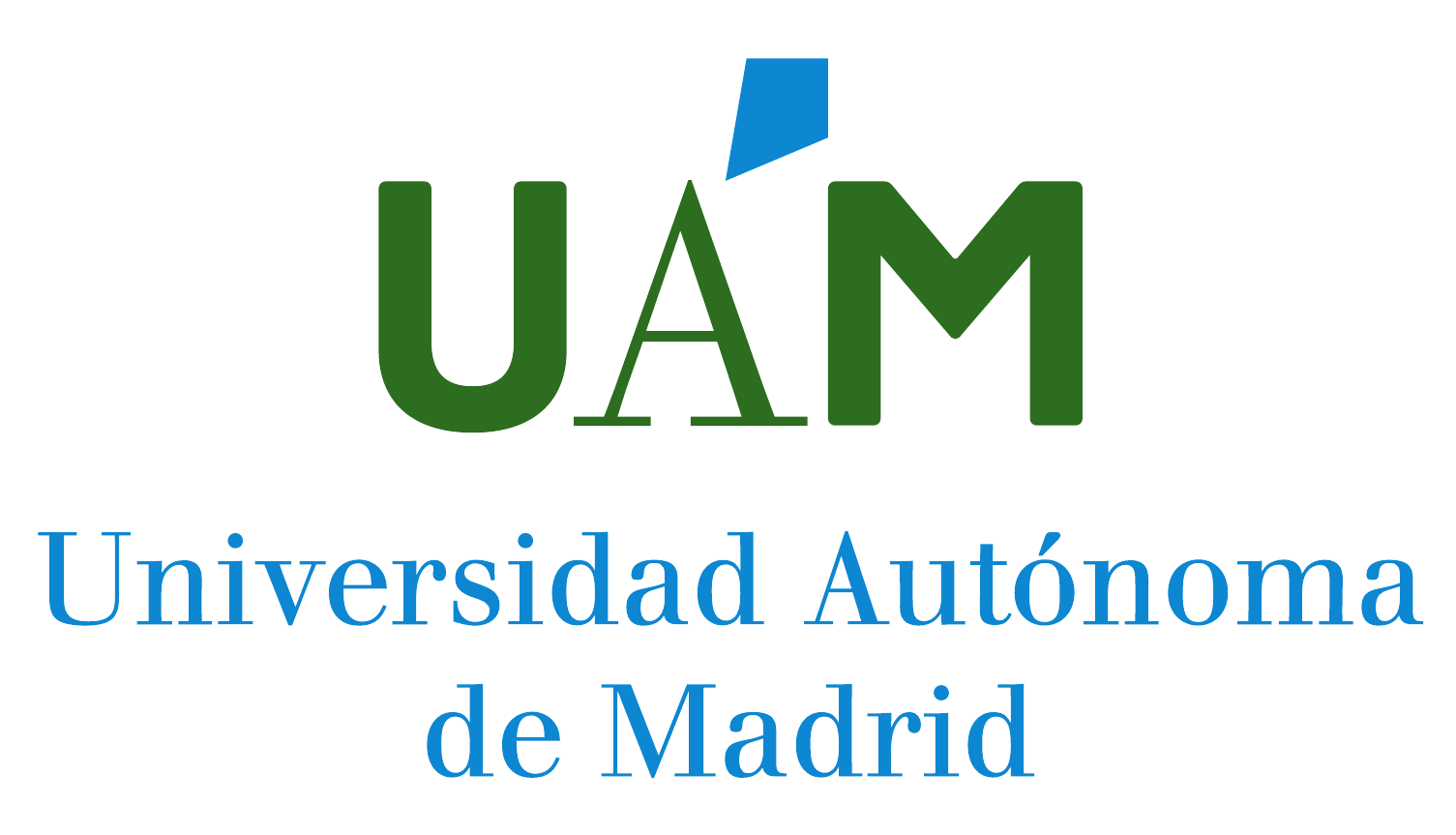}

        \large
        Departamento de F\'isica de la Materia Condensada\\
        Universidad Aut\'onoma de Madrid\\

        \vspace{0.8cm}
        Madrid, May 23, 2025
            
    \end{center}
    \restoregeometry
\end{titlepage}

\chapter*{}
\vspace{2cm}
\hfill\textit{A mi abuelita Mari}

\tableofcontents
\listoffigures
\listoftables

\chapter*{Acknowledgements}
\addcontentsline{toc}{chapter}{Acknowledgements}
\markboth{Acknowledgements}{Acknowledgements}
En cierto modo, las relaciones entre las personas forman un sistema de interacciones que bien recuerda a la f\'isica de muchos cuerpos, donde la acci\'on de uno solo de sus constituyentes tiene efectos sobre todo el conjunto (y no necesariamente de forma perturbativa). Y si bien al menos en f\'isica somos capaces hasta cierto punto de predecir el comportamiento, en las relaciones humanas no siempre es tan f\'acil establecer el grado de influencia que alguien puede tener en tu vida. En lo que sigue, intentar\'e agradecer a todas aquellas personas que han tenido un efecto en mi vida durante estos años, y que han contribuido a que hoy est\'e escribiendo estas l\'ineas.

En primer lugar, quiero dar las gracias a Juanjo, mi director de tesis, por haberme acompañado en todo este proceso. Valoro mucho su absoluta disponibilidad para discutir de física siempre que fuese necesario; no han sido pocas las ocasiones en las que comentar conceptos con él tras estar atascado durante días fuese la clave para continuar. También aprecio su \textit{laissez-faire}, del cual puede ser que abusase por momentos, su trato cercano y su humor distendido. Pero sobretodo le estoy agradecido por haberse involucrado conmigo no solo en mi formaci\'on meramente investigadora, sino tambi\'en en enseñarme a conducirme como cien\'tifico. Como mentor, ha contribuido enormemente a la visión que tengo ahora sobre cómo hacer física y ciencia, simplemente siguiendo su (buen) ejemplo. 

Dentro de la universidad, me resulta imposible olvidarme del equipo, Juanjo (Esteve), Toni y Joan. Vosotros fuisteis muchos d\'ias mi raz\'on para querer pasar el d\'ia en la universidad. Me llevo muy buenos recuerdos de todos los ratos que pasamos, tanto en los momentos cotidianos hablando sobre física o sobre la vida, con unas cervezas en la terraza o en los congresos a los que fuimos. En particular, a Juanjo le agradezco el compartir con él cultura de Internet, la afición por el \textit{gym}, que fue gracias a él que lo retomé con mayor interés que nunca, y siempre fue un placer para mi trabajar mano a mano con él. A Toni le agradezco nuestras discusiones, que aunque en alguna ocasión me desesperase un poco, en el fondo siempre fueron provechosas y siempre llegamos a un entendimiento mutuo. Toni tiene un buen hacer del que todos podemos tomar ejemplo. A Joan le agradezco los buenos ratos en el despacho, siempre me gustó discutir (o más bien aprender) QFT con él, nuestras partidas de ajedrez diarias, a pesar de mis derrotas constantes (lo cual espero cambiar algún día) y que me descubriese dónde comer un \textit{mapo tofu}. Os deseo lo mejor a los tres y espero que nuestra amistad prosiga en el futuro; con Juanjo y Toni además espero seguir colaborando y a Joan estoy seguro de que le irá muy bien vaya donde vaya.

I also want to thank the extended group, made by the latest incorporations, namely Simran, Pedro and Vinicius. It was great having you around, you made the lunchtime more interesting. I also want to thank Antonio Picón and his group, Giovanni, Mikhail, Miguel and Mauricio. For me, it was great collaborating with another group. I think that part of the PhD experience is also getting to see how things work somewhere else, trying to find points in common and ultimately sharing knowledge between the two, and I found this in the multiple reunions the two groups had. I also want to thank the rest of the people in the uni whom I shared some moments, Isidro, Simon, Manu, Nico, Jaime and more, that contributed to making these days enjoyable.

I thank Adolfo for receiving me in Grenoble. The three months I spent during my stay passed by very quickly, and during this time I had the chance to broaden my scope in what can be done. I am very grateful for the opportunity to work with you, for the support you have given me then and also the following months when we continued to finish the work started there. I am also very grateful to the rest of the people in Grenoble, especially to the people in the lab, for their help and support during my stay. To the people I met there, Justin, Selma, Mathieu. Beyond physics, I enjoyed our systematic \textit{baby foot} matches, and the beers we shared.

En Oviedo, quiero agradecer a Jaime, el que fue mi director en el pasado, por haberme motivado a hacer el doctorado. Aunque finalmente no se diese la colaboración entre Oviedo y Madrid, él es parte de la razón de que ahora escriba esto. Con él comenz\'o mi viaje por el mundo de los materiales topol\'ogicos, e igualmente tengo muy presentes sus consejos y su apoyo. Junto con Juanjo, ambos sois mis referentes en el mundo de la f\'isica.

Tambi\'en quiero dar las gracias a todos mis amigos en Xix\'on, ya sean mis amigos de toda la vida, de la universidad (que a estas alturas ya se puede considerar de buena parte de la vida), o m\'as recientes pero igualmente importantes. En orden alfa\'betico: \'Alvaro, César, Christian, Cuetos, Pablo, Pedro, Rub\'en y más gente que me dejo. Hago mención especial a Álvaro, por esas veces que bajó a Madrid ya fuese a ver al Sporting o algún concierto, y por haberme introducido al mundo de la halterofilia, y a Christian por todas las veces que vino a Madrid para ir de batalla. Estoy agracecido a todos igualmente por estar ahí, es por ellos que siempre \textit{prestóme} volver a Asturias después de los periodos en Madrid. Y luego está Elena, a la que quiero agradecer su paciencia y sus ánimos incondicionales, dándome fuerza cuando no la tenía en estos últimos días de escritura.

En Madrid, tambien estoy agradecido por el tiempo que pase con Prendes, Manu, Xavi, Andrea, Lunita, Anabel, las \textit{tech nenas} y dem\'as gente que apareci\'o en mi vida en esos a\~nos. A pesar de que nunca fui muy consistente saliendo con el grupo, siempre me acogieron los días que quise socializar, ya fuese yendo a cenar, a tomar algo, a Mondo o al Ochoymedio, y siempre me sentí parte del grupo. También en Madrid, le agradezco a Cuetos, Ropero y Paco nuestros frecuentes partidos de pádel. No está claro que dejasemos de ser unos paquetes desde aquellos comienzos los primeros años (yo al menos), pero siempre lo pasamos bien.

Obviamente estoy muy agradecido a mi familia, a mi padre Jos\'e Agust\'in, a mi madre Mar\'ia Cristina y a mi hermana Sonia que siempre me han apoyado, han estado conmigo en mis \'exitos y en mis fracasos, y que sobretodo me han aguantado en momentos en los que estaba más irascible. Sería impensable que estuviese escribiendo esta tesis si no fuese por sus constantes ánimos, mi carrera académica y laboral ha sido posible únicamente gracias a ellos. Igualmente agradezco al resto de la familia, abuelos, t\'ios, primos su apoyo durante todos estos años. Espero que todos esteis orgullosos de esta tesis.

Todas las personas aqu\'i mencionadas, y tambi\'en aquellas que no que han pasado por mi vida, han contribuido a hacerme la persona que soy hoy. Gracias a todos. Familia, amigos actuales y amigos pasados, antiguos compa\~neros de trabajo, compa\~neros de clase, compa\~neros de piso, compa\~neros de viaje, compa\~neros de fiesta, compa\~neros de deporte. No me olvido de nadie, gracias a todos.

\chapter*{List of publications}
\addcontentsline{toc}{chapter}{List of publications}
\markboth{List of publications}{List of publications}

Part of the work presented in this thesis has been published in the following articles:
\begin{itemize}
    \item \textit{Deep learning for disordered topological insulators through their entanglement spectrum}\\
    \textbf{Alejandro Jos\'e Ur\'ia-\'Alvarez}, Daniel Molpeceres-Mingo, Juan Jos\'e Palacios\\
    \href{https://journals.aps.org/prb/abstract/10.1103/PhysRevB.105.155128}{Phys. Rev. B \textbf{105}, 155128 (2022)}

    \item \textit{Efficient computation of optical excitations in two-dimensional materials with the Xatu code}\\
    \textbf{A. J. Uría-Álvarez}, J.J. Esteve-Paredes, M.A. García-Blázquez, J.J. Palacios\\
    \href{https://www.sciencedirect.com/science/article/pii/S0010465523003466}{Comput. Phys. Comms. \textbf{295}, 109001 (2024)}

    \item  \textit{tightbinder: A Python package for semi-empirical tight-binding models of crystalline and disordered solids}\\
    \textbf{Alejandro Jos\'e Ur\'ia-\'Alvarez}, Juan Jos\'e Palacios\\
    \href{https://joss.theoj.org/papers/10.21105/joss.05810}{J. Open Source Softw., 9(94), 5810 (2024)}

    \item \textit{Topologically protected photovoltaics in Bi nanoribbons}\\
    \textbf{Alejandro Jos\'e Ur\'ia-\'Alvarez}, Juan Jos\'e Palacios\\
    \href{https://pubs.acs.org/doi/10.1021/acs.nanolett.4c01277}{Nano Lett. 2024, 24, 22, 6651–6657}

    \item \textit{Amorphization-induced topological and insulator-metal transitions in bidimensional Bi$_x$Sb$_{1-x}$ alloys}\\
    \textbf{Alejandro Jos\'e Ur\'ia-\'Alvarez}, Juan Jos\'e Palacios\\
    \href{https://arxiv.org/abs/2410.16034}{arXiv:2410.16034 (2024)} (Submitted to Physical Review Research)

    \item \textit{Real-space criteria for non-crystalline fractional Chern insulators}\\
    \textbf{Alejandro. J. Ur\'ia-\'Alvarez}, Patrick Ledwith, Daniel E. Parker, C\'ecile Repellin, Adolfo. G. Grushin\\
    In preparation

\end{itemize}
Other works published during the course of the thesis that have been not been included are:

\begin{itemize}

    \item \textit{Theoretical Approach for Electron Dynamics and Ultrafast Spectroscopy (EDUS)}\\
    G. Cistaro, M. Malakhov, J. J. Esteve-Paredes, \textbf{A. J. Uría-Álvarez}, R. E. F. Silva, F. Martín, J. J. Palacios, A. Picón\\
    \href{https://pubs.acs.org/doi/full/10.1021/acs.jctc.2c00674}{J. Chem. Theory Comput. 2023, 19, 1, 333–348}

    \clearpage

    \item \textit{Shift current with Gaussian basis sets \& general prescription for maximally-symmetric summations in the irreducible Brillouin zone}\\
    M.A. García-Blázquez, J.J. Esteve-Paredes, \textbf{A. J. Uría-Álvarez}, J.J. Palacios\\
    \href{https://pubs.acs.org/doi/10.1021/acs.jctc.3c00917}{J. Chem. Theory Comput. 2023, 19, 24, 9416–9434}

    \item \textit{Excitons in nonlinear optical responses: shift current in MoS2 and GeS monolayers}\\
    J.J. Esteve-Paredes, M. A. García-Blázquez, \textbf{A. J. Uría-Álvarez}, M. Camarasa-Gómez, J.J. Palacios\\
    \href{https://doi.org/10.1038/s41524-024-01504-2}{npj Comput. Mater. 11, 13 (2025)}

    \item \textit{Strain-time engineering via exciton interactions}\\
    Maur\'icio F. C. Martins Quintela, Miguel S\'a, \textbf{Alejandro. J. Ur\'ia-\'Alvarez}, Mikhail Malakhov, Giovanni Cistaro, Jorge Quereda, Juan J. Palacios, Antonio Pic\'on\\
    \href{https://arxiv.org/abs/2501.16036}{arXiv:2501.16036 (2025)}

\end{itemize}

\chapter*{List of acronyms}
\addcontentsline{toc}{chapter}{List of acronyms}

\begin{description}[itemsep=-0.1em, style=nextline]
    \item[ANN] Artificial Neural Network
    \item[ARPES] Angle-Resolved Photoemission Spectroscopy 
    \item[BZ] Brillouin Zone
    \item[CI] Chern Insulator
    \item[CNN] Convolutional Neural Network 
    \item[DFT] Density Functional Theory 
    \item[DSM] Dirac Semimetal
    \item[FCI] Fractional Chern Insulator 
    \item[FQHE] Fractional Quantum Hall Effect
    \item[HF] Hartree-Fock
    \item[HOTI] Higher Order Topological Insulator 
    \item[HWCC] Hybrid Wannier Charge Center
    \item[IQHE] Integer Quantum Hall Effect
    \item[LL] Landau Level 
    \item[MBPT] Many-Body Perturbation Theory 
    \item[OBC] Open Boundary Conditions
    \item[PBC] Periodic Boundary Conditions  
    \item[QAHE] Quantum Anomalous Hall Effect
    \item[QSHE] Quantum Spin Hall Effect
    \item[SOC] Spin-Orbit Coupling
    \item[TI] Topological Insulator
    \item[TR-TI] Time-Reversal Topological Insulator
    \item[TRIM] Time-Reversal Invariant Momentum 
    \item[TRS] Time-Reversal Symmetry 
    \item[SK] Slater-Koster 
    \item[VCA] Virtual Crystal Approximation 
    \item[WCC] Wannier Charge Center
    \item[WF] Wannier Function 
    \item[WSM] Weyl Semimetal
\end{description}

\chapter*{Abstract}
\addcontentsline{toc}{chapter}{Abstract}
\markboth{Abstract}{Abstract}

Topological phases of matter have garnered significant interest over the past two decades for two main reasons: their identification, via topological invariants, relies on the quantum geometry of the Bloch states, bringing attention to an aspect of electronic band structure overlooked up to their discovery. Secondly, these classes of materials present electronic states with unusual properties, leading to exotic phenomena and making them relevant for potential applications. In this thesis we explore both fundamental and technological aspects of the first discovered topological phase: the topological insulator. To this end, we consider different models of topological insulators with a particular emphasis on Bismuth compounds, which have been shown to exhibit topological properties in their different forms.

The first part of the thesis deals with the optoelectronic potential of TIs. The optical response of insulators and semiconductors can be understood precisely in terms of bound electron-hole pairs, or excitons. We develop and implement a new methodology based on the tight-binding approximation, enabling us to solve the Bethe-Salpeter equation for the excitonic states several orders of magnitude faster than the ab-initio counterparts. Using hBN and MoS$_2$ as benchmark materials, we obtain exciton spectra in strong agreement with previous calculations.

We then explore the role of the most prominent feature of topological insulators in the optical response, namely the presence of gapless edge states. Breaking the appropriate symmetries of the system, we observe that it is possible to induce a finite edge charge accumulation and edge charge currents, from the transition of bulk excitons to the topological edge states. We illustrate this effect in Bi(111) nanoribbons where we estimate currents in the $\mu$A range, demonstrating the potential of TIs for photovoltaics.

The second part of the thesis focuses on the identification of topological materials and the effect of disorder on their properties. Prior research has succeeded in establishing the general framework underlying the calculation of topological invariants in crystalline systems. For those without translational invariance, we show how the entanglement spectrum, combined with deep learning, can predict the topological invariant of disordered systems. We test this methodology with a prototypical model of a topological insulator for an amorphous and a fractal lattice, accurately predicting the topological regimes even when the system is gapless.

In real materials, topologically insulating phases are typically achieved via strong spin-orbit coupling, producing band inversions which result in topologically non-trivial bands. Considering bidimensional Bi$_x$Sb$_{1-x}$ alloys, we use the entanglement spectrum to predict the critical concentrations of different allotropes. For the amorphous solid, with the aid of electronic transport calculations, we uncover a rich phase diagram where disorder, together with spin-orbit coupling, drives trivial to topological transitions and insulator to metal transitions.

Finally, we address the characterization of fractional Chern insulators. These phases, owing to the fractional quantum Hall effect, rely on the ideal flat band limit as the criterion to find candidate fractional systems. In absence of bands, we propose a real-space criterion to identify fractional Chern insulators in disordered systems. We demonstrate the connection between the real- and reciprocal-space approaches and apply the new criterion to various models, including amorphous Chern insulators, Landau levels in graphene, and Rydberg atoms in optical lattices, establishing the maximum disorder that a fractional phase can support.

\chapter*{Resumen}
\addcontentsline{toc}{chapter}{Resumen}
\markboth{Resumen}{Resumen}

Las fases topol\'ogicas de la materia han atraido un inter\'es muy significativo a lo largo de las dos \'ultimas decadas por dos razones principalmente. En primer lugar, su identificacion, por medio de invariantes topologicos, se basa en la geometr\'ia cu\'antica de los estados de Bloch, llamando la atenci\'on sobre un aspecto de la estructura electr\'onica de bandas que hab\'ia sido ignorado hasta su descubrimiento. En segundo lugar, estas clases de materiales presentan estados electr\'onicos con propiedades inusuales, dando lugar a fen\'omenos ex\'oticos y haci\'endolos relevantes por sus posibles aplicaciones. En esta tesis exploramos tanto los aspectos fundamentales como tecnol\'ogicos de la primera fase topol\'ogica descubierta: el aislante topol\'ogico. Para ello, consideramos diferentes modelos de aislantes topol\'ogicos con un \'enfasis particular en los compuestos de bismuto, los cuales se ha demostrado exhiben propiedades topol\'ogicas en sus diferentes formas.

La primera parte de la tesis trata sobre el potencial para optoelectr\'onica de los aislantes topol\'ogicos. La respuesta \'optica de aislantes y semiconductores puede entenderse de forma precisa en t\'erminos de pares electron-hueco ligados, o excitones. Hemos desarrollado e implementado una nueva metodolog\'ia basada en el m\'etodo de ligaduras fuertes, que nos permite resolver la ecuaci\'on de Bethe-Salpeter para los estados excit\'onicos con una velocidad m\'utilples ordenes de magnitud superior a la de los c\'alculos ab-initio. Usando hBN y MoS$_2$ como materiales de referencia, obtenemos espectros de excitones en fuerte acuerdo con c\'alculos previos.

A continuaci\'on exploremos el rol de la caracter\'istica m\'as notable de los aislantes topol\'ogicos en la respuesta \'optica, que es la presencia de estados de borde sin banda prohibida. Rompiendo las simetr\'ias apropiadas del sistema, observamos que es posible inducir una acumuaci\'on finita de carga en los bordes, as\'i como corrientes el\'ectricas en los mismos. Ilustramos este efecto con una nanocinta de Bi(111), donde estimamos que la intensidad de las corrientes estar\'ia en el rango de los $\mu$A, demostrando as\'i el potencial de los aislantes topol\'ogicos para la energ\'ia fotovoltaica.

La segunda parte de la tesis se centra en la identificaci\'on de materiales topol\'ogicos y el efecto del desorden en sus propiedades. Trabajos anteriores han tenido \'exito en establecer el marco general que subyace en el c\'alculo de invariantes topol\'ogicos en sistemas cristalinos. Para sistemas sin invarianza traslacional, mostramos c\'omo el espectro de entrelazamiento, cuando se combina con aprendizaje profundo, puede predecir el invariante topol\'ogico de sistemas desordenados. Probamos esta metodolog\'ia con un modelo protot\'ipico de aislante topol\'ogico para una red amorfa y una red fractal, prediciendo satisfactoriamente los reg\'imenes topol\'ogicos, incluso cuando el sistema no posee banda prohibida.

En materiales reales, los aislantes topol\'ogicos se consiguen t\'ipicamente a trav\'es de un fuerte acoplo esp\'in-\'orbita, produciendo una inversi\'on de bandas que resulta en bandas topol\'ogicamente no triviales. Tomando aleaciones bidimensionales de Bi$_x$Sb$_{1-x}$, utilizamos el espectro de entrelazamiento para predecir las concentraciones cr\'iticas de distintos al\'otropos. Para el s\'olido amorfo, con la ayuda de c\'alculos de transporte electr\'onico, encontramos un rico diagrama de fases en el que el desorden, junto al acoplo esp\'in-\'orbita, induce transiciones de aislante trivial a topol\'ogico y de aislante a metal.

En \'ultimo lugar, abordamos la caracterizaci\'on de los aislantes fraccionarios de Chern. Estas fases, a ra\'iz del efecto Hall cu\'antico fraccionario, toman el l\'imite ideal de bandas planas como criterio para identificar candidatos a sistemas fraccionarios. En ausencia de bandas, proponemos un criterio de espacio real para identificar aislantes fraccionarios de Chern en sistemas desordenados. Demostramos la conexi\'on entre los enfoques de espacio real y rec\'oproco, y aplicamos el nuevo criterio a varios modelos, concretamente aislantes de Chern amorfos, niveles de Landau en grafeno y \'atomos de Rydberg en redes \'opticas, estableciendo el m\'aximo desorden que una fase fraccionaria puede soportar.

\chapter*{Resume}
\addcontentsline{toc}{chapter}{Resume}
\markboth{Resume}{Resume}

Les fases topolóxiques de la materia xeneraron un interés bien importante a lo llargo de les dos caberes décades por dos razones principales. En primer llugar, la so identificación por mediu d’invariantes topolóxicos, basada na xeometría cuántica de los estaos de Bloch, fixo que se prestara atención a un aspeutu de la estructura electrónica de bandes que se taba inorando hasta’l so descubrimientu. En segundu llugar, estes clases de materiales tienen estaos electrónicos con propiedaes poco comunes, lo que conlleva fenómenos exóticos y failos importantes pol so potencial p'aplicaciones. Nesta tesis esploramos tanto los aspeutos fundamentales como teunolóxicos de la primer fase topolóxica descubierta: l’aislante topolóxicu. Pa dello, consideramos distintos modelos d’aislantes topolóxicos, con un énfasis particular nos compuestos de bismutu, que demostraron tener propiedaes topolóxiques nes sos distintes formes.

La primer parte de la tesis trata sobre’l potencial pa la optoelectrónica de los aislantes topolóxicos. La rempuesta óptica d’aislantes y semiconductores pue entendese con precisión en términos de pares electrón-furacu lligaos, o excitones. Desendolcamos y implementamos una nueva metodoloxía basada nel métodu de lligadures fuertes, que nos perm\'ite resolver la ecuación de Bethe-Salpeter pa los estaos excitónicos con una velocidá d’ordes de magnitú superior a la de los cálculos ab-initio. Usando hBN y MoS$_2$ como materiales de referencia, llogramos espectros d’excitones en bon alcuerdu con cálculos previos.

Dempués esploramos’l papel de la carauterística más notable de los aislantes topolóxicos na rempuesta \'optica, que ye la presencia d’estaos de borde ensin banda prohibida. Rompiendo les simetríes apropiaes del sistema, vemos que ye posible inducir una acumulación finita de carga nos bordes, asina como corrientes eléctriques nos mesmos. Ilustramos esti efeutu con una nanocinta de Bi(111), onde estimamos que la intensidá de les corrientes taría nel rangu de los $\mu$A, demostrando asina’l potencial de los aislantes topolóxicos pa la enerxía fotovoltaica.

La segunda parte de la tesis céntrase na identificación de materiales topolóxicos y l’efeutu del desorde nes sos propiedaes. Trabayos anteriores tuvieron ésitu al establecer el marcu xeneral que s’atopa na base del cálculu d’invariantes topolóxicos en sistemes cristalinos. Pa sistemes ensin invarianza tresllacional, amosamos cómo’l espectru d’entrellaciamientu, cuando se combina con aprendizaxe fondu, pue predicir l’invariante topolóxicu de sistemes desordenaos. Probamos esta metodoloxía con un modelu prototípicu d’aislante topolóxicu pa una rede amorfa y una rede fractal, anticipando con ésitu los rexímenes topolóxicos, inclusive cuando’l sistema nun tien banda prohibida.

En materiales reales, los aislantes topolóxicos consíguense típicamente al traviés d’un fuerte acoplamientu espín-órbita, produciendo una inversión de bandes que resulta en bandes topolóxicamente non triviales. Tomando aleaciones bidimensionales de Bi$_x$Sb$_{1-x}$, usamos l’espectru d’entrellaciamientu p'anticipar les concentraciones crítiques de distintos alótropos. Pal sólidu amorfu, cola ayuda de cálculos de tresporte electrónicu, atopamos un diagrama de fases ricu nel que’l desorde, xunto col acoplamientu espín-órbita, induz transiciones de trivial a topolóxicu y d’aislante a metálicu.

Finalmente, abordamos la carauterización de los aislantes fraicionarios de Chern. Estes fases, a raíz del efeutu Hall cuánticu fraicionariu, tomen el llímite ideal de bandes planes como criteriu pa identificar candidatos a sistemes fraicionarios. Na ausencia de bandes, proponemos un criteriu d’espaciu real pa identificar aislantes fraicionarios de Chern en sistemes desordenaos. Demostramos la conexón ente los enfoques d’espaciu real y recíprocu, y aplicamos el nuevu criteriu a varios modelos, concretamente aislantes de Chern amorfos, niveles de Landau en grafenu y átomos de Rydberg en redes óptiques, estableciendo’l máximu desorde que una fase fraicionaria pue soportar.

\clearpage


\pagenumbering{arabic}

\part{Optical excitations in topological insulators}
\chapter{Introduction}

The question of why things appear the way they do, or in more physical terms, why some materials are transparent, while others are opaque or reflective, is a fundamental one that traces back to very origins of physics. The study of the interaction of light with matter began with the concept of geometrical optics, establishing principles governing the propagation of light rays and lensing~\cite{newton1952opticks}. Then, the electromagnetic theory was developed, which successfully explained the reflection, refraction and diffraction of light in dielectric media~\cite{wangsness1979electromagnetic}. This explanation relies on intrinsic quantities of the materials, namely the dielectric function $\varepsilon(\omega)$. Meaning that although light behavior is well-understood, there is a gap in our knowledge as for why each material shows its specific properties. Simplified models, such as the Drude model for metals~\cite{ashcroft} or the Lorentz model for insulators~\cite{fowles1989introduction}, can describe reflectivity but in an ad hoc manner. It is with the advent of quantum mechanics that for the first time we can explain the optical properties of materials from first principles. 

Thus, optics is concerned with the dynamics and manipulation of light, either classically or quantum mechanically, after its interaction with a material, whereas the quantum mechanical description of condensed matter systems allows us to explain the behavior of the system \textit{after} its interaction with light. Returning to the initial question, our perception of objects stems from the light that is emitted, reflected and absorbed by them. Similar to the hydrogen series~\cite{eisberg1985quantum}, the absorption and emission spectra of solids correspond to the allowed transitions between its quantum states, where photons are absorbed and emitted. The study of these transitions is the study of optical excitations. 

Broadly speaking, an optical excitation refers to a state of the system above its equilibrium energy level (ground state) created by the mediation of photons. Focusing on solids, the most prominent example is the excitation of an electron-hole pair, where an electron from the valence band absorbs a photon and transitions to the conduction band. Other examples include the excitation of optical phonons~\cite{cowley1963, Deinzer2004, GENZEL19733} or plasmons~\cite{maier2007plasmonics, brown2016, Yu2019}. Note that these excitations rely on the absorption of photons, but do not involve actual photonic states $a^{\dagger}_{\mathbf{q}\lambda}$ in their description~\cite{fetter2012quantum,mahan2013many}. Instead, they can be described solely in terms of the quasiparticles of the condensed matter system, and the electric field admits a semiclassical treatment (meaning that the notion of photon is used to justify the existence of the excitations, but the electric field is mathematically treated as a classical field). One kind of excitation that involves photon states are polaritons, which arise from the coupling between photons and bosonic quasiparticles such as phonons, excitons or plasmons. Polaritons admit a fully quantum description with the quantization of the electromagnetic field, but also a classical description in terms of the macroscopic dielectric function of the material~\cite{Alvarez-Perez2020, rivera2019, zubin2014, Latini2019}.

Even though all these excitations involve photons, when describing the absorption spectrum in insulators and semiconductors, it is the electronic excitations or electron-hole pairs that are primarily responsible for the observed spectra~\cite{fox2010optical}. At finite temperature, the finite phonon density of states introduces additional processes, such as phonon-assisted pair formation or recombination or temperature-dependent absorption, resulting in a more complex picture~\cite{giustino2017, Noffsinger2012, kioupakis2010, zacharias2016}. For simplicity, however, we will focus exclusively on purely electronic optical excitations. In this context, the absorption can be obtained from the optical conductivity. The optical conductivity is defined (in an isotropic medium and in linear regime) from:
\begin{equation}
    \mathbf{J}(\omega)=\sigma(\omega)\mathbf{E}(\omega)
\end{equation}
where $\mathbf{J}(\omega)$ is the current density, $\mathbf{E}(\omega)$ is the electric field, and $\sigma(\omega)$ is the optical conductivity. Via Maxwell's equations, one can show that the dielectric function $\varepsilon(\omega)$, defined from $\mathbf{D}(\omega)=\varepsilon_0\varepsilon(\omega)\mathbf{E}(\omega)$ is related to the optical conductivity by~\cite{ashcroft}:
\begin{equation}
    \varepsilon(\omega)=1+\frac{i\sigma(\omega)}{\omega\varepsilon_0}
\end{equation}
where $\varepsilon_0$ is the vacuum permittivity and $\mathbf{D}(\omega)$ is the displacement field. Both the optical conductivity and the dielectric function are in general complex functions. From this we see that the imaginary part of the optical conductivity, or equivalently the real part of the dielectric function gives the refraction index of the material. Analogously, the real part of the optical conductivity (or the imaginary part of the dielectric function) gives the absorption spectrum of the material. The conductivity can then be obtained by means of linear-response theory, namely with the Kubo formula~\cite{coleman2015introduction}. 

Independently of its connection to the dielectric function, the optical conductivity gives us a measure of the current that the applied electric field originates in the material (i.e.\ it is a response function). This concept is fundamental to optoelectronics, which focuses on the design of devices where the incidence of light creates an electrical current, or conversely, an applied voltage results in light emission. The most common examples are photodiodes such as solar cells or light-emitting diodes (LEDs). These optical-electronic energy conversion devices have become increasingly important in the last years, due to the need for renewable energy sources and the development of information technology~\cite{rosencher2002optoelectronics, Soref1993}. 

While linear response explains absorption, non-linear optical properties~\cite{parker2019} are also of great interest, for instance for the generation of high harmonic pulses~\cite{high_harmonic} (e.g.\ second harmonic for second order), the shift current~\cite{sipe,circular_photogalvanic_weyl} (second order) which could be an alternative DC source to the conventional solar cells as it is not bound to the Shockley-Queisser limit~\cite{shockley}, the jerk current~\cite{puente-uriona2023, furchi2014} (third order), also known as photoconductivity and more. All these effects are explained in terms of single electron-hole pairs, where the order of the effect corresponds to the number of absorbed photons. Owing to the perturbative approach used when writing the current densities (in powers of the electric field, e.g. $J_{a}(2\omega)=\sum_{b,c}\sigma_{abc}(2\omega;\omega,\omega)E_{b}(\omega)E_{c}(\omega)$ for second-harmonics), the magnitude of each effect will be lower the higher its order, but potentially relevant nonetheless.

\begin{figure}[h]
    \centering
    
    \raisebox{0.4cm}{
    \begin{tikzpicture}
        \begin{feynman}[medium]
          \vertex (b1);
          \vertex [right=1.5cm of b1, dot](b2){};
          \vertex [right=1.5cm of b2, dot](b3){};
          \vertex [right=1.5cm of b3](b4);
    
          \vertex [xshift=0.75cm, yshift=0.3cm] (t1) at (b1) {\(a, \omega\)};
          \vertex [xshift=-0.75cm, yshift=0.3cm] (t2) at (b4) {\(b, \omega\)};
          \vertex [xshift=0.75cm, yshift=1.05cm] (t3) at (b2) {\(m, \omega + \omega'\)};
          \vertex [xshift=0.75cm, yshift=-1.05cm] (t4) at (b2) {\(n, \omega'\)};

          \vertex [xshift=0.1cm, yshift=1.5cm] (t5) at (b1){(a)};

          \diagram* {
            (b1) -- [boson] (b2);
            (b2) -- [fermion, half left] (b3);
            (b3) -- [fermion, half left] (b2);
            (b3) -- [boson] (b4);
          };
        \end{feynman}
      \end{tikzpicture}
    }
      \hspace{0.5cm}
      \begin{tikzpicture}
    
        \draw[black] (-2, 2.6) circle (0pt) node[below=0.1cm] {(b)};

        \draw[thick] plot[domain=-1.5:1.5, samples=100] (\x, {-0.4*\x*\x}) node[right] {\(v\)};
        
        \draw[thick] plot[domain=-1.5:1.5, samples=100] (\x, {0.4*\x*\x + 1.5}) node[right] {\(c\)};
        
        \draw[->, red, thick, decorate, decoration={snake, amplitude=1mm, segment length=5mm}] (-1, 0.4*1*1 + 0.5) -- (1 - 0.1, -0.4*1*1 + 0.1) node[midway, above=0.2cm] {\(\omega\)};
        
        \draw[->, blue, thick] (1, -0.4*1*1) -- (1, 0.4*1*1 + 1.4) node[midway, right] {};
        
        \filldraw[blue] (1, 0.4*1*1 + 1.5) circle (2pt) node[above] {\(e^-\)};
        
        \filldraw[white] (1, -0.4*1*1) circle (2pt);
        \draw[red] (1, -0.4*1*1) circle (2pt) node[below=0.5cm, right=0.2cm] {\(h^+\)};
    
        \draw[->, red, thick, dashed, decorate, decoration={snake, amplitude=1mm, segment length=5mm}] (1 + 0.2, 0.4*1*1 + 1.4) -- (3 - 0.1, -0.4*1*1 + 1.1) node[midway, above=0.2cm] {\(\omega\)};
    
        \draw[->, thick, dashed, blue] (1 + 0.1, 0.4*1*1 + 1.5 - 0.1) to[bend left] (1 + 0.1, -0.4*1*1 + 0.1);

      \end{tikzpicture}

      \begin{tikzpicture}
        \draw[thick, gray, opacity=0.2] (-2, 0) -- (2, 0);
      \end{tikzpicture}
      \vspace{0.35cm}

      \raisebox{1cm}{
        \begin{tikzpicture}
        \begin{feynman}[medium]
        \vertex (i1);
        \vertex [right=1.5cm of i1, dot](f1){};
        \vertex [below=1.5cm of i1](i2);
        \vertex [right=1.5cm of i2, dot](f2){};
        \vertex [right=1.5cm of b3](b4);
        \vertex [xshift=1.5cm, yshift=-0.75cm, dot] (f3) at (f1) {};
        \vertex [right=1.5cm of f3](i4);

        \vertex [xshift=0.75cm, yshift=0.3cm] (t1) at (i1) {\(a, \omega\)};
        \vertex [xshift=0.75cm, yshift=-0.4cm] (t2) at (i2) {\(b, \omega\)};
        \vertex [xshift=0.5cm, yshift=-0.7cm] (t3) at (i1) {\(m, \omega + \omega'\)};
        \vertex [xshift=1cm, yshift=0cm] (t3) at (f1) {\(n, \omega'\)};
        \vertex [xshift=1.4cm, yshift=0cm] (t4) at (f2) {\(r, 2\omega + \omega'\)};
        \vertex [xshift=-0.75cm, yshift=0.4cm] (t5) at (i4) {\(c, 2\omega\)};

        \vertex [xshift=0.1cm, yshift=1.5cm] (t6) at (i1){(c)};

        \diagram* {
            (i1) -- [boson] (f1);
            (i2) -- [boson] (f2);
            (f1) -- [fermion] (f2);
            (f2) -- [fermion] (f3);
            (f3) -- [fermion] (f1);
            (f3) -- [boson] (i4);
        };
        \end{feynman}
        \end{tikzpicture}
    }
    \hspace{0.35cm}
    \begin{tikzpicture}

        \draw[black] (-2, 4) circle (0pt) node[below=0.1cm] {(d)};
    
        \draw[thick] plot[domain=-1.5:1.5, samples=100] (\x, {-0.4*\x*\x}) node[right] {\(v\)};
        
        \draw[thick] plot[domain=-1.5:1.5, samples=100] (\x, {0.4*\x*\x + 1}) node[right] {\(c\)};

        \draw[thick] plot[domain=-1.5:1.5, samples=100] (\x, {0.4*\x*\x + 2.8}) node[right] {\(c'\)};
        
        \draw[->, red, thick, decorate, decoration={snake, amplitude=1mm, segment length=5mm}] (-1, 0.4*1*1 + 0.5) -- (1 - 0.1, -0.4*1*1 + 0.1) node[midway, above=0.2cm] {\(\omega\)};
        \draw[->, red, thick, decorate, decoration={snake, amplitude=1mm, segment length=5mm}] (-1, 0.4*1*1 + 2.4) -- (1 - 0.1, 0.4*1*1 + 1.1) node[midway, above=0.2cm] {\(\omega\)};
        
        \draw[->, blue, thick] (1, -0.4*1*1) -- (1, 0.4*1*1 + 0.9) node[midway, right] {};
        \draw[->, blue, thick] (1, 0.4*1*1 + 1.1) -- (1, 0.4*1*1 + 2.7) node[midway, right] {};
        
        \filldraw[white] (1, 0.4*1*1 + 1) circle (2pt);
        \draw[red] (1, 0.4*1*1 + 1) circle (2pt) node[above] {};

        \filldraw[blue] (1, 0.4*1*1 + 2.8) circle (2pt) node[above] {\(e^-\)};
        
        \filldraw[white] (1, -0.4*1*1) circle (2pt);
        \draw[red] (1, -0.4*1*1) circle (2pt) node[below=0.5cm, right=0.2cm] {\(h^+\)};
    
        \draw[->, red, thick, dashed, decorate, decoration={snake, amplitude=1mm, segment length=5mm}] (1 + 0.2, 0.4*1*1 + 2.7) -- (3 - 0.1, 0.4*1*1 + 1.4) node[midway, above=0.2cm] {\(2\omega\)};
    
        \draw[->, thick, dashed, blue] (1 + 0.1, 0.4*1*1 + 2.8 - 0.1) to[bend left] (1 + 0.1, -0.4*1*1 + 0.1);

      \end{tikzpicture}

      \caption[Diagrams for linear and non-linear optical conductivities]{(a) Feynman diagram showing the main contribution to the linear optical conductivity $\sigma_{ab}(\omega)$ in terms of free electron-hole pairs.\ (b) Band diagram illustrating the process of excitation of an electron-hole pair via absorption of a photon (solid lines) and deexcitation via emission (dashed lines).\ (c) Feynman diagram showing one of the contributions to the second-order conductivity $\sigma_{abc}(2\omega;\omega,\omega)$ for the specific case of second-harmonic generation.\ (d) Band diagram depicting the process of second-harmonic generation, where an electron-hole pair is excited to a higher conduction band via absorption of two photons. The indices $(a,b,c)$ denote light polarization, while the indices $(m,n,r)$ denote the band indices including momenta $\mathbf{k}$, and $\omega, \omega'$ are frequencies. Feynman diagrams adapted from~\cite{parker2019}.}
      \end{figure}
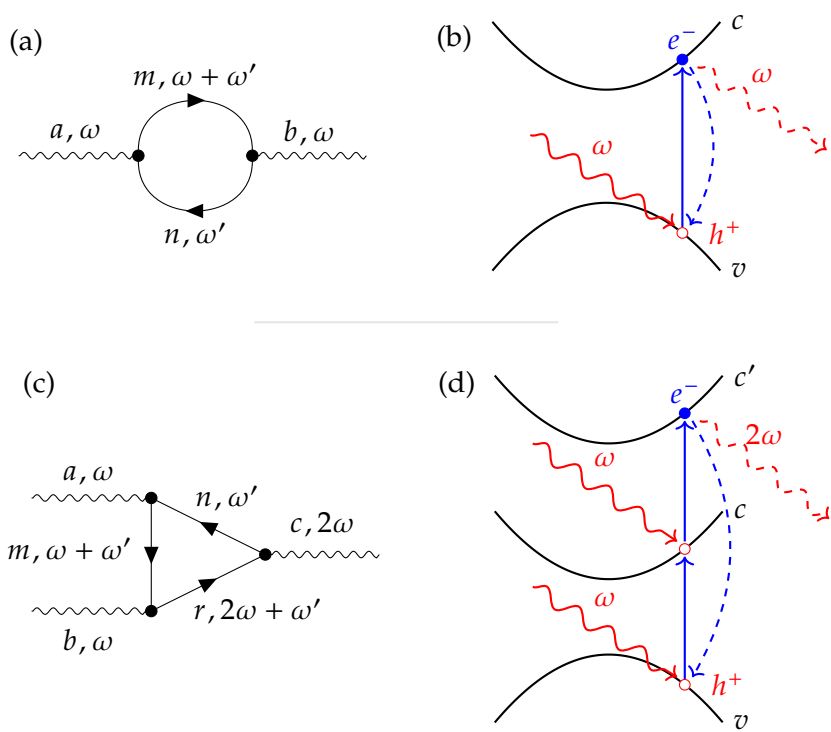

So far we have been discussing optical excitations in the form of \textit{free} electron-hole pairs, based purely on the single-particle picture. The electronic band structure is typically obtained as the mean-field solution to the interacting problem (namely the electronic Hamiltonian of the solid in the Born-Oppenheimer approximation~\cite{Kaxiras_2003}) as in density functional theory or Hartree-Fock~\cite{kohanoff2006electronic, gross1986many}. In many cases, this single-particle description is enough to describe correctly the properties of materials, including optical ones. However, there are materials where the absorption spectrum (or other non-linear properties) cannot be explained solely in terms of these free electron-hole pairs. In those cases it is necessary to move beyond the mean-field picture and take into account the Coulomb interaction between the excited electron and the hole~\cite{Hanke1980,Rohlfing1998,Rohlfing2000}. The resulting excitations of the system are \textit{bound} electron-hole pairs or excitons, which consist on a collective electron-hole pair (i.e.\ a superposition). 

\begin{figure}[h]
    \centering

    \raisebox{1cm}{
    \begin{tikzpicture}
        \begin{feynman}[medium]
        \vertex (b1);
        \vertex [right=1.5cm of b1, dot](b2){};
        \vertex [xshift=0.75cm, yshift=0.75cm] (l1) at (b2);
        \vertex [xshift=0.75cm, yshift=-0.75cm] (l2) at (b2);
        \vertex [right=1.0cm of l1](l3);
        \vertex [right=1.0cm of l2](l4);
        \vertex [xshift=0.75cm, yshift=-0.75cm, dot] (b3) at (l3){};
        \vertex [right=1.5cm of b3](b4);

        \vertex [xshift=0.5cm, yshift=-0.75cm] (text) at (l1){\(L(\omega)\)};
        \vertex [xshift=0.75cm, yshift=0.3cm] (t1) at (b1){\(a,\omega\)};
        \vertex [xshift=-0.75cm, yshift=0.3cm] (t2) at (b4){\(b,\omega\)};
        \vertex [yshift=1.5cm] (t3) at (b1){(a)};

        \diagram* {
            (b1) -- [boson] (b2);
            (b2) -- [fermion, quarter left] (l1);
            (l2) -- [fermion, quarter left] (b2);
            (l1) -- (l3);
            (l4) -- (l2);
            (l1) -- (l2);
            (l3) -- (l4);
            (l3) -- [fermion, quarter left] (b3);
            (b3) -- [fermion, quarter left] (l4);
            (b3) -- [photon] (b4);
        };
        \end{feynman}
        \end{tikzpicture}
    }
\hspace{0.5cm}
\begin{tikzpicture}

    \draw[black] (-2, 2.5) circle (0pt) node[below=0.1cm] {(b)};
    
    \draw[thick] plot[domain=-1.5:1.5, samples=100] (\x, {-0.4*\x*\x}) node[right] {\(v\)};
    
    \draw[thick] plot[domain=-1.5:1.5, samples=100] (\x, {0.4*\x*\x + 1.5}) node[right] {\(c\)};
    
    \draw[->, red, thick, decorate, decoration={snake, amplitude=1mm, segment length=5mm}] (-2, 0.75) -- (-0.6, 0.25) node[midway, above=0.2cm] {\(\omega\)};
    
    \filldraw[blue] (0, 1.5) circle (2pt) node[above] {\(e^-\)};
    
    \filldraw[white] (0, 0) circle (2pt);
    \draw[red] (0, 0) circle (2pt) node[below=0.1cm] {\(h^+\)};

    \draw[->, red, thick, dashed, decorate, decoration={snake, amplitude=1mm, segment length=5mm}] (0.6, 0.25) -- (2, 0.75) node[midway, above=0.2cm] {\(\omega\)};

    \draw[dashed, thick] (0, 0.75) ellipse (0.5cm and 1.4cm);

    \draw[->, thick, gray] (0, 1.35) -- (0, 0.1);
    \draw[->, thick, gray] (0.2, 1.3) to[bend left=20] (0.2, 0.2);
    \draw[->, thick, gray] (-0.2, 1.3) to[bend right=20] (-0.2, 0.2);
  \end{tikzpicture}
  \hspace{0.5cm}
  \raisebox{0.35cm}{
        \begin{tikzpicture}

        \draw[black] (-0.8, 2) circle (0pt) node[below=0.1cm] {(c)};
        \foreach \x in {-1, 0, 1, 2}
        \foreach \y in {-1, 0, 1, 2}
            \filldraw[black] (\x, \y) circle (1.5pt);

        \filldraw[blue] (0.4, 0.7) circle (3pt) node[below] {\(e^-\)};

        \draw[red, thick, dashed] (0.4, 0.7) circle (1cm);
        \filldraw[red] (1.4, 0.7) circle (3pt) node[right] {\(h^+\)};

        \end{tikzpicture}
        }

\caption[Diagrams for linear conductivity with excitons]{(a) Feynman diagram showing the main contribution to the linear optical conductivity $\sigma_{ab}(\omega)$ in terms of the exciton propagator $L(\omega)$, i.e.\ considering electron-hole interactions.\ (b) Band diagram illustrating the excitation of an exciton via absorption of a photon (solid line) and its deexcitation via photon emission (dashed line).\ (c) Schematic representation of an exciton in a crystal lattice, where the interacting electron-hole pair behaves similarly to a hydrogen atom.}
\end{figure}
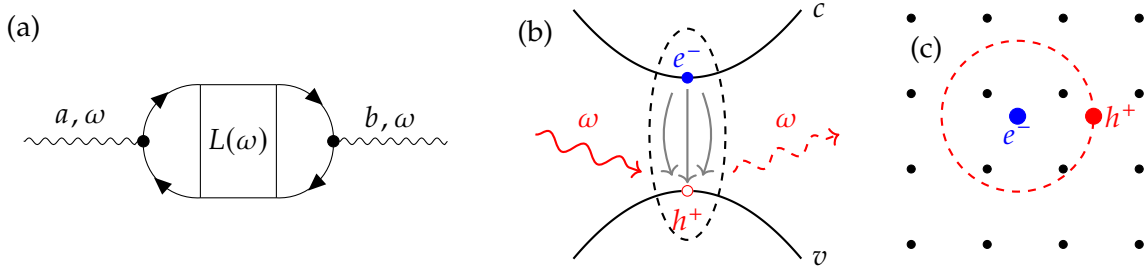

\begin{figure}
    \centering
    \raisebox{0.6cm}{
        \includegraphics[height=5.3cm]{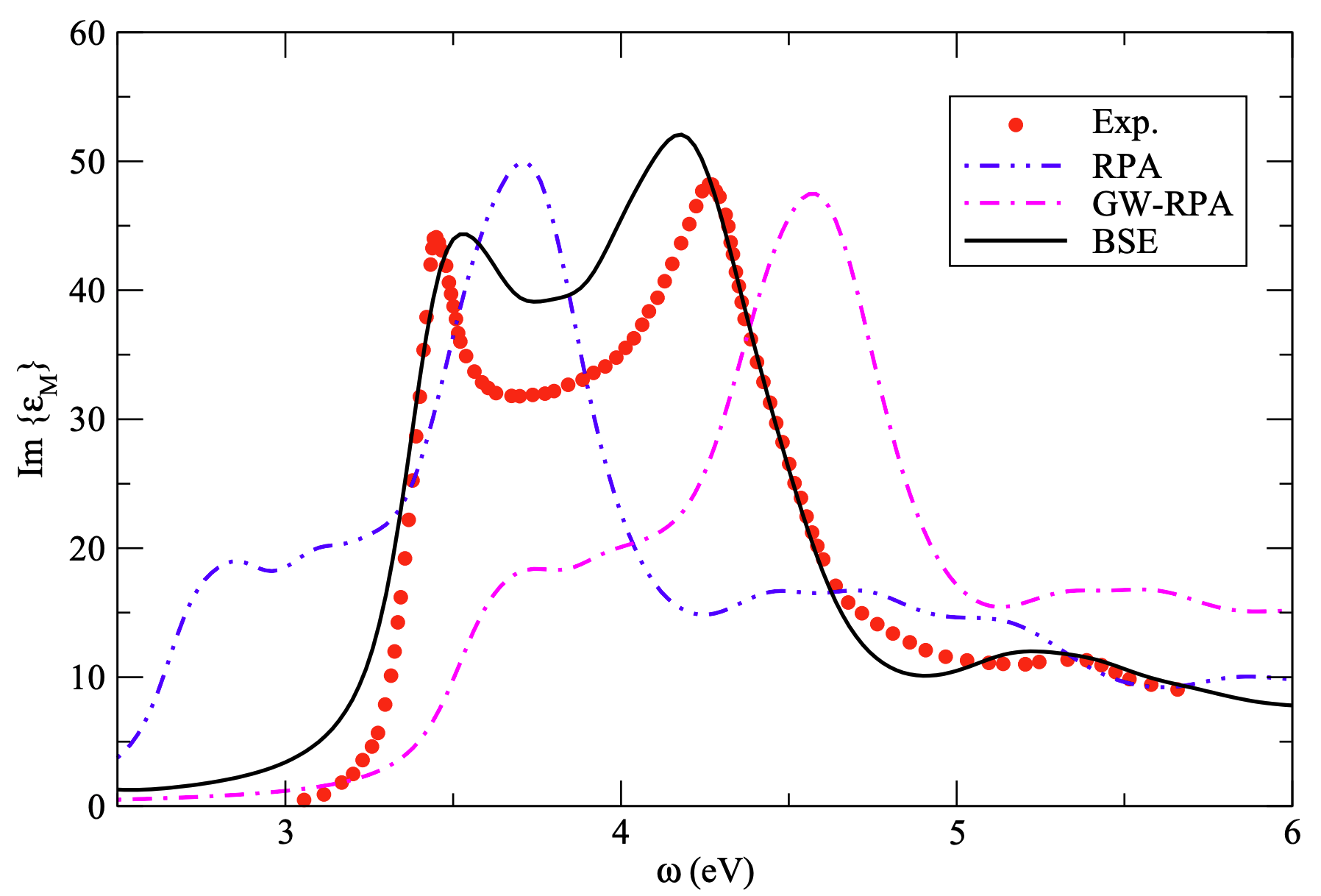}
    }
    \includegraphics[height=5.9cm]{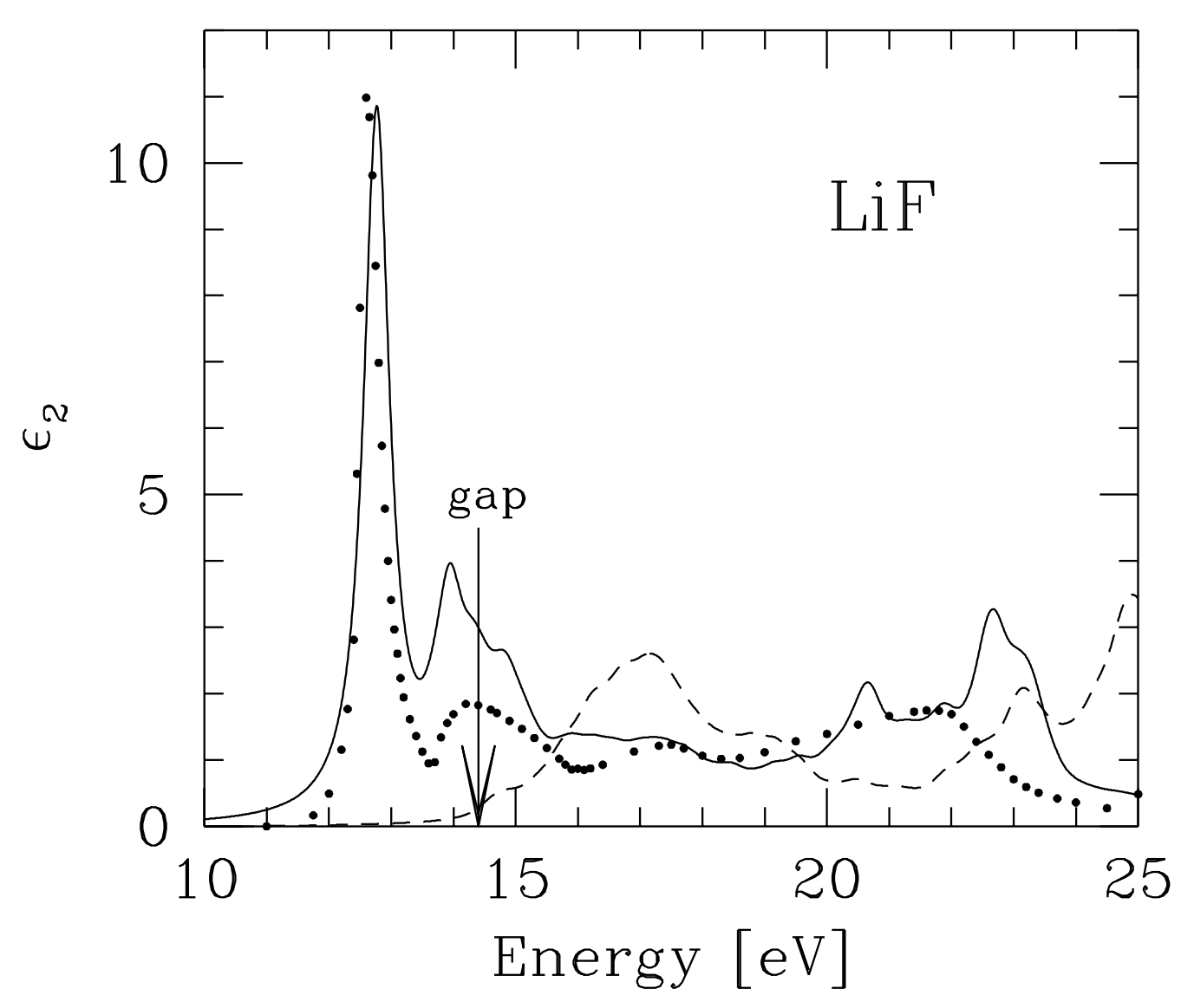}
    \caption[Absorption spectra of bulk silicon and LiF]{(Left) Absorption spectrum of bulk silicon, as obtained from DFT (RPA), GW-RPA and the BSE (excitons). The red points correspond to the experimental data, obtained from~\cite{absorption_silicon_exp}. Plot by Francesco Sottile, extracted from~\cite{sottile_thesis}. (Right) Absorption spectrum of LiF. The solid line corresponds to the BSE calculation, while the dashed one is without electron-hole excitations. The dots denote the experimental data, obtained from~\cite{Roessler67}. Extracted from~\cite{Rohlfing1998}.}\label{fig:absorption_silicon_LiF}
\end{figure}

In the limit of a strongly screened Coulomb interaction, the calculation of the conductivity approaches that of free electron-hole pairs, which is normally the case for three-dimensional materials. However, the need to include excitons in the description of optical responses became apparent with materials such as bulk Si or LiF~\cite{Hanke1980,Rohlfing1998,sottile2003} (see Fig.~\ref{fig:absorption_silicon_LiF}), where perturbative corrections beyond the mean-field approximation, such as the GW method~\cite{aryasetiawan1998gw,Golze2019}, still fail to reproduce the experimental spectrum. Only by including the Coulomb interaction between the excited electron and hole can these spectra be accurately reproduced. With the discovery of graphene~\cite{geim2009graphene} and the subsequent interest in two-dimensional materials, the study of excitons has taken on greater importance. In 2D materials, the reduced dimensionality results in less screening of the Coulomb interaction~\cite{cudazzo2011,chernikov2014}, leading to strongly bound excitons that manifest as in-gap states, producing significant signatures in the optical conductivity~\cite{galvani2016, zhang2022,wang2018,wu2015,ridolfi2018}. Note that excitonic effects are not limited to the linear conductivity; they also affect non-linear properties. Several studies have reported an enhancement of non-linear optical effects when considering excitons, for instance in second harmonic generation~\cite{trolle2014,ruan2023excitonic} or the shift current~\cite{yang-hao2021,huang2023,esteveparedes2024}.

Lastly, it is worth mentioning that in interacting systems the study of excitons can also be extended to composite states formed by a higher number of particles, such as trions~\cite{zhang2014absorption, berkelbach2013}, which involve two excited electrons and one hole (or vice versa), biexcitons~\cite{Ozfidan2015, torche2019}, consisting on two bound electron-hole pairs, as well as higher multi-particle states. While all these are less commonly observed, they can still be detected in photoluminescence experiments~\cite{vaquero2020excitons, hao2017neutral, ye2018efficient} and give a unique response that cannot be understood as the sum of the constituents. Thus, they contribute to a richer understanding of the optical behavior of materials, but similar to higher-order optical effects, their impact tends to be smaller in magnitude.

In conclusion, excitons are fundamental to understand the optical properties of materials, particularly in the context of two-dimensional materials where their role becomes critical. The study of excitons is a field of research that has been active for several decades, where on the theory side multiple theoretical approaches and computational tools have been developed~\cite{quintela2022}. With the discovery of 2D materials and their potential for optoelectronics, for instance with the manipulation of the atomic structure via strain~\cite{roldan2015strain, esteve2019, quintela2025straintime} or the ability to select electronic excitations based on the light polarization~\cite{Mueller2018}, the field has experienced a renewed impulse, with numerous experimental groups reporting their optical properties~\cite{vaquero2020excitons, sotome2021terahertz,tonndorf2015}.

In this thesis, we focus solely on the description of excitons and not so much on their impact of the optical properties of solids. In Chapter 2, we review briefly the different existing methods to compute excitonic states, and introduce a new method to determine the exciton spectrum based on an exact diagonalization (or equivalently configuration interaction) approach. We also develop the more standard approach to excitons using many-body perturbation theory, to show the similarities and differences between them. Next, in Chapter 3, we show how this new method was implemented in the Xatu code, describing the algorithms and the structure of the code, and showcasing its application to two prototypical 2D materials, hBN and MoS$_2$. For these materials we obtain results of comparable accuracy to those of ab-initio tools, but at a fraction of the computational cost. Finally, in Chapter 4 we use the code to explore the role that the topological edge states of a topological insulator play in the presence of excitons. In this case, we show that when considering the transition rates (via Fermi's golden rule) between bulk excitons and edge states in a ribbon, it is possible to generate a photovoltaic current if the appropriate symmetries are broken. Additionally, we provide and estimate of the magnitude of the resulting photocurrent, which we predict to be in the $\mu A$ range, showing the potential of topological insulators for photovoltaics.

\chapter{Mathematical description of excitons}

There are multiple ways to model excitons, depending on the degree of sophistication and accuracy desired. Historically, the first approach was developed by Wannier~\cite{wannier1937}, based on the idea that the exciton is a system analogue to a hydrogen atom, this is, one electron electrostatically bound to a positive charge, the hole. One can then write down the Schrödinger equation for the exciton, which is a two-body problem, and solve it. Specifically, the Schrödinger equation can be written in terms of the center-of-mass and relative coordinates of the exciton, and assuming that the exciton is given by $\phi(\mathbf{r}_e,\mathbf{r}_h)=\phi_e(\mathbf{r}_e)\phi_h(\mathbf{r}_h)\psi(\mathbf{r}_e-\mathbf{r}_h)$ it is possible to perform a separation of variables, finally arriving at:
\begin{equation}
\left[ -\frac{\hbar^2}{2\mu} \nabla^2 + V(\mathbf{r}) \right] \psi(\mathbf{r}) = E \psi(\mathbf{r}),
\end{equation}
which is known as the Wannier equation\index{Wannier!equation}. Here, $\mu$ is the reduced mass of the electron-hole pair (where the masses are obtained in the effective mass approximation~\cite{kittel2021introduction}), $V(\mathbf{r})$ is the Coulomb interaction, and $\mathbf{r}=\mathbf{r}_e-\mathbf{r}_h$ the relative coordinate. Thus, the Wannier equation describes the bound part or relative motion of the exciton, while the center-of-mass motion ($\phi_e(\mathbf{r}_e)$, $\phi_h(\mathbf{r}_h)$) is trivially solved by a plane wave. At this point, the Wannier equation can be solved directly in real space or in reciprocal space after a Fourier transformation~\cite{quintela2022, berkelbach2013, ninhos2024}. An alternative approach to solve it is the variational method, in which one proposes a trial wavefunction and minimizes the energy expectation value $\braket{H}$ with respect to the parameters of the wavefunction~\cite{prada2015}. 

While the Wannier equation succeeds in describing correctly the absorption by bound states, it presents multiple shortcomings. First, the Wannier model overlooks completely the microscopic details of the system, namely the atoms and orbitals that form the basis, and consequently the band structure. Likewise, the Wannier Hamiltonian has spherical symmetry, while in reality the symmetry of the exciton will be restricted to at most that of the crystal. This, together with the effects coming the band structure and the Berry curvature of the Bloch states, results in exciton energy spectrums that are non-hydrogenic~\cite{srivastava2015} and could possibly have an effect on the exciton wavefunctions themselves, for instance on the topological properties~\cite{allocca2018, maisel2023}. Also, the Wannier equation can only describe bound excitons and not the exciton continuum, which is responsible for a notable part of the absorption spectrum.

\begin{wrapfigure}{r}{0.35\textwidth}
    \begin{center}
        \begin{tikzpicture}

            \draw[->] (-2,-1.5) -- (-2,3) node[above] {E};
            
            \draw[thick] plot[domain=-1.5:1.5, samples=100] (\x, {-0.5*\x*\x});
            \node at (-1, -1.2) {VB};
            
            \draw[thick] plot[domain=-1.5:1.5, samples=100] (\x, {0.5*\x*\x + 1.5});
            \node at (-1, 2.6) {CB};
                   
            \foreach \y in {0.75, 1.125, 1.3125, 1.40625, 1.453125, 1.4765625} {
                \draw[dashed] (-0.5,\y) -- (0.5,\y);
            }
            
            \fill[gray,opacity=0.5] (-0.5,1.5) rectangle (0.5,3);
    
            \draw[decorate,decoration={brace,amplitude=10pt,mirror,raise=4pt},yshift=0pt] (0.6,0.75) -- (0.6,1.4765625) node[midway,xshift=15pt,right,rotate=-45] {Bound exc.};
            
            \draw[decorate,decoration={brace,amplitude=10pt,mirror,raise=4pt},yshift=0pt] (0.6,1.5) -- (0.6,3) node[midway,xshift=15pt,right,rotate=-45] {Continuum exc.};
            \end{tikzpicture}
    \end{center}
    \captionsetup{format=plain,singlelinecheck=false}
    \caption[Exciton spectrum diagram relative to bands]{Spectral diagram showing the usual energies for bound excitons (in-gap states) and continuum excitons (beyond the gap).}
\end{wrapfigure}
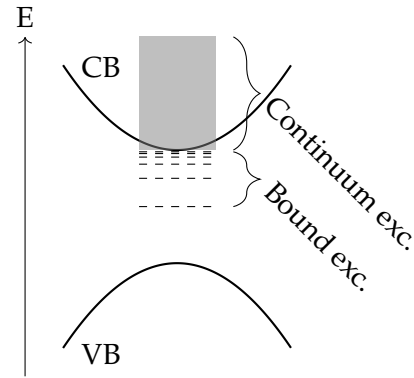

All these shortcomings can be circumvented with a many-body description of the exciton, which ultimately leads to the Bethe-Salpeter equation (BSE). The BSE has its origins in high-energy physics, for the description of bound fermion-antifermion states~\cite{bse_original}. Eventually, the equation found its way into many-body perturbation theory (i.e.\ non-relativistic quantum field theory), where it is used for bound electron-hole pairs in solids, namely excitons~\cite{Strinati_book, CSANAK1971287}. This MBPT approach, when combined with an ab-initio description of the single-particle states (also typically in MBPT, e.g. GW), leads to accurate predictions of the exciton spectrum, matching experimental absorption spectra~\cite{Hanke1980,Rohlfing1998,sottile2003,qiu2013,chang2000}. This is known as GW-BSE and is considered the state-of-the-art method to compute excitons in solids; its accuracy has prompted the development of multiple software packages~\cite{berkeleygw,yambo,Perfetto_2018,exciting}. Being a fully ab-initio calculation however, there is a high computational cost involved, making it unfeasible for large systems.

A middle ground is provided by an exact diagonalization, or in chemistry terms, configuration interaction approach~\cite{franceschetti1999,dvorak2019,bieniek2022}. In this method, one proposes a many-body basis (electron-hole pairs) in which the exciton states are expanded, and diagonalizes the interacting Hamiltonian in such finite basis. Ultimately, this simple approach leads to the same BSE\@. Although it presents some limitations, such as the lack of screening, it describes to a good extent the exciton spectrum on a qualitative, and sometimes quantitative level, and is less expensive computationally (precisely because of those limitations). Finally, it is worth mentioning that it is also possible to identify exciton resonances from time dynamics, either in DFT (TDDFT)~\cite{Onida2002}, Green's functions~\cite{attaccalite2011,Ridolfi2020} or with the density matrix~\cite{Cistaro2023}. 

In this chapter, we formally develop the theory of excitons with exact diagonalization. Using the tight-binding approximation for the orbitals, we are able to simplify the problem further, obtaining expressions that can be quickly evaluated. Next, we derive the BSE through many-body perturbation theory, highlighting the main differences between the two approaches. Specifically, we address the screening of the Coulomb interaction, showing how it can be computed in the tight-binding approximation, which can then be used as an extension of the exact diagonalization approach as it lacks screening. Finally, having established the general form of an excitonic state, we present various ways to characterize it, namely the real and reciprocal space probability densities, the spin polarization, and the optical absorption. We also discuss the symmetry transformations of excitons. \clearpage

\section{Exact diagonalization approach to excitons}

\subsection{The Bethe-Salpeter equation}
We begin by presenting the exact diagonalization approach to excitons. Exact diagonalization methods are widely used in condensed matter physics to extract the ground or excited states of strongly correlated system such as the Hubbard model, magnetic or fractional systems. The central idea is that the many-body Hamiltonian can be represented in a finite basis where it can be diagonalized directly. Thus, knowing that excitons are bound electron-hole pairs, we can expand the exciton states in a basis of electron-hole pairs and diagonalize the interacting Hamiltonian in this basis.
From a quantum chemistry perspective, for the description of excitons we consider the exact, non-relativistic electronic Hamiltonian of the solid of interest,
\begin{equation}\label{Hamiltonian}
     H = H_0 + V =\sum_{i,j}t_{ij}c^{\dagger}_{i}c_{j} + \frac{1}{2}\sum_{i,j,k,l}V_{ijkl}c^{\dagger}_ic^{\dagger}_jc_lc_k,
\end{equation}
where the indices include orbital and position degrees of freedom; we restrict to basis of localized orbitals. $H_0$ describes the kinetic and ion-electron interaction terms and $V$ is the electrostatic interaction between electrons. Diagonalization of $H_0$ yields a Bloch eigenbasis $\ket{n\mathbf{k}}$ with energies $\varepsilon_{n\mathbf{k}}$, which here will correspond to insulating or semiconducting materials. $n$ denotes the band quantum number, and $\mathbf{k}$ is the crystal momentum. The interaction term in~\eqref{Hamiltonian} contains
\begin{align}
    \label{interaction_matrix_element}
    V_{ijkl}&=\braket{i,j|V|k,l} 
    =\int d\mathbf{r} d\mathbf{r}' \varphi^*_i(\mathbf{r})\varphi^*_j(\mathbf{r}')V(\mathbf{r},\mathbf{r}')\varphi_k(\mathbf{r})\varphi_l(\mathbf{r}')
\end{align}
where $V(\mathbf{r},\mathbf{r}')$ is the two-body interaction. This can be the bare Coulomb interaction or some alternative interaction to take into account dimensionality or screening.
Since the non-interacting Hamiltonian $H_0$ describes insulating materials, it is usually a good approximation to take the ground state for the interacting Hamiltonian $H$ as the Fermi sea:
\begin{equation} \label{FermiSea}
    \ket{GS} = \prod_{n,\mathbf{k}}^{\varepsilon_{n\mathbf{k}}\leq \varepsilon_F}c^{\dagger}_{n\mathbf{k}}\ket{0}
\end{equation}
where $\ket{0}$ denotes the state with zero electrons, and $\varepsilon_F$ is the Fermi energy. Then, an electron-hole pair of center-of-mass momentum $\mathbf{Q}$ between the conduction band $c$ and the valence band $v$, and located at momentum $\mathbf{k}$ is defined as
\begin{equation}
    \ket{v,c,\mathbf{k},\mathbf{Q}} = c^{\dagger}_{c\mathbf{k} + \mathbf{Q}}c_{v\mathbf{k}}\ket{GS}
\end{equation}
meaning that one electron of momentum $\mathbf{k}$ from the valence bands is promoted to the conduction bands with momentum $\mathbf{k}+\mathbf{Q}$. Note that even though we denote these states as electron-hole pairs, we are not actually using hole quasiparticle operators, but simply refer to the hole as the absence of an electron in the Fermi sea. We will stick to the electron picture throughout this thesis, unless specified otherwise. These electron-hole pairs will serve as the basis for the exciton states, $\ket{X_n(\mathbf{Q})}$:
\begin{align}
     \nonumber\ket{X_n(\mathbf{Q})} &= \sum_{v,c,\mathbf{k}}A^n_{vc}(\mathbf{k},\mathbf{Q})\ket{v,c,\mathbf{k},\mathbf{Q}}
     = \sum_{v,c,\mathbf{k}}A^n_{vc}(\mathbf{k},\mathbf{Q})c^{\dagger}_{c\mathbf{k} + \mathbf{Q}}c_{v\mathbf{k}}\ket{GS}
\end{align}
Therefore, the exciton is expressed as a linear combination of electron-hole pairs over different bands and momenta. Note that $\mathbf{Q}$ serves as a good quantum number for the exciton states, since the interaction is momentum-conserving. The interaction only mixes electron-hole pairs with the same net momentum, which is $\mathbf{Q}$. This can be seen by computing explicitly a general interaction matrix element, $V_{ijkl}$. The quantum number $n$ here is used simply to denote different exciton states.
Next, we determine the $A^n_{vc}(\mathbf{k}, \mathbf{Q})$ coefficients that minimize the expectation value $\braket{X_n(\mathbf{Q})|H|X_n(\mathbf{Q})}$:
\begin{equation}
    \frac{\delta E[X]}{\delta X} = \frac{\delta}{\delta X}\left[\frac{\braket{X_n(\mathbf{Q})|H|X_n(\mathbf{Q})}}{\braket{X_n(\mathbf{Q})|X_n(\mathbf{Q})}}\right] = 0
\end{equation}
Performing this derivative explicitly (in practice in terms of the coefficients $A^n_{vc}(\mathbf{k}, \mathbf{Q})$) is equivalent to the problem of diagonalizing the Hamiltonian represented in the basis of electron-hole pairs,
\begin{equation}
\label{eigenproblem}
    \sum_{v',c',\mathbf{k}'}H_{vc,v'c'}(\mathbf{k}, \mathbf{k}', \mathbf{Q})A^n_{v'c'}(\mathbf{k}',\mathbf{Q}) = E^n_XA^n_{vc}(\mathbf{k}, \mathbf{Q})
\end{equation}
where $H_{vc,v'c'}(\mathbf{k}, \mathbf{k}', \mathbf{Q})=\braket{v,c,\mathbf{k},\mathbf{Q}|H|v',c',\mathbf{k}',\mathbf{Q}}$.
The expansion in electron-hole pairs of the exciton is actually an ansatz: we obtain exact eigenstates of the Hamiltonian restricted to a partition of the Hilbert space, $PHP$, where $P$ is a projector over the single electron-hole pairs,
\begin{equation}
    PHP = \sum_{\substack{v,c,\mathbf{k}\\ v',c',\mathbf{k}'}}H_{vc,v'c'}(\mathbf{k}, \mathbf{k}', \mathbf{Q})\ket{v',c',\mathbf{k}',\mathbf{Q}}\bra{v,c,\mathbf{k},\mathbf{Q}}
\end{equation}
In fact, if we only consider charge-conserving excitations, we could represent the Hamiltonian in the following way:
\begin{equation}
H = \bigoplus^{N_e}_{n=0} P_nHP_n + C, \ \text{with } 
\end{equation}
where 
\begin{align}
\nonumber &P_n=\sum_{
\substack{\{c_i\},\{v_i\}\\ \{c'_i\},\{v'_i\}}}\ket{\{c_i\}, \{v_i\}}\bra{\{c'_i\}, \{v'_i\}}\ 
\quad\text{and}\quad\ket{\{c_i\}, \{v_i\}} = \prod^n_{i=1}c^{\dagger}_{c_i}\prod^n_{i=1}c_{v_i}\ket{GS}
\end{align}
$N_e$ is the total number of electrons, $C$ the coupling between the different excitation sectors, and $P_n$ is the projector over the n-th electron-hole pairs sector.
If instead of using the Bloch states from $H_0$ we formulate the problem in terms of the Hartree-Fock (HF) solution to~\eqref{Hamiltonian}, then the coupling between the Fermi sea and the single-pair sector, $P_0HP_1$, is exactly zero according to Brillouin's theorem~\cite{szabo1996modern, gross1986many}. As we will mention later, we will assume that this always holds even when the ground state has not been calculated in the HF approximation. 
The same, however, is not true for $P_0HP_2$ or $P_1HP_2$, i.e., the interaction couples the ground state and the one electron-hole pair sector with the two electron-hole pairs sector. Thus, the proposed ground state and the exciton states are never exact but approximate eigenstates. Given that the material is insulating, we expect the coupling to be weak due to the energy differences, which justifies the ansatz. Keeping with the exact diagonalization approach, one could try to diagonalize the Hamiltonian including more excitation sectors. Although possible in principle, it becomes quickly unfeasible since the Hilbert space in many-body systems grows exponentially (constituting the main limitation of the method) and, in this case, the eigenstates would involve a mixture of excitations, losing the interpretation as a bound electron-hole pair. 

Going back to~\eqref{eigenproblem}, we compute next the Hamiltonian matrix elements in the $H_0$ basis, which are given in terms of the single particle energies and the interaction matrix elements:
\begin{align}
    \label{h_braketement}
    \nonumber 
    &H_{vc,v'c'}(\mathbf{k},\mathbf{k}', \mathbf{Q}) = \\ \nonumber&\delta_{\mathbf{k}\mathbf{k}'}\delta_{vv'}\big[\varepsilon_{c\mathbf{k}+\mathbf{Q}}\delta_{cc'}+\Sigma_{cc'}(\mathbf{k}+\mathbf{Q},\mathbf{k}'+\mathbf{Q})\big]\\
 &-\delta_{\mathbf{k}\mathbf{k}'}\delta_{cc'}\big[\varepsilon_{v\mathbf{k}}\delta_{vv'}+\Sigma_{v'v}(\mathbf{k}',\mathbf{k})\big] -(D - X)_{vcv'c'}(\mathbf{k},\mathbf{k}',\mathbf{Q})
\end{align}
where
\begin{equation}
   \begin{split}
    \label{kernel_D_X}
    D_{vc,v'c'}(\mathbf{k},\mathbf{k}', \mathbf{Q}) &= V_{c\mathbf{k}+\mathbf{Q},v'\mathbf{k}',c'\mathbf{k}'+\mathbf{Q},v\mathbf{k}} \\
     X_{vc,v'c'}(\mathbf{k},\mathbf{k}', \mathbf{Q}) & = V_{c\mathbf{k}+\mathbf{Q},v'\mathbf{k}',v\mathbf{k},c'\mathbf{k}'+\mathbf{Q}}
\end{split} 
\end{equation}
and
\begin{equation}
\begin{split}
\Sigma_{nm}(\mathbf{k},\mathbf{k}')=\sum_{j,\mathbf{k}''}^{\text{occ}}\big( V_{n \mathbf{k}, j\mathbf{k}'', m \mathbf{k}', j\mathbf{k}''} - V_{n\mathbf{k},j \mathbf{k}'', j\mathbf{k}'',m\mathbf{k}'}\big)
\end{split}
\end{equation}
$D$, $X$ correspond to the direct and exchange interactions between the electron-hole pair, whereas $\Sigma$ is the self-energy coming from the interaction of the electron/hole with the Fermi sea (hence the sum restricted to occupied states). At this point we could obtain the exciton spectrum diagonalizing (\ref{h_braketement}). Instead, it is more convenient to solve first for the ground-state of (\ref{Hamiltonian}) at the mean-field level, i.e.\ in the HF approximation~\cite{isil2014}. If we now write (\ref{h_braketement}) as an eigenvalue problem in the HF band basis, we obtain
\begin{align}
\label{bse}
     (\varepsilon_{c\mathbf{k}+\mathbf{Q}} - \varepsilon_{v\mathbf{k}})A^n_{vc}(\mathbf{k}, \mathbf{Q}) + \sum_{v',c',\mathbf{k}'}K_{vc,v'c'}(\mathbf{k},\mathbf{k}',&\mathbf{Q})A^n_{v'c'}(\mathbf{k}',\mathbf{Q})
    = E^n_XA^n_{vc}(\mathbf{k}, \mathbf{Q})
\end{align}
where $\varepsilon_{n\mathbf{k}}$ are now the HF quasiparticle energies, and $K = - (D-X)$ is the interaction kernel. Thus, the self-energies are now incorporated into the quasiparticle energies instead. Note that the Fermi sea energy has been set to zero, so that exciton energies can be compared directly with the gap of the system. This is the standard form of the Bethe-Salpeter equation for excitons using the Tamm-Dancoff approximation (TDA)~\cite{hirata1999, dreuw2005}, and it defines the starting point for any exciton calculation. As we will see in the next section, the main difference with MBPT comes from the interaction kernel, which there involves a dynamically screened interaction, usually in the random-phase approximation~\cite{hybertsen1985, adler1962, wiser1963}. The determination of the dielectric constant is a computationally intensive task~\cite{berkeleygw}, which we will avoid setting instead an effective static screening. The usage of an effective screening is one of the main reasons that makes this approach considerable faster than the ab-initio tools.

So far we have seen that it is more convenient to pose the exciton problem in terms of the HF basis, as it simplifies the problem and allows to decouple excitation sectors. 
In practice, we do not address the problem of determining the mean-field solution to~\eqref{Hamiltonian}. Instead, we start directly from equation~\eqref{bse} assuming that the initial band structure, which is already known, verifies it. Namely, for tight-binding band structures we drop the self-energy terms assuming that we are using a HF solution. Alternatively, if the band structure comes from DFT or MBPT (e.g.\ GW approximation), then we also remove the self-energy terms since the quasiparticle energies already include self-energy corrections (although they do not cancel exactly with those from~\eqref{h_braketement}). Thus, from now on we regard the starting band structure as the non-interacting Hamiltonian $H_0$.

\subsection{Interaction matrix elements}\label{sec:interaction_matrix_elements}
With Eq.~\eqref{bse} established, a practical expression for the interaction matrix elements~\eqref{interaction_matrix_element} remains to be obtained. The single-particle states, using a basis of localized orbitals, can be written as
\begin{equation}
\label{single-particle-state}
    \varphi_{n\mathbf{k}}(\mathbf{r}) = \frac{1}{\sqrt{N}}\displaystyle\sum_{\mathbf{R}}e^{i\mathbf{k}\cdot\mathbf{R}}\sum_{i,\alpha}C^{n\mathbf{k}}_{i\alpha}\phi_{\alpha}(\mathbf{r}-\mathbf{R} - \mathbf{t}_i)
\end{equation}
where $\{\phi_{\alpha}\}_{\alpha}$ denote the orbitals located at the atom $i$ of the motif and $N$ is the number of unit cells of the system. As mentioned before, this wavefunction may correspond to that of a tight-binding model (meaning that the spatial nature of the orbitals is ignored and are typically considered orthonormal), or a DFT calculation with a local orbital basis set, which are in general non-orthogonal. We write the Bloch states in the lattice gauge, and all expressions will be derived following this convention. Regarding the global phase freedom of the states, we set $\sum_{i\alpha}C^{n\mathbf{k}}_{i\alpha}\in\mathbb{R}$~\cite{Rohlfing2000, garcia-blazquez2023}. While the origin of the single-particle states can be different, for the actual calculation of the interactions we will treat them on the same footing, approximating them as point-like orthonormal orbitals.
Depending on how we treat the interaction, different working expressions for the matrix elements can be obtained. For instance, we address first the direct term, which is given by
\begin{align}
     &D_{vc,v'c'}(\mathbf{k},\mathbf{k}', \mathbf{Q})
    = \int d\mathbf{r}d\mathbf{r}'\varphi^*_{c\mathbf{k}+\mathbf{Q}}(\mathbf{r})\varphi^*_{v'\mathbf{k'}}(\mathbf{r}')V(\mathbf{r},\mathbf{r}')\varphi_{c'\mathbf{k'}+\mathbf{Q}}(\mathbf{r})\varphi_{v\mathbf{k}}(\mathbf{r}')
\label{integral}\end{align}
We substitute the single-particle Bloch states~\eqref{single-particle-state} in Eq.~\eqref{integral}. Expanding each term, we end up having to evaluate the same four-body integral, but now between the orbitals that compose each state,
\begin{equation}
\label{fourbody}
    \int d\mathbf{r} d\mathbf{r}' \phi^*_{\alpha}(\mathbf{r})\phi^*_{\beta}(\mathbf{r}')V(\mathbf{r},\mathbf{r}')\phi_{\gamma}(\mathbf{r})\phi_{\delta}(\mathbf{r}')
\end{equation}
At this point, there are two ways to compute the present four-body integral: we can evaluate directly the interaction in real space, or, instead, use its Fourier series to work in reciprocal space. In both cases we consider point-like orbitals centered at $\mathbf{R}+\mathbf{t}_i$ and orthogonality between the orbitals: 
\begin{equation}
\label{eq:approximation}
    \phi_{\alpha}(\mathbf{r} - \mathbf{R} - \mathbf{t}_i)\phi_{\beta}(\mathbf{r} - \mathbf{R'} - \mathbf{t}_j) \approx \delta_{\alpha\beta} \delta(\mathbf{r} - \mathbf{R} - \mathbf{t}_i)\delta_{ij}\delta_{\textbf{R},\textbf{R}'}
\end{equation}
Integrating in real space, after simplifying the resulting deltas, we obtain the following expression for the direct term $D$:
\begin{align}
\label{direct}
     &D_{vc,v'c'}(\mathbf{k}, \mathbf{k}', \mathbf{Q})
     = \frac{1}{N}\sum_{ij}\sum_{\alpha\beta}(C_{i\alpha}^{c\mathbf{k}+\mathbf{Q}})^*(C_{j\beta}^{v'\mathbf{k}'})^*C_{i\alpha}^{c'\mathbf{k}'+\mathbf{Q}}C_{j\beta}^{v\mathbf{k}}V_{ij}(\mathbf{k}'-\mathbf{k})
\end{align}
where
\begin{equation}
    V_{ij}(\mathbf{k}'-\mathbf{k}) = \sum_{\mathbf{R}}e^{i(\mathbf{k}'-\mathbf{k})\mathbf{R}}V(\mathbf{R} - (\mathbf{t}_j - \mathbf{t}_i)).
\end{equation}
Here $V_{ij}(\mathbf{k}'-\mathbf{k})$ can be regarded as a lattice Fourier transform centered at $\mathbf{t}_j - \mathbf{t}_i$. Since it is defined as a sum over lattice vectors and not an integral, one cannot use the shift property from the Fourier transform. Attempting to do so would result in breaking the spatial symmetries of the Hamiltonian. Then, the direct term can be interpreted as the weighted average of the Fourier transform of the interaction between the electron and the hole, over all positions and orbitals. The exchange term is computed analogously:
\begin{align}
\label{exchange}
    \nonumber X_{vc,v'c'}(\mathbf{k}, \mathbf{k}', \mathbf{Q}) &= 
    \int d\mathbf{r}d\mathbf{r}'\varphi^*_{c\mathbf{k}+\mathbf{Q}}(\mathbf{r})\varphi^*_{v'\mathbf{k'}}(\mathbf{r}')V(\mathbf{r},\mathbf{r}')\varphi_{v\mathbf{k}}(\mathbf{r})\varphi_{c'\mathbf{k'}+\mathbf{Q}}(\mathbf{r}') \\
    &= \frac{1}{N}\sum_{ij}\sum_{\alpha\beta}(C_{i\alpha}^{c\mathbf{k}+\mathbf{Q}})^*(C_{j\beta}^{v'\mathbf{k}'})^*C_{i\alpha}^{v\mathbf{k}}C_{j\beta}^{c'\mathbf{k}'+\mathbf{Q}}V_{ij}(\mathbf{Q})
\end{align}
In case that there is only one atom in the motif, then expressions~\eqref{direct},~\eqref{exchange} simplify even further since the interaction decouples from the tight-binding coefficients, yielding
\begin{equation}
\begin{split}
    &D_{vc,v'c'}(\mathbf{k}, \mathbf{k}', \mathbf{Q}) = 
    \frac{1}{N}V(\mathbf{k}'-\mathbf{k})(U^{\dagger}_{\mathbf{k}+\mathbf{Q}}U_{\mathbf{k}'+\mathbf{Q}})_{cc'}(U^{\dagger}_{\mathbf{k}}U_{\mathbf{k}'})_{v'v}\\
    &X_{vc,v'c'}(\mathbf{k}, \mathbf{k}', \mathbf{Q}) = 
    \frac{1}{N}V(\mathbf{Q})(U^{\dagger}_{\mathbf{k}+\mathbf{Q}}U_{\mathbf{k}})_{cv}(U^{\dagger}_{\mathbf{k}'}U_{\mathbf{k}'+\mathbf{Q}})_{v'c'}
\end{split}
\end{equation}
where $U_{\mathbf{k}}$ is the unitary matrix that diagonalizes the Bloch Hamiltonian $H(\mathbf{k})$~\cite{wu2015}. The evaluation of these expressions is much faster than the corresponding ones~\eqref{direct} and~\eqref{exchange} for a general case. Additionally, for $\mathbf{Q}=0$, the exchange term~\eqref{exchange} becomes exactly zero, which is not true in general, although it is usually neglected. As mentioned before, for DFT band structures we evaluate the interaction using the same point-like approximation, performing first a Löwdin orthogonalization of the basis~\cite{szabo1996modern}. This allows to improve the TB descriptions, incorporating fine details to the quasiparticle dispersion along the BZ\@. In such treatments, our interaction matrix elements are an approximation to the true ones involving ab-initio orbitals. Given that in DFT the orbitals are known (e.g.\ Gaussian-type basis in the CRYSTAL~\cite{crystal17} code), one could, in principle, evaluate the integrals (\ref{fourbody}) exactly for a closer ab-initio calculation of excitons as in~\cite{garcia-blazquez2024}.
  
The previous calculation corresponds to the evaluation of the interaction matrix elements
in real space. An alternative approach consists of writing the interaction as its Fourier series before evaluating~\eqref{fourbody}~\cite{Cistaro2023, Ridolfi2020}:
\begin{equation}
    V(\mathbf{r}-\mathbf{r'})=\frac{1}{N}\sum_{\mathbf{q}}V(\mathbf{q})e^{i\mathbf{q}\cdot(\mathbf{r}-\mathbf{r'})}
\end{equation}
where
\begin{equation}
    V(\mathbf{q}) = \frac{1}{V_{\text{cell}}}\int_{\Omega}V(\mathbf{r})e^{-i\mathbf{q}\cdot\mathbf{r}}d\mathbf{r}
\end{equation}
$\Omega=NV_{\text{cell}}$ denotes the volume of the crystal, and the Fourier series is done according to the dimensionality of the problem (e.g.\ in 2D, both $\mathbf{q},\mathbf{r}\in\mathbb{R}^2$). Usually, one takes $\Omega\to\infty$ meaning that we can evaluate the integral analytically, this is, $V(\mathbf{q})$ becomes the Fourier transform of the potential. Note, however, that $\mathbf{q}$ is not restricted to the first Brillouin Zone (BZ), and $V(\mathbf{q})$ is not periodic in the BZ\@. Therefore, in principle one has to sum over $\mathbf{q}\in\text{BZ}$, but also over reciprocal vectors $\mathbf{G}$, i.e.
\begin{equation}
    V(\mathbf{r}-\mathbf{r'})=\frac{1}{N}\sum_{\mathbf{q}\in\text{BZ}}\sum_{\mathbf{G}}V(\mathbf{q} + \mathbf{G})e^{i(\mathbf{q} + \mathbf{G})\cdot(\mathbf{r}-\mathbf{r'})}
\end{equation}
The evaluation of the integral is done in the same way, although in this case there is a plane wave instead of the electrostatic interaction. This approach is particularly useful when using a plane wave basis, since it allows to evaluate the four-body integrals exactly without need for approximation~\eqref{eq:approximation}, being the de facto methodology used by the ab-initio codes~\cite{berkeleygw,yambo}. The interaction matrix elements $D$, $X$ are now given by
\begin{equation}
\begin{split}
\label{reciprocal_interaction}
    &D_{vc,v'c'}(\mathbf{k}, \mathbf{k}', \mathbf{Q}) = \frac{1}{N}\sum_{\mathbf{G}}V(\mathbf{k}-\mathbf{k}' + \mathbf{G})I^{\mathbf{G}}_{c\mathbf{k}+\mathbf{Q},c'\mathbf{k}'+\mathbf{Q}}(I^{\mathbf{G}}_{v\mathbf{k},v'\mathbf{k}'})^* \\
     &X_{vc,v'c'}(\mathbf{k}, \mathbf{k}', \mathbf{Q}) = \frac{1}{N}\sum_{\mathbf{G}}V(\mathbf{Q} + \mathbf{G})I^{\mathbf{G}}_{c\mathbf{k} + \mathbf{Q},v\mathbf{k}}(I^{\mathbf{G}}_{c'\mathbf{k'} + \mathbf{Q},v'\mathbf{k}'})^*
\end{split}
\end{equation}
where
\begin{equation}
    I^{\mathbf{G}}_{n\mathbf{k},m\mathbf{k}'} = \sum_{i\alpha}(C_{i\alpha}^{n\mathbf{k}})^*C_{i\alpha}^{m\mathbf{k'}}e^{i(\mathbf{k}-\mathbf{k}'+\mathbf{G})\cdot\mathbf{t}_i}
\end{equation}
Usually $V(\mathbf{q})$ decays fast enough, so it suffices to sum only over $\mathbf{G}=\mathbf{0}$ for the excitons to converge in energy. As we will see in Chapter~\ref{chapter:chapter3}, our developed code \texttt{Xatu} allows to use the interactions evaluated in real-space (expressions~\eqref{direct},~\eqref{exchange}), or in reciprocal space (expressions~\eqref{reciprocal_interaction}). They are benchmarked in section~\ref{sec:xatu_validation}.

\section{Many-body perturbation theory approach to excitons}
\subsection{The particle-hole Green's function}

The Bethe-Salpeter equation was originally devised in quantum field theory to describe composite bound states made of fermions, in our case for electron-hole pairs. There are multiple ways to derive it; we will mainly follow the derivation from~\cite{Rohlfing2000}, which is also based on~\cite{CSANAK1971287,Strinati_book,strinati1984effects,strinati1982dynamical}. Since it involves two particles, we must consider the two-particle interacting Green's function,
\begin{equation}
    G(12;1'2') = (-i)^2\braket{\Psi|T[\psi(1)\psi(2)\psi^{\dagger}(1')\psi^{\dagger}(2')]|\Psi}\footnote{Throughout this section we set $\hbar=1$ for simplicity.}
\end{equation}
where $1=(\mathbf{x}_1,t_1)$, $\ket{\Psi}$ denotes the ground state of the interacting theory, $T$ is the time-ordering operator and $\psi,\psi^{\dagger}$ are the field operators (see~\eqref{eq:field_operators} for their definition). As with the one-particle Green's function, one can write the time-ordered expectation value in terms of specific time-orderings, each one with a step function. There are two specific time orderings we are interested in for the description of excitons: $1,1' > 2, 2'$ and $2, 2' > 1,1'$~\cite{Strinati_book, CSANAK1971287}. These orderings correspond to the creation of one electron and one hole, while the rest describe two-electron or two-hole processes. Also, since excitons usually appear as optical excitations, we restrict ourselves to simultaneous creation of the electron-hole pair and also simultaneous annihilation, i.e. $t_1=t_{1'}$ and $t_2=t_{2'}$\footnote{Actually, one must define the times as $t_{1'}=t_1 + \delta$, $t_{2'}=t_2+\delta$, with $\delta\to0$ for the Green's functions to be well-defined. This leads to more cumbersome expressions, so for simplicity we work directly with $\delta=0$. In the situations where there is some ambiguity, for instance in the evaluation of the $T$ operator, the underlying $\delta$ is used to solve it. We refer to~\cite{CSANAK1971287} for the details.}~\cite{Rohlfing2000}. The Green's function can be written as
\begin{equation}
    G(12;1'2') = G_1(12;1'2')\theta(t_1+t_{1'} - t_2 - t_{2'}) + G_2(12;1'2')\theta(t_2+t_{2'} - t_1 - t_{1'}) + \text{other terms}
\end{equation}
Since we are interested only in particle-hole pairs, we can define a particle-hole Green's function $G^{\text{ph}}$ which takes into account exclusively the terms described above. This propagator is defined as
\begin{equation}
    G^{\text{ph}}(12;1'2') = G_1(12;1'2')\theta(t_1- t_2) + G_2(12;1'2')\theta(t_2 - t_1)
\end{equation}
where we have already imposed the time simultaneity of the creation and annihilation of particles. Now let's examine each term individually. As one does to obtain the Lehmann representation of the one-particle interacting Green's function, we can assume that there exists a set of eigenstates of the Hamiltonian with $N$ particles $\ket{\Psi^N_n}$. Inserting a completeness relation in the first Green's function ($t_1>t_2$) we get
\begin{align}
    G_1(12;1'2') = -\sum_n \chi_n(1,1')\chi_n^*(2,2')
\end{align}
where $\chi_n(1,1') =\braket{\Psi|T[\psi(1)^{\dagger}\psi(1')]|\Psi_n^N}$ are the particle-hole (or Bethe-Salpeter) amplitudes. In the interaction picture, the time dependence of the field operators is given by $\psi(1)=e^{iHt_1}\psi(\mathbf{x}_1)e^{-iHt_1}$. With this, the particle-hole amplitudes are rewritten as
\begin{align}
    \chi_n(1,1') = -e^{i(E - E_n)t_1} \braket{\Psi|\psi^{\dagger}(\mathbf{x}_{1'})\psi(\mathbf{x}_1)|\Psi_n^N}
\end{align}
Therefore, the Green's function for $1,1' > 2,2'$ reads
\begin{equation}
    G_1(12;1'2') = -\sum_n e^{i(E-E_n)(t_1-t_2)}\chi_n(\mathbf{x}_1,\mathbf{x}_{1'})\chi^*_n(\mathbf{x}_{2'},\mathbf{x}_{2})
\end{equation}
where now we have defined the spatial particle-hole amplitudes $\chi_n(\mathbf{x}_1,\mathbf{x}_{1'}) = \braket{\Psi|\psi^{\dagger}(\mathbf{x}_{1'})\psi(\mathbf{x}_1)|\Psi_n^N}$. Proceeding in the same way, we obtain an expression of the Green's function for times $2,2'>1,1'$:
\begin{align}
     G_2(12;1'2') = -\sum_n e^{-i(E-E_n)(t_1-t_2)} \chi_n(\mathbf{x}_2,\mathbf{x}_{2'})\chi_n^*(\mathbf{x}_{1'},\mathbf{x}_{1})
\end{align}
At this point we can perform the Fourier transform to frequency domain for the particle-hole Green's function on the variable $t\equiv t_1-t_2$. Adding the convergence factors, we get:
\begin{align}
    \nonumber G^{\text{ph}}(\mathbf{x}_1\mathbf{x}_2,\mathbf{x}_{1'}\mathbf{x}_{2'};\omega) &= \int_{-\infty}^{\infty}dt\ e^{i\omega t}G^{ph}(12;1'2') \\  
    & = -i\sum_n\left[\frac{\chi_n(\mathbf{x}_1,\mathbf{x}_{1'})\chi^*_n(\mathbf{x}_{2'},\mathbf{x}_{2})}{\omega + E - E_n + i\eta} - \frac{\chi_n(\mathbf{x}_2,\mathbf{x}_{2'})\chi_n^*(\mathbf{x}_{1'},\mathbf{x}_{1})}{\omega - E + E_n - i\eta}\right]
\end{align}
This expression can be regarded as the Lehmann representation of the two-particle Green's function, whose poles denote the excitation energies of the system, namely the exciton energies $\Omega_n=E_n-E$. 

\subsection{The Bethe-Salpeter equation revisited}\label{sec:mbpt-bse}

At this point, before actually deriving the Bethe-Salpeter equation, we are going to examine further the two-particle propagator, but from a diagrammatic perspective. Expanding the particle-hole Green's function in its perturbation series in powers of $V$, the zeroth and first order diagrams are the following:

\begin{figure}[h]
\centering
\begin{tikzpicture}
  \begin{feynman}[medium]
    \vertex (f1);
    \vertex [right=0.5cm of f1](a);
    \vertex [right=of a] (b);
    \vertex [right=0.5cm of b] (i1);
    
    \vertex [below=of f1] (i2);
    \vertex [right=0.5cm of i2](c);
    \vertex [right=of c] (d);
    \vertex [right=0.5cm of d] (f2);
    
    \vertex [below=0.1cm of f1] (l1) {\(1\)};
    \vertex [below=0.1cm of i1] (l2) {\(2'\)};
    \vertex [below=0.1cm of f2] (l1p) {\(2\)};
    \vertex [below=0.1cm of i2] (l2p) {\(1'\)};

    \diagram* {
      (i1) -- [fermion] (b) -- [horizontal] (a) -- [fermion] (f1);
      (i2) -- [fermion] (c) -- [horizontal] (d) -- [fermion] (f2);
      (a) -- [vertical] (c);
      (b) -- [vertical] (d);
    };

    \node[] at (1.25,-0.75) {\(G^{\text{ph}}\)};
  \end{feynman}
\end{tikzpicture}
 \raisebox{1.25cm}{=}
\begin{tikzpicture}
  \begin{feynman}[medium]
    \vertex (i1);
    \vertex [right=of i1](f1);
    \vertex [below=of i1](f2);
    \vertex [right=of f2](i2);
    
    \vertex [below=0.1cm of i1] (l1) {\(1\)};
    \vertex [below=0.1cm of f1] (l2) {\(2'\)};
    \vertex [below=0.1cm of f2] (l1) {\(1'\)};
    \vertex [below=0.1cm of i2] (l2) {\(2\)};

    \diagram* {
      (f1) -- [fermion] (i1);
      (f2) -- [fermion] (i2);
    };
  \end{feynman}
\end{tikzpicture}
 \raisebox{1.25cm}{+}
\begin{tikzpicture}
  \begin{feynman}[medium]
   \vertex (i1);
    \vertex [right=of i1](f1);
    \vertex [below=of i1](f2);
    \vertex [right=of f2](i2);
    
    \vertex [right=0.1cm of i1] (l1) {\(1\)};
    \vertex [left=0.1cm of f1] (l2) {\(2'\)};
    \vertex [below=0.1cm of f2] (l1) {\(1'\)};
    \vertex [below=0.1cm of i2] (l2) {\(2\)};

    \diagram* {
      (f1) -- [fermion] (i2);
      (f2) -- [fermion] (i1);
    };
  \end{feynman}
\end{tikzpicture}
 \raisebox{1.25cm}{+}
\begin{tikzpicture}
  \begin{feynman}[medium]
    \vertex (i1);
    \vertex [right=0.5cm of i1, dot](a) {};
    \vertex [right=of a, dot](b) {};
    \vertex [right=0.5cm of b](f1);
    \vertex [below=of i1](f2);
    \vertex [below=of f1](i2);
    
    \vertex [below=0.1cm of i1] (l1) {\(1\)};
    \vertex [below=0.1cm of f1] (l2) {\(2'\)};
    \vertex [below=0.1cm of f2] (l1) {\(1'\)};
    \vertex [below=0.1cm of i2] (l2) {\(2\)};

    \diagram* {
      (f1) -- [fermion] (b) -- [fermion] (a) -- [fermion](i1);
      (a) -- [boson, out=-90, in=-90] (b);
      (f2) -- [fermion] (i2);
    };
  \end{feynman}
\end{tikzpicture}
 \raisebox{1.25cm}{+}
\begin{tikzpicture}
  \begin{feynman}[medium]
    \vertex (i1);
    \vertex [right=of i1](f1);
    \vertex [below=0.75cm of f1, dot](a) {};
    \vertex [right=1cm of a, dot](b) {};
    \vertex [below=of i1](f2);
    \vertex [right=of f2](i2);
    
    \vertex [right=0.1cm of i1] (l1) {\(1\)};
    \vertex [left=0.1cm of f1] (l2) {\(2'\)};
    \vertex [below=0.1cm of f2] (l1) {\(1'\)};
    \vertex [below=0.1cm of i2] (l2) {\(2\)};

    \diagram* {
      (f2) -- [fermion] (i1);
      (f1) -- [fermion] (a) -- [fermion] (i2);
      (a) -- [photon] (b);
    };
     \draw[decoration={markings, mark=at position 0.5 with {\arrow[scale=1.5]{>}}}, postaction={decorate}] (b) arc [start angle=-180, end angle=180, radius=0.4cm];
  \end{feynman}
\end{tikzpicture}
\raisebox{1.25cm}{+}
\begin{tikzpicture}
\begin{feynman}[medium]
    \vertex (i1);
    \vertex [right=1cm of i1, dot](a) {} ;
    \vertex [right=1cm of a](f1);
    \vertex [below=of i1](f2);
    \vertex [right=1cm of f2, dot](b) {};
    \vertex [right=1cm of b](i2);
    
    \vertex [below=0.1cm of i1] (l1) {\(1\)};
    \vertex [below=0.1cm of f1] (l2) {\(2'\)};
    \vertex [below=0.1cm of f2] (l1) {\(1'\)};
    \vertex [below=0.1cm of i2] (l2) {\(2\)};

    \diagram* {
      (f1) -- [fermion] (a) -- [fermion] (i1);
      (f2) -- [fermion] (b) -- [fermion] (i2);
      (a) -- [boson] (b);
    };
  \end{feynman}
\end{tikzpicture}
\raisebox{1.25cm}{+}
\begin{tikzpicture}
  \begin{feynman}[medium]
   \vertex (i1);
   \vertex [dot](a) at (0.5, -0.75) {};
   \vertex [right=of a, dot] (b) {};
    \vertex [below=of i1](f2);
    \vertex [right=2.5cm of i1](f1);
    \vertex[below=of f1](i2);
    
    \vertex [right=0.1cm of i1] (l1) {\(1\)};
    \vertex [left=0.1cm of f1] (l2) {\(2'\)};
    \vertex [below=0.1cm of f2] (l1) {\(1'\)};
    \vertex [below=0.1cm of i2] (l2) {\(2\)};

    \diagram* {
      (f2) -- [fermion] (a) -- [fermion] (i1);
      (f1) -- [fermion] (b) -- [fermion] (i2);
      (a) -- [boson] (b);
    };
  \end{feynman}
\end{tikzpicture}
\raisebox{1.25cm}{$+\quad\dotsc$}
\caption[Zero and first order contributions to the particle-hole Green's function $G^{\text{ph}}$]{Zero and first order contributions to the particle-hole Green's function $G^{\text{ph}}$. We show only some Hartree and Fock diagrams to illustrate the diagrammatic expansion.}
\end{figure}
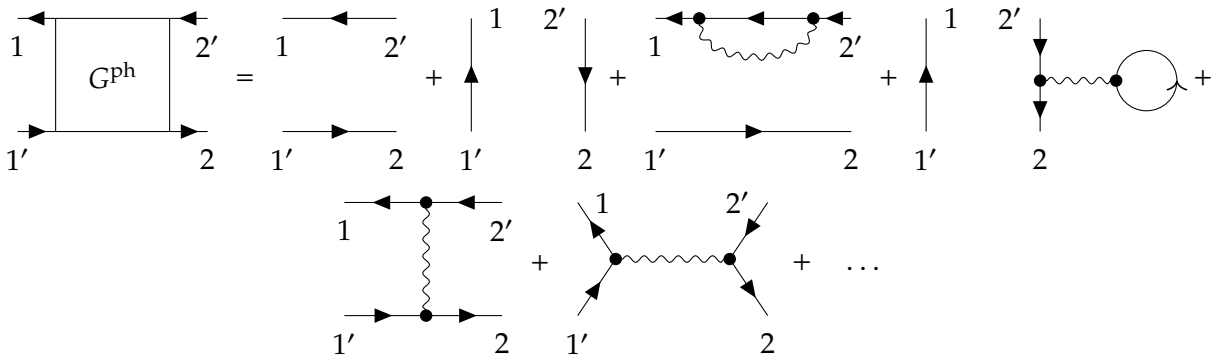

From the first diagrams we quickly observe that the interaction lines do not necessarily connect the non-interacting propagators $G_0$. Thus, if we keep adding terms to the expansion, the two-particle Green's function can be written in terms of two disconnected diagrams, which correspond to a product of interacting single-particle Green's functions plus all the remaining connected diagrams.

\begin{figure}[h]
\centering
\begin{tikzpicture}
  \begin{feynman}[medium]
    \vertex (f1);
    \vertex [right=0.5cm of f1](a);
    \vertex [right=of a] (b);
    \vertex [right=0.5cm of b] (i1);
    
    \vertex [below=of f1] (i2);
    \vertex [right=0.5cm of i2](c);
    \vertex [right=of c] (d);
    \vertex [right=0.5cm of d] (f2);
    
    \vertex [below=0.1cm of f1] (l1) {\(1\)};
    \vertex [below=0.1cm of i1] (l2) {\(2'\)};
    \vertex [below=0.1cm of f2] (l1p) {\(2\)};
    \vertex [below=0.1cm of i2] (l2p) {\(1'\)};

    \diagram* {
      (i1) -- [fermion] (b) -- [horizontal] (a) -- [fermion] (f1);
      (i2) -- [fermion] (c) -- [horizontal] (d) -- [fermion] (f2);
      (a) -- [vertical] (c);
      (b) -- [vertical] (d);
    };

    \node[] at (1.25,-0.75) {\(G^{\text{ph}}\)};
  \end{feynman}
\end{tikzpicture}
 \raisebox{1.25cm}{=}
\begin{tikzpicture}
  \begin{feynman}[medium]
    \vertex (i1);
    \vertex [right=of i1](f1);
    \vertex [below=of i1](f2);
    \vertex [right=of f2](i2);
    
    \vertex [below=0.1cm of i1] (l1) {\(1\)};
    \vertex [below=0.1cm of f1] (l2) {\(2'\)};
    \vertex [below=0.1cm of f2] (l1) {\(1'\)};
    \vertex [below=0.1cm of i2] (l2) {\(2\)};

    \diagram* {
      (f1) -- [double,double distance=0.3ex,thick,with arrow=0.5,arrow size=0.2em] (i1);
      (f2) -- [double,double distance=0.3ex,thick,with arrow=0.5,arrow size=0.2em] (i2);
    };
  \end{feynman}
\end{tikzpicture}
 \raisebox{1.25cm}{+}
\begin{tikzpicture}
  \begin{feynman}[medium]
   \vertex (i1);
    \vertex [right=of i1](f1);
    \vertex [below=of i1](f2);
    \vertex [right=of f2](i2);
    
    \vertex [right=0.1cm of i1] (l1) {\(1\)};
    \vertex [left=0.1cm of f1] (l2) {\(2'\)};
    \vertex [below=0.1cm of f2] (l1) {\(1'\)};
    \vertex [below=0.1cm of i2] (l2) {\(2\)};

    \diagram* {
      (f1) -- [double,double distance=0.3ex,thick,with arrow=0.5,arrow size=0.2em] (i2);
      (f2) -- [double,double distance=0.3ex,thick,with arrow=0.5,arrow size=0.2em] (i1);
    };
  \end{feynman}
\end{tikzpicture}
 \raisebox{1.25cm}{+}
\begin{tikzpicture}
  \begin{feynman}[medium]
    \vertex (f1);
    \vertex [right=0.5cm of f1](a);
    \vertex [right=of a] (b);
    \vertex [right=0.5cm of b] (i1);
    
    \vertex [below=of f1] (i2);
    \vertex [right=0.5cm of i2](c);
    \vertex [right=of c] (d);
    \vertex [right=0.5cm of d] (f2);
    
    \vertex [below=0.1cm of f1] (l1) {\(1\)};
    \vertex [below=0.1cm of i1] (l2) {\(2'\)};
    \vertex [below=0.1cm of f2] (l1p) {\(2\)};
    \vertex [below=0.1cm of i2] (l2p) {\(1'\)};

    \diagram* {
      (i1) -- [fermion] (b) -- [horizontal] (a) -- [fermion] (f1);
      (i2) -- [fermion] (c) -- [horizontal] (d) -- [fermion] (f2);
      (a) -- [vertical] (c);
      (b) -- [vertical] (d);
    };

    \node[] at (1.25,-0.75) {\(\delta G^{\text{ph}}\)};
  \end{feynman}
\end{tikzpicture}
\caption[Decomposition of the two-particle Green's function into disconnected and connected diagrams]{The two-particle Green's function can be written in terms of two disconnected diagrams, corresponding to the propagation of two independent self-interacting particles, plus all the connected diagrams $\delta G^{\text{ph}}$, i.e.\ all diagrams where the two particles interact. The double solid line denotes the interacting single-particle Green's function $G$.}\label{fig:disconnected_green}
\end{figure}
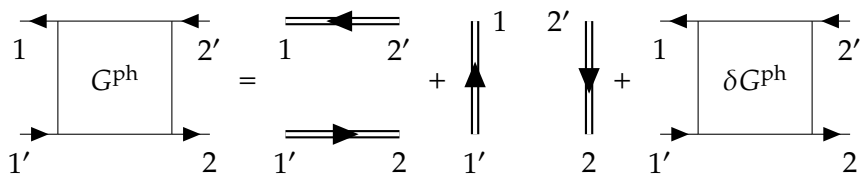
\noindent Mathematically, the above diagrams read
\begin{equation}
G^{\text{ph}}(12;1'2') = G(1,1')G(2,2') - G(1,2')G(2,1') + \delta G^{\text{ph}}(12;1'2')
\end{equation}
The excitons will be contained within the connected part or bound part of the two-particle Green's function, since it describes the interaction between two dressed particles. At this point, we can further expand the connected diagrams  $\delta G^{\text{ph}}$ in terms of their irreducible interactions, in the same way one defines the irreducible self-energy of the one-particle Green's function for the Dyson equation. Therefore, defining $\Xi$ as the irreducible interaction diagrams we can obtain a Dyson equation for the bound part of the two-particle Green's function, $\delta G^{\text{ph}}$.
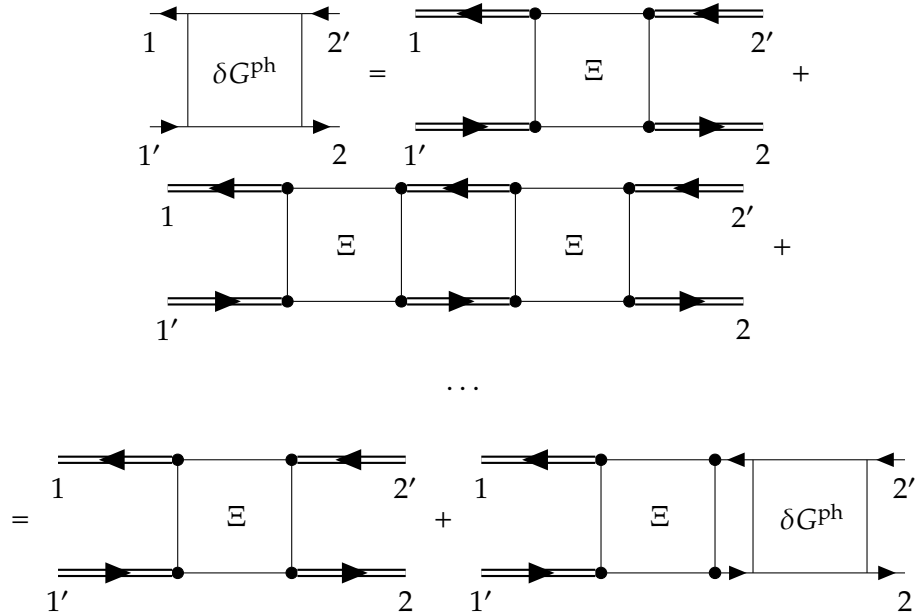
\begin{figure}[h]
\centering
\begin{tikzpicture}
  \begin{feynman}[medium]
    \vertex (f1);
    \vertex [right=0.5cm of f1](a);
    \vertex [right=of a] (b);
    \vertex [right=0.5cm of b] (i1);
    
    \vertex [below=of f1] (i2);
    \vertex [right=0.5cm of i2](c);
    \vertex [right=of c] (d);
    \vertex [right=0.5cm of d] (f2);
    
    \vertex [below=0.1cm of f1] (l1) {\(1\)};
    \vertex [below=0.1cm of i1] (l2) {\(2'\)};
    \vertex [below=0.1cm of f2] (l1p) {\(2\)};
    \vertex [below=0.1cm of i2] (l2p) {\(1'\)};

    \diagram* {
      (i1) -- [fermion] (b) -- [horizontal] (a) -- [fermion] (f1);
      (i2) -- [fermion] (c) -- [horizontal] (d) -- [fermion] (f2);
      (a) -- [vertical] (c);
      (b) -- [vertical] (d);
    };

    \node[] at (1.25,-0.75) {\(\delta G^{\text{ph}}\)};
  \end{feynman}
\end{tikzpicture}
 \raisebox{1.25cm}{=}
\begin{tikzpicture}
  \begin{feynman}[medium]
    \vertex (i1);
    \vertex [right=of i1, dot](f1) {};
    \vertex [below=of i1](f2);
    \vertex [right=of f2, dot](i2) {};

    \vertex [right=of f1, dot](a) {};
    \vertex [right=of a](b);
    \vertex [right=of i2, dot](c) {};
    \vertex [right=of c](d);
    
    \vertex [below=0.1cm of i1] (l1) {\(1\)};
    \vertex [below=0.1cm of b] (l2) {\(2'\)};
    \vertex [below=0.1cm of f2] (l1) {\(1'\)};
    \vertex [below=0.1cm of d] (l2) {\(2\)};

    \diagram* {
      (f1) -- [double,double distance=0.3ex,thick,with arrow=0.5,arrow size=0.2em] (i1);
      (f2) -- [double,double distance=0.3ex,thick,with arrow=0.5,arrow size=0.2em]  (i2);
      (f1) -- [vertical] (i2);
      (a) -- [vertical] (c);
      (f1) -- [horizontal] (a);
      (b) -- [double,double distance=0.3ex,thick,with arrow=0.5,arrow size=0.2em]  (a);
      (i2) -- [horizontal] (c) -- [double,double distance=0.3ex,thick,with arrow=0.5,arrow size=0.2em]  (d);
    };

	\node[] at (2.35,-0.75) {\(\Xi\)};
  \end{feynman}
\end{tikzpicture}
 \raisebox{1.25cm}{+}
\begin{tikzpicture}
   \begin{feynman}[medium]
    \vertex (i1);
    \vertex [right=of i1, dot](f1) {};
    \vertex [below=of i1](f2);
    \vertex [right=of f2, dot](i2) {};

    \vertex [right=of f1, dot](a) {};
    \vertex [right=of a, dot](b) {};
    \vertex [right=of b, dot](a1) {};
    \vertex [right=of a1](b1);

    \vertex [right=of i2, dot](c) {};
    \vertex [right=of c, dot](d) {};
    \vertex [right=of d, dot](c1) {};
    \vertex [right=of c1](d1);
    
    \vertex [below=0.1cm of i1] (l1) {\(1\)};
    \vertex [below=0.1cm of b1] (l2) {\(2'\)};
    \vertex [below=0.1cm of f2] (l1) {\(1'\)};
    \vertex [below=0.1cm of d1] (l2) {\(2\)};

    \diagram* {
      (f1) -- [double,double distance=0.3ex,thick,with arrow=0.5,arrow size=0.2em] (i1);
      (f2) -- [double,double distance=0.3ex,thick,with arrow=0.5,arrow size=0.2em]  (i2);
      (f1) -- [vertical] (i2);
      (a) -- [vertical] (c);
      (f1) -- [horizontal] (a);
      (b) -- [double,double distance=0.3ex,thick,with arrow=0.5,arrow size=0.2em]  (a);
      (i2) -- [horizontal] (c) -- [double,double distance=0.3ex,thick,with arrow=0.5,arrow size=0.2em]  (d);
      (b) -- [vertical] (d);
      (a1) -- [vertical] (c1);
      (b) -- [horizontal] (a1);
      (b1) -- [double,double distance=0.3ex,thick,with arrow=0.5,arrow size=0.2em]  (a1);
      (d) -- [horizontal] (c1) -- [double,double distance=0.3ex,thick,with arrow=0.5,arrow size=0.2em]  (d1);
    };

	\node[] at (2.35,-0.75) {\(\Xi\)};
	\node[] at (5.35,-0.75) {\(\Xi\)};
  \end{feynman}
\end{tikzpicture}
\raisebox{1.25cm}{$+$}
\\
\raisebox{0.75cm}{$\dotsc$}
\raisebox{1.25cm}{}
\\
 \raisebox{1.25cm}{=}
\begin{tikzpicture}
  \begin{feynman}[medium]
    \vertex (i1);
    \vertex [right=of i1, dot](f1) {};
    \vertex [below=of i1](f2);
    \vertex [right=of f2, dot](i2) {};

    \vertex [right=of f1, dot](a) {};
    \vertex [right=of a](b);
    \vertex [right=of i2, dot](c) {};
    \vertex [right=of c](d);
    
    \vertex [below=0.1cm of i1] (l1) {\(1\)};
    \vertex [below=0.1cm of b] (l2) {\(2'\)};
    \vertex [below=0.1cm of f2] (l1) {\(1'\)};
    \vertex [below=0.1cm of d] (l2) {\(2\)};

    \diagram* {
      (f1) -- [double,double distance=0.3ex,thick,with arrow=0.5,arrow size=0.2em] (i1);
      (f2) -- [double,double distance=0.3ex,thick,with arrow=0.5,arrow size=0.2em]  (i2);
      (f1) -- [vertical] (i2);
      (a) -- [vertical] (c);
      (f1) -- [horizontal] (a);
      (b) -- [double,double distance=0.3ex,thick,with arrow=0.5,arrow size=0.2em]  (a);
      (i2) -- [horizontal] (c) -- [double,double distance=0.3ex,thick,with arrow=0.5,arrow size=0.2em]  (d);
    };

	\node[] at (2.35,-0.75) {\(\Xi\)};
  \end{feynman}
\end{tikzpicture}
 \raisebox{1.25cm}{+}
\begin{tikzpicture}
   \begin{feynman}[medium]
    \vertex (i1);
    \vertex [right=of i1, dot](f1) {};
    \vertex [below=of i1](f2);
    \vertex [right=of f2, dot](i2) {};

    \vertex [right=of f1, dot](a) {};
    \vertex [right=0.5cm of a, dot](b);
    \vertex [right=of b, dot](a1);
    \vertex [right=0.5cm of a1](b1);

    \vertex [right=of i2, dot](c) {};
    \vertex [right=0.5cm of c, dot](d);
    \vertex [right=of d, dot](c1);
    \vertex [right=0.5cm of c1](d1);
    
    \vertex [below=0.1cm of i1] (l1) {\(1\)};
    \vertex [below=0.1cm of b1] (l2) {\(2'\)};
    \vertex [below=0.1cm of f2] (l1) {\(1'\)};
    \vertex [below=0.1cm of d1] (l2) {\(2\)};

    \diagram* {
      (f1) -- [double,double distance=0.3ex,thick,with arrow=0.5,arrow size=0.2em] (i1);
      (f2) -- [double,double distance=0.3ex,thick,with arrow=0.5,arrow size=0.2em]  (i2);
      (f1) -- [vertical] (i2);
      (a) -- [vertical] (c);
      (f1) -- [horizontal] (a);
      (b) -- [fermion]  (a);
      (i2) -- [horizontal] (c) -- [fermion]  (d);
      (b) -- [vertical] (d);
      (a1) -- [vertical] (c1);
      (b) -- [horizontal] (a1);
      (b1) -- [fermion] (a1);
      (d) -- [horizontal] (c1) -- [fermion]  (d1);
    };

	\node[] at (2.35,-0.75) {\(\Xi\)};
	\node[] at (4.35,-0.75) {\(\delta G^{\text{ph}}\)};
  \end{feynman}
\end{tikzpicture}
\caption[Dyson equation for the bound part of the two-particle Green's function, $\delta G^{\text{ph}}$]{Dyson equation for the bound part of the two-particle Green's function, $\delta G^{\text{ph}}$. Note that the expansion with irreducible interactions can be regarded as a ladder approximation.}
\end{figure}
Mathematically, the equation for $\delta G^{\text{ph}}$ is given by:
\begin{align}\label{bse_deltaG}
\nonumber\delta G^{\text{ph}}(12;1'2') &= \int d3456 G(1,3)G(4,1')\Xi(3,5;4,6)G(6,2')G(2,5) \\ 
&+  \int d3456 G(1,3)G(4,1')\Xi(3,5;4,6)\delta G^{\text{ph}}(62;52')
\end{align}
Equation~\eqref{bse_deltaG} is already regarded as a Bethe-Salpeter equation for the bound part of $G^{\text{ph}}$. For the description of excitons, however, it is more convenient to obtain a BSE in a different form. First, one defines the following electron-hole propagator $L$ from the two-particle Green's function,
\begin{equation}
L(12;1'2') := -G^{\text{ph}}(12;1'2') + G(1,1')G(2,2')
\end{equation}
From the previous separation of $G^{\text{ph}}$ into disconnected and connected parts, as depicted in Fig.~\ref{fig:disconnected_green}, we already see that we are effectively removing one disconnected diagram from the Green's function. An intuitive explanation is that that diagram does not contribute to the description of excitons, which are encoded in the bound part of the two-particle propagator. This can be seen explicitly evaluating the disconnected diagram $G(1,1')G(2,2')$. Fourier transforming it, we obtain
\begin{equation}
  \text{FT}[G(1,1')G(2,2')](\omega) = -i\left[\frac{\chi_0(\mathbf{x}_1,\mathbf{x}_{1'})\chi^*_0(\mathbf{x}_{2'},\mathbf{x}_{2})}{\omega + i\eta} - \frac{\chi_0(\mathbf{x}_2,\mathbf{x}_{2'})\chi^*_0(\mathbf{x}_{1'},\mathbf{x}_{1})}{\omega - i\eta}\right] 
\end{equation}
This cancels the first ($n=0$) term in $G^{\text{ph}}$, which corresponds to the ground state, and consequently $L$ strictly describes the neutral excitations of the system
\begin{align}\label{eq:L_expr}
   L(\mathbf{x}_1\mathbf{x}_2,\mathbf{x}_{1'}\mathbf{x}_{2'};\omega) = i\sum_{n\neq 0}\left[\frac{\chi_n(\mathbf{x}_1,\mathbf{x}_{1'})\chi^*_n(\mathbf{x}_{2'},\mathbf{x}_{2})}{\omega - \Omega_n + i\eta} - \frac{\chi_n(\mathbf{x}_2,\mathbf{x}_{2'})\chi_n^*(\mathbf{x}_{1'},\mathbf{x}_{1})}{\omega + \Omega_n - i\eta}\right]
\end{align}

With this definition, we can start reasoning diagrammatically like we did for $\delta G^{\text{ph}}$ to obtain a Dyson (or more accurately a Bethe-Salpeter) equation for the effective propagator $L$. If we substitute the corresponding diagrams on the definition of $L$, we see that we can write the infinite ladder expansion as a recursive equation:
\begin{equation}\label{bse_l}
L(12;1'2') = G(1,2')G(2,1') + \int d3456 G(1,3)G(4,1')\Xi(3,5;4,6)L(62;52')
\end{equation}
Equation~\eqref{bse_l} defines the Bethe-Salpeter equation for $L$, which is the most common one for the description of excitons. There exist additional forms of the Bethe-Salpeter equation, for instance in terms of the original two-particle Green's function $G^{\text{ph}}$, although we will not derive it here. So far, we have been able to write down recursive equations for the particle-hole propagators in terms of the irreducible interactions, also commonly named interaction kernel.

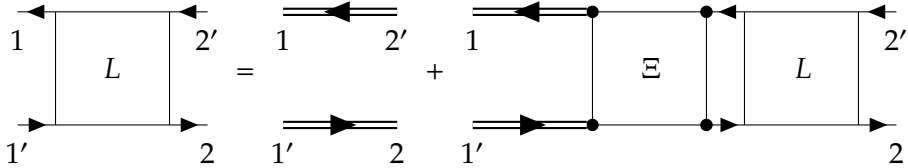
\begin{figure}[h]
\centering
\begin{tikzpicture}
  \begin{feynman}[medium]
    \vertex (f1);
    \vertex [right=0.5cm of f1](a);
    \vertex [right=of a] (b);
    \vertex [right=0.5cm of b] (i1);
    
    \vertex [below=of f1] (i2);
    \vertex [right=0.5cm of i2](c);
    \vertex [right=of c] (d);
    \vertex [right=0.5cm of d] (f2);
    
    \vertex [below=0.1cm of f1] (l1) {\(1\)};
    \vertex [below=0.1cm of i1] (l2) {\(2'\)};
    \vertex [below=0.1cm of f2] (l1p) {\(2\)};
    \vertex [below=0.1cm of i2] (l2p) {\(1'\)};

    \diagram* {
      (i1) -- [fermion] (b) -- [horizontal] (a) -- [fermion] (f1);
      (i2) -- [fermion] (c) -- [horizontal] (d) -- [fermion] (f2);
      (a) -- [vertical] (c);
      (b) -- [vertical] (d);
    };

    \node[] at (1.25,-0.75) {\(L\)};
  \end{feynman}
\end{tikzpicture}
 \raisebox{1.25cm}{=}
\begin{tikzpicture}
  \begin{feynman}[medium]
    \vertex (i1);
    \vertex [right=of i1](f1);
    \vertex [below=of i1](f2);
    \vertex [right=of f2](i2);
    
    \vertex [below=0.1cm of i1] (l1) {\(1\)};
    \vertex [below=0.1cm of f1] (l2) {\(2'\)};
    \vertex [below=0.1cm of f2] (l1) {\(1'\)};
    \vertex [below=0.1cm of i2] (l2) {\(2\)};

    \diagram* {
      (f1) -- [double,double distance=0.3ex,thick,with arrow=0.5,arrow size=0.2em] (i1);
      (f2) -- [double,double distance=0.3ex,thick,with arrow=0.5,arrow size=0.2em] (i2);
    };
  \end{feynman}
\end{tikzpicture}
  \raisebox{1.25cm}{+}
\begin{tikzpicture}
  \begin{feynman}[medium]
   \vertex (i1);
   \vertex [right=of i1, dot](f1) {};
   \vertex [below=of i1](f2);
   \vertex [right=of f2, dot](i2) {};

   \vertex [right=of f1, dot](a) {};
   \vertex [right=0.5cm of a, dot](b);
   \vertex [right=of b, dot](a1);
   \vertex [right=0.5cm of a1](b1);

   \vertex [right=of i2, dot](c) {};
   \vertex [right=0.5cm of c, dot](d);
   \vertex [right=of d, dot](c1);
   \vertex [right=0.5cm of c1](d1);
   
   \vertex [below=0.1cm of i1] (l1) {\(1\)};
   \vertex [below=0.1cm of b1] (l2) {\(2'\)};
   \vertex [below=0.1cm of f2] (l1) {\(1'\)};
   \vertex [below=0.1cm of d1] (l2) {\(2\)};

   \diagram* {
     (f1) -- [double,double distance=0.3ex,thick,with arrow=0.5,arrow size=0.2em] (i1);
     (f2) -- [double,double distance=0.3ex,thick,with arrow=0.5,arrow size=0.2em]  (i2);
     (f1) -- [vertical] (i2);
     (a) -- [vertical] (c);
     (f1) -- [horizontal] (a);
     (b) -- [fermion]  (a);
     (i2) -- [horizontal] (c) -- [fermion]  (d);
     (b) -- [vertical] (d);
     (a1) -- [vertical] (c1);
     (b) -- [horizontal] (a1);
     (b1) -- [fermion] (a1);
     (d) -- [horizontal] (c1) -- [fermion]  (d1);
   };

 \node[] at (2.35,-0.75) {\(\Xi\)};
 \node[] at (4.35,-0.75) {\(L\)};
 \end{feynman}
\end{tikzpicture}
\caption{Diagrammatic representation of the Bethe-Salpeter equation for $L$}
\end{figure}
\begin{figure}[h]
  \centering
  \begin{tikzpicture}
    \begin{feynman}
      
    \vertex (a);
    \vertex [below=1.5cm of a] (b);
    \vertex [right=1.5cm of a, dot] (bi1){};
    \vertex [right=1.5cm of b, dot] (bf1){};
    \vertex [left=0.04cm of bi1](aux1);
    \vertex [right=0.04cm of bi1](aux2);
    \vertex [left=0.04cm of bf1](aux3);
    \vertex [right=0.04cm of bf1](aux4);

    \vertex [xshift=1.5cm, yshift=-0.75cm, dot] (bi2) at (bi1){};
    \vertex [right=1.5cm of bi2, dot] (bf2){};
    \vertex [right=1cm of bf2, dot] (bi3){};
    \vertex [right=1.5cm of bi3, dot] (bf3){};

    \vertex [xshift=1.5cm, yshift=0.75cm] (c) at (bf3);
    \vertex [xshift=1.5cm, yshift=-0.75cm] (d) at (bf3);

    \vertex [left=0.1cm of a] (l1) {\(1\)};
    \vertex [left=0.1cm of b] (l2) {\(1'\)};
    \vertex [right=0.1cm of d] (l1p) {\(2\)};
    \vertex [right=0.1cm of c] (l2p) {\(2'\)};

    \diagram*{
      (bi1) -- [double,double distance=0.3ex,thick,with arrow=0.5,arrow size=0.2em] (a);
      (b) -- [double,double distance=0.3ex,thick,with arrow=0.5,arrow size=0.2em] (bf1);
      (aux1) -- [boson] (aux3);
      (aux2) -- [boson] (aux4);
      (bi2) -- [double,double distance=0.3ex,thick,with arrow=0.5,arrow size=0.2em] (bi1);
      (bf1) -- [double,double distance=0.3ex,thick,with arrow=0.5,arrow size=0.2em] (bi2);
      (bi2) -- [boson] (bf2);
      (bi3) -- [double,double distance=0.3ex,thick,with arrow=0.5,arrow size=0.2em, out=-270, in=90] (bf2);
      (bf2) -- [double,double distance=0.3ex,thick,with arrow=0.5,arrow size=0.2em, out=270, in=-90] (bi3);
      (bi3) -- [boson] (bf3);
      (c) -- [double,double distance=0.3ex,thick,with arrow=0.5,arrow size=0.2em] (bf3);
      (bf3) -- [double,double distance=0.3ex,thick,with arrow=0.5,arrow size=0.2em] (d);

    };
    \end{feynman}
  \end{tikzpicture}
  \caption[Higher order diagram in $L$]{ Example of a higher order diagram in $L$. If we keep adding loops in the exchange interaction following the ladder approximation, we are effectively screening the electrostatic interaction in the RPA, as we will see in the next section.}\label{fig:screened_exchange_bse}
\end{figure}
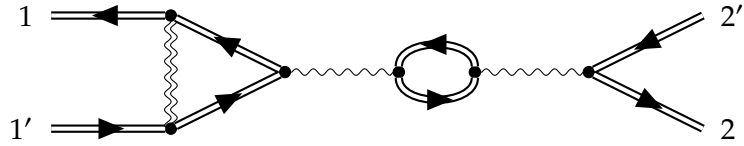

While we have done a diagrammatic derivation of the BSE, it is possible to derive it formally with Schwinger's approach of functional derivatives; in this case from the generalized equation of motion of the single-particle Green's function in presence of an external field $U$, $L=\frac{\delta G}{\delta U}$~\cite{Strinati_book}. In general, all of Hedin's equations can be derived in this way, which constitute a closed set of equations that can be solved iteratively as an alternative to perturbation theory~\cite{hedin1965new}. The same holds for the interaction kernel $\Xi$, which can be defined in terms of a functional derivative of the Hedin's equation for the self-energy, $\Xi=\frac{\delta \Sigma}{\delta G}$. Then, the interaction kernel is typically written in the GW approximation, which implies setting the vertex $\Gamma$ (another of Hedin's equations) to a delta function. The interaction kernel then reads:
\begin{equation}\label{eq:interaction_kernel}
  \Xi(3,5;4,6)=\delta(3,4)\delta(5,6)v_c(3,5) - \delta(3,6)\delta(4,5)W(3,4)
\end{equation}
where $v_c$ is the bare Coulomb interaction, and $W$ denotes the screened Coulomb interaction, which has yet to be determined. From the interaction kernel we see that only the direct interaction between the electron and the hole is screened, while the exchange term is given in terms of the bare or unscreened Coulomb interaction. This can be understood from the ladder approximation that constitutes the BSE\@; the ladder approximation itself is the responsible for generating the diagrams which would correspond to a screened exchange interaction, see for instance the diagram in Fig.~\ref{fig:screened_exchange_bse}.

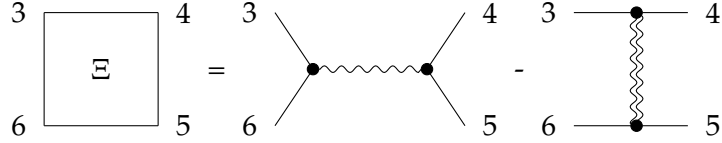
\begin{figure}[h]
  \centering
  \begin{tikzpicture}
    \begin{feynman}[medium]
      \vertex (a);
      \vertex [right=1.5cm of a] (b);
      
      \vertex [below=of a] (c);
      \vertex [right=of c] (d);
      
      \vertex [left=0.1cm of a] (l1) {\(3\)};
      \vertex [right=0.1cm of b] (l2) {\(4\)};
      \vertex [right=0.1cm of d] (l1p) {\(5\)};
      \vertex [left=0.1cm of c] (l2p) {\(6\)};
  
      \diagram* {
        (b) -- [horizontal] (a);
        (c) -- [horizontal] (d);
        (a) -- [vertical] (c);
        (b) -- [vertical] (d);
      };
  
      \node[] at (0.75,-0.75) {\(\Xi\)};
    \end{feynman}
  \end{tikzpicture}
   \raisebox{0.9cm}{=}
   \begin{tikzpicture}
    \begin{feynman}[medium]

    \vertex (a);
    \vertex [dot](i1) at (0.5, -0.75) {};
    \vertex [right=2.5cm of a] (b);
    
    \vertex [below=of a] (c);
    \vertex [right=1.5cm of i1, dot](i2){};
    \vertex [right=2.5cm of c] (d);
    
    \vertex [left=0.1cm of a] (l1) {\(3\)};
    \vertex [right=0.1cm of b] (l2) {\(4\)};
    \vertex [right=0.1cm of d] (l1p) {\(5\)};
    \vertex [left=0.1cm of c] (l2p) {\(6\)};
  
      \diagram* {
        (a) -- [horizontal] (i1) -- [horizontal] (c);
        (b) -- [horizontal] (i2) -- [horizontal] (d);
        (i1) -- [boson] (i2);
      };
    \end{feynman}
  \end{tikzpicture}
  \raisebox{0.9cm}{-}
  \begin{tikzpicture}
    \begin{feynman}[medium]
      \vertex (a);
      \vertex [right=0.75cm of a, dot](i1){};
      \vertex [right=1.5cm of a] (b);
      
      \vertex [below=of a] (c);
      \vertex [right=0.75cm of c, dot](i2){};
      \vertex [right=of c] (d);

      \vertex [left=0.04cm of i1](aux1);
      \vertex [right=0.04cm of i1](aux2);
      \vertex [left=0.04cm of i2](aux3);
      \vertex [right=0.04cm of i2](aux4);
      
      \vertex [left=0.1cm of a] (l1) {\(3\)};
      \vertex [right=0.1cm of b] (l2) {\(4\)};
      \vertex [right=0.1cm of d] (l1p) {\(5\)};
      \vertex [left=0.1cm of c] (l2p) {\(6\)};
    
        \diagram* {
          (a) -- [horizontal] (i1) -- [horizontal] (b);
          (c) -- [horizontal] (i2) -- [horizontal] (d);
          (aux1) -- [boson] (aux3);
          (aux2) -- [boson] (aux4);
          
        };
      \end{feynman}
    \end{tikzpicture}
    \caption[Diagrammatic representation of the interaction kernel in the GW approximation]{ Diagrammatic representation of the interaction kernel in the GW approximation. The first term corresponds to the exchange interaction, while the second one is the direct interaction. The double wiggle line denotes the screened Coulomb interaction $W$.}
\end{figure}

At this point we are ready to obtain the exciton eigenvalue problem from the BSE~\eqref{bse_l}. First, we define the non-interacting electron-hole propagator $L_0$ as
\begin{equation}
  L_0(1,2;1',2') = G(1,2')G(2,1')
\end{equation}
Now assuming that the interacting Green's function $G$ admits a spectral decomposition (or simply taking $G\approx G_0$), we can Fourier transform $L_0$ which yields
\begin{align}\label{eq:L0_expr}
  \nonumber &L_0(\mathbf{x}_1,\mathbf{x}_2,\mathbf{x}_{1'},\mathbf{x}_{2'};\omega) = \\
  &i\sum_{c\mathbf{k},v\mathbf{k}'}\left[\frac{\varphi_{c\mathbf{k}}(\mathbf{x}_1)\varphi^*_{v\mathbf{k}'}(\mathbf{x}_{1'})\varphi_{v\mathbf{k}'}(\mathbf{x}_2)\varphi^*_{c\mathbf{k}}(\mathbf{x}_{2'})}{\omega - (\varepsilon_{c\mathbf{k}} - \varepsilon_{v\mathbf{k}'}) + i\eta} - \frac{\varphi_{v\mathbf{k}'}(\mathbf{x}_1)\varphi^*_{c\mathbf{k}}(\mathbf{x}_{1'})\varphi_{c\mathbf{k}}(\mathbf{x}_2)\varphi^*_{v\mathbf{k}'}(\mathbf{x}_{2'})}{\omega + (\varepsilon_{c\mathbf{k}} - \varepsilon_{v\mathbf{k}'}) - i\eta}\right]
\end{align}
where $\varepsilon_{n\mathbf{k}}$ are the single-particle energies. Before substituting the expression for $L$~\eqref{eq:L_expr} in the BSE, first we need to expand the field operators within the Bethe-Salpeter amplitudes $\chi_n(\mathbf{x}_1,\mathbf{x}_{1'})$. For a detailed derivation of the amplitude, see section~\ref{sec:kwf-rswf}. In general, one obtains
\begin{equation}\label{eq:bse_amplitude_expansion}
  \chi_n(\mathbf{x}_1,\mathbf{x}_{1'}) = \sum_{vc\mathbf{k}}\left[A^n_{vc}(\mathbf{k})\varphi_{c\mathbf{k}}(\mathbf{x}_1)\varphi^*_{v\mathbf{k}}(\mathbf{x}_{1'}) + B^n_{vc}(\mathbf{k})\varphi_{v\mathbf{k}}(\mathbf{x}_1)\varphi^*_{c\mathbf{k}}(\mathbf{x}_{1'})\right]
\end{equation}
The BSE in its discrete form can now be obtained substituting~\eqref{eq:L0_expr},~\eqref{eq:L_expr} and~\eqref{eq:bse_amplitude_expansion} into the~\eqref{bse_l}. Projecting over the single-particle states $\varphi_{n\mathbf{k}}(\mathbf{x})$, one obtains the following eigenvalue problem\footnote{Here we show the eigenvalue problem obtained at $\mathbf{Q}=0$ (e.g.\ in Eq.~\eqref{eq:bse_amplitude_expansion} where the exciton states are also assumed to take $\mathbf{Q}=0$.) See~\cite{garcia-blazquez2024} for the complete treatment with $\mathbf{Q}\neq 0$.}~\cite{blase2020, garcia-blazquez2024}:
\begin{equation}\label{eq:bse_eigenvalue_antihermitian}
  \left(
  \begin{array}{cc}
    R & C \\
    -C^* & -R^*
  \end{array}
  \right)
  \left(
  \begin{array}{c}
    A \\
    B
  \end{array}
  \right)
  = \Omega_n
  \left(
  \begin{array}{c}
    A \\
    B
  \end{array}
  \right)
\end{equation}
$R$ denotes the resonant part of the BSE (creation of electron-hole pairs, $A^n_{vc}(\mathbf{k})$), while $-R^*$ is the antiresonant part (destruction of electron-hole pairs, $B^n_{vc}(\mathbf{k})$). $C$ denotes the matrix coupling the two sectors, and $A$, $B$ are the eigenvectors of the problem that appear in the expansion~\eqref{eq:bse_amplitude_expansion}. The $R$ and $C$ matrices are given by 
\begin{align}
  R_{v\mathbf{k}c\mathbf{k},v'\mathbf{k}'c'\mathbf{k}'} &= \delta_{vv'}\delta_{cc'}\delta_{\mathbf{k}\mathbf{k}'}(\varepsilon_{c\mathbf{k}} - \varepsilon_{v\mathbf{k}}) - (D_{v\mathbf{k}c'\mathbf{k}',v'\mathbf{k}'c\mathbf{k}} - X_{v\mathbf{k}c'\mathbf{k}',c\mathbf{k}v'\mathbf{k}'})\\
  C_{v\mathbf{k}c\mathbf{k},v'\mathbf{k}'c'\mathbf{k}'} &= - (D_{v\mathbf{k}v'\mathbf{k}',c'\mathbf{k}'c\mathbf{k}} - X_{v\mathbf{k}v'\mathbf{k}',c\mathbf{k}c'\mathbf{k}'})
\end{align}
where $D$ and $X$ are the direct and exchange terms of the interaction kernel, respectively. As opposed to the exact diagonalization approach, $D$ takes now a different form owing to the expression for the interaction kernel~\eqref{eq:interaction_kernel}, which involves the screened Coulomb interaction $W$. $D$ and $X$ are given by
\begin{align}
  D_{ij,kl} &= \int d\mathbf{x}d\mathbf{x}'\varphi^*_{i}(\mathbf{x})\varphi^*_{j}(\mathbf{x}')W(\mathbf{x},\mathbf{x}';\omega=0)\varphi_{k}(\mathbf{x})\varphi^*_{l}(\mathbf{x}')\\
  X_{ij,kl} &= \int d\mathbf{x}d\mathbf{x}'\varphi^*_{i}(\mathbf{x})\varphi^*_{j}(\mathbf{x}')v_c\varphi_{k}(\mathbf{x})\varphi^*_{l}(\mathbf{x}')
\end{align}
where the indices $i,j,k,l$ denote band, momentum pairs $(n,\mathbf{k})$. We observe that the direct interaction matrix element $D$ involves the screened Coulomb interaction at zero frequency, $W(\omega = 0)$. Strictly speaking, the eigenvalue problem~\eqref{eq:bse_eigenvalue_antihermitian} poses a self-consistent problem, where the exciton energies $\Omega_n$ are also present in the screened Coulomb interaction at $\omega=\Omega_n$. One way to operate the frequency dependency of $W$ is to write it in the plasmon-pole approximation. This allows to evaluate analytically the interaction kernel, which can be further simplified assuming that the plasmon energies are well separated from the exciton energies, yielding the above expression~\cite{Rohlfing2000}. Thus, with this approximation (regarded simply as static screened Coulomb interaction), often justified in insulators and semiconductors, the BSE can be solved one-shot. There is another approximation that can be made to simplify the BSE, which is the Tamm-Dancoff approximation. This approximation is made to neglect the antiresonant part of the BSE, i.e. $C=0$, and implies that the ground state of our theory is the Fermi sea. Therefore, this approximation is also justified in insulators and semiconductors, where the Fermi level is well separated from the minimum of the conduction band. With the TDA, the exciton states are given by $\ket{X_n} = \sum_{vc\mathbf{k}}A^n_{vc}(\mathbf{k})c^{\dagger}_{c\mathbf{k}}c_{v\mathbf{k}}\ket{GS}$, the same as in the exact diagonalization approach. And the BSE in the TDA reads
\begin{equation}
  (\varepsilon_{c\mathbf{k}} - \varepsilon_{v\mathbf{k}})A_{vc}^n(\mathbf{k}) + \sum_{v',c',\mathbf{k}'}K_{v\mathbf{k}c\mathbf{k},v'\mathbf{k}'c'\mathbf{k}'}A_{v'c'}^n(\mathbf{k}')
    = \Omega_n A_{vc}^n(\mathbf{k})
\end{equation}
where the interaction kernel $K=-(D-X)$ is defined using the above terms. Thus, we see that with the static screening and the TDA the problem of determining the exciton spectrum is formally the same both with ED or MBPT, the main difference lying in the presence of the screened Coulomb interaction. 

\subsection{Screening of the Coulomb interaction}
So far we have been discussing how to obtain and solve the BSE\@. To do so, however, we still need to specify the screening of the Coulomb interaction. In MBPT, one can identify an effective interaction from the perturbative expansion of the Green's functions in powers of the interaction. This is, the bare interaction plus additional diagrams where other processes take place, such as electron-hole pair creation. As with the interacting Green's function, this results in a Dyson or recursive equation for this effective, screened potential $W$:
\begin{equation}
    W(1,2) = v_c(1,2) + \int d34 v_c(1,3)P(3,4)W(4,2)
\end{equation}
\begin{figure}[h]
  \centering
  \raisebox{0.3cm}{
  \begin{tikzpicture}
    \begin{feynman}[medium, every blob={/tikz/fill=gray!30,/tikz/inner sep=2pt}]
      \vertex (bi1);
      \vertex [right=1.5cm of bi1] (bf1);

      \vertex [above=0.04cm of bi1](aux1);
      \vertex [below=0.04cm of bi1](aux2);
      \vertex [above=0.04cm of bf1](aux3);
      \vertex [below=0.04cm of bf1](aux4);
    
        \diagram* {
          (aux1) -- [boson] (aux3);
          (aux2) -- [boson] (aux4);
        };

      \end{feynman}
    \end{tikzpicture}
  }
    \raisebox{0.3cm}{=
    \begin{tikzpicture}
      \begin{feynman}[medium, every blob={/tikz/fill=gray!30,/tikz/inner sep=2pt}]
        \vertex (bi1);
        \vertex [right=1.5cm of bi1] (bf1);
      
          \diagram* {
            (bi1) -- [boson] (bf1);
          };
  
        \end{feynman}
      \end{tikzpicture}
    \ +
    }
  \begin{tikzpicture}
    \begin{feynman}[medium, every blob={/tikz/fill=gray!30,/tikz/inner sep=2pt}]
      \vertex (bi1);
      \vertex [right=1.5cm of bi1] (aux_bi2);
      \vertex [right=0.5cm of aux_bi2](a);
      \vertex [right=2.25cm of aux_bi2] (bf2);

      \vertex [above=0.04cm of aux_bi2](aux1);
      \vertex [below=0.04cm of aux_bi2](aux2);
      \vertex [above=0.04cm of bf2](aux3);
      \vertex [below=0.04cm of bf2](aux4);
    
        \diagram* {
          (bi1) -- [boson] (aux_bi2);
          (aux1) -- [boson] (aux3);
          (aux2) -- [boson] (aux4);
        };
        \vertex [right=1.5cm of bi1, blob] (bi2) {$P$};

      \end{feynman}
    \end{tikzpicture}
    \caption[Dyson series for the screened interaction $W$]{Diagrammatic representation of the Dyson series for the screened Coulomb interaction $W$, written in terms of the irreducible polarizability $P$.}
\end{figure}
where the numbers represent space-time coordinates, e.g. $1\equiv (\mathbf{x}_1,t_1)$, $P(3,4)$ corresponds to the irreducible polarizability, $v_c$ is the bare Coulomb interaction and $W$ is the screened interaction. 

This equation is one of the five Hedin's equations; it is coupled to the rest via the irreducible polarizability $P$\@. Since solving exactly Hedin's equations is an unfeasible task, one usually resorts to approximations to simplify the problem. The standard technique is to use the random-phase approximation (RPA), in which we approximate the irreducible polarizability by a free electron-hole pair $P_0$~\cite{martin2016interacting}, i.e.:
\begin{equation}
    P(3,4) \approx P_0(3,4) = -iG(3,4)G(4,3)
\end{equation}
\begin{figure}
  \centering
  \raisebox{0.45cm}{
  \begin{tikzpicture}
    \begin{feynman}[medium, every blob={/tikz/fill=gray!30,/tikz/inner sep=6pt}]
      \vertex [blob] (i1) {$P$};
    \end{feynman}
  \end{tikzpicture}
  \raisebox{0.35cm}{$\ \approx\ $}}
  \begin{tikzpicture}
    \begin{feynman}
      \vertex [dot](i1){};
      \vertex [right=2cm of i1, dot] (f1){};

      \diagram* {
        (i1) -- [double,double distance=0.3ex,thick,with arrow=0.5,arrow size=0.2em, out=-270, in=90] (f1);
        (f1) -- [double,double distance=0.3ex,thick,with arrow=0.5,arrow size=0.2em, out=270, in=-90] (i1);
      };
    \end{feynman}
  \end{tikzpicture}
  \caption[RPA approximation]{Diagram showing the approximation of the irreducible polarizability $P$ by the free electron-hole pair $P_0$, known as the random-phase approximation.}
\end{figure}
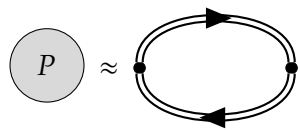
In doing so, we decouple the equation for the screened interaction from the rest of Hedin's equations. Thus, the screened interaction can be obtained solving its recursive equation. To do so, first we factor out the screened interaction:
\begin{equation}\label{eq:screened_factor_out}
    \int d4 W(4,2)\left[\delta(1 - 4) - \int d3 v_c(1,3)P_0(3,4)\right] = v_c(1,2)
\end{equation}
From this we define the microscopic dielectric function $\epsilon(1,4)$:
\begin{align}
    \epsilon(1,4) &= \delta(1 - 4) - \int d3 v_c(1,3)P_0(3,4) \\ 
    &\int d4 \epsilon(1,4) W(4,2) = v_c(1,2)
\end{align}
If we now invert the dielectric function using the relation $\int d1 \epsilon^{-1}(3,1)\epsilon(1,4)=\delta(3-4)$, we obtain the final expression of the screened interaction:
\begin{equation}
    W(3,2) = \int d1 \epsilon^{-1}(3,1)v_c(1,2)
\end{equation}
From this expression, we see the reason for denominating the effective interaction as a "screened" potential: we have identified a dielectric function (more accurately, its inverse) that is convoluted with the bare interaction, in the fashion seen in classical electrodynamics.

So far we have worked in space-time coordinates, but it is more common to Fourier transform the previous equations to frequency domain, as one obtains the Lehmann representation of the Green's function, which is more practical for solving either the Dyson or the Bethe-Salpeter equation. Doing so, Eq.~\eqref{eq:screened_factor_out} becomes:
\begin{align}\label{eq:screened_w_factor_out}
    &\int d4 W(\mathbf{x}_4,\mathbf{x}_2;\omega) 
    \left[\delta(\mathbf{x}_1 - \mathbf{x}_4) - \int d\mathbf{x}_3 v_c(\mathbf{x}_1,\mathbf{x}_3)P_0(\mathbf{x}_3,\mathbf{x}_4;\omega)\right] = v_c(\mathbf{x}_1,\mathbf{x}_2)
\end{align}
and
\begin{align}
    P_0(\mathbf{x}_1,\mathbf{x}_2;\omega)=
    \sum_{v\mathbf{k}, c\mathbf{k}'}\left[\frac{\varphi_{c\mathbf{k}'}(\mathbf{x}_1)\varphi^*_{v\mathbf{k}}(\mathbf{x}_1)\varphi_{v\mathbf{k}}(\mathbf{x}_2)\varphi^*_{c\mathbf{k}'}(\mathbf{x}_2)}{\omega - (E_{c\mathbf{k}'} - E_{v\mathbf{k}})} - \frac{\varphi_{v\mathbf{k}}(\mathbf{x}_1)\varphi^*_{c\mathbf{k}'}(\mathbf{x}_1)\varphi_{c\mathbf{k}'}(\mathbf{x}_2)\varphi^*_{v\mathbf{k}}(\mathbf{x}_2)}{\omega + (E_{c\mathbf{k}'} - E_{v\mathbf{k}})}\right]
\end{align}
Repeating the same steps as before we obtain the expression for the screened potential, which is now frequency dependent:
\begin{equation}\label{eq:screened_continuum}
    W(\mathbf{x}_3,\mathbf{x}_2;\omega) = \int d\mathbf{x}_1 \epsilon^{-1}(\mathbf{x}_3,\mathbf{x}_1;\omega)v_c(\mathbf{x}_1,\mathbf{x}_2)
\end{equation}
The idea now is to obtain an expression for both the dielectric constant $\varepsilon(\mathbf{x}_1,\mathbf{x}_3;\omega)$ and the screened potential $W(\mathbf{x}_1,\mathbf{x}_2;\omega)$. Both quantities are typically evaluated in first-principles codes, either using plane waves (meaning that all the quantities are written in reciprocal space to take advantage of the basis) or Gaussian basis. However, in both cases the evaluation of the dielectric constant is notably expensive computationally. Which is why instead, we want to exploit the tight-binding approximation, i.e.\ point-like orbitals to obtain a simplified expression for the dielectric function that allows for faster evaluation.

The starting point will be Eq.~\eqref{eq:screened_w_factor_out}. We focus first on the integral of the second term of the dielectric function:
\begin{align}
    \nonumber &\int  d\mathbf{x}_3\mathbf{x}_4 v_c(\mathbf{x}_1,\mathbf{x}_3)P_0(\mathbf{x}_3,\mathbf{x}_4;\omega)W(\mathbf{x}_4,\mathbf{x}_2;\omega) \\ 
     \nonumber & = \frac{1}{N^2}\sum_{\substack{\mathbf{x}_3\equiv(\mathbf{t}_i,\mathbf{R}) \\ \mathbf{x}_4\equiv(\mathbf{t}_j, \mathbf{R}')}} v_c(\mathbf{x}_1,\mathbf{t}_i+\mathbf{R}) \sum_{v\mathbf{k}, c\mathbf{k}'} \sum_{\alpha\beta
    }\left[\frac{C^{c\mathbf{k}'}_{i\alpha}(C^{v\mathbf{k}}_{i\alpha})^*C^{v\mathbf{k}}_{j\beta}(C^{c\mathbf{k}'}_{j\beta})^*e^{i(\mathbf{k}'-\mathbf{k})(\mathbf{R}-\mathbf{R}')}}{\omega - (E_{c\mathbf{k}'} - E_{v\mathbf{k}})} - \right.\\
    &\hspace{5cm}\left. \frac{C^{v\mathbf{k}}_{i\alpha}(C^{c\mathbf{k}'}_{i\alpha})^*C^{c\mathbf{k}'}_{j\beta}(C^{v\mathbf{k}}_{j\beta})^*e^{i(\mathbf{k}-\mathbf{k}')(\mathbf{R}-\mathbf{R}')}}{\omega + (E_{c\mathbf{k}'} - E_{v\mathbf{k}})}\right]W(\mathbf{t}_j+\mathbf{R}',\mathbf{x}_2;\omega)
\end{align}
where we have used the expression for the polarizability $P_0$ and have expanded the single-particle states in terms of the orbitals. Then, using the tight-binding approximation for point-like orbitals~\eqref{eq:approximation} we can carry out the integrations analytically.
Thus, we only take into account contributions from the same orbital in the same spatial position, and neglect all cross terms.
Then, the integral in the spatial coordinates $\mathbf{x}_3,\mathbf{x}_4$ transforms into a summation over the discrete set of lattice positions. We use the same indices as in the integral, noting now that its values are restricted to the atomic positions.
On the other hand, the first term of Eq.~\eqref{eq:screened_w_factor_out} can be rewritten in discrete form as:
\begin{align}
    W(\mathbf{x}_1,\mathbf{x}_2;\omega) &= \int d\mathbf{x}_4 W(\mathbf{x}_4,\mathbf{x}_2;\omega)\delta(\mathbf{x}_1-\mathbf{x}_4)
    =\sum_{\mathbf{x}_4\equiv\mathbf{t}_j+\mathbf{R}'}W(\mathbf{x}_4,\mathbf{x}_2;\omega)\delta_{\mathbf{x}_1\mathbf{x}_4}
\end{align}
since we only evaluate the interaction over a set of discrete positions, i.e.\ all the atomic coordinates. From now on, all coordinates correspond to pairs of motif and Bravais vectors, $\mathbf{x}\equiv (\mathbf{t}_i, \mathbf{R})$. Putting together the two previous expressions, we arrive at a discretized version of Eq.~\eqref{eq:screened_w_factor_out}, where we sum over lattice positions instead of integrating over all space:
\begin{equation}
    \sum_{\mathbf{x}_4}\bar\epsilon(\mathbf{x}_1,\mathbf{x}_4;\omega)W(\mathbf{x}_4,\mathbf{x}_2;\omega)=v_c(\mathbf{x}_1,\mathbf{x}_2)
\end{equation}
where
\begin{align}
    \nonumber &\bar\epsilon(\mathbf{x}_1,\mathbf{x}_4)=\delta_{\mathbf{x}_1\mathbf{x}_4} - \\
    &\frac{1}{N^2}\sum_{\mathbf{x}_3} v_c(\mathbf{x}_1,\mathbf{x}_3)\sum_{v\mathbf{k},c\mathbf{k}'} \sum_{\alpha\beta
    }\left[\frac{C^{c\mathbf{k}'}_{i\alpha}(C^{v\mathbf{k}}_{i\alpha})^*C^{v\mathbf{k}}_{j\beta}(C^{c\mathbf{k}'}_{j\beta})^*e^{i(\mathbf{k}'-\mathbf{k})(\mathbf{R}-\mathbf{R}')}}{\omega - (E_c - E_v)} - \right.
    \left. \frac{C^{v\mathbf{k}}_{i\alpha}(C^{c\mathbf{k}'}_{i\alpha})^*C^{c\mathbf{k}}_{j\beta}(C^{v\mathbf{k}}_{j\beta})^*e^{i(\mathbf{k}-\mathbf{k}')(\mathbf{R}-\mathbf{R}')}}{\omega + (E_c - E_v)}\right]
\end{align}
Since we are evaluating the dielectric function and the screened potential over a set of discrete positions, we may regard them as matrices. Now we can define the inverse dielectric matrix $\sum_{\mathbf{x}_1}\bar\epsilon^{-1}(\mathbf{x}_1,\mathbf{x}_4;\omega)\epsilon(\mathbf{x}_4,\mathbf{x}_3;\omega)=\delta_{\mathbf{x}_1\mathbf{x}_3}$, from which we obtain the screened potential $W$:
\begin{equation}
    W(\mathbf{x}_3,\mathbf{x}_2;\omega) = \sum_{\mathbf{x}_1} \bar\epsilon^{-1}(\mathbf{x}_3,\mathbf{x}_1;\omega)v_c(\mathbf{x}_1,\mathbf{x}_2)
\end{equation}
which is the discrete version of Eq.~\eqref{eq:screened_continuum}. We may now use this potential in the Bethe-Salpeter interaction kernel, instead of the bare Coulomb potential or the Keldysh potential. 
The interaction kernel $K$ is composed of two terms, the direct and exchange interactions $D, X$. In a typical BSE calculation, as obtained from MBPT, the direct term is given by some matrix elements of the screened interaction $W$, whereas the exchange term is a matrix element of the bare Coulomb interaction. Following the same procedure as in section~\ref{sec:interaction_matrix_elements}, we obtain the following interaction kernel:
\begin{equation}
    K_{vc,v'c'}(\mathbf{k},\mathbf{k}',\mathbf{Q}) = -( D_{vc,c'c'}(\mathbf{k},\mathbf{k}',\mathbf{Q}) - X_{vc,v'c'}(\mathbf{k},\mathbf{k}',\mathbf{Q}))
\end{equation}
where
\begin{align}
    D_{vc,v'c'}(\mathbf{k},\mathbf{k}',\mathbf{Q}) &= \sum_{ij,\alpha\beta}(C^{c\mathbf{k}+\mathbf{Q}}_{i\alpha})^*(C^{v'\mathbf{k}'}_{j\beta})^*C^{c'\mathbf{k}'+\mathbf{Q}}_{i\alpha}C^{v\mathbf{k}}_{j\beta}W_{ij}(\mathbf{k}'-\mathbf{k})\\
    W_{ij}(\mathbf{k}'-\mathbf{k}) &= \sum_{\mathbf{R}}e^{i(\mathbf{k}'-\mathbf{k})\mathbf{R}}W(\mathbf{t}_i, \mathbf{t}_j + \mathbf{R};
    \omega = 0)
\end{align}
and
\begin{equation}
    X_{vc,v'c'}(\mathbf{k},\mathbf{k}',\mathbf{Q}) = \sum_{ij,\alpha\beta}(C^{c\mathbf{k}+\mathbf{Q}}_{i\alpha})^*(C^{v'\mathbf{k}'}_{j\beta})^*C^{v\mathbf{k}}_{i\alpha}C^{c'\mathbf{k}'+\mathbf{Q}}_{j\beta}V_{ij}(\mathbf{Q})
\end{equation}
$V_{ij}(\mathbf{Q})$ has a definition analogue to that of $W_{ij}(\mathbf{k}'-\mathbf{k})$, except that we use the bare Coulomb potential $v_c$ instead of the screened potential. Also note that for the direct term, we use that $W(\mathbf{r},\mathbf{r}')=W(\mathbf{r} + \mathbf{R},\mathbf{r}' + \mathbf{R})$, $\forall \mathbf{R}\in\text{Bravais lattice}$, meaning that we can remove one summation over $\mathbf{R}'$. It is important to notice as well that we evaluate the screened interaction in the BSE at zero frequency, namely in the static limit. In doing so, we avoid performing the integration over frequencies that is required for the computation of the direct terms, which has a negligible effect as we move apart from the metallic regime, i.e.\ there are no plasmons.

\section{Properties}
\subsection{Reciprocal and real-space probability densities}\label{sec:kwf-rswf}

Once the interaction kernel is determined, Eq.~\eqref{bse} can be solved to obtain the exciton energies and wavefunctions, i.e., the coefficients $A^n_{vc}(\mathbf{k}, \mathbf{Q})$. These can be used to compute different quantities. For instance, given that the exciton is written as a linear combination of electron-hole pairs with well-defined $\mathbf{k}$ quantum number (in the Tamm-Dancoff approximation), we can define the probability density of finding the exciton in a specific pair in $\mathbf{k}$-space, or reciprocal probability density, as
\begin{equation}\label{kwavefunction}
    |\psi_X(\mathbf{k})|^2=\sum_{v,c}|A^n_{vc}(\mathbf{k}, \mathbf{Q})|^2
\end{equation}
which is the straightforward definition since all electron-hole pairs are orthonormal to each other. Note that this definition is independent of the details of the basis or the tight-binding approximation.

Plotting the probability density~\eqref{kwavefunction} is useful to extract some information about the exciton such as the wavefunction type ($s$, $p$, etc., following the hydrogenic model). The same can be argued for its real-space wavefunction, $\psi_X(\mathbf{r}_e, \mathbf{r}_h)$. However, obtaining it is not as straightforward as the $\mathbf{k}$ wavefunction. To do so, we remember the definition of the field operators,
\begin{equation}\label{eq:field_operators}
    \psi^{\dagger}(\mathbf{r}) = \sum_{n\mathbf{k}}\varphi^*_{n\mathbf{k}}(\mathbf{r})c^{\dagger}_{n\mathbf{k}},\ \psi(\mathbf{r}) = \sum_{n\mathbf{k}}\varphi_{n\mathbf{k}}(\mathbf{r})c_{n\mathbf{k}}
\end{equation}
where $\varphi_{n\mathbf{k}}(\mathbf{r})$ are the single-particle states in coordinate representation. Then, we can define the amplitude or real space wavefunction of the exciton in the following way:
\begin{equation}
    \psi_X(\mathbf{r}_e,\mathbf{r}_h) = \braket{GS|\psi(\mathbf{r}_e)\psi^{\dagger}(\mathbf{r}_h)|X_n(\mathbf{Q})}
\end{equation}
This definition is motivated by the fact that $\varphi_{n\mathbf{k}}(\mathbf{r})=\braket{GS|\psi(\mathbf{r})|n\mathbf{k}}$. We note here that this is precisely the same quantity as the Bethe-Salpeter amplitude we had defined in Section~\ref{sec:mbpt-bse}, $\chi(\mathbf{r}_e,\mathbf{r}_h)$, although here we will refer to it as the exciton wavefunction $\psi_X$. Before computing the amplitude, it is convenient to switch to the electron-hole picture. The field operator written in terms of electron and hole operators is
\begin{equation}
    \psi(\mathbf{r})=\sum_{c\mathbf{k}}\varphi_{c\mathbf{k}}(\mathbf{r})c_{c\mathbf{k}} + \sum_{v\mathbf{k}}\varphi_{v\mathbf{k}}(\mathbf{r})h^{\dagger}_{v\mathbf{-k}}\equiv \psi_e(\mathbf{r}) + \psi^{\dagger}_h(\mathbf{r})
\end{equation}
where $\psi_e(\mathbf{r})$, $\psi_h(\mathbf{r})$ are the annihilation field operators for electrons and holes respectively. Since we are switching from the electronic to the electron-hole picture, the same has to be done for the exciton state, $\ket{X_n(\mathbf{Q})}=\sum_{v,c,\mathbf{k}}A^n_{vc}(\mathbf{k}, \mathbf{Q})c^{\dagger}_{c\mathbf{k}+\mathbf{Q}}h^{\dagger}_{v,-\mathbf{k}}\ket{0}$. Evaluating the exciton amplitude in terms of the electron and hole field operators, we obtain
\begin{align}
\label{real-space-wf}
     \psi_X(\mathbf{r}_e,\mathbf{r}_h)&=\braket{GS|\psi_e(\mathbf{r}_e)\psi_h(\mathbf{r}_h)|X_n(\mathbf{Q})}
    =\sum_{v,c,\mathbf{k}}A^n_{vc}(\mathbf{k}, \mathbf{Q})\varphi_{c\mathbf{k}+\mathbf{Q}}(\mathbf{r}_e)\varphi^*_{v\mathbf{k}}(\mathbf{r}_h)
\end{align}
To obtain the first equality note that there are four cross terms containing electron and hole field operators. Two of them are zero, since they move around the electron or the hole [e.g. $\psi_e(\mathbf{r}_e)\psi^{\dagger}_e(\mathbf{r}_h)$], meaning that the final state is still orthonormal to the ground state. There is a third term consisting on the creation of an electron and a hole, $\psi^{\dagger}_e(\mathbf{r}_e)\psi^{\dagger}_h(\mathbf{r}_h)$. This term is also zero because we assume that our ground state is the Fermi sea, meaning that it does not contain excited electrons. If this were the case, then the exciton could also consist on de-excitations or antiresonant transitions. This is known as the Tamm-Dancoff approximation, and it is also usually present in GW-BSE\@. To obtain the final expression for the wavefunction, it remains to substitute the expression of the field operators. One then recovers the electron-hole pairs states of the exciton basis (up to a sign from operator permutation), and from orthonormality it results in expression (\ref{real-space-wf}).

At this point, to be able to plot the exciton real space wavefunction, we still need to evaluate~\eqref{real-space-wf} in terms of the single-particle states $\varphi_{n\mathbf{k}}(r)$. Since the exciton wavefunction depends on both the position of the electron and the hole, first we need to fix the position of either of them to be able to plot the wavefunction. Since we assume the orbitals are point-like, both the electron and hole can only be localized at the atomic positions, so we will evaluate the wavefunction and the probability density at these points only. 

We set the electron to be located at cell $\mathbf{R}_e$ and atom $\mathbf{t}_m$ of the motif, $\mathbf{r}_e=\mathbf{R}_e + \mathbf{t}_m$, while the hole is at position $\mathbf{r}_h=\mathbf{R}_h + \mathbf{t}_n$. Using the point-like approximation~\eqref{eq:approximation}, the probability density of finding the electron at a given position with the hole fixed reads
\begin{equation}
    |\psi_X(\mathbf{R}_e + \mathbf{t}_m, \mathbf{R}_h + \mathbf{t}_n)|^2 = \sum_{\alpha\beta}|\psi^{\alpha\beta}_X(\mathbf{R}_e + \mathbf{t}_m, \mathbf{R}_h + \mathbf{t}_n)|^2
\end{equation}
where
\begin{align}
    \nonumber &|\psi_X^{\alpha\beta}(\mathbf{R}_e + \mathbf{t}_m, \mathbf{R}_h + \mathbf{t}_n)|^2 = \\
    &\frac{1}{N^2}\sum_{v,c,\mathbf{k}}\sum_{v',c',\mathbf{k}'}A^n_{vc}(\mathbf{k}, \mathbf{Q})(A^n_{v'c'}(\mathbf{k}',\mathbf{Q}))^*e^{i(\mathbf{k} - \mathbf{k}')\cdot(\mathbf{R_e} - \mathbf{R_h})}
    \cdot C^{c,\mathbf{k}+\mathbf{Q}}_{m\alpha}(C^{c',\mathbf{k}'+\mathbf{Q}}_{m\alpha})^*(C^{v,\mathbf{k}}_{n\beta})^*C^{v',\mathbf{k}'}_{n\beta} 
\end{align}

\subsection{Spin polarization}

 If the single-particle basis includes spin, one can also compute the expected value of the total spin projection of the exciton, $S_z^T=S_z^e+S_z^h$. Given that we are using a fully electronic description of the exciton, we need to specify the electrons whose spin we want to measure. To this purpose, we write the total spin operator in second quantization as
\begin{equation}
\label{eq:total_spin}
    S^T_z = \sum_{c',c,\mathbf{k}}\sigma_{c'\mathbf{k}+\mathbf{Q}, c\mathbf{k}+\mathbf{Q}}c^{\dagger}_{c'\mathbf{k}+\mathbf{Q}}c_{c\mathbf{k}+\mathbf{Q}} - \sum_{v',v,\mathbf{k}}\sigma_{v\mathbf{k},v'\mathbf{k}}c_{v'\mathbf{k}}c^{\dagger}_{v\mathbf{k}}
\end{equation}
where $\sigma_{nm}=\braket{n|S_z|m}$. An alternative definition that also yields the correct result is the straightforward second-quantized version of $S_z$, namely $S^T_z=\sum_{i,j}\sigma_{ij}c^{\dagger}_ic_j$, where $i,j$ denote pairs of band index and momentum $(n,\mathbf{k})$. In~\eqref{eq:total_spin}, the labels $c,c',v,v'$ refer exclusively to the conduction and valence bands used in the definition of the excitons. Note that the second term, which corresponds to the spin of the hole, has a minus sign. This is because holes, when described as quasiparticles, have opposite momentum and spin of the corresponding electronic state, i.e. $h^{\dagger}_{n,-\mathbf{k},-\sigma} = (-1)^{\sigma}c_{n\mathbf{k}\sigma}$, for states below the Fermi energy, $\varepsilon_{n\mathbf{k}} < \varepsilon_F$~\cite{fetter2012quantum}. These $h$ operators describe creation/annihilation of holes in terms of their electronic counterpart. Although we keep $\mathbf{k}$ the same (since we are still in the electronic picture), we already incorporate this minus sign to give a correct description of the total spin of the exciton. As we will see later, this sign change is also necessary to retrieve the known singlet and triplet states when summing angular momentum. The two pictures are equivalent, and all the previous calculations can be reproduced in the electron-hole picture.
The expected value of the total spin is then given by
\begin{align}
\label{spin_total}
 \nonumber &\braket{X_n(\mathbf{Q})|S^T_z|X_n(\mathbf{Q})}=\\
 &\sum_{v,c,\mathbf{k}}A^n_{vc}(\mathbf{k},\mathbf{Q})\left[\sum_{c'}\right.(A^n_{vc'}(\mathbf{k},\mathbf{Q}))^*\sigma_{c'\mathbf{k}+\mathbf{Q}, c\mathbf{k}+\mathbf{Q}} 
      \left. - \sum_{v'}(A^n_{v'c}(\mathbf{k},\mathbf{Q}))^*\sigma_{v\mathbf{k}, v'\mathbf{k}}\right]
\end{align}
If $[H_0, S_z] = 0$, then the spin projection $S_z$ is also a good quantum number for the Bloch states. Therefore, they can be written now as $\ket{n\mathbf{k}\sigma}$, or in real space as $\varphi_{n\mathbf{k}}(\mathbf{r})\chi_{\sigma}$, where $\chi_{\sigma}$ denotes the spin part of the state. This means that the spin operator $S_z$ is diagonal, $\sigma_{nm}=\sigma_{n}\delta_{nm}$, which allows us to simplify expression~\eqref{spin_total}:
\begin{equation}
    \braket{S_z^T} = \sum_{v,c,\mathbf{k}}|A^n_{vc}(\mathbf{k},\mathbf{Q})|^2(\sigma_{c} - \sigma_{v})
\end{equation}
Another consequence of having the spin well-defined is that it also propagates to the electron-hole pairs that serve as a basis for the exciton states, i.e. $\ket{\Tilde{v}, \Tilde{c}, \mathbf{k},\mathbf{Q}}=c^{\dagger}_{\Tilde{c}\mathbf{k}+\mathbf{Q}}c_{\Tilde{v}\mathbf{k}}\ket{GS}$, where $\Tilde{v}=(v,\sigma_v)$, $\Tilde{c}=(c,\sigma_c)$. In principle, we allow the spin of the electron and the hole to be different, $\sigma_c \neq \sigma_v$. Taking into account the spin in the computation of the interaction matrix elements, we arrive at constraints on which electron-hole pairs interact. Then the direct and exchange terms read
\begin{equation}
\begin{split}
\label{spin_interaction_matrix_elements}
    D_{\Tilde{v}\Tilde{c},\Tilde{v}'\Tilde{c}'}(\mathbf{k},\mathbf{k}', \mathbf{Q}) &= \delta_{\sigma_c\sigma_{c'}}\delta_{\sigma_v\sigma_{v'}}D_{vc,v'c'}(\mathbf{k},\mathbf{k}', \mathbf{Q}) \\
    X_{\Tilde{v}\Tilde{c},\Tilde{v}'\Tilde{c}'}(\mathbf{k},\mathbf{k}', \mathbf{Q}) &= \delta_{\sigma_c\sigma_v}\delta_{\sigma_{v'}\sigma_{c'}}X_{vc,v'c'}(\mathbf{k},\mathbf{k}', \mathbf{Q})
\end{split}
\end{equation}
which can be directly obtained by substituting the single-particle states, since the spin part is not mixed with the orbital part of the states (i.e.\ $\ket{n\mathbf{k}\sigma}=\ket{n\mathbf{k}}\otimes\ket{\sigma}$). At this point, we can arrange the electron-hole pairs into four groups depending on their spin: 
\begin{equation*}
  \{\ket{++}, \ket{--}, \ket{+-}, \ket{-+}\}_e=\{\ket{\sigma_c\sigma_v}\}_e
\end{equation*}
The $e$ subindex is used to denote that this corresponds to the electronic picture. Then the Hamiltonian represented in terms of the spin groups, and taking into account~\eqref{spin_interaction_matrix_elements} becomes
\begin{equation}
    H = 
    \begin{pmatrix}
        H_0 - D + X & X           & 0       & 0 \\
        X           & H_0 - D + X & 0       & 0 \\
        0           & 0           & H_0 - D & 0 \\
        0           & 0           & 0       & H_0 - D
    \end{pmatrix}
\end{equation}
where $H_0$, $D$, $X$ are blocks which include matrix elements corresponding to different electron-hole pairs but same spin group. If we now take into account that the hole in its quasiparticle representation must have spin opposite of that of the electron vacancy, then our states are $\{\ket{+-}, \ket{-+}, \ket{++}, \ket{--}\}_{eh}$, where $eh$ denotes electron-hole picture. Therefore, the exciton spectrum would be composed of groups of three triplet states and one singlet state, as when adding angular momenta. If instead we turn off the exchange interaction, then every state should have at least four-fold degeneracy. Any additional degeneracy would come from spatial symmetries of the Hamiltonian, in particular from the irreducible representations of the little group at $\mathbf{Q}$ (see~\ref{sec:symmetry}). 

Lastly, is it interesting to note that the singlet, triplet picture for excitons can also be understood as spin conserving or spin-flip excitons respectively. Light (in the dipole approximation) only couples to the orbital part of the states, and therefore spin is conserved across transitions~\cite{xiao2012coupled}. Therefore, light can only excite singlet excitons, which corresponds to the intuitive picture of an exciton. On the other hand, the spin-flip excitons actually describe magnons, i.e.\ the lowest-energy excitations of magnetic systems, which consist precisely on a collective spin-flip. In systems with non-magnetic Fermi seas, it might be more correct to speak of magnetic excitons~\cite{wu2019physical}, but in magnetic materials (magnetic Fermi sea), the BSE provides a unified description of both excitons and magnons~\cite{olsen2021unified}.

\subsection{Optical conductivity and light absorption}

The optical conductivity can be derived from linear response theory. Let's consider the Hamiltonian of the interacting system $H$ and a weak, external perturbation consisting of the light-matter coupling,
\begin{equation}
H_{\text{tot}} = H + H_{\text{ext}}
\end{equation}
where $H_{\text{ext}}=-\int d\mathbf{r}\mathbf{j}(\mathbf{r})\cdot\mathbf{A}(\mathbf{r})=-\mathbf{j}\cdot\mathbf{A}$, obtained with the minimal coupling $\mathbf{p}\rightarrow \mathbf{p} - e\mathbf{A}$. Here the potential vector $\mathbf{A}$ is taken to be uniform in space and is monochromatic, $\mathbf{E}=i\omega\mathbf{A}$, with $\mathbf{E}=\mathbf{E}_0e^{-i\omega t}$, and $\mathbf{j}$ is the current operator, $\mathbf{j}=e\sum_{i}\mathbf{v}_i$. Computing the expected value of the total current, $\mathbf{J}=\frac{\delta H_{\text{tot}}}{\delta \mathbf{A}}$, at time $t$ and first order in the field $\mathbf{A}$, we obtain the Kubo formula for the electric conductivity~\cite{louie2006conceptual, tong2016lectures,coleman2015introduction,mahan2013many}:
\begin{equation}
    \sigma_{ab}(\omega) = \frac{ine^2}{m\omega}\delta_{ab} + \frac{1}{\hbar\omega V}\int^{\infty}_0dt\braket{[j_{a}(t),j_{b}(0)]}e^{i\omega t}
\end{equation}
which is extracted from the definition of the current $J_{a}=\sum_{b}\sigma_{ab}(\omega) E_{b}(\omega)$. The first term corresponds to the diamagnetic contribution to the current, whereas the second term is the paramagnetic contribution, both appearing directly from the minimal coupling. Assuming that we are at finite temperature, the expectation value is to be taken using the density matrix $\braket{O}=\text{Tr}(\rho O)$, where $\rho=e^{-\beta H}/Z$ and $Z = \text{Tr }e^{-\beta H}$. Also, supposed that we have a complete set of eigenstates of the unperturbed Hamiltonian, $H\ket{n}=E_n\ket{n}$, we can use the completeness relation to write the current-current correlation function as
\begin{align}
    \sigma_{ab}(\omega) &=\frac{ine^2}{m\omega}\delta_{ab} + 
    \frac{i}{\omega V}\sum_{n,m}\frac{e^{-\beta E_n}-e^{-\beta E_m}}{Z}\left[\frac{\braket{m|j_{a}|n}\braket{n|j_{b}|m}}{\hbar\omega + i\eta - (E_n - E_m)}\right] \\
    & \xrightarrow[\beta\to\infty]{} \frac{ine^2}{m\omega}\delta_{ab} + 
    \frac{i}{\omega V}\sum_n\left[\frac{\braket{0|j_{a}|n}\braket{n|j_{b}|0}}{\hbar\omega + i\eta - (E_n - E_0)} - \frac{\braket{0|j_{b}|n}\braket{n|j_{a}|0}}{\hbar\omega + i\eta - (E_0 - E_n)}\right]
    \label{eq:kubo_formula_T0}
\end{align}
where $\eta$ is an infinitesimal introduced to ensure convergence of the integral, and $E_0$ is the energy of the ground state $\ket{0}$. The first line is the conductivity at finite temperature $\beta=1/k_bT$, and the second line at $T=0$. From now on we will consider the expression at zero temperature. Taking the real part of the conductivity (dissipative part) and restricting ourselves to $\omega > 0$, we obtain the optical conductivity that reflects absorption:
\begin{equation}
    \text{Re}\ \sigma_{ab}(\omega) = \frac{\pi}{\omega V}\sum_{n\neq 0}\braket{0|j_{a}|n}\braket{n|j_{b}|0}\delta(\hbar\omega - (E_n - E))
\end{equation}
This expression is general in the sense that the many-body states $\{\ket{n}\}$ have yet to be specified. If we now particularize to exciton states (which we assume form a complete basis), the real part of the optical conductivity reads
\begin{equation}
    \label{eq:excitonkubo}
    \begin{split}
        \text{Re}\ \sigma_{ab}(\omega)=&\frac{\pi e^2 \hbar}{V} \sum_{n=1}^{N_X} \frac{1} {E_{n}}V_{n}^{a}(V_{n}^{b})^\ast \delta(\hbar\omega-E_{k})
    \end{split}
\end{equation}
which is the expression typically reported in literature~\cite{trolle2014theory,pedersen2015intraband,Rohlfing2000,ridolfi2018}. 
Here, $N_X$ is the number of exciton states, $E_n$ is the energy of the $n$-th excited state ($n=1$ being the first exciton), $V_n^{a}=\braket{GS|v_{a}|X_n}$ is the velocity matrix element (VME) transition to the ground state and $V$ is the volume of the solid under periodic boundary conditions. We have dropped $\mathbf{Q}$ in the notation, since only excitons with $\mathbf{Q}=0$ can be excited by light incidence, and instead specify only the excitation index $n$ in the exciton coefficients, $A_{vc}^{n}(\mathbf{k})$. The VME is given by
\begin{equation} \label{eq:vme_ex}
    V_n^{a}=\sum_{cv \mathbf{k}}A_{vc}^{n}(\mathbf{k})v^{a}_{vc}(\mathbf{k})
\end{equation}
where $v^{a}_{vc}(\mathbf{k}) \equiv \braket{v\mathbf{k}|v_{a}|c\mathbf{k}}=i\hbar^{-1}\braket{v\mathbf{k}|[H_0,r_{a}]|c\mathbf{k}}$ ($H_0$ is the non-interacting or mean-field Hamiltonian). Note that the same $k$-dependent phase used in the BSE must be retained here for the valence and conduction states, see discussion below Eq.~\eqref{single-particle-state}. Such procedure ensures a proper evaluation of exciton VMEs with Eq.~\eqref{eq:vme_ex}.

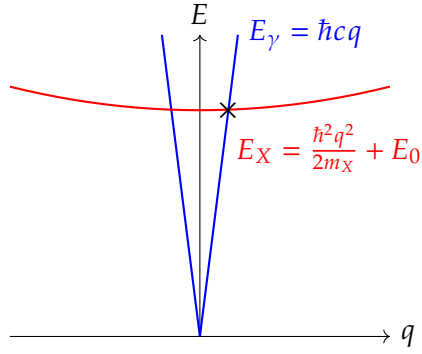
\begin{figure}
    \centering
    \begin{tikzpicture}
        \draw[->] (-2.5,0) -- (2.5,0) node[right] {$q$};
        \draw[->] (0,0) -- (0,4) node[above] {$E$};
        
        \draw[blue, thick] (0,0) -- (0.5,4) node[right] {\(E_{\gamma}=\hbar c q\)};
        \draw[blue, thick] (-0.5,4) -- (0,0) node[right] {};
        
        \draw[red, thick, domain=-2.5:2.5] plot (\x, {0.05*\x*\x + 3});
        \node[red] at (1.7, 2.5) {\(E_X=\frac{\hbar^2q^2}{2m_X} + E_0\)};
        
        \node at (0.37, 3) {\(\times \)};
    \end{tikzpicture}
    \caption[Exciton dispersion in relation to light cone]{Diagram showing the dispersion relation for light (blue) and a typical exciton (red). In emission and absorption processes, both energy and momentum must be conserved (which is the intersection of both dispersions, represented with the cross). Since photons are highly energetic, the transferred momentum is very low so in practice transitions are computed in the $q\to 0$ limit~\cite{palummo2015exciton, spataru2005theory}.}
\end{figure}

An exciton is said to be dark (or bright) if $V_n^{a}=0$ ($V_n^{a}\neq 0$), for light polarized along the $a$ direction. In general, the value of the VME will depend on the values of the single-particle velocity matrix elements $v^{a}_{vc}(\mathbf{k})$ and the exciton envelope $A_{vc}^{n}(\mathbf{k})$. Assuming a generic basis of localized states, the single-particle velocity matrix elements are given by~\cite{esteve2023comprehensive,lee2018tight}
\begin{equation} \label{eq:vme}
	\begin{split}	 &\braket{n\mathbf{k}|{\mathbf{v}}|n'\mathbf{k}}=\\
   & \sum_{\alpha \alpha'} (C^{n\mathbf{k}}_{\alpha})^*C^{n'\mathbf{k}}_{\alpha'}\nabla_{\mathbf{k}} H_{\alpha \alpha'}(\mathbf{k})
		+ i\sum_{\alpha \alpha'}(C^{n\mathbf{k}}_{\alpha})^*C^{n'\mathbf{k}}_{\alpha'}\Big[ \epsilon_{n}(\mathbf{k}) \mathbf{\xi}_{\alpha \alpha'}(\mathbf{k})-  \epsilon_{n'}(\mathbf{k})\mathbf{\xi}_{\alpha' \alpha}^{\ast}(\mathbf{k}) \Big]
	\end{split}
\end{equation}
where we have simplified the notation doing $i\alpha\to\alpha$. $\mathbf{\xi}_{\alpha \alpha'}(\mathbf{k})=i\braket{u_{\alpha\mathbf{k}}|\nabla_{\mathbf{k}}u_{\alpha' \mathbf{k}}}$ is the Berry connection between Bloch basis states. Substituting the expansion of the Bloch states in the local basis, one obtains
\begin{equation} \label{eq:quantities}
\mathbf{\xi}_{\alpha \alpha'}(\mathbf{k})=\sum_{\mathbf{R}}e^{i\mathbf{k}\cdot \mathbf{R}}\braket{\alpha \mathbf{0}|{\mathbf{r}}|\alpha' \mathbf{R}}+i\nabla_{\mathbf{k}}S_{\alpha \alpha'}(\mathbf{k})   
\end{equation}
In the case of a non-orthonormal local orbital basis, the overlap matrix $S_{\alpha \alpha'}(\mathbf{k})\equiv \braket{\alpha\mathbf{k}|\alpha'\mathbf{k}}$ is responsible for making the Berry connection non-hermitian, $\mathbf{\xi}_{\alpha \alpha'}(\mathbf{k})=\mathbf{\xi}_{\alpha' \alpha}^{\ast}(\mathbf{k})+i\nabla_{\mathbf{k}}S_{\alpha \alpha'}(\mathbf{k})$. If we have an orthogonal basis set, as in tight-binding models, the overlap matrix is a unitary matrix at all points of the Brillouin zone. Then the VMEs further simplify and read
\begin{equation} \label{eq:vmetba}
 v^{a}_{vc}(\mathbf{k})=\sum_{\alpha \alpha'}(C^{n\mathbf{k}}_{\alpha})^*C^{n'\mathbf{k}}_{\alpha'} \Bigg[\frac{\partial H_{\alpha \alpha'}(\mathbf{k})}{\partial k_a}+iH_{\alpha \alpha'}(\mathbf{k})(t^{a}_{\alpha'}-t^{a}_{\alpha}) \Bigg]
\end{equation}
This approximation is also common in the context of calculations based off maximally-localized Wannier orbitals~\cite{pizzi2020wannier90,ibanez2018ab}, where the inter-orbital position matrix elements are discarded, even though they can be obtained from the Wannier orbitals, as opposed to a TB calculation.

Finally, we can also obtain the non-interacting limit from Eq.~\eqref{eq:excitonkubo}. Turning off the interactions, for instance via infinite screening, excitons reduce to a single electron-hole pair, $A_{vc}(\mathbf{k})=\delta_{vv_0}\delta_{cc_0}\delta_{\mathbf{k}\mathbf{k}_0}$. The conductivity then becomes
\begin{align}
\label{eq:ipkubo}
    \sigma_{ab}(\omega)=\frac{\pi e^2 \hbar}{V} \sum_{cv\mathbf{k}} \frac{1} {E_{c \mathbf{k}}-E_{v\mathbf{k}}} v_{cv}^{a}(\mathbf{k})v_{vc}^{b}(\mathbf{k})
    \delta(\hbar\omega-(E_{c\mathbf{k}}-E_{v \mathbf{k}}))
\end{align}
From the frecuency-dependent optical conductivity we can obtain the absorbance, $S(\omega)=\sigma(\omega)/c\varepsilon_0$, which is the ratio of absorbed incident flux per frequency and length (considering vacuum surroundings)~\cite{wu2016theory, dresselhaus2022solid}. Note that this quantity is ill-defined for 2D lattice systems. In such case, all the absorbance is assumed to occur at $z=0$ reference plane of the material.

\subsection{Center-of-mass and relative velocity}

One could also want to determine the velocity of an excitonic state. This would be done naively simply computing the expectation value of the second quantized version of the velocity operator, similarly to the definition of the total spin:
\begin{equation}
  \mathbf{v} = \sum_{i,j}\mathbf{v}_{ij}c^{\dagger}_ic_j
\end{equation}
where the indices $i,j$ run over the band index and momentum $(n,\mathbf{k})$, and the velocity matrix elements are $\mathbf{v}_{ij}=\braket{i|\mathbf{v}|j}$. The expected velocity of an exciton state is then given by
\begin{subequations}  
\begin{align}
  \nonumber\braket{\mathbf{v}}_X&=\braket{X_n(\mathbf{Q})|\mathbf{v}|X_n(\mathbf{Q})}  \\
  &=\sum_{v,c,\mathbf{k}}A^n_{vc}(\mathbf{k}, \mathbf{Q})\left[\sum_{c'}(A^n_{vc'}(\mathbf{k},\mathbf{Q}))^*\mathbf{v}_{c'\mathbf{k}+\mathbf{Q},c\mathbf{k}+\mathbf{Q}} - \sum_{v'}(A^n_{vc'}(\mathbf{k},\mathbf{Q}))^*\mathbf{v}_{v\mathbf{k},v'\mathbf{k}}\right]  \label{eq:total_electronic_velocity} \\
  & \equiv \braket{\mathbf{v}_e}_X - \braket{\mathbf{v}_h}_X \label{eq:relative_exciton_velocity}
\end{align}
\end{subequations}
As opposed to the optical conductivity, which involved velocity matrix elements between valence and conduction bands, the velocity matrix elements in this case are between either valence or conduction bands. They are obtained generally using expression~\eqref{eq:vme}. We see that the expression we have obtained is completely analogue to that of the total projected spin of the exciton $\braket{S^T_z}_X$ in~\eqref{spin_total}. However, the interpretation differs here. The velocity of a hole is given by $\mathbf{v}_h=\hbar^{-1}\mathbf{\nabla}_{\mathbf{k}_h}\varepsilon_h(\mathbf{\mathbf{k}}_h)$, where $\mathbf{k}_h=-\mathbf{k}_e$ and $\varepsilon_h(\mathbf{k}_h)=-\varepsilon_e(-\mathbf{k}_e)$. If the system is time-reversal invariant, then using the chain rule it follows that $\mathbf{v}_h=\hbar^{-1}\nabla_{\mathbf{k}_h}\varepsilon_h(\mathbf{k}_h)=\hbar^{-1}\nabla_{\mathbf{k}_e}\varepsilon_e(\mathbf{k}_e)=\mathbf{v}_e$. Meaning that the velocity of a hole is exactly the same as the velocity of the corresponding electron~\cite{kittel2021introduction}, while for the spin it is the opposite. This implies then that in the previous expression, the total electronic velocity of the exciton can be written as the relative velocity of its components, as in~\eqref{eq:relative_exciton_velocity}.
Therefore, as we did initially for the total spin projection, we could have written directly the velocity in terms of the velocity of the conduction electron, minus the velocity of the valence hole
\begin{equation}
  \mathbf{v}= \sum_{c,c',\mathbf{k}}\mathbf{v}_{cc'}(\mathbf{k})c^{\dagger}_{c\mathbf{k}}c_{c'\mathbf{k}} - \sum_{v,v',\mathbf{k}}\mathbf{v}_{vv'}(\mathbf{k})c_{v'\mathbf{k}}c^{\dagger}_{v\mathbf{k}} \equiv \mathbf{v}_e - \mathbf{v}_h
\end{equation}
which yields the same expression as~\eqref{eq:total_electronic_velocity} after taking the expected value with an exciton $\ket{X_n(\mathbf{Q})}$. This interpretation of the total electronic velocity of the exciton as the relative velocity of its components suggests that it is possible to define a center-of-mass velocity for the exciton as well, which would be given simply by
\begin{equation}
  \label{eq:center_of_mass_exciton_velocity}
  v_{\text{CM}} \equiv \braket{\mathbf{v}_e}_X + \braket{\mathbf{v}_h}_X
\end{equation}
For $\mathbf{Q}=0$, time-reversal invariance imposes that both the center-of-mass and relative velocities of the exciton must be zero. Note that in general the center-of-mass velocity obtained this way will differ from a naive definition following from the velocity of single-particle states, $\mathbf{v}_{\text{CM}}=\hbar^{-1}\nabla_{\mathbf{Q}} E_X(\mathbf{Q})$ (see Fig.~\ref{fig:transitionsQ_velocities} in Section~\ref{sec:transitions_Q0}). Thus, the relative or total electronic velocity of the exciton gives a measure of the electric current induced by the exciton, whereas the center-of-mass velocity gives a measure of the motion of the exciton as a whole.

\subsection{Symmetry}\label{sec:symmetry}
For both the reciprocal and the real-space probability densities, one could expect them to have the symmetries of the crystal, since $[H,C]=0$, where $C$ is any symmetry operator from the space group $\mathcal{G}$. We know from group and representation theory~\cite{bir1974symmetry, dresselhaus2007group, bradley2009mathematical} that in general eigenstates of the Hamiltonian will transform according to the irreducible representations of the corresponding group. Then, if a state is degenerate (meaning that the state belongs to a representation of dim $>1$), it will not be necessarily an eigenstate of the symmetry operators and consequently the associated densities will not be invariant under symmetry transformations. Still, in this case it is possible to define a probability density that preserves the symmetry of the crystal for each degenerate subspace,
\begin{equation}
\label{wf_invariant}
    |\psi_X(\mathbf{r}, \mathbf{r}_h)|^2=\sum_n|\psi_X^{(n)}(\mathbf{r}, \mathbf{r}_h)|^2
\end{equation}
where the index $n$ runs over exciton states degenerate in energy. An analogous expression holds for the $\mathbf{k}$ wavefunction. When we introduce the code for exciton calculations, it will be a good practice to check that the resulting probability densities preserve the symmetry of the crystal to ensure that the exciton calculation was converged correctly. 
The above statement is formulated as follows: Given $|\Psi|^2=\sum_n|\psi_n|^2$, where $\psi_n$ denotes the wavefunction of the exciton states on some degenerate subspace of $PHP$, and given some symmetry operation $C$ such that $[H, C] = 0$, then
\begin{equation}
\label{symmetry_proof}
   C|\Psi|^2 = |\Psi|^2.
\end{equation}
To prove it, first we consider the action of the symmetry operator $C$ on an exciton state. Given that the eigenstates of a degenerate subspace of $H$ are not in general eigenstates of $C$, the most general action is to mix the degenerate states, i.e.
\begin{equation}
    C\psi_n=\sum_i\alpha_{in}\psi_i
\end{equation}
The coefficients $\alpha_{in}$ are the matrix elements of $C$. To prove~\eqref{symmetry_proof}, we need to know the action of $C$ on the squared amplitude, $C|\psi_n|^2$. So first we want to prove the following property:
\begin{equation}
    C|\psi_n|^2=|C\psi_n|^2
\end{equation}
This can be proven using the action of the symmetry operation on the coordinate of the wavefunction, i.e. $C\psi_n(x) = \psi_n(C^{-1}x)$:
\begin{align}
     C|\psi_n|^2(x)=|\psi_n|^2(C^{-1}x)=\psi_n(C^{-1}x)\psi_n^*(C^{-1}x)
    =C\psi_n(x)C\psi_n^*(x) = |C\psi_n|^2(x)
\end{align}
where we have also used that $C\psi_n^*(x) = (C\psi_n)^*(x)$. This last identity can be proved conjugating the action of the symmetry on the coordinates,
    $(C\psi_n)^*(x) = \psi_n^*(C^{-1}x) = C\psi_n^*(x)$.
This enables us to compute $C|\psi_n|^2$ in terms of an expansion on the states of the degenerate subspace:
\begin{equation}
    C|\psi_n|^2=|C\psi_n|^2=\left|\sum_i\alpha_{in}\psi_i\right|^2=\sum_{i,j}\alpha_{in}\alpha^*_{jn}\psi_i\psi^*_j
\end{equation}
Finally, with this expression we can prove the symmetry invariance of $|\Psi|^2 = \sum_n|\psi_n|^2$. To do so, we act with the symmetry operation $C$ on $|\Psi|^2$:
\begin{align}
     C|\Psi^2| &= \sum_n C|\psi_n|^2 = \sum_n \left[\sum_{i,j}\alpha_{in}\alpha^*_{jn}\psi_i\psi^*_j\right]
    = \sum_{ij}\left[\sum_n\alpha_{in}\alpha^*_{jn}\right]\psi_i\psi^*_j = \sum_{i}|\psi_i|^2 = |\Psi|^2
\end{align}
where we have used that $C$ is unitary, i.e. $\sum_n\alpha_{in}\alpha^*_{jn} = \delta_{ij}$. This proves that the sum of the squared amplitude of each degenerate state is invariant under the symmetry operations.
\\

So far we have considered some general symmetry $C$ such that $[H, C] = 0$. For the abstract, unrepresented Hamiltonian, given any operation $C$ of the space group of the solid, it is true that $[H, C]= 0$. However, we are not working with the total Hamiltonian, but with a sector of it. So one must actually look for symmetry operations that commute with $PHP$,
\begin{equation}
    [PHP, C] = 0
\end{equation}
Since the sectors of electron-hole pairs of different momentum are disconnected, we can define $\Tilde{H}(\mathbf{Q}) = P_{\mathbf{Q}}HP_{\mathbf{\mathbf{Q}}}$, where $P_{\mathbf{Q}}$ is the projector over electron-hole pairs of $\mathbf{Q}$ total momentum. This Hamiltonian is analogous to the Bloch Hamiltonian $H(\mathbf{k})$, and it can be shown that it transforms in the same way:
\begin{equation}
    C^{-1}\Tilde{H}(\mathbf{Q})C = \Tilde{H}(C^{-1}\mathbf{Q}) 
\end{equation}
meaning that for $\mathbf{Q}=0$ the symmetry group is the crystallographic space group, but for $\mathbf{Q}\neq 0$ the Hamiltonian is invariant only under symmetry operations of the little group of $\mathbf{Q}$, whose irreducible representations thus dictate the (unitary) transformation properties of the $\mathbf{Q}-$excitonic wavefunctions. The proof of this property is as follows: First we expand the definition of $H(\mathbf{Q})$,
\begin{align}
     C^{-1}H(\mathbf{Q})C &= C^{-1}P_{\mathbf{Q}}CC^{-1}HCC^{-1}P_{\mathbf{Q}}C = C^{-1}P_{\mathbf{Q}}CHC^{-1}P_{\mathbf{Q}}C 
\end{align}
where we have used that $[H,C]=0$. So we only need to see how the projectors transform under the symmetry operation to determine how $H(\mathbf{Q})$ transforms.
\begin{align} \label{PQtransformation}
    C^{-1}P_{\mathbf{Q}}C = \sum_{\mathbf{k},v,c}C^{-1}c^{\dagger}_{c,\mathbf{k+Q}}c_{v\mathbf{k}}\ket{GS}\bra{GS}c^{\dagger}_{v\mathbf{k}}c_{c,\mathbf{k+Q}}C
\end{align}
Inserting identities, we can transform each creation/annihilation operator according to $C^{-1}c^{\dagger}_{n\mathbf{k}}C = c^{\dagger}_{n,C^{-1}\mathbf{k}}$, up to an arbitrary phase that is cancelled in~\eqref{PQtransformation}. From the definition of the Fermi sea~\eqref{FermiSea}, it follows that it is invariant under point group operations, i.e. $C\ket{GS} = \ket{GS}$ (again, up to an arbitrary phase that is cancelled), we arrive to the following expression:
\begin{align}
     &C^{-1}P_{\mathbf{Q}}C = \sum_{\mathbf{k},v,c}c^{\dagger}_{c,C^{-1}\mathbf{k}+C^{-1}\mathbf{Q}}c_{v,C^{-1}\mathbf{k}}\ket{GS}\bra{GS}c^{\dagger}_{v,C^{-1}\mathbf{k}}c_{c,C^{-1}\mathbf{k}+C^{-1}\mathbf{Q}} = P_{C^{-1}\mathbf{Q}}
\end{align}
where we reordered the summation using the C-invariance of the BZ in $\mathbf{k}-$space to arrive to the final expression for the transformed projector. Therefore, the projected exciton Hamiltonian $H(\mathbf{Q})$ transforms in the same way as the Bloch Hamiltonian:
\begin{align}
    C^{-1}H(\mathbf{Q})C = H(C^{-1}\mathbf{Q})
\end{align}
Likewise, the application of time-reversal yields $T^{-1}H(\mathbf{Q})T=H(-\mathbf{Q})$ whenever it is a symmetry of the system. Thus, for any exciton calculation at $\mathbf{Q}$, the degeneracies (without taking into account spin) of the spectrum must coincide with the dimension of the associated irreducible representations, which we can extract from the character tables. Likewise, if we consider the total exciton wavefunction of a degenerate subspace, it must be invariant under all symmetry operations.
\chapter{Efficient computation of excitons in two-dimensional materials with the Xatu code}\label{chapter:chapter3}

After establishing the mathematical framework for excitons, we can proceed to the implementation of the algorithms required to compute the exciton spectrum. In this chapter we will cover the development of the \texttt{Xatu} code~\cite{uria_xatu}, a \texttt{C++} program and library designed to obtain the exciton spectrum in two-dimensional materials, with applicability extending to molecules, 1D and 3D systems. The code is centered around solving the Bethe-Salpeter equation, as obtained from the exact diagonalization approach together with the tight-binding approximation. As we will demonstrate, the approximations discussed in the previous chapter —namely, the absence of microscopic screening and the use of point-like orbitals— enable a rapid determination of the exciton states, offering a substantial speed advantage over existing ab-initio methods. With this approach, the code fills a gap in the field of computational optoelectronics, namely that of effective exciton calculations, given that multiple tools exist for ab-initio excitons, but none for effective, approximate calculations\footnote{During the publication of this work, another program also focusing on effective exciton computations was released~\cite{dias2023wantibexos}.}.

The code has been designed with versatility in mind: it can work with a variety of inputs, including tight-binding models that may be purely effective, based on Slater-Koster parametrizations or obtained from Wannierizations~\cite{pizzi2020wannier90}. Additionally, it can handle electronic band structures obtained from DFT calculations, provided these are based on local basis calculations (i.e. Gaussian or numerical atomic orbitals~\cite{soler2002siesta}). The first part of this chapter reviews the code's structure and the algorithms implemented to solve the BSE\@. In the second half we present the validation of the code, which was done characterizing the exciton spectrum in hexagonal boron nitride, hBN, as well as molybdenum disulfide, MoS$_2$. In both cases, the results are in agreement with available literature, highlighting the code's reliability and potential applications.

\section{Implementation and algorithms}
The programming languages of choice for the implementation of the exciton theory were \texttt{C++} and \texttt{Fortran}, which are the usual options for heavy numerical computations. In the case of \texttt{C++}, to facilitate manipulation of matrices we use the library \texttt{Armadillo}~\cite{Sanderson2016}, on top of the usual libraries for linear algebra (\texttt{BLAS} and \texttt{LAPACK}). The core of the code was written in \texttt{C++}, except the post-diagonalization calculation of the optical conductivity which is written in \texttt{Fortran}. This routine is wrapped inside the \texttt{C++} library. 

The software was designed with a hybrid approach in mind: previous packages such as DFT codes require the preparation of input files, which are then fed to the program and result in some output files which may be post-processed to extract information. We propose to use the same scheme, i.e.\ to prepare an input file which describes the system where we want to compute the excitons (namely the Hamiltonian $H_0$), and another one with the description of the excitons (participating bands, $\mathbf{k}$ mesh, etc.). However, there is an alternative usage, which is employing directly the exciton API defined to build the program. This is a common approach, where one builds a library to expose some functionality to the user (e.g.\ \texttt{Python} libraries). Therefore, one can define some system and script the computation of excitons using the API\@. This is advised whenever we are interested in performing some manipulation of the excitons, and not only obtaining the spectrum or the absorption. 

There is a third approach, consisting on using the system files to leverage the definition of the system to other programs (e.g. DFT), and then use the API instead of the exciton configuration file. The CLI option parsing has been done using the header-only library \href{https://tclap.sourceforge.net/}{TCLAP}, which is distributed with this package. The software is currently available from its \href{http://github.com/xatu-code/xatu}{repository}, where it is periodically updated with bug fixes and new features.

\subsection{Class hierarchy}

The library follows the object-oriented programming (OOP) paradigm, and as such it is structured around a set of classes to encapsulate the different parts of the exciton computation. This allows to hide the complexity of the algorithms and data structures used, and to provide a clean and more readable interface, which can then be used by either the user or the developer to build more complex programs. While a functional programming paradigm would also have suited the library, it would have resulted in more verbose code, mainly due to the need to pass around the state of the computation (i.e.\ functions with multiple arguments). With OOP, the state is stored in the objects themselves via their attributes, meaning that it suffices to pass the objects and use their methods to perform the computation, typically resulting in cleaner code. The downside of this encapsulation is that it can restrict the flexibility of use and reduce the extensibility of the library, depending on how the methods and classes are coupled. Nevertheless, we opted for the OOP approach as it results in more readable code, and ultimately with a good design it can be as flexible as a functional one.

Next we show the UML diagram with the hierarchy of classes within the code, namely the inheritance relations as well as compositions, or classes passed as arguments. For simplicity, we do not show the attributes nor the methods of each class, but only the relations between them. We detail in what follows the role of each class in the code.
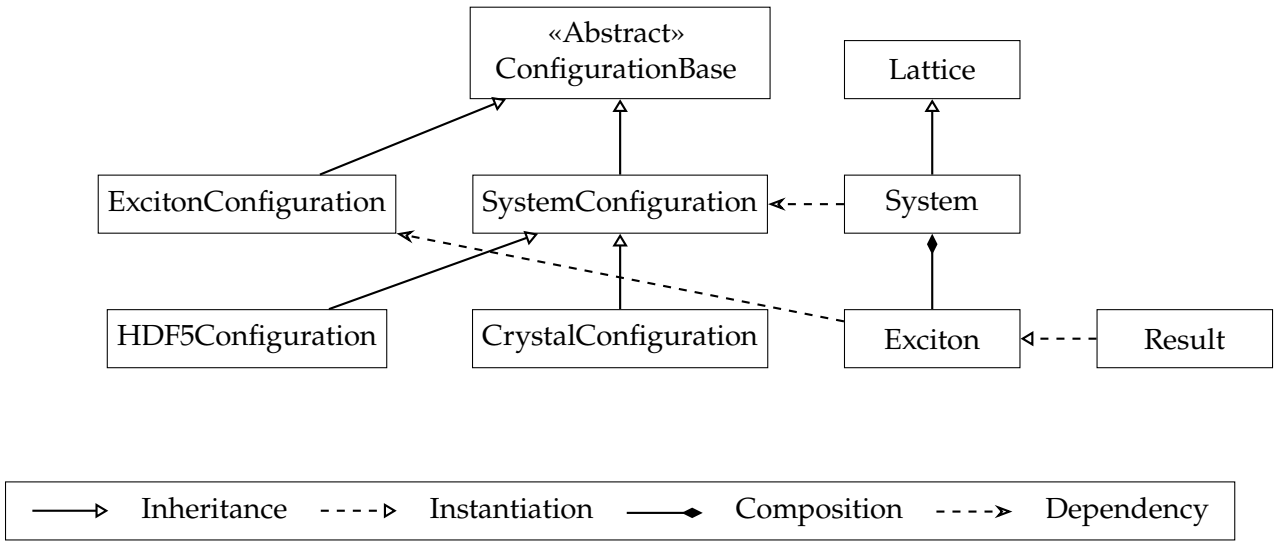
\begin{figure}[h]
  \centering
  \begin{tikzpicture}[
    class/.style={rectangle, draw=black, fill=white, text centered, minimum height=2em, minimum width=6em},
    abstractClass/.style={rectangle, draw=black, fill=white, text centered, minimum height=3em, minimum width=6em, align=center},
    inheritance/.style={draw, -{Triangle[open]}, thick},
    instantiation/.style={draw, -{Triangle[open]}, dashed, thick},
    composition/.style={draw, -{Diamond}, thick},
    dependency/.style={draw, -{Stealth[open]}, dashed, thick}
  ]
  
    \node[abstractClass] (ConfigurationBase) {
      \begin{tabular}{c}
        \textless\textless Abstract\textgreater\textgreater \\
        ConfigurationBase
      \end{tabular}
    };
    \node[class, below=of ConfigurationBase] (SystemConfiguration) {SystemConfiguration};
    \node[class, left=of SystemConfiguration] (ExcitonConfiguration) {ExcitonConfiguration};
    \node[class, below=of SystemConfiguration] (CrystalConfiguration) {CrystalConfiguration};
    \node[class, below=of ExcitonConfiguration] (HDF5Configuration) {HDF5Configuration};

    \draw[inheritance] (SystemConfiguration) -- (ConfigurationBase);
    \draw[inheritance] (ExcitonConfiguration) -- (ConfigurationBase);
    \draw[inheritance] (CrystalConfiguration) -- (SystemConfiguration);
    \draw[inheritance] (HDF5Configuration) -- (SystemConfiguration);

    \node[class, right=of SystemConfiguration] (System) {System};
    \node[class, above=of System] (Lattice) {Lattice};
    \node[class, below=of System] (Exciton) {Exciton};

    \draw[inheritance] (System) -- (Lattice);
    \draw[composition] (Exciton) -- (System);
    \draw[dependency] (System) -- (SystemConfiguration);
    \draw[dependency] (Exciton) -- (ExcitonConfiguration);

    \node[class, right=of Exciton] (Result) {Result};

    \draw[instantiation] (Result) -- (Exciton);

    \node[rectangle, draw=black, fill=white, text centered, minimum height=2em, minimum width=12em, below=of CrystalConfiguration, yshift=-0.5cm] (Legend) {
      \begin{tabular}{cccccccc}
        \begin{tikzpicture}
          \draw[inheritance] (0,0) -- ++(1,0);
        \end{tikzpicture} & Inheritance &
        \begin{tikzpicture}
          \draw[instantiation] (0,0) -- ++(1,0);
        \end{tikzpicture} & Instantiation &
        \begin{tikzpicture}
          \draw[composition] (0,0) -- ++(1,0);
        \end{tikzpicture} & Composition &
        \begin{tikzpicture}
          \draw[dependency] (0,0) -- ++(1,0);
        \end{tikzpicture} & Dependency \\
      \end{tabular}
    };
  \end{tikzpicture}
  \caption[UML diagram of the \texttt{Xatu} code]{Unified Modeling Language (UML) diagram of the \texttt{Xatu} code, showing the class hierarchy used in the design of the library. We only show the relation between the different classes and skip their attributes and methods altogether.}
\end{figure}

The library defines in the first place an abstract class named \texttt{ConfigurationBase}, which provides basic and general methods for parsing text files, as well as virtual methods that are intended to be overwritten by the specific configuration classes, for the different input files used in the code. Thus, the classes \texttt{ExcitonConfiguration} and \texttt{SystemConfiguration} extend \texttt{ConfigurationBase}, and define the parameters that are expected to be read from the input files, as well as the format that these files must have in order to parse the information correctly. \texttt{ExcitonConfiguration} is responsible for parsing the parameters relative to the BSE (e.g.\ number of bands or the screening parameters), while \texttt{SystemConfiguration} is in charge of reading the system file, which contains the single-particle Hamiltonian and the lattice information. \texttt{SystemConfiguration} specifies the necessary quantities from the single-particle model that must be present to compute the excitons, and as such, it can be used as the base class for other configuration classes to parse different file formats. Namely, the \texttt{HDF5Configuration} and \texttt{CrystalConfiguration} classes inherit from \texttt{SystemConfiguration}, in such a way that they override the methods relative to the parsing of the file format, but fill the same attributes as the base class. This allows us to use polymorphism: The \texttt{System} class will only expect a \texttt{SystemConfiguration} class, meaning that we can provide any of the derived classes and the code will work as expected, since all of them have the same attributes. The class \texttt{HDF5Configuration} is used to read the system from a HDF5 file, while \texttt{CrystalConfiguration} is used to read the Hamiltonian from a CRYSTAL calculation, i.e.\ a DFT one based on Gaussian basis sets.

Next, we define the classes responsible for the actual calculation of the BSE\@. In this case, we have separated the functionality in different classes to enhance readability. First, we define the \texttt{Lattice} class, which is responsible for storing the Bravais lattice, and holds all the methods relative to the manipulation of both the direct and reciprocal lattices, such as generating the BZ mesh.
Then, we define the \texttt{System} which inherits from \texttt{Lattice}, and is basically responsible for storing and computing all the single-particle properties of the model, mainly used to determine the bands of the model. Finally, the class \texttt{Exciton} is implemented, which defines all the methods necessary to build and solve the Bethe-Salpeter equation. While this class was originally an extension of \texttt{System}, we finally opted for a composition, in an attempt to make a clear distinction between the single-particle properties and the BSE calculation (i.e.\ we consider that the \texttt{Exciton} calculation "has" a \texttt{System} instead of "is"). Finally, the results from the diagonalization of the BSE (eigenenergies and eigenvectors) are stored and returned by the \texttt{Exciton} class in the \texttt{Result} class, which contains all the routines to perform the post-processing of the exciton states, such as computing the optical conductivity or the exciton wavefunctions.

\subsection{Bethe-Salpeter equation solver}
\begin{figure}[h]
    \centering
    \begin{tikzpicture}
        \node (start) [draw, ellipse] {Start};

        \node (bigbox) [draw, rectangle, minimum width=6cm, minimum height=1.8cm, below of=start, anchor=north] {};

        \node at (bigbox.north) [below=0.1cm, xshift=-2.3cm] {\textit{Input}};

        \node (systemfile) [draw, trapezium, trapezium left angle=75, trapezium right angle=105, below of=start, node distance=2.3cm, xshift=-1.5cm] {System file};

        \node (excitonfile) [draw, trapezium, trapezium left angle=75, trapezium right angle=105, below of=start, node distance=2.3cm, xshift=1.5cm] {Exciton file};

        \node (computeH0) [draw, rectangle, below of=bigbox, node distance=2.5cm, align=center]
            {Diagonalize $H_0(\mathbf{k})\ \forall \mathbf{k}\in$ BZ \\ Store $\varepsilon_{n\mathbf{k}}$ and $U(\mathbf{k})$};

        \node (method) [draw, diamond, aspect=2, below of=computeH0, node distance=2.5cm, align=center] {Method to use};

        \node (auxRS) [left=of method, xshift=-1.5cm] {};

        \node (computeRS) [draw, rectangle, below of=auxRS, node distance=1.5cm, align=center] {Compute and \\store $V_{ij}(\mathbf{k})$};

        \node (computeBSE) [draw, rectangle, below of=method, node distance=3cm, align=center] 
            {Compute BSE \\ $H_{vc,v'c'}(\mathbf{k},\mathbf{k}',\mathbf{Q})$};

        \node (diagonalize) [draw, rectangle, below of=computeBSE, node distance=2.cm, align=center] 
            {Diagonalize BSE \\ Store $E_X^n(\mathbf{Q})$ and $A^n_{vc}(\mathbf{k},\mathbf{Q})$};

        \node (converge_energies) [draw, diamond, aspect=2, below of=diagonalize, node distance=2.5cm, align=center] {$E^n_X(\mathbf{Q})$ converged?};

        \node (bigbox_output) [draw, rectangle, minimum width=10cm, minimum height=2cm, below of=converge_energies, node distance=3cm] {};

        \node at (bigbox_output.north) [below=0.1cm, xshift=-4cm] {\textit{Output}};

        \node (wavefunctions) [draw, trapezium, trapezium left angle=75, trapezium right angle=105, below of=bigbox_output, node distance=0.3cm, xshift=-2.3cm, align=center]
            {Wavefunctions\\ $|\psi^n_X(\mathbf{k})|^2$, $|\psi^n_X(\mathbf{r}_e,\mathbf{r}_h)|^2$};

        \node (optical_conductivity) [draw, trapezium, trapezium left angle=75, trapezium right angle=105, below of=bigbox_output, node distance=0.3cm, xshift=2.3cm, align=center]
            {Optical conductivity\\ $\sigma_{\alpha}(\omega)$};

        \node (increaseNk) [draw, rectangle, right of=converge_energies, node distance=6cm, align=center] {Increase $N_{\mathbf{k}}$ \\ and/or $N_{v/c}$};

        \node (symmetry) [draw, diamond, aspect=2, below of=bigbox_output, node distance=3cm, align=center] {Symmetries \\ preserved?};

        \node (End) [draw, ellipse, below of=symmetry, node distance=2.5cm] {End};

        \draw [->] (start) -- (bigbox);
        \draw [->] (bigbox) -- (computeH0);
        \draw [->] (computeH0) -- (method);
        \draw [->] (method) -- ++(-1.8,0) -| node[anchor=south, xshift=1.4cm] {Real space} (computeRS);
        \draw [->] (computeRS) -- ++(0,-1.5) -- (computeBSE);
        \draw [->] (method) -- node[anchor=west, yshift=0cm] {Reciprocal space} (computeBSE);
        \draw [->] (computeBSE) -- (diagonalize);
        \draw [->] (diagonalize) -- (converge_energies);
        \draw [->] (converge_energies) -- node[anchor=east] {Yes} (bigbox_output);
        \draw [->] (converge_energies) -- node[anchor=south, xshift=-0.5cm] {No} (increaseNk);
        \draw [->] (increaseNk) -- ++(0, 12.1) -- (excitonfile);
        \draw [->] (bigbox_output) -- (symmetry);
        \draw [->] (symmetry) -- node[anchor=east] {Yes} (End);
        \draw [->] (symmetry) -- node[anchor=south, xshift=-1cm] {No} ++(6,0) -- (increaseNk);
    \end{tikzpicture}
    \caption[Flow diagram of an exciton calculation with \texttt{Xatu}]{Flow diagram showing the typical steps in an exciton calculation with \texttt{Xatu}.}\label{fig:xatu_flow_diagram}
\end{figure}
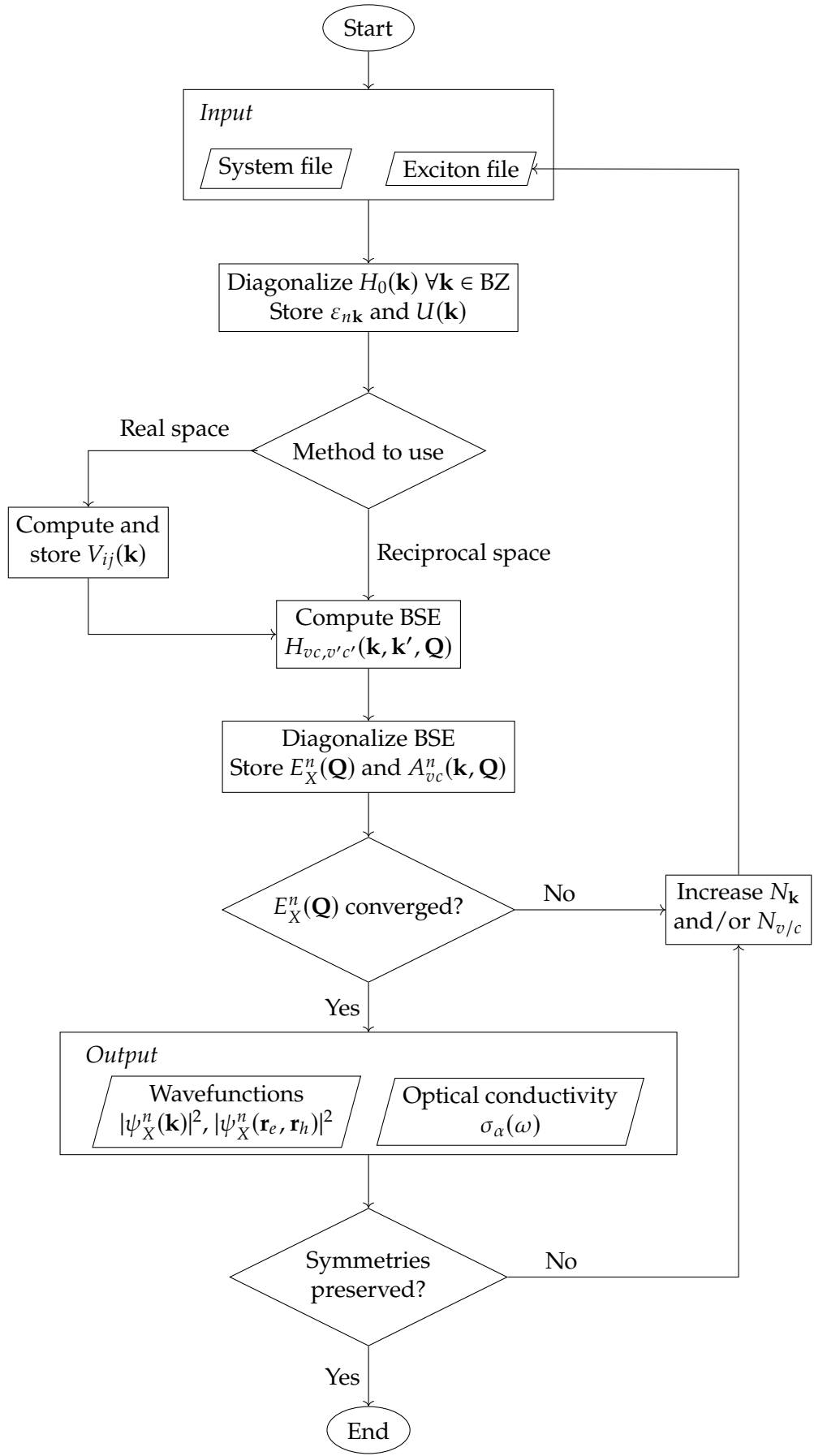

In this subsection we discuss the global algorithm behind the computation of the exciton spectrum. This algorithm is summarized in the flow diagram in Fig.~\ref{fig:xatu_flow_diagram}. Here we describe the main steps of a typical exciton calculation, as done by the \texttt{Xatu} code in its executable form (since using the library one can customize and perform alternative workflows, although the core part which is setting and solving the BSE is done in the same way).

A calculation starts by reading the system and exciton configuration files, used to define the single-particle Hamiltonian and all the parameters relative to the BSE\@. Then, the first step is to obtain the mesh of the BZ, according to the number of unit cells $N_{\mathbf{k}}$ specified in the exciton file. With all the $\mathbf{k}$ points determined, we diagonalize the Bloch Hamiltonian $H_0(\mathbf{k})$ $\forall\mathbf{k}$ and store the associated energies $\varepsilon_{n\mathbf{k}}$ and eigenstates $U(\mathbf{k})$. As we will discuss in the next section, this allows to reduce the time complexity of building the BSE matrix. Next, depending on the method chosen, the interaction matrix elements of the BSE are computed, either in real-space, which presents an intermediate step (again, to reduce time complexity), or with the reciprocal space method which is done on the fly directly. Once the BSE is built, the matrix is diagonalized using one of the multiple methods available to obtain the exciton energies $E_X^n(\mathbf{Q})$ and states $A_{vc}^n(\mathbf{k},\mathbf{Q})$. At this point, we can check whether the calculation is converged checking the exciton energies, which amounts to ensuring that degenerate states are well-identified (close enough in energies up to a user-defined threshold), and that energies do not change in general as we increase $N_{\mathbf{k}}$. If it is not the case, we increase $N_{\mathbf{k}}$ in the exciton configuration file and repeat the whole calculation. In general, the calculation should also be converged with respect to the number of bands participating, $N_{v/c}$. The number of participating bands typically determines the energy interval where excitons can be regarded as converged (for sufficiently high $N_{\mathbf{k}}$). Meaning that for the lowest excitons it typically suffices to use $N_{v/c}\sim 1-2$, whereas for more energetic excitons the number of bands required increases, which follows directly from the energy of the non-interacting electron-hole pairs associated to the furthest bands. Once the energies are converged, we can post-process the states to obtain the probability densities (in real or reciprocal space) or the optical conductivity. These quantities must preserve the symmetries of the crystal; in case that they do not (supposed that the single-particle model is well-defined) it is likely that more bands need to be included in the calculation, since a band crossing can induce this symmetry breaking of the excitons. If all quantities transform as expected, then we consider the calculation to be correct and finished.

\clearpage

\subsection{Complexity analysis}

Next we will discuss the numerical implementation of the exciton computation and related quantities. Solving the Bethe-Salpeter equation (\ref{bse}) amounts to diagonalizing the corresponding matrix $PHP$.
Diagonalization is done using the standard linear algebra libraries, meaning that the main problem is constructing $PHP$ as fast as possible. Consider a system formed by $N_{\mathbf{k}}$ unit cells in total (meaning $\sqrt{N_{\mathbf{k}}}$ along each Bravais vector for a two-dimensional system). To treat the interaction rigorously, one has to compute the excitons on a BZ mesh with the same number of $\mathbf{k}$ points as unit cells, due to the periodic boundary conditions. Therefore, one has to compute $N_{\mathbf{k}}^2$ matrix elements, and each of them requires computing the lattice Fourier transform, which involves summations over the $N_{\mathbf{k}}$ unit cells\footnote{This complexity analysis applies only to the real-space method. For the reciprocal space one, each matrix element involves instead a summation over reciprocal lattice vectors $\mathbf{G}$. This is done on the fly, meaning that the reciprocal method algorithm has an added $\mathcal{O}(N_{\mathbf{G}})$ factor, where $N_{\mathbf{G}}$ is the number of reciprocal lattice vectors used.}. This has to be done for all possible band pairs $N_B=N_vN_c$, where $N_{v/c}$ is the number of valence/conduction bands. Then, a naive implementation of~\eqref{bse} would have $\mathcal{O}(N_{\mathbf{k}}^3N_B^2)$ time complexity, on par with matrix diagonalization algorithms. Note that each interaction matrix element also requires knowing the tight-binding coefficients $\{C^{n\mathbf{k}}_{i\alpha}\}$. If the dimension of the Bloch Hamiltonian is $N_H$, then diagonalizing on the fly the system for each element of the BSE would result in time $\mathcal{O}(N_{\mathbf{k}}^2N_B^2(N_{\mathbf{k}} + N_{H}^3))$.

The easiest way to reduce the time complexity of the BSE construction is to increase the space complexity, i.e.\ to precompute and store quantities that appear multiple times, instead of computing them on the fly. This can be done for the Bloch Hamiltonian eigenvectors. Before constructing $PHP$, we diagonalize $H_0(\mathbf{k})$ $\forall\mathbf{k}\in\text{BZ}$, and store the eigenvectors. At this point, if we were to store all eigenvectors, the spatial complexity would go from $\mathcal{O}(1)$ to $\mathcal{O}(N_{\mathbf{k}}N_H^2)$. Since we only need the eigenvectors corresponding to the bands that participate in the exciton definition, it suffices to store only those, meaning that the spatial complexity would be $\mathcal{O}(N_{\mathbf{k}}N_H(N_v+N_c))$, i.e.\ we have to store $N_{\mathbf{k}}$ matrices of size $N_H\times (N_v+N_c)$. Accessing directly the eigenvectors results in a time complexity of $\mathcal{O}(N_{\mathbf{k}}^3N_B^2 + N_{\mathbf{k}}N_H^3)$. 

The same could be done for the lattice Fourier transform $V_{ij}(\mathbf{k}-\mathbf{k}')$. Since it depends on the difference between two $\mathbf{k}$ points, we could simply store $V_{ij}$ for each pair of $\mathbf{k}$ points. This implies high spatial complexity $\mathcal{O}(N_{\mathbf{k}}^2)$, but overall it does not report any speed advantage, since precomputing this would be of order $\mathcal{O}(N_{\mathbf{k}}^3)$. However, it is possible to reduce the time cost of the algorithm: as long as the $\mathbf{k}$ point mesh covers the whole BZ uniformly\footnote{Specifically, we need a BZ mesh such that it forms a vector space (modulus $\mathbf{G}$ vectors), i.e.\ it must include the identity element for addition $\mathbf{k}=0$.} (as given by Monkhorst-Pack), then we can map the $\mathbf{k}$ point difference back to a single $\mathbf{k}$ point using the periodicity of $V_{ij}(\mathbf{k}-\mathbf{k}')$:
\begin{align}\label{eq:periodic_k}
    \forall \mathbf{k},\mathbf{k}'\in\text{BZ, } \exists \mathbf{G}\in\text{Reciprocal lattice, }& \mathbf{k}''\in\text{BZ}
    \text{ s.t. } \mathbf{k}-\mathbf{k}' = \mathbf{G} + \mathbf{k}''
\end{align}
Therefore, it suffices to compute and store $V_{ij}(\mathbf{k})$ $\forall \mathbf{k}\in\text{BZ}$. Then, when initializing the matrix elements of $PHP$, one has to find the vector $\mathbf{k}''$ such that it verifies~\eqref{eq:periodic_k}. The time complexity now is $\mathcal{O}(N_{\mathbf{k}}^2)$, which is a reduction of an order of magnitude. The space complexity is also reduced, being now $\mathcal{O}(N_{\mathbf{k}})$.

With this, the algorithm for determining $PHP$ has time order $\mathcal{O}(N_{\mathbf{k}}^2N_B^2+N_{\mathbf{k}}N_H^3)$, and the memory requirements are $\mathcal{O}(N_{\mathbf{k}} + N_{\mathbf{k}}N_H(N_v+N_c))=\mathcal{O}(N_{\mathbf{k}}N_H(N_v+N_c))$. As we will see, this allows for very fast computation of the BSE matrix, meaning that the main bottleneck lies in the diagonalization, as it often happens. In some cases we might be interested in the whole spectrum, but usually it suffices to determine the lowest energy eigenstates. To address this, the code includes a custom implementation of the Davidson algorithm, which is suited to obtain the ground state of quantum chemistry Hamiltonians~\cite{DAVIDSON1975}.\\

So far the discussion has been focused on how to reduce the complexity of the algorithm, but it is equally important to comment on how to perform the actual computation of the matrix elements. The big O notation neglects all constant factors, which is fine for theoretical considerations, but these might have a considerable impact on the real behaviour of the code. The general strategy followed was to vectorize all calculations to make use of the highly optimized and parallel existing linear algebra routines. The remaining parts that do not allow vectorization, such as the matrix element initialization in $PHP$, were all parallelized with \texttt{OpenMP}. Currently, all the parallelism is shared-memory; distributed parallelism will be implemented in the future.

For instance, consider the direct interaction term which requires computing expression~\eqref{direct}. Supposed that the lattice Fourier transform of the interaction is already computed for all motif combinations $i, j$ and for all $\mathbf{k}$ points, we basically have to sum over tight-binding coefficients multiplied by the interaction. Given that the Bloch eigenstates are already stored as columns in matrices, we want to write this as matrix-vector products. Specifically, we can use $V_{ij}$ as a bilinear form, so with a well-defined matrix $\Tilde{V}$ the direct term can be written as:
\begin{equation}
    D_{vc,v'c'}(\mathbf{k},\mathbf{k}',\mathbf{Q}) = C^{T}_{cc'}\Tilde{V}(\mathbf{k}'-\mathbf{k})C_{v'v}
\end{equation}
where
\begin{equation}
    \Tilde{V} = V(\mathbf{k}-\mathbf{k}')\otimes \mathbb{I}_n, \ \ C_{nm} = C^*_n \odot C_m
\end{equation}
$\odot$ denotes element-wise array product, $C_n$ is the vector of coefficients corresponding to state $\ket{n}$ and $\mathbb{I}_n$ denotes a square matrix of ones of dimension $n$, $n$ being the number of orbitals per atom. Note that this expression is only valid if all atoms have the same number of orbitals. Otherwise, one must take into account the different number of orbitals per chemical species when performing the Kronecker's products. The exchange term $X$ can be computed in an analogous way. Note that this assumes that the order of the single-particle basis is $\{\ket{i}\otimes\ket{\alpha}\otimes\ket{\sigma}\}$, i.e.\ for each atomic position, we run over orbitals, and for each orbital we run over spin. This is also relevant for the computation of the spin of the excitons, since it follows this convention.

Lastly, another example that is worth mentioning is how to compute the probability of finding the electron on a given spatial position~\eqref{real-space-wf}. Since this requires two summations over $\mathbf{k}, \mathbf{k}'$, its cost would be $\mathcal{O}(N_{\mathbf{k}}^2)$. To obtain the whole wavefunction, a priori we have to evaluate this over each position in the crystal, meaning that the cost would be $\mathcal{O}(N_{\mathbf{k}}^3)$. However, this would be the worse case scenario in which the exciton is strongly delocalized in real-space. Usually, it will suffice to compute the real-space wavefunction on a contour of the hole position, for a few unit cells only. To actually compute the probability, we want to use the fact that we are storing the exciton coefficients as vectors. First, note that~\eqref{real-space-wf} can be written as:
\begin{align}
    |\psi_X^{\alpha\beta}(\mathbf{t}_n + \mathbf{R}_e, &\mathbf{t}_m + \mathbf{R}_h)|^2 
    = \left|\frac{1}{N}\sum_{v,c,\mathbf{k}}A_{vc}^{\mathbf{Q}}(\mathbf{k})e^{i\mathbf{k}\cdot(\mathbf{R_e} -\mathbf{R_h})}    
    C^{c,\mathbf{k}+\mathbf{Q}}_{m\alpha}(C^{v,\mathbf{k}}_{n\beta})^*\right|^2
\end{align}
which already reduces the complexity down to $\mathcal{O}(N_{\mathbf{k}})$. Then, the probability is computed as $||A\odot C||^2$, where $A$ is the vector of exciton coefficients that incorporates the exponential terms and $C$ are the tight-binding coefficients arranged such that they match the electron-hole pair ordering of the exciton.

\subsection{Effective screening, potentials and regularizations}
As we mentioned at the beginning, to compensate for the lack of screening of the theory, one typically uses the Rytova-Keldysh potential~\cite{rytova, keldysh} instead of the bare Coulomb potential in the context of two-dimensional materials. However, both interactions diverge at $r=0$. We regularize this divergence by setting $V(0)=V(a)$~\cite{FengchengWu2015}, where $a$ denotes the lattice parameter. Currently, the code implements the Keldysh potential, given by
\begin{equation}
\label{eq:keldysh}
    V(\mathbf{r})=\frac{e^2}{8\varepsilon_0\Bar{\varepsilon}r_0}\left[H_0\left(\frac{|\mathbf{r}|}{r_0}\right)- Y_0\left(\frac{|\mathbf{r}|}{r_0}\right)\right]
\end{equation}
where $\Bar{\varepsilon}=(\varepsilon_m+\varepsilon_s)/2$, with $\varepsilon_s$, $\varepsilon_m$ being the dielectric constants of the substrate and the embedding medium (usually vacuum) respectively, and $r_0$ the effective screening length. Those three parameters have to be specified for all calculations. $H_0$, $Y_0$ are Struve and Bessel functions of second kind respectively. For the reciprocal space method we employ the Fourier transform of the Keldysh potential:
\begin{equation}
    V(\mathbf{q}) = \frac{e^2}{2\varepsilon_0\Bar{\varepsilon}A}\frac{1}{|\mathbf{q}|(1+r_0|\mathbf{q}|)}
\end{equation}
where $A$ is the area of the crystal. The code also implements the standard bare Coulomb potential, which is given by
\begin{equation}
    V(\mathbf{r}) = \frac{e^2}{4\pi\varepsilon_0}\frac{1}{|\mathbf{r}|} \leftrightarrow V(\mathbf{q})\equiv \text{FT}_{2D}\left[V(\mathbf{r})\right] = \frac{e^2}{2\varepsilon_0 A}\frac{1}{|\mathbf{q}|}
\end{equation}
where $\text{FT}_{2D}$ denotes the 2D Fourier transform. While for the real space method the dimensionality of the system is specified directly by the lattice positions, for the reciprocal space method the dimensionality is specified when we write $V(\mathbf{r})$ as a Fourier series. Thus, while the code is in principle designed towards 2D systems, the real space method can be used for any dimensionality, although in the Keldysh case the potential was derived assuming a 2D system. For the reciprocal method, however, by construction it can only be used with 2D systems, due to the Fourier series. This is how the FT of the Keldysh and bare Coulomb potentials are obtained, namely via a 2D FT\@. One could extend this method to 3D systems simply writing the potential as a 3D Fourier series, which is the standard approach in all ab-initio codes, as it allows treating systems of arbitrary dimensionality. The interaction between fictitious copies of the system for dimension lower than three is usually handled ensuring that the box defined by the Bravais vectors is large enough.

Going back to the real space method, since the interaction decays quickly, we employ a radial cutoff, such that for distances $r>R_c$ we take the interaction to be zero. Then, the effective interaction is
\begin{equation}
    \Tilde{V}(\mathbf{r})=\left\{\begin{array}{cc}
        V(a) & \text{if } |\mathbf{r}| = 0 \\
        V(\mathbf{r}) & \text{ if } |\mathbf{r}| < R_c \\
        0 & \text{else}
    \end{array}\right.
\end{equation}
where $R_c$ is the cutoff radius. The cutoff has two purposes: first, it enforces the crystal symmetries in the transformed potential $V_{ij}(\mathbf{k})$ (as a function of $\mathbf{k}$). Secondly, it allows to compute the summation over lattice positions faster. Instead of evaluating the potential over all lattice positions, we restrict the sum to the lattice positions where we know the potential is different from zero. This is also helpful to avoid interactions between copies of the system when using the 3D Fourier series of the potential, although the code currently does not implement it. As with the real space method, the reciprocal method also exhibits divergences. We set $V(\mathbf{q}=0)=0$ to remove the long wavelength divergence, which is typically justified as a cancellation of the positive charge background~\cite{gross1986many}. Another way to regularize the divergence if to compute $V(\mathbf{q}=0)$ as its average over neighbouring $\mathbf{q}$ points,
\begin{equation}
    V(\mathbf{q}=0)\approx \frac{1}{N_{\mathbf{q}}}\sum_{\mathbf{q}\neq 0}V(\mathbf{q})
\end{equation}
This is also justified from the point of view of the BSE, which is ideally defined in the thermodynamic limit, such that the $\mathbf{q}=0$ divergence is integrated out. However, due to the discretization of the BZ, this divergence becomes apparent. As a middle ground between the integral and the discretization, we can discretize the integral, but take the average over neighbouring $\mathbf{q}$ points for each $\mathbf{k}$ point. The code implements for simplicity the $V(\mathbf{q}=0)=0$ method; both regularizations yield correct results in the $N_{\mathbf{k}}\rightarrow\infty$ limit, although the energy scaling with $N_{\mathbf{k}}$ will differ depending on the method used~\cite{qiu2017many}.
\\

\section{Validation and benchmarks}\label{sec:xatu_validation}

So far we have discussed the theory underlying the code and its numerical implementation. Therefore, it remains to show actual examples of the capabilities of the code. One context where excitons are relevant is valleytronics: materials with honeycomb structure which exhibit the band gap at the $\mathbf{K}, \mathbf{K}'$ points of the Brillouin zone (the "valleys"), and whose optical excitations can be tuned according to the valley~\cite{mak2018, xiao2012}. The materials most commonly used for this purpose are transition metal dichalcogenides (TMDs), with formula $\text{WX}_2$, where W is the transition metal and S some chalcogenide. Another similar material that has become highly relevant is hexagonal boron nitride (hBN), although in this case due to its good properties as an insulating substrate~\cite{Dean2010}.
These materials have become the prototypical examples to test the capabilities of an exciton code, and have been studied extensively. We will characterize the excitons in both hBN and $\text{MoS}_2$, i.e.\ obtain the exciton spectrum for $\mathbf{Q}=0$, show the associated wavefunctions and compute the optical conductivity. We will also show how a simple strain model of hBN can be used to break some crystal symmetries and modify the excitonic ground state.
All the calculations shown are done with the real-space approach to the interaction matrix elements and neglecting the exchange term, unless specified otherwise.

\subsection{Exciton spectrum of hBN}

Monolayer hexagonal boron nitride has a large quasi-particle band gap, with ab-initio calculations predicting a value of $6-8$ eV depending on the method used~\cite{zhang2022}. As we will see, the band structure of hBN is relatively flat along the $\mathbf{M}-\mathbf{K}$ path in the Brillouin zone. This, in conjunction with small screening results in excitons that are strongly delocalized in reciprocal space, but are tightly bound in real space.

This material can be described easily with a minimal 2-band tight-binding model~\cite{galvani2016}, equivalent to graphene but with opposite onsite energies for each atom of the motif (also regarded as a staggered potential). The tight-binding model for hBN reads:
\begin{equation}
    H= \sum_{i}\frac{\Delta}{2}(c^{\dagger}_ic_i-d^{\dagger}_id_i)+\sum_{\substack{\left<i,j\right> \\ i \neq j}}\left[ tc^{\dagger}_id_j + \text{h.c.}\right]
\end{equation}
where $c^{\dagger} (d^{\dagger})$ denote creation operators for $p_z$ electrons at B (N) atoms. The indices $i, j$ run over unit cells, and the summation over $\left<i,j\right>$ spans only the first neighbours.
The parameters are $t=-2.3$ eV, $\Delta/2=3.625$ eV, and the corresponding system file can be found in the code repository under the folder \texttt{examples/material\_examples}.
\begin{figure}[h]
    \centering
    \includegraphics[width=0.7\columnwidth]{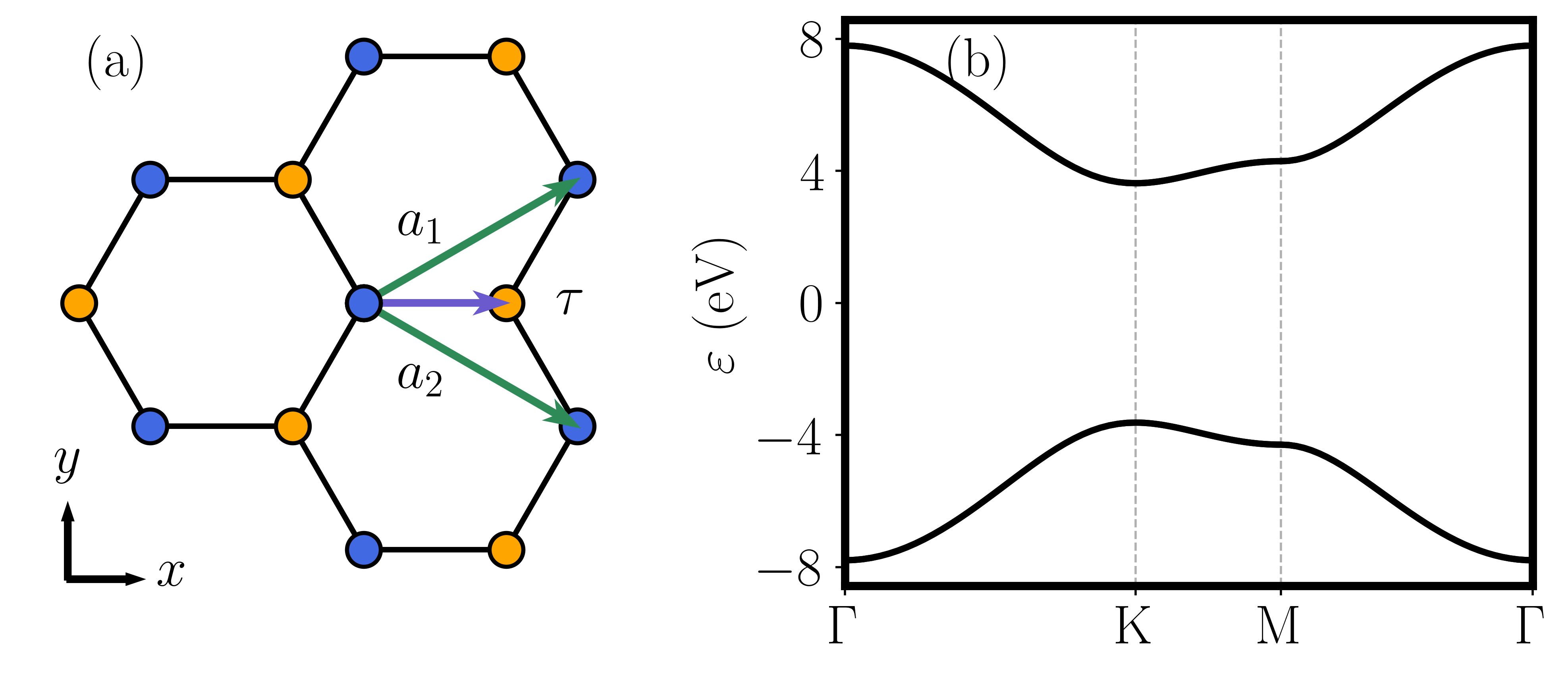}
    \caption[Crystal and band structure of hBN]{(a) hBN lattice and (b) band structure of the tight-binding model.}\label{fig:hbn_lattice}
\end{figure}

From a Slater-Koster perspective, hBN is described by $p_z$ orbitals. Taking the model to be spin-polarized for simplicity, there are only two bands and there must be only one electron per unit cell (half-filling) for it to be an insulator. Once the model is defined and the system file is appropriately constructed, we can begin setting the parameters of the calculation. First, we need to specify the constants that appear in the Keldysh potential in Eq.~\eqref{eq:keldysh}. These parameters determine the strength of the electrostatic interaction and consequently affect the exciton binding energies. Here we follow previous works to set these quantities~\cite{galvani2016}, but sometimes we will be interested in exploring the effect of tuning the dielectric constants, or instead we will want to set them to reproduce known experimental results. Nevertheless, values for typical substrates can be found in literature  and $r_0$ can also be estimated from ab-initio calculations~\cite{prada2015, berkelbach2013}.

\begin{figure}[h]
    \centering
    \includegraphics[width=0.9\columnwidth]{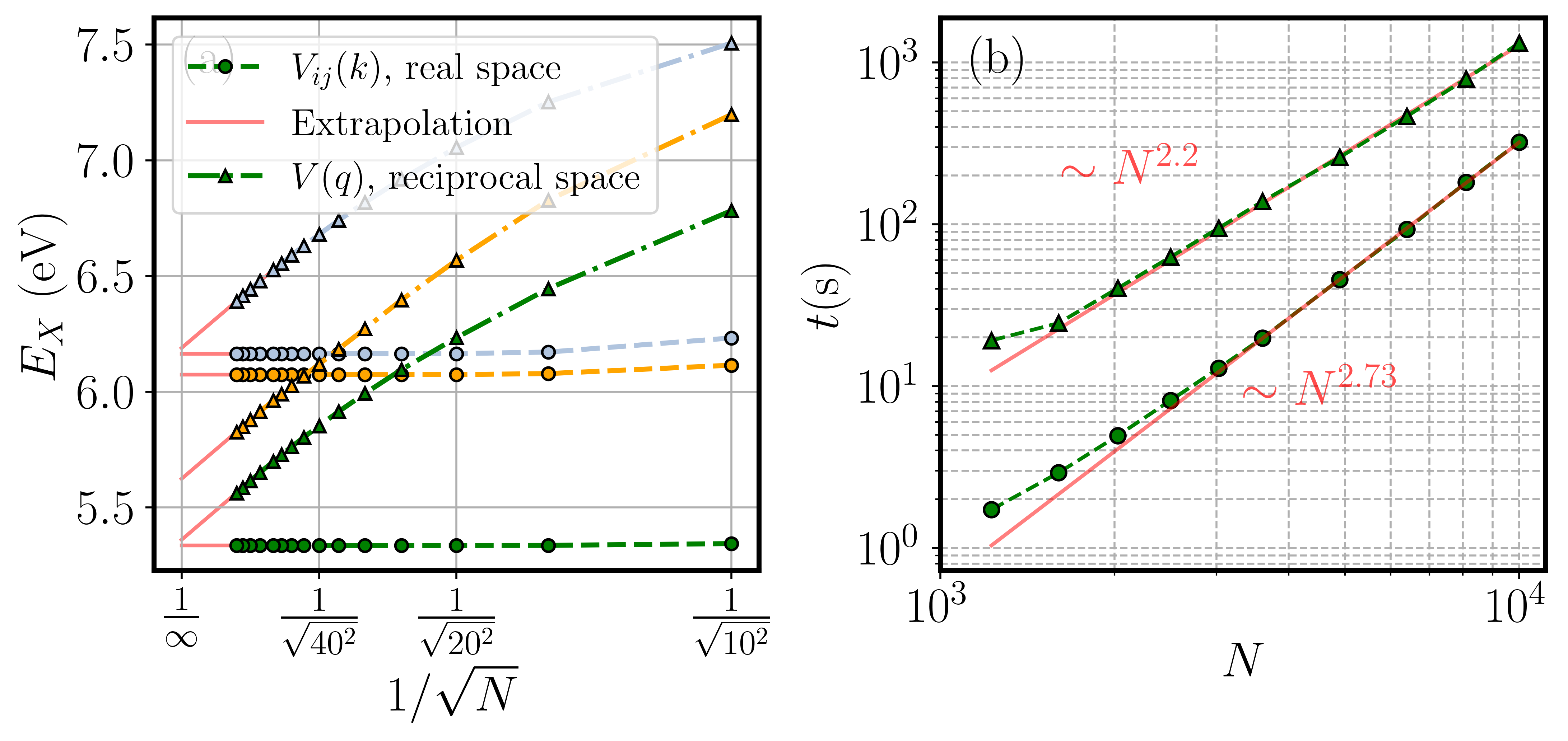}
    \caption[Exciton energy convergence and elapsed calculation time as a function of $N_{\mathbf{k}}$]{(a) Convergence of the ground state and first and second excited states for hBN as a function of the number of $\mathbf{k}$ points, $N\equiv N_{\mathbf{k}}$, computed with interactions both in real and reciprocal space. The reciprocal space calculations have to be converged also with respect to the number of reciprocal lattice vectors included, $N_{\mathbf{G}}$, with $N_{\mathbf{G}} = 25$ in this case. For $N\rightarrow\infty$, the energies in both methods are approximately the same except for the discrepancy in the second energy level, which is attributed to the real-space regularization $V(0)=V(a)$. Changing the reference length $a$ to smaller distances reduces the discrepancy between both methods.\ (b) Measured calculation time as a function of $N$. For both the real and reciprocal space calculations, the asymptotic behaviour is $\mathcal{O}(N^3)$. However, at small values of $N$ the required time is partially dominated by the BSE matrix initialization, which in both cases scales as $\mathcal{O}(N^2)$.}\label{fig:hbn_convergence}
\end{figure}

The other parameters of the exciton file are related to the convergence of the excitons themselves. Varying the number of $\mathbf{k}$ points in the mesh, $N_{\mathbf{k}}$, one obtains the convergence curves shown in Fig.~\ref{fig:hbn_convergence}(a). The convergence has been done with both the default interactions (in real-space) and with reciprocal interactions. For reciprocal interactions energies converge much slower than the real-space counterpart, on top of requiring summing over several $\mathbf{G}$ reciprocal cells. In materials with highly localized excitons in $\mathbf{k}$ space, it usually suffices to take only $\mathbf{G}=0$ (e.g.\ $\text{MoS}_2$). However, we will see  later that hBN excitons are highly delocalized in reciprocal space, which is why the interaction can see neighbouring reciprocal unit cells.

After checking convergence, we can start studying the exciton themselves. The energies of the first 8 states and their degeneracies are given in Table~\ref{tab:hbn_spectrum}. To make sense of the degeneracies, one has to check the character table of the point group of the material: hBN has the crystallographic point group $D_{3h}$, 
with both one- and two-dimensional irreducible representations.
Since the symmetry operations and their action on single-particle states are specific to each problem, the code does not address the problem of identifying the irreducible representation of each exciton, nor labeling them in terms of symmetry eigenvalues. Instead, we only check that the $\mathbf{Q}-$excitonic wavefunctions have the allowed degeneracies and~\eqref{wf_invariant} is invariant under the little group at $\mathbf{Q}$.

\begin{table}[t]
    \centering
    \begin{tabular}{|c|c|c|c|}
        \hline 
        $n$ & Energy (eV) & Binding energy (eV) & Degeneracy \\
        \hline
        \hline
        1 & 5.3357 & -1.9143 & 2  \\
        2 & 6.0738 & -1.1762 & 1  \\
        3 & 6.1641 & -1.0859 & 2  \\
        4 & 6.1723 & -1.0777 & 1  \\
        5 & 6.3511 & -0.8989 & 2  \\
        \hline
    \end{tabular}
    \caption[Exciton spectrum of hBN]{Exciton spectrum from the tight-binding model for hBN computed with $N_{\mathbf{k}}=60^2$. The binding energy $E_b$ is defined as $E_b=E_X - \Delta$, where $\Delta$ is the gap of the system.}\label{tab:hbn_spectrum}
\end{table}

\begin{figure}[H]
    \centering
    \includegraphics[width=0.49\columnwidth]{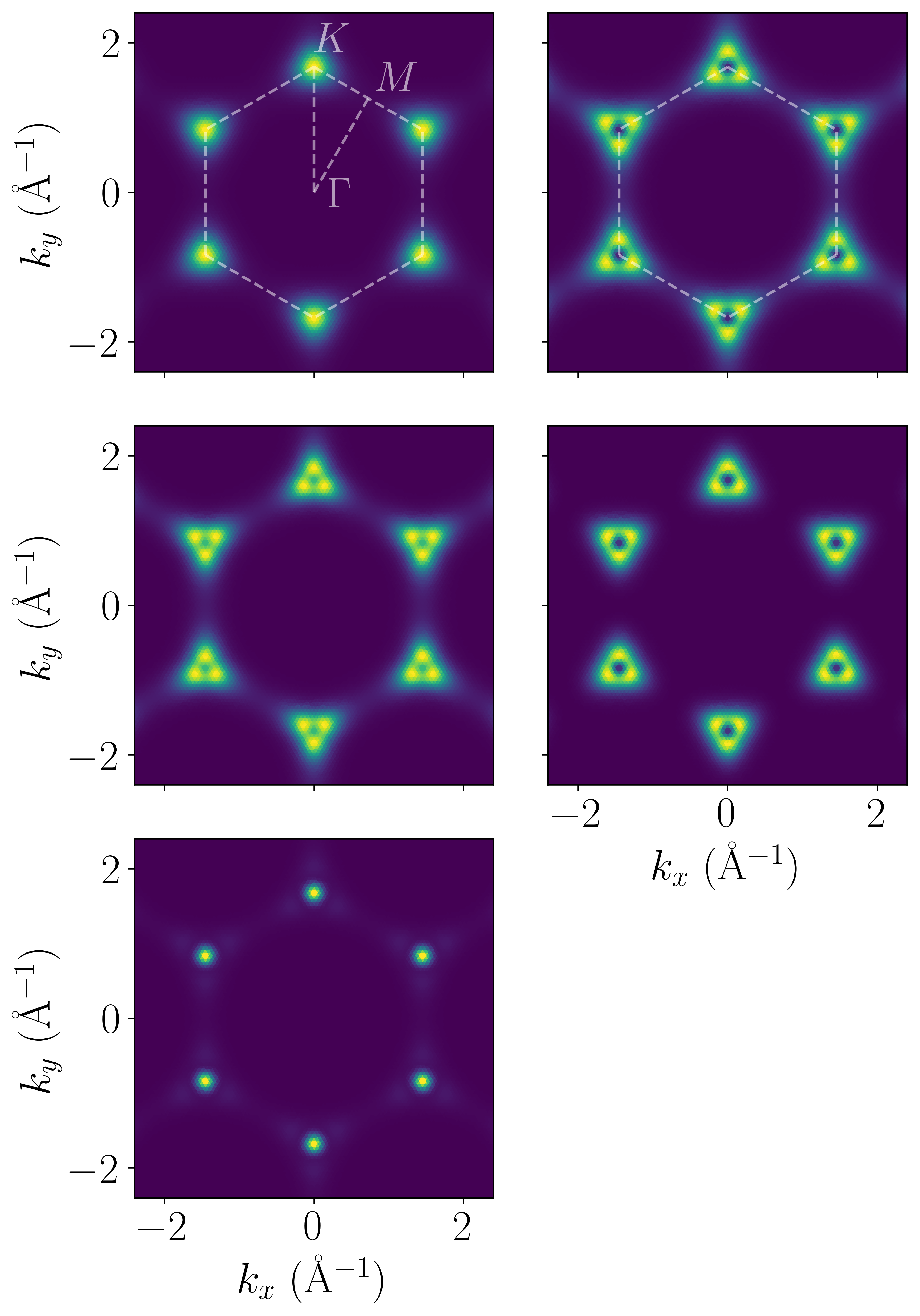}
    \raisebox{0.15cm}{
        \includegraphics[width=0.49\columnwidth]{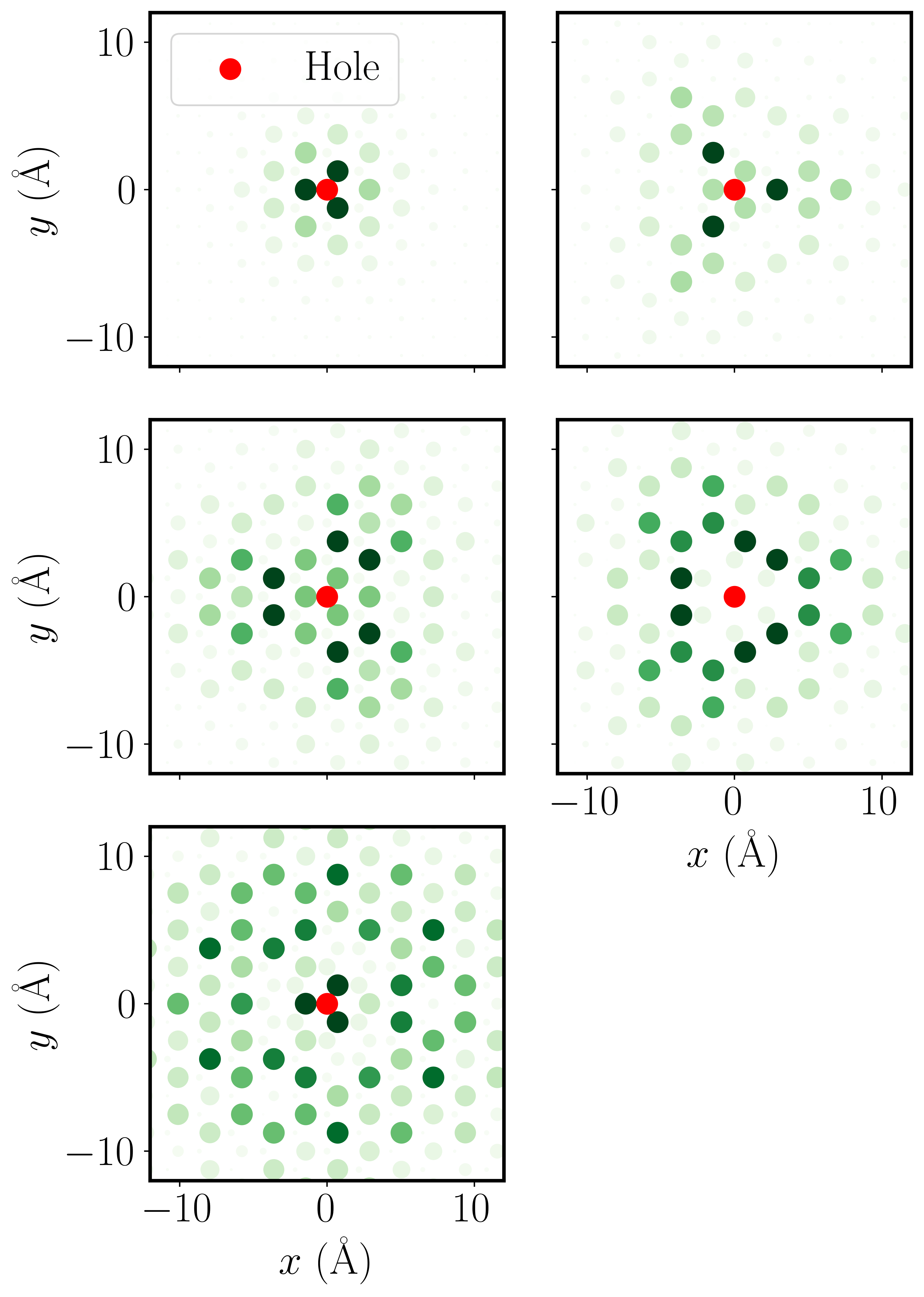}
    }
    \caption[Reciprocal and real-space probability densities of the first five excitons in hBN]{Plot of the $\mathbf{k}$ exciton probability densities (left) and electronic, real-space exciton probability densities (right) in TB hBN with $\mathbf{Q}=0$, for the first 5 energy levels (from left to right, top to bottom). For each level, we actually show $|\Psi(\mathbf{k})|^2=\sum_n|\psi_n(\mathbf{k})|^2$ for $\mathbf{Q}=0$, where the index $n$ runs over degenerate states. The red dot in the real-space probability densities shows the position of the hole.}\label{fig:hbn_kwf}
\end{figure}

The $\mathbf{k}$ probability densities of the first eight excitonic states, grouped by degenerate levels, are shown in Fig.~\ref{fig:hbn_kwf}. Each energy level has the symmetry of the lattice, as expected since we are plotting~\eqref{wf_invariant}. The additional symmetry in this case is due to time-reversal symmetry and the fact that $\mathbf{Q}=0$ is a time-reversal invariant momentum. We see that the wavefunctions peak at the valleys, although they also spread over the $\mathbf{K}-\mathbf{M}-\mathbf{K}'$ paths. This means that the excitons are formed by strongly interacting electron-hole pairs in $\mathbf{k}$ space, which explains why we need to sum over several reciprocal cells when using the reciprocal interactions. As for the shape of excitons, we find the common pattern: the first state is $s$-like in the sense that it does not have nodes. The next state would be $p$-like and so on. Note that the hydrogen analogy only concerns the shape of the wavefunctions, and not the energy spectrum, which in general differs from the hydrogen series.

\begin{figure}[t]
    \centering
    \includegraphics[width=0.5\columnwidth]{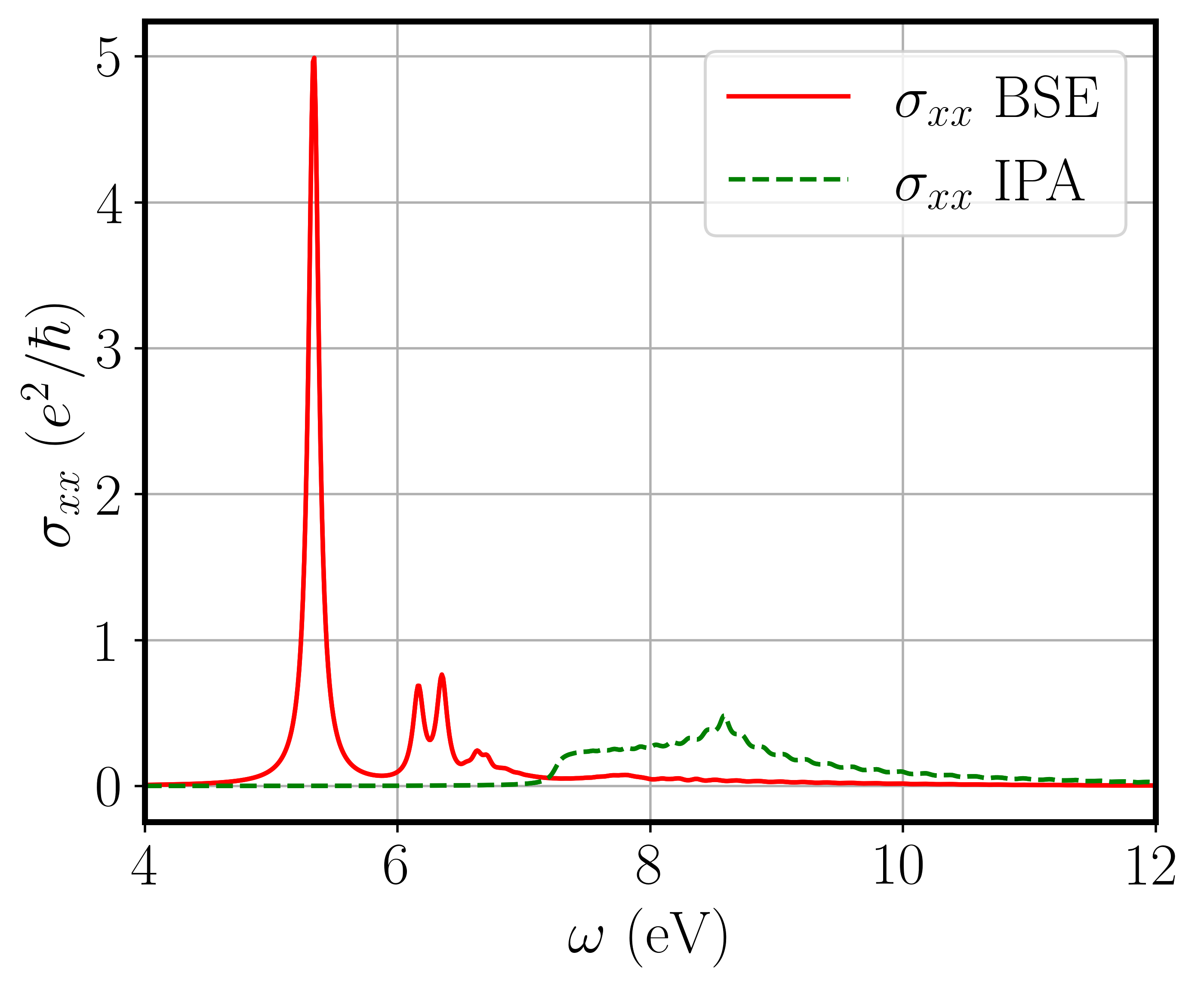}
    \caption[Optical conductivity of hBN]{Optical conductivity of monolayer hBN as a function of the incident energy. We compare the conductivity obtained with the Kubo formula in the independent particle approximation (IPA), and with the BSE, which shows a dramatic change from the inclusion of excitons. }\label{fig:hbn_absorption}
\end{figure}

Since the excitons are delocalized in reciprocal space, we expect them to be strongly localized in real space. The real space densities of each degenerate level are shown in Fig.~\ref{fig:hbn_kwf}. The hydrogenic picture makes more sense when looking at the real-space wavefunction, since it can be understood then as the problem of two interacting opposite sign charges. The spectrum and the degeneracies do not match that of hydrogen, but the wavefunctions behave radially as we would expect. 

In hBN the spin-orbit coupling is small and it suffices to compute the excitons for the spinless system, in particular given that we are also neglecting the exchange interaction. If we consider a spinful system, without exchange again, we obtain exactly the same energy levels but now four-fold degenerate (on top of the previous spatial degeneracies). The same stands for both types of wavefunctions. 

Our study of the exciton spectrum in hBN concludes with the calculation of the optical conductivity~\cite{Ridolfi2020, pedersen2015intraband, zhang2022}, which reflects the light absorbance from a source up to a constant factor. So far we have not discussed which excitons of the spectrum are bright or dark. This can be seen through the calculation of the optical oscillator strengths within Eq.~\eqref{eq:excitonkubo}, which determine the transition rate for photon emission. The frequency-dependent conductivity of monolayer hBN is given in Fig.~\ref{fig:hbn_absorption}. Electron-hole interactions move the spectral power from the continuum to pronounced sub-band gap peaks. Attending to Table~\ref{tab:hbn_spectrum} and Fig.~\ref{fig:hbn_kwf}, we see that non-degenerate excitons with mainly $s$ character are bright. The relative height of the peaks can be understood by looking at the magnitude of the wavefunctions near the $\mathbf{K}$ and $\mathbf{K'}$ points. All bright excitons can be excited with linearly polarized light along two orthonormal polarization directions, giving rise to an isotropic conductivity consistent with the $D_{3h}$ point group of the material.

It is of interest to check the validity of the results against a more refined description of the band structure of the material. This can be done with the code by using a local orbital-based DFT calculation as the starting Hamiltonian, instead of using a parametrized tight-binding model. The exciton energies will depend on the gap as estimated from the functional used, but we expect to get similar wavefunctions and conductivity. Since we consider several orbitals for each chemical species now, we have multiple valence and conduction bands so we should converge the exciton with respect to the number of bands as well. It is a proper check to do, but in this case the different bands are well separated, so their effect should be negligible.

\begin{figure*}[t]
    \centering
    \includegraphics[width=1\textwidth]{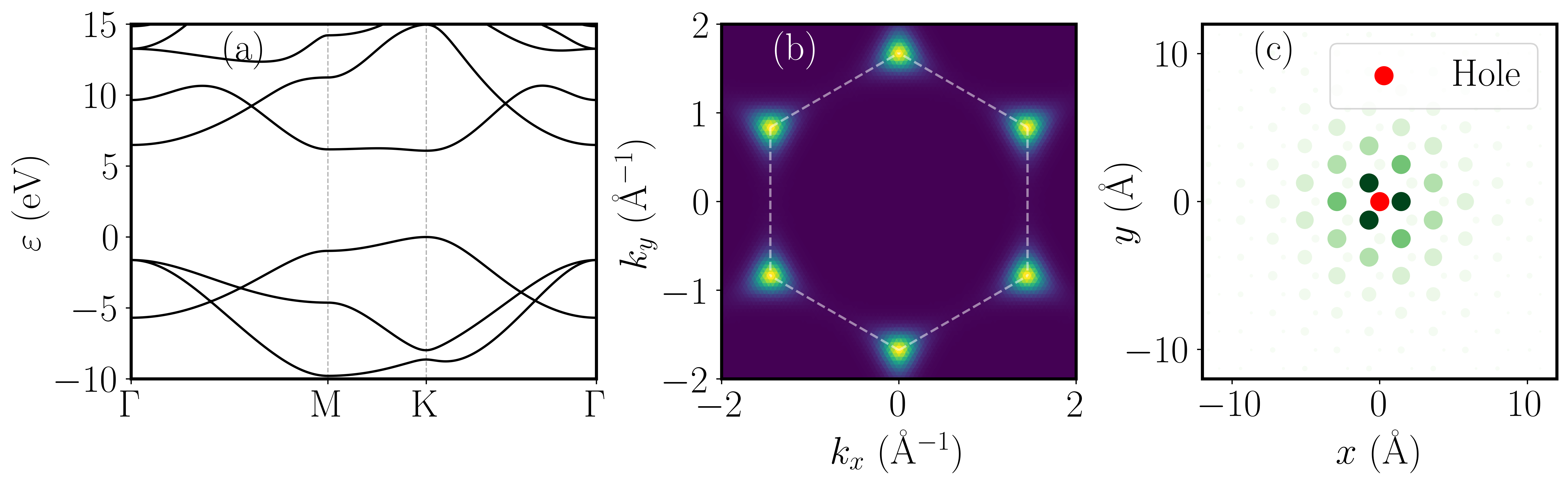}
    \caption[Exciton calculation using a DFT-HSE06 Hamiltonian for hBN]{(a) DFT band structure of monolayer hBN as obtained with the HSE06 functional.\ (b) Reciprocal and (c) real-space probability densities of the $\mathbf{Q}=0$ ground state exciton with $N_{\mathbf{k}}=60^2$, $N_c=N_v=1$. The DFT calculation involved a basis size of 36. We have run successfully exciton calculations on different systems with varying basis sizes, from 8 to 92.}\label{fig:hbn_dft}
\end{figure*}

The DFT band structure and the wavefunctions of the ground state exciton are shown in Fig.~\ref{fig:hbn_dft}. One could use standard LDA functionals, but here we opt for a hybrid functional (HSE06~\cite{krukau2006} in this case), which is efficiently implemented in \texttt{CRYSTAL}~\cite{crystal17}. This type of functional yields a better estimation of the single-particle gap due to a different treatment of the exchange-correlation term. For both LDA (not shown) and hybrid functionals such as HSE06, the wavefunctions closely resemble those obtained with TB models. For instance, we observe the same sublattice polarization present in the TB real-space densities with the HSE06 calculation Fig.~\ref{fig:hbn_dft}(c).
The energy spectrum shows the same degeneracies, although the positions of some of the levels are exchanged. 

\clearpage

\subsection{Strained hBN}

To illustrate the applicability of the code beyond standard cases, we now study the effect of strain on the exciton spectrum. If we apply some uniaxial in-plane strain along the $x$ axis, the point group of the material will change to $C_{2v}$ (with rotation axis along $x$). The degeneracy of the ground state came from the spatial symmetries, meaning that it should be broken for any strain value, given that all irreducible representations of $C_{2v}$ are of dimension 1. Therefore, we can study the energy splitting of the ground state as a function of the applied strain.

\begin{figure}[h]
    \centering
    \begin{tikzpicture}
        \node at (0,0) {\includegraphics[width=0.29\columnwidth]{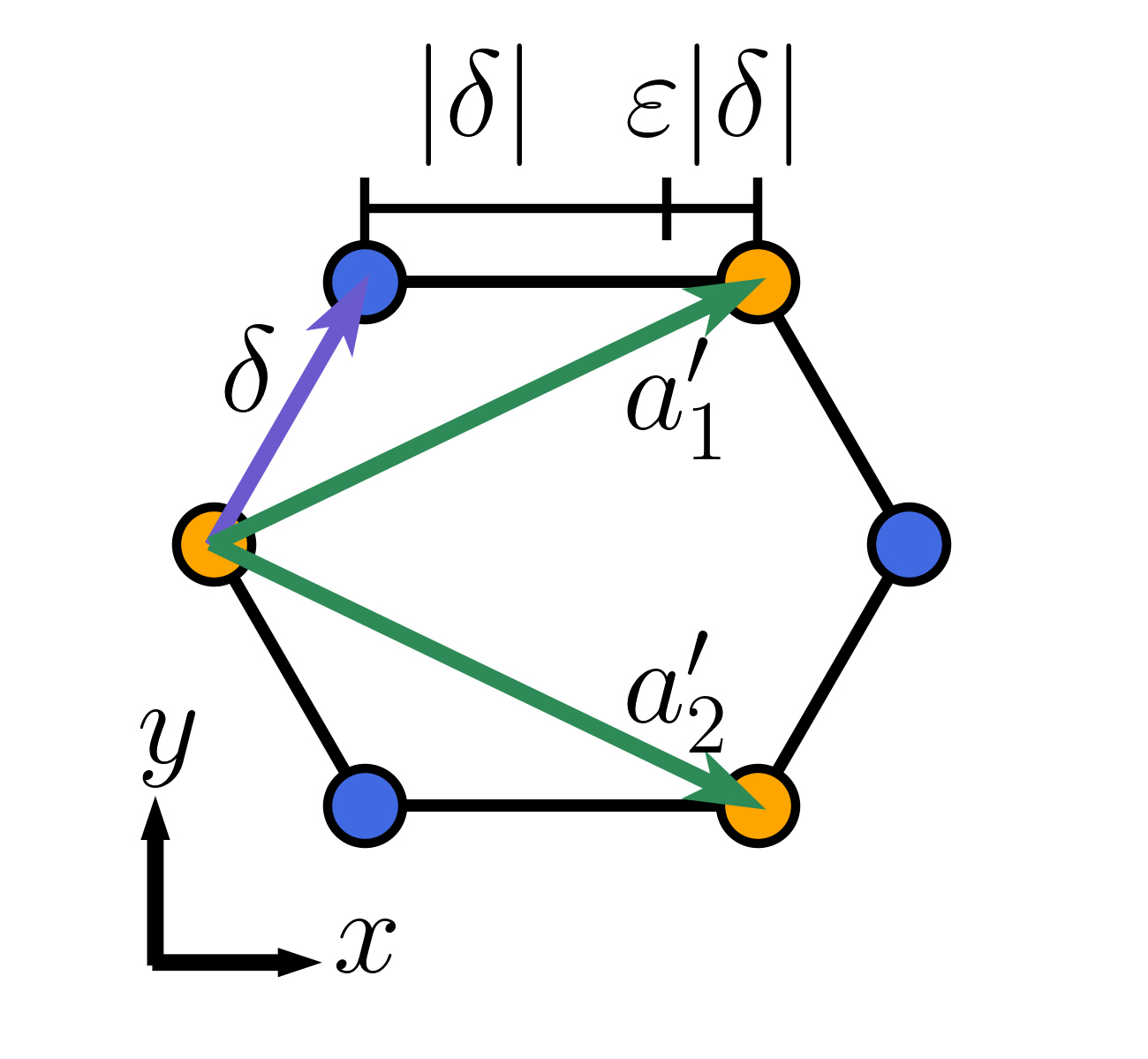}};
        \node at (-1.5,1.5) {(a)};
        
        \node at (7.5,-0.25) {\includegraphics[width=0.69\columnwidth]{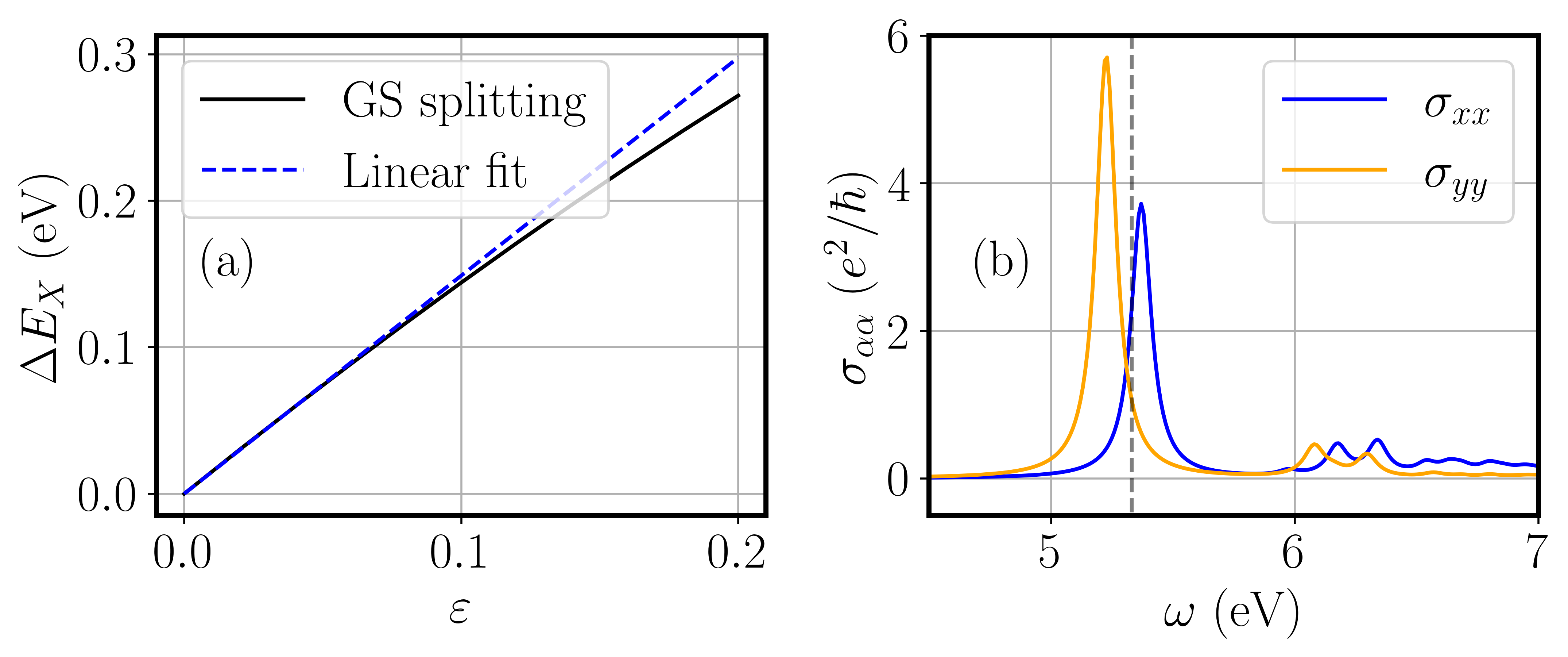}};
        \node at (2.25,1.5) {(b)};
        \node at (8,1.5) {(c)};
    \end{tikzpicture}
    \caption[Ground state exciton energy splitting and optical conductivity of strained hBN]{(a) Schematic of the distortion of the hBN lattice due to the application of uniaxial strain along the $x$ axis. Note that in reality all bonds should be distorted, due to the phenomena of Poisson contraction.\ (b) Energy splitting of the ground state exciton as a function of strain in hBN\@. We observe that the splitting is linear on strain, for small values.\ (c) Frequency-dependent conductivity on strained hBN, $\varepsilon=0.1$. The dashed line shows the position of the ground state for $\varepsilon=0$.}\label{fig:hbn_strain_exciton}
\end{figure}
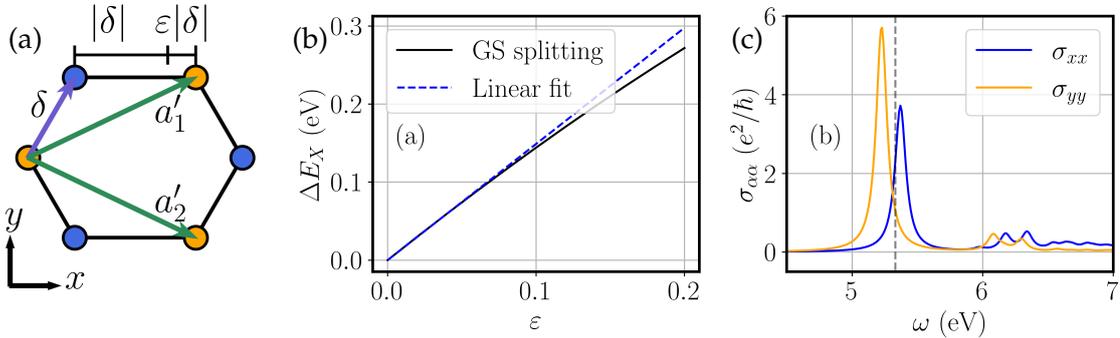

The strain model used is fairly straightforward. Based on the original tight-binding model, we now consider the hopping parameters to have an exponential dependence on the distance:
\begin{equation}
    t(r) = t_0e^{-a(r-r_0)},
\end{equation}
where $a$ is some inverse decay length, $t_0$ the original value of the hopping and $r_0$ the reference length. Additionally, the distortion of the lattice due to strain is taken to affect only bonds parallel to the strain. A rigorous approach would have to implement appropriate distortion of all atomic positions according to the stress tensor~\cite{nuno_strain}, but for our purposes this simple model suffices. This is illustrated in Fig.~\ref{fig:hbn_strain_exciton}(a).

The procedure to study the exciton spectrum as a function of strain is as follows: we generate different system files (i.e.\ different Hamiltonians) for different values of the strain, which translates into different atomic positions. Then, we run the exciton simulation for each system file, storing the energies. As we expected, now all states are non-degenerate because of the symmetry group $C_{2v}$. We can plot the ground state splitting as a function of strain, which is shown in Fig.~\ref{fig:hbn_strain_exciton}(b). In Fig.~\ref{fig:hbn_strain_exciton}(c) we plot the conductivity for some finite value of the strain, $\varepsilon=0.1$. The response is no longer isotropic due to the lattice symmetry breaking caused by strain, where the exciton peaks shift for both light polarizations.

\subsection{Exciton spectrum of MoS$_2$}

\begin{figure}[h]
    \centering
    \includegraphics[width=0.7\columnwidth]{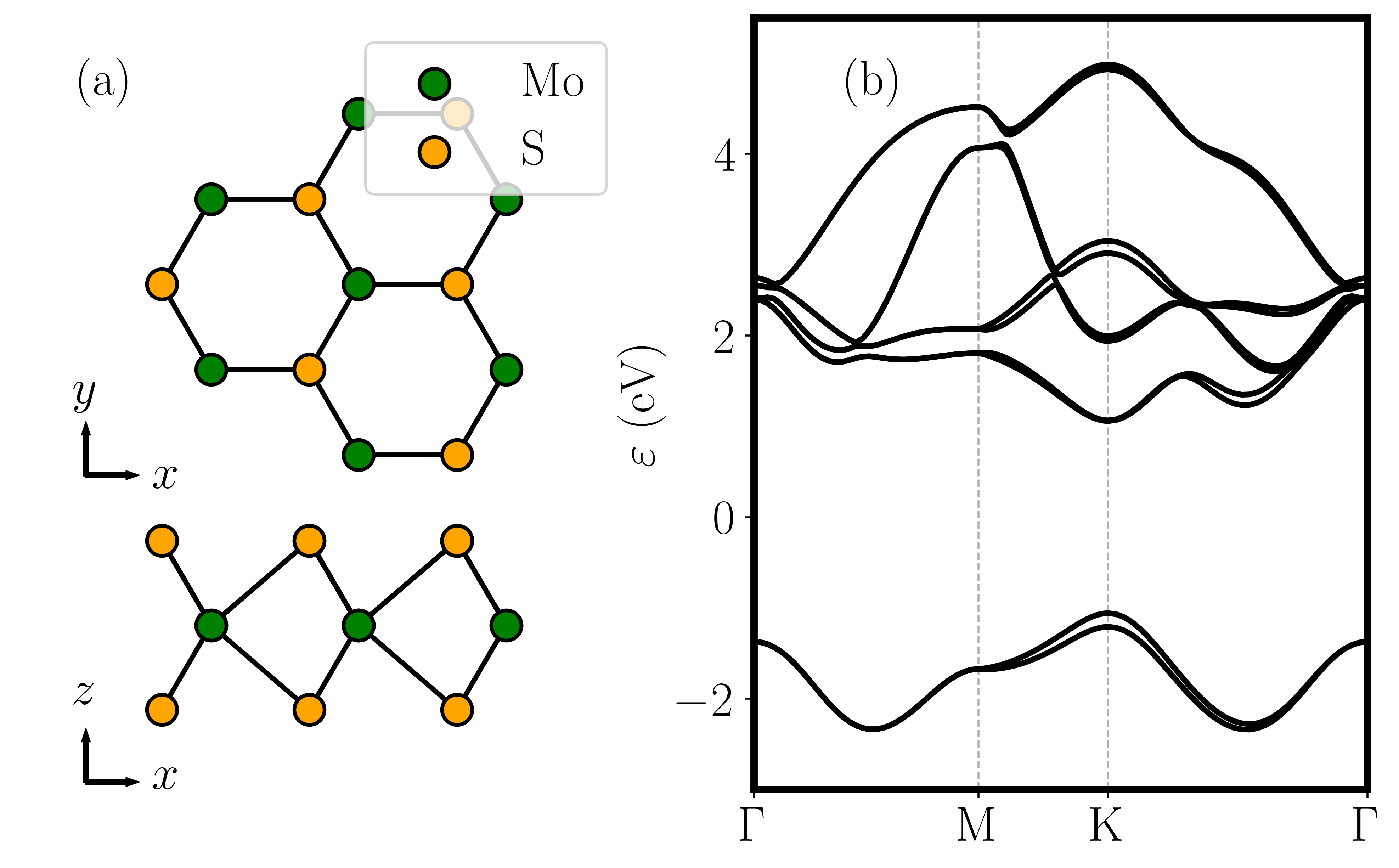}
    \caption[Crystal and band structure of MoS$_2$]{(a) Crystal and (b) tight-binding band structure of $\text{MoS}_2$.}\label{fig:mos2_bands}
\end{figure}

To conclude the validation section, we also analyze the exciton spectrum of $\text{MoS}_2$. Same as hBN, in monolayer form this material presents itself in a honeycomb lattice, although it is not planar. Instead, it is formed by three layers of composition S-Mo-S respectively. 
The description of the band structure of $\text{MoS}_2$ requires a more complex model, which is why we use it to showcase the code. We use a Slater-Koster tight-binding model~\cite{Ridolfi_2015}, where each chemical species has a different set of orbitals (Mo has $d$ orbitals, and S only $p$ orbitals). This, together with the non-negligible spin-orbit coupling results in a more complex band structure than that of hBN\@. Both the lattice and the band structure can be found in Fig.~\ref{fig:mos2_bands}.

\begin{table}[h]
    \centering
    \begin{tabular}{|c|c|c|c|}
        \hline 
        $n$ & Energy (eV) & Binding energy (eV) & Degeneracy \\
        \hline
        \hline
        1 & 1.7673 & -0.3527 & 2  \\
        2 & 1.7797 & -0.3403 & 2  \\
        3 & 1.9105 & -0.2095 & 2  \\
        4 & 1.9232 & -0.1968 & 2  \\
        5 & 1.9418 & -0.1782 & 2 \\
        6 & 1.9528 & -0.1672 & 2 \\
        \hline
    \end{tabular}
    \caption[Exciton spectrum of MoS$_2$]{Exciton spectrum from the tight-binding model for MoS$_2$ computed with $N_{\mathbf{k}}=40^2$, $N_v=N_c=2$. This model has a direct gap at $\mathbf{K}$ of 2.12 eV, used to compute the shown exciton binding energies.}\label{tab:mos2}
\end{table}

After checking convergence with the number of $\mathbf{k}$ points and the number of bands, we obtain the spectrum shown in Table~\ref{tab:mos2}. In this case, the point group of the material is again $D_{3h}$ 
and the irreducible representations realized by the wavefunctions at $\mathbf{Q}=0$ are compatible with the character table of the group~\cite{bir1974symmetry}.
As before, to ensure that the excitons were computed correctly we can plot the total densities to ensure that they have the expected symmetries.

\begin{figure}[t]
    \centering
    \includegraphics[width=0.7\columnwidth]{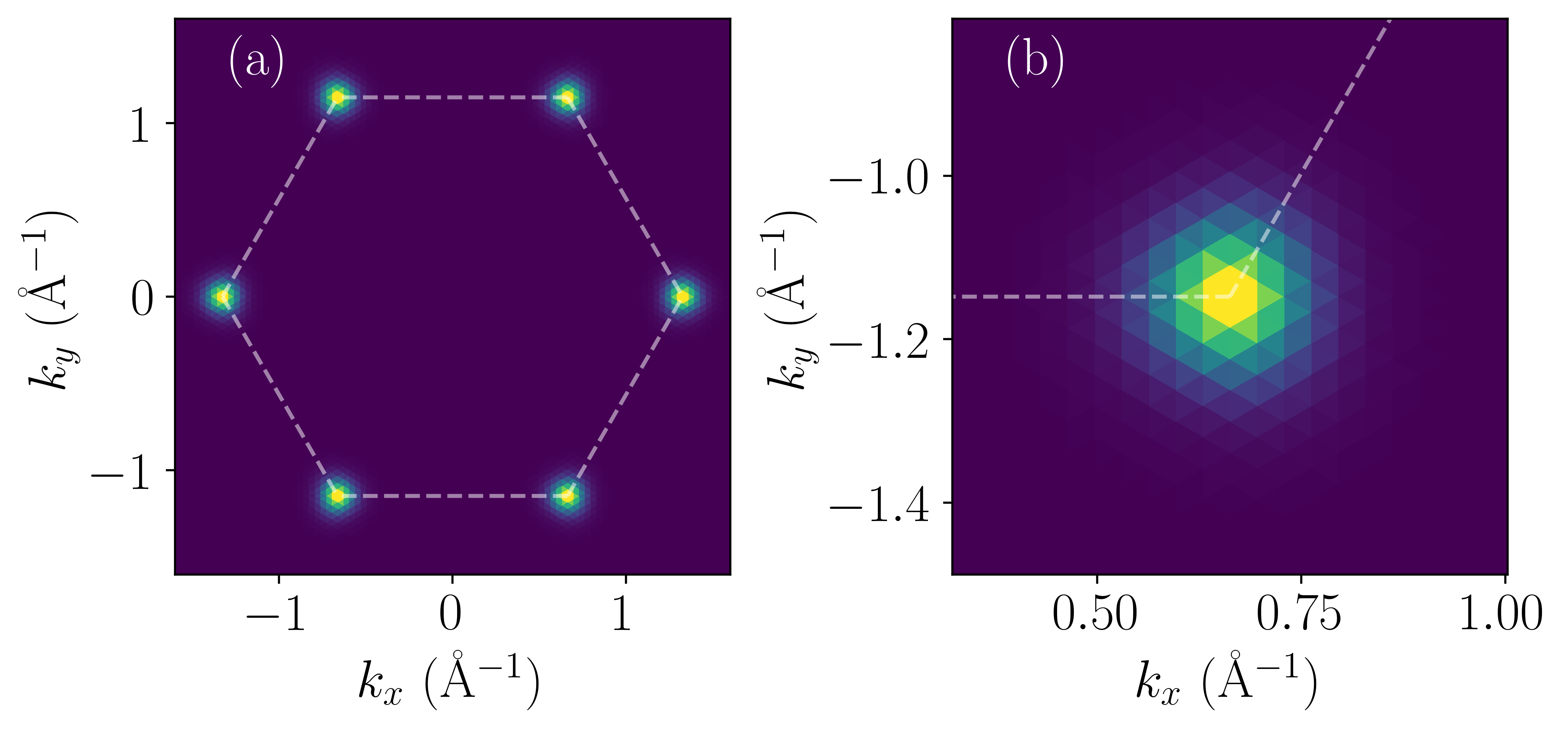}
    \caption[Reciprocal probability density of the ground state exciton in MoS$_2$]{(a) Probability density of the ground state exciton in MoS$_2$ obtained over the full BZ with $N_{\mathbf{k}} = 60^2$ for $\mathbf{Q}=0$.\ (b) Ground state exciton computed in a contour of the $\mathbf{K}$ valley with $N_{\mathbf{k}}=30^2$ with a reduction factor of 2. Both calculations were done with $N_c=N_v=2$.}\label{fig:mos2_kwf}
\end{figure}

\begin{figure}[H]
    \centering
    \begin{tikzpicture}
        \node at (0,0) {\includegraphics[width=0.41\columnwidth]{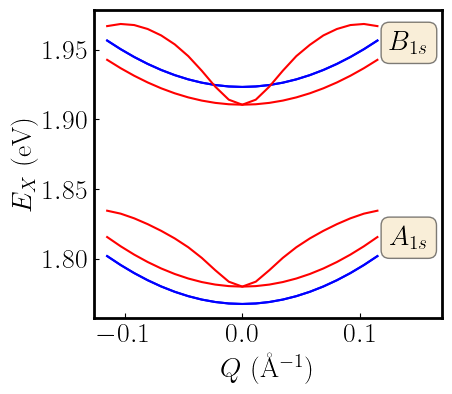}};
        \node at (-2.8, 2.7) {(a)};

        \node at (7,0.07) {\includegraphics[width=0.42\columnwidth]{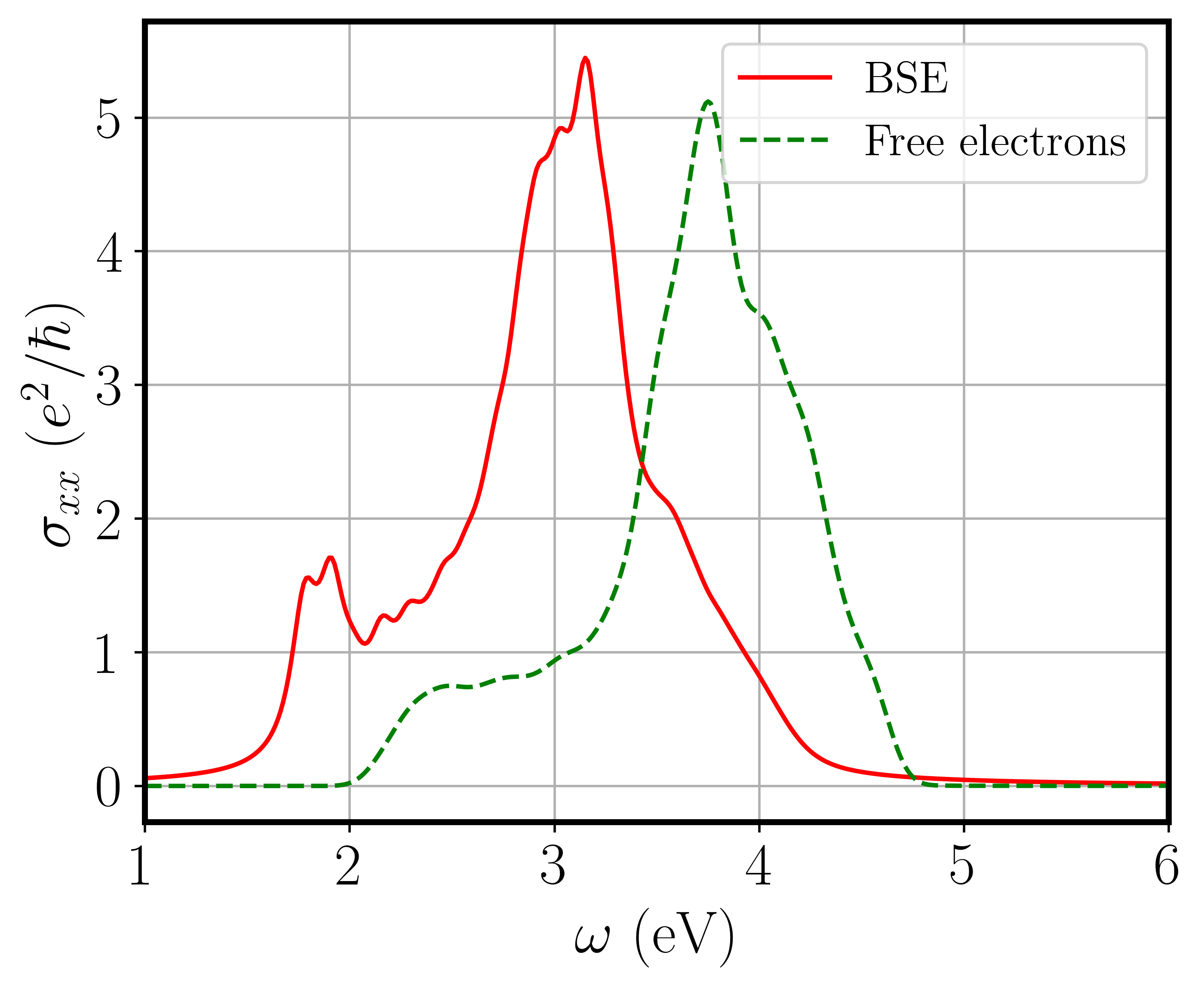}};
        \node at (3.7, 2.7) {(b)};
    \end{tikzpicture}
    \caption[Exciton band structure and optical conductivity of MoS$_2$]{(a) Exciton band structure in MoS$_2$ for $\mathbf{Q}$ near $\Gamma$ along the $\mathbf{K}'-\Gamma-\mathbf{K}$ path. The red (blue) lines correspond to excitons formed by like-spin (unlike-spin) transitions. The exchange interaction couples the valleys, resulting in a splitting of the like-spin excitons from the A, B groups with an approximately linear dispersion as reported in~\cite{nonanalyticity}.\ (b) Optical conductivity of MoS$_2$ with and without excitons. The BSE calculation was done with $N_k=34^2$, $N_v=2$ and $N_c=6$. The first two peaks correspond to the A, B excitons at the valleys, while the rest of the conductivity can be regarded as a shift of the non-interacting one.}\label{fig:mos2_exciton_bands}
\end{figure}

In Fig.~\ref{fig:mos2_kwf}(a) we show the reciprocal probability density of the first energy level. As opposed to hBN, we observe that the states are strongly localized at the valleys. 
Resolving the degeneracy by labeling each exciton with the $C_3$ eigenvalues would result in each exciton localized in a different valley~\cite{FengchengWu2015}. This shows that at least the low energy spectrum of $\text{MoS}_2$ can be studied at one valley instead of the whole BZ~\cite{bieniek2020}.  This allows to get a more precise description of the exciton since one can use a more refined mesh. The wavefunction for the exciton obtained at one valley can be seen in Fig.~\ref{fig:mos2_kwf}(b), using a feature of the code to reduce the BZ mesh by some integer factor. For higher excited states this does not hold since the states become more extended across the BZ, reaching both valleys. To illustrate a calculation with both exchange and finite $\mathbf{Q}$, we compute the exciton band structure in the vicinity of $\Gamma$, specifically in the $\mathbf{K}'-\Gamma-\mathbf{K}$ direction, represented in Fig.~\ref{fig:mos2_exciton_bands}(a). Without exchange the bands remain double degenerate, while including it results on a splitting of the excitons, following a quasi linear dispersion, due to the exchange interaction coupling both valleys~\cite{nonanalyticity}.

Since the excitons are very localized in reciprocal space, they should be delocalized in real-space, meaning that the radius of the exciton should be large (e.g.\ compared to that of hBN). To complete the characterization of the excitons, we calculate the optical conductivity as shown in Fig.~\ref{fig:mos2_exciton_bands}(b).
While the exciton energies converge quickly with $N_{\mathbf{k}}$, it is usually necessary to include more $\mathbf{k}$ points in the calculation of the optical conductivity in order to smooth unphysical oscillations derived from the discrete mesh. As it can be seen, the shape of the spectrum matches previous tight-binding studies~\cite{ridolfi2018,FengchengWu2015} and agrees well with ab-initio results~\cite{qiu2013}. At low energies, the optical conductivity of MoS$_2$ presents the characteristic A and B exciton peaks, that are understood considering the main spin-allowed electron-hole excitations at the $\mathbf{K}$ and $\mathbf{K}'$ points. The split of $\sim$$100$ meV between such peaks reflects the effect of SOC in TMD materials~\cite{liyilei2014}. At higher energies, the main feature of the spectra is a pronounced peak similar to the non-interacting case but red-shifted in energy. The excitons giving rise to such peak are often called "C" excitons and were fully characterized in Ref.~\cite{ridolfi2018}, already showing the potential of tight-binding methods for studying new exciton physics.

\section{Conclusions}

We have developed a software package that allows to solve the Bethe-Salpeter equation constructed from either tight-binding models or DFT calculations based on localized orbitals. By considering orbitals as point-like, the computation of the interactions becomes  drastically simplified. Together with an effective screening, this results in a fast determination of the BSE matrix. More specifically, our real-space implementation of the interaction matrix elements is shown to be faster and more precise than its reciprocal-space counterpart, which is the formulation more commonly used.

As in GW-BSE approximations, the starting band structure plays a crucial role in determining the resulting exciton spectrum. Therefore it is key to select the best possible functional (typically hybrids) or the most accurate tight-binding models that capture the most prominent features of the band structure. Then, by choosing appropriately the screening parameters, it is possible to reproduce the results of GW-BSE or similar first-principles codes at a fraction of the computational cost. 

The \texttt{Xatu} code currently provides all the tools needed to extract and characterize the exciton spectrum, either using the binary or via its API. Nevertheless, the package is still under development, as new functionalities and optimizations are added. Our future plans include giving support for distributed parallelism to enable bigger system sizes and calculation of different excitation types such as trions or biexcitons. The code is currently aimed at the description of 2D materials, but it can support 0D and 3D systems. Since the Keldysh potential is only adequate for 2D systems, we will implement additional potentials suitable for different dimensionalities.
We also plan to add the possibility of performing exact calculations of the interaction matrix elements when using Gaussian-based DFT codes to compute the band structure~\cite{garcia-blazquez2024}. Currently we provide an interface with the CRYSTAL code~\cite{crystal17}, and ideally more interfaces to community codes will be added over time, such as SIESTA~\cite{siesta2020}, Wannier90~\cite{pizzi2020wannier90} or PySCF~\cite{sun2018pyscf}. The project has been released under an open-source license and as such community contributions are welcome and encouraged.
\chapter{Topologically protected photovoltaics in Bi nanoribbons}

\section{Introduction}

Optoelectronics is a highly dynamic field of research, driven by the plethora of phenomena that emerge when materials interact with light, such as the multiple existing optical responses at different orders and the manipulation of the electronic structure, as demonstrated in techniques like pump-probe experiments~\cite{Cistaro2023, leone2016attosecond}. This is particularly true for 2D optoelectronics, where the confinement of the electronic states can lead to more pronounced effects, coming from exciton formation. This, combined with the high tunability of the atomic structure through strain, and the ability to select electronic excitations based on the light polarization, gives 2D optoelectronics its relevance from both fundamental and technological perspectives~\cite{Mueller2018}.

\begin{figure}[t]
    \centering
    \begin{tikzpicture}
        \node at (0,0) {\includegraphics[width=0.4\columnwidth]{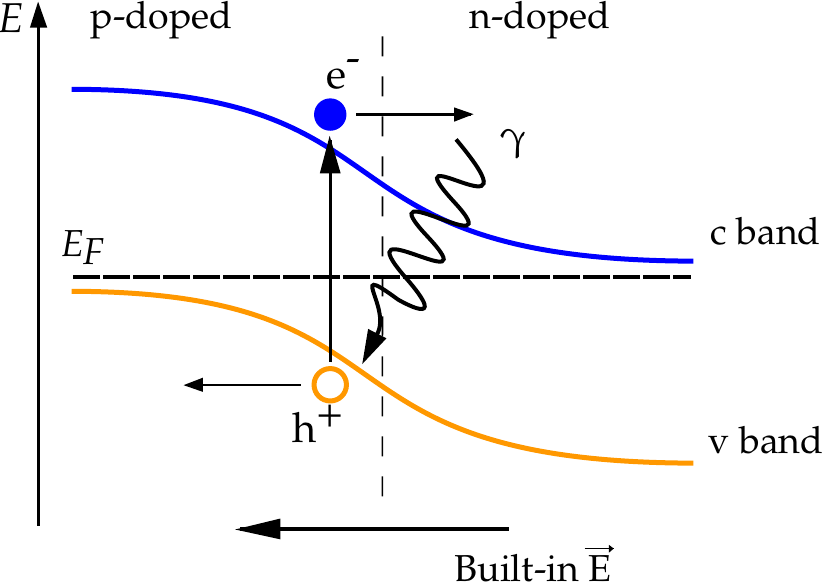}};
        \node at (-3.5,2.5) {(a)};

        \node at (7.7,0) {\includegraphics[width=0.55\columnwidth]{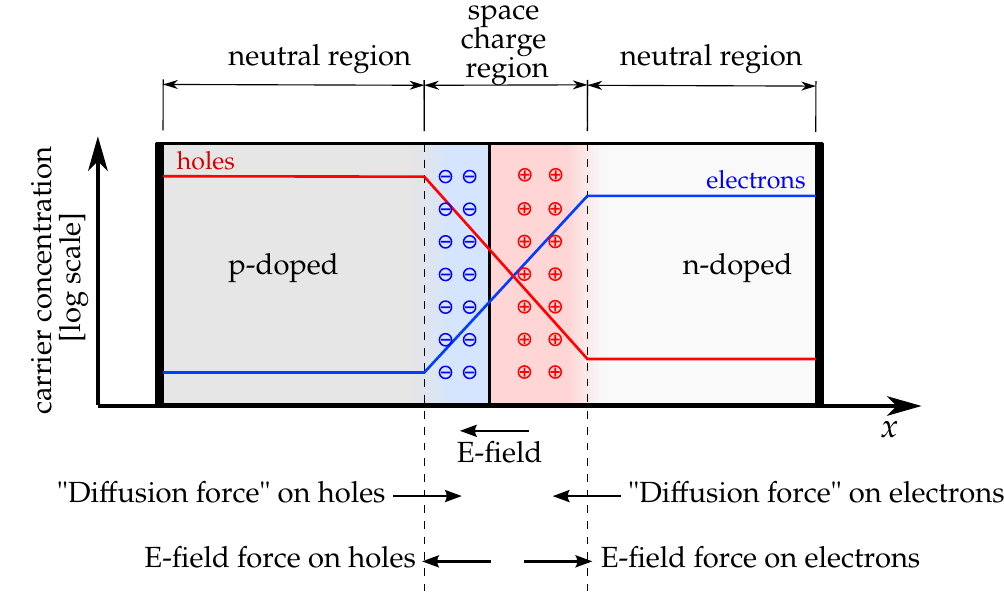}};
        \node at (4.5,2.5) {(b)};
    \end{tikzpicture}
    \caption[Schematic of charge separation in a p-n junction]{(a) Schematic diagram of the charge separation process in a p-n junction. Upon light incidence, an electron-hole pair is created in the depletion zone. Here, the built-in electric field due to the change in doping across the junction separates the carriers, generating a chemical potential difference or a current.\ (b) Representation of a p-n junction in equilibrium. The variations in the concentrations of the charge carriers across the junction produce non-neutral regions in the interface (the depletion zone), resulting in a built-in electric field. Adapted from~\cite{pn_junction_wiki}.}\label{fig:pn_junction}
\end{figure}

Of particular interest is light-energy conversion in the form of photocurrent generation, on which solar cells devices are based.
The formation of bound e-h pairs and their subsequent separation is the most common source of photocurrent generation. In conventional solar cells, based on p-n junctions, this is achieved with the built-in electric field in the depletion zone, which separates the free charge carriers generating a chemical potential difference or a current depending on the circuit scheme, as represented in Fig.~\ref{fig:pn_junction}. Since the efficiency of conventional cells is constrained by the Shockley–Queisser limit~\cite{shockley},  alternative dissociation mechanisms have been proposed as in multijunctions cells~\cite{multijunction_cells} or in excitonic solar cells, where a bound electron-hole pair is formed and diffuses to an interface where the charge separation takes place~\cite{Gregg2003, multiexciton}.

While the Shockley–Queisser limit specifies the maximum efficiency of a conventional solar cell, in practice there are many unwanted effects that hinder the performance of the device below this theoretical limit~\cite{cheng2022understanding}. These effects, depicted in Fig.~\ref{fig:recombination_mechanisms}, include radiative recombinations (photon emission) of the charge carriers before the charge separation takes place, or non-radiative recombinations, where phonons are emitted to conserve energy. One such case is the non-radiative electron-hole recombination at in-gap states coming from surface states, defects or traps. Alternative processes are non-radiative Auger recombinations, where an existing trion or biexciton partly recombines, leaving the remaining charge carriers with higher energy. 

\newcommand\ya{1}
\newcommand\yb{3.5}
\newcommand\defect{2.25}
\begin{figure}[h]
    \centering
    \begin{tikzpicture}
        \draw[->] (0,\ya - 1) -- (0,\yb + 1) node[left] {E};
        \draw[thick] (-0.1, \yb) -- (0.1, \yb) node[left, xshift=-0.1cm] {$E_c$};

        \draw[thick] (-0.1, \ya) -- (0.1, \ya) node[left, xshift=-0.1cm] {$E_v$};

        \draw[thick] (-0.1, {(\ya + \yb) / 2}) -- (0.1, {(\ya + \yb) / 2}) node[left, xshift=-0.1cm] {$E_F$};

        \shade[bottom color=blue!20, top color=white] (0.5,\yb) rectangle (3,\yb+1);
        \draw[thick] (0.5,\yb) -- (3,\yb);

        \shade[top color=orange!20, bottom color=white] (0.5,\ya) rectangle (3,\ya-1);
        \draw[thick] (0.5,\ya) -- (3,\ya);

        \node[align=center] at (1.75, -0.5) {Radiative\\recombination};
        \filldraw[blue] (1,\yb) circle (2pt) node[above] {\(e^-\)};
        \filldraw[red] (1,\ya) circle (2pt) node[below] {\(h^+\)};
        \draw[->, thick] (1,\yb) -- (1,\ya + 0.1);
        \draw[->, thick, decorate, decoration={snake, amplitude=.4mm, segment length=2mm, post length=1mm}] (1,\yb) -- (2,2) node[midway, right] {$\gamma$};

        \shade[bottom color=blue!20, top color=white] (4.5,\yb) rectangle (7,\yb+1);
        \draw[thick] (4.5,\yb) -- (7,\yb);

        \shade[top color=orange!20, bottom color=white] (4.5,\ya) rectangle (7,\ya-1);
        \draw[thick] (4.5,\ya) -- (7,\ya);

        \node[align=center] at (5.75, -0.75) {Non-radiative\\surface/trap\\recombination};
        \filldraw[blue] (5.75,\yb) circle (2pt) node[above] {\(e^-\)};
        \filldraw[red] (5.75,\ya) circle (2pt) node[below] {\(h^+\)};
        \draw[thick] (5.75 - 0.4, \defect) -- (5.75 + 0.4, \defect);
        \draw[->, thick] (5.75,\yb) -- (5.75,\defect);
        \draw[->, thick] (5.75,\ya) -- (5.75,\defect);

        \shade[bottom color=blue!20, top color=white] (8.5,\yb) rectangle (12,\yb+1);
        \draw[thick] (8.5,\yb) -- (12,\yb);

        \shade[top color=orange!20, bottom color=white] (8.5,\ya) rectangle (12,\ya-1);
        \draw[thick] (8.5,\ya) -- (12,\ya);
        \node[align=center] at (10.25, -0.5) {Auger non-radiative\\recombination};
        \filldraw[blue] (9,\yb) circle (2pt) node[above] {\(e^-\)};
        \filldraw[blue] (9.5,\yb) circle (2pt) node[below, xshift=0.2cm] {\(e^-\)};
        \filldraw[red] (9,\ya) circle (2pt) node[below] {\(h^+\)};
        \draw[->, thick] (9,\yb) -- (9,\ya);
        \draw[->, thick] (9.5,\yb) -- (9.5,\yb + 1);

        \draw[dashed, blue] (9,\yb) arc[start angle=180, end angle=360, radius=0.25];

        \filldraw[blue] (11,\yb) circle (2pt) node[above] {\(e^-\)};
        \filldraw[red] (11.5,\ya) circle (2pt) node[above, xshift=0.2cm] {\(h^+\)};
        \filldraw[red] (11,\ya) circle (2pt) node[below] {\(h^+\)};
        \draw[->, thick] (11,\yb) -- (11,\ya);
        \draw[->, thick] (11.5,\ya) -- (11.5,\ya - 1);

        \draw[dashed, red] (11,\ya) arc[start angle=180, end angle=0, radius=0.25];

        \node at (2.65, \yb + 0.6) {(a)};
        \node at (6.65, \yb + 0.6) {(b)};
        \node at (11.65, \yb + 0.6) {(c)};

    \end{tikzpicture}
    \caption[Different electron-hole recombination mechanisms in semiconductors]{Different electron-hole recombination mechanisms in semiconductors: (a) Radiative recombination, (b) non-radiative recombination via defect or surface states, (c) non-radiative Auger recombination of trions.}
    \label{fig:recombination_mechanisms}
\end{figure}
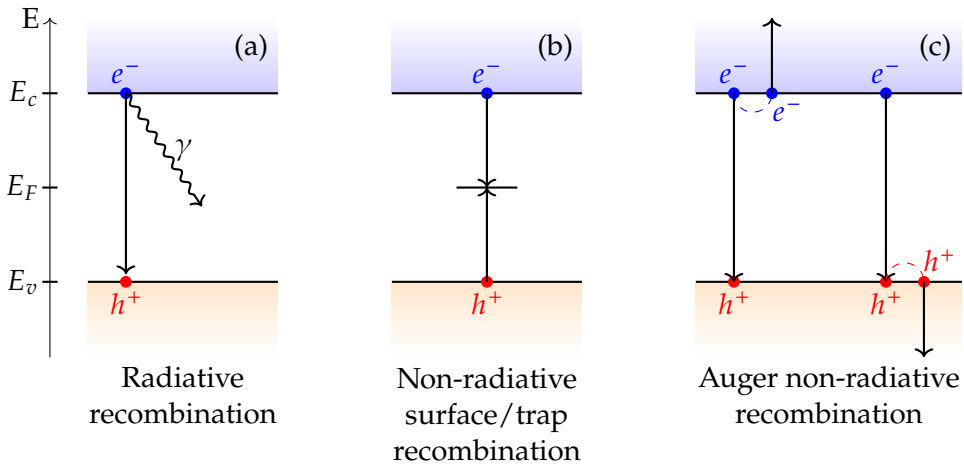

Topological insulators (TIs), on the other hand, have garnered significant attention in recent years due to their potential for use in spintronic devices, among other more fundamental reasons~\cite{spintronics}. We review in detail the properties of TIs in Chapters~\ref{chapter:intro_topology} and~\ref{chapter:topology}; here it suffices to know that TIs present topological edge states, i.e.\ when considering a semi-infinite TI such as a ribbon, its electronic band structure will exhibit edge bands connecting the valence and conduction bands, see for example Fig.~\ref{fig:unitcell} for Bi(111). These bands are helical, meaning that there is a momentum-spin locking (i.e.\ not independent), but most importantly, the edge states are topologically protected, which implies that they will always be present as long as the bulk gap of the system does not close under deformations, for instance introducing defects~\cite{hasan2010}.

Thus, TIs have been extensively studied in regard to their topological properties, identifying new topological materials or exploring the implications of the edge states for different physical phenomena. However, there has been relatively little study of TIs from an optical perspective, mostly focusing on obtaining the bulk, non-interacting optical conductivity~\cite{exp_bi_conductivity, Pandey2021, optical_abs_ti_thin_films, magneto_optical_conductivity, PhysRevB.100.195110, PhysRevB.97.081402, berry_phase_exciton, jiawei2023}, and some reporting the observation of excitons in topological insulators~\cite{Bi2Se3_exciton_exp, observation_excitons_ti}. Only recently, for instance, the exciton spectrum in Bi$_2$Se$_3$ was shown to exhibit topological properties~\cite{topological_excitons}.

While there are works addressing the role of trivial edge states in the dissociation of excitons in semiconductors~\cite{souvik2023, doi:10.1126/science.aal4211, Sui2022, Kinigstein2020, doi:10.1063/1.4968794}, the interaction between bulk excitons in TIs and their topologically protected edge states remains, however, largely unexplored. One recent work studies the interplay between bulk and topological states in the formation of excitons in Bi$_2$Se$_3$ and how these affect the optical response of the TI~\cite{bowen2023}.

According to Fermi's golden rule~\cite{cohen1986quantum}, an exciton is expected to decay elastically into a continuum of states in the presence of a given coupling, in this case the Coulomb interaction. Excitons lie within the energy gap and in a trivial insulator there are typically no pure electronic excitations accessible for the exciton to decay into. As we stated before, the usual dissociation channels would be radiative (photons) or non-radiative (phonons) recombinations~\cite{pelant2012luminescence, segall1968}, in particular recombination at the surface or defects. Topological insulators, instead, always present edge states connecting the valence and conduction bands~\cite{hasan2010} (see Fig.~\ref{fig:qshe}), meaning that in addition to light emission, the exciton can decay into e-h pairs formed by edge states. Therefore, TIs provide a situation analogue to the surface state recombination in semiconductors, where instead of having some discrete energy levels in-gap, we have topological edge states connecting continuously the valence and conduction bands. Then, excitons formed in the bulk of the material could potentially decay elastically into the edge states, instead of inelastically as in semiconductors (via phonon emission). The main difference being, in the semiconductor the exciton recombines at the surface state, while in the TI the exciton dissociates into the edge states, namely the charges remain separated.

\begin{figure}[h]
    \centering
    \begin{tikzpicture}

        \draw[->] (-2,-1.) -- (-2,2.5) node[left] {E};
 
        \draw[thick] (-2.1, 0.75) -- (-1.9, 0.75) node[left, xshift=-0.1cm] {$E_F$};

        \draw[thick] plot[domain=-1.5:1.5] (\x, {0.4*\x*\x + 1.5}) node[right] {\(c\)};
        \draw[thick] plot[domain=-1.5:1.5] (\x, {-0.4*\x*\x}) node[right] {\(v\)};

        \draw[dashed] (-0.6, 1.25) -- (0.6, 1.25) node[right] {\(\ket{X}\)};

        \draw[->, thick] (1.75, 0.75) -- (3.25, 0.75) node[midway, above] {\(\Gamma\)};

        \draw[thick] plot[domain=3.5:6.5] (\x, {0.4*(\x-5)*(\x-5) + 1.5}) node[right] {\(c\)};
        \draw[thick] plot[domain=3.5:6.5] (\x, {-0.4*(\x-5)*(\x-5) }) node[right] {\(v\)};

        \draw[thick] (3.5, -0.9) -- (5, 0.75) -- (6.5, 2.4);
        \draw[thick] (3.5, 2.4) -- (5, 0.75) -- (6.5, -0.9);

        \filldraw[blue] (5.6, 0.625 + 0.75) circle (2pt) node[right] {\(e^-\)};
        \filldraw[red]  (5.6, -0.625 + 0.75) circle (2pt) node[right] {\(h^+\)};
        \draw[dashed] (5.6, 0.625 + 0.75) -- (5.6, -0.625 + 0.75);
        
    \end{tikzpicture}
    \caption[Dissociation of a bulk exciton into an edge electron-hole pair]{Dissociation of a bulk exciton into a non-interacting electron-hole pair hosted at the topological edge bands. Since the exciton lies within the insulating gap, in a TI there is always a non-interacting edge electron-hole pair such that energy is conserved. In the left diagram, we hide the topological edge bands for clarity.}
\end{figure}
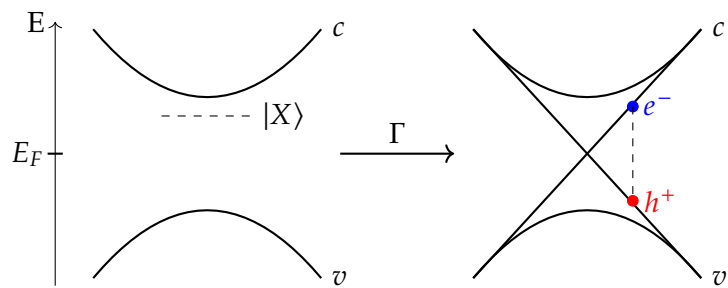

\begin{figure}[h]
    \centering
    \includegraphics[width=0.9\columnwidth]{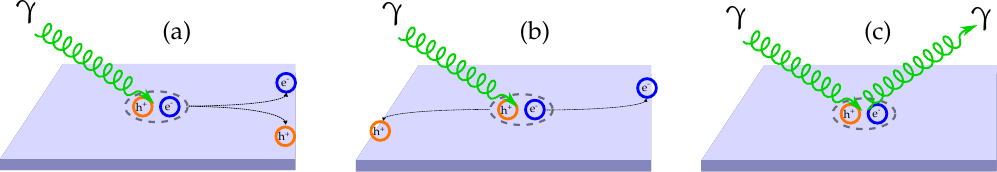}
    \caption[Exciton decay paths in a topological insulator]{Possible decay paths for an exciton in a topological insulator. The exciton can decay in principle into edge electron-hole pairs, which can then be either (a) intra-edge or (b) inter-edge.\ (c) The exciton can also decay radiatively, emitting a photon.}\label{fig:exciton_decay_processes}
\end{figure}

Therefore, in a TI we identify three main dissociation channels for the exciton, illustrated in Fig.~\ref{fig:exciton_decay_processes}: either the standard radiative recombination, or the exciton decaying into electron-hole pairs hosted at the topological edge bands, where we additionally distinguish between intra- or inter-edge transitions. Note that we disregard any phonon-assisted process as those require a more complex description, namely including an electron-phonon coupling term in the Hamiltonian. In this work, we focus on the electronic transitions only.
Our principal observation here is that, for sufficiently narrow 2D TI systems (TI ribbons), the electron and hole can decay onto opposite edges, resulting in a charge separation and eventually in a photovoltaic current~\cite{uria_topologically}. In what follows, we discuss the conditions under which these processes can be achieved. We test and evaluate the actual transition rates for the exciton decay into the edge states for a tight-binding model of Bi(111). Tuning the different parameters of the model, we are able to modify the ratios between the different dissociation channels available, making it possible for the charge separation and the current generation processes to compete with the other recombination mechanisms present.

\section{Symmetry breaking as the driving mechanism for the photovoltaic effect}

\begin{figure}[h]
    \centering
    \includegraphics[width=0.9\columnwidth]{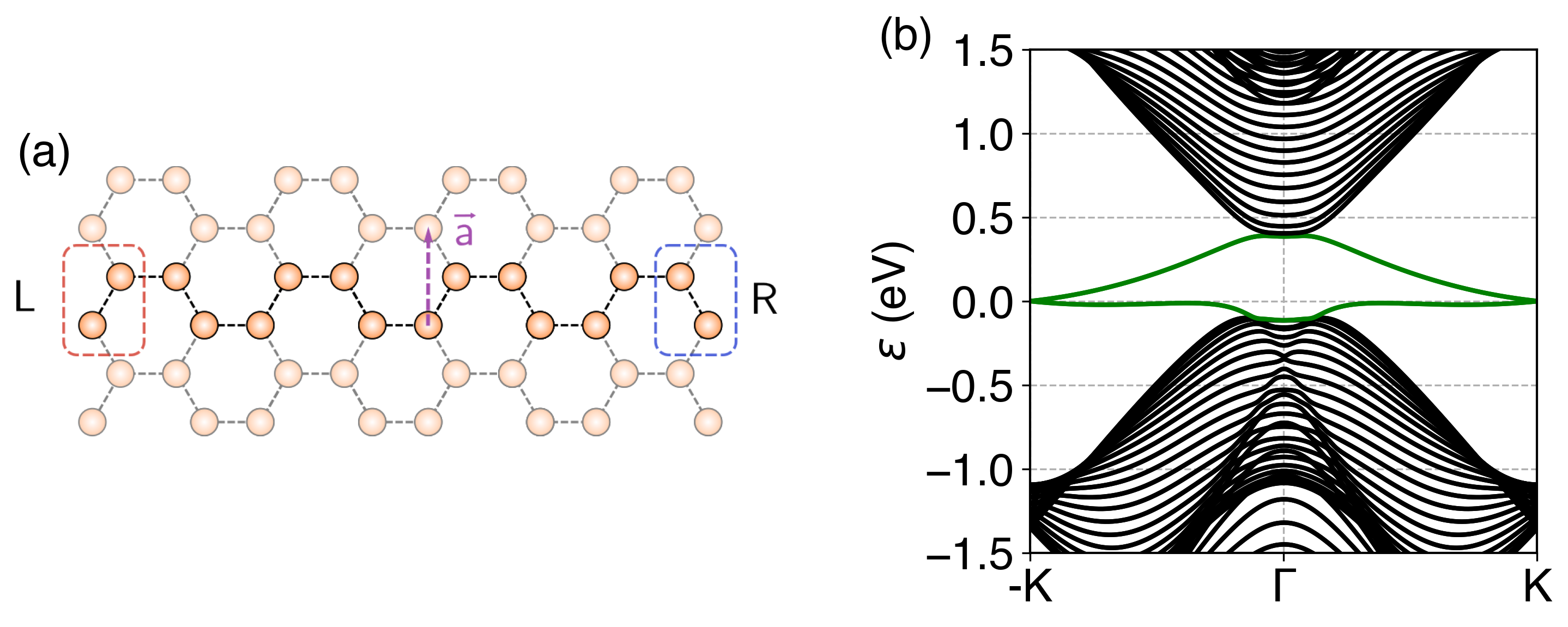}
    \caption[Unit cell and band structure of a zigzag ribbon of Bi(111)]{(a) Bi(111) zigzag nanoribbon where the dissociation process takes place. The highlighted atoms denote the unit cell, and $\mathbf{a}$ is the Bravais vector. The edge atoms are identified with the rectangles and labeled as $L$ (left) or $R$ (right). We introduce onsite energies on the left edge to split the topological edge bands.\ (b) Band structure of a Bi(111) zigzag ribbon (without edge term) for $N=20$, with the edge bands highlighted in green.}\label{fig:unitcell}
\end{figure}

A purely electronic exciton decay can take place in the form of a non-interacting e-h pair where both constituents are located on the same edge, or on opposite edges, which may result in charge transportation since edge electrons and holes have typically finite velocity. 
Due to time-reversal invariance, however, there is a ${k}\leftrightarrow -{k}$ symmetry in the electronic bands (see Fig.~\ref{fig:unitcell}(b)), meaning that the e-h pair can either be equally located at ${k}$ or $-{k}$, preventing such possibility for both inter- and intra-edge processes. On top of time-reversal invariance, note also that the system may also possess inversion symmetry, forcing any current appearing on one edge to be cancelled by the one appearing on the opposite one. 

\begin{figure}[t]
    \centering
    \includegraphics[width=0.9\columnwidth]{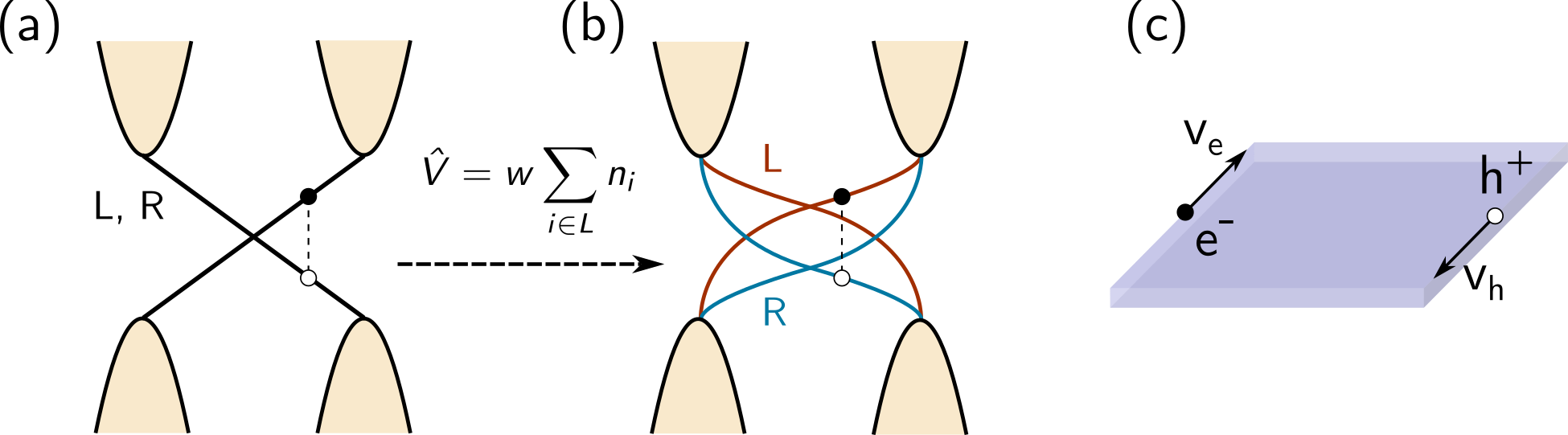}
    \caption[Splitting of the topological edge bands in presence of an edge potential]{Splitting of the edge bands.\ (a) For a topological insulator with inversion symmetry, the edge bands of both sides are degenerate, resulting in identical rates for intra-edge and inter-edge transitions.\ (b) The introduction of an edge offset potential allows to split the edge bands, producing a distinction between the different transitions.\ (c) For each edge e-h pair we can determine its total velocity as $v_{\rm{e-h}}=v_e-v_h$~\cite{kittel2021introduction} to establish whether it carries current or not. For the pair drawn in (b), we observe that $v_e>0$ and $v_h < 0$, meaning that $v_{\rm{e-h}}=v_e-v_h>0$.}\label{fig:edge_bands_splitting}
\end{figure}

Even if a priori one is unable to generate current in the presence of time-reversal symmetry (TRS), it is still possible to generate a charge imbalance between the edges. Consider that we introduce an asymmetry between the edges, via an electric field applied in the direction perpendicular to the infinite edges, or simply by some asymmetric termination.
The latter is implemented in the following Hamiltonian:

\begin{align}
     H = \sum_{i\alpha}\varepsilon_{i\alpha}c^{\dagger}_{i\alpha}c_{i\alpha} + \sum_{i\alpha,j\beta}t^{\alpha\beta}_{ij}c^{\dagger}_{i\alpha}c_{j\beta}
    + \lambda\sum_{i\alpha,j\beta}\braket{i\alpha|\mathbf{L}\cdot\mathbf{S}|j\beta}c^{\dagger}_{i\alpha}c_{j\beta} + w\sum_{i\in L} n_i,
\end{align}
where the first three terms correspond to $H_0$, which is a Slater-Koster tight-binding model of a ribbon of Bi(111)~\cite{liu_allen}, known to be a topological insulator~\cite{murakami, bi_bl_dft, bi111_thinfilms,sabater2013topologically}. 
The last term is the edge offset potential. In particular, we work with a zigzag termination~\cite{bi_bl_dft, symmorphic_ribbon}. The width of the ribbon is given by $N$, which is the number of dimers in the ribbon, taken to be even. In Fig.~\ref{fig:unitcell}(a) we show an example unit cell of the Bi(111) ribbon, and the atoms we identify as left (L) and right (R). On the left ones, we introduce additional onsite energies corresponding to the edge offset to split the edge bands. 
Then, as long as the perturbation does not close the bulk gap, the edge bands will split, as schematically shown in Fig.~\ref{fig:edge_bands_splitting}(b). The splitting is expected to produce a different transition rate depending on whether the e-h pair is localized on the left-right boundaries, respectively, or the right-left ones. 

\begin{figure}[H]
    \centering
    \includegraphics[width=0.5\columnwidth]{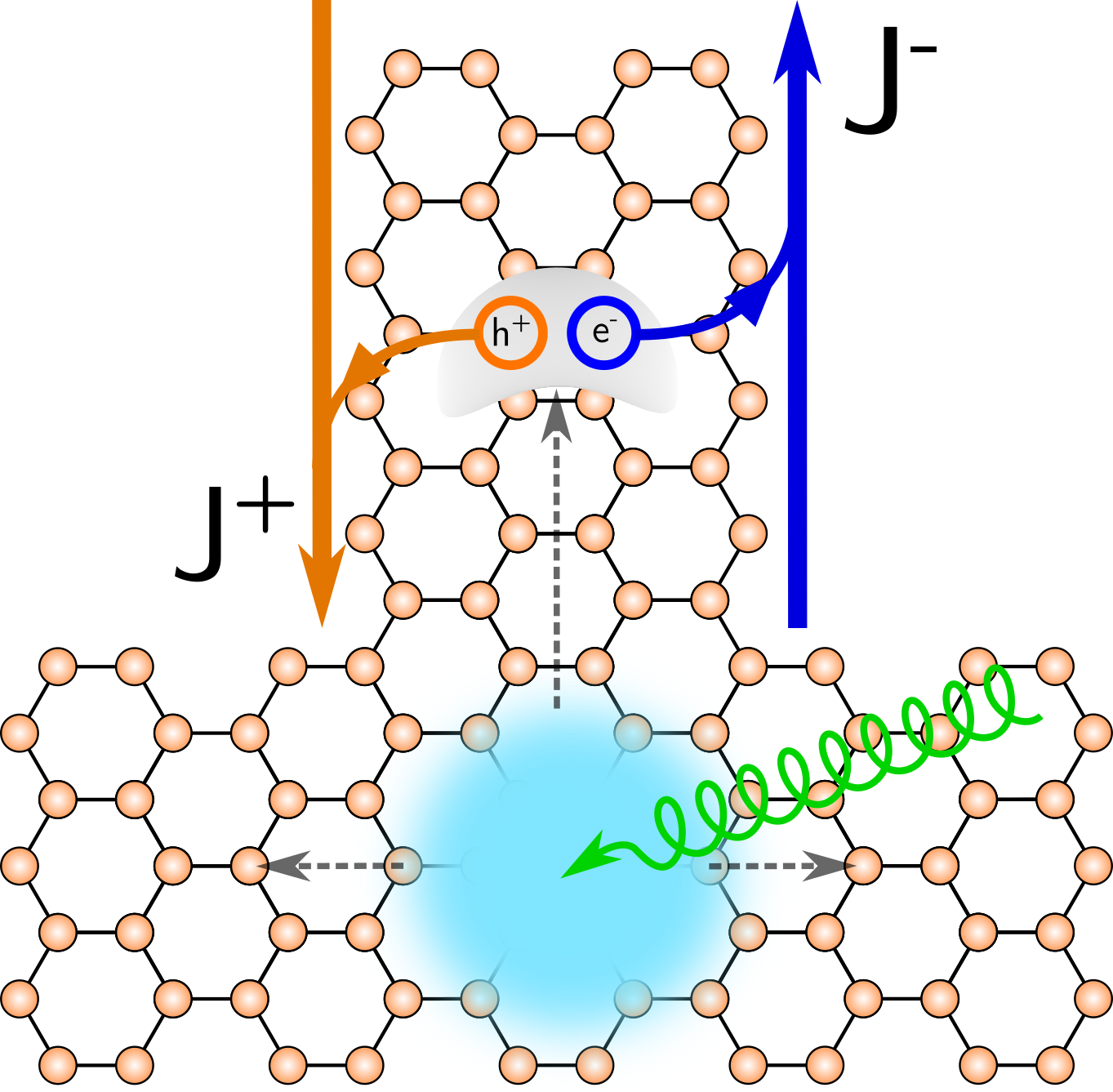}
    \caption[Device where the topological photovoltaic effect is achieved]{Schematic representation of the proposed mechanism. Device where an exciton wave packet is created at the bulk of the sample, where it will diffuse in any direction. Excitons entering the top ribbon
   present a finite momentum $Q$, giving rise to an out-of-equilibrium edge carrier population with non-zero momentum and velocity, thus forming a topologically protected current.}\label{fig:setup}
\end{figure}

In addition to charge accumulation, current generation is also possible if time-reversal symmetry is broken. This may occur by a selective population of excitons with non-zero $Q$, avoiding their time-reversal partners with opposite momentum. One possible way to achieve this is shown in Fig.~\ref{fig:setup}, where an exciton wave packet is created in the bulk of the sample. This packet is generically described by a momentum distribution $\ket{X} = \int d\mathbf{Q}f(\mathbf{Q})\ket{X(\mathbf{Q})}$. Since excitons with finite momentum also have finite velocity, some of them will propagate into the top ribbon where the edge offset is present. This populates the ribbon with excitons with finite $Q$, but not their time-reversal companions. From the dissociation of these excitons into inter-edge electron-hole pairs we expect to generate a topologically protected photocurrent.

In summary, time-reversal symmetry will prevent any current from appearing in the material, whereas inversion symmetry will impede any charge separation. Introducing an edge offset potential, we break inversion symmetry, which in principle should allow for a preferential decay direction for the charge carriers, producing an edge charge imbalance. Likewise, if we break TRS via a population of excitons that is non-time-reversal invariant, then the imbalanced charges at the edges could also potentially form a topologically protected edge current.

\section{Transition rates between bulk excitons and topological edge states}

\begin{figure}[h]
    \centering
    \includegraphics[width=0.9\columnwidth]{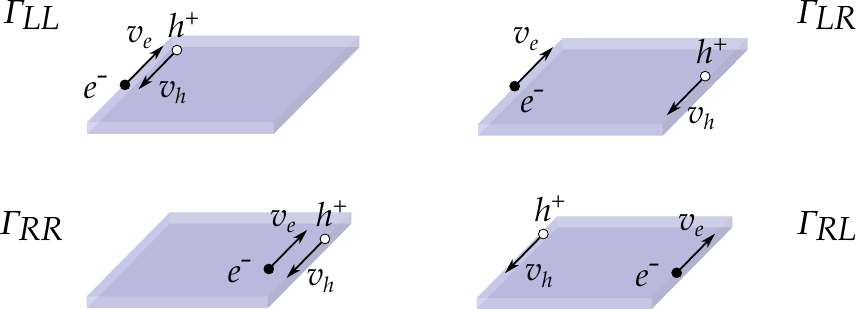}
    \caption[Possible transition rates to the edge electron-hole pairs]{Available transition rates to the edge electron-hole pairs, corresponding to the different combinations of edge indices $s,s'$. The rates can be intra-edge ($\Gamma_{RR}$ and $\Gamma_{LL}$) or inter-edge ($\Gamma_{RL}$ and $\Gamma_{LR}$).}\label{fig:rates_possibilities}
\end{figure}

To test these hypotheses, we need to evaluate the transition rate from the exciton to each one of the possible electron-hole pairs. Instead of using the band number, we denote each band by its location or edge index, $R$ (right) and $L$ (left). Thus,  for instance, an electron and a hole located on the opposite edges with momentum $k$ would be $\ket{L, R, k}$. With this notation, we want to evaluate the following transition rates:
\begin{equation}
\label{eq:fermi_rule}
    \Gamma^{\pm}_{ss'} = \sum_n\frac{2\pi}{\hbar}\left|\braket{X_n|V|s, s', \pm{k}}\right|^2\rho(E_X)
\end{equation} 
where $n$ runs over degenerate exciton states (if present), $s,s'\in\{R, L\}$ denote the edge where the electron, hole are localized respectively (see Fig.~\ref{fig:rates_possibilities}), $\rho$ is the density of states of the final continuum of states, namely the edge e-h pairs, and $V$ is the electrostatic interaction. Note that we hide from the notation the center-of-mass momentum $Q$ of both the exciton and the edge electron-hole pair, but it is implicit and can be non-zero in general (see Eqs.~\eqref{eq:exciton_chap4},~\eqref{eq:edge_eh_pair}). $E_X$ is the energy of the exciton, defined as the energy of the state relative to the Fermi sea. The initial exciton $\ket{X}$ is taken as the bulk ground state exciton:
\begin{equation}
    \ket{X_n(Q)} = \sum_{v,c,k}A^n_{vc}(k,Q)c^{\dagger}_{ck+Q}c_{vk}\ket{FS}
    \label{eq:exciton_chap4}
\end{equation}
which is a superposition of electron-hole pairs between any conduction ($c$) and valence ($v$) bands, excluding the edge bands. We write momenta quantum numbers as scalars instead of vectors, since we already assume to be working with a ribbon (one-dimensional BZ). $\ket{FS}$ denotes the Fermi sea, and the coefficients $A^n_{vc}(k,Q)$ which determine the exciton states are obtained solving the Bethe-Salpeter equation~\eqref{bse}~\cite{wu2015, bieniek2022, trolle2014theory, quintela2022}. Specifically, we use the previously developed real-space formalism for the interaction matrix element~\eqref{direct},~\eqref{exchange}, where assuming that all orbitals are point-like simplifies greatly the calculation of the exciton spectrum~\cite{uria_xatu}. Also, the real-space formalism allows taking into account the finite boundaries of the ribbon, whereas the reciprocal formalism assumes periodic boundary conditions for an infinitely sized 2D crystal. Therefore, the real-space approach is better suited to address the problem, on top of its intrinsic benefits such as faster convergence with $N_{\mathbf{k}}$ or faster calculations. As for the exchange term present in the BSE, we set $X=0$ assuming that its contribution is negligible.
Regarding screening, we use the Rytova-Keldysh potential~\eqref{eq:keldysh}, setting the material dielectric constant to $\epsilon=40$ and the environmental one to $\bar{\epsilon}=2.45$, which corresponds to a SiO$_2$ substrate. 
The edge e-h pair is defined as:
\begin{equation}
    \ket{s,s',k} = c^{\dagger}_{sk+Q}c_{s'k}\ket{FS}
    \label{eq:edge_eh_pair}
\end{equation}
where $c^{\dagger}_{sk+Q}$ creates a conduction electron such that it is located at side $s$ with the specified momentum. The same is done with $c_{s'k}$ for the valence hole.

$k$ is chosen such that, given $s,s'$, the corresponding e-h pair has the same energy as the exciton, $\braket{H}_X=\braket{H}_{e-h}$. The sign of ${k}$  must also be specified since there are two possibilities and, in principle, transitions can be asymmetric in $\pm{k}$. 
When the inversion symmetry is removed by the edge offset potential $w$, e.g\@. at the left boundary, the edge bands, as shown in Fig.~\ref{fig:transitionsQ0_splitedges_width}(a), are split. We expect now that the inter-edge transition rates $\Gamma_{RL}$ and $\Gamma_{LR}$ will be different as the inter-edge e-h pairs correspond to different $|{k}|$ points (see Fig.~\ref{fig:transitionsQ0_splitedges_width}(a)), producing an inter-edge charge imbalance in the material. This mechanism would compete with intra-edge transitions $\Gamma_{RR}$ and $\Gamma_{LL}$, where the electron and hole eventually recombine on the same edge. The intra-edge rates serve then as the baseline to estimate the efficiency of the effect.

To compute the transition rates, we simply expand the exciton states in the basis of electron-hole pairs. Then, the transition rates are cast in terms of the direct and exchange terms of the BSE, namely~\eqref{kernel_D_X}. Since we neglect the exchange term, the final expression for the rates is:
\begin{equation}
    \Gamma^{\pm}_{ss'} = \frac{2\pi}{\hbar}\sum_n\left|\sum_{v,c,k'}A^n_{vc}(k',Q)D_{vc,s's}(k,k',Q)\right|^2\rho(E_X)
\end{equation}
where $D_{vc,s's}(k,k',Q)$ is the direct term, now written in terms of the edge indices $s,s'$. Finally, it should also be noted that in case of a complex edge band structure (as it is the case for the armchair termination of the ribbon, see section~\ref{sec:armchair_bi111}), there might be multiple band quantum numbers $n$ that correspond to the same edge indices $s,s'$ and momentum $\pm k$. In this case, the transition rates are given summing over all possible final states, in analogy with the summation over degenerate exciton states.

\subsection{Exciton spectrum in Bi(111)}\label{sec:exciton_bi111}

\begin{figure}[h]
    \centering
    \includegraphics[width=0.7\columnwidth]{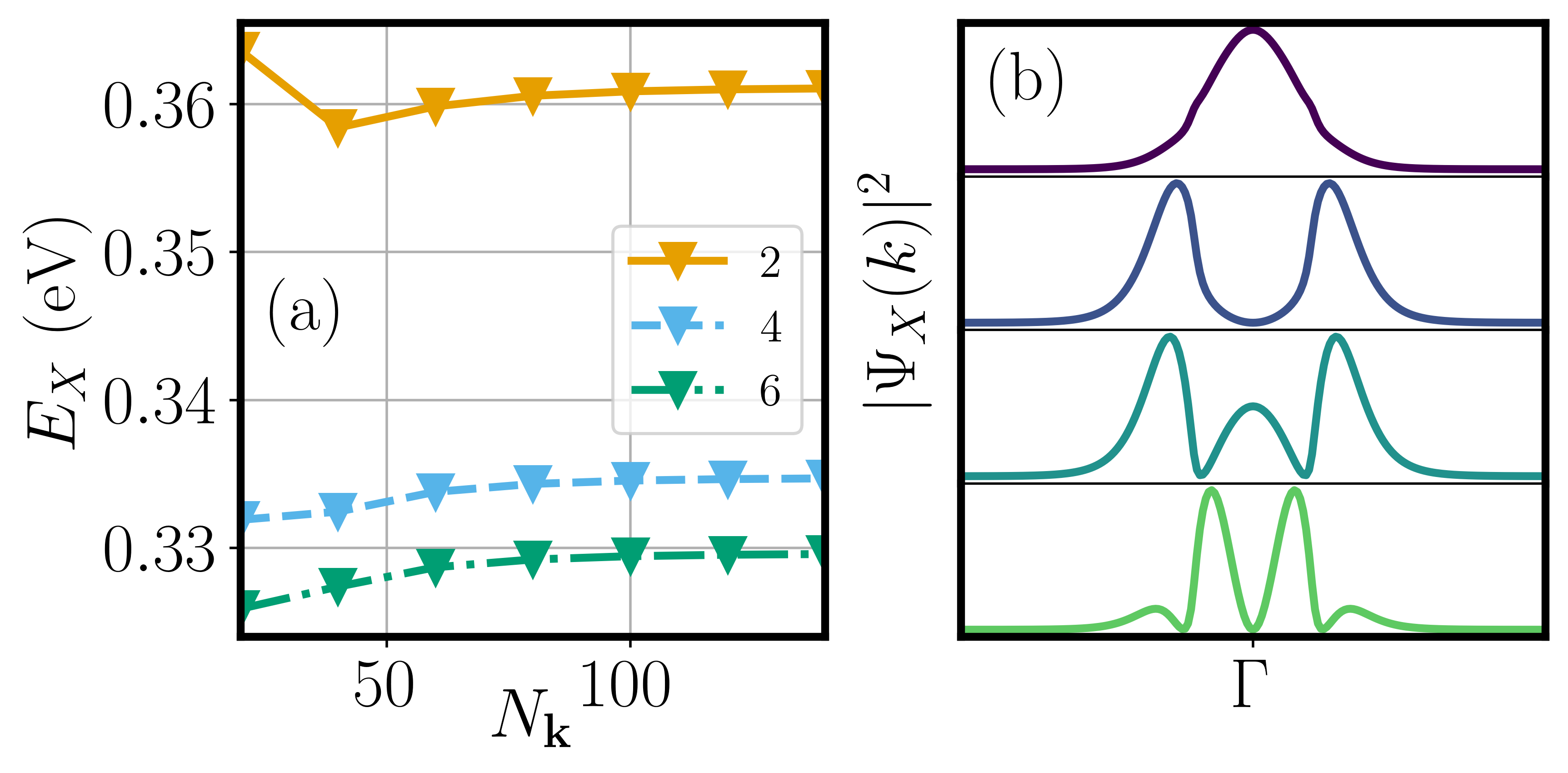}
    \caption[Energy convergence and reciprocal wavefunctions of excitons in a Bi(111) ribbon]{(a) Convergence of the ground state exciton energy in a ribbon of width $N=14$ with $N_{\mathbf{k}}$ and the number of bands included for both valence $N_v$ and conduction $N_c$, i.e. $N_c = N_v = 2$, 4 or 6.\ (b) ${k}$-probability density of the first four exciton levels for $N_v = N_c = 2$, from top ($n=1$) to bottom ($n=4$).}\label{fig:kwf_convergence}
\end{figure}

Before computing the transition rates, first we characterize the exciton spectrum of the Bi(111) ribbon. As for any exciton calculation, we begin converging the exciton energy spectrum with respect to the number of ${k}$ points, $N_{{k}}$, as well as the number of bands included in the calculation $N_{v/c}$. The convergence of the ground state exciton energy is shown in Fig.~\ref{fig:kwf_convergence}(a), which converges very quickly with $N_{{k}}$, although there is a stronger dependence with the number of bands included. Due to the limited computational resources, we cannot include an arbitrary large number of bands and ${k}$ points. As we will show later, a good compromise can be obtained using $N_v=N_c=4$ bands in the calculation of transition rates, which does not change the results qualitatively but allows for fully converged rates in $N_{{k}}$. 

\begin{figure}
    \centering
    \includegraphics[width=0.8\columnwidth]{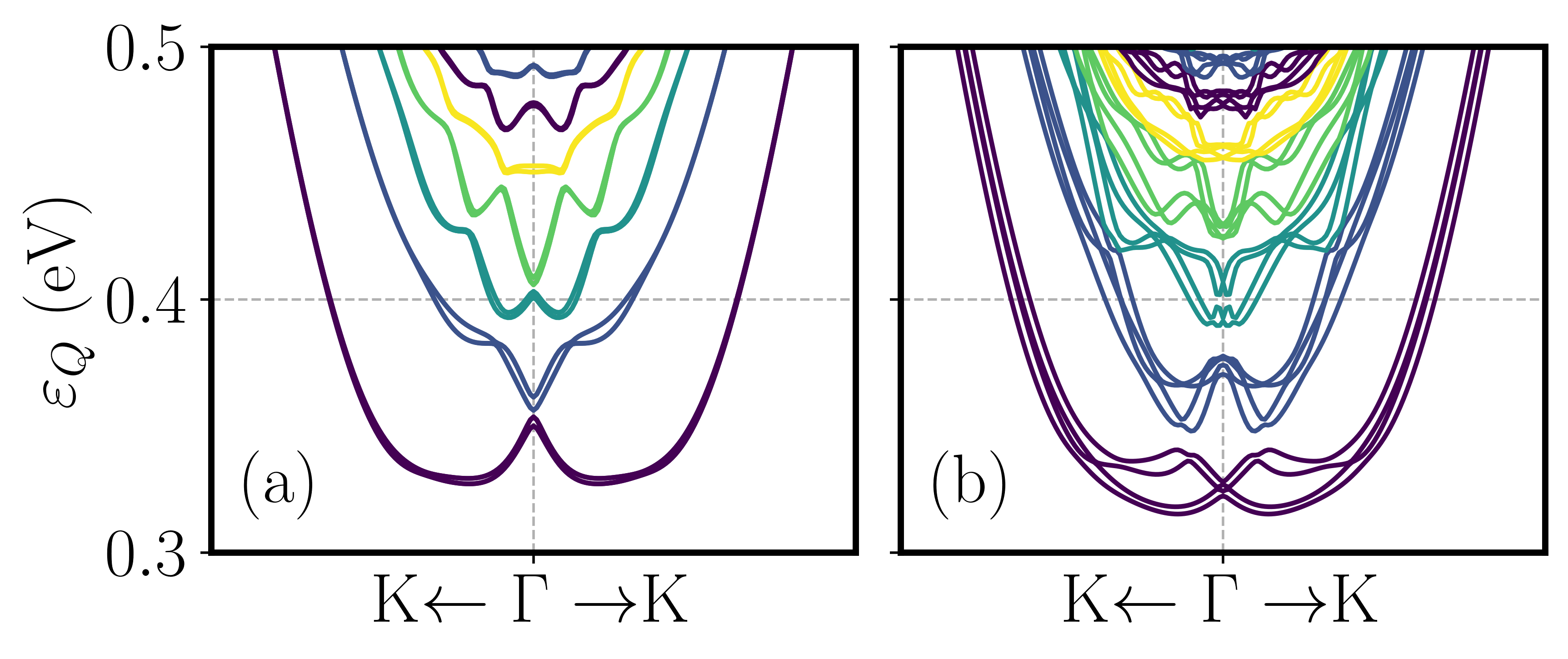}
    \caption[Exciton band structure of a Bi(111) ribbon]{Exciton band structure of a Bi(111) ribbon for $N=14$ as a function of the center-of-mass momentum $Q$, for $N_v=N_c=2$.\ (a) Band structure without inversion breaking (staggered) potential $V_{st}$, and (b) with $V_{st}=0.1$ eV. Each different color groups four consecutive exciton states.}\label{fig:exciton_bands}
\end{figure}

Regarding the energy levels, when computing the exciton spectrum with Q = 0 and without spin-orbit coupling (SOC), we observe a four-fold degeneracy of all energy levels. This is to be expected, since the point group of the ribbon is $C_{2h}$, whose irreducible representations (irreps) are all of dimension 1. For a spinless system, all states would be non-degenerate, while for a spinful system with neither SOC nor exchange, four-fold degenerate. As we turn on the SOC, the exciton spectrum acquires a fine structure. The character table of the $C_{2h}$ double group shows one-dimensional irreps~\cite{doublegroups}. However, we observe a two-fold degeneracy for some states, in particular for the ground state. The combination of inversion symmetry with time-reversal symmetry results in two-fold degenerate single-particle bands, which results then in two-fold degenerate exciton bands, even in presence of spin-orbit coupling. These two-fold degeneracies also extend to finite Q excitons. The origin of the degeneracy can be checked introducing a sub-lattice staggered potential $V_{st}$ which breaks inversion symmetry, lifting the quasiparticle band degeneracy and in turn results in a splitting of the degenerate exciton bands. The exciton bands for both $N_v=N_c=2$ with and without a staggered potential $V_{st}$ are represented in Figs.~\ref{fig:exciton_bands}(a, b) respectively.

Additionally, we examine both the reciprocal and real-space probability densities to ensure that they behave as expected. The reciprocal wavefunctions for the first four energy levels are shown in Fig.~\ref{fig:kwf_convergence}(b). They are centered around $\Gamma$, where the gap is located as seen from Fig.~\ref{fig:unitcell}(b). As we go to higher energy levels, the wavefunctions become more delocalized and show the standard oscillatory behaviour of states confined in a quantum well.

\begin{figure}[t]
    \centering
    \includegraphics[width=1\columnwidth]{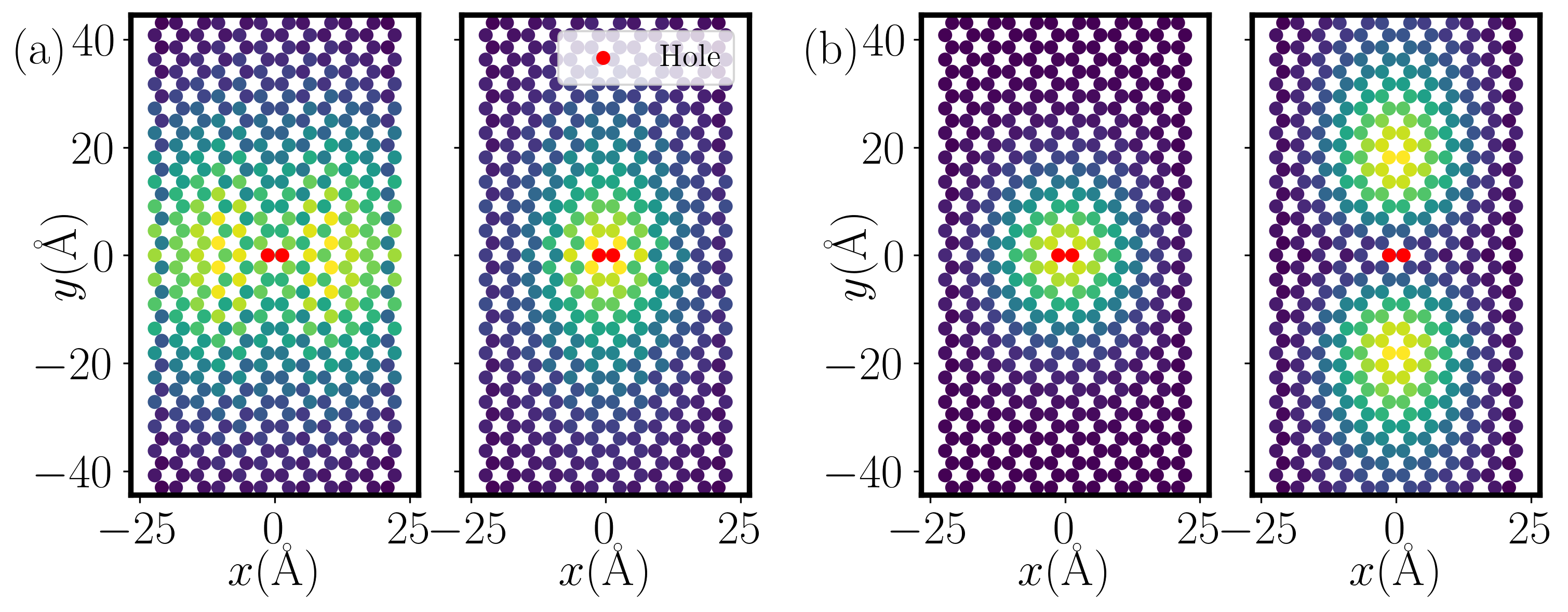}
    \caption[Real-space wavefunctions of excitons in a Bi(111) ribbon]{Real-space electronic probability density of the ground state (left) and first excited (right) excitons in a ribbon of width $N=12$.\ (a) Excitons wavefunctions with SOC and (b) without SOC.}\label{fig:rswf_exciton_ribbon}
\end{figure}

The real-space electronic probability density for the ground state and first excited exciton are shown in Fig.~\ref{fig:rswf_exciton_ribbon}. Interestingly, the ground state exciton in Fig.~\ref{fig:rswf_exciton_ribbon}(a) has a $p$-like character, instead of the usual $s$-like found in semiconductors. Instead, it is the first excited exciton the one that is $s$-like. If we turn off the spin-orbit coupling, then the wavefunction of the lowest exciton becomes $s$-like and the first excited is $p$-like but in the periodic direction, as typically expected. This could be attributed to the band inversion produced by the presence of SOC\@: without SOC, the bands are parabolic and the material corresponds to a trivial insulator, hence the observed behaviour. When including SOC, there is a band inversion at $\Gamma$ that results in the system becoming a topological insulator. 

\begin{figure}[H]
    \centering
    \includegraphics[width=0.9\columnwidth]{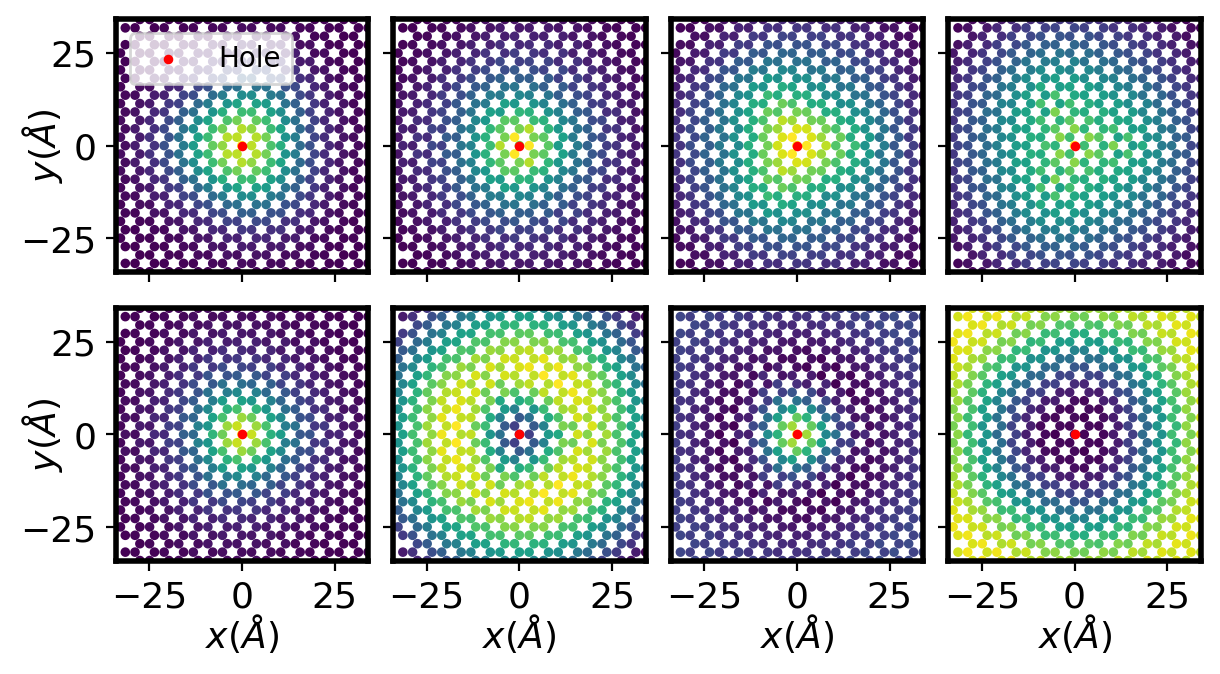}
    \caption[Real-space exciton wavefunctions in bulk Bi(111)]{Real-space electronic probability density of the first four excitonic levels in bulk Bi(111), from left to right. (Top) Excitons in Bi(111) computed without SOC and (Bottom) with SOC.}\label{fig:rswf_exciton_bulk}
\end{figure}

Since the band inversion is present both in the electronic band structure of both the ribbon and the bulk (i.e. 2D periodic), the real-space electronic probability densities should exhibit a similar behaviour to that of the ribbon. These are shown in Fig.~\ref{fig:rswf_exciton_bulk}, again with and without SOC\@. In the bulk case, however, we observe no deviation from the s-like behaviour when SOC is present for the ground state or first excited states. It is unclear therefore whether the $p$-like behaviour along the finite direction is induced by the confinement of the exciton to the ribbon, the band inversion, the intrinsic quantum geometry of the electronic bands, or a combination of these. Nevertheless, regardless of the origin of the shape found for the ground state exciton in the ribbon, its wider spread towards the edges of the ribbon could be beneficial for the dissociation process, as it could enhance the spatial overlap with the edge states, specially when compared with a purely $s$-like state.

\subsection{Transition rates for $Q=0$.}\label{sec:transitions_Q0}
\begin{figure}[t]
    \centering
    \includegraphics[width=0.95\columnwidth]{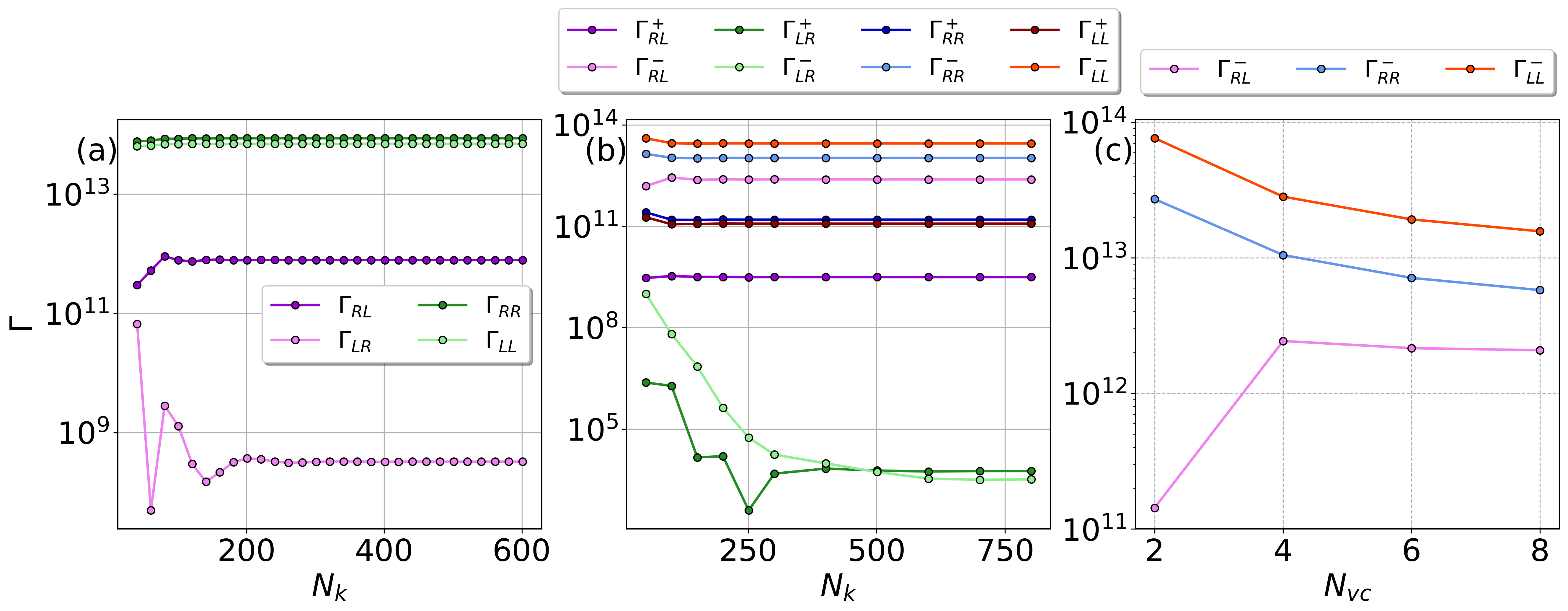}
    \caption[Convergence of the transition rates for $Q=0$ and $Q\neq 0$]{Convergence of the transition rates for the ground state exciton to the different edge electron-hole pairs.\ (a) Convergence of $Q=0$ transitions as a function of $N_k$, for width $N=12$, $N_v=N_c=2$ and $w=0.2$ eV. (b) Convergence of $Q\neq0$ transitions as a function of $N_k$, for $N=14$, $N_v=N_c=4$, $Q=0.1$ and $w=0.2$.\ (c) Convergence of the rates as a function of $N_{v/c}$ for $N=14$, $N_k=200$ and $w=0.2$ eV.}\label{fig:convergence_rates}
\end{figure}

Having established the properties of the exciton spectrum, we now proceed to compute the transition rates for the exciton decay into the edge electron-hole pairs, addressing first the $Q=0$ case.
Because of time-reversal symmetry the transition rates are symmetric in $+{k} \leftrightarrow -{k}$, i.e. $\Gamma^+_{ss'}=\Gamma_{ss'}^-$ for $Q=0$ excitons. The proof is as follows: We assume that the ground state exciton consists of a degenerate subspace of $N$ states $\{\ket{X_n}\}_n$. Then, since those excitons have $Q=0$, the time-reversal operator $\mathcal{T}$ will map those excitons onto themselves, 
\begin{equation}
    \mathcal{T}\ket{X_n}=U_{nm}\ket{X_m}
\end{equation}
where $U$ is the unitary matrix of the representation. Then, with the transition rate defined as the sum over all degenerate states, we may use that the interaction is time-reversal invariant, $[V,\mathcal{T}] = 0$ to prove the identity:
\begin{align}
    \nonumber \Gamma^+_{ss'} &= \sum_n\left|\braket{X_n|V|s,s',+k}\right|^2\rho(E_X) = \sum_n\left|\braket{X_n|\mathcal{T}^{\dagger}V\mathcal{T}|s,s',+k}\right|^2\rho(E_X) \\ 
    \nonumber & =\sum_{m,p}\sum_nU^*_{nm}U_{np}\braket{X_m|V|s,s',-k}\braket{s,s',-k|V|X_p}\rho(E_X) \\
    &= \sum_m \left|\braket{X_m|V|s,s',-k}\right|^2\rho(E_X) = \Gamma^-_{ss'}
\end{align}
where we have used that $\sum_nU^*_{nm}U_{np}=\delta_{mp}$. Consequently, we only need to compute the rates for one of the signs, for instance $\Gamma^+_{ss'}$. In what follows, we discuss the results obtained for both rates $\Gamma^{\pm}_{ss'}$. As we did with the exciton energies, first we need to ensure that the transition rates are converged with respect to $N_k$. Some examples of the convergence of the rates for $Q=0$ and $Q\neq 0$ are represented in Fig.~\ref{fig:convergence_rates}; in general, the transition rates require a higher number of $k$ points to converge than the energy spectrum. This is particularly true for the lowest magnitude rates, which are more sensitive to changes in the wavefunctions (i.e.\ the fluctuations are comparable with the converged value) and consequently require more points to converge. Regarding the convergence with respect to the number of bands, we see in Fig.~\ref{fig:convergence_rates}(c) that $N_{v/c}=4$ is the minimum number of bands required to ensure that the behaviour of the rates does not change qualitatively. For all calculations in this section and the next one, we set $N_v=N_c=4$, and use a minimum of $N_k=800$, while in some cases $N_k$ was even higher to ensure that the lowest magnitude transition rates were properly converged, up to $N_k = 1200$. 

\begin{figure}[H]
    \centering
    \includegraphics[width=0.995\columnwidth]{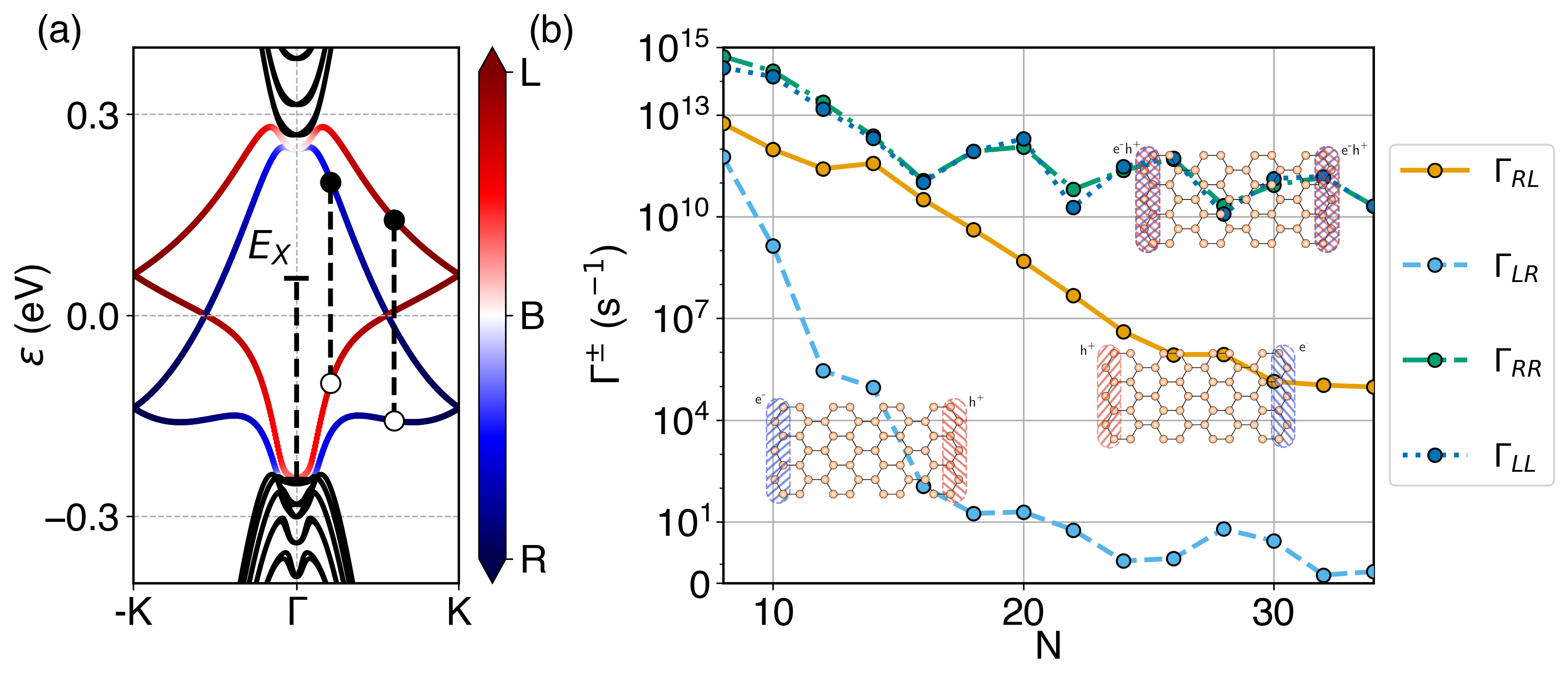}
    \caption[Edge band splitting in Bi(111) and transition rates for $Q=0$ as a function of the width of the ribbon]{{Transitions at $Q=0$.} (a) Band structure of the Bi(111) ribbon for $N=20$ and $w=0.2$ eV. The edge bands are colored according to the electronic occupation at the edges of the ribbon. We illustrate an arbitrary exciton state and the final inter-edge electron-hole pairs with the same energy.\ (b) Transition rates of the ground state exciton to the different edge electron-hole pairs as a function of the width of the ribbon $N$, for $w=0.2$ eV.}\label{fig:transitionsQ0_splitedges_width}
\end{figure}

\begin{figure}[t]
    \centering
    \includegraphics[width=0.85\columnwidth]{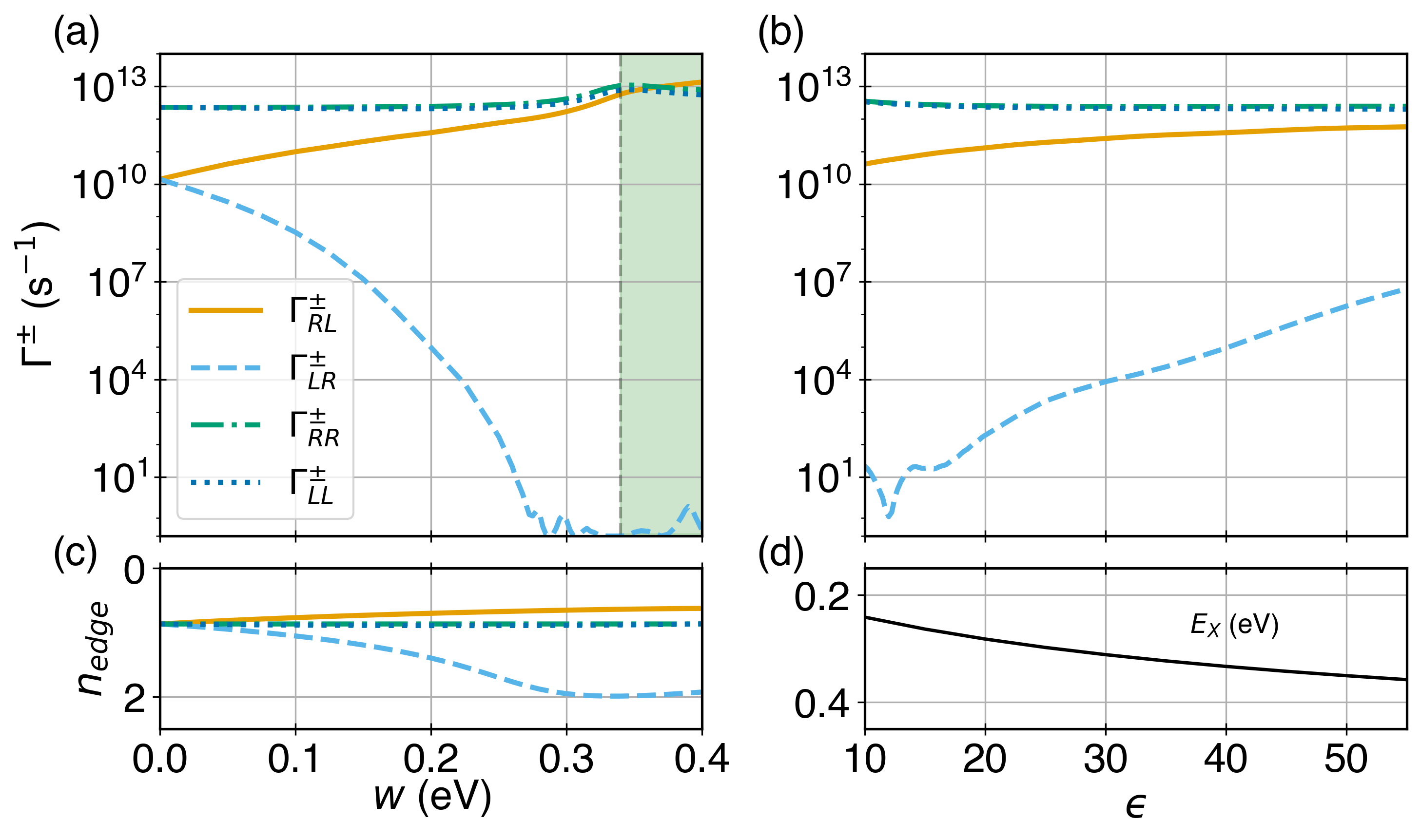}
    \caption[Transition rates for $Q=0$ as a function of the edge offset potential and the material dielectric constant]{(a, c) Transition rates and edge occupation as a function of the edge offset potential $w$ for $N=14$.\ (b, d) Transition rates and ground state exciton energy as a function of the dielectric constant $\epsilon$ for $N=14$.\ (a, b, c, d) share the same legend.}\label{fig:transitionsQ0_rates_dielectric_offset}
\end{figure}

The effect of the introduction of the offset potential can be seen in Fig.~\ref{fig:transitionsQ0_splitedges_width}(a). As discussed before, the edge bands are split which allows to identify them based on the electronic occupation at the edges. We set an offset value of $w=0.2$ eV for all calculations, except when $w$ is varied.
The rates in the presence of the onsite potential as a function of the width of the ribbon $N$ are shown in Fig.~\ref{fig:transitionsQ0_splitedges_width}(b). In general, as expected, the inter-edge rates decay faster as a function of $N$ than the intra-edge ones, with  $\Gamma^{\pm}_{RL}$ being several orders of magnitude higher than $\Gamma^{\pm}_{LR}$ for $N\sim 10 - 30$.  

Notably, for intermediate widths ($N\sim 12-16 $), $\Gamma_{RL}^\pm$ turns out to be comparable to intra-edge rates. This can be attributed, in part, to the peculiar real-space electronic probability density of the exciton, which exhibits a $p$-like character, as shown Fig.~\ref{fig:rswf_exciton_ribbon}(a). Moreover, it is possible to tune the rates to enhance the inter-edge/intra-edge ratio. In Fig.~\ref{fig:transitionsQ0_rates_dielectric_offset}(a) we show the effect of modifying the edge onsite potential $w$. For $w=0$ there is no charge imbalance since both inter-edge rates are equal. As we increase the potential, one rate $\Gamma_{RL}^\pm$ becomes enhanced as it comes closer to the intra-edge rates, while the other $\Gamma_{LR}^\pm$ decreases. The effect of the onsite potential is approximately splitting the edge bands by the same value $w$. Therefore, as we increase $w$, the corresponding edge pairs become increasingly more distant in $|{k}|$. From Fig.~\ref{fig:transitionsQ0_splitedges_width}(a) we see that those involved in $\Gamma_{LR}^\pm$ get pushed to the high-symmetry point $K$, where the wavefunctions are fully localized on the edge. On the other hand, for $\Gamma_{RL}^\pm$ the e-h pairs involved get closer to $\Gamma$ ($k=0$), where the functions have a stronger bulk component. Therefore, it is possible to improve the inter-edge/intra-edge ratio by tuning the localization of the e-h pairs on the edge, as seen in Fig.~\ref{fig:transitionsQ0_rates_dielectric_offset}(c). If the edge bands become too far apart in energies, then some electron-hole pairs will change the bands where they are hosted, which we indicate with the green region.

A similar discussion can be done with the dielectric constants of the system. We focus on the dielectric constant of the material $\epsilon$, although the same arguments apply to the substrate constant $\epsilon_s$, or the dielectric constant of the medium $\epsilon_m$. Tuning $\epsilon$ produces a change in the exciton energy, which will result in a transition to pairs with different $k$, as shown in Figs.~\ref{fig:transitionsQ0_rates_dielectric_offset}(b, d). In this case, the specific behaviour will be dependent on the form of the bands. Similarly to the onsite potential, changing the exciton energy drastically could result in pairs hosted in a different set of bands from before, although it is not the case for the range of values considered.

\clearpage

\subsection{Transition rates for $Q\neq 0$.} 

\begin{figure*}[h]
    \centering
    \includegraphics[width=1\textwidth]{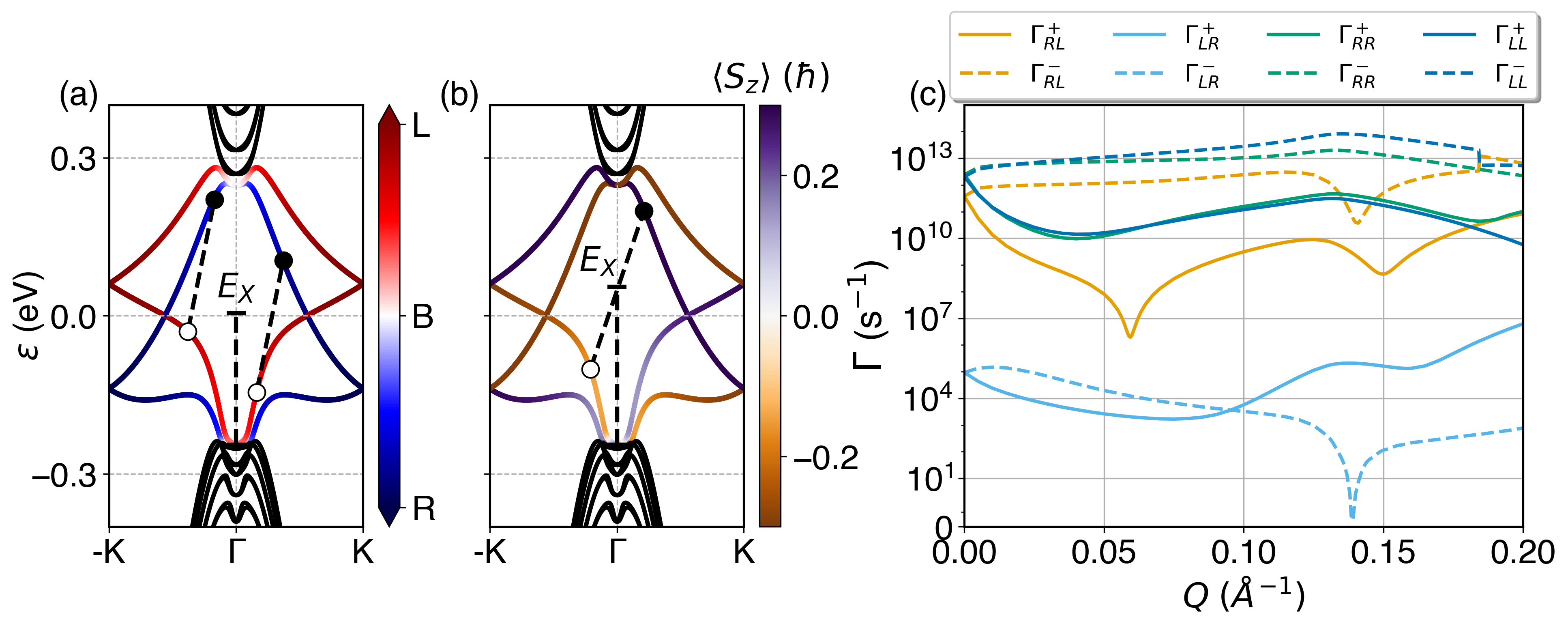}
    \caption[Split edge band structure showing the edge occupation and average spin projection]{Transitions at finite $Q$.\ (a, b) Band structure of the Bi(111) ribbon for $N=20$ and $w=0.2$ eV. The first one shows the edge occupation of the bands, while the second ones shows the average spin projection $\left<S_z\right>$ of the bands. We illustrate an exciton state with finite $Q$ and the corresponding final edge electron-hole pairs such that momentum and energy are conserved. Note that if $Q$ is too large, then the electron-hole pair might change bands, as in (b).\ (c) Transition rates of the ground state exciton as a function of $Q$ for $N=14$.}\label{fig:transitionsQ_bands_splitting}
\end{figure*}

Next we consider the transition rates for excitons with finite momentum $Q$. As for $Q=0$ excitons, the edge charge accumulation will still be present as long as we keep finite the edge offset term (we again set a fixed value of $w = 0.2$ eV). Now, the main difference with respect to the rates for $Q=0$ excitons comes from the  asymmetry in $k$. Since the initial exciton is not time-reversal invariant (as it has finite momentum $Q$), all the transition rates  $\Gamma^{\pm}_{ss'}$ $\forall s,s'\in\{R,L\}$ will be different (Figs.~\ref{fig:transitionsQ_bands_splitting}(a, b) show schematically the inter-edge processes). Both intra-edge and inter-edge pairs can carry a net current since they have a finite total velocity ${v}_{\rm{e-h}}(k)\neq 0$, but now there will be no exact cancellation between $k$ and $-k$ pairs. 

\begin{figure*}[t]
    \centering
    \includegraphics[width=1\textwidth]{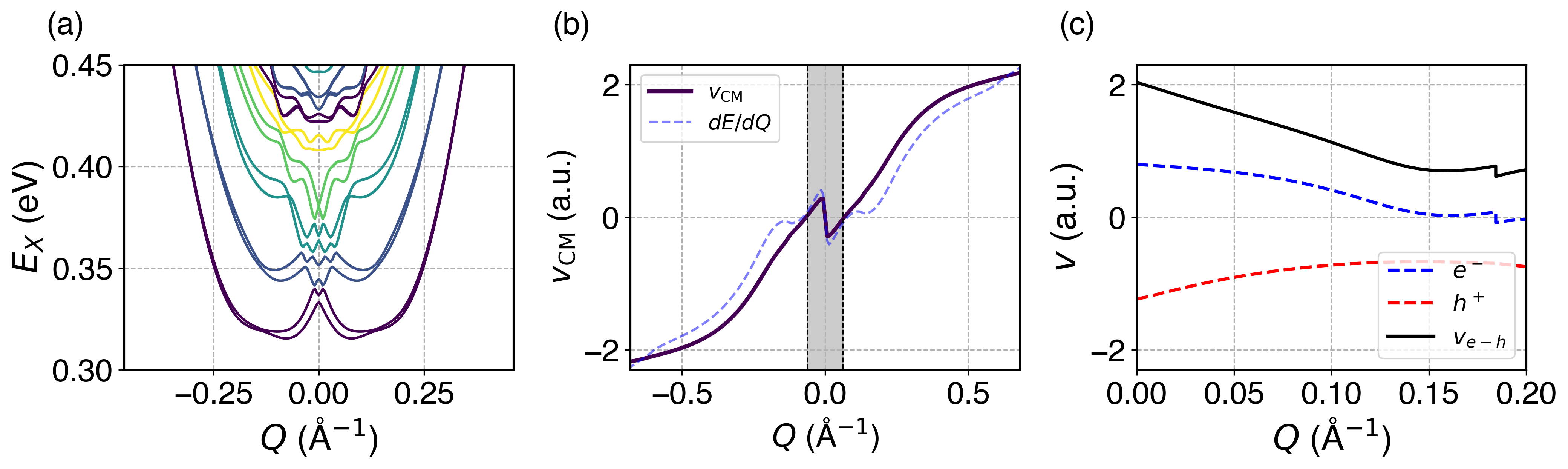}
    \caption[Exciton band structure, center-of-mass velocity of the lowest exciton and total velocity of the edge electron-hole pair as a function of Q]{(a) Low energy exciton band structure for $N_v=N_c=4$ and $N=14$. Each color corresponds to four excitonic states in total.\ (b) Center-of-mass velocity of the ground state exciton. The shadowed region denotes the fraction of excitons that do not contribute to the formation of an edge current. For comparison, we show the derivative of the lowest exciton band, which behaves similarly to $v_{\text{CM}}$, although it differs in general.\ (c) Velocity $v_{e-h}=v_e - v_h$ of the relevant electron-hole pair $\Gamma^-_{RL}$ and of each component individually, for $N=14$, $w=0.2$ eV.}\label{fig:transitionsQ_velocities}
\end{figure*}

The results, displayed in Fig.~\ref{fig:transitionsQ_bands_splitting}(c), show the expected behaviour: as $Q$ becomes non-zero, the $\pm k$ symmetry of the rates is lifted, namely $\Gamma^+_{ss'} \neq \Gamma^-_{ss'} $. We observe that, for the values of $Q$ considered, $\Gamma^{\pm}_{RL}$ and $\Gamma^{\pm}_{LR}$ rates differ by several orders of magnitude, meaning that the charge separation still takes place. We focus our attention again on these inter-edge rates since electron-hole pairs localized on the same edge are assumed not to contribute to the current as they are prone to recombination (in this case via phonon emission first). 
One inter-edge rate ($\Gamma^-_{RL}$) is close in magnitude to the intra-edge ones for all the values of $Q$ considered.
We also see that $\Gamma^-_{RL}$ differs by several orders of magnitude from $\Gamma^+_{RL}$, supporting our hypothesis that an overall edge current can develop in the material since we are inducing an electronic population imbalanced in $k$. For reference, we show in Fig.~\ref{fig:transitionsQ_velocities}(c) the total velocity of the electron-hole pair corresponding to $\Gamma^-_{RL}$, which is non-zero and positive for the values of $Q$ considered. 
Note that the plot only shows values of $Q$ up to 0.2. For higher values of $Q$, the energy of the exciton increases quadratically (see Fig.~\ref{fig:transitionsQ_velocities}(a)) and as a consequence there are no longer available edge e-h pairs. Also, as $Q$ increases, it might happen that either the electron or the hole change the band where they are hosted, as illustrated in Figs.~\ref{fig:transitionsQ_bands_splitting}(a, b). This produces the discontinuity in the rates and the velocities appearing at $Q\approx 0.18$.\\

The fraction of excitons entering the ribbon (see Fig.~\ref{fig:setup}) is determined by the sign of the velocity of these. We have thus computed the total velocity or center-of-mass velocity of the exciton ${v}_{\text{CM}}$ as a function of $Q$, as shown in Fig.~\ref{fig:transitionsQ_velocities}(b). Those with ${v}_{\text{CM}}>0 $ will enter the ribbon. For a small fraction with $Q>0$, highlighted with the gray region, the excitons  have negative velocity, i.e.\ they move away from the ribbon. For the fraction of excitons of the highlighted region with negative $Q$, they enter the channel, but contribute with opposite currents (due to time-reversal) to the ones with $Q,v_{\text{CM}}>0$. However, from the exciton bands in Fig.~\ref{fig:transitionsQ_velocities}(a), we conclude that it is more likely to have a population of excitons satisfying the latter condition, as it corresponds to a lower energy overall. 
It should be noted that for a conventional semiconductor, the exciton bands would be parabolic meaning that all velocities for positive momentum would also be positive. Thus, this fraction of excitons that hinders the performance of the device is also intrinsic to the topological insulator, but is expected to be small.

\begin{figure*}[t]
    \centering
    \includegraphics[width=0.85\textwidth]{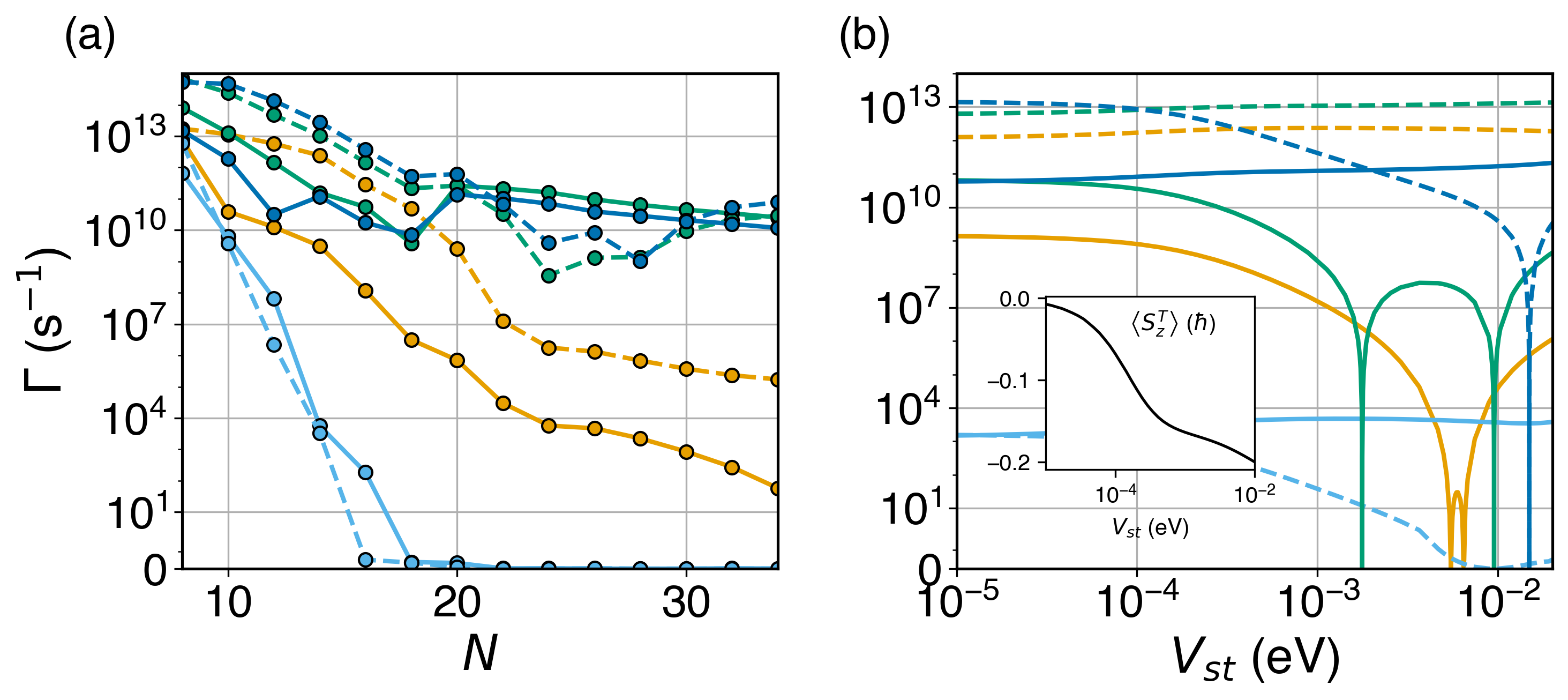}
    \caption[Transition rates for $Q\neq0$ as a function of the width of the ribbon and as a function of the sub-lattice staggered potential.]{(a) Transition rates as a function of $N$, for $Q=0.1$ \AA$^{-1}$ and $w=0.2$ eV. (b) Transition rates as a function of the staggered potential $V_{st}$ for $N=14$ and $w=0.2$ eV. The inset shows the total spin projection of the ground state exciton, $\left<S_z\right>_X$ as a function of the staggered potential.}\label{fig:transitionsQ_width_stagger}
\end{figure*}

As we did for $Q=0$, in Fig.~\ref{fig:transitionsQ_width_stagger}(a) we show the behaviour of the transition rates as we increase the width of the ribbon for $Q=0.1$. As expected, the inter-edge rates decay faster than the intra-edge rates. 
Importantly, up to $N=20$, the relevant inter-edge rate $\Gamma^-_{RL}$ is comparable to the intra-edge ones. The opposing rates $\Gamma^{\pm}_{LR}$ become completely suppressed from $N=18$, enhancing the charge separation. As for the ratio between $\Gamma^{+}_{RL}$ and $\Gamma^{-}_{RL}$, it appears to be relatively constant for the widths under consideration.

Finally, we show that it is also possible to engineer the rates by further tuning the spin of the exciton. For $Q=0$ we obtain that $\left<S_z\right>_{\rm{X}}= 0$, due to the bulk bands being degenerate. However, if the exciton had a finite value of the spin, we would expect different values for the rates $\Gamma^{\pm}_{ss'}$ given that the edge bands also present opposite spin when $k \leftrightarrow -k$, as shown in Fig.~\ref{fig:transitionsQ_bands_splitting}(b). We can induce this finite spin introducing a sub-lattice staggered potential that breaks inversion symmetry in the bulk of the material (as in hBN for instance), thereby fully splitting the bulk bands. We show in Fig.~\ref{fig:transitionsQ_width_stagger}(b) how for $Q=0.1$ this potential induces a spin in the ground state exciton (inset), and results in the relevant rates $\Gamma^+_{RL}$, $\Gamma^-_{RL}$ deviating even further from each other. Remarkably, some of the intra-edge rates, which may hinder the performance of the device, are strongly suppressed in a wide range of the staggered potential, becoming even zero at particular values.\\

Since current generation is only possible with $Q\neq 0$ excitons, we note that radiative recombination will not be present due to the finite exciton momentum. Furthermore, for the $Q=0$ excitons considered, we observe a vanishingly small oscillator strength as corresponding to dark excitons. In fact, the main limiting factor will be the exciton-phonon scattering. For reference, other bidimensional materials show exciton lifetimes due to phonon scattering between $1-1000$ fs~\cite{chen2020, chan2023}. Assuming similar lifetimes for Bi(111), the relevant rate $\Gamma^-_{RL}$ is approximately of the same magnitude for all values of $Q$ considered. We note that the effect is heavily dependent on the ribbon width, meaning that for wider systems the exciton-phonon scattering will eventually dominate.

\subsection{Effect of armchair ribbon termination}\label{sec:armchair_bi111}
\begin{figure}[h]
    \centering
    \includegraphics[width=0.85\columnwidth]{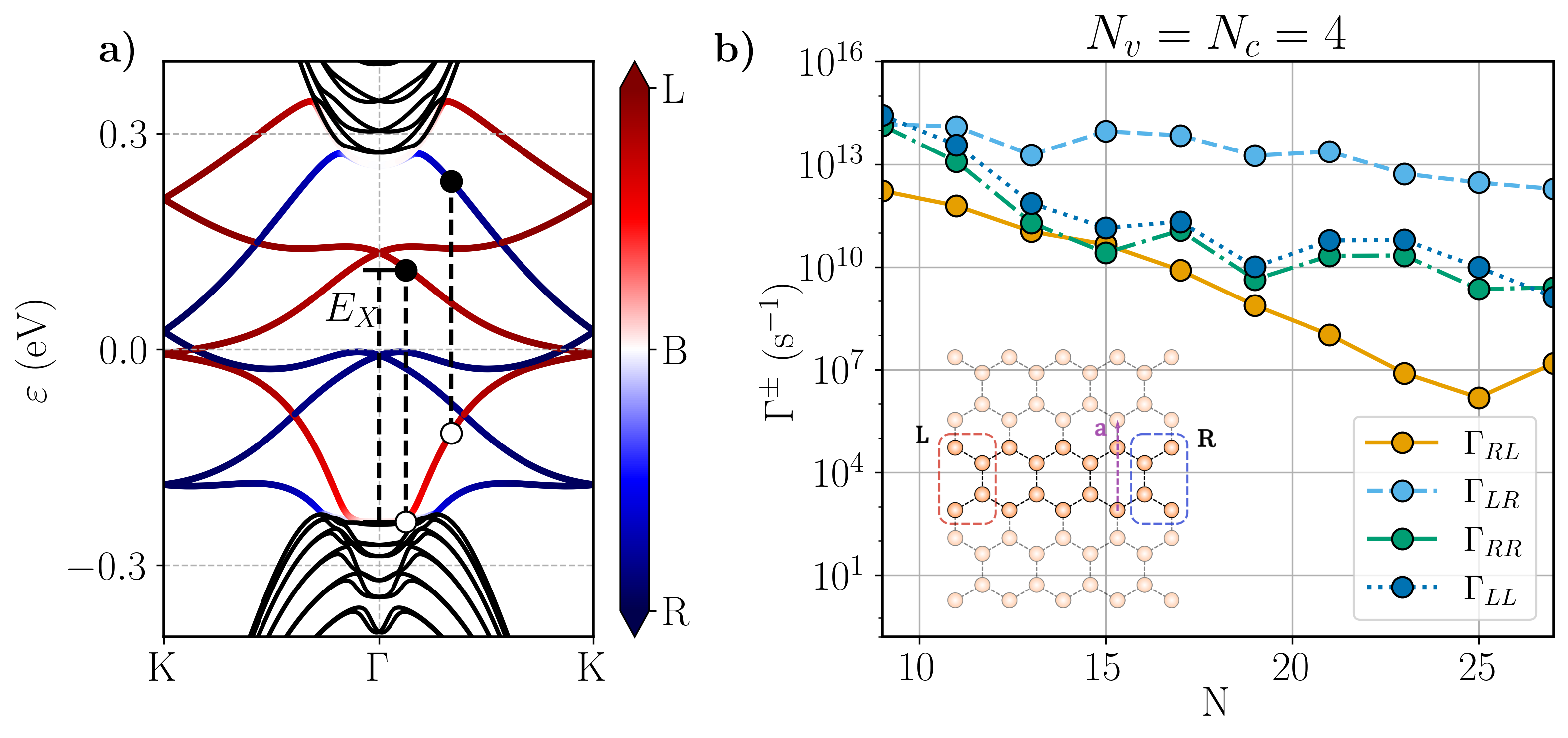}
    \caption[Edge band structure for an armchair ribbon and transition rates for $Q=0$ as a function of the width of the ribbon.]{Rates on an armchair nanoribbon.\ (a) Band structure of an armchair nanoribbon of width $N=20$ and edge offset $w=0.2$ eV. $N$ is an integer such that the unit cell has $4N + 2$ atoms.\ (b) Transition rates from the bulk excitons to the available edge electron-hole pairs as a function of $N$, for $N_{vc}=4$, $Q=0$ and $N_k = 901$.}\label{fig:armchair}
\end{figure}

For completeness, we also study the effect of the edge termination of the ribbon on the transition rates. The main effect stems from the different band structure, which for the armchair termination results in more complex edge bands. As shown in Fig.~\ref{fig:armchair}(a), for the armchair ribbon there are 4 edge bands, which in presence of the edge offset potential results in 8 different bands. Therefore, for one given rate $\Gamma^{\pm}_{ss'}$ with fixed sides $s, s'$ there can be multiple electron-hole pairs available. For simplicity, we only compute the transition rates to the electron-hole pair that gives the dominant contribution to each rate, namely the pairs closer to the $\Gamma$ point. These rates are shown in Fig.~\ref{fig:armchair}(b) as a function of the width of the armchair nanoribbon for $Q=0$ and $N_{v/c}=4$. We observe the same behaviour as in the zigzag termination, except for the fact that the dominating rate now is $\Gamma_{LR}$ instead of $\Gamma_{RL}$. This is attributed precisely to the more complex band structure, since for the exciton energies present there are two available pairs for the $\Gamma_{LR}$ rate: one closer to $K$ and another one closer to $\Gamma$ in the BZ (the latter shown in Fig.~\ref{fig:armchair}(a)), while in the zigzag ribbon there was only one close to $K$. Thus, even though the dissociation direction is reversed, the charge separation mechanism still takes place. Likewise, if we were to consider the rates as a function of $Q$, we would expect the same behaviour as in the zigzag, and more importantly, the same direction for the currents given the slope of the bands for the relevant edge electron-hole pairs.

\subsection{Estimation of the photocurrent}
We can provide a quick estimate of the current with $j = \eta n(Q) ev$, where $n(Q)$ is the linear density of excitons available for dissociation, $\eta$ is the efficiency of the effect, $e$ the charge and $v$ the velocity of the final electron once the exciton has dissociated. First, we assume that the linear density of excitons available for dissociation is steady and that the laser is intense enough so that there is always at least one exciton available in the bulk for dissociation. Then, the efficiency $\eta$ is defined as $\eta = \Gamma^-_{RL}/\sum_{ss'} \Gamma^{\pm}_{ss'}$, where $\Gamma^-_{RL}$ is the dominating rate relative to the edge current generation. Thus, $\eta n(Q)$ is the fraction of edge electron-hole pairs relevant to the effect. Since only one exciton can dissociate at a time (there can be multiple excitons simultaneously since they are approximately described as bosons, but the edge bands become saturated if there is already one electron-hole pair due to the fermionic statistics), we set $n(Q)=1$. Doing the calculation for $N=14$, $Q=0.1$ as in Fig.~\ref{fig:transitionsQ_width_stagger}(a), we estimate a current of $j\approx 0.5\ \mu A$. Summing over $Q$ would yield a higher current, but again in the order of $\mu A$. 

\section{Conclusions}

We have noted that the edge states of a 2D TI constitute an alternative dissociation path to exciton recombination. To this end, we have fully characterized the exciton spectrum in Bi(111) nanoribbons, and shown that, if one introduces an onsite edge potential to split the edge states, then one can possibly obtain an edge charge imbalance from the dissociation of excitons into non-interacting edge electron-hole pairs. Additionally, we have shown that, if we are able to generate a population of excitons in the ribbon that is not time-reversal invariant, then an edge current (topologically protected) may develop. Moreover, the corresponding transition rates can be tuned to increase or decrease the strength of the effect. Our estimates indicate that currents in the $\mu$A range can be obtained. The present arguments are not dependent on the specific shape of the bands, and we expect that they can be applied and tested both theoretically and experimentally with other 2D TIs such as Bi$_4$Br$_4$, which is a room-temperature TI~\cite{Shumiya2022}. The foundation of the effect is not exclusive of the dimensionality and can be trivially extended to three-dimensional TIs. Remarkably, in a recent work it has been reported the relaxation of photoexcited bulk electrons onto topological surface states in Bi$_2$Se$_3$~\cite{fukumoto2024}, indirectly supporting our claim.

\part{Disorder in topological insulators}
\chapter{Introduction}\label{chapter:intro_topology}

One of the primary interests in condensed matter physics is the study of the electronic transport properties of materials. Through quantum mechanics, the application of Bloch's theorem to crystalline solids led to the development of electronic band theory, which describes the electronic structure in terms of energy bands~\cite{kittel2021introduction, ashcroft}. Band theory is a cornerstone of condensed matter theory, providing a foundation for understanding the electrical, optical and magnetic properties of materials---at least as a first approximation, as interactions might induce different behaviours~\cite{fazekas1999lecture}. Importantly, band theory enables the classification of materials as insulators, semiconductors or metals, based on their band structure and the position of the Fermi level. 

In parallel, condensed matter physics also examines phase transitions, which are often characterized by a symmetry breaking and quantified using an order parameter. The Ginzburg-Landau theory provides a phenomenological framework that effectively captures the universal features of phase transitions~\cite{landau2013statistical}. It defines an order paramete $\psi$ for the system (e.g.\ the average magnetization $m=\braket{S_z}$ for ferromagnets), and uses the symmetries of the Hamiltonian to construct a free energy functional $F[T,\psi]$. This approach successfully captures the symmetry breaking of phase transitions and has been used to describe phenomena such as superconducting, ferromagnetic or liquid-gas transitions, among others~\cite{hohenberg2015introduction}.

The emergence of topological phases of matter marked a significant paradigm shift in condensed matter physics. Unlike conventional phases, these are characterized by a topological invariant instead of an order parameter. Band theory can be used to identify topological materials, but it is only upon inspection of the ground state wavefunction that their properties become apparent, and not from the energy bands as for conventional insulators and metals. Traditional concepts of phase transitions do not apply, as topological phase transitions are not characterized by symmetry breaking but by changes in the system's topological invariant. This invariant is a global property of the system; the term topological signifies that it is robust under perturbations of the system. While in second-order phase transitions the order parameter changes continuously, transitions between topological phases are abrupt and are associated to a closing of the energy gap of the system, which would correspond to a non-smooth deformation of the Hamiltonian.

The study of topological phases began with the discovery of the integer quantum Hall effect (IQHE)~\cite{klitzing1980new}. Under a strong magnetic field, the electrons of a 2D electron gas arrange into Landau levels---energy levels of flat dispersion, see Fig.~\ref{fig:qhe_chern}(a). The Hall conductivity $\sigma_{xy}$, which measures current flow perpendicular to the applied voltage, shows a quantized value in terms of the number of filled Landau levels, $n$, and the conductance quantum $e^2/h$~\cite{tong2016lectures}, 
\begin{equation}
    \sigma_{xy} = \frac{e^2}{h} n,
\end{equation}
It was later demonstrated that each Landau level carries a finite value of the topological invariant of the system, which is the Chern number~\cite{thouless1982quantized, niu1985quantized}, leading to the quantized conductivity~\cite{thouless1983quantization}. Thus, the IQHE represents the first topological phase of matter discovered. This effect was soon extended to systems without an external magnetic field, famously by Haldane who introduced a model for graphene with a staggered mass term and $\mathcal{T}$-breaking hopping terms, akin to inserted magnetic fluxes~\cite{haldane1988model}. This work led to the so-called quantum anomalous Hall effect or Chern insulators. Similar to the IQHE, Chern insulators are characterized by broken time-reversal symmetry, $[H,\mathcal{T}]\neq 0$. Under an applied electrical field, these systems develop a finite Hall conductivity, which is given in terms of the Chern number $C$ of each filled band of the system~\cite{xiao2010berry, chang2023colloquium}.

\newcommand\ybb{-1}\begin{figure}[h]
    \centering
    \begin{tikzpicture}

        \draw[->, thick] (-6,-1.) -- (-6,2.) node[left] {$E$};

        \draw[->, thick] (-6,-1) -- (-3.,-1) node[below] {$k$};
 
        \draw[thick] (-6.1, 0.5) -- (-5.9, 0.5) node[left, xshift=-0.1cm] {$E_F$};

        \draw[thick, blue] (-6, -0.4) -- (-3.1, -0.4);
        \draw[thick, blue] (-6, 0.2) -- (-3.1, 0.2);
        \draw[thick, red]  (-6, 0.8) -- (-3.1, 0.8);
        \draw[thick, red]  (-6, 1.4) -- (-3.1, 1.4);

        \draw[<->] (-3.5, -0.4) -- (-3.5, 0.2) node[midway, right] {$\hbar\omega_c$};

        \node[align=center] at (-6.7, 2.1) {(a)};

        \draw[->, thick] (-0.5, -1) -- (-1.5, -1) node[below] {$\Delta(x)$};
        \draw[->, thick] (-0.5, -1) -- (-0.5, 2) node[right] {$x$};
        \draw[thick] (-0.6, 0.5) -- (-0.4, 0.5);

        \draw[thick, red!50!blue!50] (-1.1, -1) -- (-1.1, 0.2);
        \draw[thick, red!50!blue!50] (-1.1, 0.2) .. controls (-1.05, 0.4) and (-0.6, 0.45) .. (-0.5, 0.5);

        \draw[thick, red!50!blue!50] (-0.5, 0.5) .. controls (-0.6, 0.55) and (-0.8, 0.6) .. (-0.8, 0.7);
        \draw[thick, red!50!blue!50] (-0.8, 0.7) -- (-0.8, 1.8);

        \node[align=center] at (-1.6, 2.1) {(b)};

        \shade[bottom color=yellow!20, top color=yellow!10] (0,\ybb) rectangle (3,\ybb+1.5);
        \draw[thick, dashed, opacity=0.5] (0, -1) -- (3, -1);
        \draw[thick, dashed, opacity=0.5] (0, 2) -- (3, 2);
        \draw[thick, dashed, opacity=0.5] (0, -1) -- (0, 2);
        \draw[thick, dashed, opacity=0.5] (3, -1) -- (3, 2);
        \draw[thick] (0, 0.5) -- (3, 0.5);

        \foreach \x in {0.5, 1, 1.5, 2, 2.5, 3} {
            \draw[thick, black] (\x, 0.5) arc (0:-180:0.25);
        }
        \foreach \x in {1, 2, 3} {
            \draw[thick, black, ->] (\x-0.15, 0.25) -- (\x-0.149, 0.25);
        }


        \node[align=center, scale=0.8] at (0.8, -0.5) {CI/QHE};
        \node[align=center, scale=0.8] at (2.5, -0.5) {$C\neq 0$};

        \node[align=center, scale=0.8] at (0.8, 1.5) {Insulator};
        \node[align=center, scale=0.8] at (2.5, 1.5) {$C=0$};

        \fill[bottom color=red!30, top color=red!30] (4.5, 2) -- plot[domain=4.5:7.5, samples=100] (\x, {0.8 + 0.5*(1 + cos(2*180*(\x-4.5)/3))}) -- (7.5, 2) -- cycle;
        \draw[thick, domain=4.5:7.5, samples=100] plot (\x, {0.8 + 0.5*(1 + cos(2*180*(\x-4.5)/3))});

        \fill[top color=blue!30, bottom color=blue!30] (4.5, -1) -- plot[domain=4.5:7.5, samples=100] (\x, {0.2 - 0.5*(1 + cos(2*180*(\x-4.5)/3))}) -- (7.5, -1) -- cycle;
        \draw[thick, domain=4.5:7.5, samples=100] plot (\x, {0.2 - 0.5*(1 + cos(2*180*(\x-4.5)/3))});

        \draw[->, thick] (4.5,-1.) -- (4.5,2.) node[left] {$E$};

        \draw[->, thick] (4.5,-1) -- (7.5,-1) node[below] {$k$};

        \draw[thick] (4.4, 0.5) -- (4.6, 0.5) node[left, xshift=-0.1cm] {$E_F$};
        \draw[thick, dashed] (4.5, 0.5) -- (7.5, 0.5);
        \fill[black] (6, 0.5) circle (2pt);

        \draw[thick] (5.5, -0.05) -- (6.5, 1.05);
        \node[align=center, scale=0.8] at (6, -0.5) {VB};
        \node[align=center, scale=0.8] at (6, 1.5) {CB};

        \node[align=center] at (3.8, 2.1) {(c)};
    \end{tikzpicture}
    \caption[Energy bands of Landau levels and Chern insulator. Schematic of the edge state in the interface between a Chern insulator and a conventional insulator]{(a) Energy levels of a 2D electron gas in a strong magnetic field. Each Landau level carries a finite Chern number $C=1$. The separation between consecutive levels is given in terms of the cyclotron frequency $\omega_c$. (b) Schematic of the edge state appearing in the interface between a Chern insulator or quantum Hall state and vacuum/trivial insulator. The edge state is represented as skipping cyclotron orbits that would clasically appear under an external magnetic field. The local energy gap $\Delta(x)$ has to close at the interface due to the change in the topological invariant. (d) Energy bands of a semi-infinite Chern insulator, showing the dispersion relation for the chiral edge state, connecting the valence and conduction bands. Figures adapted from \cite{hasan2010}.}
    \label{fig:qhe_chern}
\end{figure}
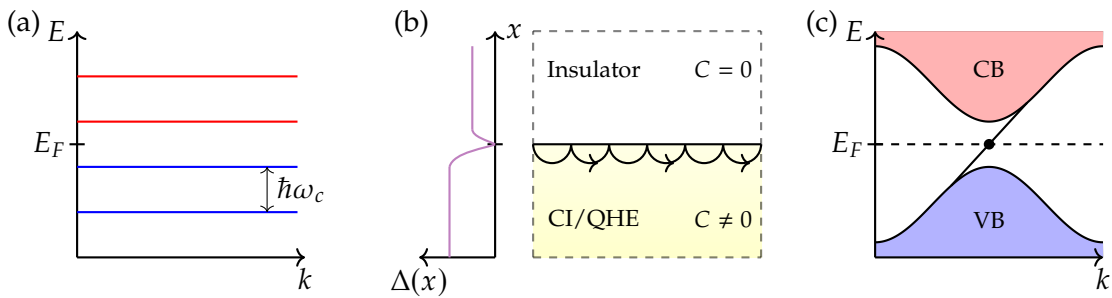

In the absence of external magnetic fields, Chern insulators can be realized in nature if the system intrinsically breaks time-reversal symmetry, as it is the case of magnetic topological insulators~\cite{chang2013experimental, chang2015high}. Alternatively, they can be engineered using quantum simulators~\cite{jotzu2014experimental, aidelsburger2013realization}. However, most materials in nature are non-magnetic, raising whether topological phases can exist in materials that preserve time-reversal symmetry. The answer was given by Kane and Mele, who extended Haldane's model to spinful graphene with an effective time-reversal invariant spin-orbit term~\cite{kane2005quantum}. The resulting phase, termed quantum spin Hall insulator, is not characterized by the total Chern number of the system, which is zero $C_T=C_{\uparrow} + C_{\downarrow} = 0$, but by the spin Chern number, $C_s=(C_{\uparrow}-C_{\downarrow})/2$, under the condition $[H, S_z]=0$. For more general spinful, time-reversal invariant systems where $[H, S_z]\neq 0$, the topological nature is governed by the $\mathbb{Z}_2$ invariant~\cite{kane2005z, fu2006time}. These systems are the so-called time-reversal topological insulators, or simply topological insulators.

\begin{figure}
    \centering
    \begin{tikzpicture}

        \shade[bottom color=yellow!20, top color=yellow!10] (2.5,\ybb) rectangle (5.5,\ybb+1.5);
        \draw[thick, dashed, opacity=0.5] (2.5, -1) -- (5.5, -1);
        \draw[thick, dashed, opacity=0.5] (2.5, 2) -- (5.5, 2);
        \draw[thick, dashed, opacity=0.5] (2.5, -1) -- (2.5, 2);
        \draw[thick, dashed, opacity=0.5] (5.5, -1) -- (5.5, 2);
        \draw[thick] (2.5, 0.5) -- (5.5, 0.5);

        \draw[thick, blue, ->] (2.6, 0.35) -- (4.5, 0.35);
        \draw[thick, blue] (4.5, 0.35) -- (5.4, 0.35);
        \draw[thick, green!40!black!60] (2.6, 0.2) -- (3.5, 0.2);
        \draw[thick, green!40!black!60, <-] (3.5, 0.2) -- (5.4, 0.2);

        \node[align=center] at (2.1, 2.1) {(a)};

        \node[align=center, scale=0.8] at (3.3, -0.5) {Top. Ins.};
        \node[align=center, scale=0.8] at (5., -0.5) {$\nu=1$};

        \node[align=center, scale=0.8] at (3.3, 1.5) {Trivial Ins.};
        \node[align=center, scale=0.8] at (5., 1.5) {$\nu=0$};

        \draw[thick, blue] (8, -0.05) -- (9, 1.05);
        \draw[thick, green!40!black!60] (8, 1.05) -- (9, -0.05);

        \fill[bottom color=red!30, top color=red!30] (7, 2) -- plot[domain=7:10, samples=100] (\x, {0.8 + 0.5*(1 + cos(2*180*(\x-7)/3))}) -- (10, 2) -- cycle;
        \draw[thick, domain=7:10, samples=100] plot (\x, {0.8 + 0.5*(1 + cos(2*180*(\x-7)/3))});

        \fill[top color=blue!30, bottom color=blue!30] (7, -1) -- plot[domain=7:10, samples=100] (\x, {0.2 - 0.5*(1 + cos(2*180*(\x-7)/3))}) -- (10, -1) -- cycle;
        \draw[thick, domain=7:10, samples=100] plot (\x, {0.2 - 0.5*(1 + cos(2*180*(\x-7)/3))});

        \draw[->, thick] (7,-1.) -- (7,2.) node[left] {$E$};

        \draw[->, thick] (7,-1) -- (10,-1) node[below] {$k$};

        \draw[thick] (6.9, 0.5) -- (7.1, 0.5) node[left, xshift=-0.1cm] {$E_F$};
        \draw[thick, dashed] (7, 0.5) -- (10, 0.5);
        \fill[black] (8.5, 0.5) circle (2pt);

        \node[align=center, scale=0.8] at (8.5, -0.5) {VB};
        \node[align=center, scale=0.8] at (8.5, 1.5) {CB};

        \draw[thick, blue, ->] (9.2, 0.6) -- (9.2, 1);
        \draw[thick, green!40!black!60, <-] (7.8, 0.6) -- (7.8, 1);

        \node[align=center] at (6.3, 2.1) {(b)};

        \draw[->, thick] (2.5,-5.5) -- (2.5, -2.5) node[left] {$E$};

        \draw[->, thick] (2.5,-5.5) -- (5.5,-5.5) node[below] {$k$};

        \draw[thick] (2.4, -4) -- (2.6, -4) node[left, xshift=-0.1cm] {$E_F$};
        \draw[thick, dashed] (2.5, -4) -- (5.5, -4);
        \fill[black] (2.8, -4) circle (2pt);
        \fill[black] (4, -4) circle (2pt);

        \fill[bottom color=red!30, top color=red!30] (2.5, -2.5) -- plot[domain=2.5:5.5, samples=100] (\x, {-3.3 + 0.3*(1 + cos(2*180*(\x-7)/6 + 90))}) -- (5.5, -2.5) -- cycle;
        \draw[thick, domain=2.5:5.5,, samples=100] plot (\x, {-3.3 + 0.3*(1 + cos(2*180*(\x-7)/6 + 90))});

        \fill[top color=blue!30, bottom color=blue!30] (2.5, -5.5) -- plot[domain=2.5:5.5, samples=100] (\x, {-4.7 - 0.25*(1 + cos(2*180*(\x-7)/6 + 90))}) -- (5.5, -5.5) -- cycle;
        \draw[thick, domain=2.5:5.5,, samples=100] plot (\x, {-4.7 - 0.25*(1 + cos(2*180*(\x-7)/6 + 90))});

        \node[align=center, scale=0.8] at (2.8, -3) {CB};
        \node[align=center, scale=0.8] at (2.8, -5) {VB};

        \draw[thick] (2.5, -3.8) -- (5.5, -4.2);
        \draw[thick] (2.5, -3.8) .. controls (3.5, -4.5) and (4.5, -4.5) .. (5.5, -4.2);

        \draw[thick] (4, -3) -- (5.5, -3);
        \draw[thick] (3.5, -3.15) .. controls (4.5, -3.5) and (5, -3.2) .. (5.5, -3);

        \draw[->, thick] (2.5,-5.5) -- (2.5, -2.5) node[left] {$E$};

        \draw[->, thick] (2.5,-5.5) -- (5.5,-5.5) node[below] {$k$};

        \node[align=center] at (2.1, -2.1) {(c)};

        \draw[->, thick] (7,-5.5) -- (7, -2.5) node[left] {$E$};

        \draw[->, thick] (7,-5.5) -- (10,-5.5) node[below] {$k$};

        \draw[thick] (6.9, -4) -- (7.1, -4) node[left, xshift=-0.1cm] {$E_F$};
        \draw[thick, dashed] (7, -4) -- (10, -4);
        \fill[black] (7.8, -4) circle (2pt);

        \fill[bottom color=red!30, top color=red!30] (7, -2.5) -- plot[domain=7:10, samples=100] (\x, {-3.3 + 0.3*(1 + cos(2*180*(\x-11.5)/6 + 90))}) -- (10, -2.5) -- cycle;
        \draw[thick, domain=7:10, samples=100] plot (\x, {-3.3 + 0.3*(1 + cos(2*180*(\x-11.5)/6 + 90))});

        \fill[top color=blue!30, bottom color=blue!30] (7, -5.5) -- plot[domain=7:10, samples=100] (\x, {-4.7 - 0.25*(1 + cos(2*180*(\x-11.5)/6 + 90))}) -- (10, -5.5) -- cycle;
        \draw[thick, domain=7:10, samples=100] plot (\x, {-4.7 - 0.25*(1 + cos(2*180*(\x-11.5)/6 + 90))});

        \node[align=center, scale=0.8] at (7.3, -3) {CB};
        \node[align=center, scale=0.8] at (7.3, -5) {VB};

        \draw[thick] (8.7, -5) -- (10, -4.7);
        \draw[thick] (7, -4.3) -- (10, -4.7);
        \draw[thick] (7, -4.3) -- (10, -3.2);
        \draw[thick] (8.45, -3.) -- (10, -3.2);

        \draw[->, thick] (7,-5.5) -- (7, -2.5) node[left] {$E$};

        \draw[->, thick] (7,-5.5) -- (10,-5.5) node[below] {$k$};

        \node[align=center] at (6.6, -2.1) {(d)};

    \end{tikzpicture}

    \caption[Band structure and schematic of the edge states in the interface between a topological insulator and a conventional insulator. Distinction between topological and trivial edge states]{(a) Schematic of the helical edge states appearing at the interface between a $\mathbb{Z}_2$ topological insulator and a conventional insulator. (b) Edge band structure of a topological insulator. In (c, d) we compare the edge band structure of a trivial and a topological insulator respectively in half BZ. A conventional insulator may present trivial edge states (e.g. from defects), which may be gapped by smooth deformations, while the topological edge states cannot be gapped. Figures adapted from \cite{hasan2010, fu2007topological}.}
    \label{fig:qshe}
\end{figure}
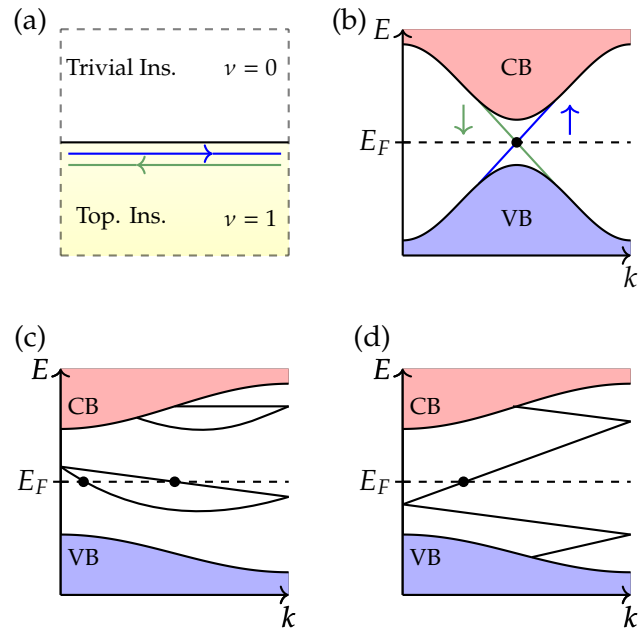

Both the Chern number and the $\mathbb{Z}_2$ invariant can be understood from the bulk wavefunction of the system. However, the topological character of the system can also be intuitively explained from its edge states, through the bulk-boundary correspondence~\cite{hasan2010, qi2011topological}. The topological invariant is a property of the ground state of gapped systems and remains unchanged as long as the bulk gap does not close. If we now consider a finite topological system, the gap must locally close at its boundary to transition to the trivial insulating state, leading to the emergence of topological edge states, as represented in Figs~\ref{fig:qhe_chern}(b, c) and~\ref{fig:qshe}(a, b). The bulk-boundary correspondence then relates the invariant to the number of edge states at the boundary: for Chern insulators, one has $\Delta C=N_+-N_-$, where $N_+ (N_-)$ is the number of chiral edge states with positive (negative) velocity, and $\Delta C$ is the change in the Chern number across the interface. For topological insulators, the relation is $\nu=N_K \text{ mod } 2$, $N_K$ being the number of pairs of helical edge states and $\nu$ the value of the $\mathbb{Z}_2$ index, as illustrated in Fig.~\ref{fig:qshe}(c, d).

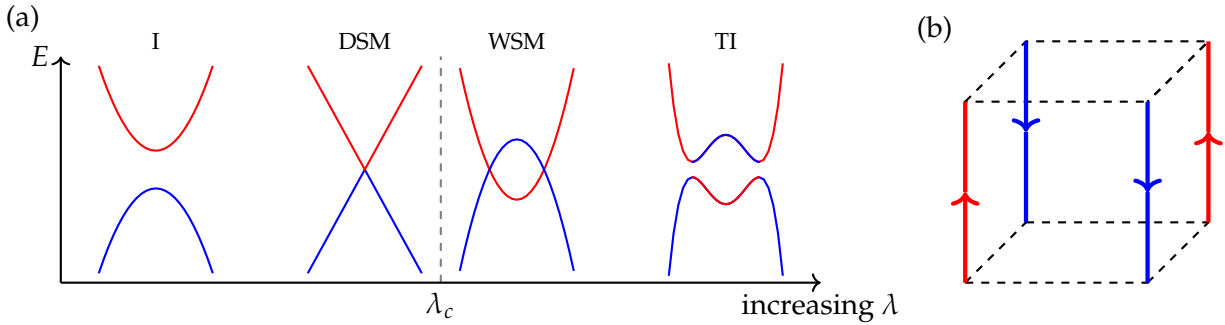
\begin{figure}[h]
    \centering
    \begin{tikzpicture}
        \draw[->, thick] (0,0) -- (0,3) node[left] {$E$};

        \draw[->, thick] (0,0) -- (10,0) node[below] {increasing $\lambda$};

        \draw[thick, red, domain=0.5:2.] plot (\x, {1.75 + 2*(\x-1.25)^2});
        \draw[thick, blue, domain=0.5:2.] plot (\x, {1.25 - 2*(\x-1.25)^2});

        \draw[thick, blue] (3.25, 0.125) -- (4, 1.5) -- (4.75, 0.125);
        \draw[thick, red] (3.25, 2.875) -- (4, 1.5) -- (4.75, 2.875);
        \draw[thick, dashed, opacity=0.5] (5., 0) -- (5., 3);
        \node[align=center] at (5, -0.3) {$\lambda_c$};

        \draw[thick, red, domain=5.25:6.75] plot (\x, {1.1 + 3.1*(\x-6.)^2});
        \draw[thick, blue, domain=5.25:6.75] plot (\x, {1.9 - 3.1*(\x-6.)^2});

        \draw[thick, red, domain=8:9.5] plot (\x, {1.6 + (3.1*(\x - 8.75)^2 - 0.6)^2});
        \draw[thick, blue, domain=8:9.5] plot (\x, {1.4 - (3.1*(\x - 8.75)^2 - 0.6)^2});

        \draw[thick, blue, domain=8.3:9.2] plot (\x, {1.6 + (3.1*(\x - 8.75)^2 - 0.6)^2});
        \draw[thick, red, domain=8.3:9.2] plot (\x, {1.4 - (3.1*(\x - 8.75)^2 - 0.6)^2});

        \node[align=center, scale=0.8] at (1.25, 3.2) {I};
        \node[align=center, scale=0.8] at (4, 3.2) {DSM};
        \node[align=center, scale=0.8] at (6., 3.2) {WSM};
        \node[align=center, scale=0.8] at (8.75, 3.2) {TI};

        \node[align=center] at (-0.5, 3.5) {(a)};

        \begin{scope}[shift={(11.9,0)}, scale=0.8]
            \draw[thick, dashed] (0, 0, 0) -- (3, 0, 0) -- (3, 3, 0) -- (0, 3, 0) -- cycle;
            \draw[thick, dashed] (0, 3, 0) -- (1, 4, 0) -- (4, 4, 0) -- (3, 3, 0);
            \draw[thick, dashed] (3, 0, 0) -- (4, 1, 0) -- (4, 4, 0) -- (3, 3, 0);
            \draw[thick, dashed] (0, 0, 0) -- (1, 1, 0);
            \draw[thick, dashed] (1, 1, 0) -- (4, 1, 0);
            \draw[thick, dashed] (1, 1, 0) -- (1, 4, 0);

            \draw[ultra thick, ->, red] (0, 0, 0) -- (0, 1.5, 0);
            \draw[ultra thick, red] (0, 1.5, 0) -- (0, 3, 0); 
            \draw[ultra thick, blue] (3, 0, 0) -- (3, 1.5, 0);
            \draw[ultra thick, <-, blue] (3, 1.5, 0) -- (3, 3, 0);

            \draw[ultra thick, blue] (1, 1, 0) -- (1, 2.5, 0);
            \draw[ultra thick, <-, blue] (1, 2.5, 0) -- (1, 4, 0); 
            \draw[ultra thick, ->, red] (4, 1, 0) -- (4, 2.5, 0);
            \draw[ultra thick, red] (4, 2.5, 0) -- (4, 4, 0);

            \node[align=center] at (-0.5, 4.2) {(b)};
        \end{scope}

    \end{tikzpicture}
    \caption[Different topological phases as a function of spin-orbit coupling and edge states of second-order HOTI]{(a) Evolution of the band structure of a system as a function of spin-orbit coupling $\lambda$. Starting from a conventional insulator (I), at a critical SOC $\lambda_c$ the bands close at a Dirac point, resulting in a Dirac semimetal (DSM) or Weyl semimetal (WSM) if either $\mathcal{P}$ or $\mathcal{T}$ are broken. For higher SOC, the gap reopens giving a TI. (b) Hinge edge states of a second-order higher-order topological insulator.}
    \label{fig:weyl}
\end{figure}

Topological materials are commonly realized in nature through spin-orbit coupling (SOC), inducing band inversions that lead to topological phases~\cite{bansil2016topological, narang2021topology}. This band inversion is responsible for producing a charge polarization (for Chern insulators) or a time-reversal polarization (for TIs), which determines the value of the topological invariant. In between conventional or inverted bands, SOC can also result in gapless systems~\cite{murakami2007phase}, as represented in Fig.~\ref{fig:weyl}(a). This is the case of Dirac semimetals~\cite{wang2013three, wang2012dirac}, characterized by the presence of two degenerate Dirac points in the band structure $H=\hbar v_F(k_x\sigma_x+k_y\sigma_y+k_z\sigma_z)\otimes\tau_z$, such as graphene. When either time-reversal symmetry $\mathcal{T}$ or inversion symmetry $\mathcal{P}$ or both are broken, the degenerate Dirac point splits into two separate points, resulting in a Weyl semimetal~\cite{yan2017topological}. Each Weyl point carries a different value of the Chern number, and like topological insulators, these materials present their own edge states called Fermi arcs connecting the Weyl points~\cite{wan2011topological, turner2013bandinsulatorstopologysemimetals}.

Dirac and Weyl semimetals can exist not only as isolated points in the BZ but also as nodal lines, where spatial symmetries enforce lines of degeneracy~\cite{burkov2011topological, fang2015topological}, hence identified as topological nodal line semimetals. Symmetries can also give rise to chiral multifold fermions, namely points of higher degeneracy with a finite Chern number~\cite{martinez2023linear, Robredo_2024}. In insulators, additional topological phases exist, such as higher-order topological insulators, defined by corner or hinge states~\cite{frank2018higher} (see Fig.~\ref{fig:weyl}(b)), and topological crystalline insulators, where crystalline symmetries provide the topological protection~\cite{fu2011topological}. Remarkably, the theory developed for Chern and topological insulators also applies to superconductors, owing to the Bogliubov-de Gennes formalism, leading to topological superconductors~\cite{qi2011topological}. Furthermore, any system whose states are parametrized by a continuous parameter $\lambda$, $\ket{u(\lambda)}$ fall under the same framework, enabling non-electronic topological systems such as photonic~\cite{lu2014topological}, phononic~\cite{wang2015topological} or mechanical modes~\cite{huber2016topological}. 

Thus far, the topological phases discussed correspond to non-interacting systems. However, interactions can lead to new types of topological phases, as exemplified by the fractional quantum Hall effect~\cite{tsui1982two}. In this case, beyond the topological character of the Landau levels, the interacting ground state exhibits topological order. This is characterized by ground-state degeneracy due to the fractionalization of electrons into quasiparticles, which act as anyonic excitations~\cite{arivas1984fractional, tong2016lectures}. The degeneracy depends on the system's filling, e.g. $\nu=1/(2m+1)$, $m\in\mathbb{N}$ for Laughlin states~\cite{laughlin1999nobel} and arises because of the topology of the manifold where the ground state is defined, typically a torus with genus $g=1$~\cite{wen1989vacuum, wen1990ground}. Similarly to the IQHE and Chern insulators, strong interactions in partially filled Chern bands produce fractional Chern insulator---fractionalized topological phases occurring without Landau levels~\cite{regnault2011fractional}. 

Since the discovery of the IQHE and the conception of the Chern insulator, the study of topological phases has expanded rapidly, with the discovery of new phases and the development of new theoretical tools for their description. In crystalline systems, the concepts of Berry connection and curvature are central in determining the Chern number~\cite{vanderbilt2018berry}. For the $\mathbb{Z}_2$ invariant, its computation is more intricate and has evolved to the current Wilson loop method~\cite{soluyanov2011computing, yu2011equivalent, alexandradinata2014wilson}. The Wilson loop, which generalizes the Berry phase, measures charge pumping in the system, allowing to determine both Chern and the $\mathbb{Z}_2$ invariant~\cite{z2pack}. Remarkably, a different framework to that of topological band theory and charge pumping is the use of representation theory to describe topological phases. Named topological quantum chemistry, this approach relies on the representations labeling the system's states and their compatibility relations to distinguish between topological phases~\cite{kruthoff2017topological, po2017symmetry, bradlyn2017topological, elcoro2021magnetic}.

While the theory for determining whether a system is topological is well-established, it assumes that the material is crystalline and is formulated in momentum space. This creates a challenge when dealing with disordered systems, as reciprocal space is not inherently suitable for describing them. Even though a topological phase is guaranteed to be protected against disorder, it is necessary to develop alternative tools, for instance real-space quantities like the Chern marker~\cite{bianco2011mapping}, to accurately quantify the effect of disorder on the topological properties. Momentum-space theory has been successful enough to allow high-throughput calculations of topological materials, using either the Wilson loop~\cite{olsen2019discovering, mounet2018two} or the topological quantum chemistry approach~\cite{vergniory2019complete, xu2020high, yuanfeng2024catalog}. However, given the ubiquity in nature of disordered materials such as amorphous solids or alloys, there is a strong interest in developing methodologies to extend the topological classification to these materials~\cite{Corbae_2023}.

In this thesis, we focus on the study of two-dimensional disordered topological insulators, including both Chern and $\mathbb{Z}_2$ topological insulators. First, in Chapter 7 we address the problem of identifying disordered time-reversal topological insulators, proposing a novel approach. Our methodology leverages the entanglement spectrum of the solid, which measures the entanglement between two halves of the system, serving as a proxy for the topological invariant. We combine this with deep learning to classify the topological phase, illustrating it with a toy model of topological fermions in an amorphous lattice. In Chapter 8, we extend this methodology to a realistic model of 2D Bi$_x$Sb$_{1-x}$ alloys, determining the maximum disorder the material can tolerate while remaining topological and identifying the different phase transitions that occur varying alloy concentration and disorder. Finally, in Chapter 9 we explore non-crystalline fractional Chern insulators. Following the strategy of Landau level mimicry in reciprocal space, we introduce a set of real-space criteria to determine if a disordered system could potentially host a fractional Chern insulator. We test these with different models, establishing the maximum degree of disorder they can withstand before losing the topological character.

\chapter{Theory of topological invariants}\label{chapter:topology}

\section{Introduction}

Topology is the branch of mathematics concerned with the properties of objects preserved under continuous deformations. The foundational area of topology is point set topology, which introduces the concept of topological spaces. A topological space, denoted as $(X, \tau)$, is the most general type of mathematical structure that formalizes the notions of continuity and connectivity. Then, one can define continuous functions between topological spaces. When such a function $f$ is bijective and its inverse $f^{-1}$ is continuous, it establishes an isomorphism known as homeomorphism~\cite{munkres2013topology}.

\begin{figure}[h]
    \centering
    \includegraphics[width=0.6\textwidth]{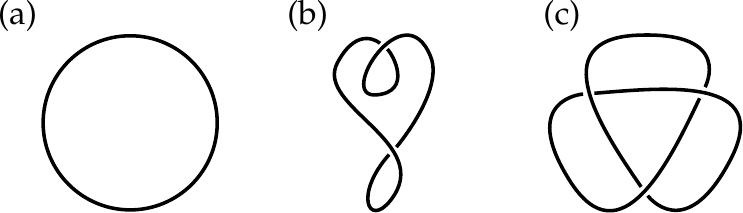}
    \caption[Topologically inequivalent knots]{Examples of topologically inequivalent knots.\ (a) and (b) correspond both to the unknot, (b) being a smooth deformation of (a).\ (c) is the trefoil knot, which is the simplest non-trivial knot that cannot be smoothly deformed into the unknot without cutting it.}\label{fig:knots}
\end{figure}

Properties of topological spaces that remain invariant under homeomorphisms are what we know as topological properties or invariants. Examples include abstract properties, like whether a space satisfies the Kolmogorov separation condition $( T_0 )$, and structures such as the fundamental group  $(\pi_1(X), +)$, from algebraic topology. The fundamental group measures the connectivity of loops within a topological space and is closely related to the genus g of orientable manifolds~\cite{massey1967algebraic}. The genus, also a topological invariant, represents the number of holes in a closed surface; for example, a sphere has $g = 0$, while a torus has $g = 1$. This is given by the Gauss-Bonnet theorem,
\begin{equation}
    \frac{1}{2\pi}\int_M K dS = \chi(M) = 2-2g
\end{equation}
stating that the integral of the Gaussian curvature $K$ of the manifold is equal to the Euler characteristic $\chi(X)$, which is linked to the genus of the surface~\cite{do2016differential}. Another classic example of topological invariants comes from knot theory, where these invariants are used to classify different types of knots, as in Fig.~\ref{fig:knots}~\cite{crowell2012introduction}.

In physics, gauge theories are described by the mathematical theory of fiber bundles~\cite{wu1975concept}. In particular, this also holds for the Berry connection and curvature, which can be interpreted as a gauge field and its associated field strength, and with their topological properties explored under this framework~\cite{simon1983holonomy}. A key concept is the Chern number, a topological invariant associated with the first Chern class of the valence bundle~\cite{Freed_2013}, which plays a central role in the classification of topological phases of matter. One of the main results from the purely mathematical approach is the so-called "ten-fold way", which provides the topological classification of Hamiltonians solely from their Altland-Zirnbauer symmetry classes~\cite{chiu2016classification, altland1997nonstandard}. Here, we will not delve into the formal mathematical theory, see for details~\cite{nakahara2018geometry}. Instead, we will focus on the physical implications of these invariants, how they relate to the physical system and why they are topological in nature.

To address the topological properties of disordered insulators, it is paramount to first develop a solid grasp of the physical origin of topological invariants in crystalline systems and the methods used to compute them. This chapter specifically reviews the theory concerning conventional topologically insulating phases, namely Chern and $\mathbb{Z}_2$ topological insulators. Nonetheless, the framework discussed  here is also applicable to other topological phases, such as WSMs or HOTIs, albeit with phase-specific nuances that distinguish them. Notably, the Wilson loop which is central in determining topological invariants, is also instrumental in characterizing the topological properties of 3D TIs, WSMs and HOTIs~\cite{fu2007topological, weng2015weyl, saini2022wloopphi, benalcazar2017quantized, franca2018anomalous}.

We begin introducing the concepts of Berry phase, Berry connection and Berry curvature, which are fundamental to understanding topological insulators. The Berry phase appears in multiple contexts in physics, and is defined in terms of the Berry connection and Berry curvature, which are the quantities used to express the topological invariants, in particular the Chern number. We show how to compute the Chern number, with both continuum and discrete formulations, the latter being more convenient numerically as it is gauge invariant. Additionally, we illustrate how the Hall conductivity in insulators is written directly in terms of the Chern number. 

With the role of the Berry connection in defining the Chern number established, we extend the discussion to the calculation of the $\mathbb{Z}_2$ invariant, starting from the early methods devised for its computation. With the original picture laid out, we shift to an alternative approach based on the Wannier representation of the Bloch states. From this perspective, the topological properties of both types of insulators are tied to a non-trivial charge pumping inherent to the systems~\cite{thouless1983quantization, fu2006time}. This charge pumping is directly related to the evolution of the Wannier charge centers~\cite{soluyanov2011wannier}, which can be tracked and whose trajectories reveal the value of the invariants. Specifically, this evaluation is done via the Wilson loop~\cite{z2pack, soluyanov2011computing}, which is the generalization of the Berry phase to the multiband case, in case of band degeneracies.

After setting the framework of topological invariants, we turn our attention to the entanglement spectrum. Among the different forms of entanglement spectrum, we consider the spatial entanglement spectrum across two halves of the system, which measures the entanglement of the many-body ground state between the two regions. Notable, it can be shown that the reduced density matrix for one region is equivalent to a flattened version of the Hamiltonian~\cite{hughes2011inversion}. Consequently, if the original Hamiltonian is topological, edge states will appear in the spectrum of the reduced density matrix, providing a direct way to define topological indices from the spectrum~\cite{alexandradinata2011trace}. As we will show in the following chapters, a combination of the standard techniques for invariants with the entanglement spectrum will provide a powerful mean to study topological phases in disordered systems.

\section{Berryology}
The Berry phase was originally introduced by Berry~\cite{berry1984quantal} and by Wilczek and Zee for the non-abelian case~\cite{wilczek1984appearance}. It was then noted that the Berry phase is actually ubiquitous in physics, appearing in multiple contexts both classical and quantum~\cite{xiao2010berry, shapere1989geometric}. Importantly, its role in condensed matter physics was originally noted by Zak~\cite{zak1989berry}, which gave the topological invariant for 1D systems, the Zak phase, relevant for instance for the SSH model~\cite{su1979solitons}. In the context of quantum mechanics, the Berry phase is particularly interesting due to the $U(1)$ gauge invariance (i.e.\ phase invariance) of states, namely under the gauge transformation
\begin{equation}
    \ket{\psi} \longrightarrow e^{i\theta}\ket{\psi}
\end{equation}
all observables $\braket{\mathcal{O}}$ related to the state remain invariant. Thus, one would not expect a priori that the Berry phase would have a relevant role in the physics of the system. There are three factors, however, that make it significant: while for individual states phases are not relevant, they are as soon as we consider a superposition of them, as for instance when time-evolving states with different energies. This is also the case of the Ahanorov-Bohm effect~\cite{ahanorov1959significance}, where the system picks up a phase due to the presence of the magnetic potential $A_{\mu}$ (but zero field), which can be measured in the interference pattern of electrons. The second factor is that the Berry phase is a geometric phase, meaning it is related to the intrinsic geometric and topological properties of the system. And lastly, it is gauge invariant which already hints that it will be related to physical observables~\cite{xiao2010berry}.

\subsection{Discrete formulation}\label{sec:discrete_berryology}

The Berry phase is typically defined in the continuum limit and as we will see, it appears naturally when considering the adiabatic evolution of a particle, as it was originally derived. However, to introduce it is convenient to start from a discrete formulation~\cite{asboth2016short, vanderbilt2018berry}. Additionally, the discrete formulation is manifestly gauge invariant, which will be convenient to obtain practical expressions to compute invariants. Given two states $\ket{\psi_1}$ and $\ket{\psi_2}$, we can define the relative phase between them as
\begin{equation}
    e^{-i\varphi_{12}} = \frac{\braket{\psi_1|\psi_2}}{|\braket{\psi_1|\psi_2}|} \longleftrightarrow \varphi_{12} = -\text{Im}\ln\braket{\psi_1|\psi_2}
\end{equation}
where the logarithm $\ln$ is defined with the branch cut $-\pi < \varphi \leq \pi$. Note that $\text{Im}\ln z$ is equivalent to taking the argument $\varphi$ of the complex number $z=|z|e^{i\varphi}$ and discarding the magnitude. If we now consider a Hilbert space of $N\geq3$ states, we can define a closed path in the space of states, $\{\ket{\psi_1}, \ket{\psi_2}, \dots, \ket{\psi_N}, \ket{\psi_1}\}$, see Fig.~\ref{fig:berry_fluxes}(a). The Berry phase over the closed path is then defined as
\begin{equation}
    \label{eq:berry_phase_discrete}
    \phi = -\text{Im}\ln\braket{\psi_1|\psi_2}\braket{\psi_2|\psi_3}\ldots\braket{\psi_{N-1}|\psi_N}\braket{\psi_N|\psi_1}
\end{equation}
This phase can be written in a manifestly gauge invariant form writing it terms of the projector for each state $P_n=\ket{\psi_n}\bra{\psi_n}$ and the trace:
\begin{equation}
    \label{eq:berry_phase_discrete_gauge}
    \phi = -\text{Im}\ln\text{Tr}[P_1P_2\ldots P_NP_1]
\end{equation}
where the additional phases introduced by the trace cancel out as they correspond to complex conjugate terms. Both expressions~\eqref{eq:berry_phase_discrete} and~\eqref{eq:berry_phase_discrete_gauge} are gauge invariant under the transformations $\ket{\psi_n}\longrightarrow e^{i\theta_n}\ket{\psi_n}$ $\forall n$. The gauge invariance is present due to the closed path; extending the above definition to arbitrary paths it can be easily seen that the phase would not be gauge invariant. Eq.~\eqref{eq:berry_phase_discrete} already has a peculiarity regarding the branch cut of the logarithm. We can write Eq.~\eqref{eq:berry_phase_discrete} as 
\begin{equation}
    \phi = -\sum_{n=1}^N\text{Im}\ln\braket{\psi_n|\psi_{n+1}}
\end{equation}
where $\ket{\psi_{N+1}}=\ket{\psi_1}$. Since the value of each $\ln$ is restricted to $-\pi < \varphi \leq \pi$, from the sum over logarithms we could obtain a value potentially different from that of Eq.~\eqref{eq:berry_phase_discrete} as we add phases. Therefore, the Berry phase is actually a gauge invariant quantity defined modulo $2\pi$, consistent with the fact that it appears exponentiated.

\input{diagrams/berry_phase.tex}

Now, we consider a Hilbert space such that our states are labelled by two different quantum numbers $n,m\in \mathbb{N}$ with $n\in[1,N]$ and $m\in[1,M]$, i.e.\ the parameter space is two-dimensional. Assuming that the labels $n,m$ denote the vertices of squares (1 and $N,M$ being the outermost), we can compute the Berry phase over a closed path in the parameter space, as in Fig.~\ref{fig:berry_fluxes}(b). The Berry phase is then given by

\begin{align}
   \nonumber \phi = -\text{Im}\ln\exp\left[-i\left(\sum_{n=1}^{N-1}\right.\right.&\varphi_{(n,1),(n+1,1)} + \sum_{m=1}^{M-1} \varphi_{(N,m),(N,m+1)} \\ 
    &\left.\left.+ \sum_{n=1}^{N-1} \varphi_{(n + 1,M),(n,M)} + \sum_{m=1}^{M-1} \varphi_{(1,m+1),(1,m)}\right)\right]
\end{align}
Although globally gauge independent, each individual phase is gauge dependent. It is more convenient instead to formulate the Berry phase in terms of Berry fluxes, which involve computing the phase around each plaquette of the lattice. The Berry flux $F_{nm}$ is defined as
\begin{align}
    \nonumber F_{nm} = -\text{Im}\ln\exp\left[-i\left(\varphi_{(n,m),(n+1,m)} \right.\right.&+ \varphi_{(n+1,m),(n+1,m+1)} \\
    &\left.\left.+ \varphi_{(n+1,m+1),(n,m+1)} + \varphi_{(n,m+1),(n,m)}\right)\right]
\end{align}
The Berry flux is itself a Berry phase, and consequently it is gauge invariant. A manifestly gauge invariant formulation is the following:
\begin{align}
    \label{eq:berry_flux}
    F_{nm} = -\text{Im}\ln\left(\braket{\psi_{n,m}|\psi_{n+1,m}}\braket{\psi_{n+1,m}|\psi_{n+1,m+1}}\braket{\psi_{n+1,m+1}|\psi_{n,m+1}}\braket{\psi_{n,m+1}|\psi_{n,m}}\right)
\end{align}
Next, if we consider the product of all plaquette phase factors $e^{-iF_{nm}}$, we obtain the Berry phase as
\begin{align}
    \label{eq:discrete_stokes_theorem}
    \prod_{n=1}^{N-1}\prod_{m=1}^{M-1}e^{-iF_{nm}} = \exp\left[-i\sum_{n=1}^{N-1}\sum_{m=1}^{M-1}F_{nm}\right] = e^{-i\phi}
\end{align}
since all internal edges in the plaquettes cancel out except the external edges corresponding to path $C$ which are not shared, as illustrated in Fig.~\ref{fig:berry_fluxes}(b). Therefore, the sum of all plaquette phase factors is equal to the Berry phase factor, $e^{-i\phi}$, as in Fig.~\ref{fig:berry_fluxes}(c). This can be regarded as a softer, discretized version of Stokes theorem, where the left term corresponds to the curl of a vector field (the Berry connection we will introduce later) on a surface, and the right term is the line integral of the vector field along the boundary $C$ of the surface. It should be noted that the above equality does not hold directly in terms of the phases, but rather in terms of the exponentiated phases, consistent with the fact that the Berry phase is only well-defined modulo $2\pi$, $\sum_{n,m}F_{nm}\equiv\phi \mod2\pi$, since the logarithm maps the phases to the $(-\pi, \pi]$ interval.

Having established the connection between the Berry flux and the Berry phase, next we assume that the states $\ket{\psi_{nm}}$ of our Hilbert space are arranged in cyclic order, i.e. $\ket{\psi_{N+1,m}}=\ket{\psi_{1,m}}$ and $\ket{\psi_{n,M+1}}=\ket{\psi_{n,1}}$. This is the same as assuming that the underlying parameter space corresponds to a discretized torus. In this case, the product of all Berry flux phase factors gives
\begin{equation}
    \label{eq:berry_fluxes_torus}
    \prod_{n=1}^{N}\prod_{m=1}^{M}e^{-iF_{nm}} = 1
\end{equation}
since all edges are shared and therefore cancel out. Now the Berry phase takes the trivial value $\phi=0$ since there is no boundary in the parameter space. We can associate a Chern number $Q$ to this Hilbert space, defined from the sum of all Berry fluxes:
\begin{equation}
    Q = \frac{1}{2\pi}\sum_{n,m}F_{nm}\in \mathbb{Z}
\end{equation}
It follows trivially from~\eqref{eq:berry_fluxes_torus} that the Chern number must be an integer. Since each Berry flux is gauge invariant as seen from~\eqref{eq:berry_flux}, the Chern number is also gauge invariant. Note that each Berry flux, being a Berry phase, is only well-defined modulo $2\pi$, which may imply that the above Chern number can take arbitrary values. This is solved computing systematically the Berry flux using expression~\eqref{eq:berry_flux}, where the logarithm restricts the value to the $(-\pi,\pi]$ interval. Its relevance can be made explicit if we consider the following modified Berry fluxes
\begin{equation}
    \tilde{F}_{nm} = \varphi_{(n,m),(n+1,m)} + \varphi_{(n+1,m),(n+1,m+1)} + \varphi_{(n+1,m+1),(n,m+1)} + \varphi_{(n,m+1),(n,m)}
\end{equation} 
directly in terms of the phases that conform the plaquette. Since all edges are shared, the sum of all phases must cancel:
\begin{equation}
    \sum_{n=1}^{N}\sum_{m=1}^{M}\tilde{F}_{nm} = 0
\end{equation}
If $-\pi < \tilde{F}_{nm} \leq \pi$, then $\tilde{F}_{nm} = F_{nm}$. However, in general $\tilde{F}_{nm}$ may be outside this interval and the logarithm in~\eqref{eq:berry_flux} takes it back to the $(-\pi,\pi]$ interval, adding an integer multiple of $2\pi$ to the phase (e.g.\ suppose that there are two fluxes $\tilde{F}_{nm}=2\pi$ and $\tilde{F}_{n'm'}=-2\pi$. Then, $F_{nm}=F_{n'm'}=0$). Therefore, each plaquette contains a number $Q_{nm}\in\mathbb{Z}$ of vortices
\begin{equation}
    Q_{nm} = \frac{F_{nm} - \tilde{F}_{nm}}{2\pi} \in \mathbb{Z}
\end{equation}
and consequently the Chern number measures the number of vortices in the Berry fluxes of the closed surface,
\begin{equation}
    \label{eq:chern_number_discrete}
    Q = \frac{1}{2\pi}\sum_{n,m}F_{nm} = \sum_{n,m}Q_{nm} \in \mathbb{Z}
\end{equation}
This is the discrete formulation of the Chern number, which as we will see later gives the topological invariant for Chern insulators. Expression~\eqref{eq:chern_number_discrete} together with~\eqref{eq:berry_flux} provide the standard numerical recipe that is used to compute the Chern number in a gauge invariant manner~\cite{fukui2005chern}.

\clearpage

\subsection{Continuum formulation}

\begin{figure}[h]
    \centering
    \includegraphics[width=0.6\textwidth]{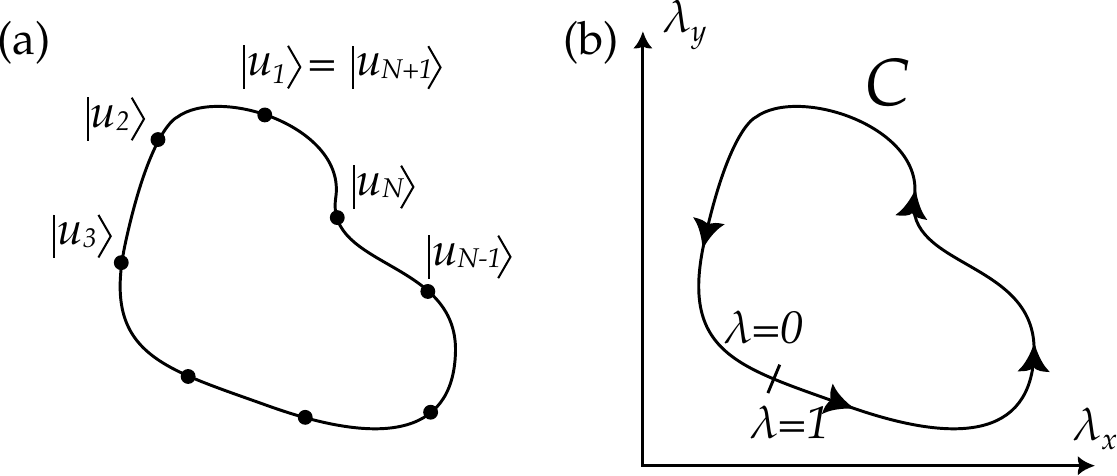}
    \caption[Continuum limit in the number of states around a closed loop for the Berry phase]{(a) Calculation of the Berry phase along a discrete closed loop of $N$ states.\ (b) Continuum limit increasing the density of states around the loop. The states are labelled by a continuous parameter $\lambda\in[0,1]$ with $\ket{u_{\lambda=0}} = \ket{u_{\lambda=1}}$.}\label{fig:berry_continuum}
\end{figure}

The definition of the Berry phase~\eqref{eq:berry_phase_discrete} is also well-defined in the continuum limit, which may also hint that the Berry phase is a physically meaningful quantity. 
From now on, we denote the states of the Hilbert space as \{$\ket{u_{\lambda}}\}$, where $\lambda$ is a continuous parameter that labels the states with $\lambda\in[0, 1]$, and such that $\ket{u_{\lambda=0}} = \ket{u_{\lambda=1}}$. Connecting with the discrete formulation of the Berry phase, we may assume that we have a discrete set of states around a closed loop, as in Fig.~\ref{fig:berry_continuum}(a). The states in the loop would be $\{\ket{u_{\lambda}}, \ket{u_{\lambda+d\lambda}}, \ldots, \ket{u_{\lambda+(N-1)d\lambda}}, \ket{u_{\lambda + Nd\lambda}}=\ket{u_{\lambda}}\}$, where $d\lambda$ is the separation between states. The Berry phase is then given by
\begin{align}
   \nonumber \phi &= -\text{Im}\sum_{\lambda}^N\ln\braket{u_{\lambda}|u_{\lambda+d\lambda}}=-\text{Im}\sum_{\lambda}^N\ln\bra{u_{\lambda}}\left(\ket{u_{\lambda}} + \partial_{\lambda}\ket{u_{\lambda}}d\lambda + \mathcal{O}(d\lambda^2)\right)\\
    \nonumber &= - \text{Im}\sum_{\lambda}^N\ln\left(1 + d\lambda\braket{u_{\lambda}|\partial_{\lambda}u_{\lambda}} + \mathcal{O}(d\lambda^2)\right) = -\text{Im}\sum_{\lambda}^N \left[d\lambda\braket{u_{\lambda}|\partial_{\lambda}u_{\lambda}} + \mathcal{O}(d\lambda^2)\right] \\
    &\xrightarrow[d\lambda\to0]{} -\text{Im}\oint \braket{u_{\lambda}|\partial_{\lambda}u_{\lambda}}d\lambda
\end{align}
where we have done a Taylor expansion to first order in $d\lambda$ for both $\ket{u_{\lambda+d\lambda}}$ and $\ln$. Note that we also assume that $\ket{u_{\lambda}}$ are smooth on $\lambda$ to perform the expansion. We observe that the quantity $\braket{u_{\lambda}|\partial_{\lambda}u_{\lambda}}$ is purely imaginary,
\begin{equation}
    2\text{Re}\braket{u_{\lambda}|\partial_{\lambda}u_{\lambda}} = \braket{u_{\lambda}|\partial_{\lambda}u_{\lambda}} + \braket{\partial_{\lambda}u_{\lambda}|u_{\lambda}} = \partial_{\lambda}\braket{u_{\lambda}|u_{\lambda}} = 0
\end{equation}
which allows us to write the Berry phase as
\begin{equation}
    \label{eq:berry_connection_definition}
    \phi = \oint i\braket{u_{\lambda}|\partial_{\lambda}u_{\lambda}}d\lambda \equiv \oint {A}(\lambda)d\lambda
\end{equation}
where $A(\lambda)=i\braket{u_{\lambda}|\partial_{\lambda}u_{\lambda}}$ is the so-called Berry connection. While the Berry phase $\phi$ must be gauge invariant modulo $2\pi$ as before, the Berry connection we have just introduced is not gauge invariant. Under a gauge transformation of the form
\begin{equation}
    \ket{u_{\lambda}} \longrightarrow e^{-i\beta(\lambda)}\ket{u_{\lambda}}
\end{equation}

\noindent the Berry connection transforms as
\begin{equation}
    A(\lambda) \longrightarrow A(\lambda) + \partial_{\lambda}\beta(\lambda)
\end{equation}
Since $\ket{u_{\lambda=0}}=\ket{u_{\lambda=1}}$, then necessarily $\beta(0)=\beta(1) + 2\pi m$. Taking this into account, the gauge transformation of the Berry phase is:
\begin{equation}
    \phi \longrightarrow \oint (A(\lambda) + \partial_{\lambda}\beta(\lambda))d\lambda = \phi + \int_0^1 \partial_{\lambda}\beta(\lambda)d\lambda = \phi + 2\pi m
\end{equation}
Namely, the Berry phase is gauge invariant modulo $2\pi$ as in the discrete case, as expected. So far we have considered states arranged in a one-dimensional parameter space, parametrized by $\lambda$. Now, as in the discrete case we will consider a two-dimensional parameter space, with states $\ket{u_{\bm{\lambda}}}$ where $\bm{\lambda}=(\lambda_x,\lambda_y)$. The Berry connection defined in~\eqref{eq:berry_connection_definition} is then generalized to a vector,
\begin{equation}
    A_{\mu}(\bm{\lambda}) = i\braket{u_{\bm{\lambda}}|\partial_{\mu}u_{\bm{\lambda}}}
\end{equation}
where $\partial_{\mu}\equiv\partial/\partial{\lambda_{\mu}}$, and $\mu=x,y$. The Berry phase is expressed as the line integral of the Berry connection along a closed path $P$ in the parameter space,
\begin{equation}
    \phi = \oint_P \bm{A}(\bm{\lambda})\cdot d\bm{\lambda}
\end{equation}
Connecting back with the discrete formulation, in Eq.~\eqref{eq:discrete_stokes_theorem} we saw that the sum of all Berry flux factors is equal to the Berry phase factor, $e^{-i\phi}$. Next we want to write the continuous version of this expression. So far we have seen that
\begin{equation}
    \oint_{\partial\mathcal{F}}\bm{A}(\bm{\lambda})\cdot d\bm{\lambda} = \lim_{\Delta\lambda_x,\Delta\lambda_y\rightarrow 0}\phi(\partial\mathcal{F})
\end{equation}
where $\partial\mathcal{F}$ is the closed path in parameter space, and $\phi(\partial\mathcal{F})$ is the discretized version of the Berry phase computed along this loop. From expression~\eqref{eq:discrete_stokes_theorem} it follows that
\begin{equation}
    \exp\left[-i\oint_{\partial\mathcal{F}}\bm{A}(\bm{\lambda})\cdot d\bm{\lambda}\right] = \lim_{\Delta\lambda_x,\Delta\lambda_y\rightarrow 0}\exp\left[-i\sum_{n,m}F_{nm}\right]
\end{equation}
Therefore, we need to determine a continuous expression for the Berry fluxes $F_{nm}$. Assuming that we have a smooth gauge in the vicinity of the plaquette $(n,m)$ (which can be assumed without loss of generality, as we can always choose a gauge where this is the case due to the gauge invariance $e^{-iF_{nm}}=e^{-iF'_{nm}}$), we can write the Berry flux as
\begin{equation}
    F_{nm} = \left[\partial_xA_y(\bm{\lambda}) - \partial_yA_x(\bm{\lambda})\right]\Delta\lambda_x\Delta\lambda_y
\end{equation}
This expression is obtained Taylor expanding to first order in $\Delta\lambda_x$, $\Delta\lambda_y$ multiple times, first in the states that appear in the plaquette, then in the logarithm, and finally to identify the finite differences that give the derivatives of the Berry connection~\cite{asboth2016short, soluyanov2012topological}. The Berry curvature then is defined as
\begin{equation}
    \label{eq:berry_curvature_2d}
    B(\bm{\lambda}) = \lim_{\Delta\lambda_x,\Delta\lambda_y\rightarrow 0}\frac{F_{nm}}{\Delta\lambda_x\Delta\lambda_y} = \partial_xA_y(\bm{\lambda}) - \partial_yA_x(\bm{\lambda})
\end{equation}
With this, the continuum version of equation~\eqref{eq:discrete_stokes_theorem} is
\begin{equation}
    \label{eq:berry_curvature_stokes}
    \exp\left[-i\oint_{\partial\mathcal{F}}\bm{A}(\bm{\lambda})\cdot d\bm{\lambda}\right] = \exp\left[-i\int_{\mathcal{F}}B(\bm{\lambda})d\lambda_xd\lambda_y\right]
\end{equation}
where $\mathcal{F}$ is the surface enclosed by the path $\partial\mathcal{F}$. While the Berry connection was not gauge invariant, now in the two-dimensional case as $\bm{A}\longrightarrow\bm{A} + \nabla \beta$, under the transformation $\ket{u_{\bm{\lambda}}}\longrightarrow e^{i\beta(\bm{\lambda})}\ket{u_{\bm{\lambda}}}$, it can be easily seen that Berry curvature is gauge invariant. Expression~\eqref{eq:berry_curvature_stokes} is general and does not depend on the gauge chosen for the states. The exponents, however, are not necessarily the same, given that the Berry phase is only well-defined modulo $2\pi$. Therefore, one has
\begin{equation}
    \oint_{\partial\mathcal{F}}\bm{A}(\bm{\lambda})\cdot d\bm{\lambda} = \int_{\mathcal{F}}B(\bm{\lambda})d\lambda_xd\lambda_y + 2\pi m
\end{equation}
Stokes' theorem corresponds to the case where $m=0$, which only applies if the manifold of states $\ket{u_{\bm{\lambda}}}$ is globally smooth on $\mathcal{F}$. As we will see, the inability to define a global gauge for the states is what gives rise to the topological invariants in the first place. The discretized Chern number~\eqref{eq:chern_number_discrete} reads in the continuum limit
\begin{equation}
    Q = \frac{1}{2\pi}\int_{\mathcal{P}}B(\bm{\lambda})d\lambda_xd\lambda_y
\end{equation}
where the integral now takes place over the whole parameter space $\mathcal{P}$. It is worthwhile to examine the behaviour of the above integral when considering an underlying torus, as in the discrete case. We may split the integral into two separate ones, along each direction of the torus. We assume for simplicity that the torus is defined on the unit square $[0,1]\times[0,1]$. Then, the Chern number reads:
\begin{align}
    \label{eq:winding_chern}
    \nonumber Q &= \frac{1}{2\pi}\int_{0}^{1}d\lambda_x\int_{0}^{1}d\lambda_y(\partial_xA_y - \partial_yA_x) = \frac{1}{2\pi}\int_{0}^{1} \partial_x\phi^{(y)}d\lambda_x \\
    &= \frac{1}{2\pi}(\phi^{(y)}(1) - \phi^{(y)}(0)) = m
\end{align}
where we have used that $\int^1_0 A_{\mu}d\lambda_{\mu}=\phi^{(\mu)}$, i.e.\ the integral of one component of the Berry connection along the $\mu$ direction of the torus gives a Berry phase, denoted as $\phi^{(\mu)}$. The second term vanishes because we suppose that we are using a periodic gauge for the states, $\ket{u_{\lambda_x,\lambda_y=0}}=\ket{u_{\lambda_x,\lambda_y=1}}$. Also, since we are considering a torus, the Berry phases at the end points $x=0$ and $x=1$ must match modulo $2\pi$, or equivalently, the Berry phase $\phi^{(y)}$ must have evolved by $2\pi m$. Therefore, the Chern number is quantized to an integer, as we had seen in the discrete case. Interestingly, we arrive at a different interpretation for the Chern number: the Chern number corresponds to the number of times $m$ the Berry phase of one coordinate winds around the torus, as a function of the other coordinate. This interpretation as a winding number will appear multiple times in the following sections, as it is a common feature of the topological invariants.

\begin{figure}[t]
    \centering
    \includegraphics[width=0.6\textwidth]{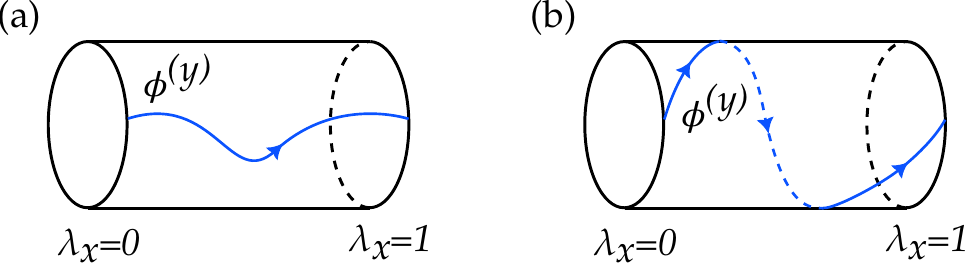}
    \caption[Winding of the Berry phase in a cylinder/torus]{Examples of windings of the Berry phase in a cylinder/torus.\ (a) No winding, corresponding to a trivial phase $m=C=0$.\ (b) Winding by $2\pi$, corresponding to $m=C=1$.}
\end{figure}

Lastly, for completeness we also discuss the case of a three-dimensional parameter space. We consider again states labeled by $\ket{u_{\bm{\lambda}}}$, where $\bm{\lambda} = (\lambda_x,\lambda_y,\lambda_z)$. For simplicity, we consider from the beginning the case where the gauge chosen for the states $\ket{\bm{\lambda}}$ is smooth on a two-dimensional open surface $\mathcal{F}$ embedded in the parameter space. Then, we can apply Stokes' theorem directly, writing the line integral of the Berry connection as the surface integral of the Berry curvature,
\begin{equation}
    \oint_{\partial\mathcal{F}}\bm{A}(\bm{\lambda})\cdot d\bm{\lambda} = \int_{\mathcal{F}}\bm{B}(\bm{\lambda})\cdot d\bm{S}
\end{equation}
where now the Berry curvature is defined as a (pseudo)vector via 
\begin{equation}
    \bm{B}(\bm{\lambda}) = \nabla\times\bm{A}(\bm{\lambda})
\end{equation}
As in the two-dimensional case, the Berry curvature is gauge invariant, while the Berry connection is not. Note that in case the gauge is not smooth, the above expression does not hold. Instead, Eq.~\eqref{eq:berry_curvature_stokes} is still valid but adapted to the 3D case. This generalization can be done to arbitrary dimensions; the three-dimensional case is particularly relevant because it provides a close analogy with the electromagnetic field, where the Berry curvature plays the role of the magnetic field $\bm{B}$ and the Berry connection $\bm{A}$ the role of the vector potential, hence the naming of the variables so far.

\subsection{Adiabatic evolution}

So far we have been discussing the properties of the Berry phase and connection in a rather abstract way, solely from the definition of the Berry phase. However, in the original derivation by Berry~\cite{berry1984quantal}, this geometric phase arises naturally when considering the adiabatic time evolution of a quantum state. We consider a Hamiltonian $H(\bm{\lambda})$ that depends on a set of parameters $\bm{\lambda}$, which are slowly varied in time, $\bm{\lambda}=\bm{\lambda}(t)$. For a given $\bm{\lambda}$, the eigenstates of the Hamiltonian are
\begin{equation}
    H(\bm{\lambda})\ket{n({\bm{\lambda}})} = E_n(\bm{\lambda})\ket{n({\bm{\lambda}})}
\end{equation}
where $n$ labels the eigenstates. Next, we consider that the system is prepared at $t=0$ in a state $\ket{\psi(t=0)}=\ket{n(\bm{\lambda}_0)}$, such that the state is gapped from the rest of the spectrum. We want to determine the solution to the time-dependent Schrödinger equation
\begin{equation}
    i\partial_t\ket{\psi(t)} = H(\bm{\lambda}(t))\ket{\psi(t)}
\end{equation}
In general, we would assume that the solution at time $t$ would be given by a linear superposition of the instantaneous eigenstates of the Hamiltonian at time $t$,
\begin{equation}
    \ket{\psi(t)} = \sum_n c_n(t)\ket{n(\bm{\lambda}(t))}
\end{equation}
However, by virtue of the adiabatic approximation~\cite{sakurai2020modern}, as long as the variation in the Hamiltonian parameters $\bm{\lambda}$ is slow enough, which implies that the variation in the Hamiltonian $\dot{H} \ll |E_n(\bm{\lambda}) - E_{n\pm 1}(\bm{\lambda})|$, where $n$ labels the initial state, then the system will remain in an instant eigenstate of the Hamiltonian. This leads to taking the Ansatz
\begin{equation}
    \ket{\psi(t)} = e^{i\phi_n(t)}e^{-i\int_0^t E_n(\bm{\lambda}(t'))dt'}\ket{n(\bm{\lambda}(t))}
\end{equation}
Substituting this Ansatz into the Schrödinger equation, we may determine the phase $\phi_n(t)$. Ultimately, after some algebra one obtains the following differential equation:
\begin{equation}
    -\partial_t \phi_n(t)\ket{n(\bm{\lambda})} + i\ket{\partial_tn(\bm{\lambda})} = 0
\end{equation}
Taking the inner product with $\bra{n(\bm{\lambda})}$ and integrating in time, we obtain the adiabatic phase
\begin{equation}
    \phi_n(t) = i\int_0^t\braket{n(\bm{\lambda})|\partial_tn(\bm{\lambda})}dt = i\int_{\mathcal{C}}\braket{n(\bm{\lambda})|\nabla_{\bm{\lambda}}n(\bm{\lambda})}\cdot d\bm{\lambda}
\end{equation}
where $\mathcal{C}$ is the curve traced by the parameters $\bm{\lambda}(t)$ in the parameter space. When the curve $\mathcal{C}$ is closed, the above expression gives the Berry phase,
\begin{equation}
    \phi_n(\mathcal{C}) = i\oint_{\mathcal{C}}\braket{n(\bm{\lambda})|\nabla_{\bm{\lambda}}n(\bm{\lambda})}\cdot d\bm{\lambda}
\end{equation}
We see here one of the reasons why it is identified as a geometrical phase: the phase picked up in the adiabatic evolution of the state does not depend on the dynamics of the system, but only on the geometry of the path travelled by the state in parameter space. This is in contrast with the second phase appearing in the Ansatz, $\exp\left(-i\int_0^t E_n(\bm{\lambda}(t'))dt'\right)$ which is precisely a dynamic phase that depends on the rate of change of the system. 

\clearpage

\section{Topological insulators}

Up to this point we have discussed the properties of the Berry phase in terms of an abstract parameter space $\bm{\lambda}$. Now we will particularize to the case of condensed matter physics, for the description of electrons in solids. The Berry phase, as introduced originally, was shown to appear in arbitrary quantum mechanical systems. It was Zak~\cite{zak1989berry} who noted that the Berry phase would also appear in the periodic motion of electrons in solids, and moreover, that a non-zero value arises because of the torus geometry. The electrons are then described by Bloch states
\begin{equation}
    H\ket{n\mathbf{k}} = E_{n\mathbf{k}}\ket{n\mathbf{k}}
\end{equation}
Or in terms of the cell-periodic part of the Bloch states, $\ket{u_{n\mathbf{k}}}$,
\begin{equation}
    H(\mathbf{k})\ket{u_{n\mathbf{k}}} = E_{n\mathbf{k}}\ket{u_{n\mathbf{k}}}
\end{equation}
where $H(\mathbf{k})$ is the Bloch Hamiltonian, $n$ denotes the band index and $\mathbf{k}$ is the crystal momentum. The first question that arises is whether we should use the complete Bloch states $\ket{n\mathbf{k}}$ or the periodic part $\ket{u_{n\mathbf{k}}}$ in the definition of the Berry connection. The answer is that one should use the periodic part~\cite{vanderbilt2018berry}, as it ensures that the inner products that appear for instance in the computation of the discretized Berry phase are well-defined. The Berry connection associated to band $n$ is then defined as
\begin{equation}
    A_{n,\mu}(\mathbf{k}) = i\braket{u_{n\mathbf{k}}|\partial_{\mu}u_{n\mathbf{k}}}
\end{equation}
where $\partial_{\mu}=\partial/\partial k_{\mu}$, and $\mu$ is a Cartesian index. Similarly, the Berry curvature of band $n$ reads
\begin{equation}
    \Omega_{n,\mu\nu}(\mathbf{k}) = \partial_{\mu}A_{n,\nu}(\mathbf{k}) - \partial_{\nu}A_{n,\mu}(\mathbf{k})
\end{equation}
Note that from now on we change the notation and denote the Berry curvature as $\Omega$ instead of $B$ to match the existing literature. So far both quantities have been defined generally and can be used either in 2D or 3D. In 2D, the Berry curvature is a scalar, $\Omega_n\equiv\Omega_{n,xy}$ and the Chern number of a band $C_n$ is given by the integral over the BZ which is a torus:
\begin{equation}
    \label{eq:chern_number_band}
    C_n = \frac{1}{2\pi}\int_{\text{BZ}}\Omega_n(\mathbf{k})d^2\mathbf{k}
\end{equation}
In this section we will discuss the bulk properties of Chern and $\mathbb{Z}_2$ topological insulators, from which the topological invariants are derived, namely the Chern number and the $\mathbb{Z}_2$ invariant respectively, both expressed in terms of the Berry connection. Nevertheless, aside from their relevance in the determination of the topological nature of materials, the Berry quantities also appear in other contexts, such as semi-classical transport. In this case, the Berry curvature plays the role of an anomalous velocity~\cite{sundaram1999wave, chang1995berry, xiao2010berry} in the equations of motion,
\begin{align}
    \left\{
    \begin{aligned}
        \dot{\bm{x}} &= \frac{1}{\hbar}\nabla_{\mathbf{k}}\varepsilon_n(\mathbf{k}) + \dot{\bm{k}}\times\bm{\Omega}(\bm{k}) \\
        \dot{\bm{k}} &= -\frac{e}{\hbar}\left(\bm{E} + \dot{\bm{x}}\times\bm{B}\right)
    \end{aligned}
    \right.
    \end{align}
The Lorentz term $\dot{\bm{x}}\times \bm{B}$ in the force is responsible for the (classical) Hall effect~\cite{ashcroft}, whereas the anomalous velocity $\dot{\bm{k}}\times\bm{\Omega}(\bm{k})$ induces the anomalous Hall effect~\cite{nagaosa2010anomalous}. As we will see next, the Berry curvature is also responsible for the quantization of the Hall conductivity in Chern insulators, leading to the AQHE\@.

\subsection{Chern insulators}\label{sec:chern_insulators}

A Chern insulator can be simply regarded as a two-dimensional insulating material with broken time-reversal symmetry $\mathcal{T}$, $[H,\mathcal{T}]\neq 0$ and such that the total Chern number of the filled bands, $C=\sum_nC_n$ is non-zero, where $C_n$ denotes the individual Chern number of band $n$, as defined in~\eqref{eq:chern_number_band}. As we are anticipating now, Chern insulators are topological phases of matter characterized by a non-zero Chern number, which is the topological invariant of the system.

In the introduction, we mentioned that Chern insulators appeared originally in the context of the IQHE as a realization of the quantized Hall conductivity but in absence of external magnetic fields. The natural way to introduce the Chern number therefore is to derive the general expression for the Hall conductivity for insulators. The Hall conductivity can be obtained by means of the Kubo formula~\eqref{eq:kubo_formula_T0} at finite frequency and taking the $\omega\rightarrow 0$ limit. Setting $a=x$, $b=y$, we have
\begin{equation}
    \sigma_{xy}(\omega) = \frac{ine^2}{m\omega}\delta_{xy} + \frac{i}{\omega V}\sum_n\left[\frac{\braket{0|j_{x}|n}\braket{n|j_{y}|0}}{\hbar\omega + i\eta - (E_n - E_0)} - \frac{\braket{0|j_{y}|n}\braket{n|j_{x}|0}}{\hbar\omega + i\eta - (E_0 - E_n)}\right]
\end{equation}
Since we are taking the $\omega\rightarrow 0$ limit, we may also set directly $\eta=0$ given that all the deltas coming from the imaginary part of the denominators involve a finite energy difference, $\delta(\omega - \Delta E)=0$ at $\omega=0$ with $\Delta E=E_n - E_0\neq 0$. We rewrite the fractions as
\begin{equation}
    \label{eq:fraction_expansion}
    \frac{1}{\hbar\omega \pm (E_n - E_0)} = \frac{1}{E_n - E_0}\left(1 \mp \frac{\hbar\omega}{E_n - E_0 \pm \hbar\omega}\right)
\end{equation}
Note that we kept the diamagnetic term above, even though we are already evaluating the $xy$ component of the conductivity. This is because the first term arising from~\eqref{eq:fraction_expansion} corresponds to a $f$-sum rule and cancels the diamagnetic contribution~\cite{ming2013linear, pines2018theory}. Expanding the current operator as $j_a=e\sum_{ij}v^a_{ij}c^{\dagger}_ic_j$, and taking into account that the states $\ket{n}$ denote Slater determinants (as we are in the non-interacting case), the Hall conductivity $\sigma_{xy}$ is given by
\begin{align}
    \label{eq:hall_conductivity_derivatives}
    & \sigma_{xy} \equiv \nonumber \sigma_{xy}(\omega = 0) = \\ 
    &\frac{ie^2}{V\hbar}\sum^{E_{n\mathbf{k}}< E_F<E_{m\mathbf{k}}}_{n,m,\mathbf{k}}\left[\frac{\braket{u_{n\mathbf{k}}|\partial_xH_{\mathbf{k}}|u_{m\mathbf{k}}}\braket{u_{m\mathbf{k}}|\partial_y H_{\mathbf{k}}|u_{n\mathbf{k}}} - \braket{u_{n\mathbf{k}}|\partial_yH_{\mathbf{k}}|u_{m\mathbf{k}}}\braket{u_{m\mathbf{k}}|\partial_x H_{\mathbf{k}}|u_{n\mathbf{k}}}}{(E_{m\mathbf{k}} - E_{n\mathbf{k}})^2}\right]
\end{align}
where the $\omega\rightarrow 0$ limit was trivially taken, and the identity $\braket{n\mathbf{k}|v_a|m\mathbf{k}}=\hbar^{-1}\braket{u_{n\mathbf{k}}|\partial_a H_{\mathbf{k}}|u_{m\mathbf{k}}}$~\cite{esteve2023comprehensive} was used, with $\partial_a \equiv \partial/\partial k_a$ and $H_{\mathbf{k}}\equiv H(\mathbf{k})$. This is the expression for the Hall conductivity that was originally reported in the TKNN paper~\cite{thouless1982quantized}, and one can already identify the Berry curvature in a gauge independent formulation~\cite{berry1984quantal, bernevig2013topological}. Now using that $\braket{u_{n\mathbf{k}}|\partial_aH_{\mathbf{k}}|u_{m\mathbf{k}}} = (E_{m\mathbf{k}} - E_{n\mathbf{k}})\braket{u_{n\mathbf{k}}|\partial_a u_{m\mathbf{k}}} = -(E_{m\mathbf{k}} - E_{n\mathbf{k}})\braket{\partial_au_{n\mathbf{k}}| u_{m\mathbf{k}}}$, we get
\begin{equation}
    \sigma_{xy} = \frac{ie^2}{V\hbar}\sum^{E_{n\mathbf{k}}< E_F<E_{m\mathbf{k}}}_{n,m,\mathbf{k}}\left[\braket{\partial_xu_{n\mathbf{k}}|u_{m\mathbf{k}}}\braket{u_{m\mathbf{k}}|\partial_yu_{n\mathbf{k}}} - \braket{\partial_yu_{n\mathbf{k}}|u_{m\mathbf{k}}}\braket{u_{m\mathbf{k}}|\partial_xu_{n\mathbf{k}}}\right]
\end{equation}
Finally, using the completeness relation $\sum_{n\mathbf{k}}\ket{u_{n\mathbf{k}}}\bra{u_{n\mathbf{k}}} = 1$ and switching from a sum to an integral over the BZ, we can write the Hall conductivity as
\begin{align}
    \nonumber \sigma_{xy} &= \frac{ie^2}{\hbar}\sum_{n}^{E_{n\mathbf{k}}<E_F}\int_{\text{BZ}}\frac{d^2\mathbf{k}}{(2\pi)^2}\left[\braket{\partial_xu_{n\mathbf{k}}|\partial_{y}u_{n\mathbf{k}}} - \braket{\partial_yu_{n\mathbf{k}}|\partial_yu_{n\mathbf{k}}}\right] \\
    &= \frac{e^2}{2\pi\hbar}\sum_{n}^{E_{n\mathbf{k}}<E_F}\int_{\text{BZ}}\frac{d^2\mathbf{k}}{2\pi}\Omega_n(\mathbf{k}) = \frac{e^2}{h}\sum_{n}^{\text{filled}}C_n = \frac{e^2}{h}C
\end{align}
where we have used an alternative way to express the Berry curvature,
\begin{align}
 \Omega_n(\mathbf{k}) &= \partial_{k_x}A_{n,y}(\mathbf{k}) - \partial_{k_y}A_{n,x}(\mathbf{k}) = i\left[\braket{\partial_{k_x}u_{n\mathbf{k}}|\partial_{k_y}u_{n\mathbf{k}}} - \braket{\partial_{k_y}u_{n\mathbf{k}}|\partial_{k_x}u_{n\mathbf{k}}}\right] 
\end{align}
Thus, the Hall conductivity of any insulator is given by the sum of the Chern number of each filled band~\cite{tong2016lectures}. Those insulators with a total non-zero Chern number are the class of materials regarded as Chern insulators. As we saw in the previous section, the Chern number must be an integer, and importantly, it is a topological invariant of the system~\cite{avron1983homotopy, thouless1983quantization, nakahara2018geometry, niu1985quantized, kohmoto1985topological}. Being a topological invariant implies that the Chern number is robust under deformations of the system, such that the bands remain gapped. But why is this the case? From the mathematical picture, we know the Chern number corresponds to the integral of the first Chern class, which is a topological invariant and in fact a generalization of the Gauss-Bonnet theorem, the Chern-Gauss-Bonnet theorem~\cite{nakahara2018geometry}. An intuitive explanation is that under adiabatic deformations, namely variations of the bands such that we can always track the same one if there are no closings, the Chern number cannot change since it corresponds to the same state.

An attempt in formalizing this statement is the following: consider an infinitesimal, adiabatic perturbation of the Hamiltonian $H(\mathbf{k})\rightarrow H(\mathbf{k}) + \delta H(\mathbf{k})$. We take the unperturbed state $\ket{u_{\mathbf{k}}}$; after perturbing the system the state evolves to $\ket{\tilde{u}_{\mathbf{k}}}$, such that $\ket{\tilde{u}_{\mathbf{k}}} = \ket{u_{\mathbf{k}}} + \ket{\delta u_{\mathbf{k}}}$. Note that we can write the state in terms of the previous one since we assume that the perturbation is adiabatic and the gap does not close. Since the perturbation is infinitesimal, we may write it as a phase:
\begin{equation}
    \ket{\tilde{u}_{\mathbf{k}}} = (1 + i\phi(\mathbf{k}))\ket{u_{\mathbf{k}}} \approx e^{i\phi(\mathbf{k})}\ket{u_{\mathbf{k}}}
\end{equation}
The variation in the Berry connection is simply the term arising from its gauge dependence as seen before:
\begin{equation}
    \delta A_{\mu}(\mathbf{k}) \approx -\partial_{\mu}\phi(\mathbf{k})
\end{equation}
And consequently, the variation of the Berry curvature is zero, 
\begin{equation}
    \delta \Omega(\mathbf{k}) = \partial_{x}\partial_{y}\phi(\mathbf{k}) - \partial_{y}\partial_{x}\phi(\mathbf{k}) = 0
\end{equation}
which is nothing else than the statement of gauge invariance of the Berry curvature. Thus, under consecutive adiabatic deformations of the system, the Chern number will remain invariant as long as there are no closings of the gap. If there is a closing of the gap, the perturbed state will no longer be adiabatically connected to the original state, and the Chern number will change. Note that while the individual Chern numbers are not conserved after a closing of the bands, the total Chern number of the crossing bands is conserved~\cite{avron1983homotopy}.

We have seen that a physical observable, the Hall conductivity, is expressed in terms of the topological invariant of the system, which is the Chern number. We saw previously that while~\eqref{eq:berry_curvature_stokes} always holds, Stokes' theorem does not necessarily always apply. Suppose that we can define a smooth gauge for the states $\ket{u_{n\mathbf{k}}}$. Then, according to Stokes' theorem,
\begin{equation}
    C_n = \frac{1}{2\pi}\int_{\text{BZ}}\Omega_n(\mathbf{k})d^2\mathbf{k} = \oint_{\partial(\text{BZ}) = \emptyset}\bm{A}_n(\mathbf{k})\cdot d\bm{k} = 0
\end{equation}
Since the Chern number is obtained integrating the Berry curvature over the whole BZ, the boundary of the manifold is empty, $\partial(\text{BZ})=\partial(\mathbb{T}^2)=\emptyset$, and consequently the application of Stokes' theorem yields a zero Chern number. Thus, for Chern insulators we conclude that it must not be possible to define a smooth gauge over the whole BZ, as otherwise it would result in $C=0$. Hence, the Chern number represents a topological obstruction to defining a globally smooth gauge for the states~\cite{bernevig2013topological}. 

At this point, we may use our knowledge of the topological obstruction to operate the expression of the Chern number. Assume that we know the positions of the points in the BZ where the initial gauge $\ket{u^\text{I}_{n\mathbf{k}}}$ is undefined, $k_s, s\in\{1,\ldots,N\}$. We may define a region, $R^\epsilon_s=\{\bm{k}\in \text{BZ} \text{ s.t. }|\bm{k}-\bm{k}_s|<\epsilon\}$ where the gauge $\ket{u^{\text{II}}_{n\mathbf{k}}}$ is smooth but undefined outside~\cite{hatsugai1993chern, bernevig2013topological}. Then, both gauges are related at the boundary of the region $R_s^{\epsilon}$ by
\begin{equation}
    \ket{u^\text{I}_{n\mathbf{k}}} = e^{i\beta(\mathbf{k})}\ket{u^{\text{II}}_{n\mathbf{k}}}
\end{equation}
Since each gauge its smooth in its respective region, we can apply Stokes' theorem to each region separately, and sum the results to obtain the Chern number. The Chern number is then given by
\begin{align}
    \label{eq:chern_obstruction}
    C_n =\frac{1}{2\pi}\left[\int_{\partial(\mathbb{T}^2\setminus R_s^{\varepsilon})}\bm{A}^{\text{I}}_n(\mathbf{k})\cdot d\bm{k} + \int_{\partial(R_s^{\epsilon})}\bm{A}^{\text{II}}_n(\mathbf{k})\cdot d\bm{k}\right]
\end{align}
Now using that both boundaries are the same, $\partial(\mathbb{T}^2\setminus R_s^{\varepsilon}) = -\partial(R_s^{\epsilon})$, the above expression is written as
\begin{equation}
    C_n = \frac{1}{2\pi}\int_{\partial(R_s^{\epsilon})}\left[\bm{A}^{\text{II}}_n(\mathbf{k}) - \bm{A}^{\text{I}}_n(\mathbf{k})\right]\cdot d\bm{k} = \frac{1}{2\pi}\int_{\partial(R_s^{\epsilon})}\nabla\beta(\bm{k})\cdot d\bm{k}
\end{equation}
which again corresponds to the winding number of the gauge transformation $\beta(\bm{k})$ similarly to~\eqref{eq:winding_chern}, but around the boundary of the region $R_s^{\epsilon}$. We can see this explicitly: consider the boundary $\partial(R_s^{\epsilon})=\{\mathbf{k}_s + \varepsilon e^{i\theta} \text{ s.t. }\theta\in[0, 2\pi)\}$. Then,
\begin{equation}
    C_n = \frac{1}{2\pi}\int_{0}^{2\pi}\partial_{\theta}\beta(\mathbf{k}_s + \varepsilon e^{i\theta})d\theta = \frac{1}{2\pi}\left[\beta(\mathbf{k}_s + \varepsilon e^{i2\pi}) - \beta(\mathbf{k}_s + \varepsilon e^{i0})\right] = m
\end{equation}
Since the states $\ket{u^\text{I}_{n\mathbf{k}}}$, $\ket{u^\text{II}_{n\mathbf{k}}}$ must be monovalued, the gauge transformation necessarily verifies $\beta(\mathbf{k}_s + \varepsilon e^{i2\pi}) - \beta(\mathbf{k}_s + \varepsilon e^{i0}) = 2\pi m$, $m$ being the winding number. It is possible to use this procedure to determine the Chern number of analytically solvable models, such as the Haldane model or general two-band models~\cite{bernevig2013topological}. However, the Chern number is most generally computed numerically using the Berry flux procedure detailed in~\eqref{eq:chern_number_discrete}.

\subsection{Charge pumping and Wannier representation}\label{sec:charge_pumping}

We conclude the discussion of the Chern insulator addressing an aspect that was not yet mentioned: the presence of a charge pumping in the system. Beyond the interpretation of the Chern number as a topological invariant and a topological obstruction to a smooth gauge, it also possesses a physical meaning. It is related to the charge pumped through the system when the Hamiltonian is adiabatically driven through the BZ\@. This was originally shown by Laughlin's argument of quantized charge transport in the IQHE~\cite{laughlin1981quantized}, and subsequently shown in general by Thouless~\cite{thouless1983quantization, thouless1982quantized, niu1994quantised}.

\begin{figure}[h]
    \centering
    \includegraphics[width=1\textwidth]{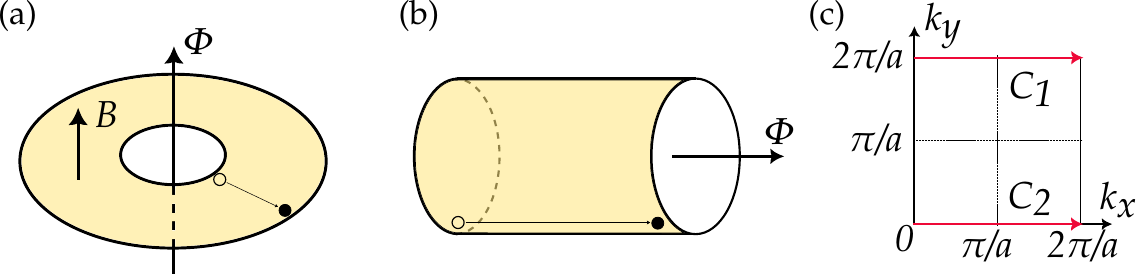}
    \caption[Charge pumping in the IQHE and Chern insulators]{(a) Laughlin's experiment of quantized charge pumping in a Corbino geometry. A magnetic field $B$ is applied on the disc, producing a IQH state, while a pure gauge flux $\Phi$ is threaded through the disc, inducing charge transport across the disc.\ (b) Hall cylinder, corresponding either to a Chern insulator or a IQH state (equivalent to the Corbino geometry). Adiabatically varying the threaded flux $\Phi$ results in charge transport across the cylinder.\ (c) Paths $C_1,\ C_2$ in the BZ corresponding to the calculation of the polarization at the beginning and end of a flux cycle.}\label{fig:charge_pumping}
\end{figure}

In the case of the IQHE and Laughlin's thought experiment, depicted in Figs.~\ref{fig:charge_pumping}(a, b), the quantized charge pumping can be formally shown solving the Schrödinger equation for the 2D electron gas. Consider periodic boundary conditions in the $y$ direction, with an applied vector potential $\bm{A}=(Bx + \Phi/L)\hat{\bm{y}}$ such that it induces Landau levels and threads a flux through the cylinder. Then, after the flux changes adiabatically from $\Phi(0)=0$ to $\Phi(T)=\Phi_0=h/e$, $\Phi_0$ being the quantum of flux, the center of the wavefunctions shifts rigidly, resulting in a charge pump $\Delta Q = ne$~\cite{laughlin1981quantized, laughlin1999nobel, tong2016lectures}. Note that the fluxes must always be changed adiabatically in order to track the same state at every flux value, otherwise charge quantization is not guaranteed. In general, flux insertions can be understood directly from the minimal coupling,
\begin{equation}
    \hbar\bm{k} \rightarrow \hbar\bm{k} - e\bm{A}
\end{equation}
which in practice, for Bloch Hamiltonians implies that $H(\bm{k})\rightarrow H(\bm{k} - e\bm{A}/\hbar)$~\cite{landau2013statistical}, namely inserting a flux through the torus is equivalent to a translation of the $\bm{k}$ vector. For this reason, in what follows we will consider directly $H(\bm{k})$ instead of $H(\Phi)$, knowing that $\bm{k}$ can be adiabatically modified from the external application of a magnetic flux.

The concept of charge pumping also applies to Chern insulators, although in this case the simplest way to introduce it is through the modern theory of polarization in solids. It was originally noted by Resta and Vanderbilt~\cite{resta2007theory, resta1992theory, kingsmith1993theory, resta1994macroscopic} that the electronic polarization in a crystalline solid can be formulated in terms of Berry phases, and more specifically in terms of Wannier functions $w_{n}(\bm{r} - \bm{R})=\braket{\bm{r}|n\bm{R}}$~\cite{wannier1937structure}. Consider a $d$-dimensional system, with Bloch functions $\psi_{n\bm{k}}(\bm{r}) = \braket{\bm{r}|n\bm{k}}$. The Wannier states are defined as the Fourier transform of the Bloch states,
\begin{equation}
    \ket{n\bm{R}} = \frac{V}{(2\pi)^d}\int_{\text{BZ}} d^d\bm{k} e^{-i\bm{k}\cdot\bm{R}}\ket{n\bm{k}}
\end{equation} 
where $V$ is the volume of the unit cell. Like Bloch states, these states form an orthogonal basis of the Hilbert space, $\braket{n\bm{R}|m\bm{R}'} = \delta_{nm}\delta_{\bm{R}\bm{R}'}$. The main difference lies in that Wannier states are localized in real space within cell $\bm{R}$, while Bloch states are extended. This difference makes them convenient to define the polarization of a solid. For each Wannier state, we may define its Wannier charge center (WCC) as
\begin{equation}
    \bar{\bm{r}}_n = \braket{n\bm{0}|\bm{r}|n\bm{0}}
\end{equation}
The total electronic polarization $\bm{P}$ is then given by the sum of the charge centers of all the Wannier states~\cite{vanderbilt1993electric},
\begin{equation}
    \label{eq:total_polarization}
    \bm{P} = -e\sum_n^N \bar{\bm{r}}_n
\end{equation}
where $N$ denotes the active subspace of bands we are considering, which may be the filled bands or an isolated subset of this. A central result shown by Blount~\cite{blount1962formalisms} is that the WCC can be written in terms of the Berry connection as
\begin{equation}
    \label{eq:blount_formula}
    \bar{\bm{r}}_n = i\frac{V}{(2\pi)^d}\int_{\text{BZ}}d^d\bm{k}\braket{u_{n\bm{k}}|\nabla_{\bm{k}}u_{n\bm{k}}}
\end{equation}
which is the expression originally used in~\cite{vanderbilt1993electric, kingsmith1993theory} to be able to write the electronic polarization~\eqref{eq:total_polarization} in terms of the WCCs. Upon $U(1)$ gauge transformations, the WCCs and the polarization are gauge independent modulo a lattice vector~\cite{marzari1997maximally}. However, at this point we may consider a different set of gauge transformations. Given the isolated set of $N$ bands, we may also consider $U(N)$ gauge transformations, i.e.\ unitary transformations that rotate the space of $N$ bands, $\ket{n\bm{k}}\longrightarrow\sum_{m}U_{nm}(\bm{k})\ket{m\bm{k}}$. For a given band $n$, these gauge transformations do not preseve energy nor the WCC, however the total polarization is gauge invariant, again modulo a lattice vector. The polarization being defined only $\bm{P}\rightarrow\bm{P}+e\bm{R}$ is consistent with the definition of the Wannier states, which are all connected to the state at the origin via the translation operator $T_{\bm{R}}\ket{n\bm{0}}=\ket{n\bm{R}}$. These gauge transformations appear naturally when considering the non-abelian Berry connection~\cite{wilczek1984appearance}, which we will introduce in the next section. In the context of Wannier functions, the gauge freedom present in the definition of the states can be exploited, for instance, to obtain maximally localized Wannier functions (MLWF). These are obtained minimizing the total spread of the manifold under consideration,
\begin{equation}
    \label{eq:wannier_spread}
    \Omega = \sum_{n=1}^N\left[\braket{n\bm{0}|\bm{r}^2|n\bm{0}} - \braket{n\bm{0}|\bm{r}|n\bm{0}}^2\right].
\end{equation}
MLWFs are famously used to derive tight-binding models for a subset of bands coming from a DFT calculation~\cite{pizzi2020wannier90, marzari2012maximally}. In the context of topological insulators, as we will see MLWFs appear naturally in the description of the topological nature of the system, specifically in the charge pumping.

As we see in Blount's formula~\eqref{eq:blount_formula}, the WCC can be written as an integral of the Berry connection over the whole BZ\@. If the system under consideration is 1D, then WCCs correspond precisely to Berry phases o alternatively Zak phases,
\begin{equation}
    \bar{r}_n = i\frac{a}{2\pi}\int_{-\pi/a}^{\pi/a}\braket{u_{n\bm{k}}|\partial_{k}u_{n{k}}}dk
\end{equation}
where $a$ is the length of the unit cell. One interest property of 1D MLWFs is that the set of WFs that minimizes the spread $\Omega$ is equivalent to the set of eigenfunctions of the projected position operator $PxP$ (here, $r=x$)~\cite{kivelson1982wannier}. Consider a Wannier state $\ket{n0}$ such that $PxP\ket{n0}=\bar{x}_n\ket{n0}$. Such a state can be shown to minimize the spread of the Wannier state~\eqref{eq:wannier_spread} and consequently is maximally localized. Namely, if $\Omega=\Omega_I + \Omega_D$, where $I/D$ denotes gauge invariant or dependent respectively, then these states verify $\Omega_D=0$~\cite{marzari1997maximally, soluyanov2012topological}. Likewise, if $\Omega_{\min}=\Omega_I$, then necessarily $\braket{mR|x|n0}=\bar{x}_n\delta_{mn}\delta_{R0}$, meaning that the WFs are eigenstates of $PxP$. This property will come up in the next section when we show the practical method for computing the topological invariants.

The appearance of Berry phases when considering 1D systems is highly suggestive of the potential relevance of WCCs in the description of topological phases. At this point we introduce a new kind of Wannier states: hybrid Wannier states~\cite{sgiarovello001electron, soluyanov2011wannier}, which correspond to Bloch states (extended) along some coordinates, and Wannier states (localized) along the others. In 2D, these states are defined as
\begin{equation}\label{eq:hybrid_wannier_state}
    \ket{nR_xk_y} = \frac{a_x}{2\pi}\int_{-\pi/a_x}^{\pi/a_x}dk_x e^{-ik_xR_x}\ket{nk_xk_y}
\end{equation}
where $R_x=ma_x$, $m\in\mathbb{N}$ and $a_x$ being the unit cell length along the $x$ direction. From this, we may define the hybrid Wannier charge centers (HWCCs) as
\begin{align}
    \label{eq:hybrid_wannier_center}
    \nonumber\bar{x}_n(k_y) = \braket{n0k_y|x|n0k_y} &= i\frac{a_x}{2\pi}\int_{-\pi/a_x}^{\pi/a_x}dk_x\braket{nk_xk_y|\partial_{x}nk_xk_y} \\
    &= \frac{a_x}{2\pi}\int_{-\pi/a_x}^{\pi/a_x}dk_x A_{n,x}(k_x,k_y).
\end{align}
Again, the HWCC can be written as an integral of the Berry connection. Note that since the integral is one-dimensional, it actually corresponds to a loop over the BZ, and therefore it describes a Berry phase. As before, one can define HWFs that are maximally localized in the $x$ direction from the eigenstates of the $PxP$ operator; assuming that a hybrid Wannier state verifies this one can show that it minimizes the spread in the $x$ direction.
This definition of HWFs can be extended to higher dimensions, e.g. 3D insulators. Focusing on the 2D case, note that we can interpret the system as a 1D system in the $x$ direction, coupled to an external parameter $k_y$, i.e. $H(k_y)$, which as mentioned before could be connected for instance to a flux insertion. We can define the partial polarization $P_x$,
\begin{equation}
    P_x(k_y) = -e\sum_n^N\bar{x}_n(k_y)
\end{equation}
Then, assuming that the gauge is smooth on $k_y$, we can track the evolution of the HWCCs $\bar{x}_n(k_y)$ and the partial polarization $P_x$ as a function of $k_y$. The evolution of the polarization along an adiabatic, complete cycle in the BZ, from $k_y=-\pi/a_y$ to $k_y=\pi/a_y$ is
\begin{align}
    \label{eq:polarization_cycle}
    &P_x(k_y=\pi/a_y) - P_x(k_y=-\pi/a_y) = -e\sum_n^N\left[\bar{x}_n(k_y=\pi/a_y) - \bar{x}_n(k_y=-\pi/a_y)\right] 
\end{align}
Now using Eq.~\eqref{eq:hybrid_wannier_center} we can write the change in the partial polarization $\Delta P_x$ as
\begin{align}
    \nonumber &P_x(k_y=\pi/a_y) - P_x(k_y=-\pi/a_y) = \\ 
    \nonumber &-ei\frac{a_x}{2\pi}\sum_n^N\left[\int^{\pi/a_x}_{-\pi/a_x}dk_xA_{n,x}(k_x,k_y=\pi/a_y) - \int^{\pi/a_x}_{-\pi/a_x}dk_xA_{n,x}(k_x,k_y=-\pi/a_y)\right] \\ 
    \label{eq:polarization_chern}& = -ea_x\sum^N_{n}\int_{\text{BZ}}\frac{d^2\bm{k}}{2\pi}\Omega_n(\bm{k}) = -ea_x C
\end{align}
where in the last line we have used Stokes' theorem to write the line integrals as the integral of the Berry curvature on the cylinder, assuming that the gauge is continuous in $k_y$ in the interval $[-\pi/a_y,\pi/a_y]$. However, since the ends are connected, we recover the integral over the BZ and consequently the Chern number, as shown in Fig.~\ref{fig:charge_pumping}(c). Note here the distinction between the cylinder and the torus: while in the cylinder we may assume a smooth gauge, this does not translate into the loop (or the torus). Namely, as in~\eqref{eq:chern_obstruction}, a finite Chern number implies that a smooth gauge for $k_y$ does not exist in the loop as otherwise the line integrals would cancel directly, corresponding to the same boundary. Equivalently, the gauge discontinuity is apparent in that the WCCs do not return to the original values after evolving across the BZ\@. While in the $x$ direction we may assume a smooth gauge (e.g.\ if using the WCCs $\bar{x}_n$ coming from the eigenstates of $PxP$), a finite Chern number implies an obstruction to a smooth gauge in $y$, and consequently an obstruction to defining 2D exponentially localized Wannier states, since the integral would be ill-defined~\cite{thouless1984wannier, brouder2007exponential}.

\begin{figure}[h]
    \centering
    \includegraphics[width=0.9\columnwidth]{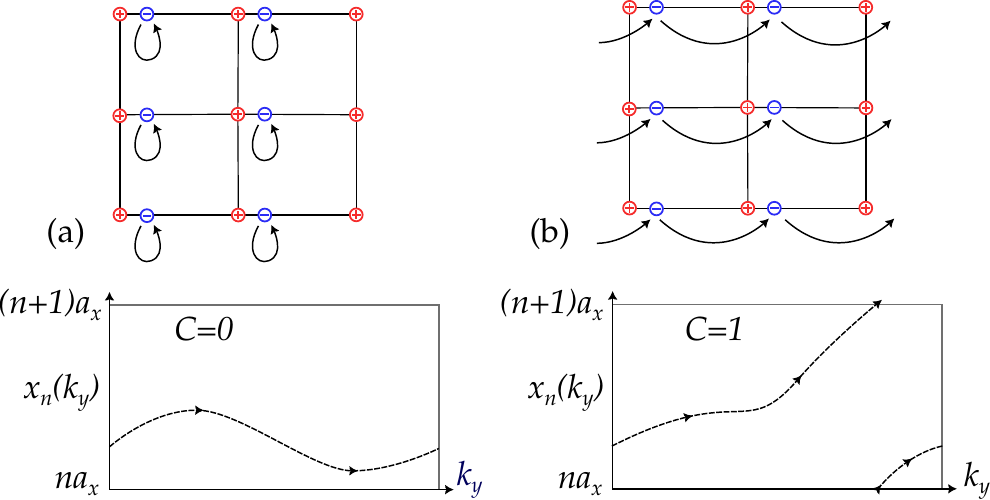}
    \caption[Depiction of the charge pumping in Chern insulators across unit cells as a function of momentum]{Schematic of the charge pumping process in Chern insulators.\ (a) In a trivial insulator the HWCCs move and stay within the unit cell as a function of $k_y$, which corresponds to $C=0$.\ (b) For a Chern insulator, here with $C=1$, the HWCCs move from one unit cell to the next after a complete pumping cycle. In the top diagrams, the positive charges denote the ions of the lattice, while the negative charges are the electrons. The bottom diagrams show the evolution of the HWCCs $x_n(k_y)$.}\label{fig:charge_pumping_chern}
\end{figure}

Additionally, beyond the previous interpretations of the Chern number, we also see from Eq.~\eqref{eq:polarization_chern} that the Chern number gives the change in the polarization $\Delta P_x$, i.e.\ the number of electrons that is pumped through the torus from one unit cell to the next, in one adiabatic cycle of the Hamiltonian, from $H(k_y=-\pi/a_y)$ to $H(k_y=\pi/a_y)$. 
Expression~\eqref{eq:polarization_cycle} already provides a recipe to extract the Chern number, where tracking the evolution of the HWCCs one can infer directly the value of the invariant, counting the number of times they loop around the circle, as depicted in Fig.~\ref{fig:charge_pumping_chern}.

In summary, the Chern number is a topological invariant that represents a topological obstruction to defining a smooth gauge for the states, or equivalently is an obstruction to defining localized Wannier states in both directions. From a physical perspective, the Chern number represents the number of electrons that is pumped from one unit cell to the next as the Hamiltonian is adiabatically driven along the BZ, and is directly related to an observable, the Hall conductivity, being responsible for giving a quantized response to an applied DC electric field.

\subsection{$\mathbb{Z}_2$ topological insulators}

All the knowledge obtained from Chern insulators sets the stage to describe time-reversal topological insulators. As we will see, the same interpretations we derived in the case of the Chern insulator, such as the topology representing an obstruction to a smooth gauge or the presence of a charge pumping in the system (although of a differen kind) will also apply to the $\mathbb{Z}_2$ topological insulator.

$\mathbb{Z}_2$ topological insulators, or equivalently time-reversal topological insulators, can be regarded as the most common family of topological insulators, in the sense that they are naturally appearing since most materials in nature respect time-reversal symmetry. As their name implies, these systems are invariant under time-reversal, $[\mathcal{T}, H]=0$. For the Berry curvature, this means that $\Omega(\bm{k})=-\Omega(-\bm{k})$~\cite{bernevig2013topological, grushin2021introduction}, and consequently the Chern number for a time-reversal topological insulator is always zero, $C=0$. The question that arises then is whether it is possible to define a topological invariant for these systems, namely if there exist different families of $\mathcal{T}$-invariant materials that cannot be adiabatically connected. As stated in the introduction, this question has an affirmative answer, which was first given by Kane and Mele~\cite{kane2005quantum}. In their case, they define a spin Chern number which is only valid if $[H,S_z]=0$. Since TIs in nature typically appear precisely because of spin-orbit coupling, which mixes the spin projection, it should be possible to define some topological invariant that generalizes the spin Chern number. This will be the $\mathbb{Z}_2$ invariant\footnote{The notation $\mathbb{Z}_2$ signifies $\mathbb{Z}_2=\mathbb{Z}/2\mathbb{Z}$, which denotes the quotient group $(\mathbb{Z}/2\mathbb{Z}, +)$, $\mathbb{Z}/2\mathbb{Z}=\{\bar{0}, \bar{1}\}$, where $\bar{n}=\{m\in\mathbb{N} \text{ s.t. }m\equiv n \mod 2\}$ are the corresponding equivalence classes. Namely, the $\mathbb{Z}_2$ is an index that can only take two values, 0 (trivial) or 1 (topological).}. Unlike Chern insulators, where the topological invariant is directly related to an observable quantity, TR-TIs do not have an observable associated to the $\mathbb{Z}_2$ invariant. However, as we will see, the value of the index can be associated to a non-trivial charge flow, or more simply to the presence of topological edge states in the sense shown in Fig.~\ref{fig:qshe}. The theory underlying $\mathbb{Z}_2$ TIs is arguably more abstract than that of CIs; in what follows we will review the different developments that led the definition of the $\mathbb{Z}_2$ index, concluding again with a practical recipe to extract the invariant.

First, we give a quick overview of the time-reversal operator $\mathcal{T}$; for a detailed review see~\cite{bernevig2013topological, asboth2016short}. From its expected action on the position and momentum operators, $\mathcal{T}\bm{x}\mathcal{T}^{-1}=\bm{x}$ and $\mathcal{T}\bm{p}\mathcal{T}^{-1}=-\bm{p}$, the time-reversal operator must correspond to $\mathcal{T}=U{K}$, where ${K}$ is the complex conjugation operator and $U$ is some unitary operator. Because of the complex conjugation, $\mathcal{T}$ is an anti-unitary operator. If one additionally requests that $\mathcal{T}\bm{S}\mathcal{T}^{-1}=-\bm{S}$, then for spin-1/2 systems the time-reversal operator is given by $\mathcal{T}=i\sigma_yK$, where $\sigma_y$ is the second Pauli matrix. It can be seen then that $\mathcal{T}^2=-1$, which is central in proving Kramers' theorem. This theorem states that in a fermionic time-reversal invariant system, for each state there is a different time-reversal state with the same energy. Using that $\mathcal{T}^2=-1$, one can prove that $\braket{\psi|\mathcal{T}\psi}=0$, namely that one state $\ket{\psi}$ and its time-reversal companion $\mathcal{T}\ket{\psi}$ (i.e.\ a Kramers' pair) are different states. Since $[H,\mathcal{T}]=0$, both states must have the same energy.

For Bloch Hamiltonians, Kramers' theorem has the following implication. From time-reversal symmetry, it follows that the Bloch Hamiltonian must transform as $\mathcal{T}H(\bm{k})\mathcal{T}^{-1}=H(-\bm{k})$. Then, one has
\begin{equation}
    H(-\bm{k})\mathcal{T}\ket{u_{\bm{k}}} = \mathcal{T}H(\bm{k})\ket{u_{\bm{k}}}=E(\bm{k})\mathcal{T}\ket{u_{\bm{k}}} \equiv E(-\bm{k})\ket{u_{-\bm{k}}}
\end{equation}
Thus, in a fermionic translational and time-reversal symmetric system, $E(-\bm{k})=E(\bm{k})$. If we now restrict $\bm{k}$ to time-reversal invariant momenta (TRIM), which are $\bm{k}$-points such that $\bm{k}=-\bm{k}+\bm{G}$, for some $\bm{G}\in\text{Reciprocal lattice}$, we have $\mathcal{T}H(\bm{k})\mathcal{T}^{-1}=H(\bm{k})$. Consequently, at each TRIM, because of Kramers' degeneracy, every eigenstate of the Bloch Hamiltonian $H(\bm{k_{\text{TRIM}}})$ is at least two-fold degenerate. As we will see, the degeneracy at these special points will be central in establishing the value of the invariant.

Beyond the spin Chern number, Kane and Mele also introduced the first formula for the $\mathbb{Z}_2$ invariant~\cite{kane2005z}. The first thing they noted is that according to the theory of fiber bundles, manifolds with the time-reversal symmetry constraint follow a $\mathbb{Z}\times\mathbb{Z}_2$ classification, $\mathbb{Z}$ being the dimension of the manifold and $\mathbb{Z}_2$ the desired index. Their definition of the $\mathbb{Z}_2$ invariant is given in terms of a function $P(\bm{k})$, defined by
\begin{equation}
    P(\bm{k}) = \text{Pf}\left[\braket{u_{i\bm{k}}|\mathcal{T}|{u_{j\bm{k}}}}\right]
\end{equation}
where $\text{Pf}$ denotes the Pfaffian and $i,j\in\{1,\ldots,N\}$. Since $m_{ij}=\braket{u_{i\bm{k}}|\mathcal{T}|{u_{j\bm{k}}}}$ is an antisymmetric matrix, it holds that $\text{Pf}(m)^2 = \det(m)$, i.e.\ the Pfaffian is given by a polynomial of the matrix elements in the upper triangular portion.
The idea here is that the time-reversal operator distinguishes two different subspaces within the Hamiltonian. Considering for simplicity $N=2$ occupied bands, $\ket{u_{i=1,2,\bm{k}}}$, one distinguishes the even subspace, in which $\{\mathcal{T}\ket{u_{i\bm{k}}}\}$ spans the same space as $\{\ket{u_{i\bm{k}}}\}$ (up to a $U(N=2)$ transformation). The alternative is the odd subspace, in which the space spanned by $\{\mathcal{T}\ket{u_{i\bm{k}}}\}$ is orthogonal to that of $\{\ket{u_{i\bm{k}}}\}$. Consequently, at each $\bm{k}$, $P(\bm{k})$ measures whether the occupied bands correspond to the even or the odd subspaces. In the even subspace, $|P(\bm{k})|=1$, whereas in the odd one we have $P(\bm{k})=0$.

The claim made now is that the number of zeros of $P(\bm{k})$ in half BZ gives the $\mathbb{Z}_2$ index of the system. Each zero carries a vorticity; if there is one zero at $\bm{k}$ there is also a zero with opposite vorticity at $-\bm{k}$. If the number of zeros is even, perturbations to the Hamiltonian can annihilate the zeros. However, if they are odd, perturbations again can annhilate the vortices except for the last one, which is forbidden from annihilating with its time-reversal partner at $-\bm{k}$, since this would have to occur at a TRIM, which always belong to the even subspace~\cite{bernevig2013topological, fruchart2013introduction}. The index in this case is defined as
\begin{equation}\label{eq:first_z2_expression}
    \Delta = \frac{1}{2i\pi}\oint_{\partial\tau}d\bm{k}\cdot \nabla_{\bm{k}}\log[P(\bm{k})]\mod2
\end{equation}
which is the winding number of the phase of $P(\bm{k})$ evaluated on a contour along half BZ\@.
\begin{figure}[h]
    \centering
    \includegraphics[width=0.6\textwidth]{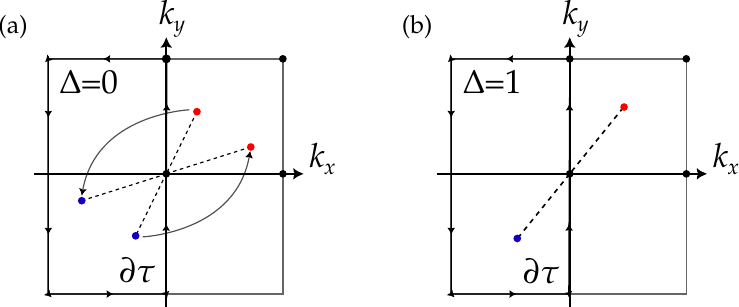}
    \caption[Distinction between trivial and topological $\mathbb{Z}_2$ phase according to the zeros of the Pfaffian]{(a) In a trivial phase (atomic limit), the number of zeros in half BZ is even. By means of perturbations in the Hamiltonian, zeros of positive vorticity (red dots) can be brought together with the zeros of negative vorticity (blue dots), annihilating.\ (b) In the odd phase (topological), there is at least one zero in $P(\bm{k})$ that cannot be annihilated via perturbations, since it would necessarily take place at a TRIM (black dots), which belongs to the even subspace. Adapted from~\cite{fruchart2013introduction}.}
\end{figure}

An alternative approach was proposed by Fu and Kane~\cite{fu2006time}, following the theory of charge pumping and polarization for Chern insulators. This was already suggested in~\cite{kane2005z}, but the idea of a $\mathcal{T}$-polarization was not formally developed. For a Chern insulator, we have seen that the change in polarization corresponds to a non-trivial flow of the WCCs across the BZ\@. For a $\mathbb{Z}_2$ TI, the polarization state at the beginning and end of a pumping cycle must be exactly the same, since $C=0$. It is possible, however, that a non-trivial flow of the charge exists within the material. To illustrate this, it is easier to consider a 2D TI such that $[H,S_z]=0$. Since $S_z$ is a good quantum number, it allows to resolve the degeneracy of the states (e.g.\ if the system also has inversion symmetry). Then, we may compute the Chern number for each subspace separately, $C_{\uparrow}$ and $C_{\downarrow}$. Each sector is the time-reversal conjugate of the other, and consequently each one may carry finite $C$, verifying $C_{\uparrow}=-C_{\downarrow}$ (since the total Chern must be zero because of TRI). 

\begin{figure}[h]
    \centering
    \includegraphics[width=0.7\columnwidth]{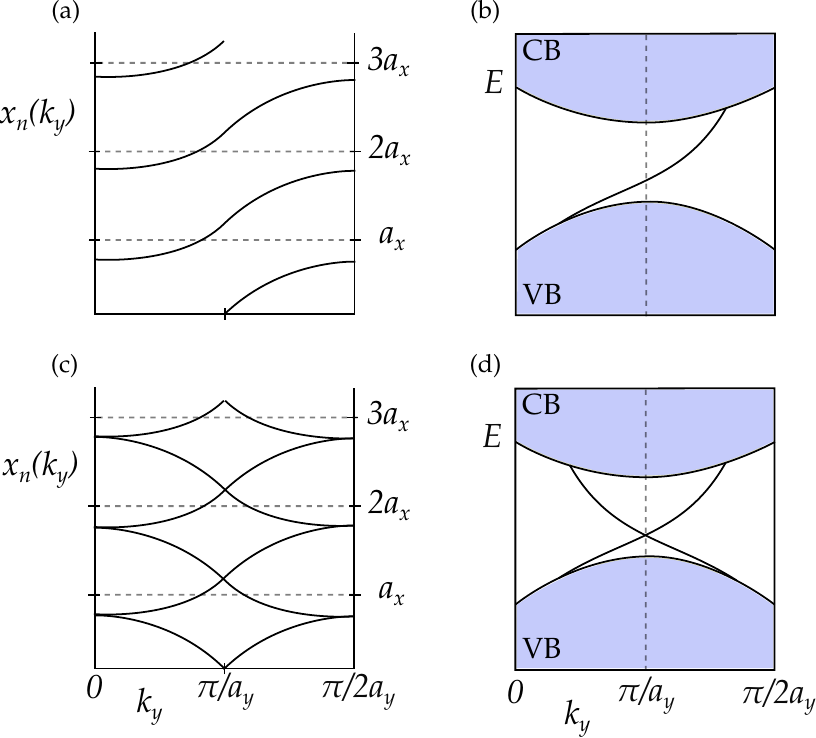}
    \caption[Comparison of the WCC flows and edge states for a CI and a $\mathbb{Z}_2$ TI]{(a) HWCC flow and (b) edge state dispersion for a Chern insulator with $C=1$.\ (c) HWCC flow and (d) edge state dispersion for a $\mathbb{Z}_2$ topological insulator. The TR-TI can be regarded as the union of the $C_{\uparrow} $ sector and the $C_{\downarrow}$ sector. Adapted from~\cite{vanderbilt2018berry}.}\label{fig:spin_chern_flow_edge_states}
\end{figure}

Assume now that $(C_{\uparrow},C_{\downarrow})=(1,-1)$. If we consider solely the WCC evolution of the spin-up sector, we would observe the characteristic charge pumping of a Chern insulator, with the WCCs moving from one unit cell to the next. If we now consider both sectors, we know that one state and its time-reversal companion must have the same HWCC, since $\braket{(n,\uparrow) 0k_y|x|(n,\uparrow) 0k_y}=\braket{(n,\downarrow) 0,-k_y|x|(n,\downarrow) 0,-k_y}$, using that $[x,\mathcal{T}]=0$. Consequently, the WCC flow for the spin-down sector amounts to reflecting the spin-up flow. This situation is illustrated in Fig.~\ref{fig:spin_chern_flow_edge_states}, for the $C_{\uparrow}$ sector and for both sectors. What the observe then is that there are charges that flow in both directions, which is why the polarization state at both $k_y=0, 2\pi/a$ is the same. Note however, that if we track the evolution of the WCCs starting from $k_y=0$, we see that at $k_y=\pi/a$ the WCCs have exchanged partners. Namely, at $k_y=0$ we have two degenerate hybrid Wannier states. Then, as $k_y$ evolves, one of these degenerate states becomes degenerate not with the previous partner, but with a different state. This is the key observation that allows to define the $\mathbb{Z}_2$ invariant.

\begin{figure}[h]
    \centering
    \includegraphics[width=0.8\textwidth]{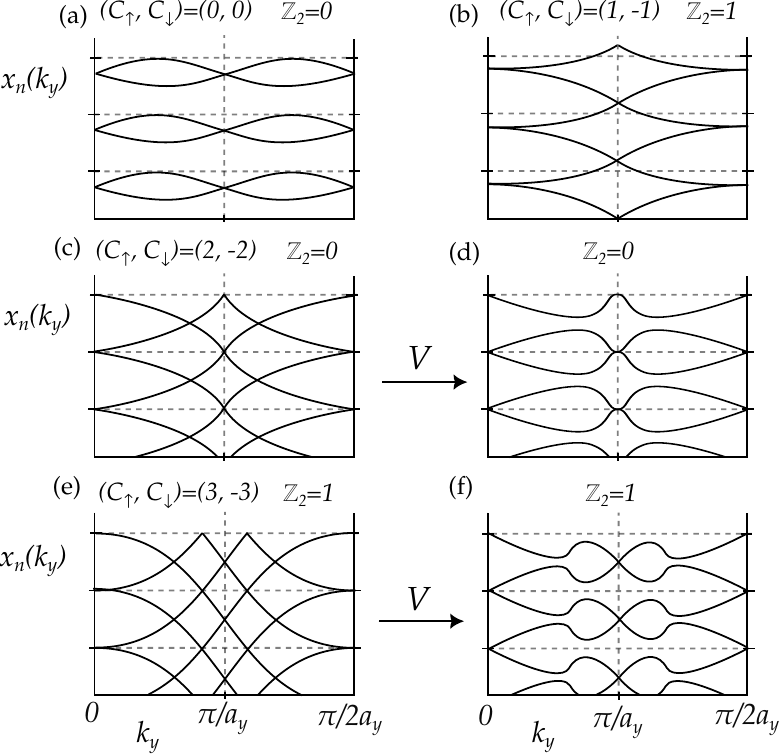}
    \caption[Examples of HWCC evolutions for different $(C_{\uparrow}, C_{\downarrow})$]{Examples of HWCC evolutions for different pairs $(C_{\uparrow}, C_{\downarrow})$ if the 2D TI conserves $S_z$, or equivalently in terms of the $\mathbb{Z}_2$ invariant. In (c, e), the introduction of a perturbation $V$ (e.g.\ a spin mixing term such as spin-orbit coupling) opens trivial gaps in the HWCC evolution, leading to an odd or even $\mathbb{Z}_2$ index as shown in (d, f). Adapted from~\cite{vanderbilt2018berry}.}\label{fig:wcc_flows_cs}
\end{figure}

This exchange of partners is then quantified defining the time-reversal polarization, which is simply the difference between the polarization for one sector and the other, $P_{\uparrow}-P_{\downarrow}$ (still sticking to the $S_z$ picture). From Fig.~\ref{fig:spin_chern_flow_edge_states}, we see that this explicitly measures the exchange of WCCs, since some flow up while the other flow down. Thus, for $k_y=0$ we would have one time-reversal polarization state, and for $k_y=\pi/a$ where the exchange takes place, a different one. For comparison, see in Fig.~\ref{fig:wcc_flows_cs}(a) the flow for a trivial case, where the time-reversal polarization is clearly the same at both $k_y=0$ and $k_y=\pi/a$.

And what happens if the WCCs exchange by more than one partner? Say for instance they exchange with the $n$-th consecutive partner. This situation would be described by $(C_{\uparrow},C_{\downarrow})=(n,-n)$. Specific examples for different values of $n$ are shown in Fig.~\ref{fig:wcc_flows_cs}. Apart from the degeneracies at the TRIMs, we also get accidental crossings at arbitrary points within the BZ\@. These crossings are not protected by time-reversal symmetry since they do not take place between time-reversal companions (and are outside TRIMs), and consequently they can be gapped by perturbations as they hybridize. Only if there is an even number of crossings in half BZ (including TRIMs) the flow will be non-trivial as it will be impossible to eliminate all crossings. Which ultimately implies that the topological invariant can only take two values: for the spin Chern it would be $C_s=(C_{\uparrow}-C_{\downarrow})\mod 2$, and likewise the $\mathbb{Z}_2$ invariant is defined modulo 2.

As an aside before formally deriving the $\mathbb{Z}_2$ invariant within this approach, the discussion of the WCC flow is also convenient to justify the existence of the topological edge states for both CIs and TR-TIs, as illustrated in Fig.~\ref{fig:spin_chern_flow_edge_states}. Beyond the argument given in the introduction, where we reason that the gap must locally close in order to transition to the trivial state, Vanderbilt~\cite{vanderbilt2018berry, taherinejad2014wannier} argues that the bulk-boundary correspondence is a consequence of the non-trivial charge pumping and the conservation of charge. Considering for simplicity a CI, if we were to cut the torus along the $x$ direction, then from the WCC flow we see that for some $k_y$ values the WCCs must be localized at the edges. We know however, that at fluxes $k_y=0,2\pi/a$ the system remains the same, so in particular charge at the edges must be conserved. This implies that as $k_y$ evolves and a charge is pumped from one edge to the other, there must exist one state injecting an electron from the bulk into the initial edge, and also a different state must extract it from the final edge. The edge state corresponding to the initial edge must be up-crossing the Fermi energy to remove the charge (so that it becomes empty as we increase $k_y$), whereas the edge state corresponding to the final edge must be down-crossing the Fermi energy to inject the charge (to become occupied as we increase $k_y$), consequently closing the gap.

In what follows we show briefly how to formalize the above ideas to define the $\mathbb{Z}_2$ invariant in systems that do not preserve $S_z$. The main distinction with the $[H,S_z]=0$ case is that individual Chern numbers cannot be computed due to the band degeneracy, which is why one formulates a theory directly for the $\mathbb{Z}_2$ invariant. The above ideas transfer directly to this case, stemming in the end from the degeneracy of the states and the WCCs at the TRIMs: depending on their connectivity we distinguish the non-trivial flow. Following~\cite{fu2006time, bernevig2013topological}, assuming for simplicity that there are no degeneracies beyond those required by TR symmetry, we can label each Kramers' pair by a label $\alpha=1,\ldots,N/2$. Then the states of each Kramers' pairs are related in general by
\begin{equation}\label{eq:gauge_t}
    \left\{\begin{aligned}
        \ket{u^{\text{I}}_{\alpha,-\bm{k}}}&=-e^{i\chi_{\alpha,\bm{k}}}\mathcal{T}\ket{u^{\text{II}}_{\alpha,\bm{k}}} \\
        \ket{u^{\text{II}}_{\alpha,-\bm{k}}}&=\ \ \ e^{i\chi_{\alpha,\bm{k}}}\mathcal{T}\ket{u^{\text{I}}_{\alpha,\bm{k}}}
    \end{aligned}\right.
\end{equation}
where the labels I, II are used to denote the two different states that form each pair. We know from~\eqref{eq:hybrid_wannier_center} that the total polarization is given in terms of the Berry connections (now we take $a_x=1$ and $e=1$ for simplicity),
\begin{equation}
    P_x(k_y) = \frac{1}{2\pi}\int_{-\pi}^{\pi}dk_x A_x(\bm{k})
\end{equation}
where $A_x(\bm{k})=i\sum_{n}\braket{u_{n\bm{k}}|\partial_{x}u_{n\bm{k}}}$ is the total Berry connection, summed over the $N$ occupied states. We may now decompose the total Berry connection in terms of the sectors I and II,
\begin{equation}
    A_x(\bm{k}) = A^{\text{I}}_x(\bm{k}) + A^{\text{II}}_x(\bm{k})
\end{equation}
where 
\begin{equation}
    A^{S}_x(\bm{k}) = i\sum_{\alpha}\braket{u^S_{\alpha\bm{k}}|\partial_{x}u^S_{\alpha\bm{k}}}
\end{equation}
and $S=\text{I},\text{II}$. This allows to define the partial polarizations $P_x^S$,
\begin{equation}
    P_x^S(k_y) = \frac{1}{2\pi}\int_{-\pi}^{\pi}dk_x A^S_x(\bm{k})
\end{equation}
which trivially verify $P_x = P^{\text{I}}_x + P^{\text{II}}_x$. As mentioned before, the $\mathbb{Z}_2$ invariant reflects the exchange of partners during the evolution of the WCCs, which is quantified by the time-reversal polarization $P_x^{\theta}$,
\begin{equation}
    \label{eq:time_reversal_polarization}
    P_x^{\theta}(k_y) = P_x^{\text{I}}(k_y) - P_x^{\text{II}}(k_y)
\end{equation}
Just like the total or partial polarizations, the time-reversal polarization is not gauge invariant~\cite{fu2006time}. Following the previous arguments, the $\mathbb{Z}_2$ invariant is then given by the difference of time-reversal polarizations at $k_y=0$ and $k_y=\pi$,
\begin{equation}
    \Delta = P_x^{\theta}(k_y=\pi) - P_x^{\theta}(k_y=0) \mod 2
\end{equation}
For comparison, the Chern number was obtained as $P_x(k_y=2\pi) - P_x(k_y=0) = C$. For $\Delta=0$, the WCCs do not exchange partners and the system is in the trivial state, whereas for $\Delta=1$ the exchange takes place and the system is in the topological state. At this point, it still remains to obtain a practical equation to explicitly compute the value of the invariant. Still following Fu and Kane (see~\cite{fu2006time, bernevig2013topological} for the detailed derivation), the time-reversal polarization may be written in terms of the sewing matrix
\begin{equation}
    w_{mn}(\bm{k})=\braket{u_{m,-\bm{k}}|\mathcal{T}|u_{n,\bm{k}}}
\end{equation}
which is simply the matrix that describes the action of the time-reversal operator, $\mathcal{T}\ket{u_{m,\bm{k}}}=\sum_{n}B_{mn}(\bm{k})\ket{u_{n,-\bm{k}}}$. The basic idea is that one may relate the Berry connections $A^{\text{I}}$ and $A^{\text{II}}$ by means of the transformations in~\eqref{eq:gauge_t}. Then, the gauge chosen $\chi_{\alpha,\bm{k}}$ which will appear when transforming between states can be written in terms of the sewing matrix. In the end, the time-reversal polarization reads
\begin{align}
    P_x^{\theta}(k_y) &=\frac{1}{2\pi i}\left(\int_0^{\pi}dk_x\partial_x\log\det [w(\bm{k})] - 2\log\left[\frac{\text{Pf }[w(\pi,k_y)]}{\text{Pf }[w(0,k_y)]}\right]\right) \\
    &= \frac{1}{\pi i}\log\left[\frac{\sqrt{\det [w(\pi,k_y)]}}{\text{Pf }[w(\pi,k_y)]}\frac{\text{Pf } [w(0,k_y)]}{\sqrt{\det [w(0,k_y)]}}\right]
\end{align}
or equivalently
\begin{eqnarray}
    (-1)^{P_x^{\theta}(k_y)} = \frac{\sqrt{\det [w(\pi,k_y)]}}{\text{Pf }[w(\pi,k_y)]}\frac{\sqrt{\det [w(0,k_y)]}} {\text{Pf } [w(0,k_y)]}
\end{eqnarray}
where one uses that $\det [w] = \text{Pf }^2 [w]$, and the election of the branch of the determinant $\pm\sqrt{\det[w(\bm{k})]}$ is done ensuring a smooth gauge~\cite{fu2006time}. Note that due to the branch ambiguity of the log, the time-reversal polarization is only defined modulo 2, as we had reasoned before solely from the possible evolutions of the WCCs\@. From the evaluation of the change in the time-reversal polarization in half a pumping cycle we obtain the invariant as
\begin{eqnarray}\label{eq:second_z2_expression}
    (-1)^{\Delta} = \prod^{4}_{i=1}\frac{\sqrt{\det[w(\bm{\Gamma}_i)]}}{\text{Pf }[w(\bm{\Gamma}_i)]}
\end{eqnarray}
where $\bm{\Gamma}_i$ denote the four TRIM points that are present in the 2D BZ\@. Eq.~\eqref{eq:second_z2_expression} provides an improvement over the previous expression~\eqref{eq:first_z2_expression} in the sense that it only requires the evaluation of the sewing matrix at the TRIMs. In practice however, it requires setting a globally smooth gauge for the states to enforce a relation between all four special points~\cite{soluyanov2011wannier}, making it impractical numerically. There is a third form to determine the $\mathbb{Z}_2$ invariant, which was also introduced by Fu and Kane in the same paper~\cite{fu2006time}. Without formally deriving it (see again~\cite{fu2006time, bernevig2013topological} for details), the index can be expressed as~\cite{soluyanov2011wannier}
\begin{equation}\label{eq:z2_obstruction}
    \Delta = \frac{1}{2\pi}\left[\oint_{\partial\tau}\bm{A}(\bm{k})\cdot d\bm{k} - \int_{\tau}\Omega(\bm{k})d^2\bm{k}\right]\mod 2
\end{equation}
where $\tau$ denotes half BZ, and $\partial\tau$ is its boundary. This expression is the analogue of Eq.~\eqref{eq:chern_number_band} for the $\mathbb{Z}_2$ invariant, although for $N$ states instead of a single band, $\bm{A}(\bm{k})=i\sum_{n=1}^N\braket{u_{n\bm{k}}|\bm{\nabla}_{\bm{k}}|u_{n\bm{k}}}$ and $\Omega(\bm{k})=\bm{\nabla}_{\bm{k}}\times \bm{A}(\bm{k})$; to see it explicitly it suffices to write the total Chern number as
\begin{equation}
    C = \frac{1}{2\pi}\int_{\text{BZ}}\Omega(\bm{k})d^2\bm{k} = \frac{1}{2\pi}\left[\int_{\text{BZ}}\Omega(\bm{k})d^2\bm{k} - \oint_{\partial\text{BZ}}\bm{A}(\bm{k})\cdot d\bm{k}\right]
\end{equation}
where the second term is zero since the BZ does not have a boundary. Therefore, if both $\mathbf{A}$ and $\Omega$ are constructed from a smooth gauge over $\tau$, which is always possible for a TR-TI since by definition ($C=0$) there is no obstruction to a smooth gauge over the whole BZ, then the index is always $\Delta=0$ by virtue of Stokes' theorem~\cite{soluyanov2012smooth}. Thus, Eq.~\eqref{eq:z2_obstruction} can only give meaningful results if we require that the gauge respects time-reversal symmetry~\cite{fu2006time, bernevig2013topological, soluyanov2011wannier}, 
\begin{equation}\label{eq:gauge_t_smooth}
    \left\{\begin{aligned}
        \ket{u^{\text{I}}_{\alpha,-\bm{k}}}&=-\mathcal{T}\ket{u^{\text{II}}_{\alpha,\bm{k}}} \\
        \ket{u^{\text{II}}_{\alpha,-\bm{k}}}&=\ \ \ \mathcal{T}\ket{u^{\text{I}}_{\alpha,\bm{k}}}
    \end{aligned}\right.
\end{equation}
which is the same as in Eq.~\eqref{eq:gauge_t} but now requiring specifically $\chi_{\alpha,\bm{k}}=0$. Then, for the non-trivial $\mathbb{Z}_2$ state, it is not possible to find a gauge which is simultaneously smooth over half BZ and verifies Eq.~\eqref{eq:gauge_t_smooth}~\cite{soluyanov2011wannier, fu2006time}. Thus, under the time-reversal constraint, from Eq.~\eqref{eq:z2_obstruction} we see that in the $\mathbb{Z}_2$ non-trivial state, the $\mathbb{Z}_2$ index represents a topological obstruction to finding a smooth gauge over half BZ (with the additional time-reversal constraint) or equivalently an obstruction to Stokes' theorem, in the same way the Chern number represented a topological obstruction to finding a smooth gauge over the whole BZ\@.

This method of computing the invariant is particularly useful for numerical calculations, since as opposed to the previous methods, it requires very little gauge fixing. Namely, it is necessary to ensure the time-reversal constraint at the boundary of the integral $\partial\tau$~\cite{soluyanov2011wannier}, while the rest of points of the BZ do not require fixing since when discretizing, the formulation becomes gauge invariant, as it happened in the calculation of the Chern number. The discretization procedure for individual bands was introduced in section~\ref{sec:discrete_berryology}; now we show the general procedure when having $N$ bands~\cite{fukui2007quantum, soluyanov2011wannier}. First, one defines the link matrices $M_{\mu,nm}(\bm{k})=\braket{u_{n\bm{k}}|u_{m,\bm{k}+\bm{s}_{\mu}}}$ and link variables $L_{\mu}(\bm{k})=\det M_{\mu}/|\det M_{\mu}|$, where $\bm{s}_{\mu}$, $\mu=1,2$ are steps in the BZ mesh in each reciprocal lattice direction. Then, the discretized Berry connection reads $A_{\mu}(\bm{k})=\log L_{\mu}(\bm{k})$, and the Berry fluxes (curvature) are
\begin{equation}
    \Omega(\bm{k}) = \log[L_{1}(\bm{k})L_{2}(\bm{k}+\bm{s}_1)L^{-1}_{1}(\bm{k}+\bm{s}_2)L^{-1}_{2}(\bm{k})]
\end{equation}
The $\mathbb{Z}_2$ index is then computed on the lattice as
\begin{align}
    \Delta &= \frac{1}{2\pi i}\left[\sum_{\bm{k}\in\partial\tau}A_2(\bm{k}) - \sum_{\bm{k}\in\tau}\Omega(\bm{k})\right]\mod 2 \\
    & = \frac{1}{2\pi i}\sum_{\bm{k}\in\tau}\left[A_{1}(\bm{k}) + A_2(\bm{k} + \bm{s}_1) - A_1(\bm{k} + \bm{s}_2) - A_{2}(\bm{k}) - \Omega(\bm{k})\right]\mod 2
\end{align}
where the second line is obtained using that the sum of the phases $A_{\mu}(\bm{k})$ cancel as we sum over plaquettes, except at the boundary $\partial\tau$. This is exactly the same situation we had in~\eqref{eq:chern_number_discrete} when we first attributed a meaning to the Chern number as the number of vortices within the BZ\@. Here, we are counting the number of vortices of the Berry curvature, appearing from the branch cut of the logarithm, i.e. $\Omega(\bm{k})\in(-\pi,\pi]$ whereas the sum of the phases $A_{\mu}(\bm{k})\in(-\pi,\pi]$ over a plaquette can be outside this interval. The difference between the two divided by $2\pi$ is the vorticity of the plaquette, which summed over half BZ and modulo 2 give the $\mathbb{Z}_2$ index.

\begin{figure}[h]
    \centering
    \includegraphics[width=0.7\textwidth]{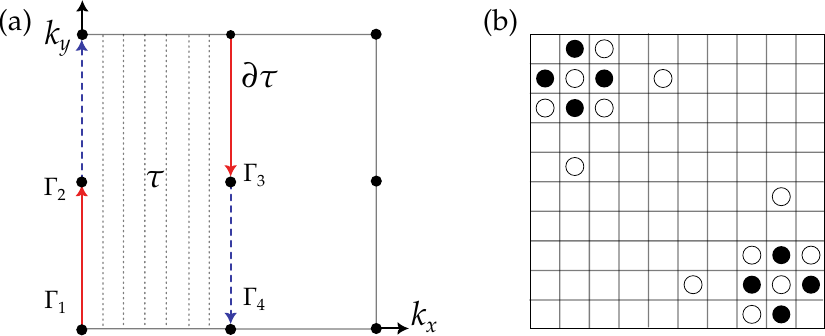}
    \caption[Calculation of the $\mathbb{Z}_2$ invariant as an obstruction to Stokes' theorem]{(a) Sketch of the BZ\@. The integral of the Berry curvature is evaluated in the interior of the half BZ $\tau$, while the Berry connection is evaluated over its boundary $\partial\tau$. The states defined over $\partial\tau$ require gauge fixing: we choose an arbitrary gauge on the red arrows, and then impose via time-reversal that gauge over the blue dashed arrows.\ (b) Discretization of the BZ into plaquettes, showing an example of the calculation of the $\mathbb{Z}_2$ invariant using the discretized Berry connection and curvatures. Black and white circles denote vorticities of $1$ and $-1$ respectively, which in this case amount to $\Delta = 0$. Adapted from~\cite{soluyanov2011wannier} and~\cite{fukui2007quantum} respectively.}
\end{figure}

Lastly, there is yet another way to compute the $\mathbb{Z}_2$ invariant which goes back to the time-reversal polarization. All our previous reasoning was done based on the behaviour of the HWCCs, which we would measure with the formulation introduced for the index in terms of the sewing matrix. However, this is not necessary since we know that the partial and total polarizations can be written explicitly in terms of the HWCCs (which is what allowed our reasoning in the first place). Since the partial polarizations are
\begin{equation}
    P_x^{S}(k_y) = \frac{1}{2\pi}\int_{-\pi}^{\pi}dk_x A_x^{\text{I}}(\bm{k}) = \sum_{\alpha}\bar{x}_{\alpha}^{S}(k_y),
\end{equation}
the invariant written from the difference in time-reversal polarizations in half a pumping cycle is~\cite{soluyanov2011computing, soluyanov2012topological}
\begin{equation}\label{eq:z2_hwccs}
    \Delta = \sum_{\alpha}\left[\bar{x}_{\alpha}^{\text{I}}(\pi) - \bar{x}_{\alpha}^{\text{II}}(\pi)\right] - \sum_{\alpha}\left[\bar{x}_{\alpha}^{\text{I}}(0) - \bar{x}_{\alpha}^{\text{II}}(0)\right]
\end{equation}
Note that this quantity is intrinsically modulo 2 since $\bar{x}^{\text{I}}_{\alpha}(k_y) = \bar{x}^{\text{II}}_{\alpha}(-k_y)\mod 1$. To exemplify, if the HWCCs exchange partners, at $k_y=\pi$ we would have $\bar{x}^{\text{I}}_{\alpha}=0.5=-\bar{x}^{\text{II}}_{\alpha}$, while at $k_y=0$ we would have $\bar{x}^{\text{I}}_{\alpha}=\bar{x}^{\text{II}}_{\alpha}=0$, which gives $\Delta=1$. If the HWCCs exchange every two states, then the change in $\bar{x}$ would be 1 and consequently $\Delta = 0$ due to the HWCCs being defined modulo the lattice parameter (here $a_x=1$). 

Therefore, if we are able to numerically extract the HWCCs, in principle we can simply track their evolution along half BZ to determine the $\mathbb{Z}_2$ invariant. In the next section, we will introduce the Wilson loop, which generalizes the Berry phase and as we will see allows to compute the HWCCs in a gauge invariant way. There exists an additional form to compute the invariant if the system exhibits inversion symmetry~\cite{fu2007inversion}, but which we will not discuss here since we focus exclusively on the most general formulations possible of the index.

To conclude, we have not discussed so far the topological nature of the invariant, this is, its robustness against perturbations of the system. From the $(C_{\uparrow},C_{\downarrow})$ picture, it is clear that the invariant defined as $C_s=(C_{\uparrow}-C_{\downarrow})\mod 2$ must be topologically protected, since each Chern number is protected from the arguments presented in section~\ref{sec:chern_insulators}. If we consider the $\mathbb{Z}_2$ invariant in the general case, the simplest way to argue about its topological nature is to consider the effect of a perturbation on the $\mathbb{Z}_2$ edge band structure. Say we introduce, for instance an impurity as the perturbation, $V=wd^{\dagger}d$. Then, using degenerate perturbation theory we can evaluate the effect of the perturbation on the energy of the edge states. Focusing on in-gap states, such as those from Fig.~\ref{fig:qshe}(d) or Fig.~\ref{fig:spin_chern_flow_edge_states}, we can obtain the energy corrections to first order in the perturbation series diagonalizing the following matrix:
\begin{equation}
    \begin{pmatrix}
        \braket{\alpha\text{I},k|V|\alpha\text{I},k} & \braket{\alpha\text{I},k|V|\alpha\text{II},k} \\
        \braket{\alpha\text{II},k|V|\alpha\text{II},k} & \braket{\alpha\text{II},k|V|\alpha\text{II},k}
    \end{pmatrix} = \begin{pmatrix}
        \braket{Ik|V|Ik} & 0 \\
        0 & \braket{Ik|V|Ik}
    \end{pmatrix} 
\end{equation}
where $\text{I, II}$ denote one edge state and its time-reversal companion and $\alpha$ is the label of the Kramer pair. As long as the perturbation is time-reversal invariant, $[V,\mathcal{T}]=0$, one can prove that there is no scattering between one edge state and its time-reversal companion,
\begin{equation}
    \braket{\alpha\text{I},k|V|\alpha\text{II},k} = \braket{\alpha\text{I},k|\mathcal{T}^{\dagger}V\mathcal{T}|\alpha\text{II},k} = -\braket{\alpha\text{I},k|V|\alpha\text{II},k}
\end{equation}
since the off-diagonal matrix elements are zero. Therefore, a time-reversal invariant perturbation results in a rigid shift of the edge bands. In particular, this means that at the TRIMs the degeneracy is always preserved, and consequently it is impossible to open a gap in the edge band structure, from what follows the $\mathbb{Z}_2$ topological nature of the system. If we were to consider an edge band structure with an even number of Kramers' pairs, then there would be scattering between non-time-reversal partners, which could potentially open a gap.

For completeness, a similar argument can be formulated for the evolution of the WCCs. If we consider a variation $\ket{n0k_y} + \delta\ket{n0k_y}$ coming from a $\mathcal{T}$-invariant perturbation, then $\mathcal{T}$ will also connect the variations. Consequently, both perturbed states will share the same WCC and in particular the degeneracy at the TRIMs will remain unmodified. Note that this is deeply connected to the argument used before to show how a gap can open in the WCC evolution; only for time-reversal companions degeneracies are guaranteed to be present at the TRIMs, while any other degeneracy between arbitrary states is purely accidental and can be gapped.

\subsection{The Wilson loop}\label{sec:wilson_loop}

Finally, we conclude the discussion of the topological invariants by introducing the Wilson loop, which constitutes the most practical method to extract the Chern number, the $\mathbb{Z}_2$ invariant, and can be used in fact also to characterize other topological phases such as HOTIs or Weyl semimetals~\cite{weng2015weyl, saini2022wloopphi, benalcazar2017quantized, franca2018anomalous}. The Wilson loop appears naturally when considering the adiabatic evolution of a degenerate manifold of states~\cite{kato1950adiabatic, wilczek1984appearance, mead1992geometric, budich2013from}, in the same way that the Berry phase appears for a single state, which is why it is regarded as the generalization of the Berry phase. 

So far we have been working with the $U(1)$ abelian Berry connection, defined for a single band. However, in practice bands in most materials present crossings and degeneracies, which forces us to consider the complete manifold of states simultaneously since we cannot resolve the degeneracies. In this context, the natural objects to work with are the non-abelian Berry connection and its corresponding Berry curvature.
First, let us introduce the $U(N)$ non-abelian Berry connection, defined by
\begin{equation}
    A_{nm,\mu}(\bm{k}) = i\braket{u_{n\bm{k}}|\partial_{\mu}u_{m\bm{k}}}
\end{equation}
where $n,m$ denote the bands in the manifold of $N$ states under consideration (which may involve states of different energies, not necessarily degenerate everywhere but with crossings at some points). Thus, the Berry connection is a matrix; it is denoted as the $U(N)$ connection since the natural gauge transformation now involves rotations of the whole space of states $\ket{u_{n\bm{k}}}\rightarrow\sum_{m}U_{nm}(\bm{k})\ket{u_{m\bm{k}}}$. As before, the Berry connection is not gauge invariant under such transformations,
\begin{equation}
    A_{\mu}(\bm{k})\rightarrow U^{\dagger}(\bm{k})A_{\mu}(\bm{k})U(\bm{k}) + iU^{\dagger}(\bm{k})\partial_{\mu}U(\bm{k})
\end{equation}
where $A_{\mu}(\bm{k})=[A_{nm,\mu}(\bm{k})]$ denotes the matrix Berry connection for the $\mu$ component. The non-abelian nature becomes clear now: since matrices do not commute in general we cannot simplify the first term as we did for the $U(1)$ Berry connection. The Berry curvature is also now a matrix, reading
\begin{equation}
    \Omega_{\mu\nu}(\bm{k}) = \partial_{\mu}A_{\nu}(\bm{k}) - \partial_{\nu}A_{\mu}(\bm{k}) - i\left[A_{\mu}(\bm{k}),A_{\nu}(\bm{k})\right].
\end{equation}
Notice the additional term $i[A_{\mu},A_{\nu}]$ that is present in the definition of the Berry curvature. It is necessary to ensure that the Berry curvature is gauge covariant (and not invariant in this case)~\cite{vanderbilt2018berry}. Namely, under a $U(N)$ gauge transformation, the Berry curvature transforms as
\begin{equation}
    \Omega_{\mu\nu}(\bm{k})\rightarrow U^{\dagger}(\bm{k})\Omega_{\mu\nu}(\bm{k})U(\bm{k})
\end{equation}
Just like gauge invariance, gauge covariance is a relevant property since it allows defining gauge invariant quantities simply by tracing them. Another relevant property of the $U(N)$ Berry curvature is that its trace coincides with the trace of the non-gauge covariant Berry curvature,
\begin{equation}
    \text{Tr }\Omega_{\mu\nu}(\bm{k}) = \sum_{n=1}^N\Omega_{n,\mu\nu}(\bm{k})
\end{equation}
where $\Omega_{n,\mu\nu}(\bm{k})=\partial_{\mu}A_{n,\nu}-\partial_{\nu}A_{n,\mu}$ is the Berry curvature for the $n$-th band. Therefore, the total Chern number can be written in terms of the gauge covariant Berry curvature as
\begin{equation}
    C = \frac{1}{2\pi}\int_{\text{BZ}}d^2\bm{k}\text{Tr }\Omega_{xy}(\bm{k})
\end{equation}

The Wilson loop was originally introduced by Wilson to describe quark confinement~\cite{wilson1974confinement}, in that case written in terms of gauge fields. As stated before, when considering the adiabatic evolution of a degenerate manifold of bands, the Wilson loop appears in the time-evolved state together with the dynamical phase~\cite{leone2011geometry}. 

For the characterization of the topological nature, we are interested not in a degenerate subspace but in all occupied bands. Then, one form to extend the original formulation is to deform the Bloch Hamiltonian $H(\bm{k})$ into a flattened version, where all occupied states fall into a single constant energy band, and likewise for the unoccupied states. While topological properties of individual bands are not preserved~\cite{avron1983homotopy}, the topological invariant of the complete valence bundle is preserved as long as the gap does not close and the stabilizing symmetries of the topological phase are not broken~\cite{alexandradinata2014wilson}. Denoting the projection onto occupied states by $P^{\text{occ}}_{\bm{k}}=\sum_n^{N} P_{\bm{k}}^n=\sum_{n}^N\ket{u_{n\bm{k}}}\bra{u_{n\bm{k}}}$, the flat-band Hamiltonian $H_F$ is 
\begin{equation}\label{eq:flattened_hamiltonian}
    H_F(\bm{k}) = (\varepsilon_- - \varepsilon_+)P_{\bm{k}}^{\text{occ}} + \varepsilon_+I
\end{equation}
where $\varepsilon_-$ denotes the energy of the occupied band of $N$ states, $\varepsilon_+$ is the energy of the flat-band for the unoccupied states, and $I$ is the identity operator. We assume that the temporal dependence is introduced from the momentum $\bm{k}=\bm{k}(t)$. As in the abelian case, we expand the time-evolved state in the basis of instantaneous eigenstates of the Hamiltonian~\cite{alexandradinata2014wilson},
\begin{equation}
    \ket{\psi(t)}=\exp\left(-i\int_{t_0}^t\varepsilon_-(t')dt'\right)\sum_{n}\psi_n(t)\ket{u_{n\bm{k}}}
\end{equation}
where the exponential is the dynamical factor. Note that the expansion is done only in the degenerate subspace by virtue of the adiabatic approximation, so that the time-evolved state is still also an instantaneous eigenstate at time $t$. Acting with the time-dependent Schrödinger equation on this state, one finally arrives at a parallel transport equation for the coefficients $\psi(t)$~\cite{leone2011geometry}:
\begin{equation}
    \dot{\psi}_n(t) - i \sum_{m,\mu}A_{nm,\mu}\dot{k}_{\mu}\psi_m = 0
\end{equation}
where one uses that $\partial_t=\sum_{\mu}\dot{k}_{\mu}\partial_{\mu}$ to arrive to the final expression. This differential equation has the following solution,
\begin{equation}
    \psi_n(t)=\sum_{m}W_{nm}[\gamma]\psi_m(0)
\end{equation}
where $\psi_m(0)$ are the coefficients at the initial time $t=0$ and $W_{nm}[\gamma]$ denotes the matrix elements of the Wilson line, computed on an arbitrary path on parameter space $\gamma$,
\begin{equation}
    W[\gamma] = \mathcal{P}\exp\left(i\int_{\gamma}\bm{A}(\bm{k})\cdot d\bm{k}\right)
\end{equation}
$\mathcal{P}$ denotes the path-ordering operator. Note now that the Wilson line involves a line integral of the non-abelian Berry connection, where each component corresponds to a matrix. In case that $\gamma$ describes a closed loop in parameter space, we refer to the Wilson loop,
\begin{equation}\label{eq:wilson_loop}
    W[\gamma] = \mathcal{P}\exp\left(i\oint_{\gamma}\bm{A}(\bm{k})\cdot d\bm{k}\right)
\end{equation}
Thus, we time-evolved state finally reads
\begin{align}
    \ket{\psi(t)}=\exp\left(-i\int_{t_0}^t\varepsilon_-(t')dt'\right)\sum_{n,m}W_{nm}[\gamma]\psi_m(0)\ket{u_{n\bm{k}}}
\end{align}
At this point, we may specify that the initial state was $\psi_m(0)=\delta_{mp}$, which also allows us to label the final state, $\ket{\psi(t)}\equiv\ket{\psi_p(t)}$:
\begin{equation}\label{eq:time_evolution_non_abelian}
    \ket{\psi_p(t)}=\exp\left(-i\int_{t_0}^t\varepsilon_-(t')dt'\right)\sum_{n}W_{np}[\gamma]\ket{u_{n\bm{k}}}
\end{equation}
from which we see that the Wilson loop is an operator that acts on the manifold of states, transporting it across the specified trajectory. Note that it is also possible to write a parallel transport equation directly for $W[\gamma]$~\cite{alexandradinata2014wilson}, if we had started directly from Eq.~\eqref{eq:time_evolution_non_abelian}. It is clear that when considering a single occupied band $N=1$, the Wilson loop reduces to the Berry phase as expected. 

The idea of the Wilson loop being a generalization of the Berry phase already hints that it will be relevant to address the topological properties of systems, but before discussing this first let's explore some properties of the Wilson loop. So far we have seen its continuous formulation, but in practice for us, it will be used it in its discrete form, which also serves to emphasize its operator nature. Upon a discretization of the BZ, we may approximate the Berry connection as a finite difference. Then, the Wilson loop is simply written in terms of the projectors over occupied states along the loop:
\begin{equation}\label{eq:discrete_wilson_loop}
    W[\gamma] = \prod_{\bm{k}\in\gamma}P^{\text{occ}}_{\bm{k}}=\prod_{\bm{k}\in\gamma}\left(\sum_{n}^{{N}}\ket{u_{n\bm{k}}}\bra{u_{n\bm{k}}}\right).
\end{equation}
Note that the loop $\gamma$ now refers to a discrete set of $\bm{k}$ points, $\gamma=\{\bm{k}_0,\bm{k}_1,\ldots,\bm{k}_{n-1},\bm{k}_0\}$. This expression is not surprising since it is the same as the original one used to define the discrete Berry phase in~\eqref{eq:berry_phase_discrete_gauge}, although here considering cross inner products between states. The matrix elements of the Wilson loop are 
\begin{equation}
    W[\gamma]_{nm} = \braket{u_{m\bm{k}_0}|W[\gamma]|u_{n\bm{k}_0}}
\end{equation}
where $\bm{k}_0$ is the initial and final point of the trajectory $\gamma$. Either from~\eqref{eq:wilson_loop} or from~\eqref{eq:discrete_wilson_loop} it can be seen that under a $U(N)$ gauge transformation of the occupied manifold, the Wilson loop is gauge covariant:
\begin{equation}
    W[\gamma] \longrightarrow U^{\dagger}(\bm{k}_0)W[\gamma]U(\bm{k}_0)
\end{equation}
As it happened with the Berry phase, which is only gauge invariant modulo $2\pi$ when computed over a closed loop, the Wilson line is only gauge covariant if computed on a loop. Then, the trace of the Wilson loop will be gauge invariant and as we will see, it is in fact related to the electronic polarization. In the spirit of the HWCCs, we may now consider loops $\gamma$ in the BZ such that they leave (without loss of generality) $k_y$ constant. For such a path $\gamma(k_y)$, its Wilson loop is given by
\begin{equation}
    W[\gamma(k_y)] = \exp\left(i\oint_{\gamma(k_y)}dk_xA_{x}(k_x,k_y)\right)
\end{equation}
These loops have a very special property: the spectrum of $\displaystyle\frac{1}{2\pi}\text{Im }\log W[\gamma(k_y)]$ matches the spectrum of $PxP$~\cite{bradlyn2022lecture, yu2011equivalent, alexandradinata2014wilson}\footnote{Here, the most general expression would be the spectrum of $\frac{a_x}{2\pi}\text{Im }\log W[\gamma(k_y)]$, since as seen in~\eqref{eq:hybrid_wannier_center} the Berry phases are related to the WCCs by $\bar{x}_n=\frac{a_x}{2\pi}\phi_n$. Namely, the eigenvalues of $PxP$ should be within the unit cell. However, for simplicity, we set $a_x=1$ throughout the section, implying that all WCCs will be in the interval $\bar{x}_n\in[0,1)$.}. Consider a general eigenstate $\lambda_n$ of the Wilson loop $W[\gamma(k_y)]$. As a complex number, it can be written in polar form as $\lambda_n=|\lambda_n|e^{i\theta_n}$. It can be proven then that the phases, $\theta_n(k_y) = \text{Im }\log \lambda_n(k_y)$ belong to the spectrum of $PxP$. To do so, consider the following eigenvalue problem:
\begin{equation}\label{eq:eigenproblem_wilson_loop}
    \left(PxP - \frac{\theta(k_y)}{2\pi} - R_x\right)\ket{\psi}=0
\end{equation}
where $R_x$ denotes the position of some unit cell along the $x$ axis. Thus, by finding one state $\ket{\psi}$ such that the above equation is fulfilled, we will have found an eigenstate of $PxP$ with eigenvalue $\theta/2\pi + R_x$. See~\cite{bradlyn2022lecture, alexandradinata2014wilson} for the detailed proof of this statement, or alternatively~\cite{yu2011equivalent} where it is proven in terms of the projected periodic position operator $PXP$ with $X=\sum_{i\alpha}e^{-i\delta k_xR_i}\ket{i\alpha}\bra{i\alpha}$, $\delta k_x=2\pi/N_xa_x$. In the end, one arrives at the following eigenstates of $PxP$:
\begin{equation}
    \ket{\psi}\equiv\ket{jR_xk_y}=\sum_{n=1}^N\int dk_xe^{-ik_x(\theta + R_x)}Q_{nj}(\bm{k})\ket{n\bm{k}}
\end{equation}
This is, the eigenstates of~\eqref{eq:eigenproblem_wilson_loop} actually correspond to hybrid Wannier states, as can be seen comparing with their definition in Eq.~\eqref{eq:hybrid_wannier_state}, also including a $U(N)$ transformation of the Bloch states via $Q_{nj}(\bm{k})$ which is the matrix that diagonalizes the Wilson loop. This result is not surprising, since we had seen previously that eigenstates of $PxP$ correspond to maximally localized WFs in the $x$ direction. The relation shown here between the HWCCs and the line integral of the Berry connection can be seen as the generalization of expression $\eqref{eq:hybrid_wannier_center}$, where now we have
\begin{equation}
    \bar{x}_n = \frac{1}{2\pi}\text{Im }\log \lambda_n
\end{equation}

In summary, we have just seen that when considering Wilson loops defined along reciprocal lattice vectors, for instance $W[\gamma(k_y)]$ in which the loop would be $\bm{k}_0\rightarrow\bm{k}_0 + \bm{G}_x$, the phases of its eigenvalues correspond to HWCCs. In particular, the eigenstates correspond to maximally localized Wannier states in the $x$ direction, as they correspond to eigenstates of the $PxP$ operator. This means that we have obtained a gauge covariant way to determine the HWCCs; while for the Chern number we already had a discrete formulation that was manifestly gauge invariant, all the formulas for the $\mathbb{Z}_2$ required some form of gauge fixing, except when writing the invariant $\Delta$ directly in terms of the HWCCs. As a reminder, each invariant can be computed in terms of the HWCCs as follows:
\begin{equation}\label{eq:invariants_together}
    \begin{aligned}
        C &= \sum_{n=1}^N\left[\bar{x}_n(k_y=2\pi/a_y) - \bar{x}_n(k_y=0)\right] \\ 
        \Delta &= \sum_{\alpha=1}^{N/2}\left[\bar{x}_{\alpha}^{\text{I}}(k_y=\pi/a_y) - \bar{x}_{\alpha}^{\text{II}}(k_y=\pi/a_y)\right] - \sum_{\alpha}\left[\bar{x}_{\alpha}^{\text{I}}(k_y=0) - \bar{x}_{\alpha}^{\text{II}}(k_y=0)\right]
    \end{aligned}
\end{equation}
We know that the WCCs in general are not gauge independent, and likewise under a gauge transformation the WCCs as obtained from the Wilson loop will change due to its gauge covariance. However, the trace of Wilson loop is gauge invariant, which corresponds to the sum of all WCCs. The advantage of the Wilson loop is that, even if the WCCs are gauge dependent, we do not care about their particular value: we only need to look at their spectral flow properties. Therefore, the Wilson loop can be computed with any arbitrary choice of gauge, for instance that obtained from numerical diagonalizations.

In practice, we do not evaluate expressions~\eqref{eq:invariants_together} numerically since the WCCs outputted by the computer are only well-defined modulo $a_x$, which can results in errors when evaluating for instance the Chern number (e.g.\ if there is only one occupied band and $\bar{x}_n\in[0,1)$, then we always have $C=0$ even for a topological system). Instead, what we do is simply track the evolution of the HWCCs along the BZ, and from their flow we determine the value of the invariant based on the considerations introduced in the previous sections.

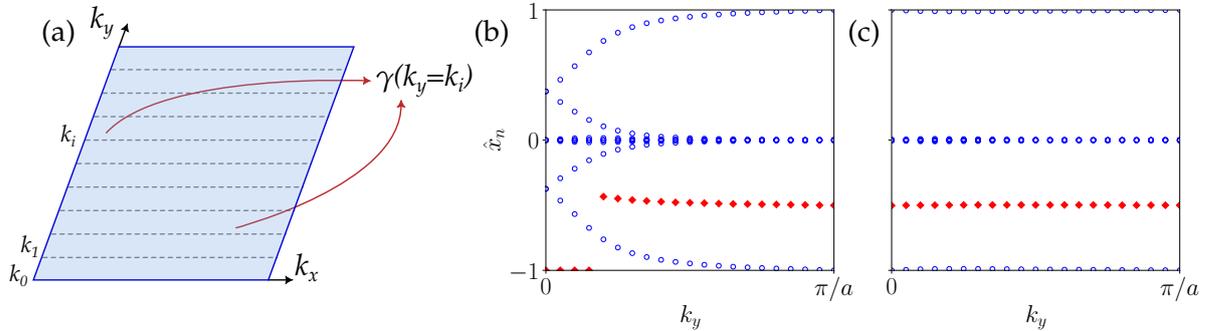
\begin{figure}[h]
    \centering
    \begin{tikzpicture}
        \node at (0,0) {\includegraphics[width=0.38\columnwidth]{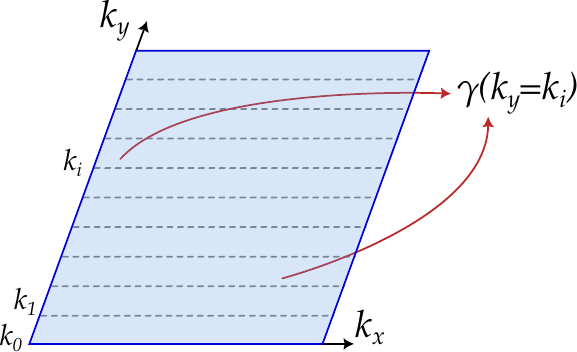}};
        \node at (-2.4,1.5) {(a)};
        
        \node at (7.9,-0.25) {\includegraphics[width=0.6\columnwidth]{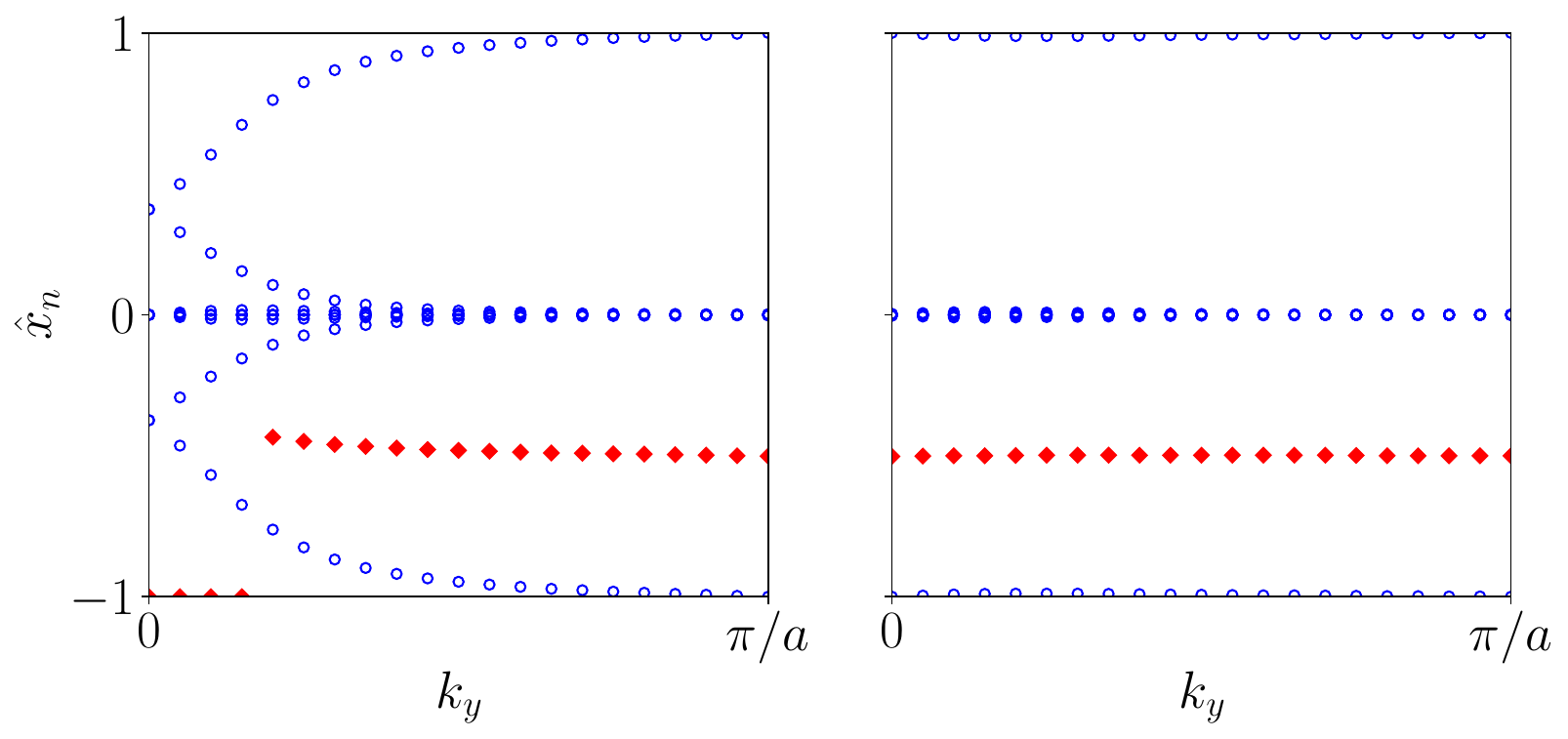}};
        \node at (3.3,1.5) {(b)};
        \node at (8.2,1.5) {(c)};
    \end{tikzpicture}
    \caption[Wannier charge center flow in Bi(111) and Sb(111)]{Example calculations of the $\mathbb{Z}_2$ invariant from the Wannier charge center flow or equivalently the Wilson loop eigenvalues in (b) Bi(111), which is topologically non-trivial, $\Delta=1$ and (c) Sb(111) which is topologically trivial, $\Delta=0$. The blue dots denote the HWCCs, while the red dots correspond to the largest gaps defined from~\eqref{eq:largest_gaps_z2}.\ (a) Sketch of the discretization of the BZ into paths of constant $k_y$ (dashed gray lines), $\gamma(k_y)$. The Wilson loop is computed along each path $W[\gamma(k_y)]$, and then diagonalized to obtain the HWCCs.}\label{fig:wannier_bi_sb}
\end{figure}

Of course, while ocular inspection might suffice for simple cases, we are interested in an algorithm to extract the invariants automatically. Soluyanov and Vanderbilt~\cite{soluyanov2011computing, z2pack} proposed a simple algorithm in which one tracks the midpoint of the largest gap between all HWCCs. Counting the total number of times these midpoints cross HWCCs in half BZ we obtain the invariant. Given the set of WCCs $\{\bar{x}_n^i\equiv \bar{x}_n(k_i),\text{s.t. }n\in\{1,\ldots,N\},i\in\{1,\ldots,M\}\}$, one determines at each $k_i$ the largest gap position $g^i\equiv g(k_i)$ such that it maximizes the distance to the closest WCC (equivalently the midpoint of the largest gap), which is given by
\begin{equation}\label{eq:largest_gaps_z2}
    \max_{g^i}\min_{n} d(g^i,\bar{x}_n^i)
\end{equation}
where $d(x,y)$ is a periodic distance since the WCCs are only defined modulo 1, $\bar{x}_n^i\in[0,1)$. Then, counting the number of crossings $n_i$ between the WCCs and the largest gap at each step,
\begin{equation}
    \min(g^i,g^{i+1}) \leq \bar{x}_n^{i+1} < \max(g^i, g^{i+1})
\end{equation}
we determine the value of the invariant, which is given by
\begin{equation}
    \Delta = \sum_{i=1}^{M-1} n_i \mod 2.
\end{equation}
See Fig.~\ref{fig:wannier_bi_sb} for an example calculation. A similar algorithm can be used to extract the Chern number. Thus, the Wilson loop provides us with a tool that can be easily evaluated and unifies the extraction of the invariant for different kinds of topological systems.

Lastly, it is interesting to mention another procedure that has been proposed multiple times in literature, mainly by Vanderbilt et al.~\cite{marzari1997maximally, z2pack, soluyanov2011computing, taherinejad2014wannier, vanderbilt2018berry}, which is closely linked to the Wilson loop. In order to achieve a smooth gauge in $\bm{k}$, which is necessary to build Wannier states and compute the charge center, Vanderbilt argues that one can follow a parallel transport procedure, which is more precisely described as an optimal phase alignment. Assuming a discretized BZ, for a single band this implies transforming the states $\{\ket{u_{\bm{k}}}\}$ to a new basis $\{\ket{\tilde{u}_{\bm{k}}}\}$ such that $\braket{\tilde{u}_{\bm{k}_i}|\tilde{u}_{\bm{k}_{i+1}}}\in\mathbb{R}^+$. In the multiband case, one instead considers the overlap matrices $M^{(\bm{k}_i,\bm{k}_{i+1})}_{mn}=\braket{u_{m\bm{k}_i}|u_{n\bm{k}_{i+1}}}$ and performs a rotation of the states at $\bm{k}_{i+1}$ such that $M^{(\bm{k}_i,\bm{k}_{i+1})}_{mn}$ is a hermitian matrix. From the singular value decomposition (SVD) $M=V\Sigma W^{\dagger}$, the rotation can be done as $VW^{\dagger}$~\cite{marzari1997maximally}. At the end, the final states at $\bm{k}_0$ are rotated by
\begin{equation}
    \Lambda = V_{n-1}W^{\dagger}_{n-1}\ldots V_0W^{\dagger}_0
\end{equation}
which is precisely the non-abelian phase picked up during an adiabatic evolution along a closed path (notice that we are taking solely the phases, and dropping the magnitude). In both cases, the Wannier functions built from this gauge election will have charge centers corresponding to the Berry phase, $\bar{x}_n=\frac{1}{2\pi}\text{Im }\log \lambda_n$, where $\lambda_n$ are the eigenvalues of $\Lambda$~\cite{vanderbilt2018berry}. So while the procedure is useful if the specific gauge and the states are needed, for the purpose of establishing the topological nature of a system, the Wilson loop as presented above is better suited as it is more direct: compute the overlap matrices (without needing to take their unitary part from the SVD), and diagonalize the loop to obtain the WCCs.

\clearpage

\section{Topology from the entanglement spectrum}
After having established successfully the main ideas concerning the description of topological phases in terms of the Berry phase and the Wilson loop, we now turn our attention to the entanglement spectrum. The entanglement spectrum and entropy are commonly used in quantum information theory~\cite{nielsen2010quantum}, as they measure the degree of non-local correlations between quantum states, but they have become increasingly relevant in condensed matter physics for the characterization of quantum many-body systems~\cite{amico2008entanglement, islam2015measuring}. For instance, the area law scaling of the spatial entanglement entropy $S\propto l^{d-1}$ of gapped ground states can be used to detect quantum phase transitions~\cite{vidal2003entanglement, eisert2010colloquium}, and enables the use of matrix product states and the density-matrix renormalization group technique to study many-body systems~\cite{hauschild2018efficient}.

Amidst the developments in the theory concerning topological insulators, it was noticed that the entanglement spectrum could also be used to characterize the topological nature of a system~\cite{li2008entanglement, pollmann2010entanglement, regnault2017entanglement}. Regarding topological order, such as the one present in fractional phases, the entanglement entropy presents a subleading constant term $S\propto \gamma$, named topological entanglement entropy~\cite{kitaev2006topological, levin2006detecting}, which can be used also to establish numerically the topological behaviour of a fractional system~\cite{grushin2015characterization}.
There exist different forms of entanglement spectrum depending on the degree of freedom that is partitioned, such as the spatial entanglement spectrum, but also momentum, orbital or particle entanglement spectrum~\cite{wu2012zoology, mondragon2013characterizing, legner2013relating}. For non-interacting topological phases of matter, it was shown that the spatial entanglement spectrum captures the topological nature of the system for a variety of reasons: degeneracies in the entanglement spectrum can be linked to the presence of edge states~\cite{fidkowski2010entanglement}, and more importantly, if the system is topological, then the entanglement spectrum will show a non-trivial flow, akin to topological edge states in the energy spectrum~\cite{turner2010entanglement, hughes2011inversion, alexandradinata2011trace}.

In this section we review the properties of the entanglement spectrum and entropy, and demonstrate how they are computed in practice for non-interacting fermion systems. Using this simplified picture for non-interacting systems, regarded as single-particle entanglement spectrum, we show why it can be used to identify topological phases of matter, as an alternative to the standard invariants seen previously. The combination of the Wilson loop to evaluate topological invariants and the entanglement spectrum will provide a powerful tool to characterize topological insulators, which we will use in the following chapters.

\subsection{Entanglement spectrum and entropy}\label{sec:entanglement_spectrum}
Before we introduce the entanglement spectrum and entropy, it is good to review the concept of density matrix and some of its properties, since it is fundamental to define the entanglement entropy. In general, given a Hilbert space $\mathcal{H}$ and an orthonormal basis $\{\ket{\phi_i}\}_i$, a density operator (or density matrix) takes the form:
\begin{equation}
    \rho = \displaystyle\sum_{ij} p_{ij}\ket{\phi_i}\bra{\phi_j}
\end{equation}
We introduce now the distinction between pure and mixed quantum states: pure states are regular states from the Hilbert space, written as a superposition in general. On the other hand, mixed states correspond to a statistical ensemble of quantum states, meaning that each state of the ensemble has a certain probability of appearing when measuring. While it sounds similar, there is a fundamental difference: pure states are in superposition, and therefore are subject to quantum interference, while for mixed states the probabilities come from experimental uncertainties.
Then, if we consider any pure quantum state $\ket{\psi}$, the density operator is defined simply by the projector onto this state:
\begin{equation}
    \rho = \ket{\psi}\bra{\psi}
\end{equation}
If we were to expand $\ket{\psi}$ in the basis, then it would be written in the general form we saw at the beginning. A key property of the density operator for pure state is the following: it can always be written as the projector of a single state, whereas for mixed states it is not possible. This property is also denoted as idempotency, $\rho^2=\rho$. 
For mixed states, the density operator is constructed specifying the probability $p_i$ of measuring a state $\ket{\psi_i}$ of the ensemble. Then, the density operator is:
\begin{equation}
    \rho = \displaystyle\sum_i p_i \ket{\psi_i}\bra{\psi_i}
\end{equation}
Note that we are using general states $\ket{\psi_i}$ of the Hilbert space, and not necessarily elements from the basis. So the density matrix for mixed states can be thought of as a linear combination of density matrices for pure states. It is easy to check that the density matrix for a mixed state is not idempotent, which also implies that it cannot be written as a projector for a single state, $\ket{\Psi}\bra{\Psi} \neq \sum_i p_i \ket{\psi_i}\bra{\psi_i}$.

The entanglement entropy is well-known within quantum information theory, as it allows to measure the degree of entanglement of a bipartite system. This quantity is also present in other contexts such as high energy physics, or as we outlined in the introduction, in condensed matter physics since it proves useful to identify topological systems. First, we introduce the von Neumann entropy, which is defined as:
\begin{equation}
    S(\rho) = -\mathrm{Tr}(\rho \ln\rho)
\end{equation}
Using the spectral decomposition of an operator, the density matrix is written in general as $\rho = \sum_i\lambda_i\ket{\nu_i}\bra{\nu_i} $, where $\{\lambda_i\}_i$ are its eigenvalues, and $\{\ket{\nu_i}\}_i$ are its eigenstates. In this representation it is immediate to carry out the trace, and the von Neumann entropy becomes:
\begin{align*}
    S(\rho) &= -\sum_{i}\lambda_i\ln\lambda_i
\end{align*}
The von Neumann entropy measures the mixture of a state. It can be seen that the entropy for a pure state is zero:
$S(\rho) = S(\ket{\psi}\bra{\psi}) = 0$. This follows since $\rho$ is by construction already in its spectral representation, namely its eigenvalues are $1$ (with multiplicity 1) and $0$ (with multiplicity $D-1$, $D$ being the dimension of the basis), which is to be expected since the eigenvalues of the density matrix also measure the occupation of states. 
The von Neumann entropy is used to define the entanglement entropy. Consider now a bipartite Hilbert space for a many-body system, i.e.:
\begin{equation}
    \mathcal{H}=\mathcal{H}_A\otimes\mathcal{H}_B
\end{equation}
This bipartite space can be constructed either specifying the subspaces $\mathcal{H}_A, \mathcal{H}_B$, or simply by considering a partition along some of the quantum numbers intrinsic to our system, which will be the case generally for us. 
\begin{figure}[h]
    \centering
    \includegraphics[width=0.4\textwidth]{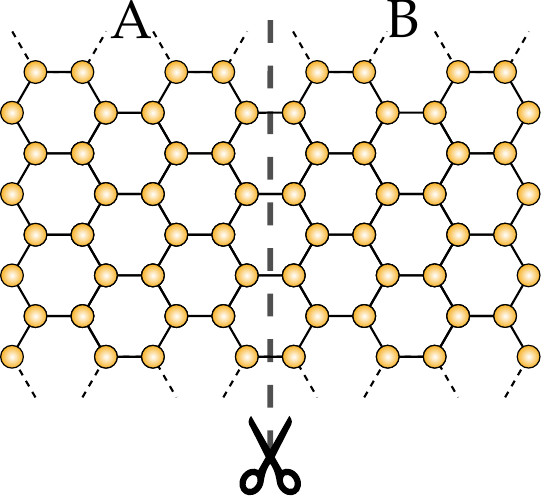}
    \caption[Illustration of a spatial entanglement cut in a ribbon]{Illustration of a spatial entanglement cut in a ribbon. The black dashed line denotes the cut, in which the system is divided into two subsystems $A$ and $B$. The smaller dashed lines are used to denote the PBC of the ribbon.}\label{fig:spatial_entanglement_cut}
\end{figure}
Next, consider a general state $\ket{\psi}\in\mathcal{H}$. According to the Schmidt decomposition theorem~\cite{peres1997quantum}, it can be written as
\begin{equation}
    \ket{\psi} = \sum_i\alpha_i\ket{\psi_i^A}\otimes\ket{\psi^B_i}
\end{equation}
where the coefficients $\{\alpha_i\}_i$ are uniquely determined, and the states $\{\ket{\psi_{A,B}^i}\}_i$ form an orthonormal basis (of a subspace) of the Hilbert spaces $\mathcal{H}_{A,B}$ respectively. Now, instead of computing the entropy for the associated density matrix (which we now it is zero as it is a pure state), we introduce the reduced density matrix associated to subsystem $\mathcal{H}_A$:
\begin{equation}
    \rho_A = \mathrm{Tr}_B(\rho)
\end{equation}
where $\mathrm{Tr}_B$ denotes the trace over $\mathcal{H}_B$ basis states, i.e.\ we trace out the degrees of freedom coming from subsystem $B$. Performing this calculation, one gets
\begin{align}
    \rho_A &= \mathrm{Tr}_B(\ket{\psi}\bra{\psi})=\sum_i|\alpha_i|^2\ket{\psi_i^A}\bra{\psi_i^A}
\end{align}
It is interesting to mention that the reduced density matrix we have obtained now corresponds to a mixed state, so its von Neumann entropy will not be zero. Since $\{\ket{\psi_i^A}\}_i$ forms a basis of $\mathcal{H}_A$, it is immediate to compute its entropy:
\begin{equation}
    S(\rho_A) = -\sum_i|\alpha_i|^2\ln|\alpha_i|^2    
\end{equation}
The entropy associated to $\rho_A$ is called entanglement entropy, and allows one to measure the degree of entanglement of the original state $\ket{\psi}$ across the partition. This quantity is well-defined in the sense that if we were to compute instead $\rho_B$ and its associated entropy, we would obtain the same quantity:
\begin{equation}
    \rho_B = \sum_i|\alpha_i|^2\ket{\psi_i^B}\bra{\psi_i^B} \Rightarrow S(\rho_B) = -\sum_i|\alpha_i|^2\ln|\alpha_i|^2
\end{equation}
As for the entanglement spectrum, it corresponds to the eigenvalues of the reduced density matrix, which in this case would be the set $\{\alpha_i\}_i$. Thus, while the Schmidt decomposition of the ground state allows to compute both quantities, in practice we do not know it, or it is hard to obtain.

\subsection{Single-particle entanglement spectrum}

So far we have considered general states $\ket{\psi}\in\mathcal{H}$. From now on, we will discuss the entanglement properties of the many-body ground state of our Hamiltonians, which in the non-interacting case is simply the Fermi sea. While performing the Schmidt decomposition for the ground state might prove difficult, it was proven that for non-interacting fermion systems one can obtain the reduced density matrix corresponding to a spatial entanglement cut (as in Fig.~\ref{fig:spatial_entanglement_cut}) by computing single-particle correlation functions~\cite{peschel2003calculation}. Then, from the reduced density matrix we determine both the entanglement spectrum and entropy. To show this, consider a tight-binding Hamiltonian for free fermions hopping between lattice sites:
\begin{equation}
    H = \sum_{i,j}t_{ij}c^{\dagger}_ic_j
\end{equation}
where $t_{ij}$ are the hopping amplitudes, and $c^{\dagger}_i(c_i)$ are the creation (destruction) operators at lattice site $i$. Any possible orbital or spin degree of freedom is also included under the indices $i,j$. Upon diagonalization, we can construct the many-body ground state which is simply the Fermi sea, $\ket{\Omega} = \prod_{n<F}d^{\dagger}_n\ket{0}$, 
where $d^{\dagger}_n$ are creation operators in the diagonal basis. Knowing the ground state, we can compute different correlation functions. Since the ground state is a determinant, by virtue of Wick's theorem all $n$-particle correlation functions can be written in terms of the one-particle correlation function,
\begin{equation}
    C_{ij} = \braket{c^{\dagger}_ic_j}
\end{equation}
For example, the two-particle correlation function factorizes as $\braket{c^{\dagger}_nc^{\dagger}_mc_ic_j} = \braket{c^{\dagger}_nc_j}\braket{c^{\dagger}_mc_i} -  \braket{c^{\dagger}_nc_i}\braket{c^{\dagger}_mc_j}$. At this point we introduce the entanglement cut of our Hilbert space, in which we partition all the system sites into two separate halves $A$ and $B$. If we now restrict the indices $i,j$ to one partition, $i,j\in A$, we can use the reduced density matrix $\rho_A$ associated to the partition to write the correlation matrix as
\begin{equation}
    C_{ij} = \mathrm{Tr}_A\left(\rho_A c^{\dagger}_ic_j\right),\quad i,j\in A
\end{equation}
where the trace is to be taken over the Fock space associated to $\mathcal{H}_A$. The previous expression holds because the reduced density matrix by definition must reproduce all expectation values in the subsystem. The same holds for higher order functions, and since these can be factored using Wick's theorem, for $\rho_A$ to behave the same way necessarily it must be an operator of the form:
\begin{equation}
    \rho_A=\mathcal{K}e^{-H_e},\quad H_e = \sum_{i,j}H_{ij}c^{\dagger}_ic_j,\ i,j\in A
\end{equation}
where $\mathcal{K}$ is a normalization constant, and $H_e$ is necessarily a free-fermion operator, named the entanglement Hamiltonian. The exponential form for the reduced density matrix corresponds to a gaussian state. We know that if a state is gaussian, then Wick's theorem holds~\cite{gaudin1960demonstration}, e.g. $\text{Tr }(\rho_A c^{\dagger}_ic^{\dagger}_jc_mc_n)$ factors into one-particle correlation functions. In our case, we already know that Wick's theorem holds for both the system and the subsystem (due to the ground state being a Slater determinant). If we then assume that the reduced density matrix $\rho_A$ is gaussian, then Wick's theorem applies, and we recover the same expectations values as for the full system. Since a state is uniquely determined by all its expectation values (all $n$-particle correlators), then the state must be gaussian.

Note now that the matrix elements $H_{ij}$ are not hoppings, but rather the most general one-particle operator. At this point, we want to determine the eigenvalues of $H_e$ as that would give us the entanglement spectrum and entropy. To do so, first we diagonalize the entanglement Hamiltonian transforming the operators to a new basis $c_i = \sum_kU_{ik}a_k$:
\begin{equation}
    \rho_A = \mathcal{K}\ \text{exp}\left(-\sum_k\varepsilon_ka^{\dagger}_ka_k\right)
\end{equation}
Before substituting this into the expression for $C_{ij}$, first it is convenient to determine the normalization constant $\mathcal{K}$. Imposing $\text{Tr} \rho_A=1$, we find
\begin{equation}
    \text{Tr}_A \left(\mathcal{K}\exp\left(-\sum_k\varepsilon_ka^{\dagger}_ka_k\right)\right) = 1 \Rightarrow \mathcal{K} = \prod_{i}\left(1 + e^{-\varepsilon_i}\right)^{-1}
\end{equation}
where the trace is to be taken over the Fock space of subsystem $A$. We may substitute this into $C_{ij}$. Tracing over all possible state configurations (different occupations numbers) in subsystem $A$ we arrive at 
\begin{align}\label{eq:correlation_matrix_trace}
    C_{ij} &= \mathrm{Tr}_A\left( \mathcal{K}\ \text{exp}\left(-\sum_k\varepsilon_ka^{\dagger}_ka_k\right)\sum_{n,m}U_{in}^*U_{jm}a^{\dagger}_na_m\right)  = \sum_k U_{ik}^*U_{jk}\frac{1}{1+e^{\varepsilon_k}}
\end{align}
Analogously, we can obtain the matrix elements of $H_e$ in the lattice basis, $H^e_{ij}=\sum_{k}U_{ik}U_{jk}^*\varepsilon_k$. Consequently, both $C$ and $H_e$ are diagonal in the new basis. Denoting the eigenvalues of $C$ by $\{\xi_k\}$, they are related to the eigenvalues of $H_e$ by:
\begin{align}
    \xi_k = \frac{1}{1+e^{\varepsilon_k}}
\end{align}
This same relation can also be expressed in matrix form:
\begin{equation}
    H^T_e = \ln\left[\frac{1 - C}{C}\right]
\end{equation}
The key point here is that the eigenvalues of the correlation matrix are isomorphic to the entanglement Hamiltonian eigenvalues, and from these we can obtain the eigenvalues of the reduced density matrix. Therefore, in practice it suffices to obtain the eigenvalues of the correlation matrix:
\begin{equation}\label{eq:restricted_correlation_matrix}
    C_{ij} = \braket{c^{\dagger}_ic_j},\ i,j\in A
\end{equation}
in the conventional way (diagonalize the Hamiltonian and compute the expectation value with respect to the Fermi sea) to obtain the entanglement spectrum and entropy. We refer to the eigenvalues of the entanglement Hamiltonian $H_e$ as the single-particle entanglement spectrum. Given the one-to-one relation between the eigenvalues of $C$ and $H_e$, we will also refer to the eigenvalues of $C$ as the single-particle entanglement spectrum. 

For completeness, we show how to compute the entanglement spectrum and entropy from the single-particle entanglement spectrum. The entanglement spectrum, eigenvalues of the reduced density matrix $\rho_A$, are obtained simply considering all different possible occupations of the states in subsystem $A$. Given a configuration of occupations $\{n_i\}$ defining a state, its associated eigenvalue is:
\begin{equation}
    \lambda_{\{n_i\}} = \prod_{i} \xi_i^{n_i}(1-\xi_i)^{1-n_i}
\end{equation}
To compute the entanglement entropy, in this case it is simpler to consider its definition instead of substituting the above entanglement spectrum. Since $S(\rho_A)=-\text{Tr}_A(\rho_A\ln\rho_A)$, we may use that $\rho_A = \mathcal{K}e^{-H_e}$ to write:
\begin{equation}
    S(\rho_A) =  \text{Tr}_A(\rho_A H_e) - \ln\mathcal{K}
\end{equation}
Performing the trace as in~\eqref{eq:correlation_matrix_trace} and simplifying, one arrives at the expression for the entanglement entropy:
\begin{equation}
    S(\rho_A) = -\sum_k\left[ \xi_k\ln\xi_k + (1-\xi_k)\ln(1-\xi_k) \right]
\end{equation}
which corresponds to the sum of the binary cross entropy for each single-particle entanglement spectrum eigenvalue.

\subsection{Non-trivial flow of the entanglement spectrum}

So far we have discussed how to obtain the entanglement spectrum and entropy from the single-particle entanglement spectrum in a non-interacting fermionic system. However, it still remains to see the most relevant property for our purposes: the flow of the (single-particle)\footnote{From now on, we will denote the single-particle spectrum directly as entanglement spectrum to abbreviate, since the many-body entanglement spectrum is not used in this section.} entanglement spectrum in topological systems. We will show that a non-trivial entanglement spectrum flow is a direct consequence of the topological nature of the system, and so it can be used to identify topological phases of matter.

As before, we assume to be working with a Hamiltonian of non-interacting 2D fermions in tight-binding representation, whose ground state is the Fermi sea. Considering a spatial partition of the system into two halves $A$ and $B$, we want to evaluate the restricted one-particle correlation matrix~\eqref{eq:restricted_correlation_matrix}. While doing so, it will be instructive to consider different boundary conditions for the system. First, we assume to be working with open boundary conditions. Switching to the diagonal basis, the correlation matrix is written as
\begin{equation}c^{\dagger}_i=\sum_{n}U_{in}c^{\dagger}_n \Longrightarrow C_{ij}=\sum_{n,m}^FU_{in}U^{\dagger}_{mj}\braket{c^{\dagger}_nc_m}=\sum_{n}^FU_{in}U^{\dagger}_{nj} = P_{ij} 
\end{equation}
where $P$ is the projector over the ground state, and $U$ the matrix that diagonalizes $H$. The indices $i,j$ here specify exclusively atomic position and orbital, while $n,m$ denote eigenstate indices and $F$ is the filling (Fermi level). Here, we already see that the one-particle correlation matrix (which in fact is also the one-particle density matrix), corresponds simply to the projector over occupied states,
\begin{equation}
    C \equiv P = \sum_{n}^F\ket{n}\bra{n}
\end{equation}
If we instead have periodic boundary conditions, the eigenstates of $H$ can be described with two quantum numbers, $n, \textbf{k}=(k_x, k_y)$. The relation between the original basis and the Bloch basis is: 
\begin{equation}
    c^{\dagger}_i=\frac{1}{\sqrt{N}}\sum_{\textbf{k}}e^{-i\textbf{k}\textbf{R}_i}c^{\dagger}_{i\textbf{k}}=\frac{1}{\sqrt{N}}\sum_{n,\textbf{k}}e^{-i\textbf{k}\textbf{R}_i}U_{in}(\textbf{k})c^{\dagger}_{n\textbf{k}}
\end{equation}
where $N$ is the number of unit cells. Note that the indices $i, j$ in $c^{\dagger}_i$ now include also the cell index $\bm{R}_i$ in addition to the atomic position and orbital, whereas in $c^{\dagger}_{i\bm{k}}$ it only includes atomic position and orbital. The matrix elements of $C$ are obtained similarly:
\begin{equation}
    C_{ij}=\frac{1}{N}\sum^F_{n, \textbf{k}}e^{i\textbf{k}(\bm{R}_j-\bm{R}_i)}U_{in}(\textbf{k})U^{\dagger}_{nj}(\textbf{k})=
    \frac{1}{N}\sum_{\textbf{k}}e^{i\textbf{k}(\bm{R}_j-\bm{R}_i)}P_{ij}(\textbf{k})
\end{equation}
where $P_{ij}(\bm{k})=\sum_{n}^{\text{occ}}\ket{n\bm{k}}\bra{n\bm{k}}$ is the momentum-resolved projection operator. 
Alternatively, if we define the entanglement cut parallel to one of the reciprocal axis, say $k_y$, then $k_y$ is still a good quantum number in the description of the entanglement spectrum (we cannot consider $k_x$ as it would involve positions outside the partition). Thus, we can define the one-particle correlation matrix to be a function of $k_y$, so that in analogy with the HWCC evolution, we can evaluate the entanglement spectrum as a function of $k_y$:
\begin{equation}
    C_{ij}(k_y)=\braket{c^{\dagger}_{ik_y}c_{jk_y}}=\frac{1}{N_x}\sum_{k_x}e^{ik_x(x_j-x_i)}P_{ij}(\textbf{k})
\end{equation}
where the operators $c^{\dagger}_{ik_y}$ are the Fourier transform of $c^{\dagger}_i$ only in the $y$-axis. Finally, if we instead consider a ribbon of a two-dimensional material, that would mean having one an open boundary in one axis, and periodicity in the other one. Like before, setting the entanglement cut parallel to our momenta, we may resolve the correlation matrix as a function of $k$:
\begin{equation}
    C_{ij}(k)=\braket{c^{\dagger}_{ik}c_{jk}}=P_{ij}(k)=\sum_{n}^{\text{occ}}\ket{n{k}}\bra{n{k}}
\end{equation}
In all four cases, the one-particle correlation matrix corresponds to the projector over the ground state, which ultimately is related to a flattened version of the Hamiltonian, as in~\eqref{eq:flattened_hamiltonian}. For instance, for the open system we would simply have $H_F = (\varepsilon_- - \varepsilon_+)P+ \varepsilon_+I$. In the case of the ribbon, we can write directly the Bloch flat-band Hamiltonian in terms of the correlation matrix, 
\begin{align}
    \nonumber H_F(k) &= (\varepsilon_- - \varepsilon_+)P_{k}^{\text{occ}} + \varepsilon_+I \\ 
    &= (\varepsilon_- - \varepsilon_+)C(k)+ \varepsilon_+I
\end{align}
where $\varepsilon_{\pm}$ denote arbitrary energies for the occupied (+) and unoccupied (-) bands.
We know that the occupied manifold of the Hamiltonian retains its topological properties when its flattened; as long as it is gapped and the symmetries that protect the topology are there, the system will still be topological. Thus, when we restrict the indices $i, j\in A$ in the correlation matrix to a half of the system, we are effectively creating a virtual boundary in the system. Since the correlation matrix is proportional to the Hamiltonian, if the system is topological then a non-trivial flow will appear in the entanglement spectrum as a function of $k$, connecting the occupied and unoccupied bands, analogous to the topological edge states in the energy spectrum~\cite{turner2010entanglement, hughes2011inversion, alexandradinata2011trace}. 

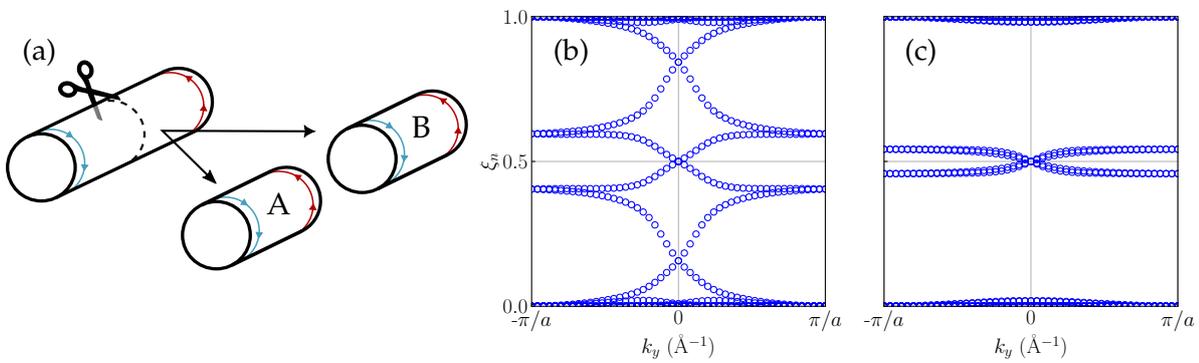
\begin{figure}[h]
    \centering
    \begin{tikzpicture}
        \node at (0,0) {\includegraphics[width=0.38\columnwidth]{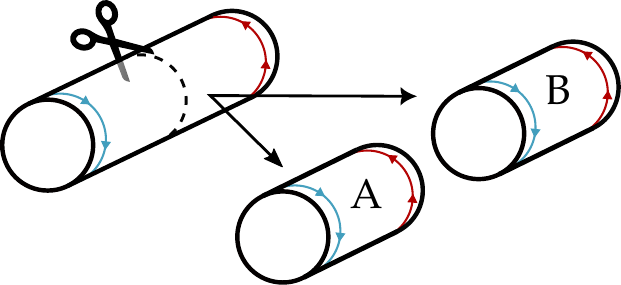}};
        \node at (-2.6,1.5) {(a)};
        
        \node at (7.9,-0.25) {\includegraphics[width=0.6\columnwidth]{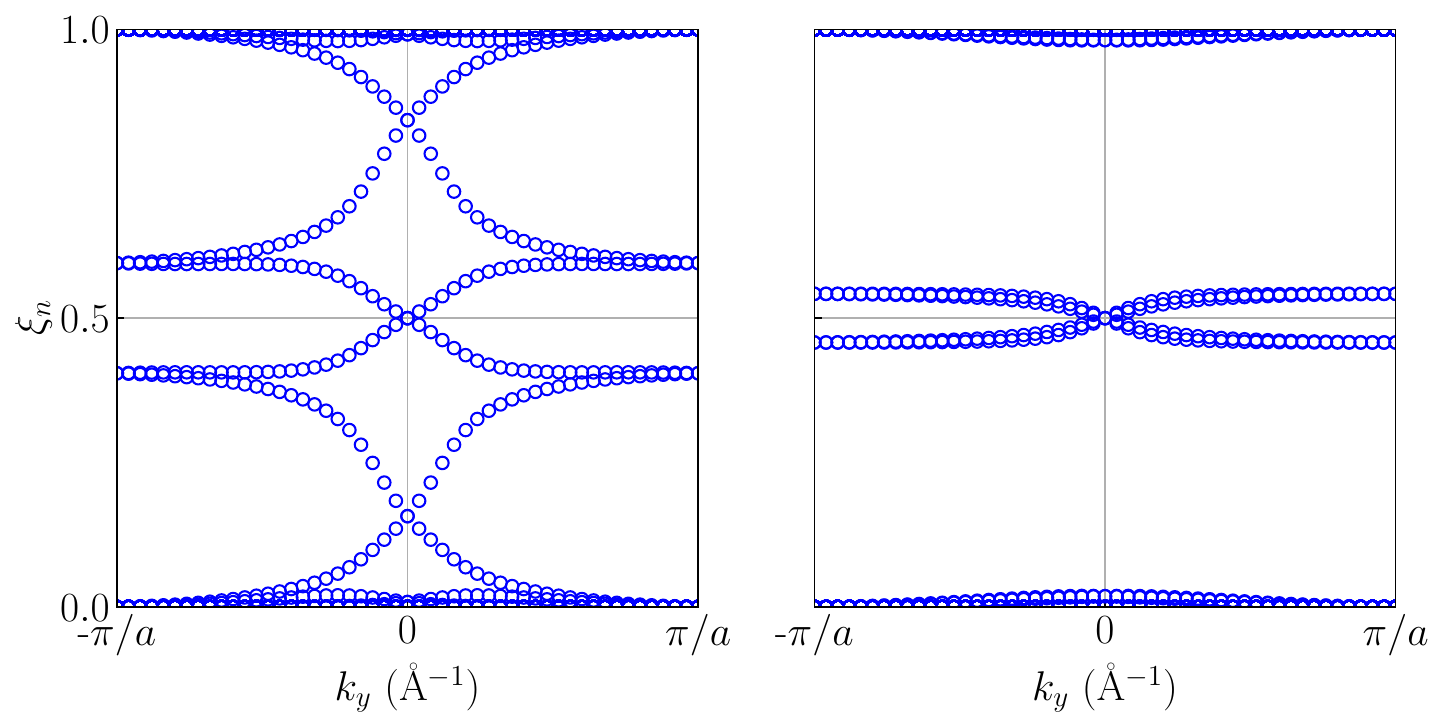}};
        \node at (4.4,1.5) {(b)};
        \node at (9.0,1.5) {(c)};
    \end{tikzpicture}
    \caption[Entanglement spectrum flow in Bi(111) and Sb(111)]{(a) Sketch of the entanglement cut on a ribbon or nanotube, which already presents topological edge states. After the cut into partitions $A$ and $B$, new topological edge states will appear on the boundary of the cut, which will be reflected in the entanglement spectrum.\ (b, c) Example calculations of the single-particle entanglement spectrum flow in Bi(111), which is topologically non-trivial, $\Delta=1$ and Sb(111) which is topologically trivial, $\Delta=0$.}\label{fig:entanglement_flows}
\end{figure}

An example of both the non-trivial and trivial flows of the entanglement spectrum is shown in Fig.~\ref{fig:entanglement_flows}, which is highly reminiscent of the flow of the WCCs we have seen before, for instance in~\ref{fig:wannier_bi_sb}. It was shown by some authors that in fact the entanglement spectrum can also be regarded as coarse grained Wannier charge centers, and consequently the entanglement spectrum would also reflect the charge pumping across the partition~\cite{lee2014position, lee2015free, lee2014position}.

\chapter{Deep learning for disordered topological insulators through their entanglement spectrum}\label{chapter:deep_learning}

\section{Introduction}\label{sec:introduction_chapter8}

In Chapter~\ref{chapter:topology} we learnt that the identification of topological materials requires the computation of Berry phases, which ultimately are related to the Chern number~\cite{thouless1982quantized} or the $\mathbb{Z}_2$ index~\cite{kane2005z}, the so-called topological invariants. The evaluation of Wilson loops~\cite{z2pack}, the most common methodology used to unveil the presence of invariants, works well for crystalline systems since they are performed in reciprocal space, to the point of allowing for high-throughput screening of materials~\cite{olsen2019discovering, mounet2018two}. There exist other approaches based on reciprocal space to condensed matter topology, such as topological quantum chemistry~\cite{bradlyn2017topological, vergniory2019complete}, symmetry indicators~\cite{fu2007inversion, po2017symmetry, song2018quantitative}, or the scattering invariant approach~\cite{fulga2012scattering}. Alternatively, one can resort to the bulk-boundary correspondence: if the system is topological, we expect the presence of conducting surface states. By calculating different observables, such as the conductance or the density of states, one strives to find evidence of the topology, without directly computing any invariant~\cite{yang2019topological}.

\begin{figure}[h]
    \centering
    \includegraphics[width=0.9\textwidth]{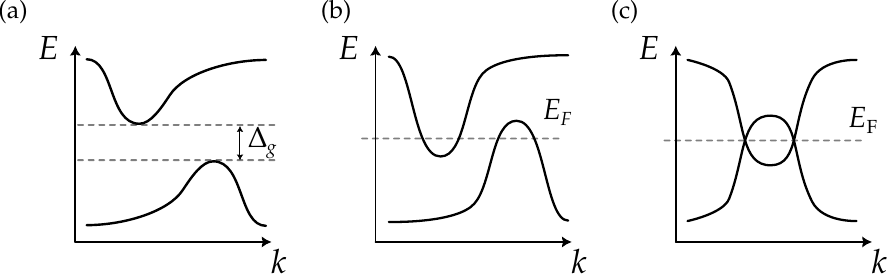}
    \caption[Comparison between a global gap, an overall gap and a band crossing]{Comparison between different types of gaps in materials.\ (a) Indirect global gap in the BZ, $\Delta_g$.\ (b) There is no global gap, but an overall gap since at each $\bm{k}$ the bands are gapped.\ (c) Band crossing, where the bands touch at some $\bm{k}$. For (a, b), since the bands are smooth in $\bm{k}$ we can compute the topological invariant associated to the lowest band even if the Fermi energy crosses two bands, as in (b). In (c), there are crossings of the bands which prevent the computation of the invariant, unless the crossings are resolved.}\label{fig:gap_types}
\end{figure}

The use of the Wilson loop technique requires the existence of a direct band gap everywhere in the Brillouin zone (e.g.\ no band crossings) near the Fermi level, since the valence manifold must be smooth in $\bm{k}$. An overall gap does not need to exist and the system may still be topologically non-trivial, as it could be the case in a semimetal, see Fig.~\ref{fig:gap_types}. For disordered or non-translationally invariant systems where the bands are not well-defined (e.g.\ in an open system), or are defined in a very small Brillouin zone (a large supercell  where the spectrum becomes essentially discrete) the direct gap concept is lost. In the latter case, if an overall gap is clearly visible, the Wilson loop can still be calculated~\cite{costa2019toward}. However, the absence of a spectral gap does not preclude a non-trivial topology, as it happens in topological Anderson insulators~\cite{leung2012effect, li2009topological}.

The introduction of topological markers such as the Chern marker~\cite{bianco2011mapping} and the Bott index~\cite{loring2011disordered, hastings2011topological}, has changed our view that working in reciprocal space is essential since they manage to provide information on the topology of the system based exclusively on real-space computations, making them particularly suitable for the description of disordered Chern insulators.
For time-reversal topological insulators, on the other hand, much less has been reported in this regard, except for a variant of the Bott index which enables a real-space study of the $\mathbb{Z}_2$ number~\cite{loring2011disordered} or a spin Bott index~\cite{huang2018theory} in analogy with the spin Chern number~\cite{sheng2006quantum}. Consequently, the systematic characterization of disordered topological insulators remains elusive.

The aim of this chapter is to develop an alternative methodology to the Wilson loop to establish the topological nature of time-reversal invariant disordered materials, such as amorphous solids or materials with impurities. Here we will show that one can use the entanglement spectrum for disordered systems where no momentum component is conserved  with the aid of artificial neural networks (ANNs)~\cite{uria2022}. Machine learning (ML) algorithms, and ANNs in particular, have been shown to accurately predict topological phases on a wide range on inputs, such as wavefunctions~\cite{mano2019application, holanda2020machine, scheurer2020unsupervised}, density matrices~\cite{carvalho2018real, che2020topological}, Berry curvature~\cite{molignini2021supervised} or Hamiltonians~\cite{zhang2018machine, sun2018deep}. It was also demonstrated that the entanglement spectrum can be used to train ML algorithms to identify topology in translationally invariant systems~\cite{tsai2020deep, tsai2021deep}, as a function of disorder in one-dimensional AIII models~\cite{zhuang2022classification}, or localization phases in interacting systems~\cite{hsu2018machine}.
In our case, we consider disordered two-dimensional (2D) time-reversal topological insulators. By training an ANN to differentiate between topological and trivial entanglement spectra in models whose invariant is known or can be computed through the Wilson loop technique, we show that we can predict the topology of the system without resorting to the calculation of momentum-space flows. More importantly, our ANN is blind to the absence or existence of a gap in the system.

\clearpage

\section{The single-point entanglement spectrum as a proxy for the topological invariant}

In section~\ref{sec:entanglement_spectrum}, we introduced yet more quantities that were shown to be related to the topology of the system, which are the entanglement spectrum and entropy~\cite{prodan2010entanglement, brzezinska2018entanglement, budich2014topological}. These magnitudes measure the degree of entanglement of our ground state between two halves of the system.  In the presence of translational symmetry, the $\mathbb{Z}_2$ index can be defined from the entanglement spectrum~\cite{alexandradinata2011trace} through its $\bm{k}$ flow, much in analogy with the HWCC flow~\cite{soluyanov2011computing}. Resorting to the insulating picture, it is known that a topologically non-trivial ground state presents in-gap single-particle states that circulate around the material, wrapping it. The flow of the entanglement spectrum thus reflects the appearance of surface states upon the separation of the two halves. 

Namely, we showed that when considering a spatial partition of the Hilbert space into two subspaces $\mathcal{H}=\mathcal{H}_A\otimes\mathcal{H}_B$, then the entanglement spectrum of the reduced density matrix $\rho_A=\text{Tr }_B\rho$ for the ground state $\ket{\Psi}=\sum_i e^{-E_i}\ket{\alpha_i}\otimes\ket{\beta_i}$ is connected to the topological nature of the system. Specifically, for non-interacting fermion systems, which are the systems under consideration, it suffices to evaluate the one-particle correlation matrix $C_{ij}$,
\begin{equation}\label{eq:correlation_matrix_no_k}
    C_{ij} = \braket{\Psi|c^{\dagger}_ic_j|\Psi}
\end{equation}
whose eigenvalues, named single-particle entanglement spectrum, can be used to construct the many-body entanglement spectrum. In presence of translational symmetry, one can choose the entanglement cut parallel to one of the two reciprocal directions, e.g. $k_y$, meaning that the correlation matrix can be written as a function of $k_y$ instead:
\begin{equation}
    C_{ij}(k_y) = \braket{\Psi|c^{\dagger}_{ik_y}c_{jk_y}|\Psi} = \sum_{k_x}e^{ik_x(x_j - x_i)}P_{ij}(\bm{k})
\end{equation}
This allows to obtain the entanglement spectrum as a function of $k_y$, which then reveals a non-trivial flow of its eigenvalues if the system is topological, akin to topological edge states, as we discussed previously. While also applicable to a disordered system, this method virtually has no advantage over the Wilson loop, given that both require diagonalizing the supercell Hamiltonian $H_{\text{SC}}(\bm{k})$ on a complete grid of the BZ of $N_{\bm{k}}$ points. From a computational standpoint, both methods would scale $\mathcal{O}(N^3N_{\bm{k}})$, where $N$ denotes the dimension of the supercell Hamiltonian $H_{\text{SC}}(\bm{k})$. Additionally, the two techniques are essentially based on the same idea, tracking a non-trivial flow of the spectrum (i.e.\ absence of gaps), which for big system sizes might prove difficult due to the relatively high number of eigenvalues compared with primitive unit cells. This could then require increasing the number of points $N_{\bm{k}}$ to resolve the flow, making the computation even more expensive.

Therefore, for disordered systems where translational symmetry is lost, ideally we would avoid resorting to inconveniently large supercells (the $N_{\bm{k}}$ copies of the supercell), and work only with $H_{\text{SC}}(\bm{k}=0)$ if considering PBC, or simply $H_{\text{SC}}$ for OBC\@. In this case, we argue that the entanglement spectrum of either the finite system or the $\bm{k}=0$ supercell, which we denote as single-point entanglement spectrum, contains the information necessary to determine the topological invariant of the system. This is understandable since in the crystalline case, an unfolding procedure of the eigenvalues would reveal the flow the spectrum. The problem then, is how to extract the topological invariant from the spectrum of the single-particle correlation matrix $C_{ij}$ in Eq.~\eqref{eq:correlation_matrix_no_k}, since it is not obvious how to unfold the eigenvalues to recover the $\bm{k}$ dependence and reveal the flow. 

\begin{figure}[h]
    \centering
    \includegraphics[width=0.7\textwidth]{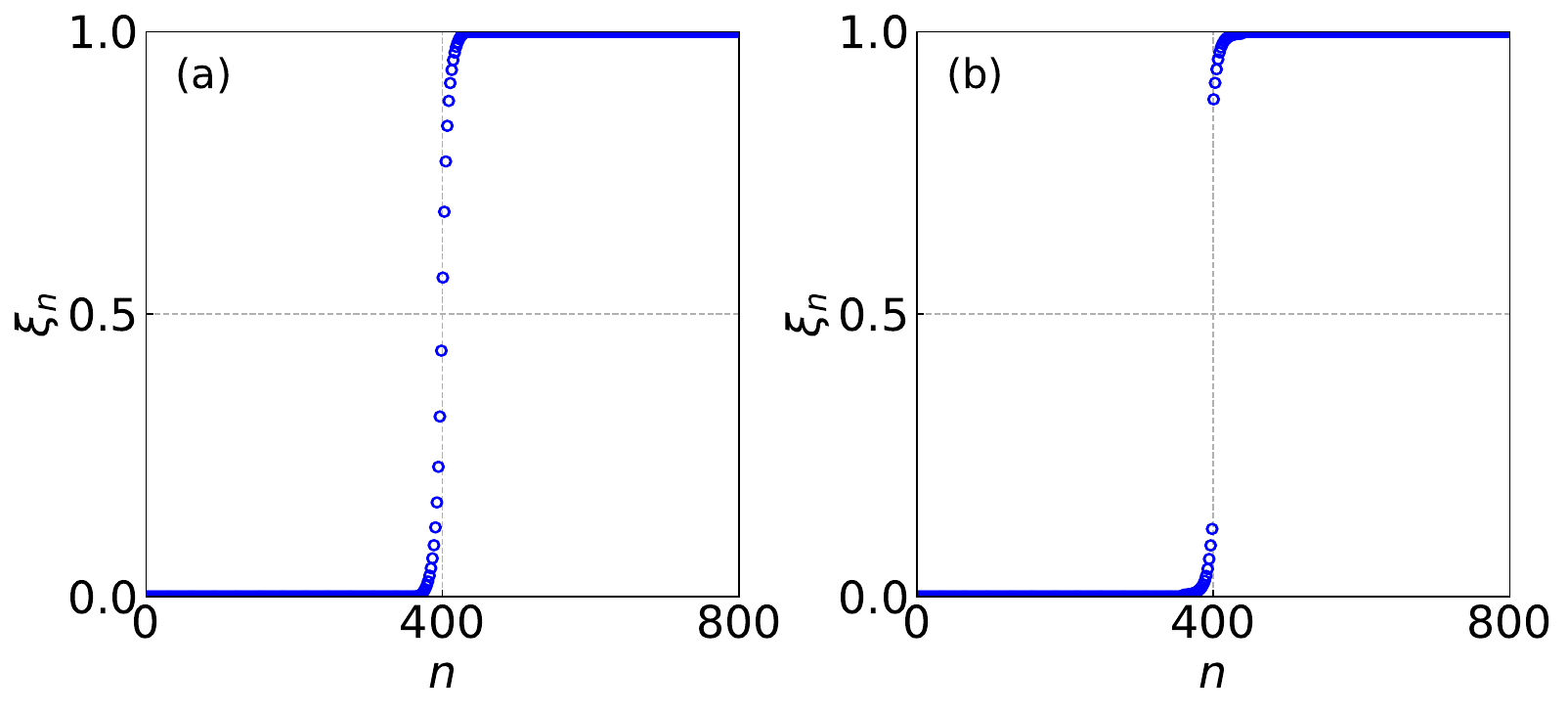}
    \caption[Single-point entanglement spectrum samples of the Wilson-Dirac lattice model]{Samples of the single-point entanglement spectrum for the Wilson-Dirac lattice model in 2D, defined in Eq.~\ref{eq:wilson_dirac_amorphous}, for (a) $M=2.5$ (topological) and (b) $M=5.1$ (trivial). The distinction between the two spectra at first glance seems apparent, although it does not have to be the case for all systems. Here, one advantage of the ANN is also the automatic classification of these spectra.}\label{fig:single-point_entanglement}
\end{figure}

Based on this, we expect the single-particle entanglement spectrum to encode features that can be used to tell apart the system from being trivial or topological, see Fig.~\ref{fig:single-point_entanglement} for examples of such spectra. To be able to do so, on top of the automated invariant computation which is necessary to establish the phase diagrams, we leverage the capabilities of an artificial neural network (ANN) to predict the topological invariant solely from the single-point entanglement spectrum. For topological materials one would naively expect states evenly scattered across the gap of the entanglement spectrum, coming from the folding into $\bm{k}=0$, although flattened Hamiltonians might present midgap energies as well without being necessarily topological~\cite{hughes2011inversion}. Otherwise, the spectrum does not present intricate features, meaning that a simple network should suffice to classify correctly the phases. 

The training procedure will be specified in the next section, when we introduce the details of the network architecture used, as well as the dataset employed to train the ANN and the results from the training. Thus, once the ANN is trained, this method will prove advantageous for disordered systems when compared with the Wilson loop, as it will only require one diagonalization for the single-point entanglement spectrum. Furthermore, its use will be mandatory when the system becomes gapless, as it usually happens for strong disorder. These arguments apply to the single-point entanglement spectrum, obtained either with PBC or OBC\@. In what follows for the rest of the chapter, we will assume to be working with the $\bm{k}=0$ supercell, namely PBC with only $N_{\bm{k}}=1$. 

\clearpage

\section{Application to the amorphous Wilson-Dirac lattice fermions model}

\subsection{Characterization of the crystalline model}

We test the above ideas with a simple model of a time-reversal topological insulator, the Wilson-Dirac lattice fermion model~\cite{fradkin2013field}. This model is obtained from the discretization of the Dirac equation on a cubic lattice, with an additional mass term corresponding to a second derivative of the field~\cite{montvay}, named the Wilson term. The purpose of the term is to remove unphysical states when comparing with the continuum Hamiltonian~\cite{wilsonchandra}. The Hamiltonian in real-space reads:

\begin{align}
    H &= \underbrace{\sum_{i,\mu}\left[ i\frac{t}{2}c^{\dagger}_{i+\mu}\alpha_{\mu}c_{i}  + \text{h.c.}\right] + \sum_i mc^{\dagger}_i\beta c_i}_{\text{Dirac eq.\ discretization}}  + \underbrace{\sum_{i,\mu} r\left(\frac{1}{2}c^{\dagger}_i\beta c_{i + \mu}  + \frac{1}{2}c^{\dagger}_{i+\mu}\beta c_{i} - c^{\dagger}_i\beta c_i\right)}_{\text{Wilson term}} \\ 
    &= \sum_{i,\mu}\left[ i\frac{t}{2}c^{\dagger}_{i+\mu}\alpha_{\mu}c_{i} + \frac{r}{2}c^{\dagger}_{i+\mu}\beta c_{i} + \text{h.c.}\right]
        + (M - 3r)\sum_i c^{\dagger}_i\beta c_i
    \label{wilsondiracrealspace}
\end{align}
where the index $i$ sums over lattice positions, and $\mu$ sums over spatial coordinates ($\mu=x,y,z$). Therefore, we are specifying hoppings only between first neighbours along the Cartesian axis (since it is a cubic lattice). $\{\alpha_{\mu}\}_{\mu}, \beta$ denote gamma matrices. They are given by:
\begin{align*}
    \alpha_{\mu} &= \sigma_{x}\otimes\sigma_{\mu} = 
    \left(\begin{array}{cc}
    0 & \sigma_{\mu} \\
    -\sigma_{\mu} & 0
    \end{array}\right), \ \ \ 
    \beta = \sigma_z\otimes I = 
    \left(\begin{array}{cc}
    I & 0 \\
    0 & -I
    \end{array}\right)
\end{align*}
Therefore $c^{\dagger}_i$ $(c_i)$ denote four-component spinors. In the following, we will fix $r=1$, which is a conventional value when working with the Wilson-Dirac model. Setting also $t=1$, if we consider a periodic cubic lattice, the Bloch Hamiltonian is given by:
\begin{equation}
    H =\overrightarrow{\sin p}\cdot\vec{\alpha} + M(\vec{p})\beta
\end{equation}
where $p_{\mu} = k_{\mu}a$, $a$ being the lattice spacing and
\begin{align}
    \nonumber \overrightarrow{\sin p} &= (\sin p_x, \sin p_y, \sin p_z), \\
    \vec{\alpha} &= (\alpha_x, \alpha_y, \alpha_z), \\
    \nonumber M(\vec{p}) & = \cos p_x + \cos p_y + \cos p_z + M - 3.
\end{align}
Depending on the value of $M$, this model describes different topological phases in three dimensions. Specifically, for $0>M>2$ and $4>M>6$ the model is in the strong topological insulator phase. If instead $2>M>4$, then the model is in the weak topological insulator phase. For any other value of $M$, the model is in the topologically trivial phase. 

\begin{figure}[t]
    \centering
    \includegraphics[width=0.9\columnwidth]{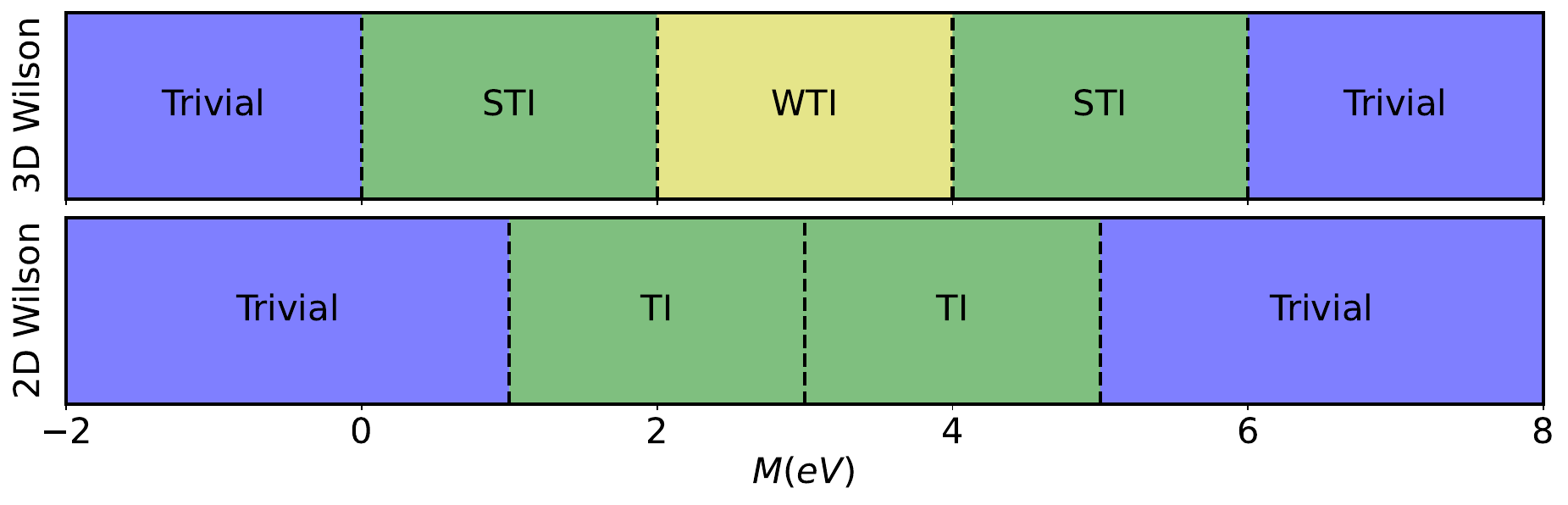}
    \caption[Topological phase diagram of the crystalline Wilson-Dirac fermions model]{Diagram with the topological phases present in the Wilson-Dirac model in three dimensions (top) and two dimensions (bottom) as a function of the mass $M$.}\label{fig:crystaldiagram}
\end{figure}

\begin{figure}[h]
    \centering
    \includegraphics[width=1\columnwidth]{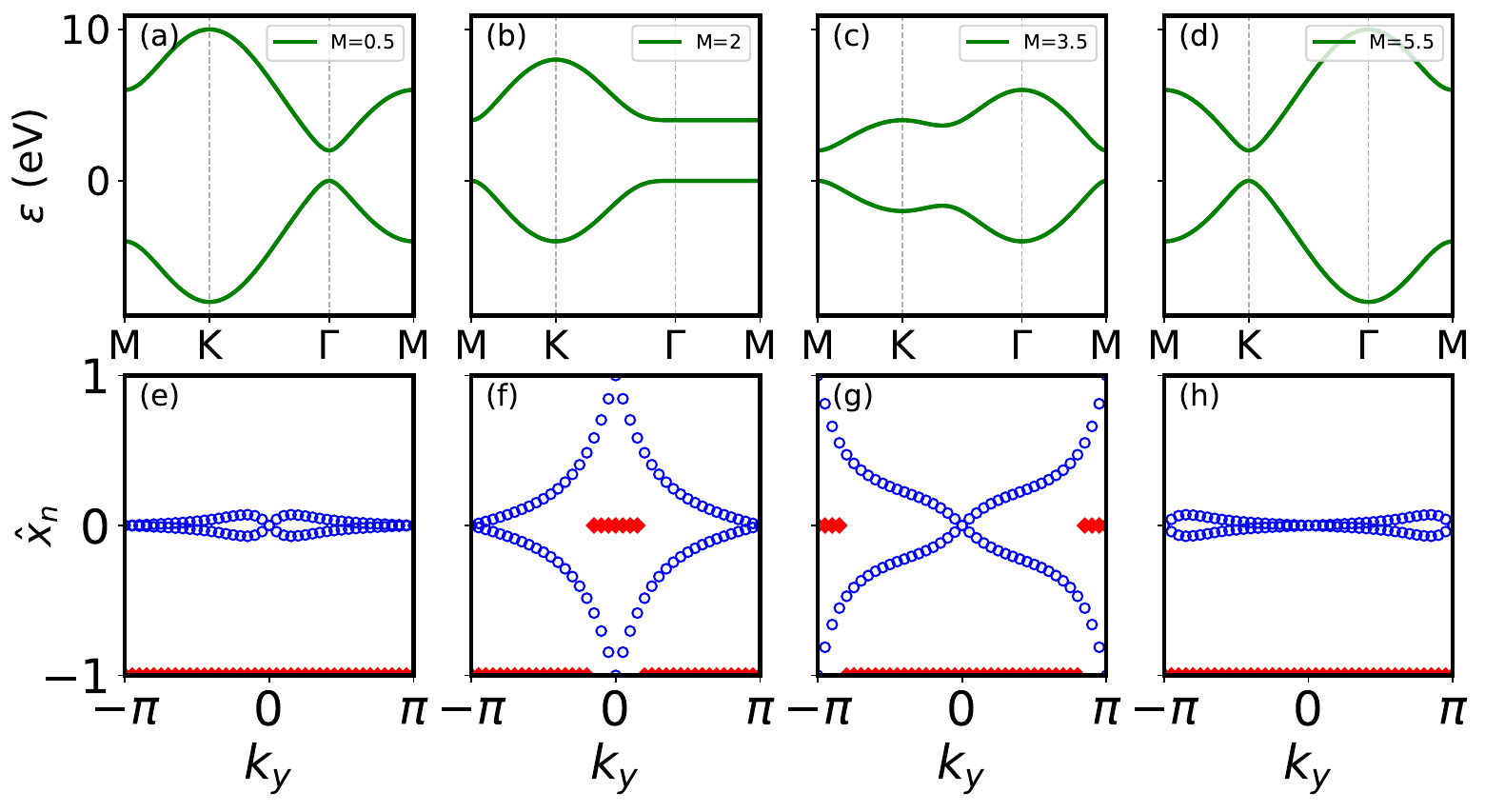}
    \caption[Examples of band structures and HWCCs for different values of $M$ in the Wilson-Dirac model]{Bands and Wilson loops for different values of $M$ for the Wilson-Dirac fermion model in 2D. The top row corresponds to the bands, while the Wilson loops appear in the bottom row.\ (a), (e) $M=0.5$.\ (b), (f) are for $M=2$.\ (c), (g) correspond to $M=3.5$, and (d), (h) to $M=5.5$.}\label{bands_wilson_loops}
\end{figure}

So far we have considered the model in three dimensions, but we are only interested in the two-dimensional model. While Eq.~\eqref{wilsondiracrealspace} holds independently of the dimension, for a square lattice the reciprocal Hamiltonian has the same shape, but the vectors are now two-dimensional:
\begin{align*}
    \overrightarrow{\sin p} &= (\sin p_x, \sin p_y) \\
    \vec{\alpha} &= (\alpha_x, \alpha_y) \\
    m(\vec{p}) & = \cos p_x + \cos p_y + M - 3
\end{align*}
The resulting topological phase diagram for the two-dimensional system differs from that in three dimensions, which is to be expected since in two dimensions there can not be weak phases. In the 2D model, we distinguish the following phases: For $1 > M > 3$ and $3 > M > 5$ it is a topological insulator, with a gap closing at $M=3$, while for any other values of $M$, it corresponds to a trivial insulator. Both phase diagrams for 2D and 3D are summarized in Fig.~\ref{fig:crystaldiagram}. This can be readily checked by computing the $\mathbb{Z}_2$ invariant with the help of the HWCCs or eigenvalues of the Wilson loop $W(k_y)=\prod_{k_i\in\text{path}}M_{k_i,k_{i+1}}$, where $M_{k_i, k_{i+1}}=U^{\dagger}(k_i, k_y)U(k_{i+1}, k_y)$ and $U(k_i, k_y)$ is the unitary matrix that diagonalizes the Bloch Hamiltonian in the atomic gauge~\cite{vanderbilt2018berry}. Some examples of the electronic bands for different values of $M$ and the corresponding Wilson loops are shown in Fig.~\ref{bands_wilson_loops}. \\

It has also been shown that this model can also realize non-trivial topology in disordered systems~\cite{mano2019application, agarwala2017topological, kobayashi2013disordered, grushin2023topological}. Following~\cite{agarwala2017topological}, we introduce a generalized version of the Wilson-Dirac model~\eqref{wilsondiracrealspace} to describe crystalline disorder or amorphous solids:
\begin{align}\label{eq:wilson_dirac_amorphous}
    H = \sum_{i,j}& \frac{i}{2}t(R)c^{\dagger}_i(\sin\phi\sin\theta\alpha_x + \sin\phi\cos\theta\alpha_y +  
    \cos\theta\alpha_z - i\beta)c_j+ \sum_i\beta (M - 3)c^{\dagger}_ic_i.
\end{align}
which can be obtained simply considering hoppings in directions $(\sin\phi\sin\theta, \sin\phi\cos\theta,\cos\theta)$. Here, the variables $(R, \phi, \theta)$ denote the spherical coordinates of the vector determined by the relative position between lattice sites $i, j$, which for the amorphous lattice will be randomly placed near their original crystal positions. The degree of disorder is characterized by the parameter $\Delta r$, which measures the characteristic distance of the site from its crystal position. As for the hopping amplitude $t\equiv t(R)$, we introduce a dependence with the bond length $R$ through an exponential law,
\begin{equation}
    t(R) = \exp\left({\frac{a-R}{a}}\right)\theta_H(R - R_{c})
\end{equation}
where $a$ is some reference lattice spacing (which here will be set to $a=1$).
Thus, when $R=a$, we recover $t=1$ as in the original model. Note that there is also a Heaviside step function $\theta_H$, which serves as a cut-off for bonds between atoms that are too far apart, $R_c$ being the cut-off distance.
It can be seen that when the lattice sites are restricted to the cubic or square lattice, we recover the same Hamiltonian in (\ref{wilsondiracrealspace}). To conclude with the tight-binding model, it can be written in more compact form grouping the angular terms into a single matrix:
\begin{equation}
    H = \sum_{i,j}t(R)c^{\dagger}_iM_{ij}(\theta,\phi)c_j
\end{equation}
where $M_{ij}$ is:
\begin{equation}
    M_{ij}(\theta, \phi) = \frac{1}{2}
    \left(\begin{array}{cccc}
    1 & 0 & -i\cos\theta & -ie^{-i\phi}\sin\theta \\
    0 & 1 & -ie^{i\phi}\sin\theta & i\cos\theta \\
    -i\cos\theta & -ie^{-i\phi}\sin\theta & -1 & 0 \\
     -ie^{i\phi}\sin\theta & i\cos\theta & 0 & -1 
\end{array}\right)
\end{equation}
Two different systems will be studied with this methodology: first, a square lattice which is increasingly deformed by random displacements of the atoms, to the point of the crystal becoming amorphous. Second, a fractal lattice, specifically the Bethe lattice. We will show that in both cases we are able to predict the presence of topological phases by means of the entanglement spectrum, as confirmed by the direct observation of edge states.

\subsection{Neural network architecture and training procedure}

Once we have defined the model, we proceed to train the ANN to predict the topological phase of the system based on the single-point entanglement spectrum. The idea behind the ANN is to feed it the single-point entanglement spectrum of the amorphous lattice, and let the ANN predict its invariant. Consequently, the training set for the ANN will be composed of pairs of $(\{\xi_i\},\nu)$, where $\{\xi_i\}$ is the single-point entanglement spectrum and $\nu$ is the topological invariant of the system. 

Based on the shape of the spectrum, as seen in Fig.~\ref{fig:single-point_entanglement}, our election here was a one-dimensional convolutional neural network. The main reason for this is the fact that we want to extract qualitative features from an ordered set of values, namely the spectrum, although a standard fully-connected neural network would have worked as well. We follow the standard architecture of CNNs, consisting on a series of convolutional layers that act as filters on the input data to extract features, and max-pooling layers to simplify the output of each layer. The extracted features are then fed to the fully-connected part of the network, which interpolate the features. The output layer corresponds to a sigmoid, since for two-dimensional materials we can only distinguish between topological and trivial phases, i.e.\ this is a binary classification problem. For three-dimensional materials, which can be strong or weak topological insulators, we must use a softmax as the output layer to be able to classify the three different phases.

\begin{figure}[h]
    \centering
    \includegraphics[width=0.95\textwidth]{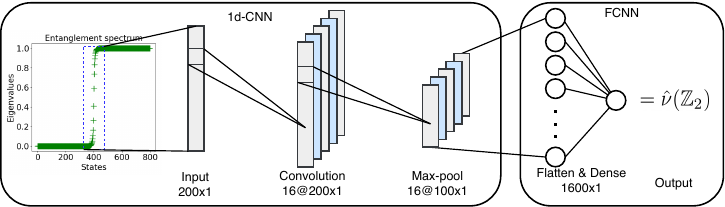}
    \caption[Architecture of the CNN trained with the Wilson-Dirac lattice fermions model]{Architecture of the one-dimensional convolutional neural network (1d-CNN) used. From the entanglement spectrum we extract only the part that is relevant to the classifier, allowing to build a smaller network. Then, a single convolution and max-pool layers are applied to extract features. These are passed onto a fully-connected neural network (FCNN) with a single dense layer to classify the spectrum as trivial or topological using a sigmoid activation function.}\label{fig:cnn_architecture}
\end{figure}

Due to the simplicity of the input data in this particular case, it suffices to use only one layer of each type, as shown in Fig.~\ref{fig:cnn_architecture}. Training with bigger networks (e.g.\ more convolutional layers and/or more dense layers) did not improve performance, which is why we use with the simplest convolutional one. The specific details of each layer used in the network can be seen in Table~\ref{tab:cnn_wilson}. Interestingly, one should be aware that inversion-symmetric materials can present non-trivial spectral flows~\cite{turner2010entanglement}, which could result in false positives. Since we are dealing with disordered topological insulators, generally breaking inversion symmetry, this will not be problematic.

\begin{table}[h]
    \centering
    \begin{tabular}{ccccc}
    \hline
    \hline
    Type of layer & Kernel & Filters & Shape & Activation function \\
    \hline 
        Input &  - & - & 1$\times$200$\times$1 & - \\
        1d convolutional & 2$\times$1 & 16 & 16$\times$200$\times$1 & ReLU \\
        Max-pool & 2$\times$1 & 1 & 16$\times$100$\times$1  & -  \\
        Fully-connected & - & - & 1600 & ReLU\\
        Output & - & - & 1 & Sigmoid \\
    \hline
    \hline
    \end{tabular}
    \caption[Architecture of the CNN used for the Wilson-Dirac fermions model]{Structure of the one-dimensional convolutional neural network used. Shape is written in format $c\times n_r\times n_c$, where $c$ is the number of channels, $n_r$ the number of rows and $n_c$ the number of columns. Since we are working with a 1D CNN, we have $n_c=1$.}\label{tab:cnn_wilson}
\end{table}

\begin{figure*}[h]
    \centering
    \includegraphics[width=0.75\textwidth]{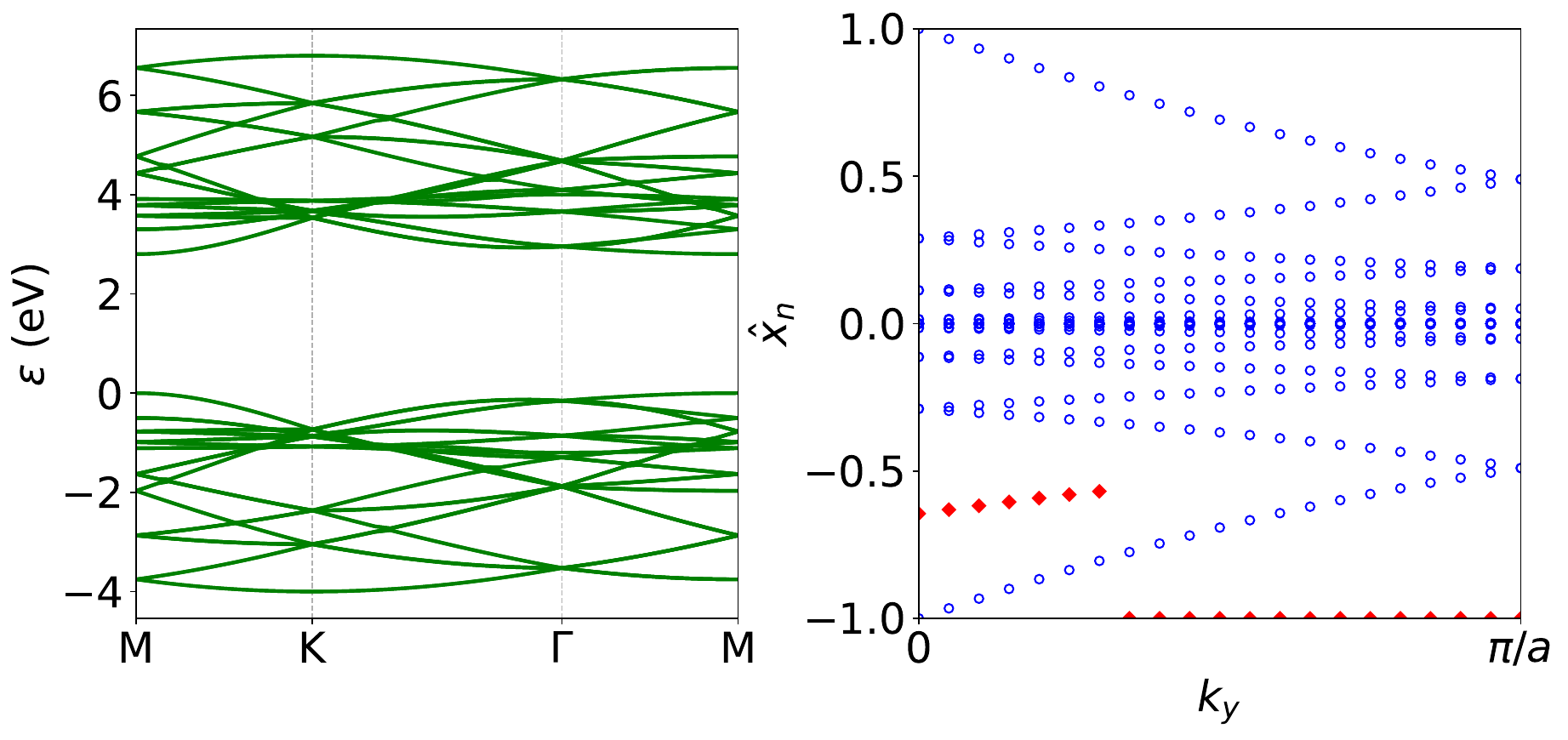}
    \caption[Band structure and HWCC flow of a supercell of the Wilson-Dirac model]{(Left) Bands of the model with $M=2.3$, $R_c=1.1$, $N_c=5\times 5$ unit cells and $\Delta r=0.01$. (Right) HWCC evolution for a particular disorder realization, corresponding to $\nu=1$.}\label{fig:wilson_supercell}
\end{figure*}

The training of the ANN is done by explicitly computing the $\mathbb{Z}_2$ invariant (when possible) and associating it with its corresponding entanglement spectrum.
We first use data from the crystalline regime (zero disorder), whose invariant is easy to compute, and then associate it with the entanglement spectrum corresponding to a supercell big enough so that it is representative of the actual samples for which we want to determine their topology. For completeness, we also use spectra corresponding to the low disorder region ($\Delta r < 0.05$) to train the ANN\@. The only difficulty here is that the computation of the topological invariant is expensive since a supercell is needed and cannot be done for strong disorder where the gap is lost. Nevertheless, as we show below, non-trivial topology can be predicted beyond the training set.
The neural network is able to extrapolate to the strong disorder region since it is learning the shape of the entanglement spectrum, rather than an explicit dependence on the phase diagram parameters.

The dataset we have used to train and assess the performance of the ANN is made of 3000 samples, of which 2500 correspond to the crystalline square lattice (i.e.\ zero disorder) for values of $M$ in $M\in\left[-1, 7\right]$, sampled uniformly. The remaining 500 samples correspond to the system with little disorder introduced, to introduce some variety in the dataset, up to $\Delta r = 0.05$. Then, the dataset was split randomly into 90\% for the training set, and 10\% for the test set. Interestingly, removing the low-disorder data from the dataset made no difference to the network accuracy, when compared with the full dataset. To visualize the dataset, some examples of the entanglement spectrum obtained for two values of $M$ are shown in Fig.~\ref{fig:single-point_entanglement} for the crystalline regime, as well as an example of the Wilson loop calculation for a slightly disordered supercell in Fig.~\ref{fig:wilson_supercell}.

The training was performed using the ADAM optimizer, with a learning rate of $\alpha=1E-5$. With higher learning rates the neural network was usually stuck in some local minimum, making random guesses (50\% for each phase). Running the training for 20 epochs we quickly reached an accuracy close to $100\%$ on both training and test sets, which is understandable since as seen in Fig.~\ref{fig:single-point_entanglement} the trivial and topological spectra can be easily distinguished.

\subsection{Topological phase diagram of the amorphous square lattice}
The specific disorder model we use is set in the following way.
Given a maximum displacement value $\Delta r$, which we take as the disorder parameter, we define the following random variables:
\begin{equation}
 r \sim U(0, \Delta r), \
 \theta \sim U(0, \pi), \
 \phi \sim  U(0, 2\pi)
\end{equation}    
where $U(a,b)$ denotes a uniform distribution between $a, b$, $a < b$. Note that out-of-plane displacements are allowed.
Out of one sampling of these variables, we generate a displacement vector given by 
$\Delta \textbf{r} = r(\sin\theta\sin\phi, \sin\theta\cos\phi, \cos\theta)$, so the final position of each atom in the supercell is $\textbf{r}_i=\textbf{r}^0_i + \Delta \textbf{r}$,
where $\textbf{r}^0_i$ denotes the crystalline lattice position. As we increase the value of $\Delta r$, the lattice becomes more disordered until long-range order is lost.

\begin{figure}[h]
    \centering
    \includegraphics[width=0.7\columnwidth]{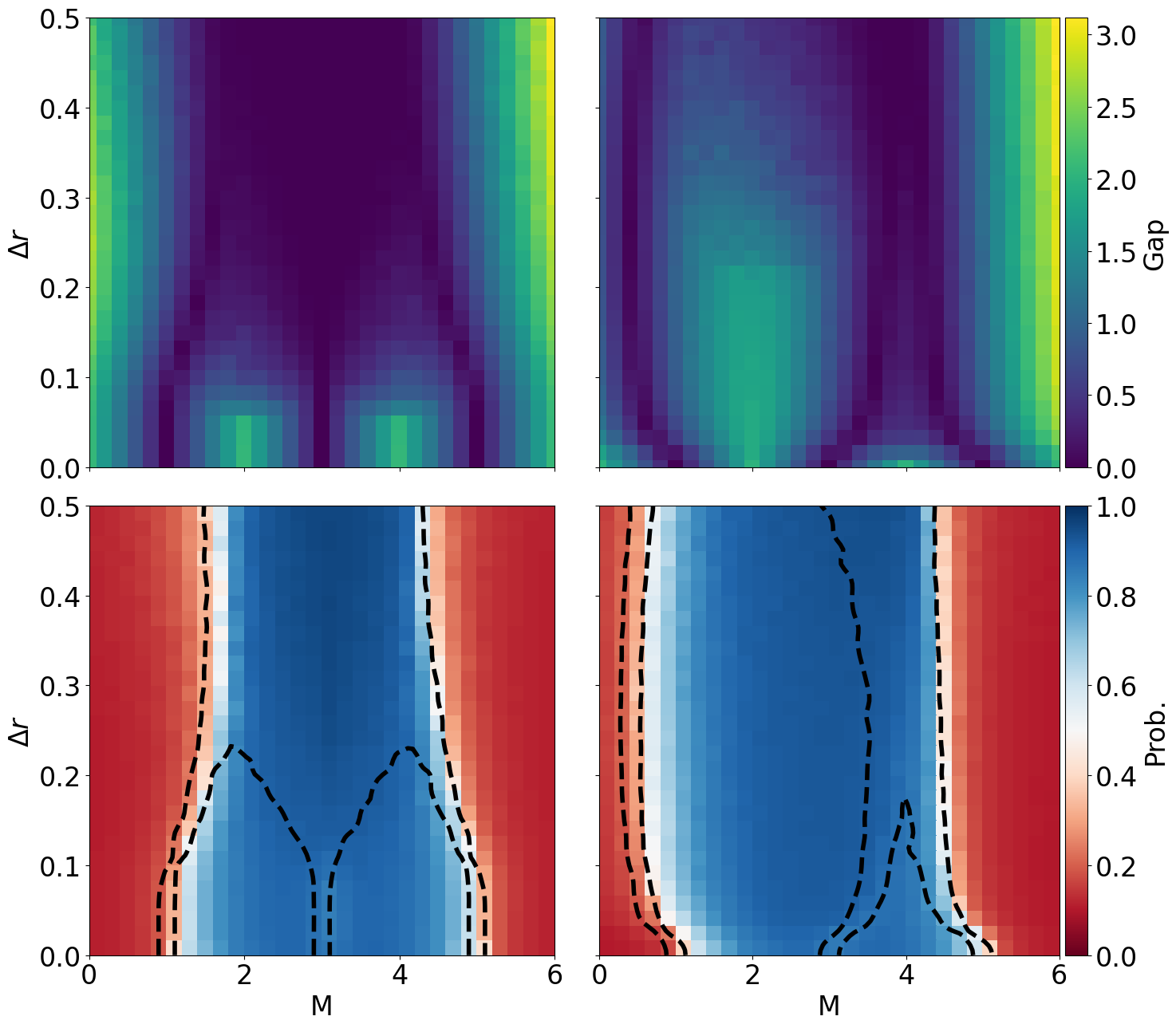}
    \caption[Gap diagram and topological phase diagram as a function of disorder and $M$]{(Top) Gap diagrams for the Wilson-Dirac fermion model on a square lattice as a function of the mass parameter $M$ and the displacement $\Delta r$ of the atoms, for two different cutoff distances. (Bottom) Topological phase diagrams predicted by the ANN, in terms of the outputted probability. Black lines correspond to contour lines
    from the gap diagram for $0.1$ eV. (Left) $R_c=1.1$, (Right) $R_c=1.4$. The cell size is $N=30$ (30 unit cells in each direction).}\label{fig:wilson_gap}
\end{figure}
\begin{figure}[h]
    \centering
    \includegraphics[width=0.7\columnwidth]{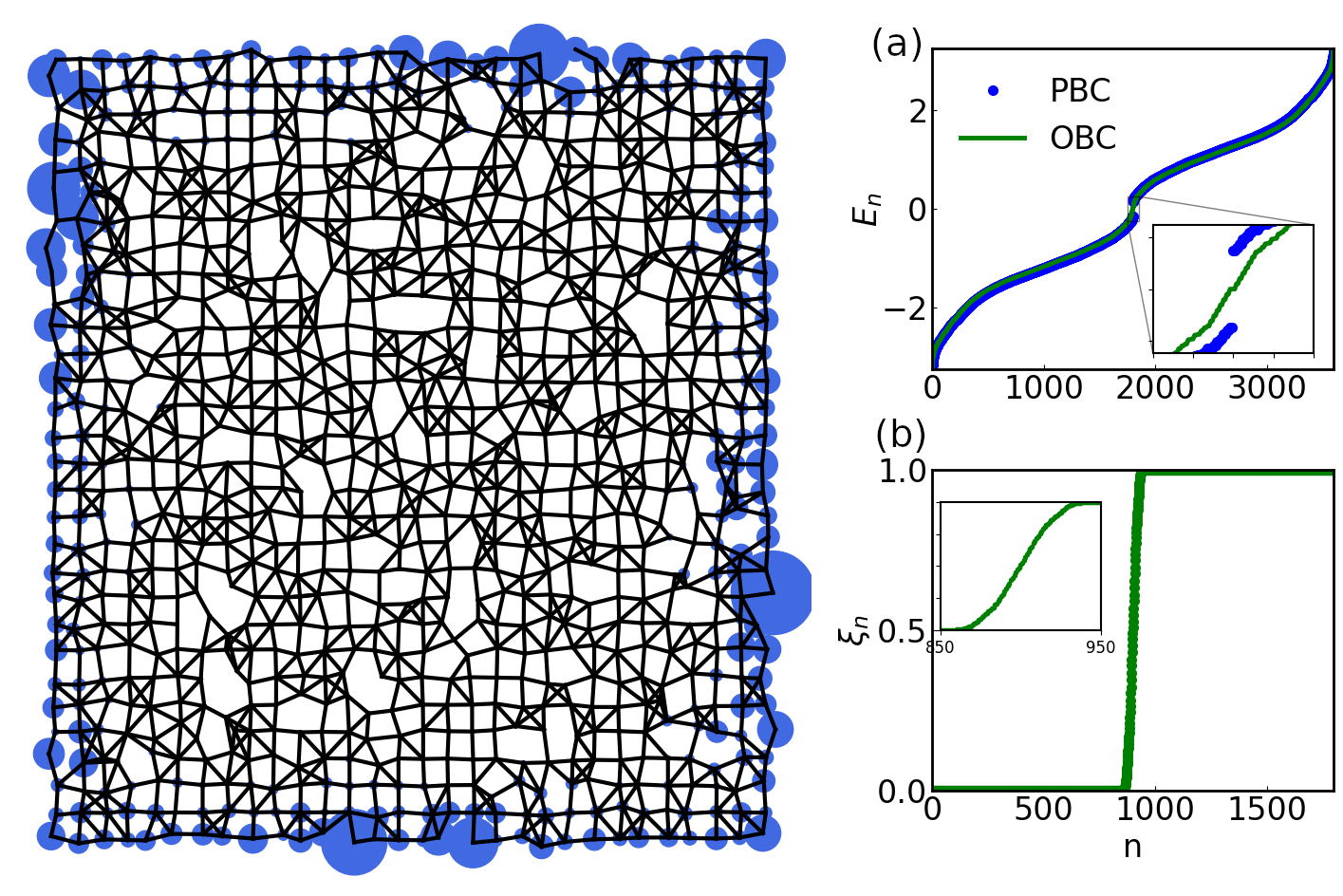}
    \caption[Topological edge state for $R_c=1.4$]{(Left) Edge state, (a) energy spectrum for both open and periodic boundary conditions and (b) entanglement spectrum for $M=3$, $R_c=1.4$, $\Delta r = 0.5$.}\label{fig:edge_state_r14}
\end{figure}
\begin{figure}[h]
    \centering
    \includegraphics[width=0.7\columnwidth]{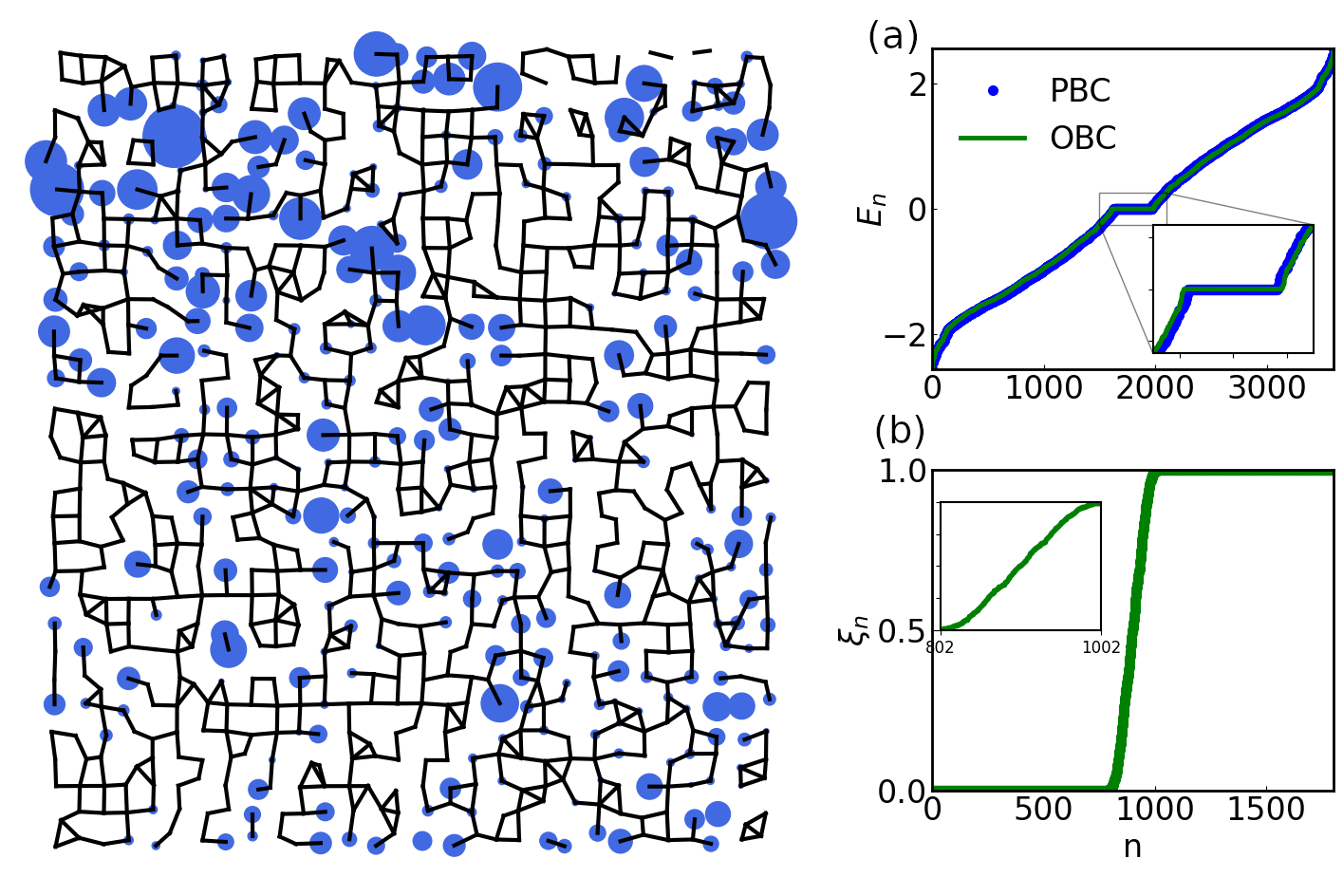}
    \caption[Topological edge state for $R_c=1.1$]{(Left) Edge state, (a) energy spectrum for both OBC and PBC and (b) entanglement spectrum for $M=3$, $R_c=1.1$, $\Delta r = 0.5$.}\label{fig:edge_state_r11}
\end{figure}

The topological phase diagram will be obtained as a function of the maximum displacement $\Delta r$ and the mass parameter $M$.
In the following we will work with one supercell only, imposing periodic boundary conditions. To obtain them, first we have to generate data, both for training and prediction. We compute the entanglement spectrum from Eq.~\eqref{eq:correlation_matrix_no_k} for the combinations of $M$ and $\Delta r$ specified before (see Figs.~\ref{fig:edge_state_r14}(c) and~\ref{fig:edge_state_r11}(c) for additional examples of entanglement spectra). The training set will be given by the points corresponding to zero or very low disorder for $R_c=1.1$. As long as the functional form of the Hamiltonian remains the same, we expect the neural network to be valid even if it has not seen data from the system with $R_c=1.4$.

\begin{figure}[h]
    \centering
    \includegraphics[width=0.74\columnwidth]{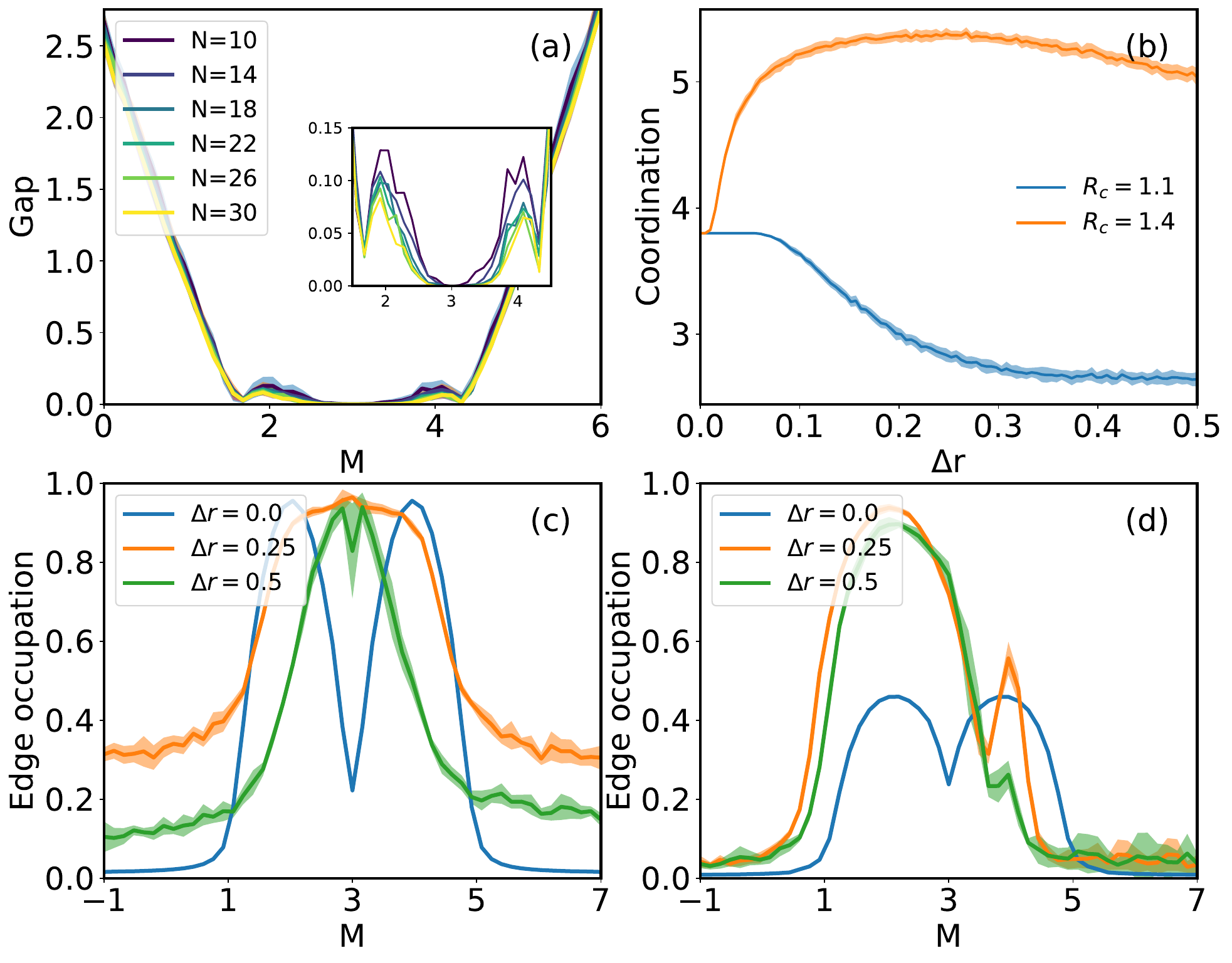}
    \caption[Gap and edge occupation as a function of $M$. Coordination number as a function of disorder]{(a) Gap for $\Delta r = 0.25$ and different system sizes.\ (b) Coordination number of the solid for different cutoff distances as the disorder increases.\ (c) Average edge occupation of the lowest 21 eigenstates in energy for $R_c=1.1$ (d) Same as (c) for $R_c=1.4$.}\label{fig:wilson_characterization}
\end{figure}

In Fig.~\ref{fig:wilson_gap} we show the topological phase diagrams obtained with the trained model for the largest systems studied (lower panels). We also show the gap (upper panels), which shows a weak dependence with the system  size [see Fig.~\ref{fig:wilson_characterization}(a)]. Here, we choose to plot the output of the ANN as an estimator of the probability of being in the topological phase. As commonly accepted, for probabilities higher than $0.5$ the system is considered to be topological. 
The model predicts the existence of topological states even in regions where the gap has vanished.

To verify the predictions of the ANN, we look for edge states near the Fermi energy where the ANN predicts non-trivial topology for disorder values outside the training set. In Fig.~\ref{fig:edge_state_r14}(a) we show an edge state obtained from one instance of the model near zero energy for $R_c=1.4$ and disorder $\Delta r = 0.5$. The gapful spectrum for this particular case is shown in Fig.~\ref{fig:edge_state_r14}(b) for periodic and open boundary conditions. We see a standard edge state in the sense that it is delocalized around the edge of the solid, as it would be expected. 
If we take a look at the bonds between atoms, the crystal has a high percolation due to more bonds appearing as we increase disorder.
However, for the smaller cutoff distance $R_c=1.1$, the solid starts to break, as shown by the diminishing coordination number in Fig.~\ref{fig:wilson_characterization}(b). This means that there are fewer paths available for a state to spread along, or equivalently that it has a lower percolation. If we now take a look at some edge state in the regions predicted to be non-trivial by the ANN, as in Fig.~\ref{fig:edge_state_r11}(a), we see that the occupation is not what we would have expected for an edge state, that is, around the borders of the solid. Still, looking closely we see that the electronic density appears mainly at the end of chains, which is the behaviour we would expect for one-dimensional topological systems. This clearly indicates that the system has undergone a transition from 2D to quasi-1D as it becomes increasingly disordered (due to the imposed cutoff between neighbours), while keeping a non-trivial topological nature. In this case, as Fig.~\ref{fig:edge_state_r11}(b) shows, there is no gap.

Finally, we can quantify the edge character in the transition from trivial to topological by looking at the average edge localization of  eigenstates near zero energy as a function of $M$, as shown in Fig.~\ref{fig:wilson_characterization}(c) and (d). As we approach the boundary between phases predicted by the ANN, there is a drastic change in edge localization, which is indicative of the phase transition.

\subsection{Topological phase diagram of the Bethe lattice}
\begin{figure}[h]
    \centering
    \includegraphics[width=0.8\columnwidth]{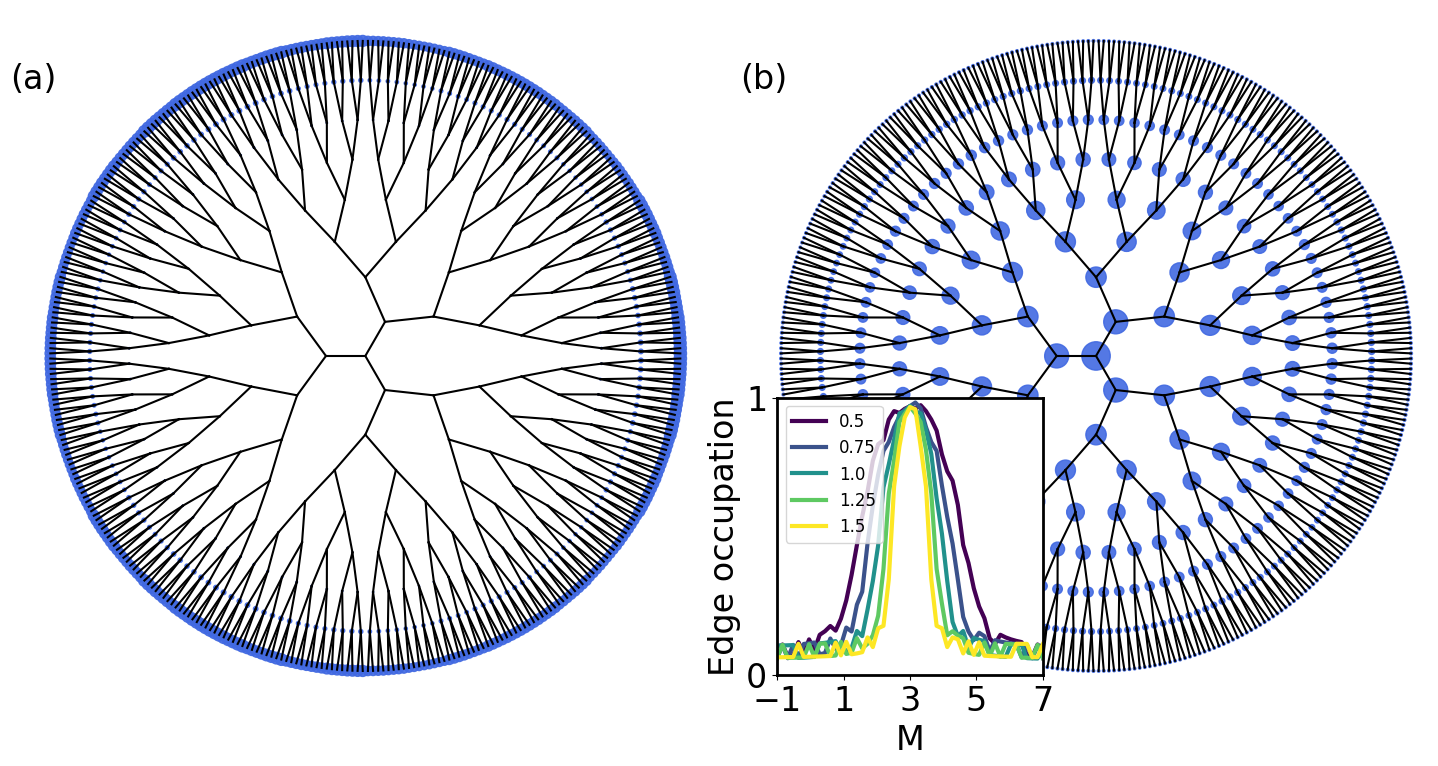}
    \caption[Topological and trivial edges state on the Bethe lattice]{(a) Edge state for $M=2.5$, $l=0.7$.\ (b) Lowest absolute energy eigenstate for $M=1$, $l=1.4$. Inset: Edge occupation as a function of $M$ for different bond lengths $l$.}\label{fig:bethe_states}
\end{figure}

We consider next a different type of system without translational symmetry, namely a Bethe lattice, which is a type of fractal lattice where the $\mathbb{Z}_2$ invariant cannot be computed by standard means. Since the underlying model Hamiltonian is the same, we expect the previously trained neural network to predict the topological phase diagram of this system as well. The Bette lattice is defined by a coordination number $z$, which specifies the number of neighbours each atom has. Then, starting from a central node, the number of nodes in each consecutive layer is given by:
\begin{equation}
    N_k = z(z-1)^{k-1},\ k>0
\end{equation}
where $k$ denotes the $k$-th layer, e.g.\ in layer 1 there are three nodes. 
It is important to mention that, mathematically speaking, the Bethe lattice is realized if the above equation is fulfilled. For us, however, the specific arrangement of the atoms is relevant since the generalized Wilson model depends explicitly on the angles between the atoms. Thus, we arrange the atoms of each layer such that the angular spacing between them is uniform. Also, the distance between every connected pair of atoms is fixed. These two constraints, with the coordination number, reproduce the lattice shown in Fig.~\ref{fig:bethe_states}.

\begin{figure}[h]
    \centering
    \includegraphics[width=0.78\columnwidth]{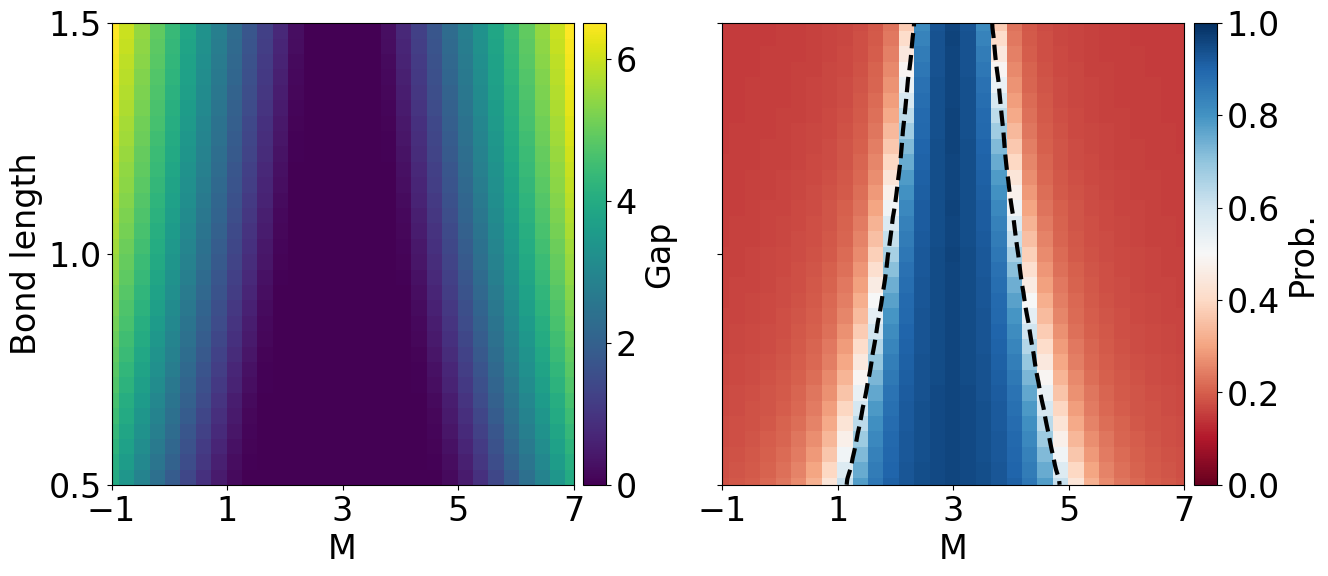}
    \caption[Gap and topological phase diagrams for the Wilson-Dirac model on the Bethe lattice]{(Left) Gap diagram for the Bethe lattice with $z = 3$ and $\text{depth}=8$. (Right) Topological phase diagram for the same model as predicted by the ANN trained with the Wilson-Dirac fermions model. Black lines correspond to contour lines from the gap diagram for 0.1 eV.}\label{fig:gap_bethe}
\end{figure}

With the model established, we proceed in an analogous fashion as with the amorphous lattice. To obtain an equivalent topological phase diagram, first we must choose some parameters that define the parameter space. As opposed to the amorphous model, for the Bethe lattice there is not a disorder parameter since the lattice is fixed. Instead, we choose the bond length. The bond length affects the hopping amplitude, effectively changing the electronic structure. The corresponding gap diagram is shown in Fig.~\ref{fig:gap_bethe}(a). 

The entanglement spectrum for different combinations of the mass and bond length parameters is fed into the neural network, which predicts the phase diagram shown in Fig.~\ref{fig:gap_bethe}(b). In this case, the whole topological region has a negligible gap. To verify that the neural network is predicting correctly the different phases, we may represent eigenstates near the Fermi energy. In Fig.~\ref{fig:bethe_states}(a) we see how an edge state appears, similar to the ones in the amorphous system for $R_c=1.1$ (note that the probability density is located mainly at the end of the different branches, which in this case happens to be also the outermost atoms). For comparison, a trivial state is also shown in Fig.~\ref{fig:bethe_states}(b), as well as the average edge occupation for several eigenstates close to the Fermi level. In all cases the results are consistent with the diagram predicted by the ANN\@.

\section{Conclusions}
The calculation of topological invariants for crystalline systems is well understood in reciprocal space, and while working in the reciprocal space may still be possible for disordered materials, computations will become too expensive due to the mandatory increase in cell size. On top of that, the most popular techniques such as the Wilson loop are only well-defined in insulating materials, difficulting the study of gapless systems such as Anderson insulators or metals. Consequently, there is presently a need for the development of techniques that ease the analysis of disordered $\mathbb{Z}_2$ TIs.

In this chapter, we address this showing that it is possible to use the entanglement spectrum of a non-translationally invariant system to train a neural network to predict topological and trivial phases~\cite{uria2022}. By training the neural network with spectra obtained from crystalline or weakly disordered phases, we can predict the topological phase diagram for generic disordered systems. We have applied it, in particular, to the case of a disordered, even amorphous lattice and to a Bethe lattice with an underlying Wilson-Dirac fermion model. This method reduces the computational time for the calculation of the invariant once we have an ANN already trained, as opposed to using, e.g., the Wilson loop technique. More importantly, it can be used with gapless systems where no other method is currently available. Additionally, this methodology could also be applied to interacting systems, where it has been shown to work in the many-body localization context~\cite{hsu2018machine}.

Lastly, one disadvantage of the present technique is that is still required one matrix diagonalization, which ultimately still limits the size of the system under study due to the $\mathcal{O}(N^3)$ scaling. We note that the training procedure introduced here is general in nature, and could potentially be applied to the deep learning of different quantities, for instance Hamiltonian deep learning, avoiding diagonalizations all together. Nevertheless, we expect that this method will allow exploring realistic models of disordered topological insulators, such as alloys as a function of the composition, topological metals and disorder-induced phase transitions in general.

\clearpage

\chapter{Amorphization-induced topological and insulator-metal transitions in bidimensional Bi$_x$Sb$_{1-x}$ alloys}
\chaptermark{Amorphization-induced topological and insulator-metal transitions in 2D Bi$_x$Sb$_{1-x}$ alloys}

\section{Introduction}

With the advent of topological phases of matter, significant effort has been devoted to the characterization of these phases, particularly into the identification of quantities that distinguish these materials, i.e.\ topological invariants and markers. As we have seen over the previous chapters, multiple approaches have been developed, being two of the most successful ones the Wilson loop, based on the behaviour of the bulk states, or the topological quantum chemistry approach, based instead on representation theory. From these efforts, the topological classification of crystalline materials is becoming complete. Consequently, the focus is shifting to disordered phases, i.e.\ systems where these techniques become less effective or entirely fail in the case of the Wilson loop if the system becomes gapless. The ultimate goal being achieving a complete classification of topological phases that includes disordered systems.

In practice, the question of how to reliably establish the topological nature of a disordered $\mathbb{Z}_2$ topological insulator remains open. 
There exist multiple methods already; in addition to the more established techniques mentioned in section~\ref{sec:introduction_chapter8}, newer methods have been introduced recently, such as different takes on the spin Chern marker~\cite{favata2023single, bau2024theory}, or the spectral localizer~\cite{cerjan2022local, franca2024topological} which could be used to assess metallic topological behaviour. Other measures that reflect topological features without directly computing the system's invariant include the spillage~\cite{munoz2023structural}, which uses the projector onto the ground state to detect band inversions.
In Chapter~\ref{chapter:deep_learning} we addressed this question, proposing our own solution to the problem. It consisted in using a conjunction of the entanglement spectrum of the system, together with ANNs to predict the $\mathbb{Z}_2$ index. In this regard, machine learning (ML) techniques, especially artificial neural networks (ANNs), have been shown to successfully determine topological invariants from a variety of inputs. Overall, ANNs have found extensive applications in condensed matter physics~\cite{carrasquilla2020machine, bedolla2020machine} and have kept up with the more modern developments of deep learning, such as attention mechanisms and transformers~\cite{li2024deep, viteritti2023transformer, luo2022autoregressive}.

\begin{figure}[h]
    \centering
    \includegraphics[width=1\textwidth]{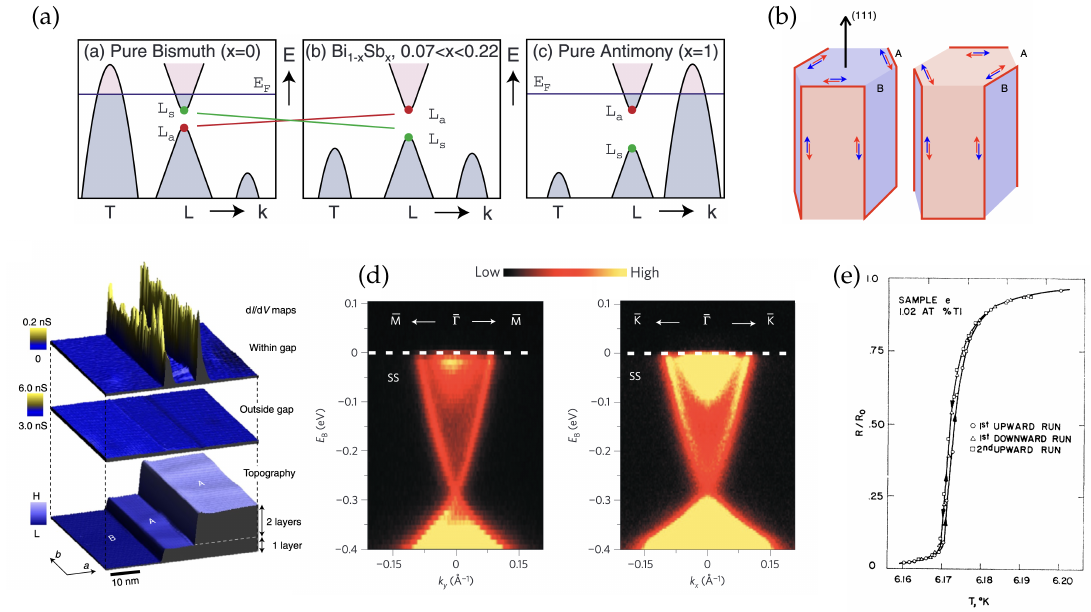}
    \caption[Examples of different topological compounds of Bi]{(a) Diagram for 3D Bi$_x$Sb$_{1-x}$ alloys showing the band inversion as a function of the concentration $x$, resulting in a strong topological insulator.\ (b) Sketch of the hinge modes appearing on 3D bulk Bi.\ (c) STM measurement of the hinge edge states appearing in stacked layers of Bi$_4$Br$_4$, which form a HOTI.\ (c) ARPES measurement of the surface states (SS) of Bi$_2$Se$_3$, a strong topological insulator.\ (d) Resistivity measurement of a-Bi, showing the superconducting transition. Figures adapted from~\cite{hasan2010, schindler2018higher, Shumiya2022, xia2009observation, shier1966superconducting}}\label{fig:bi_compounds}
\end{figure}

Among disordered systems, alloys and amorphous systems hold particular significance due to their experimental accessibility and presence in multiple technologies~\cite{Corbae_2023}. Only recently their topological properties began to be explored~\cite{grushin2023topological, agarwala2017topological, yang2019topological, marsal2020topological, tao2023average}. One notable group of materials are Bi-based compounds, which are known to be topological in crystalline form, see Fig.~\ref{fig:bi_compounds}. For instance, Bi(111), $\beta$-bismuthene and $\alpha$-bismuthene exhibit topological behaviour in crystalline and amorphous states~\cite{murakami2006quantum, liu2011stable, costa2019toward, sabater2013topologically, syperek2022observation, bai2022doubled}, bulk Bi is a HOTI~\cite{schindler2018higher} and the three-dimensional alloy Bi$_x$Sb$_{1-x}$ is a strong topological insulator~\cite{teo2008surface, hsieh2009observation}. Other well-known materials include Bi$_2$Se$_{3}$ or Bi$_2$Te$_{3}$ which are strong topological insulators~\cite{xia2009observation,zhang2009topological}. This motivates us to study two-dimensional alloys of Bi$_x$Sb$_{1-x}$, in both crystalline and amorphous form, as they could potentially reveal topological behaviours. While the topological properties of the crystalline 2D alloy have already been addressed in previous works~\cite{brzezinska2018entanglement, nouri2018topological, corbae2024hybridization}, the amorphous form remains unexplored. On top of the topological properties of the different compounds, amorphous bismuth a-Bi is known to be superconducting~\cite{shier1966superconducting, mata2016superconductivity}. If proven to be topological, it could lead to exotic physics as those arising from topological insulator-superconductor junctions~\cite{fu2008superconducting}.

In this chapter, we extend the methodology developed in the previous Chapter~\ref{chapter:deep_learning}~\cite{uria2022} to determine the topological phase diagrams of BiSb alloys. Specifically, we consider a Slater-Koster tight-binding model of Bi and Sb, which we use to describe both crystalline and amorphous alloys. Then, using a suitably trained ANN we map the topological phase diagram of the $\beta$ alloy (buckled lattice) and also explore the $\alpha$ lattice (puckered), typically overlooked for Bi. Furthermore, we examine the phase diagram as a function of the structural disorder, revealing a trivial to topological transition as disorder increases for a fixed concentration~\cite{uria2024amorphization}. This finding confirms a result that was previously reported for stanane~\cite{wang2022structural}. With the aid of electronic transport calculations, we investigate the phase diagram in more depth and uncover an insulator-metal transition in the strongly disordered alloy, which may reflect the behaviour of 3D amorphous Bi, known to be superconducting. Thus, we test the viability of the technique to address the topological properties of realistic models of disordered materials.

\section{Characterization of the $\beta$ and $\alpha$ allotropes}
\subsection{Slater-Koster parametrizations and band structures}\label{sec:sk_parameters}

Bi and Sb are described each one individually by a Slater-Koster (SK) tight-binding model, with the onsite energies, hopping amplitudes and spin-orbit couplings (SOC) taken from Ref.~\cite{liu_allen} (see parameter tables~\ref{tab:sk_beta},~\ref{tab:sk_alpha} at the end of the section). These parameters, while originally devised for the three-dimensional semimetals, also reproduce correctly the band structure of the 2D compounds, as we will see later. In particular, for bismuth it captures the topological behaviour of Bi(111) monolayers, which exhibit a quantum spin Hall state. The Hamiltonian of the tight-binding models is written in general as
\begin{align}
      H &= \sum_{i\alpha}\varepsilon_{i\alpha}c^{\dagger}_{i\alpha}c_{i\alpha} + \sum_{i\alpha,j\beta}t^{\alpha\beta}_{ij}c^{\dagger}_{i\alpha}c_{j\beta} + \lambda\sum_{i\alpha,j\beta}\braket{i\alpha|\mathbf{L}\cdot\mathbf{S}|j\beta}c^{\dagger}_{i\alpha}c_{j\beta}
\end{align}
where the indices $i,j$ run over lattice positions, and $\alpha,\beta$ run over orbital degrees of freedom, including spin. $\lambda$ denotes the strength of the SOC\@. The previous Hamiltonian is written either for pure Bi or pure Sb; to describe the alloys, we simply build the crystal considering different atomic substitutions in each lattice site, and mix the hopping amplitudes between Bi and Sb whenever there is a hopping involving atoms of different species, i.e.
\begin{equation}
    t^{\alpha\beta}_{\text{Bi-Sb}} = \frac{1}{2}\left(t^{\alpha\beta}_{\text{Bi}} + t^{\alpha\beta}_{\text{Sb}}\right), \quad \forall\alpha,\beta
\end{equation}
The rest of the Hamiltonian is built such that each atom retains the onsite energies and SOC intrinsic to the chemical element.

\begin{figure}[h]
    \centering
    \includegraphics[width=0.7\columnwidth]{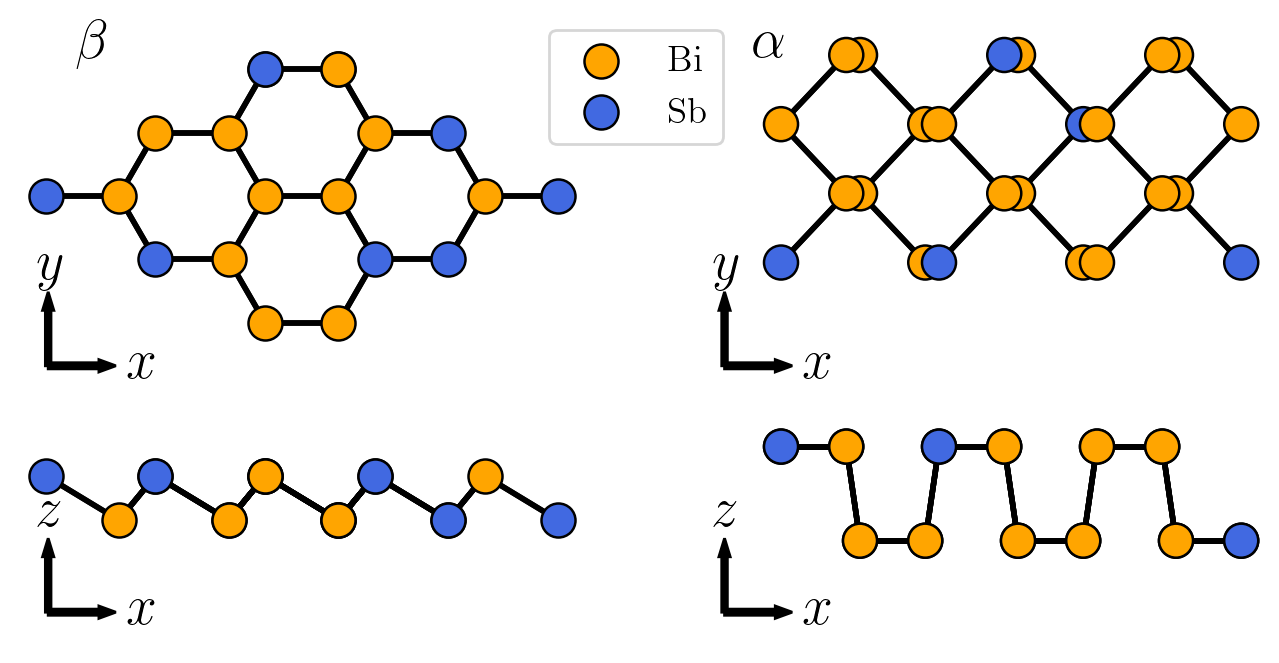}
    \caption[$\beta$ and $\alpha$ crystal structures of Bi/Sb]{Two different crystalline allotropes of $\text{Bi}_x\text{Sb}_{1-x}$ alloys: the $\beta$ crystal, which corresponds to a buckled honeycomb lattice, and the $\alpha$ crystal for the puckered lattice.}\label{fig:crystal-alloys}
\end{figure}
On this work we focus first on two different crystalline allotropes of the Bi$_x$Sb$_{1-x}$ alloys, the $\alpha$ and $\beta$ crystals. The $\alpha$ allotrope corresponds to a puckered lattice, whereas the $\beta$ one is buckled, as shown in Fig.~\ref{fig:crystal-alloys}. The lattice parameters for the two crystals have been taken from~\cite{akturk2016single, nouri2018topological}, and we use the same for both Bi and Sb. For the crystalline alloys, we will use the mixing approach to obtain the interspecies hopping amplitudes. 

The SK parameters provided in~\cite{liu_allen} can be used to describe both Bi and Sb in their different forms. However, for the tight-binding models to correctly describe the DFT band structures, we may need to consider hoppings beyond those specified in the original model. This is particularly true for the $\alpha$ crystal, where we need to consider hoppings up to fourth neighbours to correctly reproduce the band structure. The hopping parameters for the $\alpha$ and $\beta$ crystals are shown in Tables~\ref{tab:sk_beta} and~\ref{tab:sk_alpha} respectively. We only show the hopping amplitudes for Bi; the same holds for Sb. 

\renewcommand{\arraystretch}{1.5}
\begin{table}[h]
    \centering
    \begin{tabular}{ccccc}
    \hline
    \hline
        $n$-th nn  & $V_{ss\sigma}$ & $V_{sp\sigma}$ & $V_{pp\sigma}$ & $V_{pp\pi}$ \\
    \hline
    1nn & -0.608 & 1.32 & 1.854 & -0.6 \\
    \hline
    \hline
    \end{tabular}
    \caption[SK parameters of $\beta$-Bi]{Hopping amplitudes up to the $n$-th next neighbour ($n$-th nn) for the $\beta$ crystal. We show the specific values for Bi; the same holds for the description of Sb with the corresponding hopping parameters. Units are eV.}\label{tab:sk_beta}
\end{table}

\begin{table}[h]
    \centering
    \begin{tabular}{ccccc}
    \hline
    \hline
        $n$-th nn  & $V_{ss\sigma}$ & $V_{sp\sigma}$ & $V_{pp\sigma}$ & $V_{pp\pi}$ \\
    \hline
    1nn & -0.608 & 1.32  & 1.854 & -0.6   \\
    2nn & -0.453 & 0.984 & 1.382 & -0.447 \\
    3nn &   0    &   0   &   0   &    0   \\
    4nn &   0    &   0   & 0.156 &    0   \\
    \hline
    \hline
    \end{tabular}
    \caption[SK parameters of $\alpha$-Bi]{Hopping amplitudes (in eV) up to the $n$-th next neighbour ($n$-th nn) for the $\alpha$ crystal. We show the specific values for Bi; the same holds for the description of Sb with the corresponding hopping parameters.}\label{tab:sk_alpha}
\end{table}

The onsite energies in both cases correspond directly to those from~\cite{liu_allen}, so we do not show them. For the $\beta$ crystal, we only use the hopping parameters up to first neighbours (even though the original model also provides them up to second), to use the model as the starting point for the amorphous case. For the $\alpha$ alloy, as stated before we needed more complex hopping parameters to reproduce the band structures. We introduce hopping parameters to second neighbours, obtained from the hopping parameters to 1-nn but scaled with $t(r)=t_0(r_0/r)^{1.5}$, where $t_0$, $r_0$ are the reference hopping and bond length. For the third and fourth nn hopping parameters, we used those to 2-nn from the original model such that they gave similar bands to DFT\@. All three models are implemented using the \texttt{tightbinder} code~\cite{uria_tightbinder}.

\begin{figure}
    \centering
    \begin{tikzpicture}
        \node at (0,0) {\includegraphics[width=0.56\columnwidth]{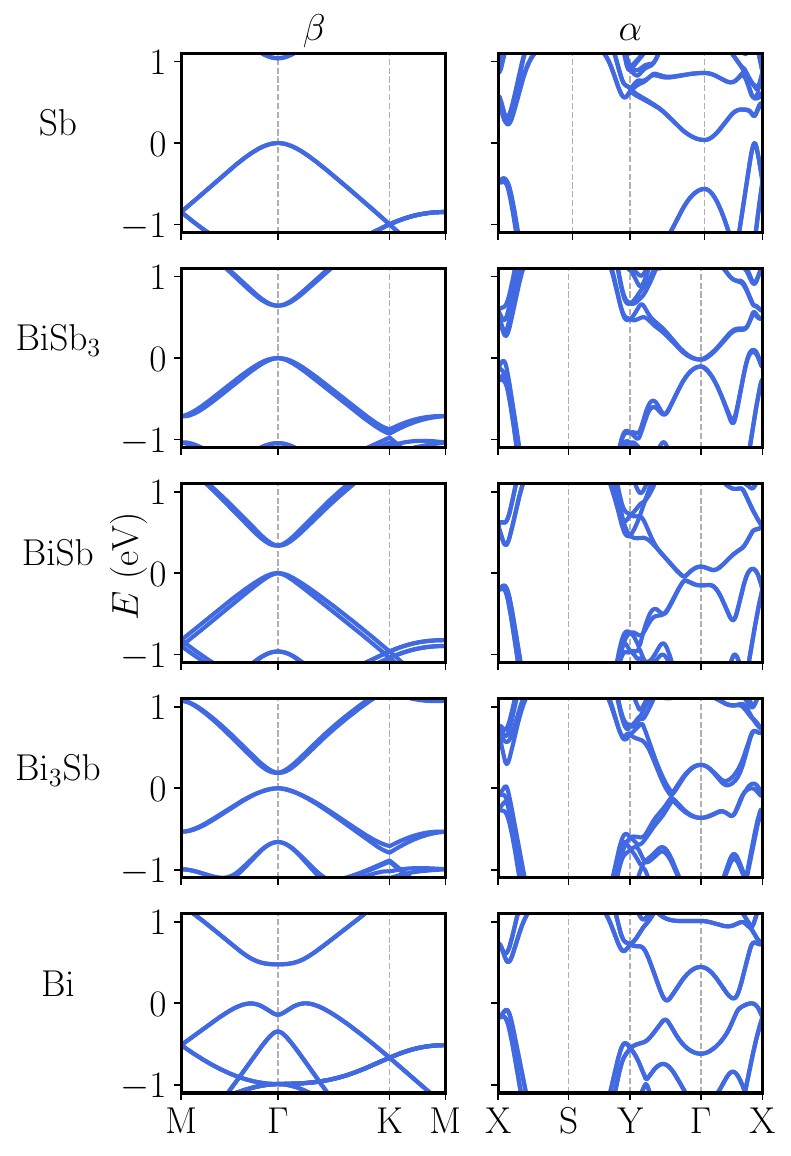}};
        \node at (-3,6.3) {(a)};
        
        \node at (7.9,0.02) {\includegraphics[width=0.42\columnwidth]{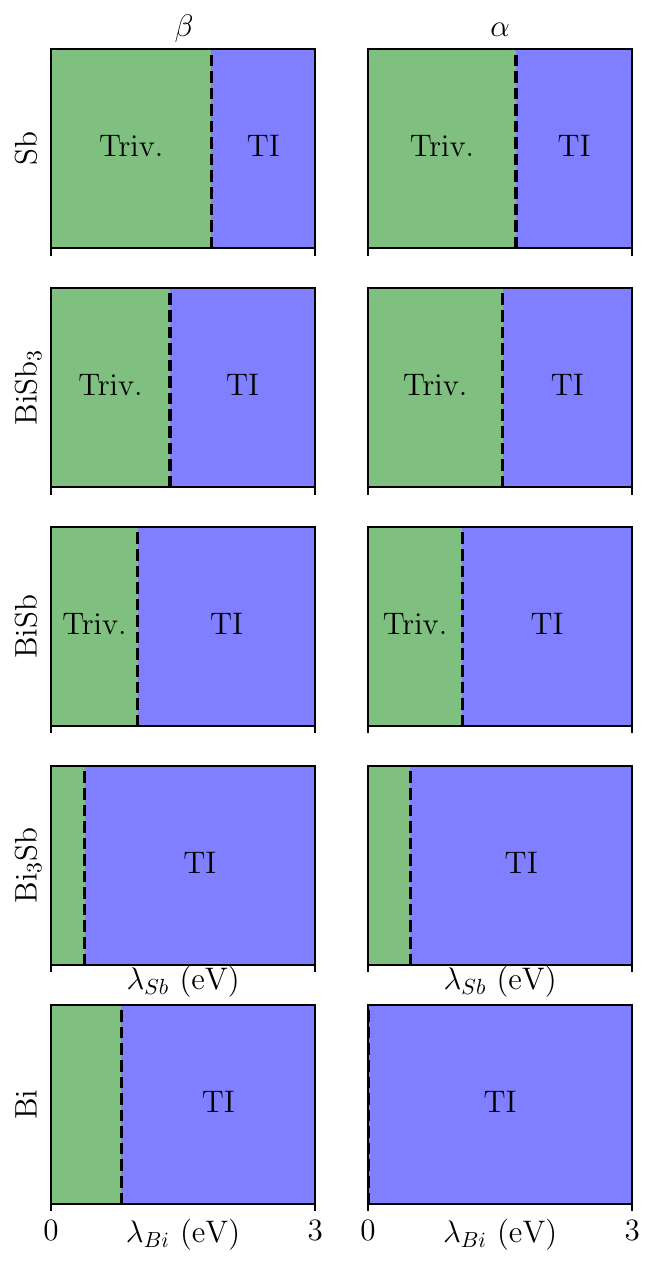}};
        \node at (4.6,6.3) {(b)};
    \end{tikzpicture}
    
    \caption[Band structure and topological phase diagrams for different $\beta$ and $\alpha$ BiSb compounds]{(a) Band structures and (b) topological phase diagrams as a function of $\lambda_{Sb}$ ($\lambda_{Bi}=\lambda_{Sb}+0.9$), for five different concentrations of the Bi$_x$Sb$_{1-x}$ alloy, for the $\beta$ structure (left column) and $\alpha$ structure (right column).}\label{fig:band_structures}
\end{figure}

The band structure for some realizations of the alloys for concentrations $x=i/4$, $i\in\{0,\ldots,4\}$ are shown in Fig.~\ref{fig:band_structures}(a), which with the SK parametrization used, we find to be in good agreement with those from DFT~\cite{singh2019low, zhou2021atomic, wang2015atomically, bai2022}. As we increase the Bi concentration, we undergo a transition from a trivial system (pure Sb) to a topological one (pure Bi). This can be seen from the fact that as $x$ increases, the gap diminishes until it starts increasing again, but now with a band inversion present. This band inversion appears at the $\Gamma$ point and can be seen in both allotropes. Thus, only from the band structures one could estimate the critical concentrations $x_c$ to be between $0.75 < x_c < 1$ for the $\beta$ crystal and between $0.25<x_c<0.5$ for the $\alpha$ one. However, as we will see these estimates are incorrect; one needs to take into account a bigger crystal and remove artificial periodicities to correctly estimate it.

For all the above concentrations we also compute the topological phase diagram as a function of the SOC $\lambda$, determined by means of the Wilson loop, shown in Fig.~\ref{fig:band_structures}(b). We observe that as we increase the concentration of Bi, the critical SOC $\lambda_c$ where the transition takes place moves to lower values, i.e.\ the effect of the concentration is to renormalize the critical SOC value. As we see from the band structures in Fig.~\ref{fig:band_structures}, the effect of the concentration is mainly related to the closing and inversion of the band gap and therefore can be understood as an effective SOC, and vice versa, the SOC can be understood as an effective concentration parameter.

\subsection{Critical concentrations and edge states in the topological regime}\label{sec:crystalline_alloys_results}

We begin applying the methodology developed in Chapter~\ref{chapter:deep_learning} to obtain the topological phase diagrams of the $\alpha$ and $\beta$ Bi$_{x}$Sb$_{1-x}$ alloys. To do so, we consider nine different stoichiometries corresponding to concentrations $x=i/8$, $i\in\{0,\ldots,8\}$. These ratios allow us to consider the smallest possible unit cells to extract easily the Wilson loop. 

\begin{figure}[h]
    \centering
    \includegraphics[width=0.7\columnwidth]{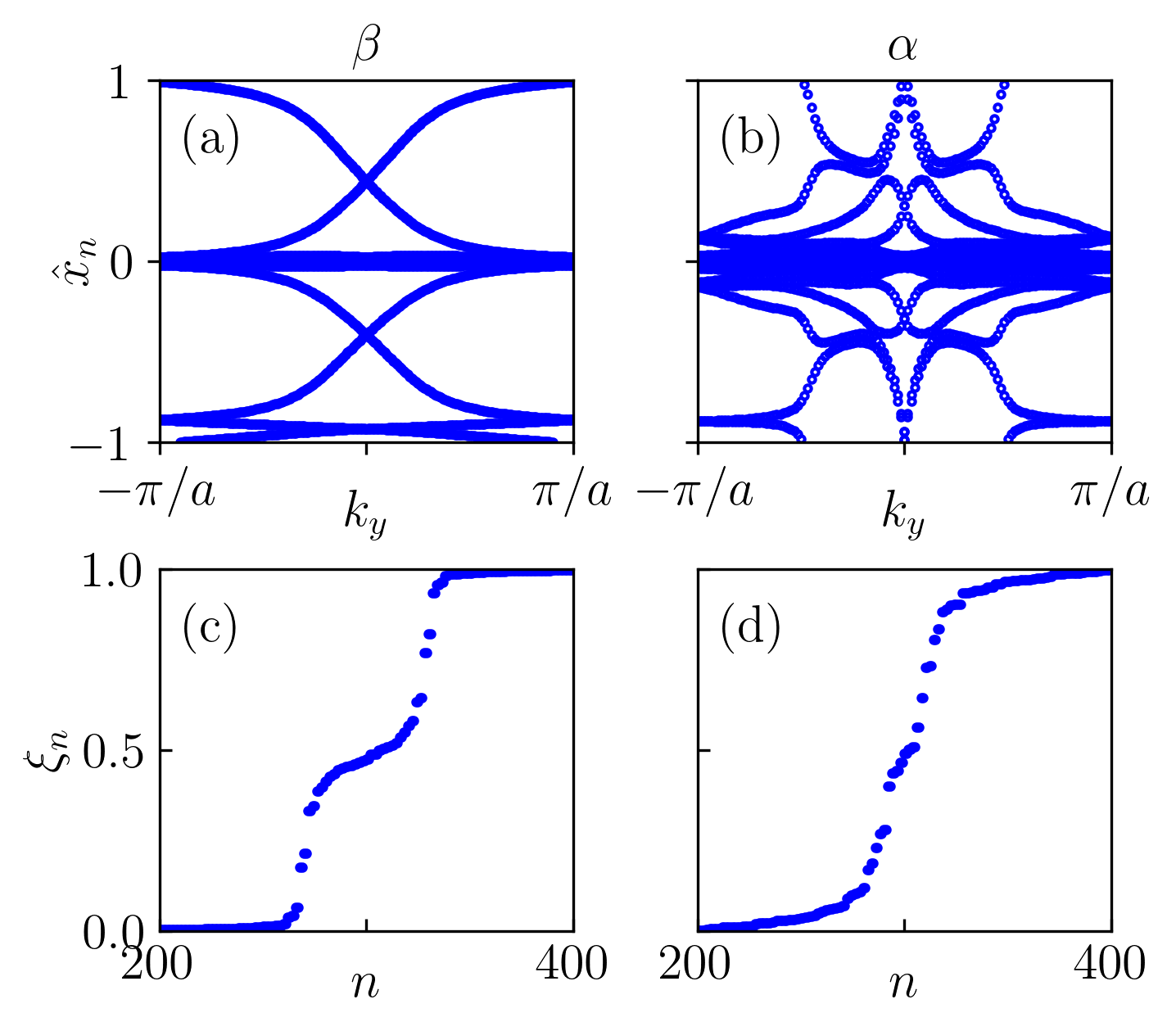}
    \caption[Wannier charge center flow and entanglement spectrum for topological samples of the $\beta$ and $\alpha$ alloys]{Wannier charge center evolution for (a) $\beta$-Bi$_{0.875}$Sb$_{0.125}$ and (b) $\alpha$-Bi$_{0.875}$Sb$_{0.125}$, both exhibiting topological behaviour. The entanglement spectrum for each system is shown in (c, d) respectively.}\label{fig:wilson_loop_entanglement}
\end{figure}

\begin{figure}[t]
    \centering
    \includegraphics[width=0.7\columnwidth]{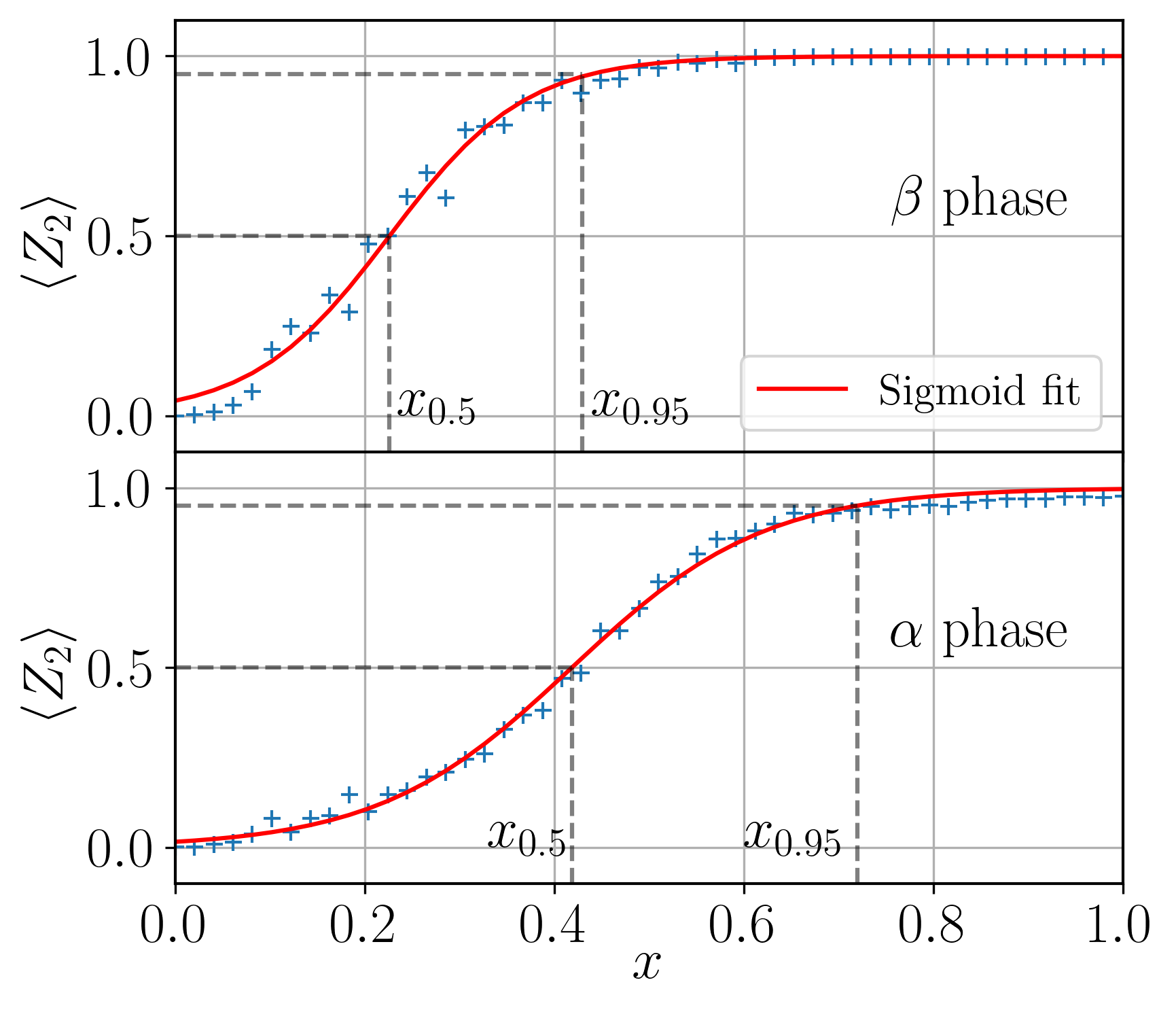}
    \caption[Topological phase diagrams as a function of the concentration for the $\beta$, $\alpha$ Bi$_x$Sb$_{1-x}$ alloys]{Topological phase diagrams for the (top) beta and (bottom) alpha alloys. The blue points correspond to the prediction of the ANN averaged over $N_s=50$ samples, while the red line is obtained fitting the points to a sigmoid function.}\label{fig:crystals_phase_diagrams}
\end{figure}

To obtain the critical concentration in a precise way, we generate training samples for the mentioned nine stoichiometries, varying the SOC in order to increase the size of the dataset. The resulting datasets are shown in section~\ref{sec:training}, Fig.~\ref{fig:training_data}. 
An example of the samples used can be seen in Figs.~\ref{fig:wilson_loop_entanglement}(c, d) for both crystals, and for a stoichiometry corresponding to the topological regime. Thus, following the workflow previously described and showed in Fig.~\ref{fig:nn_training}, in which we associate the supercell entanglement spectrum to the primitive unit cell, we create the datasets. We then train two ANNs, one for each crystal and use them to predict each topological phase diagram. These are shown in Fig.~\ref{fig:crystals_phase_diagrams}, where we now plot the average prediction of the neural networks for an intermediate set of concentrations together with a fit to a sigmoid function, $f(x)=(1+e^{-b(x - x_0)})^{-1}$, which typically describes phase transitions. We highlight two different critical concentrations: $x_{0.5}$ which is the standard definition for critical concentrations as it signals that it is more likely to be in the topological region ($p>0.5$). Additionally, we define $x_{0.95}$ in order to establish a critical concentration that indicates a \textit{global} transition of the alloy from trivial to topological. Given how the entanglement spectrum works, it suffices to make the spatial cut across a \textit{locally topological} region to mark the system as topological, while it could be trivial elsewhere. Namely, we take the probability $p$ of the neural network as the probability of the entanglement cut being at a topological region. For this reason, only for $p > 0.95$ we assume that the alloy is globally topological.

\begin{figure}[h]
    \centering
    \includegraphics[width=0.8\columnwidth]{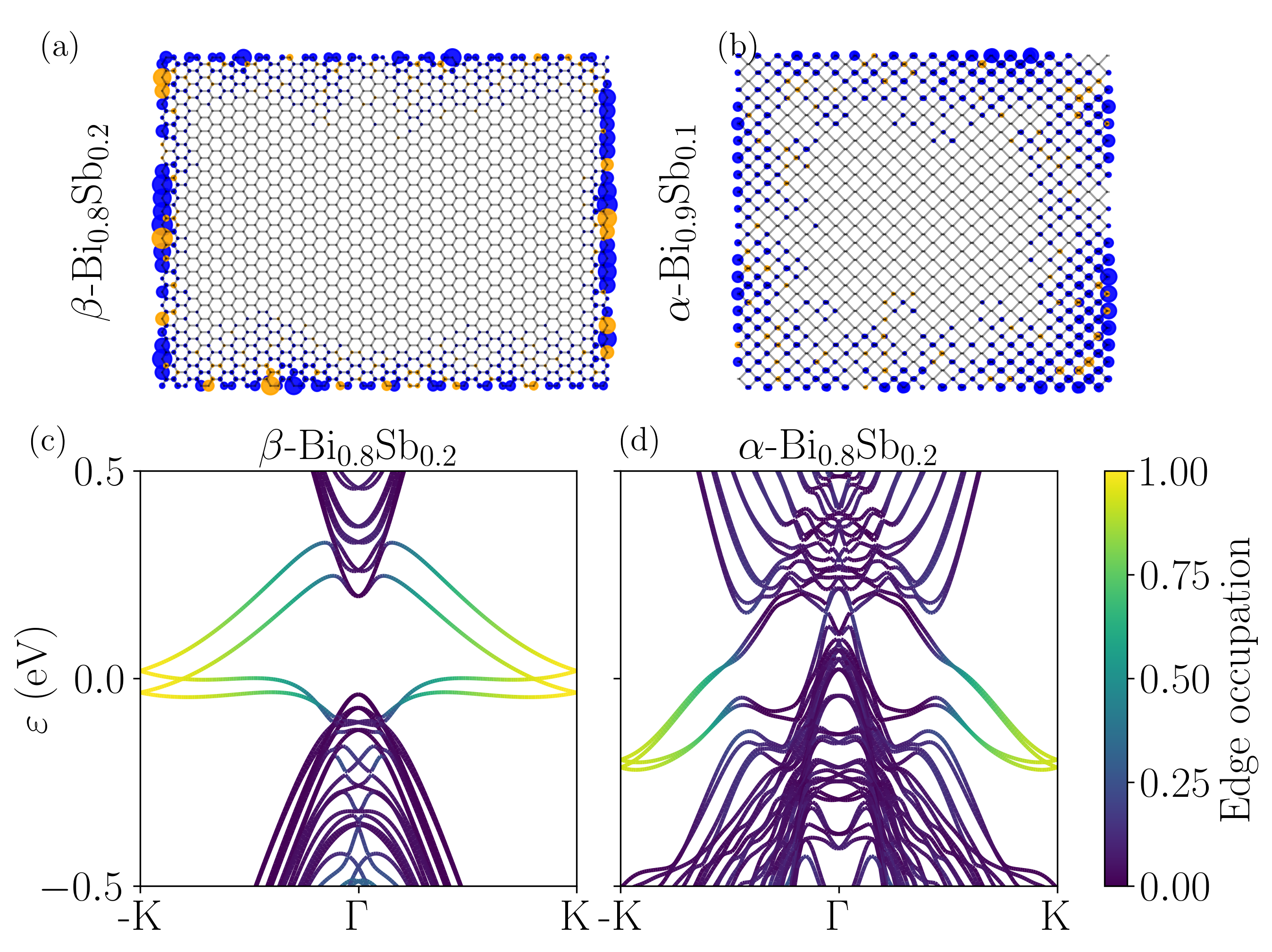}
    \caption[Topological edge states and edge bands in the crystalline alloys]{Examples of (a) topological edge state of the $\beta$-Bi$_{0.8}$Sb$_{0.2}$ alloy and (b) edge state of $\alpha$-Bi$_{0.9}$Sb$_{0.1}$. Electronic band structure of a ribbon of (c) $\beta$-Bi$_{0.8}$Sb$_{0.2}$ (50 atoms wide) and (d) $\alpha$-Bi$_{0.8}$Sb$_{0.2}$ (52 atoms wide). The color of the bands represents the probability of the state being localized at the edges of the ribbon.}\label{fig:crystals_edge_states_bands}
\end{figure}

\begin{figure}[H]
    \centering
    \includegraphics[width=0.7\columnwidth]{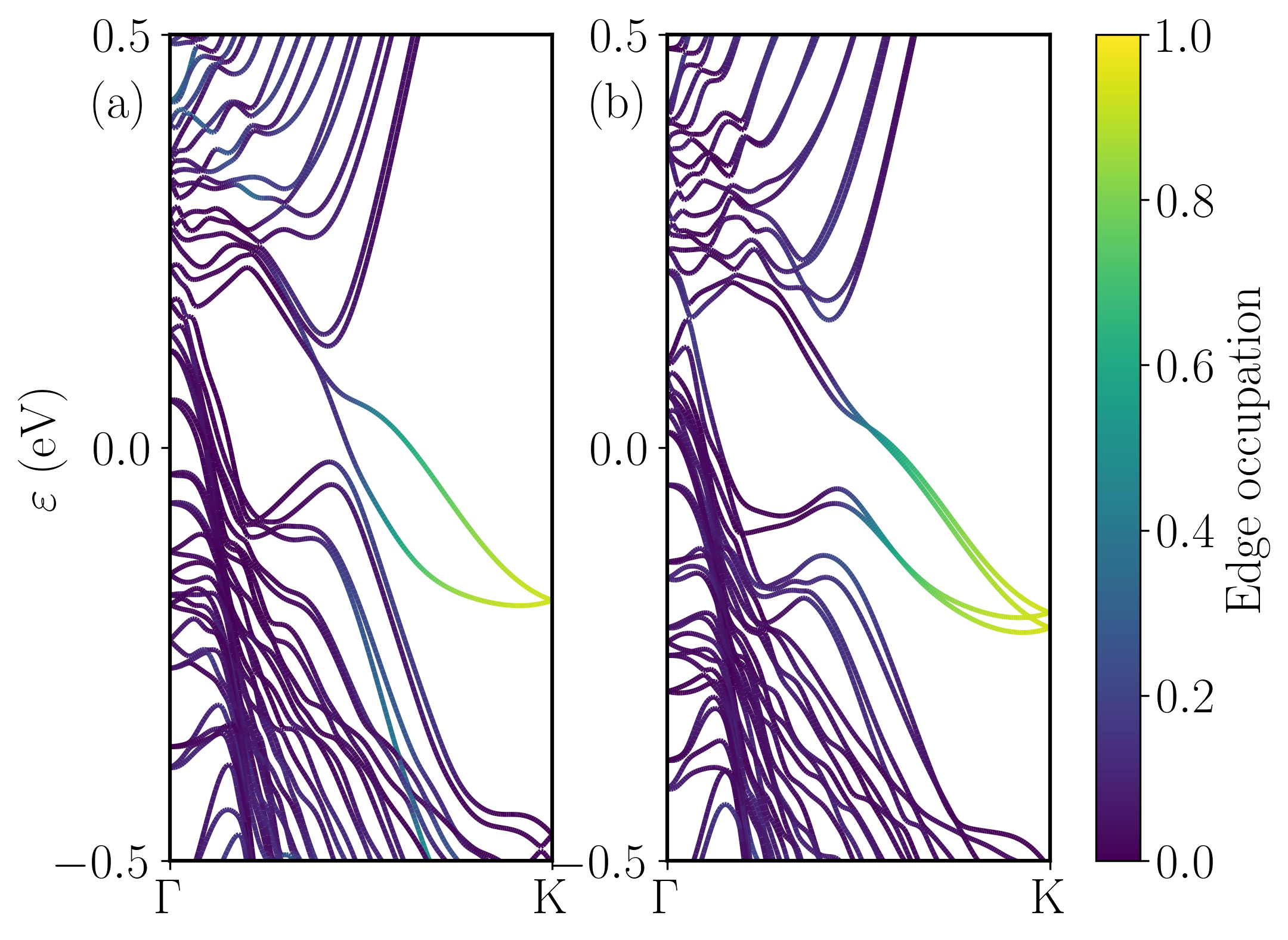}
    \caption[Comparison between trivial and topological edge bands in the $\alpha$-Bi$_x$Sb$_{1-x}$ alloy]{Comparison of the band structure of a ribbon of $\alpha$-Bi$_{0.8}$Sb$_{0.2}$ (52 atoms wide) for (a) topologically trivial sample and (b) non-trivial.}\label{fig:alpha_bands_trivial_topological}
\end{figure}

To verify the findings of the ANN, we compute the band structure of a ribbon of both alloys for some concentration beyond $x_{0.95}$. Then, if the system is topological, we should see topological edge bands connecting the valence and conduction bands. In Fig.~\ref{fig:crystals_edge_states_bands}(c) we show the bands of a ribbon of $\beta$-Bi$_{0.8}$Sb$_{0.2}$, where each band is colored according to the weight of the wavefunction at the boundaries of the ribbon. Then, we can identify perfectly the topological edge bands. In the case of the $\alpha$ alloy, in Fig.~\ref{fig:crystals_edge_states_bands}(d) we show the bands of an instance of the alloy that is topological. In this case, as opposed to the $\beta$ alloy, there is not a gap in the ribbon as a consequence of the bulk band structure, as depicted in Fig.~\ref{fig:band_structures}, where there is a direct gap in the system, but not a global gap across the whole BZ\@. To ensure that the edge bands are indeed topological, we can compare them with those of the trivial case, as shown in Fig.~\ref{fig:alpha_bands_trivial_topological}. The SK model we are considering presents trivial edge bands in the trivial regime (for both crystals), which could be confused for the topological ones. From Fig.~\ref{fig:alpha_bands_trivial_topological} their distinction becomes clear, as the topological ones still connect valence and conduction bands. 

As an additional check of topological behaviour, we also look for topological edge states in a finite sample of both alloys, again for a concentration beyond the critical one $x_{0.95}$. In Fig.~\ref{fig:crystals_edge_states_bands}(a, b) we plot an instance of a topological edge state for the $\beta$-Bi$_{0.8}$Sb$_{0.2}$ and $\alpha$-Bi$_{0.9}$Sb$_{0.1}$ alloys respectively. In both cases we identify the standard behaviour of a topological edge state, with the majority of its weight located at the boundary of the solid. Note that due to the small gap, for the $\beta$ alloy the edge state penetrates into the solid. This is more prominent in the $\alpha$ alloy, which exhibits a longer penetration length.

\section{Amorphous Bi$_x$Sb$_{1-x}$ alloys}

Having characterized the crystalline alloys, we now turn our attention to the amorphous phase. So far we have been using a more realistic approach to the modelling of the alloy, where we mix the hopping parameters of Bi and Sb to obtain the interspecies hopping amplitudes. Alternatively, the alloy can be modelled using the virtual crystal approximation (VCA), where we modify all the tight-binding parameters in a continuous way from pure Bi to pure Sb~\cite{brzezinska2018entanglement}:
\begin{align}
    \nonumber \varepsilon^{\alpha}_x &= x \varepsilon^{\alpha}_{\text{Bi}} + (1-x)\varepsilon^{\alpha}_{\text{Sb}} \\
        t^{\alpha\beta}_{x} &= x t^{\alpha\beta}_{\text{Bi}} + (1-x)t^{\alpha\beta}_{\text{Sb}},\quad x\in[0,1],\quad \forall\alpha,\beta \\
    \nonumber \lambda_x &= x \lambda_{\text{Bi}} + (1-x)\lambda_{\text{Sb}}
\end{align}
where $x$ denotes the concentration of the alloy Bi$_x$Sb$_{1-x}$. Note that this approach is unable to capture spatial fluctuations in the lattice, as opposed to the mixing approach, arguably making the latter more realistic for the description of the alloy. Namely, in the VCA we homogenize all the parameters, whereas in the mixing approach we only change the hopping amplitudes, leaving the corresponding onsite energies and spin-orbit strength of the chemical species.

For the amorphous lattice we drop the mixing approach in favour of a reduced version of the VCA to simplify the generation of the structures, where the hopping parameters are fixed and only SOC changes. This is justified, since as seen in section~\ref{sec:sk_parameters} the effect of the concentration is basically to renormalize the critical SOC\@. Consequently, we may expect the same physics fixing all the parameters, and modifying exclusively the SOC $\lambda$, taking it as an effective concentration parameter. The generation of the amorphous structure will be based for simplicity on the $\beta$ structure. Starting from the crystalline position, we introduce structural disorder via random displacements of the atomic positions. The magnitude of the displacement, $|\mathbf{\delta r}|$, is sampled from a uniform distribution:
\begin{align}
    \nonumber&\mathbf{r}' = \mathbf{r} + \mathbf{\delta r},\quad  \text{where}\\
    &|\mathbf{\delta r}|\sim U(0,\sigma r_0),\ \theta\sim U(0,2\pi),\ \varphi\sim U(0,\pi)
\end{align}
$\theta$, $\varphi$ being the angles of the displacement vector $\mathbf{\delta r}$ in spherical coordinates. $\sigma$ is the maximum displacement possible, i.e.\ the disorder strength, given in terms of the reference length $r_0$ (first neighbours in the crystalline case). Since the distances between the atoms will change as we increase $\sigma$, we need to modify the hopping parameters accordingly to properly capture the amorphous lattice. We introduce this dependency on the distance via an exponential law,
\begin{equation}
    t'^{\alpha\beta}(\mathbf{r})= t^{\alpha\beta}e^{-C(r-r_0)}\theta_H(R_c-r),\quad\forall\alpha,\beta
    \label{eq:hoppings_amorphous}
\end{equation}
where $r=|\mathbf{r}|$ and $C$ is the inverse decay length, $C=1$. Note that we have also introduced a cutoff distance $R_c$ via a Heaviside step function $\theta_H(r)$, which we set to $R_c=1.4r_0$. Thus, in the amorphous case, instead of defining a set of hopping parameters up to the $n$-th neighbour, the hopping parameters are defined up to a maximum distance, based on those to first neighbours in the crystal.

\subsection{Amorphization-induced topological transition}

Following the characterization of the crystalline alloys, we now address the topological properties\ of amorphous Bi$_x$Sb$_{1-x}$, this is, as a function of the disorder strength $\sigma$. In the previous sections we established that the concentration of the alloy can be effectively substituted by the SOC $\lambda$. For this reason, we drop the distinction between the two species, and instead fix the hopping parameters and onsite energies, and modify only the SOC strength $\lambda$. Also, given the way the hopping parameters are defined in the amorphous solid (see Eq.~\ref{eq:hoppings_amorphous}), the $\beta$ crystal will be the reference structure, as its SK model is simpler than the one for the $\alpha$ crystal (namely, the $\beta$ crystal can be described in the crystalline limit of the amorphous model, while this is not possible for the $\alpha$ crystal due to its hopping parameters to different neighbours).

\begin{figure}[h]
    \centering
    \includegraphics[width=0.75\linewidth]{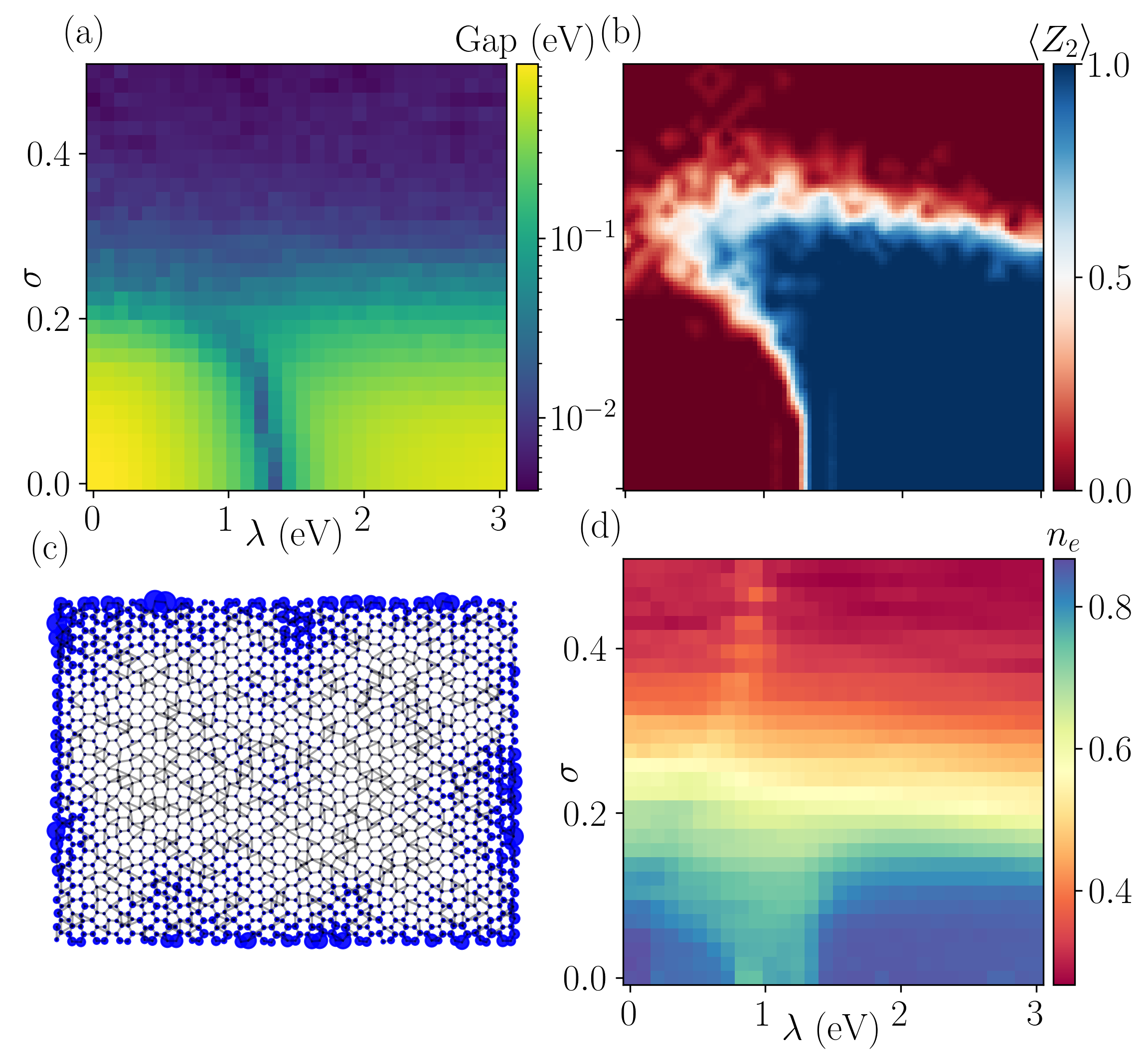}
    \caption[Gap diagram, topological phase diagram, edge occupation diagram and topological edge state of the amorphous alloy]{Diagrams for the amorphous alloy as a function of the SOC strength $\lambda$ and the disorder strength $\sigma$ for (a) gap, for a system with $N_c=16^2$ unit cells and averaged over 10 samples, and (b) $\mathbb{Z}_2$ invariant (averaged over 20 samples).\ (c) Topological edge state for the amorphous solid at $\sigma=0.25$, $\lambda=2.5$ eV. (d) Average edge occupation of the lowest $21$ state energy states as a function of $\sigma$ and $\lambda$, with $N_s=10$ samples and $N_c=20\times 15$ unit cells. At high disorder, there is a decrease in the edge occupation of the lowest states for all values of $\lambda$, possibly signaling a global topological phase transition to a trivial state.}\label{fig:gap_topo_diagrams_amorphous}
\end{figure}

A topological phase transition from a topological to a trivial insulator can only take place following a closing of the gap. For this reason we begin the characterization of the amorphous model obtaining the gap of the system to identify potential transitions. The gap as a function of both disorder and SOC strength is shown in Fig.~\ref{fig:gap_topo_diagrams_amorphous}(a). Most notably, we distinguish two different regions: at low disorder ($\sigma < 0.3$) there is a closing of the gap that moves with disorder, i.e. $\lambda_c=\lambda_c(\sigma)$, potentially showing a topological phase transition. At high disorder ($\sigma > 0.3$) the system becomes gapless $\forall \lambda$. This region can correspond potentially to four different phases: either an Anderson insulator or a metal, which could then be trivial or topological~\cite{groth2009theory, li2009topological}.

To establish the topological nature of all three regions, we resort again to the ANN\@. We follow the same procedure as before: we generate the dataset to train a new neural network, which consists of pairs of entanglement spectrum plus topological invariant determined via the Wilson loop. In this case, we want to obtain the topological phase diagram as a function of both disorder and SOC\@. For this reason, we generate training samples at zero and finite disorder, for all the SOC values considered ($\lambda\in[0,3]$). The distribution of the training data can be seen in section~\ref{sec:training}, in Fig.~\ref{fig:training_data}. Once the ANN is trained, we use it to predict the topological phase diagram of the amorphous solid, shown in Fig.~\ref{fig:gap_topo_diagrams_amorphous}(b). As we guessed, at low disorder there is a trivial to topological transition. Remarkably, the critical strength $\lambda_c$ depends on the disorder. This implies that for fixed SOC, say $\lambda=1$, if we increase disorder we undergo a transition from trivial to topological. In other words, it is an amorphization-induced topological transition. At high disorder we find the opposite behaviour: for high SOC ($\lambda > 1.3$) disorder destroys the topological phase, transitioning to a gapless trivial region. For low SOC ($\lambda < 1.3$), the system was already in a trivial region, so it does not change in that regard, although it becomes gapless.

\begin{figure}[t]
    \centering
    \includegraphics[width=0.8\linewidth]{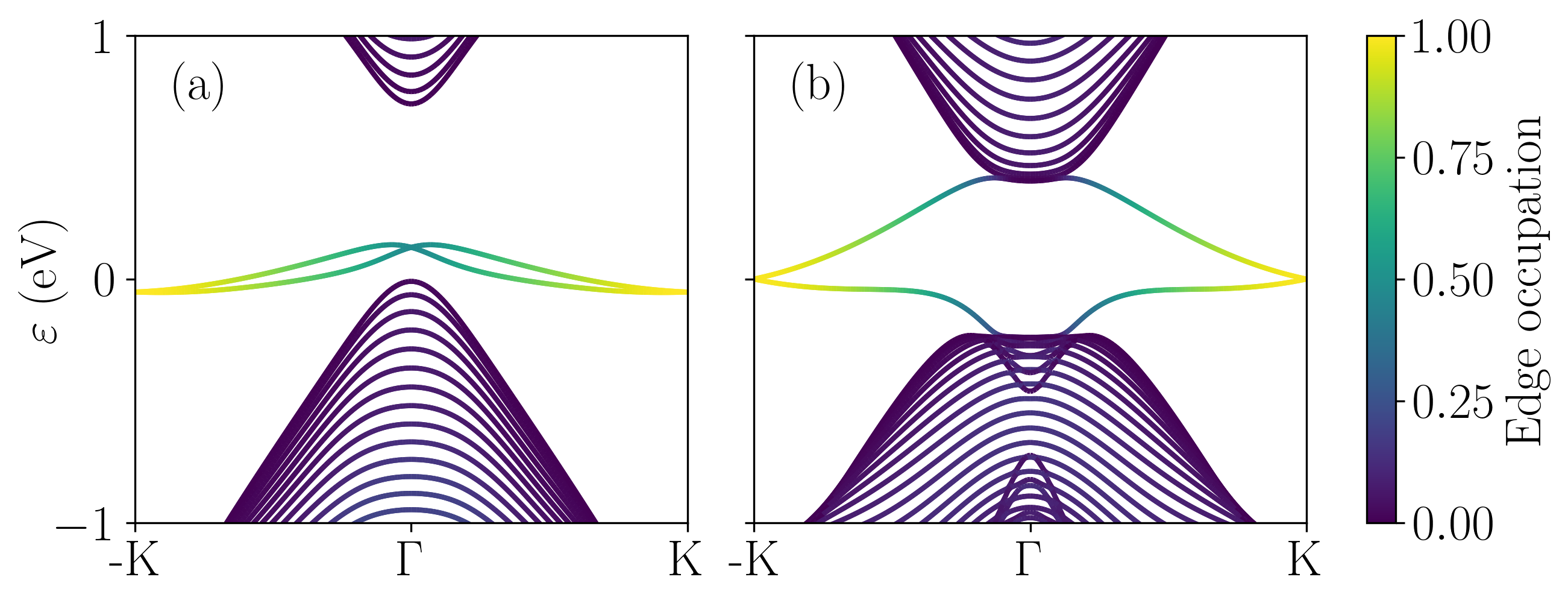}
    \caption[Comparison between trivial and topological edge bands in the $\beta$ alloy]{Bands of a $\beta$ (zigzag) ribbon with a width of 50 atoms using the parameters of the amorphous alloy, for two different SOC values: (a) $\lambda=0.5$ eV, which exhibits trivial edge states, and (b) $\lambda=2.5$ eV, showing the topological edge bands.}\label{fig:beta_ribbon_bands}
\end{figure}

To verify the results of the ANN, we look for signatures of the topological edge states. First, we evaluate the average edge occupation $\overline{\braket{n_e}}=1/N\sum_n^N\sum_{i\in\partial\Omega}\braket{c^{\dagger}_ic_i}_n$ for the first $N$ states closer to the Fermi energy, where $\partial\Omega$ denotes the boundary of the solid $\Omega$. This quantity gives a direct measure of the degree of localization of the states, which for topological edge states should be close to 1. The average edge occupation is shown in Fig.~\ref{fig:gap_topo_diagrams_amorphous}(d). There, we observe that both trivial and topological regions at low disorder exhibit high edge occupations. For the trivial region, this is due to the presence of trivial edge states, whereas in the topological one it is due to the topological edge states, as depicted in Fig.~\ref{fig:beta_ribbon_bands}. The transition between the regions is characterized by a lower edge occupation, as it is expected from the increasingly higher bulk component of the edge states as the gap closes. At high disorder, the edge occupation drops significantly for all values of $\lambda$. This is to be expected due to the appearance of localized bulk states coming from the disorder, which will reduce the average even if there are proper topological edge states present. 

Next, we look explicitly for the presence of topological edge states. One instance of such a state is shown in Fig.~\ref{fig:gap_topo_diagrams_amorphous}(c), at the limit of the topological region with $\sigma=0.25$. In this case, the edge state shown already shows some bulk component, which will increase as disorder increases (gap decreases). Looking for edge states at high disorder ($\sigma > 0.3$), we are unable to find a well-defined edge state since all of them exhibit strong bulk components and in general do not wrap the solid as one would expect.


\subsection{Electronic transport setup}\label{sec:transport}

Finally, to ensure that the topological edge states do not play any role in the high disorder region, we set up electronic transport calculations. The aim is two-fold: to establish whether there are topological edge states contributing to the conductance and to determine the transport nature of the bulk system, this is, insulating or metallic. For this, we consider two different setups: with open boundary conditions (i.e.\ transport along a ribbon), and with periodic boundary conditions (transport along a nanotube). Examples of both transport setups are shown in Fig.~\ref{fig:transmission}(a, b) respectively. The leads are taken to be of the same material as the sample, but doped to ensure that they are always metallic irrespective of the sample behaviour. 

\begin{figure}[h]
    \centering
    \includegraphics[width=0.8\linewidth]{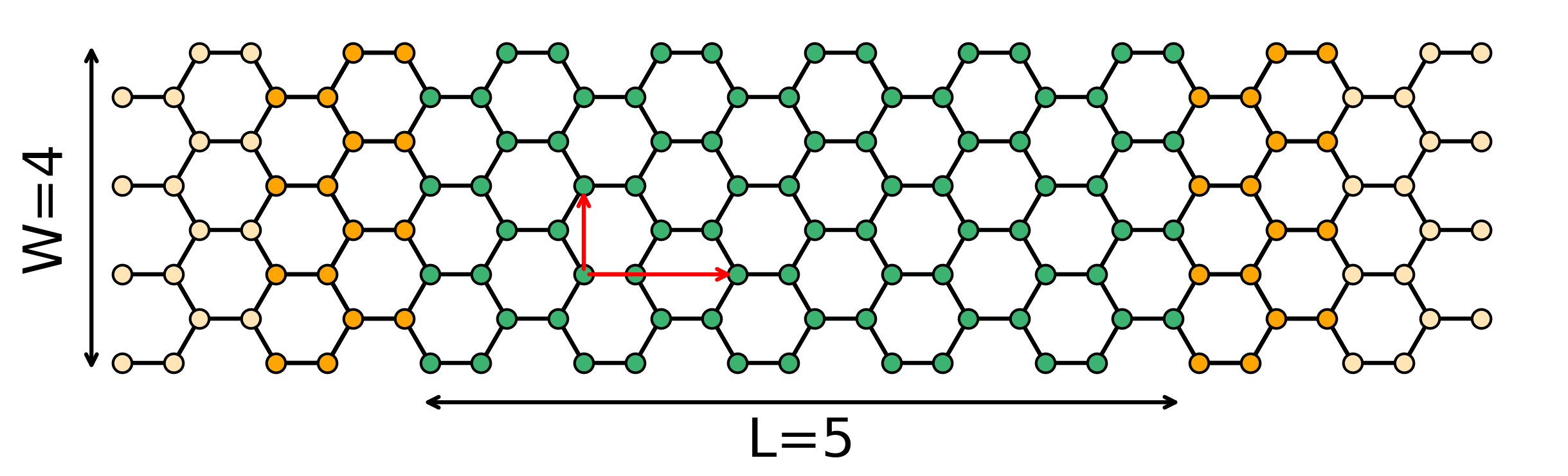}
    \caption[Setup of a two-terminal transport calculation in a ribbon]{Transport setup for an OBC calculation. This setup corresponds to a sample (green atoms) with $L=5$ and $W=4$ unit cells. The orange and yellow atoms denote the leads attached to the sample. The red arrows indicate the unit cell of the ribbon. The device $D$ refers to the sample together with the first unit cell of the lead (green and orange atoms).}\label{fig:transport_setup_obc}
\end{figure}

\begin{figure}[h]
    \centering
    \includegraphics[width=0.7\linewidth]{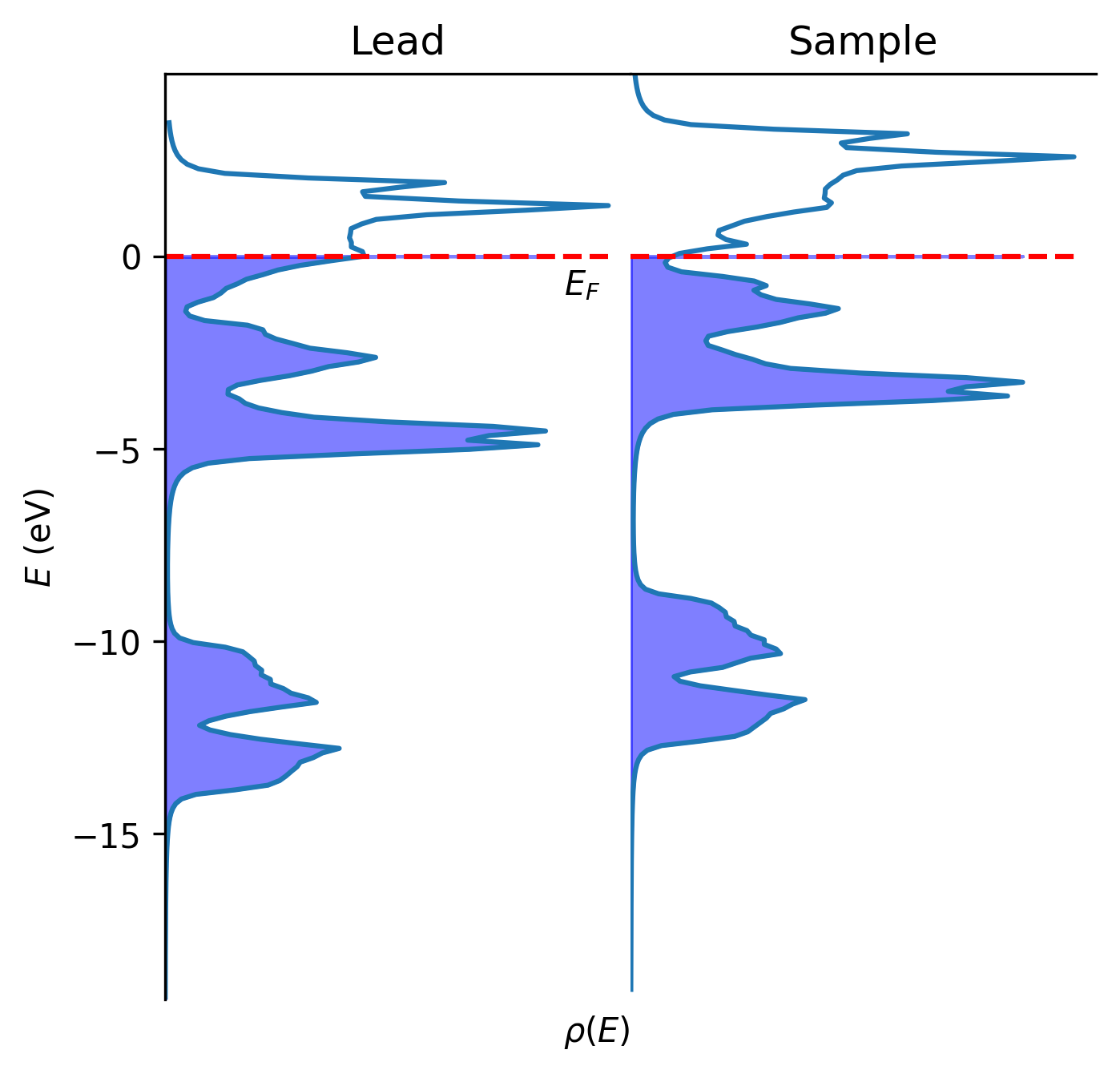}
    \caption[Density of states of the leads and the sample used in the transport calculations]{Density of states of (left) leads, with an applied gate voltage to impose a specific doping and (right) sample.}\label{fig:dos_sample_leads}
\end{figure}

We use Landauer's formalism to determine the electronic transport properties of a sample of the amorphous alloy~\cite{datta1997electronic}. The conductance of the sample is given by Landauer's formula~\cite{landauer1957spatial}:
\begin{equation}
    G = \frac{e^2}{h}T(E_F)
\end{equation}
where $E_F$ is the Fermi energy of the sample, and $T$ is the transmission, which is computed with the Caroli formula~\cite{caroli1971direct}:
\begin{equation}
    T(E) = \text{Tr}\left[\Gamma_L(E)G_D^-(E)\Gamma_R(E)G_D^+(E)\right]
\end{equation}
$\Gamma_{L/R}$ denotes the coupling of the leads with the device (sample plus unit cell of the lead, see Fig.~\ref{fig:transport_setup_obc}), given by $\Gamma_{L/R}=i(\Sigma_{L/R}^+ - \Sigma^-_{L/R})$ with $\Sigma_{L/R}^{+/-}$ being the selfenergy of lead $L/R$, and $G_D$ is the Green's function of the device. For the definition of these quantities we refer to previous works~\cite{jacob2011critical}.

The conductance is obtained at the Fermi level of the sample. To ensure that the leads are always metallic and provide the same current for different instances of disorder, we consider a gate potential applied to the leads to ensure constant charge across calculations. The gate potential for a fixed charge in the leads is obtained integrating the density of states of the leads,
\begin{equation}
    \int_{-\infty}^{E_F}\rho_{L/R}(E - V)dE = N_e
\end{equation}
where $N_e$ is the desired charge. In all calculations, we have set $N_e=N + N_a/2$, where $N$ is the number of electrons of the lead at charge neutrality, and $N_a$ is the number of atoms of the lead. This level of doping ensures that the Fermi level of the sample is aligned with the middle of the conduction band of the lead, resulting in constant transport across the leads. The density of states of both the sample and the lead are shown in Fig.~\ref{fig:dos_sample_leads}, with the gate voltage already applied to the leads. The transport calculations were done with the \texttt{tightbinder} library~\cite{uria_tightbinder}.

\subsection{Insulator-metal transition at high disorder}

\begin{figure}[t]
    \centering
    \includegraphics[width=0.75\textwidth]{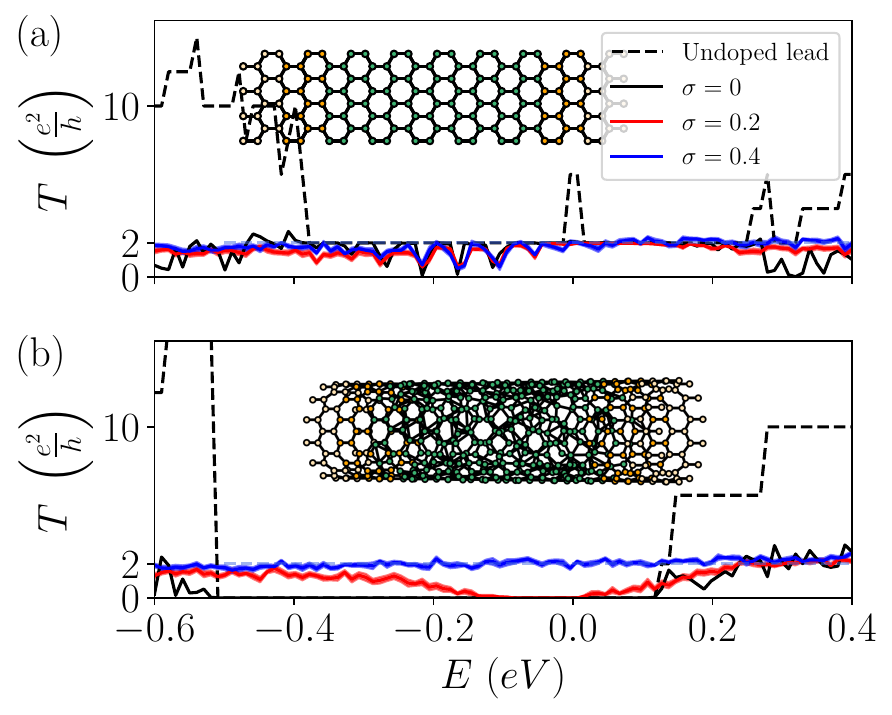}
    \caption[Transmission of a nanoribbon and a nanotube of a-Bi$_x$Sb$_{1-x}$]{(a, b) Transmission as a function of disorder of the Bi$_x$Sb$_{1-x}$ alloy for a nanoribbon (OBC) and a nanotube (PBC) respectively, for $\lambda=2.5$ and averaged over 10 samples. Calculations done for samples of size $W=16$, $L=9$.}\label{fig:transmission}
\end{figure}

\begin{figure}[!h]
    \centering
    \includegraphics[width=0.8\textwidth]{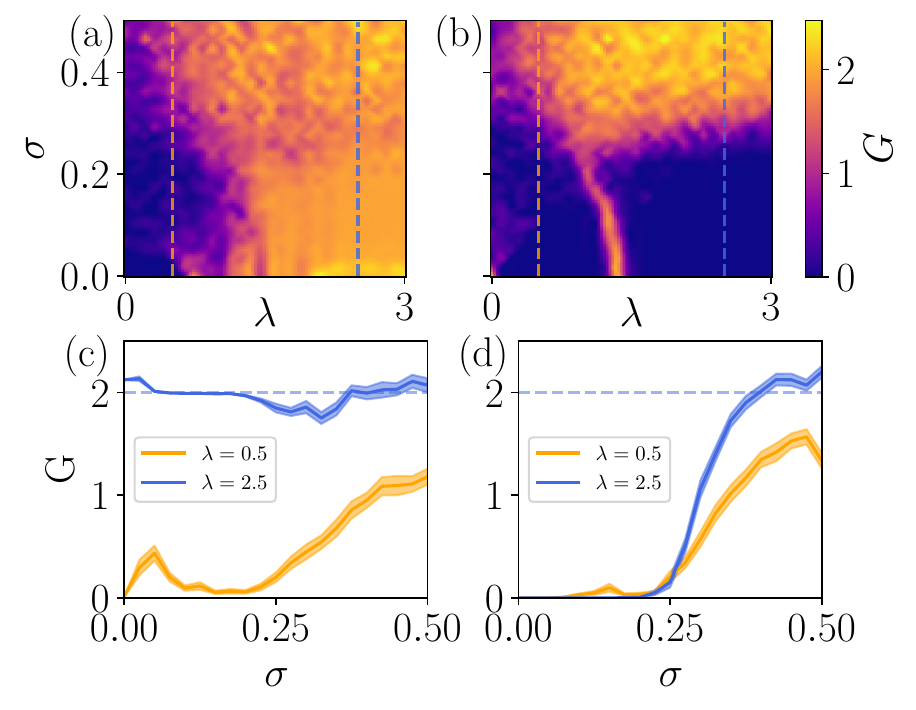}
    \caption[Conductance diagrams for PBC and OBC as a function of disorder and SOC for the amorphous alloy]{(a) Conductance diagrams as a function of SOC strength $\lambda$ and disorder strength $\sigma$ for OBC and (b) PBC respectively, averaged over 10 samples. The dashed lines correspond to the cuts that are represented on subplots (c, d), which show the conductance as a function of disorder for different $\lambda$ for OBC and PBC respectively, averaged over 50 samples (shaded areas denote standard error of the mean). All calculations have been done with system size $W=16$, $L=9$.}\label{fig:conductance}
    \includegraphics[width=0.75\columnwidth]{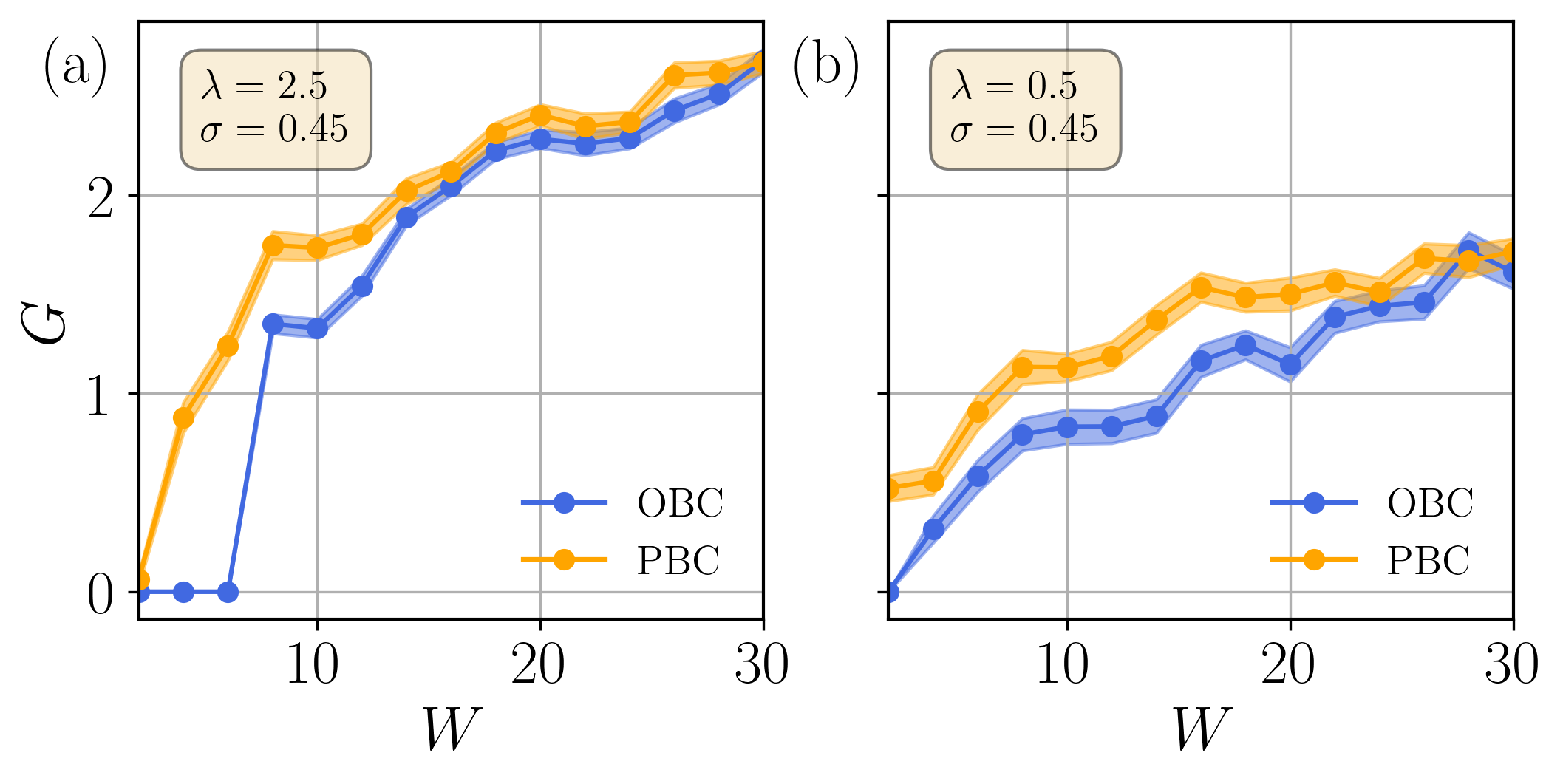}
    \caption[Scaling analysis of the conductances at high disorder in the amorphous alloy]{Scaling of the conductances for high disorder $\sigma=0.45$ as a function of $W$ ($=L$) for (a) $\lambda=2.5$ and (b) $\lambda=0.5$ for both OBC and PBC\@. We have chosen pairs $(W,L)$ such that the aspect ratio of the ribbon is closest to 1. All points are averaged over $N_s=50$ samples, and the shaded areas denote the standard error of the mean $\sigma_{\mu}=\sigma/\sqrt{N_s}$.}\label{fig:scaling}
\end{figure}

In the first place, we determine the transmission with both OBC and PBC as a function of disorder for a topological sample with $\lambda=2.5$, shown in Fig.~\ref{fig:transmission}(a, b). For the nanoribbon, we observe that for all values of disorder the transmission at the Fermi level is approximately two (the Fermi level is set at zero). On the contrary, for the nanotube at low disorder the sample is insulating as expected, but at high disorder its transmission also increases to two, meaning that the bulk states contribute to the conduction. This is further verified with the conductance diagrams in Fig.~\ref{fig:conductance}(a, b), also for OBC and PBC respectively. For OBC at low disorder, we can identify the topological region as it exhibits a quantized conductance $G=G_0=2e^2/h$. At high disorder the conductance starts fluctuating and increases beyond the quantum of conductance $G_0$. For the PBC diagram, we observe the same: the trivial and topological regions at low disorder are insulating as expected, but in the high disorder region the sample becomes metallic. The spurious conductances seen at low disorder in both cases are attributed to the trivial edge states in the OBC case, and to the proximity to the valence band in the PBC case (see Fig.~\ref{fig:conductance}(c, d)).

To conclude the metallic nature of the system, we perform a scaling analysis in the spirit of the scaling theory of localization~\cite{abrahams1979scaling}. Namely, as we increase simultaneously the width $W$ and the length $L$ of the sample and the leads, the bulk system must show one of two different behaviours: either its conductance increases with the sample size (metallic, more channels available for transport), or the conductance decays to zero (insulating, due to exponentially localized states). We examine the scaling of the system for both OBC and PBC at high disorder, $\sigma=0.45$. From the diagrams in Fig.~\ref{fig:conductance}(a, b) we also observe a difference in the conductance at low SOC ($\lambda =0.5$) and high SOC ($\lambda=2.5$). For this reason we address the scaling for both values, to detect any possible difference between these two metallic regions. The scaling analysis is illustrated in Fig.~\ref{fig:scaling}. For both values of $\lambda$ we observe an increase of the conductance with the system size, in principle indicating a metallic scaling. Regarding the difference between OBC and PBC, while the conductances show some differences for small system sizes, these discrepancies decrease as the system size increases, hinting that there is not any contribution from topological edge states in the OBC case. In the particular case of $\lambda=0.5$, it appears the PBC conductance could be plateauing, which begs the question of what happens for bigger system sizes, namely, if it would still be possible that the PBC conductance drops to an insulating state whereas the OBC one plateaus due to topological edge states. Unfortunately, those system sizes are beyond our reach for the present study. With this we conclude that the system undergoes an insulator-metal transition with disorder, which was previously observed in the context of Anderson insulators and dubbed inverse Anderson insulators~\cite{goda2006inverse, zuo2024topological}.

\section{Neural networks and training data}\label{sec:training}

As we mentioned throughout this chapter, the topological properties of the $\beta$, $\alpha$ alloys and the amorphous alloy are mainly assessed through an ANN trained to predict the topological invariant based on the entanglement spectrum of the system. As a reminder, the topological invariant for the crystalline structures is obtained from the Wilson loop (see section~\ref{sec:wilson_loop}), defined as $W(k_y)=\prod_{i\in\gamma}\left(\sum_n\ket{u_{n\mathbf{k}}}\bra{u_{n\mathbf{k}}}\right)$, where $\gamma$ denotes a closed path in the BZ, and $\mathbf{k}=(k_i,k_y)$~\cite{vanderbilt2018berry}. Tracking the evolution of its eigenvalues, which correspond to HWCCs, as a function of $k_y$, one can extract the topological invariant of the system, in this case the $\mathbb{Z}_2$ invariant~\cite{soluyanov2011computing}. For the disordered systems, be it the crystalline alloy or the amorphous alloy we use the entanglement spectrum to predict the topological invariant, as explained in Chapter~\ref{chapter:deep_learning}, by means of an ANN\@. The entanglement spectrum is obtained from the eigenvalues of the correlation matrix $C_{ij}=\braket{\Psi|c^{\dagger}_ic_j|\Psi}$, with $i,j$ restricted to one half of the system. 

\begin{figure}[!h]
    \centering
    \includegraphics[width=1\textwidth]{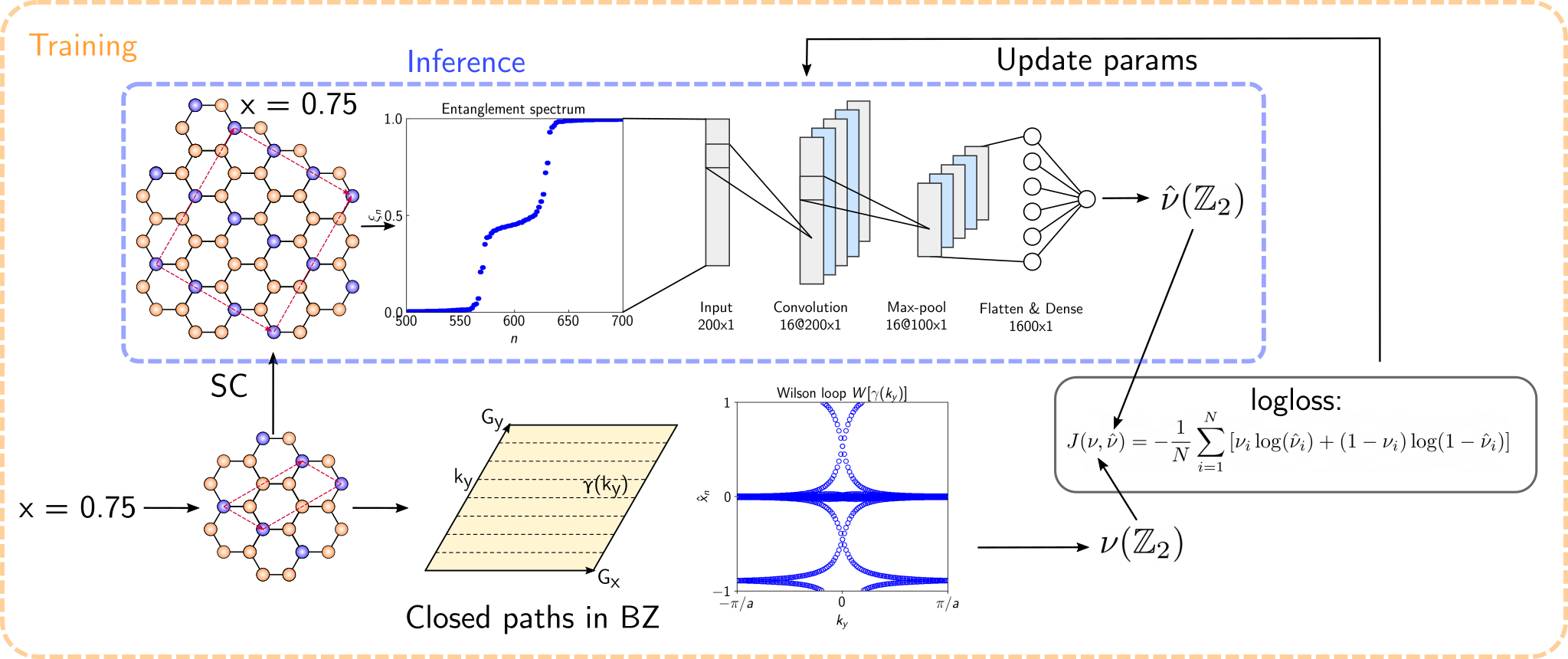}
    \caption[Workflow with the artificial neural networks to train and predict the invariants]{Description of the workflow with the artificial neural network. For training, we consider a small cell matching some concentration. We compute its invariant with the Wilson loop, and then extract the entanglement spectrum from a supercell. The neural network is trained using the (entanglement, invariant) pair. For inference, we consider a supercell for one specific concentration, obtain its entanglement spectrum and feed it to the ANN, which will predict its invariant.}\label{fig:nn_training}
\end{figure}

Here, we consider a different neural network for each of the three systems: the $\beta$ alloy, the $\alpha$ alloy and the amorphous alloy. To train each one, we create a dataset formed by pairs of the actual invariant, computed by means of the Wilson loop, and the corresponding entanglement spectrum for a supercell. For the crystalline alloys, as explained in section~\ref{sec:crystalline_alloys_results} we generate data considering a finite set of concentrations and changing the value of the SOC\@. The datasets for the $\beta$ and $\alpha$ alloy contain a total of 4500 samples each, corresponding to 500 per concentration. After balancing the datasets to ensure the same number of samples per category, we get approximately a total of 2900 samples for the $\beta$ alloy, and 3300 samples for the $\alpha$ alloy.

\begin{figure}[t]
    \centering
    \includegraphics[width=0.7\linewidth]{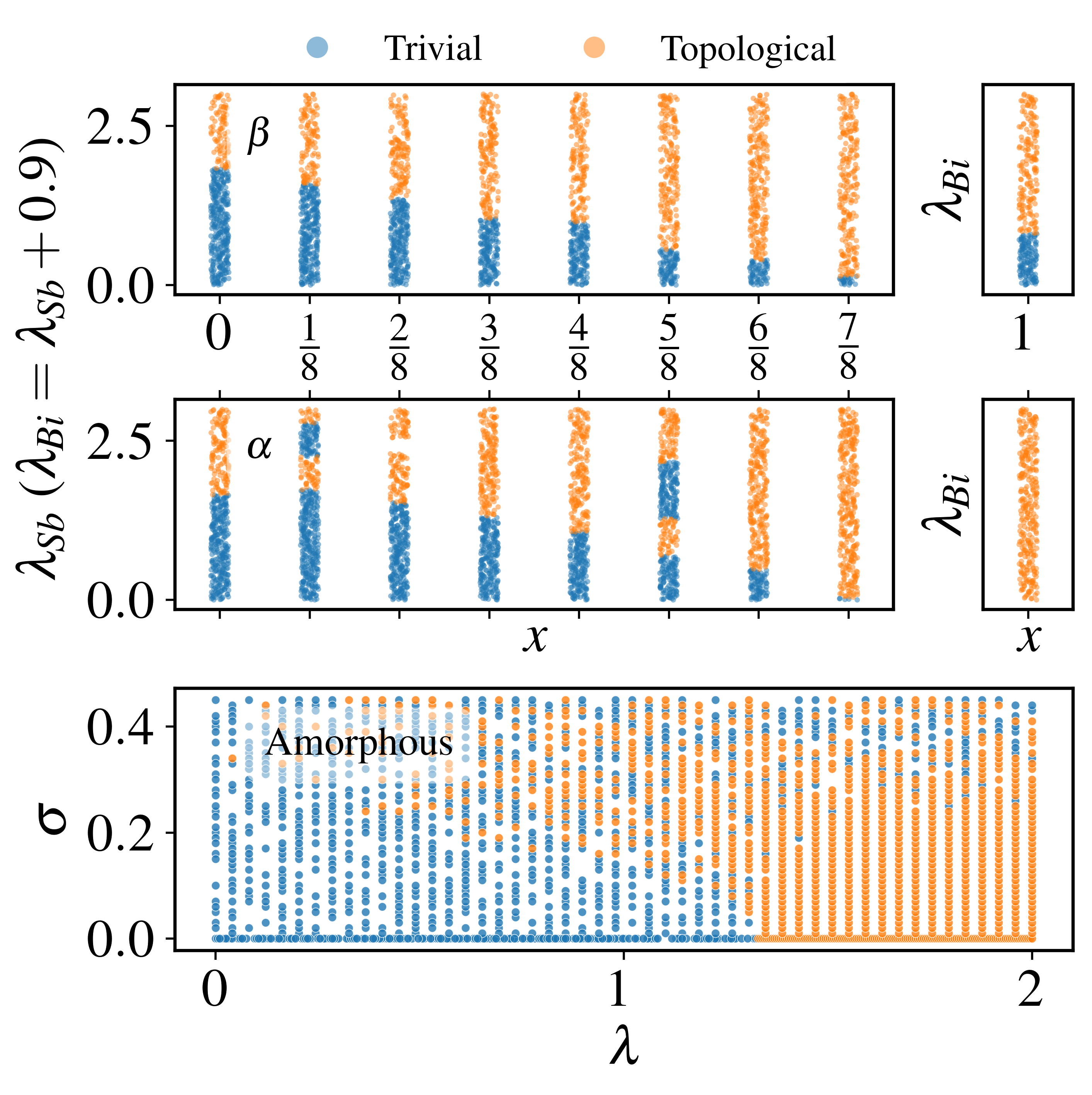}
    \caption[Datasets used in the training and testing of the ANN]{Datasets used in the training of the neural networks. (Top) Dataset for the $\beta$ alloy, (middle) dataset for the $\alpha$ alloy and (bottom) dataset for the amorphous alloy. In the crystalline cases, the samples are represented as a function of the SOC of Sb, $\lambda_{\text{Sb}}$; the SOC for Bi is always defined as  $\lambda_{\text{Bi}}=\lambda_{\text{Sb}}+0.9$. For $x=1$ there is no Sb, so the points are represented directly as a function of $\lambda_{\text{Bi}}$.}\label{fig:training_data}
\end{figure}

For the amorphous alloy, we take directly the SOC as an effective concentration parameter, and change the disorder strength to generate samples for both crystalline and disordered cases. In this case, we generated a total of 2600 samples, which after balancing resulted in 2100 samples. The datasets used to train and test the three neural networks are represented in Fig.~\ref{fig:training_data}. The training and inference procedures are schematically represented in Fig.~\ref{fig:nn_training}.

Regarding the neural network architectures, we used two different ones, one for the crystalline case and a different one for the amorphous case. These architectures are shown in Table~\ref{tab:arch_crystalline} and Table~\ref{tab:arch_amorphous} respectively. The idea is to use convolutional layers to extract the main features of the entanglement spectrum, followed by dense layers to perform the interpolation. While in the crystalline case it achieved high accuracy with the most simply network possible (one layer of each type), in the case of the amorphous network it seemed to benefit from the inclusion of additional convolutional and dense layers, increasing its accuracy. All layers use a ReLU activation function, except for the last layer which uses a sigmoid.

In training, we used a train-test split of $10\%$. For the $\beta$ network, we achieve an accuracy on the test set of $99\%$, while for the $\alpha$ network we achieve $85\%$. Both networks were trained for 50 epochs, with a learning rate $\alpha=10^{-4}$ with the ADAM optimizer. The amorphous network was trained for 200 epochs, also with $\alpha=10^{-4}$, achieving an accuracy on the test set of $85 \%$.

\renewcommand{\arraystretch}{1.2}
\begin{table}[h]
    \centering
    \begin{tabular}{ccc}
        \hline
        \hline
        Layer type & Kernel & Output shape \\
        \hline
        1d convolutional & 16 &  $32\times200\times1$ \\
        Max-pool & 2 & $32\times100\times1$ \\
        Dense & - & 1000 \\
        Dense & - & 1 \\
        \hline
        \hline
    \end{tabular}
    \caption[Architecture of the ANN used for the crystalline alloys]{Architecture of the ANN used to predict the topological phase diagrams of the $\beta$ and $\alpha$ alloys.}\label{tab:arch_crystalline}
\end{table}

\begin{table}[h]
    \centering
    \begin{tabular}{ccc}
        \hline
        \hline
        Layer type & Kernel & Output shape \\
        \hline
        1d convolutional & 32 &  $32\times200\times1$ \\
        Max-pool & 2 & $32\times100\times1$ \\
        1d convolutional & 64 &  $64\times100\times1$ \\
        Max-pool & 2 & $64\times50\times1$ \\
        1d convolutional & 128 &  $128\times50\times1$ \\
        Max-pool & 2 & $128\times25\times1$ \\
        Dense & - & 256 \\
        Dense & - & 10 \\
        Dropout ($p=0.1$) & - & - \\
        Dense & - & 10 \\
        Dense & - & 1 \\
        \hline
        \hline
    \end{tabular}
    \caption[Architecture of the ANN used for the amorphous alloy]{Architecture of the ANN used to determine the phase diagram of the amorphous alloy.}\label{tab:arch_amorphous}
\end{table}

\subsection{Entanglement spectrum from the kernel polynomial method}

In the previous chapter, we mentioned that one of the main shortcoming of the proposed method to predict the invariant was that it requires diagonalizing the Hamiltonian. We show here that this diagonalization can be bypassed by means of the kernel polynomial method (KPM)~\cite{weisse2006kernel}. To do so, we follow the approach taken in previous works~\cite{varjas2020computation, carvalho2018real} where the projector over the Fermi sea can be written in terms of a function of the Hamiltonian $H$. This allows to rewrite the correlation matrix directly in terms of this function of $H$, which can then be expanded in powers of $H$ according to the KPM\@.
\begin{align}
     C_{ij}&=\braket{\Psi|c^{\dagger}_ic_j|\Psi}=\sum_{n }^{E_n\leq E_F}\braket{n|i}\braket{j|n}  
    =\braket{j|\left(\sum_n^{E_n\leq E_F}\ket{n}\bra{n}\right)|i} = \braket{j|\theta(E_F - H)|i}
\end{align}
where $E_F$ is the Fermi energy, and $\theta(E)$ is the Heaviside step function. Thus, we identify the function of the Hamiltonian $P(E, H)=\theta(E - H)$. Normalizing the Hamiltonian such that its eigenvalues lie in the interval $[-1, 1]$, we can expand this function in Chebyshev polynomials:
\begin{equation}
    P(E,H) = \sum^M_{m=0}g_m\mu_m(E)T_m(H)
\end{equation}
where $g_m$ are the Jackson kernel coefficients, $\mu_m(E)$ are the moments and $T_m(H)$ are the Chebyshev polynomials, defined in terms of a recursive relation:
\begin{align}
    \nonumber T_0(H) &= I \\
    T_1(H) &= H \\
    \nonumber T_m(H) &= 2HT_m(H) - T_{m-1}(H)
\end{align}
The moments $\mu_m(E)$ can be evaluated knowing that $P(E,H)=\theta(E - H)$ and read~\cite{varjas2020computation}:
\begin{align}
    \mu_m(E) &= \frac{2}{\pi}\frac{1}{1+\delta_{m0}}\int_{-1}^1\frac{P(E,x)T_m(x)}{\sqrt{1 - x^2}}dx \\ 
    &=\left\{
    \begin{array}{cc}
            1 - \frac{1}{\pi}\text{arccos(E)} & m=0 \\
        \frac{-2}{m\pi}\sin[{m\ \text{arccos}(E)}] & m\neq 0 
    \end{array}\right.
\end{align}
Finally, taking matrix elements of $P(E,H)$ in the orbitals basis $\{\ket{i}\}$ we obtain the correlation matrix $C_{ij}$. The eigenvalues of this matrix form the entanglement spectrum used to train the neural network and predict the topological nature of the alloys. Thus, given that we still need to diagonalize the correlation matrix, the speedup provided by the KPM is restricted by this diagonalization of a matrix of dimension $N/n$, where $N$ is the dimension of the complete system and $n$ is the reduction factor coming from the restriction to a partition of the system. Therefore, the time complexity is still $\mathcal{O}(N^3)$, although in practice it leads to faster calculations.

\section{Conclusions}

Using the entanglement spectrum as a proxy for the $\mathbb{Z}_2$ topological invariant in conjunction with artificial neural networks, we were able to determine the topological phase diagram of the crystalline Bi$_x$Sb$_{1-x}$ alloys in their $\beta$ and $\alpha$ forms, estimating the critical concentrations. Applying this same methodology to the amorphous alloy, we observe that for low disorder strength, this is in the gapped region, the structural disorder renormalizes the critical SOC, resulting in trivial to topological transitions. This result was already observed in SK models of stanane~\cite{wang2022structural}, and we conjecture that it extends to all families of topological insulators that accept the same description. 

Additionally, we go beyond previous works to study the high disorder region, i.e.\ where the system becomes gapless. With the ANN predicting a global transition to a trivial region, we resort to electronic transport calculations. Remarkably, we observe that the system undergoes a transition from an insulator to a metal with disorder, although it is possible that the metallic phase exhibits different regimes. While in principle we have discarded any topological behaviour in the $\mathbb{Z}_2$ sense, it would be interesting to explore other forms of topology, namely HOTIs, as other forms of Bi exhibit, and in general a more detailed study of the scaling would benefit the characterization of the metallic phase to ensure the distinction between trivial and topological. We also highlight the performance of the neural network, which was able to predict the phase diagram independently of the gap of the system.

Regarding the applicability of the methodology to more realistic models, we observe that it is in principle restricted only to tight-binding models, i.e.\ models where one can tune the topological behaviour in a continuous fashion. This is so because one needs to be able to generate enough training samples for the neural network to accurately predict the phase diagram, and it can be regarded in general as a weakness of machine learning based methods. Therefore, while for DFT calculations conceptually the methodology may be applied identically, the lack of tunable parameters prevents us from training the neural network. Nevertheless, we note that it is possible to sort this problem downfolding the DFT model to a tight-binding description, which would then allow to generate enough training samples. In this spirit, we believe that the methodology presented here can be applied to a wide range of materials, namely to all materials accepting a Slater-Koster description or an effective tight-binding model, and in those cases it can be used to predict successfully their topological phase diagram.

\chapter{Real-space criteria for non-crystalline fractional Chern insulators}
\newcommand{\olsi}[1]{\,\overline{\!{#1}}} 
\newcommand\norm[1]{\left\lVert#1\right\rVert}

\section{Introduction}
The discovery of the fractional Quantum Hall effect without external magnetic field in twisted MoTe$_2$~\cite{Cai2023,Zeng2023,Park2023,Xu2023} and few layer graphene~\cite{Lu2024,Xie2024,Waters2024} realizes a long-sought prediction of the many-body fractional Chern insulator (FCI) state~\cite{Neupert2011,Sun2011,Sheng2011,regnault2011fractional}.
In these experiments, an incommensurate potential with respect to lattice translations, or moir\'{e} potential, determines the filling at which such fractional many-body state is stabilized.
Confounded with the existence of Chern insulator states in quasicrystals~\cite{Kraus:2012iqa,Tran:2015cj,huang2018theory,Fuchs:2018dd,Varjas2019,Chen2019,Schirmann2024} and amorphous systems~\cite{agarwala2017topological,mitchell_amorphous_2018,mano2019application,costa2019toward,marsal2020topological,Sahlberg2020,agarwala_higher-order_2020,grushin2023topological,wang_structural-disorder-induced_2021,spring_amorphous_2021,wang2022structural,uria2022,munoz2023structural,marsal_obstructed_2022, uria2024amorphization}
this discovery raises the more general question: what are the real-space criteria for single-particle bands that favor non-crystalline fractional Chern insulators states?

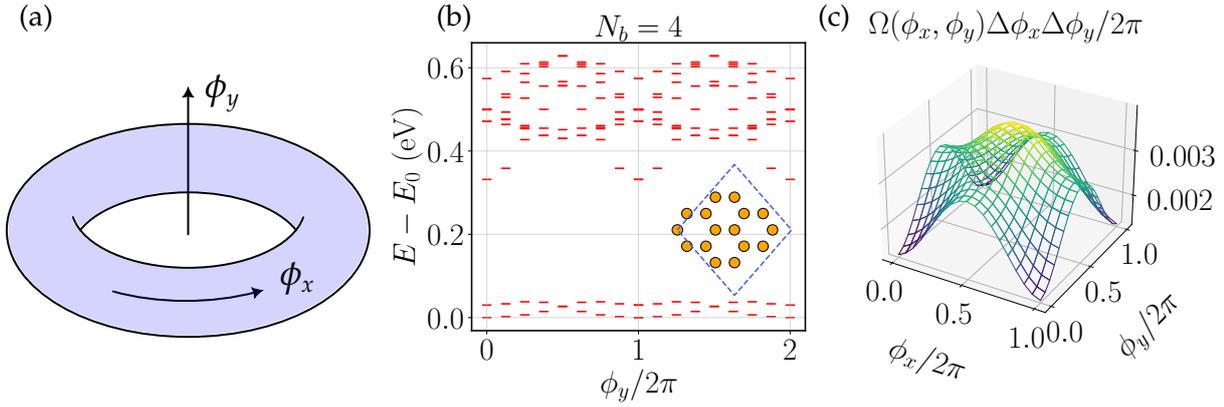
\begin{figure}[h]
    \centering
    \begin{tikzpicture}
        \node at (0,0) {\includegraphics[width=0.3\textwidth]{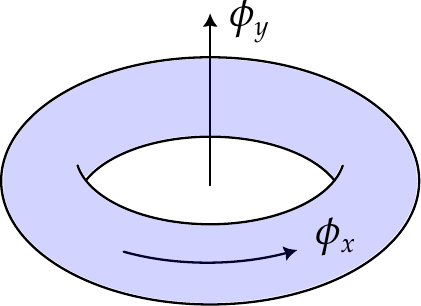}};
        \node at (-2,2.5) {(a)};

        \node at (8.1,0) {\includegraphics[width=0.69\textwidth]{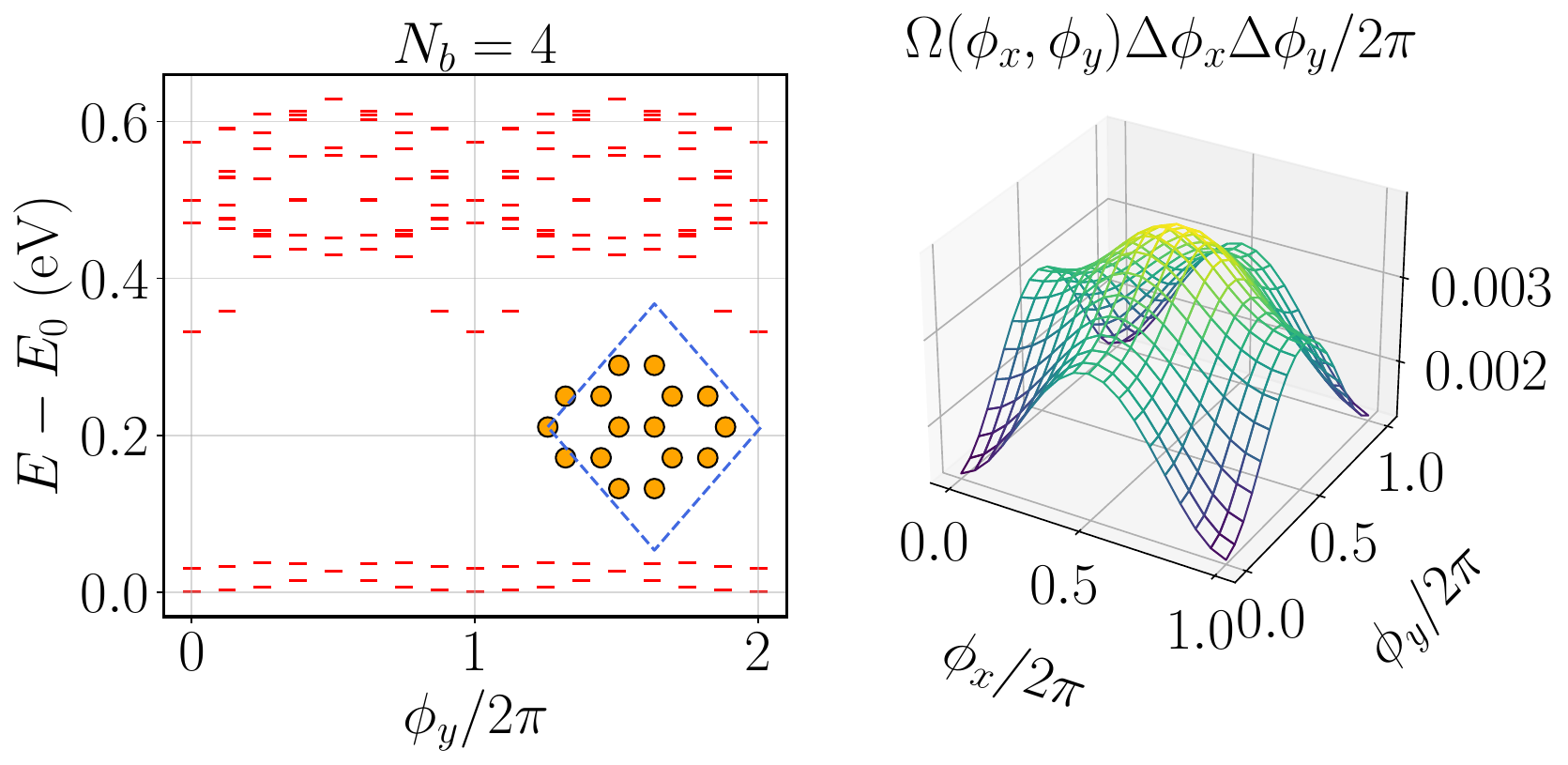}};
        \node at (3.5,2.5) {(b)};
        \node at (8.5,2.5) {(c)};
    \end{tikzpicture}
    \caption[Example calculation of the fractional Chern insulator ground state for $\nu=1/2$ in a $C=1$ band]{Example calculation of a bosonic fractional Chern insulator ground state for filling $\nu=1/2$ in a $C=1$ band.\ (a) Flux insertions in the torus.\ (b) Energy spectrum flow as a function of the incident flux $\phi_y$. One can identify the two-fold degenerate ground state, which returns to the original state after two flux insertions.\ (c) Discretized Berry curvature associated to the quasi-degenerate ground state, with $C=1$. These results correspond to a finite system of model~\eqref{eq:a-coldmodel} with $N_b=4$ bosons.}\label{fig:fci_examples}
\end{figure}

Many-body calculations are typically necessary to evaluate the emergence and stability of FCIs in microscopic models, which typically resort to the insertion of magnetic fluxes through the torus, very much in the style introduced back before in section~\ref{sec:charge_pumping}, to examine the degeneracy of the ground state, the charge transport or directly evaluating the many-body Chern number, see Fig.~\ref{fig:fci_examples}. The Chern number is given by
\begin{equation}
    C = \frac{1}{2\pi}\int_{[0,2\pi]^2} d^2\bm{\phi}\ \Omega(\bm{\phi}),
\end{equation}
where the integral is performed over the incident fluxes $\mathbb{T}^2=[0,2\pi]\times[0,2\pi]$, $\Omega(\bm{\phi}) = \partial_xA_y(\bm{\phi}) - \partial_yA_x(\bm{\phi})$ is the many-body Berry curvature with corresponding Berry connection $A_{\mu}(\bm{\phi}) = \text{Tr }\Phi^{\dagger}(\bm{\phi})\partial_{\mu}\Phi(\bm{\phi})$ and $\Phi(\bm{\phi})=(\ket{G_1(\bm{\phi})},\ldots,\ket{G_q(\bm{\phi})})$ is the ground state multiplet of degeneracy $q$~\cite{kudo2019many}.
Interestingly, candidate models may be identified based on the value of single-particle indicators~\cite{Parameswaran2012,Claassen2015,Jackson2015,Lee2017,Ledwith2020,Mera2021,Mera2021b,Wang2021,Ozawa2021,Mera2022,ledwith2022vortexability,Varjas2022}. 
Up to a few notable exceptions~\cite{Simon_2015, Zhu_2016, Kourtis_2018, andrewsPRB2024}, FCIs nominally emerge in Chern bands whose properties most closely resemble those of the lowest Landau level. 
Namely, a small band dispersion and small inhomogeneities of the Berry curvature and Fubini-Study metric in momentum space emerge as favorable criteria for single-particle bands to stabilize FCIs. 
These properties favor that all single-particle states contribute equally to the total Chern number $C$ which equals $C=1$ for each Landau level.

In addition to these criteria, it has been understood that bands that favor fractional Chern insulator states are close to saturating the so-called trace condition of an ideal Chern band~\cite{Parameswaran2013,Claassen2015,Roy2014,Ledwith2020,Wang2021}.
The trace condition is an inequality that expresses the non-wannierizability of a Chern band, and follows from the Souza-Wilkens-Martin optical sum-rule~\cite{Souza2000,OnishiFu2024}.
It states that the trace of the integrated quantum metric $g^{ij}_{\mathbf{k}}$, which is roughly the mean squared displacement of a given Wannier function, must be larger or equal to the Chern number.
When saturated, as occurs for the lowest Landau level, the band is called fully vortexable, or ideal, in the sense that single-particle states are holomorphic wavefunctions in the particle coordinate, and many-body states admit an exact vortex attachment~\cite{Ledwith2020}. 

The challenge to identify ideal single-particle states prone to exhibit fractional Chern insulators is that the existing criteria, namely band flatness, homogeneous Berry curvature and the trace condition, are all defined in momentum-space, and may not be employed in the absence of translation invariance.
In this chapter we introduce an analogue set of criteria for ideal Chern bands based on real-space quantities. The criterion of small fluctuations of the Berry curvature in momentum space translates into small fluctuations of the local Chern marker in real-space.
We formulate the trace inequality in terms of the real-space vortexability, which measures the degree to which many-body states admit a vortex attachment, in terms of the local Chern marker.  
We then compute these criteria for several disordered Chern insulating models: a model of an amorphous Chern insulator, disordered graphene in a magnetic field and a Chern insulator realized in ultra-cold atomic systems. We study their phase diagram as a function of disorder and indicate the most favorable regions, according to these real-space criteria, where many-body fractional Chern insulator ground states could appear upon adding strong electron-electron interactions. 

\section{Real-space criteria for non-crystalline fractional Chern insulators}\label{sec:criteria}


\subsection{Fluctuations of the Berry curvature in terms of the local Chern marker} 

The real-space quantities we wish to introduce hinge on the definition of the local Chern marker~\cite{bianco2011mapping}. We recall that the Chern number of a two-dimensional system is a global property of filled single-particle states:
%
\begin{equation}\label{eq:chern_number}
    C = \frac{1}{2\pi}\int_{\text{BZ}} \frac{d^2\mathbf{k}}{A_{\text{BZ}}}\ \Omega(\mathbf{k}) = 2\pi i \olsi{\text{Tr}}[PxP, PyP].
\end{equation}
The first equality is familiar in the context of crystalline topological phases, expressing the Chern number $C$ as the flux of the Berry curvature, $\Omega(\mathbf{k})$, of a filled band over the Brillouin zone (BZ)~\cite{bernevig_quantum_2006}.

The second equality, put forward by Bianco and Resta~\cite{bianco2011mapping}, is formally equivalent to the first if $x, y$ are the position operators in two perpendicular real-space directions, $P=\sum_{n\in \mathrm{occ}}\ket{n}\bra{n}$ is the projector onto the occupied eigenstates $\ket{n}$, and $\olsi{\text{Tr}}$ denotes the trace over the bulk of the system per unit area, $\olsi{\text{Tr}}\ \mathcal{O}=\text{Tr}\ \mathcal{O}/A$, $A$ being the bulk area.

From the commutator of the projected position operators it is straightforward to define the local Chern marker:
\begin{align}
     C &= 2\pi i\olsi{\text{Tr}}[PxP, PyP] = \frac{2\pi i}{A}\sum_{i,\alpha} \braket{i\alpha|[PxP, PyP]|i\alpha} \equiv \frac{1}{N_c}\sum_{i}C(\mathbf{r}_i),
\end{align}
where we have defined the local Chern marker as:
\begin{align}
\label{eq:LCM}
    C(\mathbf{r}_i) &=\frac{2\pi i}{A_c}\sum_{\alpha}\braket{i\alpha|[PxP, PyP]|i\alpha},
\end{align}
with $A=N_cA_c$, $A_c$ being the area of the unit cell, and $N_c$ the number of unit cells. The index $i$ runs over lattice positions, whereas $\alpha$ runs over internal degrees of freedom, for example the orbital or spin degree of freedom.  
For an insulator, averaging $C(\mathbf{r})$ over the bulk of a sufficiently large finite system results in an integer equal to the total Chern number of the filled states. Thus, $C(\mathbf{r})$ identifies a topologically non-trivial Chern insulator if $C\neq0$. Such real-space representation of the Chern number is particularly useful for amorphous or disordered systems, where the absence of lattice translational symmetry hinders standard techniques in momentum space.

Minimizing the fluctuations of the Berry curvature favours the eventual stability of a fractional Chern insulator state~\cite{Varjas2019,moralesduran2023}. 
Fluctuations of the Berry curvature in momentum space would constitute a deviation from Landau level physics, where the Berry curvature is flat throughout all the Brillouin zone. These fluctuations can be quantified with the standard deviation associated to the Berry curvature~\cite{moralesduran2023}:
\begin{equation}\label{eq:fluctuations_berry}
    F_m = \left[\int_{\text{BZ}}\frac{d^2{\mathbf{k}}}{A_{\text{BZ}}}\left(\frac{\Omega(\mathbf{k})}{2\pi} - C\right)^2\right]^{1/2}
\end{equation}
The dual description of the Chern number given in Eq.~\eqref{eq:chern_number} suggests that we can also compute the Berry curvature fluctuations working in real space.  
To express the fluctuations in real space we first define the Chern operator $\mathcal{C}$ as
\begin{gather}
    C = \frac{1}{N_b}\text{Tr}\ \mathcal{C} \equiv \braket{\mathcal{C}}, \\
    \mathcal{C} = \frac{2N_b\pi i}{N_cA_c}[PxP,PyP],
\end{gather}
where $N_b= N_c N_\mathrm{orb}$ is the number of basis states in the real-space traced region, given by the number of unit cells $N_c$ times the number of internal degrees of freedom $N_\mathrm{orb}$.
From this definition we define the standard deviation relative to the Chern operator:
\begin{align}\label{eq:chern_fluctuations}
     F_r &= \left[\frac{1}{N_b}\text{Tr}\left(\mathcal{C} - C\right)^2\right]^{1/2} = \braket{(\mathcal{C} - C)^2}^{1/2} = \left(\braket{\mathcal{C}^2} - \braket{\mathcal{C}}^2\right)^{1/2},
\end{align}
where we have used that $\text{Tr}\ \mathbb{I}=N_b$. 
The trace is intended to be taken in the infinite crystal limit $N_c\rightarrow\infty$.
Hence, in all calculations we separate the trace into edge and bulk parts, $A = \partial A \cup B$, and take the trace exclusively over the bulk part $B$. In the same way, we take $N_b$ to be the number of basis states corresponding to the bulk part.

Writing explicitly the trace we arrive to the following expression for the Berry curvature fluctuations:
\begin{equation}
\label{eq:fluctuationsBerryreal}
    F_r = \left[\frac{1}{N_b}\sum_{\substack{i,\alpha \\ i',\alpha'}}\left|\braket{i\alpha|\mathcal{C}|i'\alpha'}\right|^2 - \left(\frac{1}{N_b}\sum_{i\alpha}\braket{i\alpha|\mathcal{C}|i\alpha}\right)^2 \right]^{1/2},
\end{equation}
where we used that $\mathcal{C}^{\dagger}=\mathcal{C}$. We see that it is the off-diagonal elements of the Chern operator $\mathcal{C}$ that contribute to the fluctuations of the Berry curvature. Thus, if the off-diagonal elements of $\mathcal{C}$ are zero and $\mathcal{C}=C\mathbb{I}$, then $F_r=0$. Equivalently, in the reciprocal formulation, the fluctuations are only zero if $\Omega(\mathbf{k})\equiv\Omega_0\ \forall \mathbf{k}\in\text{BZ}$, as in a Landau level. In what follows, we show the precise mathematical relation between the real-space Berry curvature fluctuations~\eqref{eq:fluctuationsBerryreal} and the momentum-space Berry curvature fluctuations~\eqref{eq:fluctuations_berry}.

\subsection{Relation between $F_r$ and $F_m$}\label{app:Berry}
We have introduced the real-space Berry curvature fluctuations, $F_r$, to measure the degree of fluctuations of the Berry curvature in situations where the Berry curvature in momentum space is not accessible. As we will see in the next section, for all our models both quantities predict the same parameter regions with the lowest fluctuations. We justify now this mathematically by explicitly relating both quantities. Instead of following the original derivation of the Chern marker~\cite{bianco2011mapping}, we take a different approach. First, consider the action of the operator $Px_iP$ over a generic state $\ket{f}=\sum_{n,\mathbf{k}}f_{n\mathbf{k}}\ket{n\mathbf{k}}$. Then, it can be shown that a general matrix element is~\cite{bradlyn2022lecture}:
\begin{equation}
    \braket{n\mathbf{k}'|Px_iP|f} = i\partial_if_{m\mathbf{k}'} + \sum_{m\mathbf{k}}A_{nm}(\mathbf{k}')f_{m\mathbf{k}},
\end{equation}
where $\partial_i\equiv \partial/\partial k_i$. From this we see that $Px_iP=iP\partial_i P$, and we can also write directly the action of the operator over the state:
\begin{equation}\label{eq:action_PxP}
    Px_iP\ket{f} = \sum_{n\mathbf{k}}\left(i\partial_if_{n\mathbf{k}} + \sum_{m\mathbf{k}}A_{nm}(\mathbf{k})f_{m\mathbf{k}}\right)\ket{n\mathbf{k}}.
\end{equation}
Next, we consider the action of the commutator of the projected position operators over the same state $\ket{f}$, i.e. $[Px_iP,Px_jP]\ket{f}$. Using equation (\ref{eq:action_PxP}) we can compute the resulting state:
\begin{align}
     [Px_iP,&Px_jP]\ket{f}  = \sum_{n,m}\sum_{\mathbf{k}}\left[i\partial_iA^j_{nm}(\mathbf{k}) - i\partial_jA^i_{nm}(\mathbf{k})\right. \left. + \sum_l A^i_{nl}A^j_{lm} - \sum_l A^j_{nl}A^i_{lm}\right]f_{m\mathbf{k}}\ket{n\mathbf{k}}.
\end{align}
Here we identify the non-abelian Berry curvature $\Omega^{ij}_{nm}(\mathbf{k})$:
\begin{equation}
    \frac{\Omega^{ij}_{nm}(\mathbf{k})}{A_{\text{BZ}}} = \partial_iA^j_{nm}(\mathbf{k}) - \partial_jA^i_{nm}(\mathbf{k}) - i[A^i,A^j]_{nm}.
\end{equation}
Note the additional factor $A^{-1}_{\text{BZ}}$ to match the definition of the Berry curvature used in Eq.~\eqref{eq:chern_number}. Thus, the action of the commutator on a generic state is given in terms of the non-abelian Berry curvature:
\begin{equation}
    [Px_iP,Px_jP]\ket{f} = \frac{i}{A_{\text{BZ}}}\sum_{n,m,\mathbf{k}}\Omega^{ij}_{nm}(\mathbf{k}) f_{m\mathbf{k}}\ket{n\mathbf{k}}.
\end{equation}
From this expression we can write the general form of the commutator of projected positions. The action over a specific Bloch state $\ket{m\mathbf{k}'}$ can be obtained setting $f_{n\mathbf{k}}=\delta_{nm}\delta_{\mathbf{k}\mathbf{k}'}$. With this we obtain a general matrix element in the Bloch eigenstate basis, and one arrives at:
\begin{equation}
\label{eq:commutator_berry}
    [Px_iP,Px_jP] = \frac{i}{A_{\text{BZ}}}\sum_{n,m,\mathbf{k}}\Omega_{nm}^{ij}(\mathbf{k})\ket{n\mathbf{k}}\bra{m\mathbf{k}}.
\end{equation}
Equation~\eqref{eq:commutator_berry} is a key identity. From it, we will re-derive Eq.~\eqref{eq:chern_number} and also prove the relation between the two fluctuations. 

To re-derive Eq.~\eqref{eq:chern_number} note that it is expressed in terms of the abelian Berry curvature, whereas Eq.~\eqref{eq:commutator_berry} is written in terms of the non-abelian Berry curvature. 
%
Since the trace over bands of the abelian and non-abelian Berry curvatures is the same~\cite{vanderbilt2018berry}, i.e. $\sum_n\ \Omega^{ij}_{nn} = \sum_n\ \Omega^{ij}_n$, we can define the Berry curvature as $\Omega(\mathbf{k})=\sum_n \Omega^{xy}_n$, to rewrite the trace of~\eqref{eq:commutator_berry} to arrive to Eq.~\eqref{eq:chern_number}.

For the fluctuations, starting also from the non-abelian case, we can prove an inequality between the two types of fluctuations that holds both in the non-abelian and abelian cases. Consider first the square of the commutator:
\begin{align}
     &[Px_iP,Px_jP]^2 = -\frac{1}{A^2_{\text{BZ}}}\sum_{n,m,m'}\sum_{\mathbf{k}}\Omega^{ij}_{nm}\Omega^{ij}_{mm'}(\mathbf{k})\ket{n\mathbf{k}}\bra{m'\mathbf{k}}.
\end{align}
The first term of Eq.~\eqref{eq:chern_fluctuations} can be computed directly:
\begin{align}
    \nonumber \braket{\mathcal{C}^2}& =\frac{i^2}{N_b}\left(\frac{2\pi N_b}{N_cA_c}\right)^2\text{Tr} \ [PxP,PyP]^2 = \frac{(2\pi)^2N_b}{N_c^2A^2_cA^2_{\text{BZ}}}\sum_{n,m,\mathbf{k}}|\Omega_{nm}|^2(\mathbf{k}) \\ 
    \nonumber &= \frac{(2\pi)^2N_b}{N_c^2A^2_cA^2_{\text{BZ}}}\left[\sum_{n,\mathbf{k}}\Omega_{nn}^2(\mathbf{k}) + \sum_{n,m\neq n,\mathbf{k}}|\Omega_{nm}|^2(\mathbf{k})\right] \\ 
    \nonumber & \geq \frac{(2\pi)^2N_b}{N_c^2A^2_c}\frac{N_c}{A^3_{\text{BZ}}} \sum_n\int d^2\mathbf{k}\ \Omega^2_{nn}(\mathbf{k}) \\
    & \geq \frac{(2\pi)^2N_b}{N_{\text{occ}} N_cA^2_cA^3_{\text{BZ}}}\int d^2\mathbf{k}\ \Omega^2(\mathbf{k}) \equiv \alpha \int \frac{d^2\mathbf{k}}{A_{\text{BZ}}}\frac{\Omega^2(\mathbf{k})}{4\pi^2}.
\end{align}
In the second line we have used that $\Omega_{nm}=\Omega_{mn}^*$, and in the last equality we have used the Cauchy-Schwarz inequality to write the integral in terms of the square of the Berry curvature.
We have introduced $\alpha$ to group all the constants, $\alpha=N_{\text{orb}}/N_{\text{occ}}$, and $N_{\text{occ}}$ is the number of occupied bands. With this, we already recognize the first term of Eq.~\eqref{eq:fluctuations_berry} which we repeat here:
\begin{equation}
    F_m^2 =  \int \frac{d^2\mathbf{k}}{A_{\text{BZ}}}\frac{\Omega^2(\mathbf{k})}{4\pi^2} - C^2
\end{equation}
Thus, if we subtract $C^2$ in both sides we recover the fluctuations:
\begin{align}
    \braket{\mathcal{C}^2} - C^2 \geq \alpha \int \frac{d^2\mathbf{k}}{A_{\text{BZ}}}\frac{\Omega^2(\mathbf{k})}{4\pi^2} - C^2 = \alpha F_m^2 + (\alpha - 1)C^2.
\end{align}
With this we arrive at a lower bound for the real-space Berry curvature fluctuations written in terms of the Berry curvature fluctuations in momentum space:
\begin{equation}
    F_r^2 \geq \alpha F_m^2 + (\alpha - 1)C^2.
\end{equation}
The same inequality holds in the abelian case, simply taking $\Omega_{nm} = \Omega_n\delta_{nm}$ and following similar steps. In the particular case that $N_{\text{occ}}=1$, then the equality holds, $F^2_r=\alpha F_m^2 + (\alpha - 1)C^2$. The constant $\alpha$ amounts to $\alpha = N_\mathrm{orb}$, where $N_\mathrm{orb}$ is the number of internal degrees of freedom within the unit cell, such that $N_b=N_c N_\mathrm{orb}$ with $N_b$ the total number of basis states. Since $N_\mathrm{orb} > 0$, we see that if $F_m = 0$, then $F_r = (\alpha - 1)C^2 > 0$ in a topological region. In general $F_r$ will have the lower bound $F^2_r \geq (\alpha - 1) C^2$ since $F^2_m>0$. Only for $N_\mathrm{orb}=1$ it would be possible for both fluctuations to be zero simultaneously, which would correspond to the trivial case since the Chern number for the complete Bloch manifold is trivial~\cite{panati2007triviality}.

\begin{figure}
    \centering
    \includegraphics[width=0.8\columnwidth]{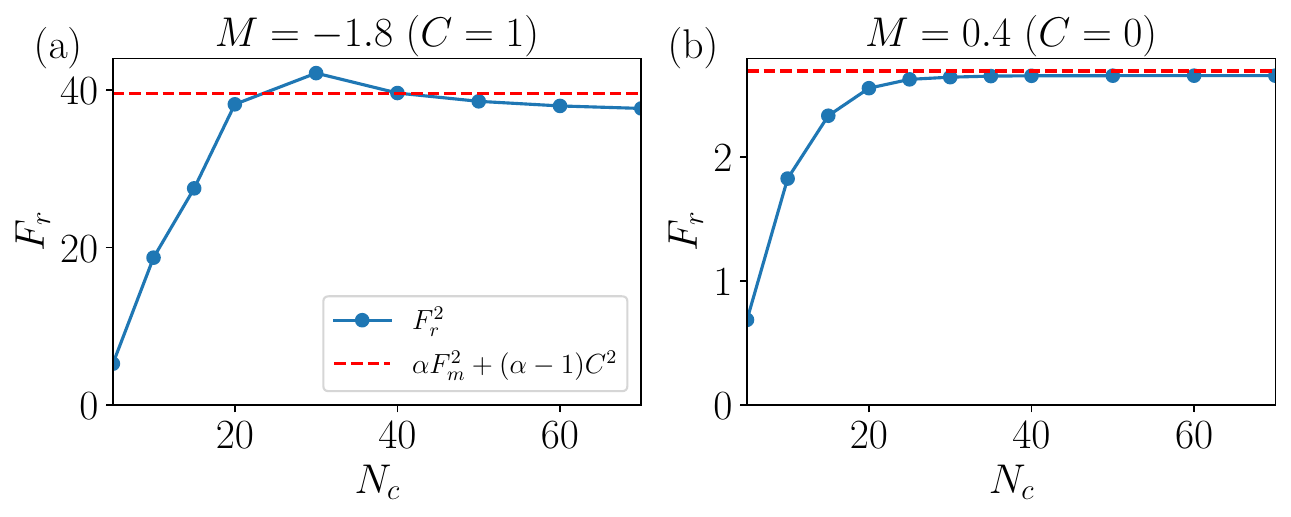}
    \caption[Scaling of the real-space Berry curvature fluctuations as a function of system size]{Real-space Berry curvature fluctuations $F_r$ compared with the Berry curvature fluctuations $F_m$ as a function of system size $N_c$ (number of unit cells along one axis) for the amorphous Chern insulator model.\ (a) $F_r$ in a topological region ($M=-1.8$) and (b) in a trivial region ($M=0.4$).}\label{fig:fluctuations_size_dependence}
\end{figure}

Lastly, it is instructive to reorder the inequality to see why $F_r$ works as an alternative criterion to $F_m$:
\begin{equation}
    F^2_m \leq \frac{1}{\alpha}F_r^2 - \frac{\alpha - 1}{\alpha}C^2.
\end{equation}
Written this way we see that the real-space Berry curvature fluctuations $F_r$ give an upper bound on the value of the momentum-space Berry fluctuations $F_m$. The lower $F_r$ the lower the Berry fluctuations $F_m$ are guaranteed to be, explaining why $F_r$ works as an alternative criterion to $F_m$. In Fig.~\ref{fig:fluctuations_size_dependence} we check whether the equality holds for the amorphous Chern insulator model. We observe that as $N_c\rightarrow\infty$ the real-space Berry curvature fluctuations do converge to the expected value.

\subsection{Trace condition and vortexability} 

The notion of \textit{vortexable} bands~\cite{ledwith2022vortexability} stems from the following definition: a band of states $\{\ket{n}\}$, defined by the projector $P=\sum_{n} \ket{n}\bra{n}$ is said to be vortexable if
\begin{equation}\label{eq:vortexability_condition}
    z\ket{n}=Pz\ket{n},
\end{equation}
for any $n$ in the band. Namely, $\ket{n}$ is still within the band defined by $P$ after attaching a vortex operator $z(\mathbf{r})$, $z:\mathbb{R}^2\to\mathbb{C}$, to the state.
Attaching a vortex operator like $z=x+iy$ physically amounts to shifting single particle states by one unit of angular momentum.
The vortexability condition allows to attach an holomorphic function of $z$ to states within the vortexable band.
In particular, we can attach the holomorphic function required to build a Laughlin-like many-body state to states in a vortexable band~\cite{ledwith2022vortexability}:
\begin{equation}
\label{eq:Laugh}
    \ket{\Psi^{m}}=\prod_{i<j}(z_i-z_j)^{m}\ket{\Psi}.
\end{equation}
Eq.~\eqref{eq:Laugh} is the ground-state wave-function of a fractional quantum Hall state for filling $\nu = 1/m$, where $m$ is either an even or odd integer for bosons or fermions, respectively. 

It is possible to write Eq.~\eqref{eq:vortexability_condition} fully in terms of operators. Taking the exterior product with a bra for the same state $\ket{n}$, and then summing over all states we arrive at
\begin{equation}
    zP=PzP.
\end{equation}
Now we can use the completeness relation $Q = 1 - P$ to arrive at a simpler expression of the vortexability condition:
\begin{equation}
    QzP=0,
\end{equation}
i.e., the operator $QzP$ has to be identically zero. One form to check this condition is using the Frobenius or trace norm, defined as $\norm{A}_F=\sqrt{\text{Tr}(A^{\dagger}A)}$~\cite{golub2013matrix}. 
From the norm property $\norm{A} = 0 \Leftrightarrow A=0$ it follows that if we define $A\equiv QzP$, then the vortexability condition can be finally written as
\begin{equation}\label{eq:vortexability_norm}
    \norm{A}^2_F= \text{Tr}(A^{\dagger}A) = \text{Tr}[P\bar{z}QzP] = 0,
\end{equation}
where $\bar{z}$ is the complex conjugate of $z$. We then define the vortexability itself in real space as
\begin{equation}
\label{eq:vortexabilitygen}
    T_r = \text{Tr}[P\bar{z}QzP].
\end{equation}
Deviations from zero of the vortexability quantify how close a given set of filled states is from being an ideal Chern band.
This is one of the criteria to assess how likely it is for this set of states to support a fractional Chern insulator upon partial filling and under strong interactions. 
Through Eq.~\eqref{eq:vortexabilitygen} we have reformulated the vortexability condition in real space, similar to the definition of the Chern number in Eq.~\eqref{eq:chern_number}.
We emphasize that this is a real-space quantity by using the label $r$ of $T_r$.

It is also possible, and useful, to relate the vortexability and the local Chern marker explicitly. 
To derive this relationship we consider the simplest vortex function $z=x+iy$, although in principle vortexable bands can be defined in terms of any vortex function $z$.
%
Then, the Chern number~(\ref{eq:chern_number}) can be rewritten after some algebra as
\begin{align}
        C = 2\pi i\text{Tr}[PxP, PyP] = \pi \text{Tr}[P\bar{z}P, PzP]
\end{align}
Now using that $[z,\bar{z}]=0$ and $P = 1 - Q$, the Chern number is written as
\begin{align}
\label{eq:vortantivort}
     C &= \pi\left(\text{Tr}[PzQ\bar{z}P] - \text{Tr}[P\bar{z}QzP]\right)\equiv \pi(\bar{T}_r - T_r)
\end{align}
where we have defined $\bar{T}_r$ as the \textit{antivortexability}, i.e.\ the vortexability corresponding to the opposite vortex function $\bar{z}=x-iy$. Eq.~\eqref{eq:vortantivort} shows that the Chern number can be split into two separate contributions, namely two vortexabilities with opposite chirality. 

The relationship between vortexability and local Chern marker allows us also to make contact with the trace condition.
For a crystal, this condition is expressed via the quantum geometric tensor, which is given by:
\begin{align}
    \eta_{\mu\nu} &= \sum_n \braket{\partial_{k_{\mu}}u_{n\mathbf{k}}|Q(\mathbf{k})|\partial_{k_{\nu}}u_{n\mathbf{k}}} \equiv g_{\mu\nu}(\mathbf{k}) - \frac{i}{2}\Omega(\mathbf{k})\varepsilon_{\mu\nu},
\end{align}
where 
$Q(\mathbf{k}) = I - \sum_{n}\ket{u_{\mathbf{k}n}}\bra{u_{\mathbf{k}a}}$, 
$\ket{u_{\mathbf{k}n}}$ is the periodic part of Bloch's function,
$g_{\mu\nu}(\mathbf{k})=\text{Re}\ \eta_{\mu\nu}$ is the quantum metric, $\Omega(\mathbf{k})=-2\text{Im}\ \eta_{\mu\nu}$ is the Berry curvature, and $\varepsilon_{\mu\nu}$ is the antisymmetric tensor.
Then, the trace condition relates the quantum metric $g$ with the Berry curvature in the following way:
\begin{equation}
\label{eq:vortexmom}
    T_m=\int d^2\mathbf{k} \left(\text{Tr}\ g_{\mu\nu}(\mathbf{k}) - \Omega(\mathbf{k})\right) \geq 0.
\end{equation}
The inequality is saturated if and only if the bands are vortexable with $z=x+iy$~\cite{ledwith2022vortexability}. 
To emphasize that this is a momentum-space quantity we use the label $m$ in $T_m$. 

For crystalline systems, deviations from the trace condition are used to signal parameter regions where a model might favor a fractional Chern insulator. To work in real space we can alternatively monitor deviations from the vortexability condition~\eqref{eq:vortexability_norm}. Following similar steps as in our definition of a local Chern marker in Eq.~\eqref{eq:LCM}, we can define a local vortexability simply taking the trace over real-space positions
\begin{equation}
\label{eq:vortexabilityrealspace}
    T_r = \sum_{i}T(\mathbf{r}_i),\quad T(\mathbf{r}_i)\equiv\sum_{\alpha}\braket{i\alpha|P\bar{z}QzP|i\alpha}.
\end{equation}
To contrast with Eq.~\eqref{eq:vortexmom} and to emphasize that this is a real-space quantity we use the label $r$ in $T_r$.
The closer $T_r$ is to zero in the bulk, the closer the trace condition Eq.~\eqref{eq:vortexability_norm} is of being satisfied.

\begin{figure}
    \centering
    \includegraphics[width=0.9\columnwidth]{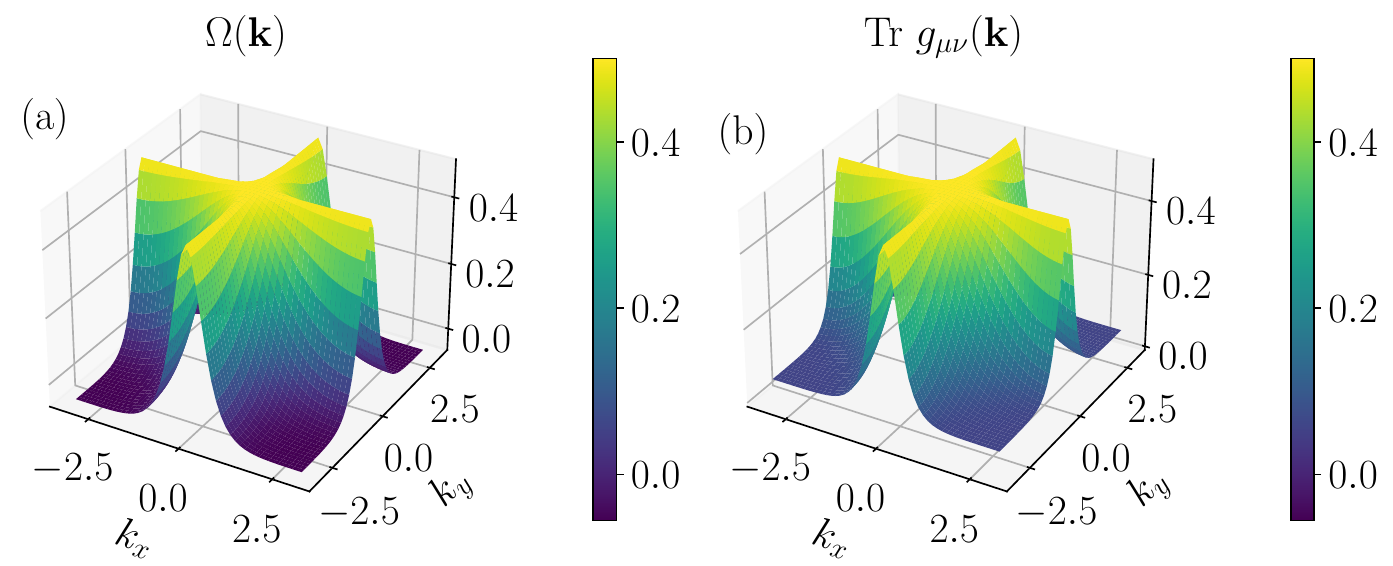}
    \caption[Example calculation of the Berry curvature $\Omega(\bm{k})$ and traced quantum metric $\text{Tr }g_{\mu\nu}(\bm{k})$]{(a) Berry curvature $\Omega(\mathbf{k})$ and (b) trace of the quantum metric $\text{Tr }g_{\mu\nu}(\mathbf{k})$ for the amorphous Chern insulating model, for $M=-1$, $t_2=\lambda=0$.}\label{fig:metric_curvature_agarwala}
\end{figure}

The calculation of the Berry curvature in Eqs.~\eqref{eq:chern_number},~\eqref{eq:fluctuations_berry} and~\eqref{eq:vortexmom} is done using the gauge invariant formulation of Fukui~\cite{fukui2005chern}, as implemented in the \texttt{tightbinder} library~\cite{uria_tightbinder}. For the calculation of the quantum metric $g_{\mu\nu}(\mathbf{k})$ in the deviation from the trace condition~\eqref{eq:vortexmom}, we require a gauge invariant formulation as well. If we take the derivative $\partial_{\mu}\equiv \partial_{k_{\mu}}$ of the eigenvalue problem $H(\mathbf{k})\ket{u_{m\mathbf{k}}} = E_m(\mathbf{k})\ket{u_{m\mathbf{k}}}$ for a reference, non-degenerate band $m$ and then take the scalar product with a general state $\ket{u_{n\mathbf{k}}}$, we arrive at the Feynman-Hellman equations~\cite{cheng2010quantum}:
\begin{gather}
\braket{u_{m\mathbf{k}}|\partial_{\mu}H(\mathbf{k})|u_{m\mathbf{k}}} = \partial_{\mu}E_m(\mathbf{k}),\ \text{if }n=m \\
    \braket{u_{n\mathbf{k}}|\partial_{\mu}u_{m\mathbf{k}}} = \frac{\braket{u_{n\mathbf{k}}|\partial_{\mu}H(\mathbf{k})|u_{m\mathbf{k}}}}{E_n(\mathbf{k})-E_m(\mathbf{k})},\ n\neq m
\end{gather}
Then the second relation specifically allows to compute the quantum geometric tensor, from which we can extract both the Berry curvature and the quantum metric for band $m$:
\begin{align}
    \nonumber \eta_{\mu\nu}(\mathbf{k}) &= \braket{\partial_{\mu}u_{m\mathbf{k}}|Q(\mathbf{k})|\partial_{\nu}u_{m\mathbf{k}}} = 
    \sum_{n\neq m}\braket{\partial_{\mu}u_{m\mathbf{k}}|u_{n\mathbf{k}}}\braket{u_{n\mathbf{k}}|\partial_{\nu}u_{m\mathbf{k}}} \\
    & = \sum_{n\neq m}\frac{\braket{u_{m\mathbf{k}}|\partial_{\mu}H(\mathbf{k})|u_{n\mathbf{k}}}\braket{u_{n\mathbf{k}}|\partial_{\mu}H(\mathbf{k})|u_{m\mathbf{k}}}}{(E_n(\mathbf{k})-E_m(\mathbf{k}))^2}.
\end{align}
From this expression we can extract both the quantum metric $g_{\mu\nu}=\text{Re}\ \eta_{\mu\nu}$ and the Berry curvature $\Omega(\mathbf{k})\varepsilon_{\mu\nu}=-2\text{Im}\ \eta_{\mu\nu}$, see Fig.~\ref{fig:metric_curvature_agarwala} for an example calculation corresponding to model~\eqref{eq:a-CImodel}. Here we have addressed the quantum geometric tensor for only one band, but it can be generalized to the non-abelian case where the quantum geometric tensor becomes a matrix, $\eta_{\mu\nu}^{mm'}$. 

\subsection{Band flatness}


The flatness of a band with respect to the adjacent energy gaps is a common criterion to enhance the role of interactions that might favor a fractional Chern insulator ground state. For crystalline systems it is defined relative to the gap between consecutive bands
\begin{equation}\label{eq:flatness_m}
    f_m =   \frac{\displaystyle\min_{\mathbf{k}\in\text{BZ}}\varepsilon_{\mathbf{k}n+1} - \max_{\mathbf{k}\in\text{BZ}}\varepsilon_{\mathbf{k}n}}{
        \displaystyle\max_{\mathbf{k}\in\text{BZ}}\varepsilon_{\mathbf{k}n} - \min_{\mathbf{k}\in\text{BZ}}\varepsilon_{\mathbf{k}n}}\equiv \frac{\Delta}{W},
\end{equation}
where $\Delta$ is the gap between two consecutive bands $n$ and $n+1$ and $W$ the bandwidth of a given band. 
The higher the ratio $f_m$, the stronger the effect of interactions. 
Specifically, large interactions favor interacting many-body ground states like the fractional Chern insulator. 

In non-crystalline systems, disorder in general closes the spectral gap. 
Topological properties are determined by the existence of a mobility gap between states that are non-wannierizable. However, in certain situations it is still useful to define a quantity that can track how the spectral gap closes in disordered systems, a real-space analogue of $f_m$, or $f_r$. This quantity is useful to track how the spectral gap closes as we increase disorder in models with a well-defined crystalline limit. It can also be useful in disordered models that retain hard-spectral gaps, as is the cased in certain fixed-coordination models with only geometric disorder~\cite{marsal2020topological,marsal_obstructed_2022}.

With these caveats in mind, the momentum space criterion of a small $f_m$ has a direct translation in the disordered case.
Consider a subspace of eigenstates of interest, instead of a specific band, with the eigenenergies $\{\varepsilon\}$ sorted in ascending order.
We can identify sets of eigenstates that correspond to well-defined groups of eigenvalues, those whose gaps between sets are larger than the average energy separation within a set.
Close to the crystalline limit, these sets will match the energy regions where bands are defined. 
Referring to the eigenenergies $\varepsilon_i$ via their indices $i$,
    one such set is $\mathcal{B}=\{i_1,\dots,i_f|i_n\in\mathbb{N},\ \forall n\}$, for instance $\{0,\ldots,(N_b - 1)/2\}$ for a half-filled model; then the flatness in the disordered case is defined as
\begin{equation}\label{eq:flatness_r}
    f_r = \frac{\displaystyle\varepsilon_{i_f+1} - \max_{i\in\mathcal{B}}\varepsilon_{i}}{\displaystyle\max_{i\in\mathcal{B}}\varepsilon_{i} - \min_{i\in\mathcal{B}}\varepsilon_{i}}.
\end{equation}
Note that $f_r$ is still well-defined if we use periodic boundary conditions, which we can thus call the supercell band flatness.
This is helpful because with open boundary conditions a topological band will appear with in-gap edge states, interfering with the flatness calculation. 

This completes the set of real-space criteria in relation to the standard criteria for Landau level mimicry. Because they are defined in real space, they are applicable to non-crystalline systems such as disordered, quasi-crystalline and amorphous systems. We summarize the different criteria in Table~\ref{tab:fci_criteria}, showing both real-space and momentum-space definitions.

\begin{table*}[t]
    \centering
    \begin{tabular}{|c||c|c|}
    \hline
        & \text{Crystalline} & \text{Disordered} \\
        \hline 
        \hline
        \rule{0pt}{0.8cm}   
        Band flatness & 
        $\displaystyle f_m=\frac{\displaystyle\min_{\mathbf{k}\in\text{BZ}}\varepsilon_{\mathbf{k}n+1} - \max_{\mathbf{k}\in\text{BZ}}\varepsilon_{\mathbf{k}n}}{
        \displaystyle\max_{\mathbf{k}\in\text{BZ}}\varepsilon_{\mathbf{k}n} - \min_{\mathbf{k}\in\text{BZ}}\varepsilon_{\mathbf{k}n}}$ & 
        $\displaystyle f_r=\frac{\displaystyle\varepsilon_{i_f+1} - \max_{i\in\mathcal{B}}\varepsilon_{i}}{\displaystyle\max_{i\in\mathcal{B}}\varepsilon_{i} - \min_{i\in\mathcal{B}}\varepsilon_{i}}$
        \\[0.5cm]
        \hline
        \makecell{Berry curvature \\ fluctuations}
        \rule{0pt}{0.8cm} 
        &\quad\quad  $\displaystyle F_m = \left[\int_{\text{BZ}}d^2{\mathbf{k}}\left(\frac{\Omega(\mathbf{k})}{2\pi} - C\right)^2\right]^{1/2}$\quad\quad & \quad\quad  
        $\displaystyle F_r=\left(\braket{\mathcal{C}^2} - \braket{\mathcal{C}}^2\right)^{1/2}\quad\quad  $
        \\[0.5cm]
        \hline
        \rule{0pt}{0.8cm} 
        \makecell{Deviation from \\ trace condition} & $\displaystyle T_m=\int d^2\mathbf{k} \left(\text{Tr}\ g_{\mu\nu}(\mathbf{k}) - \Omega(\mathbf{k})\right)$  &  $T_r=\displaystyle\text{Tr}[P\bar{z}QzP]$ \\[0.5cm]
        \hline
    \end{tabular}
    \caption[Comparison between the momentum- and real-space criteria for fractional Chern insulators]{Criteria used to find favorable bands to stabilize a fractional Chern insulator. We compare their definitions in momentum space for crystals and their real-space equivalent introduced in this work for disordered systems. We use the subscripts $m$ and $r$ to label the \textit{momentum} or \textit{real-space} version of each quantity. To identify fractional Chern insulator candidates in real-space, we look for parameter regions where the band flatness $f_r$ is maximal, while keeping the Berry curvature fluctuations, $F_r$, and the deviations from the trace condition, the vortexability $T_r$, to a minimum, ideally zero. We refer to the text for the definition of each quantity.}\label{tab:fci_criteria}
\end{table*}

\section{Real-space criteria of non-crystalline Chern insulators}\label{sec:examples}

Here we apply our real-space criteria to three examples of disordered tight-binding models in two-dimensions: an amorphous Chern insulator~\cite{agarwala2017topological}, disordered Landau levels in graphene generated by an applied magnetic field, and a four-orbital model for ultra-cold atomic systems predicted to host a fractional Chern insulator~\cite{zhao2023}.

Starting from the crystalline structure for all three models, we introduce the structural disorder sampling the displacement of the atoms from a uniform distribution, such that the maximum displacement $\eta$ is the parameter characterizing the degree of disorder of the solid:
%
\begin{equation}\label{eq:disorder_sampling}
    \mathbf{r}'=\mathbf{r} + \mathbf{\delta r}\quad |\mathbf{\delta r}| \sim U(0,\eta).
\end{equation}
We also restrict the displacements to be in-plane, and allow any arbitrary direction $\theta\sim U(0,2\pi)$.
Unless otherwise noted, the hopping amplitude between sites will depend on the distance and angles between sites.
Hence, displacing the site positions will introduce hopping disorder.

\subsection{Amorphous Chern insulator}

We start by defining an amorphous Chern insulator model introduced in Ref.~\cite{agarwala2017topological}, that defines the Qi-Wu-Zhang model of a Chern insulator in an amorphous point-set~\cite{Qi2006}.
%
The Hamiltonian reads
\begin{equation}
\label{eq:aCI}
    H =\sum_{i,\alpha,\beta} \varepsilon_{\alpha\beta} c^{\dagger}_{i\alpha}c_{i\alpha} + \sum_{\substack{i,j\neq i \\ \alpha,\beta}} t_{\alpha\beta}(\mathbf{r}_{ij})c^{\dagger}_{i\alpha}c_{j\beta},
\end{equation}
where
    \begin{subequations}\label{eq:a-CImodel}
\begin{eqnarray}
t_{\alpha\beta}(\mathbf{r}_{ij})&=&T_{\alpha\beta}(\hat{\mathbf{r}}_{ij})e^{-(|\mathbf{r}_{ij}|-a)}\theta_H(R_c - |\mathbf{r}_{ij}|) \\
    \left[\varepsilon_{\alpha\beta}\right] &=& 
    \begin{pmatrix}2 + M & (1 - i)\lambda \\ 
    (1 + i)\lambda & -(2+M)\end{pmatrix} \\
\left[T_{\alpha\beta}\right](\hat{\mathbf{r}}_{ij}) &=& \frac{1}{2}
\begin{pmatrix} 
-1 & -ie^{-i\theta} + \lambda[\sin^2{\theta}(i+1)-1] \\ 
-ie^{i\theta} + \lambda[\sin^2{\theta}(i-1)-1] & 1 
\end{pmatrix}
\end{eqnarray}
\end{subequations}
The indices $i,j$ run over lattice positions with two orbitals per site. The indices $\alpha,\beta$ denote the orbital degree of freedom. In the crystalline limit of the model, $\eta=0$ in Eq.~\eqref{eq:disorder_sampling}, and $i$ and $j$ label the vertices of a square lattice. The angle $\theta$ is set by the direction of the vector $\mathbf{r}_{ij}=\mathbf{r}_i-\mathbf{r}_j$, $\hat{\mathbf{r}}_{ij}$ being the associated unitary vector. $a$ is the nearest-neighbor distance in the square lattice, which we set to $a=1$. The hopping cutoff distance $R_c$ sets the number of nearest-neighbors to which hopping is permitted. We set it to $R_c=1.4$, such that in the crystalline case with $\eta=0$ we recover the original model (hoppings up to first neighbours). 

To create the amorphous lattice for a given disorder strength $\eta$ we draw random displacements of the sites in a square lattice following~\eqref{eq:disorder_sampling}. 
For each realization, these define the disordered site positions and the set of $\mathbf{r}_{ij}$. We then input these into Eq.~\eqref{eq:a-CImodel} to define the hopping amplitudes.

\begin{figure}[!h]
    \centering
    \includegraphics[width=0.7\columnwidth]{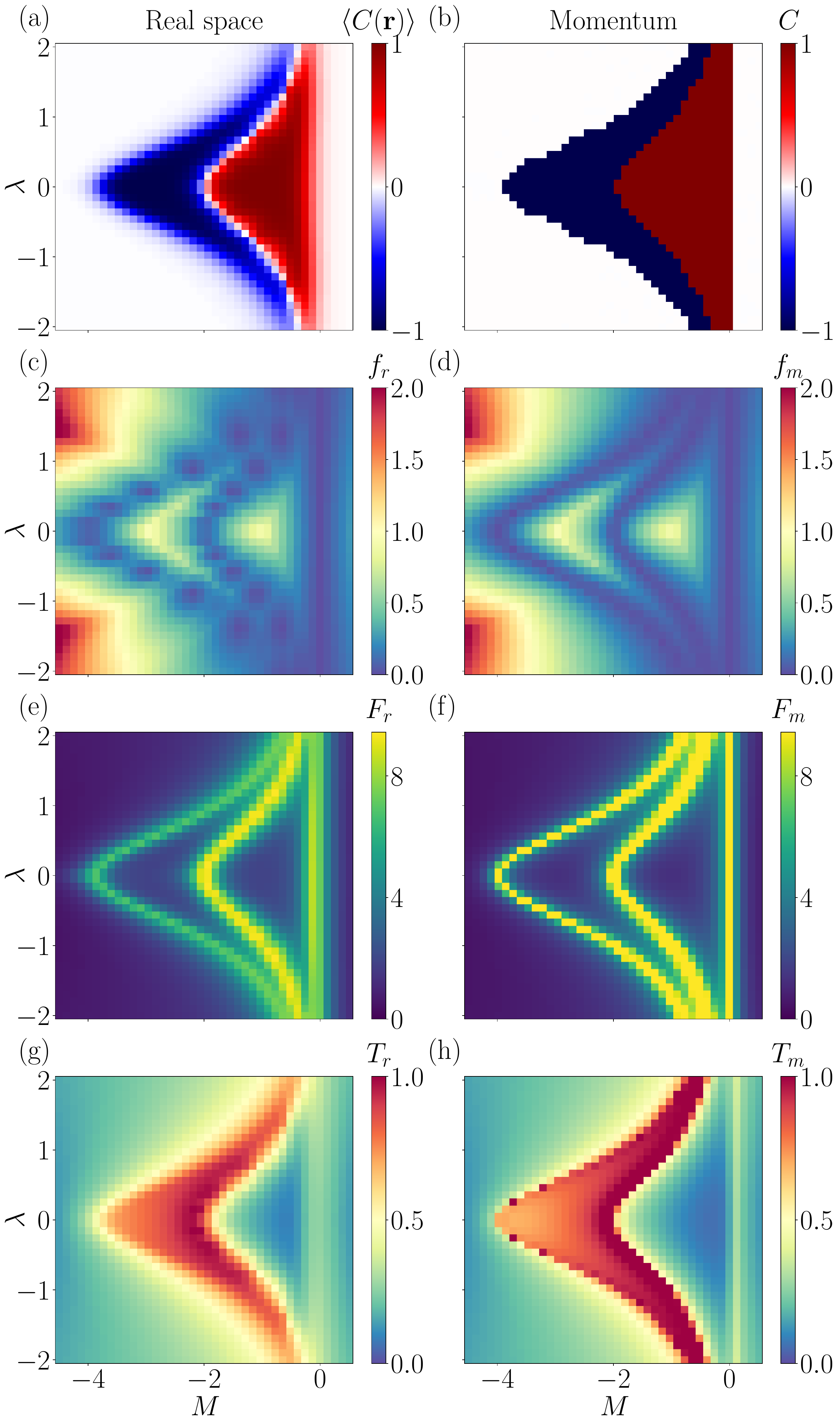}
    \caption[Comparison between the real- and momentum-space criteria for the amorphous Chern insulator model]{Phase diagrams for the amorphous Chern insulator model Eq.~\eqref{eq:aCI} in the crystalline limit as a function of $M$ and $\lambda$. We compare the quantities calculated in real space (left column) and momentum space (right column).\ (a) Bulk-averaged local Chern marker for $N_c=10\times 10$ for the lowest band.\ (b) Chern number of the lowest band computed by integrating the Berry curvature in momentum space.\ (c) Supercell band flatness $f_r$ for $N_c=15\times 15$.\ (d) Crystalline band flatness $f_m$.\ (e) Real-space Berry curvature fluctuations $F_r$ for $N_c=20\times 20$ and (f) Berry curvature fluctuations $F_m$ calculated in momentum space.\ (g) Real-space bulk vortexability $T_r$ for $N_c=10\times 10$ and (h) deviation from the trace condition $T_m$.
    All momentum quantities were obtained with a grid of $N_k=50\times 50$ points in the Brillouin zone.
    }\label{fig:bhz_characterization}
\end{figure}

To benchmark our real-space criteria against the momentum space criteria as listed in Table~\ref{tab:fci_criteria}  we start with the crystalline case. 
Fig.~\ref{fig:bhz_characterization}(a) shows the topological phase diagram as a function of $M$ and $\lambda$ calculated using the local Chern marker~\eqref{eq:LCM}, where the sum is taken over the bulk of a finite system, of size $N_c=10\times 10$. There are three phases with Chern numbers $C=0$ (white), $C=1$ (red) and $C=-1$ (blue).
Fig.~\ref{fig:bhz_characterization}(b) shows the corresponding phase diagram by integrating the Berry curvature in momentum space using the first expression in Eq.~\eqref{eq:chern_number}.

The flatness of the lowest band is computed using Eq.~\eqref{eq:flatness_r} for Fig.~\ref{fig:bhz_characterization}(c) and Eq.~\eqref{eq:flatness_m} for Fig.~\ref{fig:bhz_characterization}(d). We observe a good agreement between both calculations, with both phase diagrams showing the same broad features. The band flatness can be optimized to be a maximum of $f_r\sim 1$ within the regions of finite Chern number.

Next we compare the fluctuations of the Berry curvature for the crystalline system calculated in real space using Eq.~\eqref{eq:fluctuationsBerryreal}, shown in Fig.~\ref{fig:bhz_characterization}(e), with those calculated in 
momentum space using Eq.~\eqref{eq:fluctuations_berry}, shown in Fig.~\ref{fig:bhz_characterization}(f). We observe that both formulas match well as a function of the model parameters.
We note that the magnitude of the fluctuations is similar in Figs.~\ref{fig:bhz_characterization}(e) and (f) but not exactly the same, likely due to finite-size effects. Nonetheless, their similarity suggests that we can use the fluctuations in the Chern operator given by Eq.~\eqref{eq:fluctuationsBerryreal} as a proxy of the fluctuations of the Berry curvature in the disordered amorphous case.

Analogously, Figs.~\ref{fig:bhz_characterization}(g) and (h) show similar behaviour for both the deviation from the trace condition calculated as the vortexability in real space, $T_r$, and momentum space, $T_r$, predicting similar regions where those quantities are lower. From the phase diagrams in Fig.~\ref{fig:bhz_characterization}, we conclude that a region that is likely to stabilize a fractional Chern insulator corresponds to the region with $C=1$, as it minimizes both the vortexability and the curvature fluctuations, and maximizes the flatness ratio. We will focus specifically on the line $\lambda = 0$ when considering disorder.

\begin{figure}[h]
    \centering
    \includegraphics[width=0.7\columnwidth]{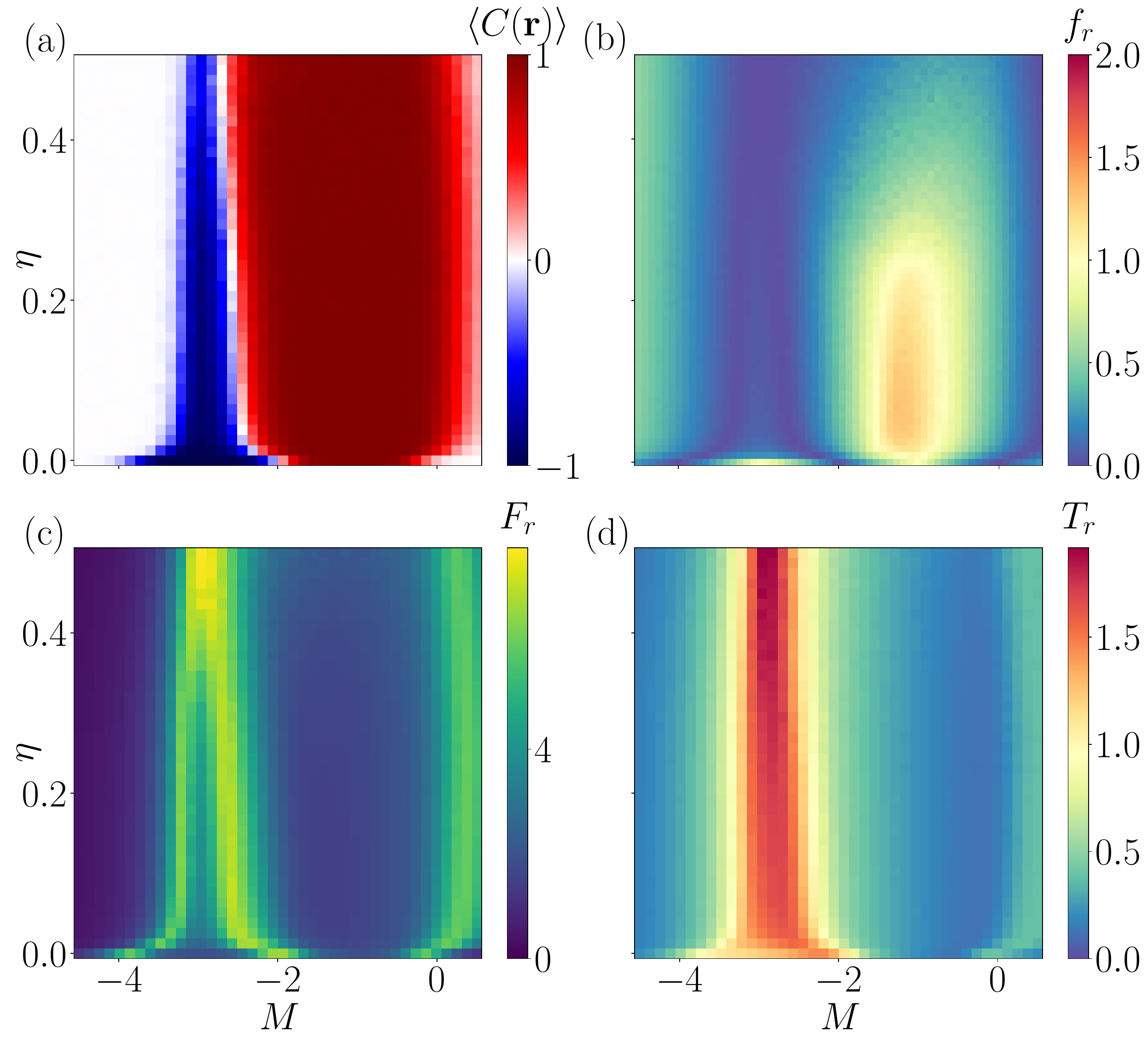}
    \caption[Real-space criteria as a function of disorder for the amorphous Chern insulator model]{Phase diagram of the amorphous Chern insulator as a function of the disorder strength $\eta$ and $M$. We fix $\lambda = 0$ as in the crystalline case it corresponds to the most likely region to host a fractional Chern insulator.\ (a) Chern marker, (b) flatness ratio, (c) real-space Berry curvature fluctuations and (d) vortexability. One can identify a candidate region that minimizes both the fluctuations and the vortexability, and maximizes the flatness ratio. The calculations are done for a system size $N_c=10\times 10$ and averaged over $N_s=50$ disorder realizations.}\label{fig:bhz_disordered_characterization}
\end{figure}

After benchmarking the real-space criteria, we address the effect of structural disorder on the phase diagram of Fig.~\ref{fig:bhz_characterization}. The phase diagrams for all four quantities as a function of the disorder strength $\eta$ and the onsite energy $M$ are displayed on Fig.~\ref{fig:bhz_disordered_characterization}. Focusing first on the topological regions in Fig.~\ref{fig:bhz_disordered_characterization}(a), we see that increasing $\eta$ results in the $C=-1$ phase disappearing in favor of the $C=1$ region. The band flatness $f_r$, shown in Fig.~\ref{fig:bhz_disordered_characterization}(b), appears to increase for low disorder within the $C=1$ region. Both fluctuations and vortexability, shown in Figs.~\ref{fig:bhz_disordered_characterization}(c) and (d), respectively, stay relatively constant for all values of disorder considered. 

These results combined suggest that the increase in the flatness could signal the stability of a fractional phase with disorder. 
However, note that optimizing these criteria does not guarantee a fractional Chern insulator state. They should be taken as indicators of promising regions in parameter space for performing many-body calculations.

\begin{figure}[t]
    \centering
    \includegraphics[width=1\columnwidth]{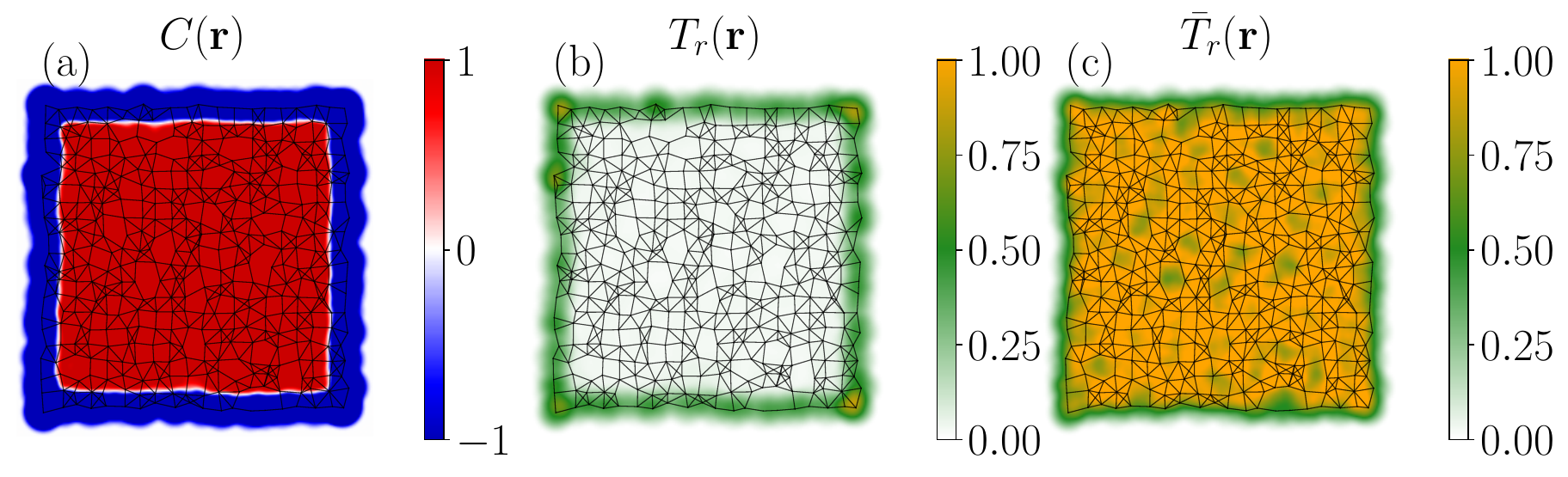}
    \caption[Real-space dependence of the local Chern marker, vortexability and antivortexability]{Real-space dependence of the (a) local Chern marker $C(\mathbf{r}_i)$ from Eq.~\eqref{eq:LCM}, (b) vortexability $T_r(\mathbf{r}_i)$ from Eq.~\eqref{eq:vortexabilityrealspace} and (c) antivortexability $\bar{T}_r(\mathbf{r}_i)$ for the amorphous Chern insulator with $w=0.4$, $N_c=20\times 20$, $M=-1$ and $t_2=\lambda=0$.}\label{fig:bhz_chern_vortex_markers}
\end{figure}

Lastly, the trace formulation for the Chern number and the vortexability allows defining local markers and hence real space maps for both quantities. We represent these in Fig.~\ref{fig:bhz_chern_vortex_markers}. For the local Chern marker, shown in Fig.~\ref{fig:bhz_chern_vortex_markers}(a), we reproduce the known behaviour where the marker takes the approximately quantized value in the bulk of the system, while it takes the opposite value at the boundary since for a finite system the total Chern number must be zero~\cite{bianco2011mapping}. The local vortexability $T_r(\mathbf{r}_i)$ and local antivortexability $\bar{T}_r(\mathbf{r}_i)$ are shown in Figs.~\ref{fig:bhz_chern_vortex_markers}(b) and (c) respectively. The vortexability is always correlated with the Chern number ($\Xi C > 0$, where $\Xi$ is the vortex chirality)~\cite{ledwith2022vortexability}. This means that in the $C=1$ region it will take lower values, while for $C=-1$ its value will increase. For the local antivortexability the opposite takes place: it takes higher values in the bulk than in the edge, since it correlates with the opposite Chern number $C=-1$.

The real-space fluctuations of the Chern marker have been recently studied as a signature of the topological Anderson transitions~\cite{Mildner2023,Assun2024}. 
It is reasonable to expect that the vortexability and antivortexability can also be used to characterize further Anderson transitions, as they are related to both the Berry curvature and the metric.

\subsection{Dirac Landau levels in graphene}

\begin{figure}[t]
    \centering
    \includegraphics[width=0.7\columnwidth]{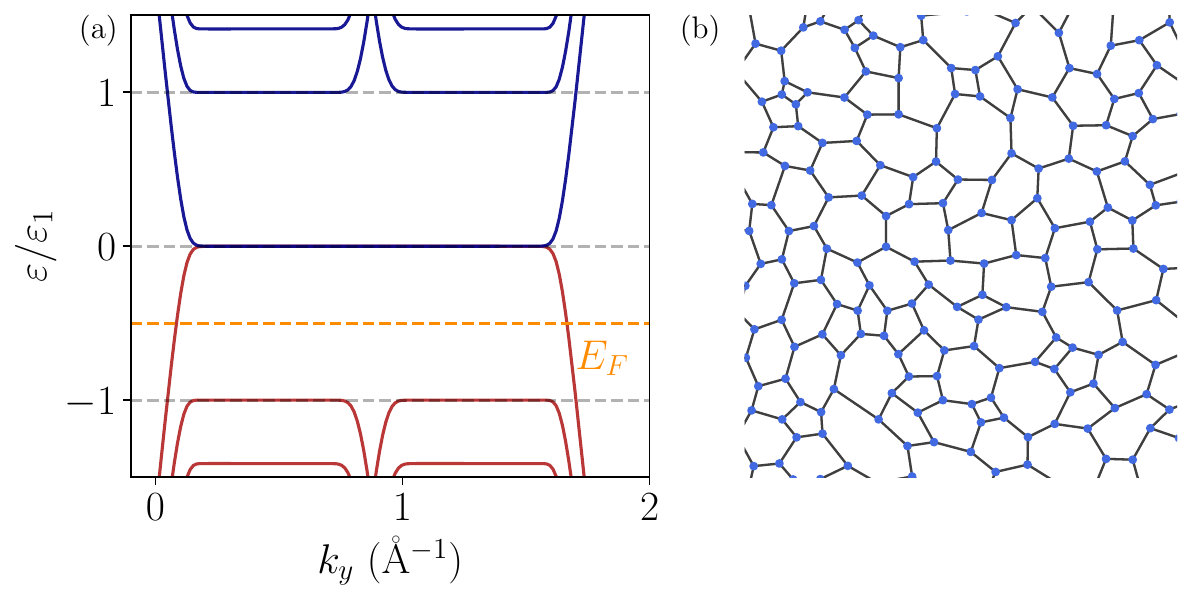}
    \caption[Dirac Landau levels in graphene and voronoization of the lattice]{(a) Landau levels for $B\approx 100$ T in a zigzag ribbon of graphene of 500 atoms width. The energy is measured relative to the first Landau level $\varepsilon_1=v_F\sqrt{2eB}/\hbar$.\ (b) Amorphous graphene system obtained with the Voronoization procedure for $\eta=2$. This procedure enforces a disordered lattice with three-fold coordination.}\label{fig:landau_levels_voronoi}
\end{figure}

We next apply the real-space criteria for the study of Landau levels of amorphous graphene. We start with the tight-binding description of $p_z$ orbitals of graphene~\cite{CastroNeto2009}, with an applied magnetic field perpendicular to the plane, $\mathbf{B}=B_0\mathbf{z}$. 
Using the Peierls substitution, the Hamiltonian of this model can be written as
\begin{equation}
    H = \sum_{\left<i,j\right>}t_{ij}'c^{\dagger}_ic_j + \hspace{0.2cm} \mathrm{h.c.},
\end{equation}
where
\begin{equation}
    t_{ij}'=t\cdot\text{exp}\left(i\frac{e}{\hbar}\int_i^j \mathbf{A}\cdot\mathbf{dl}\right).
\end{equation}
$\mathbf{A}$ is the magnetic potential vector, which in this case corresponds to $\mathbf{A}=B_0x\mathbf{\hat{y}}$, and $t$ is the real hopping amplitude without magnetic field ($B_0=0$). The indices $\left<i,j\right>$ run over pairs of nearest neighbours of a honeycomb lattice. It is more convenient to write the magnetic field in terms of the magnetic flux per unit cell $\phi/\phi_0=BA_c/\phi_0$, where $\phi_0=e/h$ is the flux quantum, and $A_c=\sqrt{3}a^2/2$ is the unit cell area, with $a=2.46$\AA~in graphene. Using the above expression for the vector potential along the nearest-neighbor paths we can write the hoppings as
\begin{equation}\label{eq:hopping-amplitude-peierls}
    t'=t\cdot \text{exp}\left(i\frac{\pi}{A_c}\frac{\phi}{\phi_0}(x_j+x_i)(y_j-y_i)\right).
\end{equation}
where now the applied magnetic field is taken into account in units of the flux quantum per unit cell. In Fig.~\ref{fig:landau_levels_voronoi}(a) we see the first Landau levels for a crystalline zigzag ribbon of graphene~\cite{CastroNeto2009}. 

We are interested in the effect of structural disorder on the Landau levels. To continuously control the number of non-hexagonal plaquettes we follow the procedure introduced in Refs.~\cite{munoz2023structural,Grushin2023}.
It starts with the observation that a honeycomb lattice can be created as the dual lattice of a triangular lattice. This means that the honeycomb lattice can be constructed by finding the area closer to each point that forms the triangular lattice. Since all vertices in the triangular lattice are equally spaced, this procedure creates the hexagonal cells of the honeycomb lattice. 

This procedure, called Voronization, can be conveniently modified to introduce structural disorder, i.e.\ non-hexagonal plaquettes~\cite{munoz2023structural,Grushin2023}. 
Starting from a triangular lattice, we disorder it by introducing random displacements of the positions sampling from a uniform distribution $U(0,\eta)$. Then, the distorted lattice is Voronoized, i.e.\ we find the area closest to a given site, called Voronoi cells. 
The vertices of each Voronoi cell define the sites of the amorphous system, and the cell edges define its connectivity. The connectivity is such that each vertex has exactly three neighbours, which defines the amorphous graphene system, as shown in Fig.~\ref{fig:landau_levels_voronoi}(b). The advantage of this procedure is that is keeps the coordination number constant and equal to three, mimicking what is observed in solid-state~\cite{toh_synthesis_2020} and synthetic experimental realizations of amorphous graphene~\cite{Mitchell2021}. 

\begin{figure}[h]
    \centering
    \includegraphics[width=0.65\columnwidth]{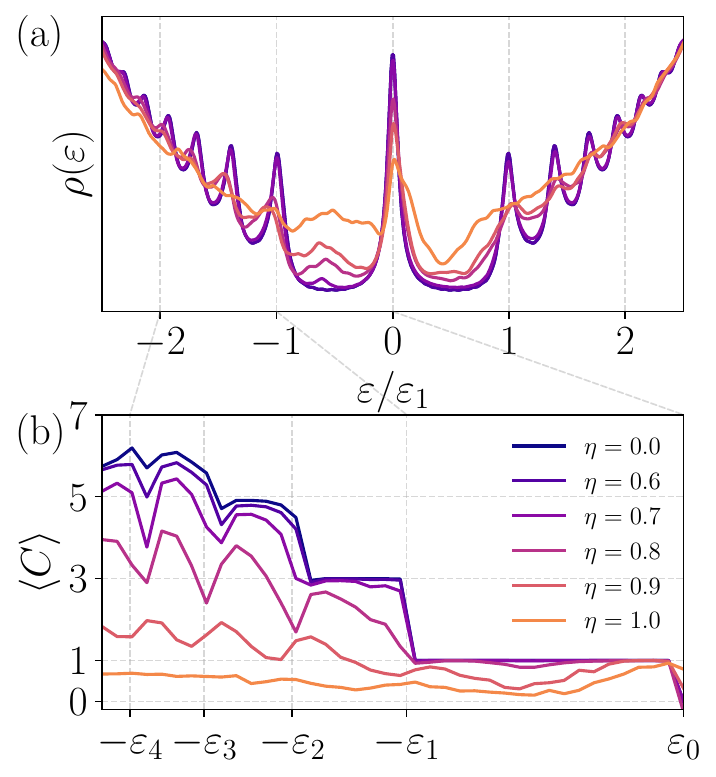}
    \caption[Effect of disorder on the density of states and Chern number of Dirac Landau levels]{Effect of different disorder strengths on the stability of the Landau levels for $N_c=30\times 45$ rectangular unit cells, $\phi/\phi_0=1/100$ averaged over 20 samples.\ (a) Total density of states for the finite size sample.\ (b) Bulk average of the local Chern marker as a function of energy. The labels $\varepsilon_n$ denotes the energy of the $n-$th Landau level in the clean system, given by $\varepsilon_{\pm n}=\pm v_F\sqrt{2neB}/\hbar$.}\label{fig:landaulevels-disorder}
\end{figure}

This amorphous three-fold coordinated lattice will contain in general a density of non-hexagonal plaquettes proportional to $\eta$~\cite{munoz2023structural,Grushin2023}. An example of the resulting amorphous graphene system is shown~\ref{fig:landau_levels_voronoi}(b) for $\eta=2$. The hopping amplitudes $t_{ij}'$ are modified due to the changes in distances and areas, as reflected by Eq.~\eqref{eq:hopping-amplitude-peierls}. The original hoppings $t$ remain constant.
Keeping $t$ constant as a function of disorder allows us to separate the effects coming from the connectivity of the lattice from the more often studied effects associated to disordering the hoppings or on-site energies~\cite{Peres2010}. 
A similar problem was studied with different methods in a structurally-disordered square lattice in Ref.~\cite{sahlberg_quantum_2023}.

The effect of disorder on the Landau level spectrum is summarized in Fig.~\ref{fig:landaulevels-disorder}. We plot the bulk density of states in panel (a) and the local Chern marker averaged in the bulk of the system in panel (b). The energy dependence in (b) is captured via the projector $P=\theta(E - H)$, that enforces that all states with energy below $E$ are filled. The Chern number is then averaged over disorder for $N_s=20$ realizations.

For small disorder, $\eta < 0.6$, the changes in the lattice do not produce any noticeable effect on the stability of the Landau levels. This can be likely attributed to the fact that the coordination number is fixed and that the unit cell area is only slightly modified. For stronger disorder, as soon as non-hexagonal polygons appear, the stability of the higher Landau levels is rapidly compromised, as seen in Fig.~\ref{fig:landaulevels-disorder}(b). For $\eta=0.8\sim 0.9$ only the first Landau level still shows quantization, and for stronger disorder the topological behaviour disappears.

\begin{figure}[t]
    \centering
    \includegraphics[width=0.75\columnwidth]{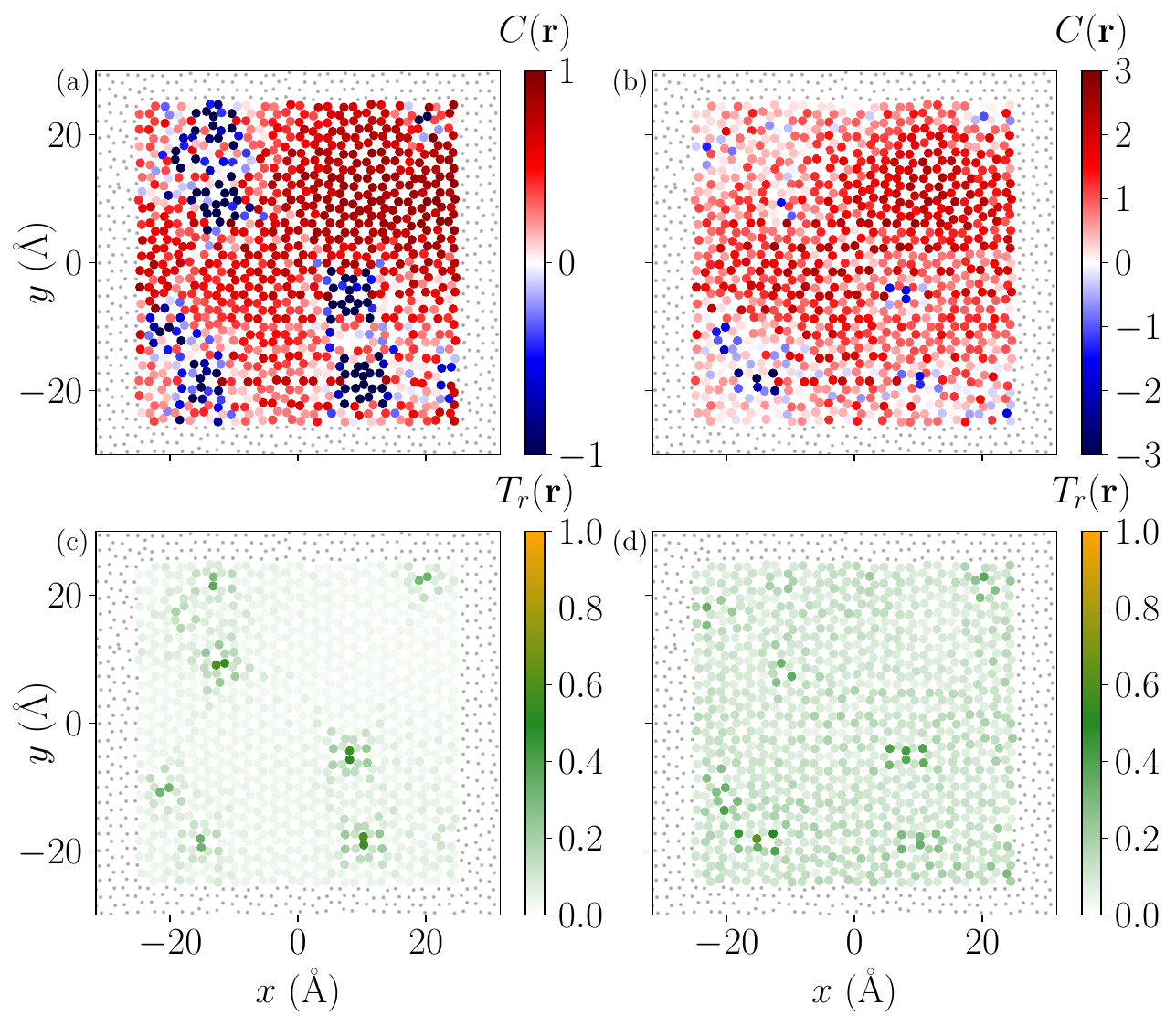}
    \caption[Local Chern marker and local vortexability for the first and second Dirac Landau levels]{(a, b) Chern marker and (c, d) local vortexability computed in the bulk region of a slab of $N_c=25\times 40$ (rectangular) unit cells, with $\phi/\phi_0=1/100$ and $\eta=0.9$ for two different Landau levels, (a, c) the $n=-1$ Landau level at $E=-\varepsilon_1/2$, and (b, d) the $n=-2$ Landau level at $E=-(\varepsilon_2+\varepsilon_1)/2$. To ensure that each LL is completely filled, one selects an energy in-between levels.}\label{fig:chern-marker-landaulevels}
\end{figure}

In what follows, we will consider the first and second Landau levels, referring to quantum numbers $n=-1$ and $n=-2$ respectively, according to the energy formula for Dirac Landau levels $\varepsilon_{\pm}=\pm v_F\sqrt{2neB}\hbar$. Fig.~\ref{fig:landau_levels_voronoi}(a) illustrates filling the system up to the first Landau level.
The effect of disorder can also be understood by mapping the local Chern marker $C(\mathbf{r})$ in real space. In Fig.~\ref{fig:chern-marker-landaulevels} we show $C(\mathbf{r})$ for the first (a) and second (b) Landau levels at $\eta=0.9$. Concomitant with non-hexagonal polygons we observe that boundaries between trivial and topological regions appear. 
These regions surround non-hexagonal plaquettes which have a local negative Chern marker. For the first Landau level in Fig.~\ref{fig:chern-marker-landaulevels}(a), quantization is still robust in the neighborhood of a pentagon or heptagon. In contrast, the second Landau level in Fig.~\ref{fig:chern-marker-landaulevels}(b) appears to be more sensitive to the presence of non-hexagonal polygons showing values smaller than the expected quantization $C=3$.  

These effects can also be seen plotting the local vortexability for the first and second Landau levels, shown in Figs.~\ref{fig:chern-marker-landaulevels}(c) and (d), respectively. The vortexability deviates from zero around non-hexagonal plaquettes. Comparing these two figures, we observe that this effect is once again more prominent for the second Landau level.

\begin{figure}
    \centering
    \includegraphics[width=0.85\columnwidth]{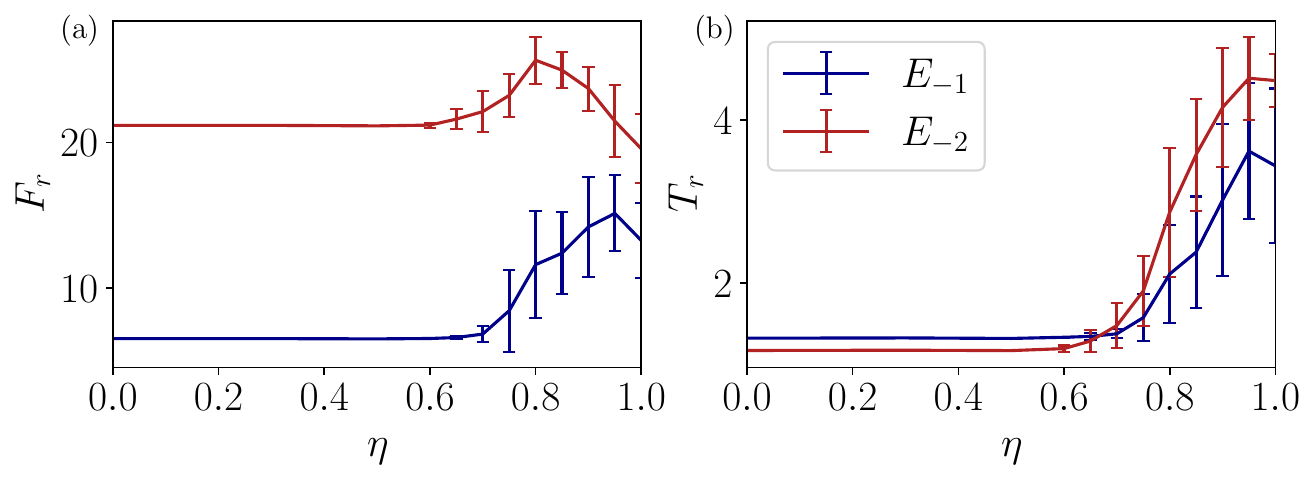}
    \caption[Real-space Berry curvature fluctuations and vortexability for the first and second Landau levels as a function of disorder]{(a) Real-space Berry curvature fluctuations and (b) vortexability in a finite system of size $N_c=25\times 35$ with flux $\phi/\phi_0=1/100$ as a function of the disorder strength $\eta$. Both quantities have been computed for two different Landau levels, $E_{-1}=-\varepsilon_1/2$ and $E_{-2}=-(\varepsilon_2+\varepsilon_1)/2$, and averaged over $N_s=20$ samples.}\label{fig:landaulevels-fci-markers}
\end{figure}

We now discuss how the difference between Landau levels becomes apparent through the Berry curvature fluctuations $F_r$.
In particular, large
real-space Berry curvature fluctuations and vortexability showcase deviations from the ideal Landau level. The results are shown in Fig.~\ref{fig:landaulevels-fci-markers}. For the first Landau level, shown in blue, we observe that both the fluctuations of the Berry curvature and the vortexability are small. As disorder is increased both quantities increase as soon as the first non-hexagonal plaquettes appear. 
The increase in both $F_r$ and $T_r$ reflect the localization transition of a Landau level as disorder is increased, which we do not analyze further.

A more pronounced behaviour is observed for the second Landau level, shown in red. The Berry curvature fluctuations, $F_r$, start from a higher value compared to the first Landau level. This is consistent with the fact that the associated total Chern number is higher ($C=3$) and therefore the variance of the local Chern marker is expected to be higher as well.
Interestingly, for disorder $\eta\geq 0.8$ the Berry curvature fluctuations decrease, most likely due to the transition to a localized phase. The vortexability also increases to values beyond those of the first Landau level, signaling that the transition to a topologically trivial state occurs faster than for the first Landau level.

\begin{figure}
    \centering
    \includegraphics[width=0.65\columnwidth]{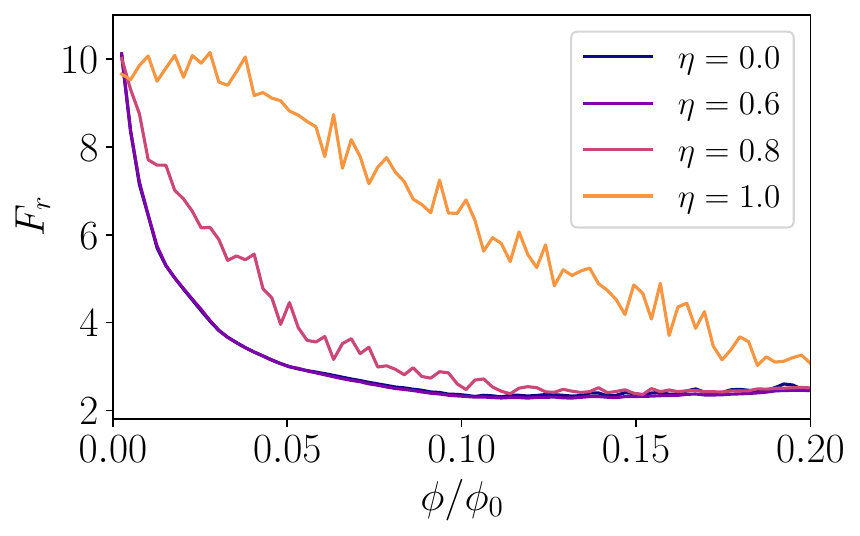}
    \caption[Real-space Berry curvature fluctuations as a function of the incident flux for the first Dirac Landau level]{Real-space Berry curvature fluctuations of the $n=-1$ Landau level of graphene at energy $E=-\varepsilon_1/2$ as a function of applied magnetic flux $\phi/\phi_0$ on a finite system of size $N_c=15\times 25$, for different disorder strengths and averaged over 40 realizations. Berry curvature fluctuations decrease until the magnetic length is comparable to the lattice spacing, $l_B\approx a$. After this scale, the Landau levels become Chern bands, whose Berry fluctuations may vary with flux.}\label{fig:dirac_fluctuations_v_phi}
\end{figure}

Lastly, in Fig.~\ref{fig:dirac_fluctuations_v_phi} we represent the value of the real-space Berry curvature fluctuations as a function of the incident flux, for different disorder strengths. Up to $\eta\sim 0.8$, we observe that the value of the fluctuations $F_r$ decreases as $\phi$ increases. This is consistent with the fact that, as the Landau level is increasingly better defined (wider flat band), its Berry curvature fluctuations approach zero, as that of an ideal flat-band Landau level. The Berry curvature fluctuations start increasing mildly at fluxes larger than $\phi/\phi_0\sim 0.15$. At this flux value, the magnetic length $l_B$ becomes comparable to the lattice parameter $a$. This point signals the transition from the Landau level regime to the Hofstadter regime. Consequently, the bands deviate from Landau levels, and realize increasingly dispersive Chern bands, whose Berry curvature fluctuations can vary.



In short, structural disorder drives Landau levels in graphene away from the ideal limit of zero vortexability, large band-flatness and homogeneously quantized bulk Chern marker. These markers show that increasing structural disorder affects Landau levels at higher energy more efficiently than those at lower energies. The increasing fragility of higher Landau levels might be expected on the grounds that the energy-level spacing for Landau levels in graphene is not constant, but rather decreasing the higher the Landau level index is~\cite{CastroNeto2009}. Hence, disorder is more efficient in broadening and mixing higher Landau levels than lower Landau levels, driving them faster to a non-ideal limit. 

So far we have discussed the effect of structural disorder on Landau levels in graphene close to half-filling. It is possible to perform a similar analysis on the Landau levels emerging from a parabolic band, such as those emerging at the $\Gamma$ point in graphene's valence band, as discussed next.

\subsection{Landau levels at $\Gamma$ in graphene}\label{app:Landaugraphene}

In addition to the Dirac Landau Levels discussed before, which are arguably the relevant ones as they take place at the Fermi energy, the tight-binding model of graphene can be pushed to exhibit Landau Levels at the bottom of the valence band at $\Gamma$.
These correspond to Landau levels of nearly free electrons, because they follow a parabolic dispersion with an effective mass. These Landau Levels can also be used to benchmark the real-space markers we have introduced. 

\begin{figure}[t]
    \centering
    \includegraphics[width=0.75\columnwidth]{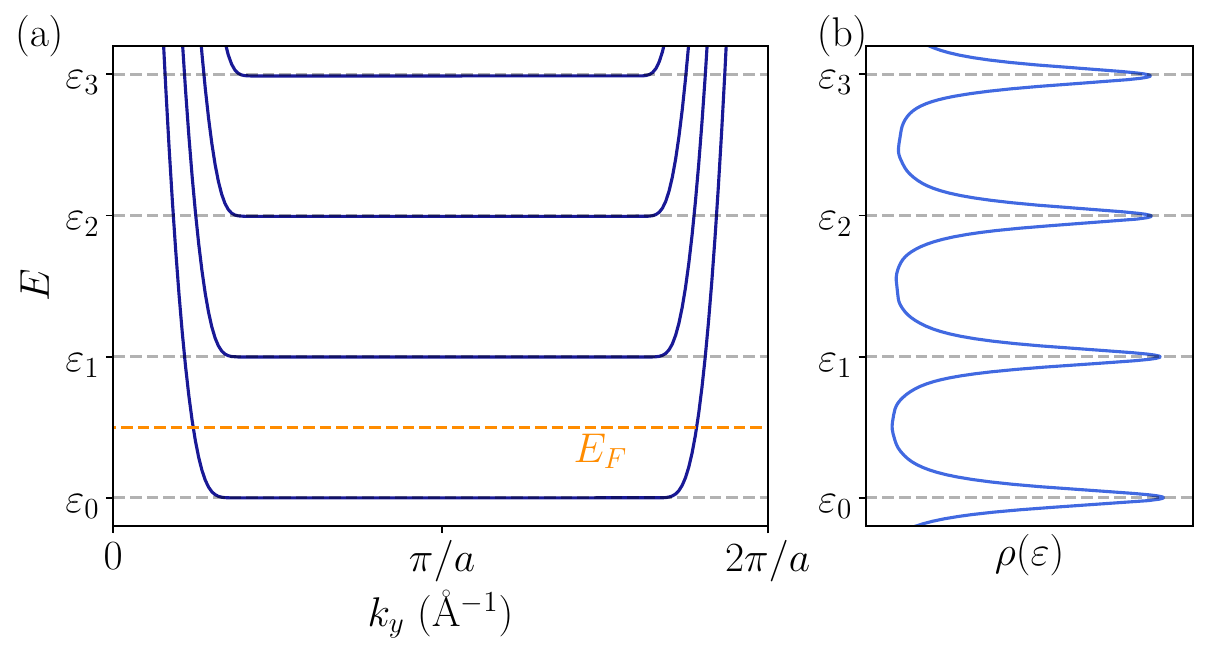}
    \caption[Band structure and density of states of Landau levels for free electrons]{(a) Landau levels associated to the parabolic band dispersion at the $\Gamma$ point in a zigzag graphene ribbon of 500 atoms width and applied magnetic field $B\approx 250$ T. (b) Density of states.}\label{fig:landaulevels_parabolic}
\end{figure}

\begin{figure}[t]
    \centering
    \includegraphics[width=0.8\columnwidth]{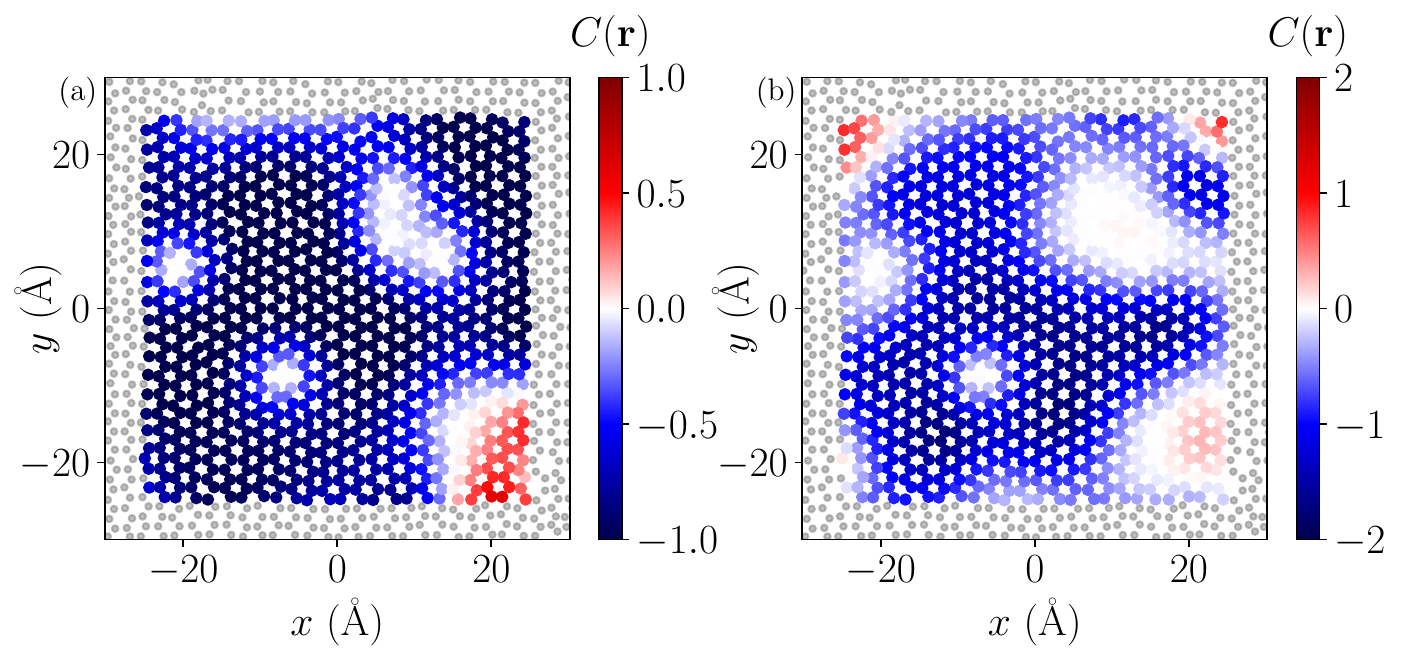}
    \caption[Local Chern marker for the first and second parabolic Landau levels]{Chern marker in the bulk of a finite graphene island of size $N=25\times 40$, with applied flux $\phi/\phi_0=1/100$ and disorder strength $\eta=0.8$, evaluated at (a) the first parabolic Landau level at energy $E=\frac{3}{3}\hbar \omega_c$ and (b) second LL at $E=\frac{5}{2}\hbar\omega_c$.}\label{fig:chernmarker_parabolic}
\end{figure}

Their band structure and the corresponding density of states are shown in Fig.~\ref{fig:landaulevels_parabolic}. We check the stability against disorder of these Landau Levels; to this end we consider for the first and second Landau Levels the local Chern marker in the bulk of the systems, which are represented in Fig.~\ref{fig:chernmarker_parabolic}. Their behaviour follows closely that of the Dirac Landau Levels, namely changes in the connectivity of the honeycomb lattice (i.e.\ the introduction of polygons different from hexagons) result in regions with $C=0$, or if close the edge, with $C=-1$. This effect is more prominent for the second LL, which can be possibly attributed to the fact that the corresponding flat band is smaller, or equivalently that is more sensitive to finite size effects.

\begin{figure}[!h]
    \centering
    \includegraphics[width=0.65\columnwidth]{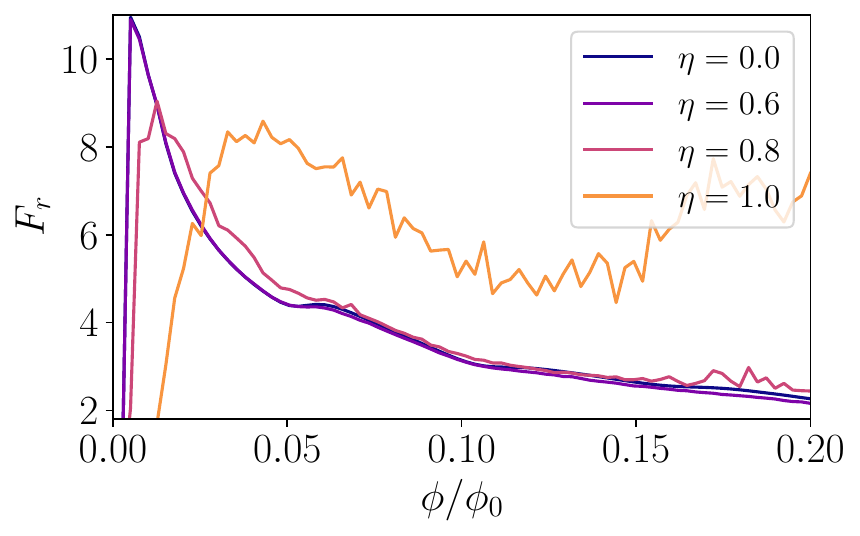}
    \caption[Real-space Berry curvature fluctuations of the first parabolic Landau level as a function of the incident flux]{Fluctuations of the Chern marker associated to the lowest Landau level at $\Gamma$ as a function of the applied magnetic flux $\phi/\phi_0$ for different disorder strengths and averaged over 40 samples.}\label{fig:fluctuations_parabolic}
\end{figure}

Using the real-space Berry curvature fluctuations we address the stability of the first LL, which we show in Fig.~\ref{fig:fluctuations_parabolic}. In this case, up to $\eta=0.8$ the LL mostly retains its characteristics, as can be seen also from the spatially resolved LCM, where it is still mostly topological in the bulk. For higher disorder strengths however, the system strongly departs from the ideal LL behaviour, which is reflected in the fluctuations generally taking higher values.

\subsection{Chern insulator model for ultra-cold atomic lattices}

The last model we consider is a model conceived for ultra-cold atoms~\cite{weber2022,zhao2023}. It was introduced as a basis to  pursue fractional Chern insulator ground states of interacting bosonic nuclei that are allowed to hop on a regular lattice. Here we focus on its single-particle properties relative to our local criteria, similarly to our previous models. The Hamiltonian for this model is defined on a honeycomb lattice, and reads
\begin{equation}
    H =\sum_{i,\alpha,\beta} \varepsilon_{\alpha\beta} c^{\dagger}_{i\alpha}c_{i\alpha} + \sum_{\substack{i,j\neq i \\ \alpha,\beta}} t_{\alpha\beta}(\mathbf{r}_{ij})c^{\dagger}_{i\alpha}c_{j\beta},
\end{equation}
where
\begin{subequations}\label{eq:a-coldmodel}
\begin{eqnarray}
    \left[\varepsilon_{\alpha\beta}\right] &=& 
    \begin{pmatrix}\Delta & 0 \\ 
    0 & 0\end{pmatrix}, \quad\quad\\
\left[t_{\alpha\beta}\right](\mathbf{r}_{ij}) &=& \frac{a^3}{|\mathbf{r}_{ij}|^3}
\begin{pmatrix} -t_a & \omega e^{-i2\theta} \\ 
\omega e^{i2\theta} & -t_b 
\end{pmatrix}
\theta_H(R_c - |\mathbf{r}_{ij}|).
\end{eqnarray}
\end{subequations}

\begin{figure}[t]
    \centering
    \includegraphics[width=0.8\columnwidth]{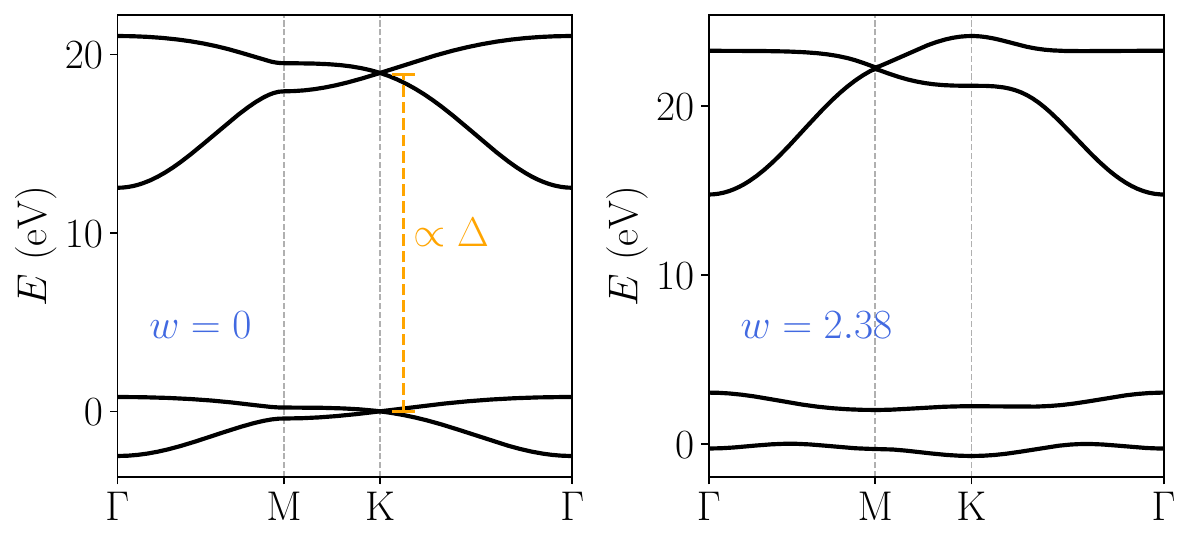}
    \caption[Band structure of the cold atoms model]{Band structure of the model defined by Eq.~\eqref{eq:a-coldmodel} for two different values of $w$. The rest of the parameters are chosen to be $\Delta=18.52$, $t_a=1.26$, $t_b=0.49$ and $R_c=2.1$.}\label{fig:cold_atoms_bands}
\end{figure}

This model is similar to the amorphous Chern insulator model introduced in Eq.~\eqref{eq:a-CImodel}. The two main differences are that, in its crystalline form, $i,j$ run over a honeycomb lattice instead of the square lattice, and the exponential mixing of orbitals in the off-diagonal matrix elements has an additional factor of $2$. Additionally, while the Chern insulator model Eq.~\eqref{eq:a-CImodel} allows hopping between first-nearest neighbours, for this model we set the cutoff distance to $R_c=2.1$, which corresponds to hoppings up to third-nearest neighbours in the crystalline limit (with $a=1$). For $w=0$, the model realizes two decoupled models of the $p_z$ orbitals of graphene, as shown in the band structure in Fig.~\ref{fig:cold_atoms_bands}(a). For $w=2.38$, the lowest bands open a topological gap that assigns the lowest band a Chern number $C=1$. In what follows we focus on this band by setting the total filling to $\nu=1/4$, fully filling the lowest band. The band structure for this value of $w$ is shown in Fig.~\ref{fig:cold_atoms_bands}(b).

\begin{figure}[h]
    \centering
    \includegraphics[width=0.8\columnwidth]{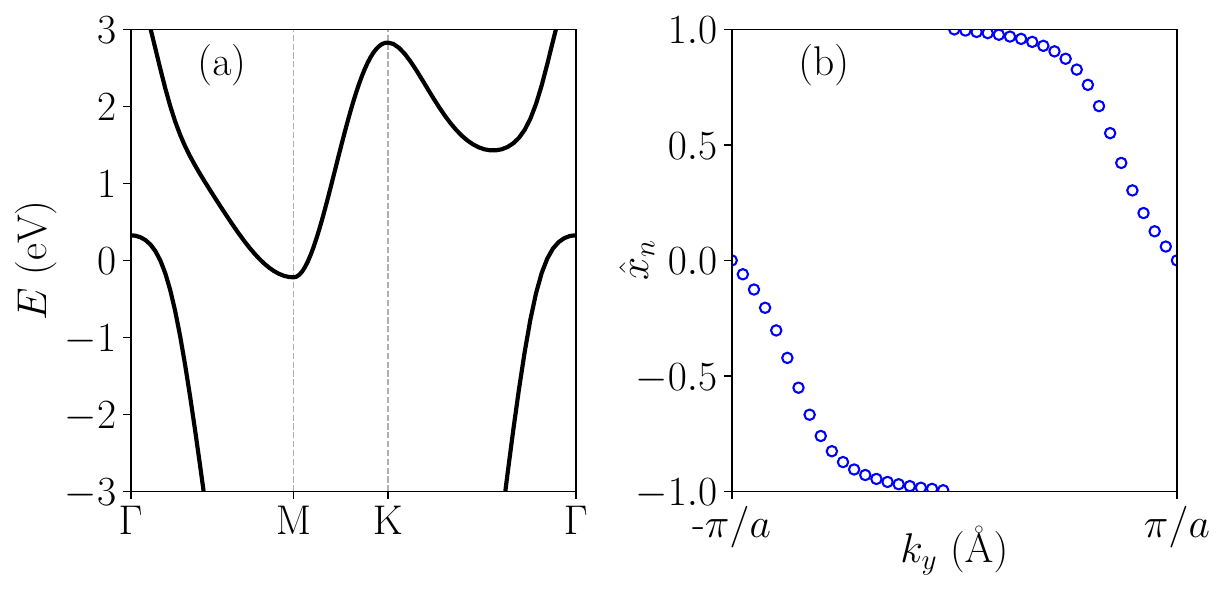}
    \caption[Topological semimetallic behaviour of the cold atoms model]{(a) Band structure of the model defined by Eq.~\eqref{eq:a-coldmodel} for $\Delta=10$ and $w=5.5$.\ (b) Wannier charge center flow of the lowest occupied band, corresponding to $C=-1$. For some regions of parameters, highlighted in Fig.~\ref{fig:rydberg-crystalline}, the model transitions from insulating to semimetallic behaviour, where it still displays topological properties.}\label{fig:rydberg_semimetal}
\end{figure}

\begin{figure}
    \centering
    \includegraphics[width=0.697\columnwidth]{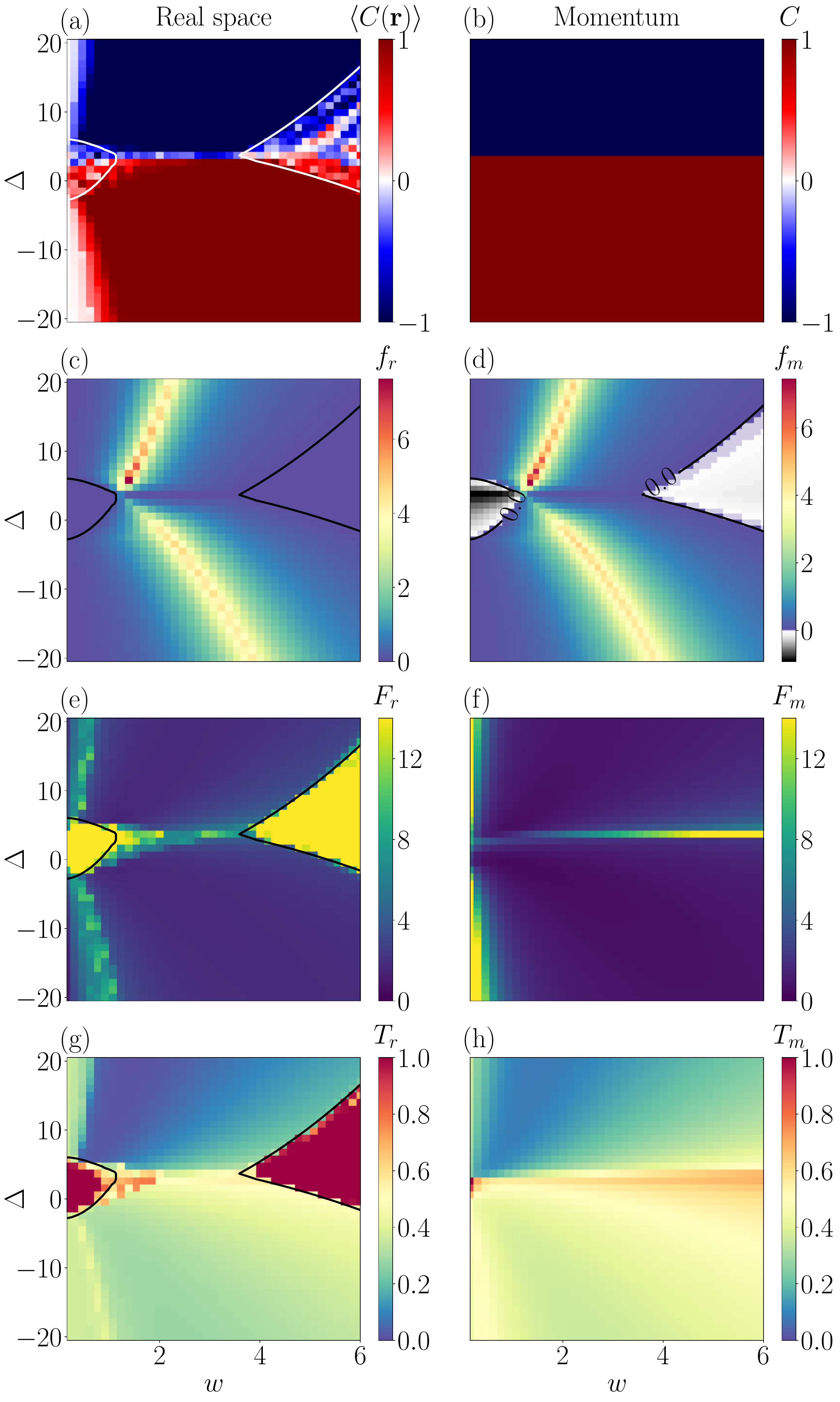}
    \caption[Comparison between the real- and momentum-space criteria for the cold atoms model]{
    Phase diagram of the crystalline Chern insulator~\eqref{eq:a-coldmodel} as a function of $\Delta$ and $w$. We compare the quantities calculated in real space (left column) and momentum space (right column).\ (a) Bulk-averaged local Chern marker for the lowest band.\ (b) Chern number of the lowest band computed by integrating the Berry curvature in momentum space.\ (c) Supercell band flatness $f_r$ for $N_c=15\times 15$.\ (d) Crystalline band flatness $f_m$.\ (e) Real-space Berry curvature fluctuations and (f) Berry curvature fluctuations calculated in momentum space.\ (g) Real-space bulk vortexability.\ (h) Deviation from the trace condition. The solid white lines in panel (a), and solid black lines in panels (c, e, g) correspond to contour lines obtained from the condition $f_m=0$ in panel (d). The regions colored in gray scale in (d) correspond to semimetallic behaviour (see Fig.~\ref{fig:rydberg_semimetal}), which implies $f_m\leq 0$ according to the definition in Eq.~\eqref{eq:flatness_m}.
    All momentum quantities were obtained with a grid of $N_k=50\times 50$ points in the Brillouin zone. The real-space quantities have been obtained with a system size of $N_c=18\times 18$ unit cells.}\label{fig:rydberg-crystalline}
\end{figure}
Fig.~\ref{fig:rydberg-crystalline} compares the real-space criteria for the flatness, Berry curvature fluctuations and vortexability for the crystalline version of the model~\eqref{eq:a-coldmodel}. First, the topological phase diagrams in Figs.~\ref{fig:rydberg-crystalline}(a) and (b), obtained via the real-space local Chern marker and the momentum-space Berry curvature respectively, see Eq.~\eqref{eq:chern_number}, yield fundamentally the same phases. 

The main difference is that, in the real-space calculation, there are two regions outlined by white lines in Fig.~\ref{fig:rydberg-crystalline}(a) where the averaged local Chern marker $\left<C(\mathbf{r})\right>$ deviates from quantized values. In those regions, the gap is indirect, and the system becomes semimetallic, see  Fig.~\ref{fig:rydberg_semimetal}(a). It is still possible to assign a topological Chern number to each band in momentum space. In Fig.~\ref{fig:rydberg_semimetal}(b) we show a representative Wannier charge center flow within this semimetallic phase for $\Delta=10$, $w=5.5$. 
Aside from these semimetallic regions, where the ability of the Chern marker to capture topology is expected to breakdown, Figs.~\ref{fig:rydberg-crystalline}(a, b) display the same behaviour.

These deviations between real and momentum space criteria due to semimetallic regions is also seen in other quantities. Comparing the flatness plots in Figs.~\ref{fig:rydberg-crystalline}(c) and (d), we observe a region with negative band flatness, per our definition in Eq.~\eqref{eq:flatness_m}, as a consequence of this semimetallic band structure. Once again, discarding these semimetallic regions, Figs.~\ref{fig:rydberg-crystalline}(c, d) display the same behaviour.

Figs.~\ref{fig:rydberg-crystalline}(e) and (f) compare the Berry curvature fluctuations calculated in real and momentum space, which mimic the features seen in Figs.~\ref{fig:rydberg-crystalline}(a) and (b). In general, we observe good agreement between both quantities, aside from the semimetallic region we already discussed. Deviations occur because the projector onto occupied states, $P$, 
mixes states from different bands when computed for finite systems, resulting in regions with large real-space Berry curvature fluctuations $F_r$. We see how this also affects the real and momentum space comparison of vortexabilities in Figs.~\ref{fig:rydberg-crystalline}(g) and (h). Overall, for both the real-space vortexability and Berry curvature fluctuations, the regions where we find their minimum values coincide with those predicted by the momentum markers.

\begin{figure}[h]
    \centering
    \includegraphics[width=0.7\columnwidth]{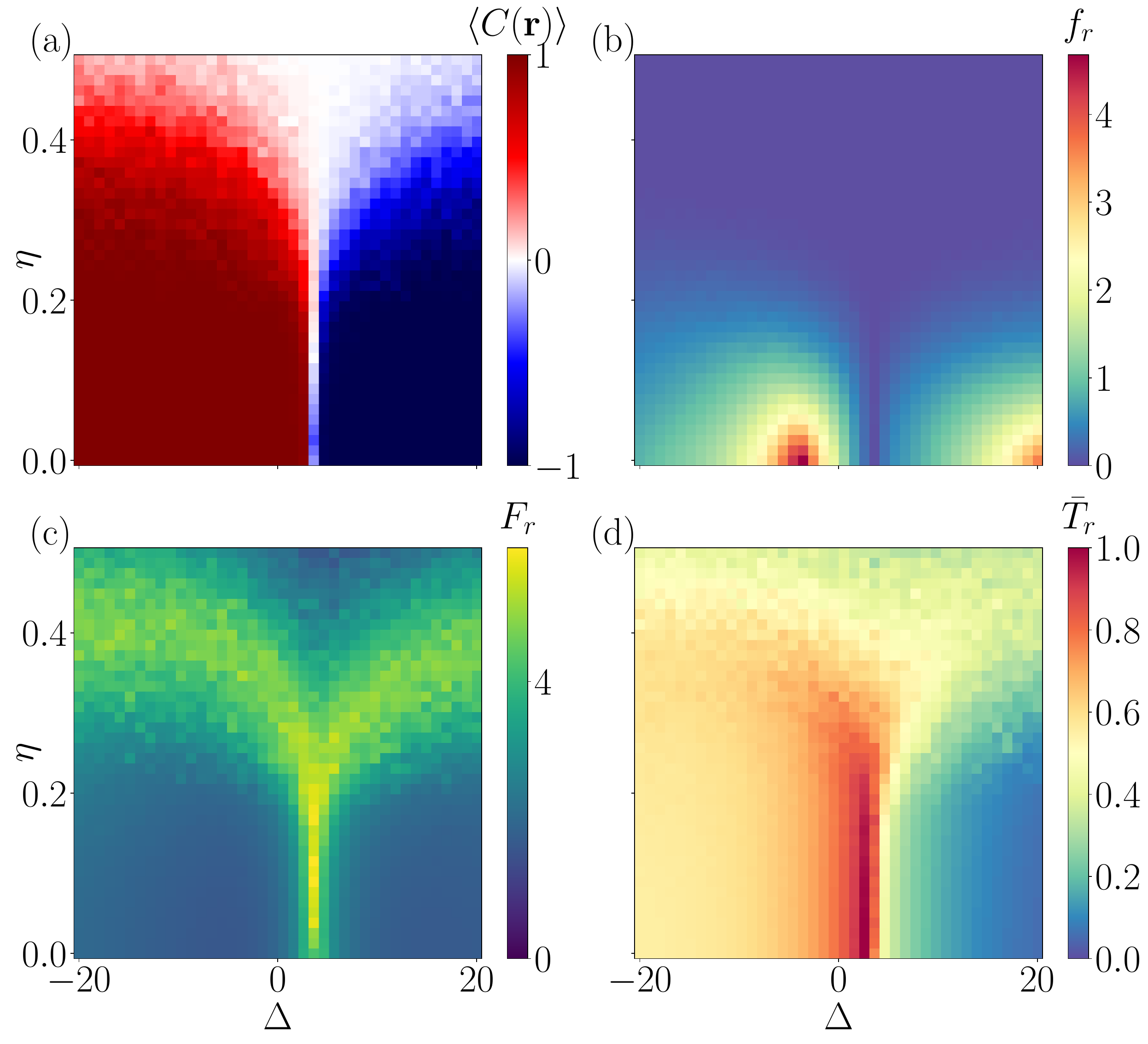}
    \caption[Real-space criteria as a function of disorder for the cold atoms model]{Characterization of the cold atoms model as a function of $\Delta$ and the disorder strength $\eta$, for $N_c=10\times 10$. We fix $w = 2.38$ as in the crystalline case it corresponds to the most likely region to host an FCI\@. (a) Chern marker, (b) flatness ratio, (c) real-space Berry curvature fluctuations and (d) (anti)vortexability. From these quantities we can estimate the maximum value of disorder that could withstand a fractional Chern insulator phase, which lies around $\eta\sim 0.2-0.3$. All plots have been averaged over $N_s=40$ samples.}\label{fig:rydberg-amorphous}
\end{figure}

Next, we apply the real-space criteria to the disordered model. 
These are shown in Fig.~\ref{fig:rydberg-amorphous}, where we plot the averaged local Chern marker, band flatness, real-space Berry curvature fluctuations and vortexability as a function of disorder $\eta$ and $\Delta$. The following discussion proceeds similarly if we would vary $w$ instead of $\Delta$.

The effect of disorder is apparent by plotting the average Chern marker as a function of the parameters, as shown in Fig.~\ref{fig:rydberg-amorphous}(a). Upon increasing disorder we eventually destroy the topological character of both Chern number $\pm1$ regions. We observe a similar behaviour when plotting the Berry curvature fluctuations and vortexability, in Figs.~\ref{fig:rydberg-amorphous}(c) and (d), respectively. Both quantities are essentially constant with disorder as long as the Chern number remains well-quantized. 

As disorder is increased and the average Chern marker decreases to zero, the Berry curvature fluctuations and the vortexability also increase. When we enter the topologically trivial region both markers drop in value. This is reasonable because we expect a trivial (Anderson) insulator to have vanishing Berry curvature, and hence vanishing fluctuations, and a small vortexability~\cite{ledwith2022vortexability}. Finally, the flatness phase diagram, shown in Fig.~\ref{fig:rydberg-amorphous}(b) shows that the introduction of disorder very quickly reduces band flatness. 

To plot Fig.~\ref{fig:rydberg-amorphous} we chose $w>0$, where a fractional Chern insulator exists in the crystalline case~\cite{zhao2023}. Taken together, the results in Fig.~\ref{fig:rydberg-amorphous} indicate that even when disorder is present, $\eta>0$, there is a region where the criteria to stabilize a fractional Chern insulator are still favorable. From vortexability and fluctuations, one could estimate a maximum disorder strength to stabilize a fractional Chern insulator to be $\eta\sim0.2-0.3$, while from the flatness this threshold might be even lower, $\eta\sim 0.15-0.2$. However, we emphasize that flatness alone is not necessarily a useful criterion in disordered models, due to the presence of localized states induced by disorder. These can increase the spectral bandwidth without introducing a change in topological character.


\section{Conclusions}\label{sec:conclusions}

We have defined and benchmarked three real-space criteria of ideal Chern bands: band-flatness, real-space Berry curvature fluctuations and the real-space vortexability. We have calculated these markers for three different models hosting Chern bands: an amorphous Chern insulator model defined from a disordered square lattice, structurally disordered graphene in a quantizing magnetic field, and an amorphous Chern insulator model defined from a disordered honeycomb lattice. 

We have checked that our real-space definitions reproduce their momentum-space counterparts for the crystalline models. 
The only exceptions occur at parameter regimes where the gap is indirect and the system becomes semimetallic. In this case, the real-space projectors cease to be as informative regarding the stability of fractional Chern insulators, because the projectors needed to define real-space criteria inevitably mix bands.

Once we benchmarked these criteria, we have applied them to study the effect of structural disorder on realizing ideal conditions to stabilize fractional Chern insulators. Structural disorder typically drives the system away from ideal Chern bands in a way that can be quantitatively captured by our real-space criteria.

%

It is worth mentioning that ultimately, as with the momentum-space markers, our real-space criteria only provide promising guidelines as to where to find fractional Chern insulators. Thus, optimizing all markers is a necessary but not sufficient condition to find a fractional Chern insulators in real space. Ultimately it is still necessary to perform a many-body calculation to establish the nature of the system.

The definitions of the real-space markers we present here open the possibility to come up with models where disorder acts in favor of creating ideal Chern bands, rather than acting against it. Finding these models is similar in spirit to finding topological Anderson insulator models~\cite{groth_theory_2009}, where disorder helps to stabilize topology, rather than acting to destroy it. The real-space criteria we have defined can be used to assess how ideal is a Chern band in systems without lattice periodicity, and in the presence of disorder. Therefore, we believe these criteria can become particularly useful in the context of finding suitable regions of parameter space that may favor fractional Chern insulator states in moir\'{e}, amorphous and quasicrystalline materials.
\chapter{Conclusions}
Throughout the first part of the thesis we have developed, implemented and applied a new method for computing excitons in two-dimensional materials, which allowed us to identify and propose a new mechanism to generate photovoltaic currents in topological insulators. The field of optoelectronics and excitonic physics has experienced a renaissance with the advent of two-dimensional materials, and the work developed in the past chapters contributes to the advancement of the field. Below, we summarize the key results from the preceding chapters, discuss their main implications and outline future directions.

We begin providing in Chapter 2 a complete review of the mathematics involved in the description of excitons. The main contribution from this chapter is the development of the real-space formulation for the interaction matrix elements present in the Bethe-Salpeter equation. This method extends previously developed theory to a more general case, and offers an alternative to the standard reciprocal-space formulation derived through Fourier transformation. Additionally, the review connects two seemingly unrelated approaches to the calculation of excitons, namely the exact diagonalization and many-body perturbation theory approaches. Coupled with a detailed characterization of exciton properties, this chapter provides a comprehensive foundation of theoretical methods for excitonic physics.

One area that is not intrinsic to condensed matter physics but has become increasingly relevant is the development of software packages to address its different challenges systematically. The origin of this interest could be traced back to the first DFT packages, such as \texttt{Quantum ESPRESSO}, \texttt{fireball}, \texttt{SIESTA}, \texttt{CRYSTAL}, among others. Their remarkable success can be attributed to the accuracy of the technique itself and the convenience of well-developed tools instead of scripts (whose reusability is limited at best), allowing researchers to focus on the underlying physics of the problem rather than the technicalities of the implementation. As the computational resources and our knowledge continues to grow, new programs appear to tackle increasingly complex problems. In this thesis we concern ourselves with the development of a software solution for the calculation of excitons, guided as far as possible by established principles of software engineering.  

The results of these efforts condensed in Chapter 3, where we present the \texttt{Xatu} code, a program and library tailored for exciton calculations, with a particular emphasis on two-dimensional materials. Building on the theoretical development present in Chapter 2, \texttt{Xatu} implements both real-space and reciprocal-space interaction matrix elements within the exact diagonalization framework. We show that this approach, which is greatly simplified owing to the tight-binding approximation, achieves results comparable to those of existing ab-initio codes, but at a fraction of the computational cost. Consequently, the code makes it possible to explore much larger systems, such as those involving Moir\'e physics, that were previously beyond the reach of conventional computational methods.

Currently, the code includes all the essential features to compute and characterize the excitonic spectrum of two-dimensional materials, and it is ready to be used by the community. However, it remains under active development, with ongoing efforts to maintain the code, address emerging bugs, and incorporate new features. On the technical front, a key planned enhancement is \texttt{MPI} parallelization, distributing the different parts of the calculation, particularly diagonalization, which often constitutes the bottleneck both time- and memory-wise. From a physics perspective, future upgrades include enabling the computation of the dielectric function in the RPA, allowing the use of the screened Coulomb interaction instead of effective ones. Additional features include various screened potentials for different systems, leveraging basis details to compute the interaction matrix elements, and extending the code's functionality to handle excitations like trions or biexcitons. 

To conclude the first half of the thesis, in Chapter 4 we propose a novel mechanism to generate photovoltaic currents in topological insulators through the dissociation of excitons into edge electron-hole pairs. We demonstrate that, under appropriate conditions---specifically the breaking of inversion and time-reversal symmetries---it would be theoretically possible to achieve an imbalanced population of charge carriers, both spatially and in momentum space, which would form a topologically protected edge current. While a direct calculation of these populations is not provided, we compute the transition rates as a proxy, highlighting the potential of this mechanism. Moreover, this chapter also serves to showcase the usefulness of \texttt{Xatu} in addressing unconventional problems. Experimentally observing the proposed effect might prove to be challenging, partly due to the difficulty in synthesizing purely two-dimensional topological insulators with sufficiently large gaps, along with the required geometrical and edge asymmetry constraints for the effect. Extending the mechanism to three-dimensional topological insulators, which are more readily available, may offer a more experimentally viable alternative.

The second section of the thesis is dedicated to the study of disorder in two-dimensional topological insulators, encompassing both Chern insulators and time-reversal topological insulators. Here we address primarily two different problems. First, it explores the development of tools, either numerical or mathematical, to determine the topological properties of disordered systems. Second, it applies these tools to extend the existing classification of disordered topological materials, focusing on one family of materials known for displaying topological properties in crystalline form. In what follows, we highlight the main results from each chapter in this section and discuss their significance.

This half of the thesis opens up with Chapter 6, with an in-depth review of the physical theory underlying topological insulators. Initially, we developed a good understanding of the mathematical quantities used to express topological invariants, namely the Berry phase, connection and curvature. These concepts are then applied to describe the Chern insulator, followed by the $\mathbb{Z}_2$ topological insulator as an extension of the Chern insulator. A key insight from this chapter, which summarizes the general understanding of topological materials, is that their topological properties are attributed to non-trivial Berry phases flows within the Brillouin zone. Additionally, these flows can be described in a unified framework via the Wilson loop, which generalizes the Berry phase to the multiband case. Appropriately evaluating the Wilson loop in the BZ, we can extract the Berry phase flows relevant for the topological classification of the material. The material presented in this chapter is an attempt to provide a cohesive and comprehensive description of topological insulators, introducing necessary concepts step-by-step to build a solid theoretical foundation for the subsequent chapters.

Once we have established the theory behind crystalline topological materials, Chapter 7 addresses for the first time the problem of evaluating the topological properties of disordered systems. The difficulty arises because the theoretical framework introduced earlier is based on reciprocal space, hence the emphasis on crystalline materials, and consequently can not be applied to systems lacking translational invariance. This issue is particularly prominent for time-reversal topological insulators, where the absence of an observable quantity linked to the topological invariant, most likely precludes us from a direct, real-space evaluation of the invariant, unlike Chern insulators. We circumvent this problem using the entanglement spectrum of the material, which can be readily computed in non-interacting systems. By feeding this to artificial neural networks, we successfully predict the $\mathbb{Z}_2$ invariant of a structurally disordered topological insulator, even when the system becomes gapless.

With the methodology for assessing topological properties of disordered systems developed, Chapter 8 is concerned with applying this technique to extend the topological classification of materials. Specifically, we use it to predict the topological phase diagram of Bi$_x$Sb$_{1-x}$ alloys, in both crystalline and amorphous phases. With a combination of electronic transport calculations and the careful examination of the single-particle states in the alloy, we confirm the predictions of the neural network. Notably, we observed a disorder-driven transition from a trivial to topological phase, a phenomenon that was previously reported, and which we conjecture is a common feature of topological insulators induced by spin-orbit coupling. In such systems, the perturbative addition of disorder serves to favor band inversion, renormalizing the critical spin-orbit coupling.

Chapters 7 and 8 yield several insights regarding the applicability of the technique. First, compared with the Wilson loop, it significantly reduces the amount of diagonalizations required to determine the topological invariant once the neural network is trained. The training requirement, as with most machine learning algorithms, is the main limitation of the method, as the generation of a sufficient amount of data can prove to be by itself a computationally demanding task. This is particularly challenging for amorphous DFT compounds, where sample generation is not as straightforward as with a tight-binding model, due to the increased computational cost and the absence of tunable parameters that facilitate a balanced sampling of trivial and topological cases. One possible solution is to derive a Wannierized model from DFT, which could mitigate these issues. While the technique shows promise, being one of its biggest successes its applicability to gapless systems, it is not without its limitations and remains far from an ideal solution for addressing disorder in $\mathbb{Z}_2$ topological insulators, namely the elusive local $\mathbb{Z}_2$ marker.

In Chapter 9 we revisit the problem of evaluating the topological properties in disordered systems, this time focusing on fractional Chern insulators. The characterization of fractional Chern insulators typically requires applying the ideal Landau level criteria to identify regions likely to host an interacting topological ground state, before performing the actual many-body calculations.
However, similarly to non-interacting topological insulators, all these criteria are formulated in reciprocal space, making them unsuitable for systems without translational symmetry. We solve this problem introducing a set of real-space criteria based on the local Chern marker. We then mathematically relate these real- and momentum-space criteria, demonstrating through several models that both predict the same parameters regions in the crystalline limit. By evaluating the real-space criteria as a function of disorder, we identify candidate regions where amorphous fractional Chern insulators might exist. Nevertheless, these criteria are not definitive and many-body calculations are still required to confirm the presence of a topological ground state.

Finally, it is worth mentioning that the results presented in this part of the thesis were obtained using the \textsl{tightbinder} library, a Python package developed during this thesis. This library is designed to build, modify and characterize tight-binding models, with a particular emphasis on their topological properties. In addition to all the standard capabilities of a tight-binding code, it includes a complete suite of topology-related quantities, such as the Chern number, local Chern marker, the Wilson loop, the quantum geometric tensor or the entanglement spectrum. It also features a complete implementation of the non-equilibrium Green's function formalism to obtain the conductance of a sample in a two-terminal setup.

\chapter{Conclusiones}

A lo largo de la primera parte de la tesis hemos desarrollado, implementado y aplicado un nuevo método para el cálculo de excitones en materiales bidimensionales, lo cual nos ha permitido identificar y proponer un nuevo mecanismo para la generación de correntes fotovoltaicas en aislantes topológicos. El campo de la optoelectrónica y la física excitónica ha experimentado un resurgir con la llegada de los materiales bidimensionales, y el trabajo desarrollado en los capítulos anteriores contribuye al avance del campo. En lo que sigue, resumimos los resultados clave de los capítulos de la primera parte, discutimos sus implicaciones y mostramos futuras direcciones de investigación.

Comenzamos en el Capítulo 2 con una revisión completa de las matemáticas involucradas en la descripción de excitones. La principal contribución de este capítulo es el desarrollo de la formulación en espacio real de los elementos de matriz de interacción presentes en la ecuación de Bethe-Salpeter. Este método extiende teoría previamente desarrollada a un caso más general, y ofrece una alternativa a la formulación estándar basada en espacio recíproco obtenida mediante transformadas de Fourier. Además, esta revisión conecta dos aproximaciones aparentemente distintas al cálculo de excitones, esto es, la diagonalización exacta y la teoría de perturbaciones de muchos cuerpos. Junto con una caracterización detallada de las propiedades de los excitones, este capítulo proporciona unos fundamentos sólidos de métodos teóricos para física excitónica.

Un área que no es intrínseca a la física de la materia condensada, pero que se ha vuelvo cada vez más relevante es la del desarrollo de software para atacar problemas complejos. El origen de este interés probablemente se pueda trazar a los primeros códigos de DFT, tales como \texttt{Quantum ESPRESSO}, \texttt{fireball}, \texttt{SIESTA}, \texttt{CRYSTAL}, entre otros. Su éxito puede ser atribuido a la propia precisión de la técnica, pero también a la conveniencia de tener herramientas bien desarrolladas en lugar de scripts (cuya reusabilidad es limitada en el mejor de los casos), permitiendo a los investigadores centrarse en la física del problema en lugar de en los tecnicismos de la implementación. A medida que los recursos computaciones siguen aumentado, nuevos programas aparecen para abordar problemas aún más complejos. En esta tesis nos centramos en el desarrollo de una solución de software para el cálculo de excitones, guiados en la medida de lo posible por principios establecidos de la ingeniería de software.

El resultado de estos esfuerzos se condensa en el Capítulo 3, donde presentamos $\texttt{Xatu}$, un programa y librería diseñado para el cálculo de excitones, con un énfasis particular en materiales bidimensionales. Tomando como base los desarrollos teóricos del Capítulo 2, \texttt{Xatu} implementa el método de diagonalización exacta, con los elementos de matriz de interacción tanto en espacio real como en espacio recíproco. Este método, que se simplifica en gran medida gracias a la aproximación de ligaduras fuertes, logra resultados comparables con los obtenidos mediante códigos de primeros principios, pero a una fracción del coste computacional. Consecuentemente, el código hace posible explorar sistemas mucho mas grandes, tales como aquellos con física de Moiré, los cuales típicamente están fuera del alcance de los métodos computacionales estándar.

Actualmente, el código incluye todas las características esenciales para calcular y caracterizar el espectro de excitones en materiales bidimensionales, y está listo para ser usado por la comunidad. Sin embargo, sigue desarroll\'andose activamente, con los esfuerzos centrados en mantener el código, solucionar errores y añadir nuevas funcionalidades. En el plano técnico, una mejora clave planeada es añadir paralelización con MPI, para distribuir diferentes partes del cálculo, en particular la diagonalización que es con frecuencia la parte más costosa computacionalmente, tanto en tiempo como en memoria. En el plano de la física, futuras actualizaciones incluyen el cálculo de la constante dielécetrica en la RPA, permitiendo el uso de la interacción de Coulomb apantallada en lugar de interacciones efectivas. Características adicionales serían la inclusión de varios potentiales apantallados para distintos sistemas, usar los detalles de la base para el cálculo de los elementos de matriz de interacción, y extender la funcionalidad del código para incluir otros tipos de excitaciones como triones o biexcitones.

Para concluir la primera parte de la tesis, en el Capítulo 4 proponemos un nuevo mecanismo para generar corrientes fotovoltaicas en aislantes topológicos por medio de la disociación de excitones en pares electron-hueco de borde. Demostramos que, bajo las condiciones apropiadas, lo cual implica la rotura de las simetrías de inversión espacial y temporal, sería posible en teoría lograr una población desbalanceada de portadores de carga, tanto espacialmente como en momentos, lo que daría pie a una corrienta de borde protegida topológicamente. Mientras que no proporcionamos un cálculo de estas poblaciones, en su lugar calculamos la tasa de transición como sustituto, mostrando el potencial de este mecanismo. Este capítulo también sirve para poner de manifiesto la utilidad de \texttt{Xatu} al tratar problemas no convencionales. Observar experimentalmente este efecto podría ser un reto, en parte debido a la dificultad de sintetizar aislantes topológicos bidimensionales con bandas prohibidas suficientemente grandes, junto con los requisitos geométricos y de asimetría necesarios para el efecto. La extensión de este efecto a aislantes topológicos tridimensionales, cuya disponibilidad es mucho mayor, podría ofrecer una alternativa más viable experimentalmente.

La segunda sección de la tesis está dedicada al estudio del desorden en aislantes topológicos bidimensionales, abarcando tanto los aislantes de Chern como los aislantes topológicos con inversión temporal. Aquí abordamos principalmente dos problemas distintos. En primer lugar, se explora el desarrollo de herramientas, ya sean numéricas o matemáticas, para determinar las propiedades topológicas de sistemas desordenados. En segundo lugar, se aplican estas herramientas para extender la clasificación actual de materiales topológicos desordenados, centrándose en una familia de materiales conocida por exhibir propiedades topológicas en su forma cristalina. A continuación, destacamos los principales resultados de cada capítulo en esta sección y discutimos su importancia.

Esta parte de la tesis comienza con el Capítulo 6, que presenta una revisión detallada de la teoría física subyacente a los aislantes topológicos. Inicialmente, desarrollamos una comprensión sólida de las cantidades matemáticas utilizadas para expresar los invariantes topológicos, es decir, la fase de Berry, la conexión y la curvatura. Estos conceptos se aplican luego para describir el aislante de Chern, seguido del aislante topológico $\mathbb{Z}_2$ como una extensión del aislante de Chern. Una idea clave de este capítulo, que resume la comprensión general de los materiales topológicos, es que sus propiedades topológicas se atribuyen a flujos no triviales de las fases de Berry dentro de la zona de Brillouin. Además, estos flujos pueden describirse en un marco unificado a través del bucle de Wilson, que generaliza la fase de Berry al caso de múltiples bandas. Al evaluar adecuadamente el bucle de Wilson en la zona de Brillouin, podemos extraer los flujos de la fase de Berry relevantes para la clasificación topológica del material. El contenido presentado en este capítulo es un intento de ofrecer una descripción cohesiva y exhaustiva de los aislantes topológicos, introduciendo los conceptos necesarios paso a paso para construir una base teórica sólida para los capítulos posteriores.

Una vez establecida la teoría detrás de los materiales topológicos cristalinos, el Capítulo 7 aborda por primera vez el problema de evaluar las propiedades topológicas de sistemas desordenados. La dificultad surge porque el marco teórico introducido anteriormente se basa en el espacio recíproco, de ahí el énfasis en los materiales cristalinos, y, por lo tanto, no puede aplicarse a sistemas que carecen de invariancia traslacional. Este problema es particularmente relevante en los aislantes topológicos con inversión temporal, donde la ausencia de una cantidad observable vinculada al invariante topológico probablemente impide una evaluación directa en el espacio real del invariante, a diferencia de los aislantes de Chern. Para sortear este obstáculo, utilizamos el espectro de entrelazamiento del material, que puede calcularse fácilmente en sistemas no interactuantes. Al proporcionar estos datos a redes neuronales artificiales, logramos predecir con éxito el invariante $\mathbb{Z}_2$ de un aislante topológico estructuralmente desordenado, incluso cuando el sistema no tiene banda prohibida.

Con la metodología desarrollada para evaluar las propiedades topológicas de sistemas desordenados, el Capítulo 8 se centra en la aplicación de esta técnica para extender la clasificación topológica de materiales. Específicamente, la utilizamos para predecir el diagrama de fases topológico de las aleaciones de Bi$_x$Sb$_{1-x}$, tanto en sus fases cristalinas como amorfas. Mediante una combinación de cálculos de transporte electrónico y un análisis detallado de los estados de una partícula en la aleación, confirmamos las predicciones de la red neuronal. En particular, observamos una transición impulsada por el desorden de una fase trivial a una topológica, un fenómeno que ya había sido reportado anteriormente y que conjeturamos es una característica común de los aislantes topológicos inducidos por el acoplamiento espín-órbita. En estos sistemas, la adición perturbativa de desorden favorece la inversión de bandas, renomalizando el acoplamiento espín-órbita crítico.

Los Capítulos 7 y 8 proporcionan varias ideas sobre la aplicabilidad de la técnica. En primer lugar, en comparación con el bucle de Wilson, se reduce significativamente la cantidad de diagonalizaciones necesarias para determinar el invariante topológico una vez que la red neuronal ha sido entrenada. Sin embargo, el requisito del entrenamiento, como ocurre con la mayoría de los algoritmos de aprendizaje automático, es la principal limitación del método, ya que la generación de una cantidad suficiente de datos puede resultar en sí misma una tarea computacionalmente exigente. Esto es particularmente desafiante para compuestos amorfos calculados con teoría del funcional de densidad, donde la generación de muestras no es tan sencilla como en un modelo de ligaduras fuertes, debido al mayor costo computacional y a la falta de parámetros ajustables que faciliten un muestreo equilibrado de casos triviales y topológicos. Una posible solución es derivar un modelo de Wannier a partir de la teoría del funcional de densidad, lo que podría mitigar estos problemas. Aunque la técnica es prometedora, siendo uno de sus mayores logros su aplicabilidad a sistemas sin gap, no está exenta de limitaciones y dista de ser una solución ideal para abordar el desorden en aislantes topológicos $\mathbb{Z}_2$, que sería el esquivo marcador local $\mathbb{Z}_2$.

En el Capítulo 9 se retoma el problema de evaluar las propiedades topológicas en sistemas desordenados, esta vez centrándose en los aislantes de Chern fraccionarios. La caracterización de estos sistemas suele requerir la aplicación de los criterios de niveles de Landau ideales para identificar regiones que potencialmente alberguen un estado fundamental topológico interacturante, antes de realizar los cálculos de muchos cuerpos. Sin embargo, al igual que ocurre con los aislantes topológicos no interactuantes, todos estos criterios están formulados en espacio recíproco, lo que los hace inadecuados para sistemas sin simetría traslacional. Resolvemos este problema introduciendo un conjunto de criterios en espacio real basados en el marcador de Chern local. Luego, relacionamos matemáticamente los criterios en espacio real y en espacio de momentos, demostrando a través de varios modelos que ambos predicen las mismas regiones de parámetros en el límite cristalino. Al evaluar los criterios de espacio real en función del desorden, identificamos regiones candidatas donde podrían existir aislantes de Chern fraccionarios amorfos. No obstante, estos criterios no son definitivos y los cálculos de muchos cuerpos siguen siendo necesarios para confirmar la presencia de un estado fundamental topológico.

Finalmente, cabe mencionar que los resultados presentados en esta parte de la tesis se obtuvieron utilizando la librería \textsl{tightbinder}, un paquete de Python desarrollado durante esta tesis. Esta biblioteca está diseñada para construir, modificar y caracterizar modelos de ligaduras fuerte, con un énfasis particular en sus propiedades topológicas. Además de incluir todas las capacidades estándar de un código de ligaduras fuertes, cuenta con un abánico completo de cantidades relacionadas con la topología, como el número de Chern, el marcador de Chern local, el bucle de Wilson, el tensor geométrico cuántico y el espectro de entrelazamiento. También incorpora una implementación completa del formalismo de funciones de Green fuera del equilibrio para calcular la conductancia de una muestra en una configuración de dos terminales.


\clearpage
\phantomsection
\addcontentsline{toc}{chapter}{Bibliography}
\bibliographystyle{unsrt}
\bibliography{bibliography}


\end{document}